\documentclass[final,5p,times,twocolumn]{elsarticle}

\usepackage{amsmath,amssymb,latexsym,amsthm}
\usepackage{bm}
\usepackage{CJKutf8}
\usepackage{subcaption}
\usepackage{xcolor}
\newcommand{\iparallel}{\mathord{/ \mkern-4mu /}}

\usepackage[maxfloats=256]{morefloats}
\journal{Atomic Data and Nuclear Data Tables}

\begin{document}

\begin{frontmatter}
    \title{Photoabsorption Cross Sections studied within the axially deformed Relativistic Quasiparticle Finite Amplitude Framework}

    \author[inst1,inst2]{C. Chen (\begin{CJK}{UTF8}{gbsn}陈晨\end{CJK})}

    \author[inst1,inst2]{Y. F. Niu (\begin{CJK}{UTF8}{gbsn}牛一斐\end{CJK})\corref{cor1}}
    \cortext[cor1]{Corresponding author at: 
    Frontiers Science Center for Rare isotope, Lanzhou University, Lanzhou, 730000, China 
    and School of Nuclear Science and Technology, Lanzhou University, Lanzhou, 730000, China}
    \ead{niuyf@lzu.edu.cn}

    \affiliation[inst1]{
        organization={Frontiers Science Center for Rare isotope, Lanzhou University},
        city={Lanzhou},
        postcode={730000},
        country={China}}
    \affiliation[inst2]{
        organization={School of Nuclear Science and Technology, Lanzhou University},
        city={Lanzhou},
        postcode={730000},
        country={China}}
    
    \author[inst3]{R. Xu\corref{cor2}}
    \ead{xuruirui@ciae.ac.cn}

    \author[inst3]{Y. Tian\corref{cor2}}
    \cortext[cor2]{Corresponding authors at: 
    China Nuclear Data Center, China Institute of Atomic Energy, P.O. Box 275(41), Beijing, 102413, China}
    \ead{tiany@ciae.ac.cn}
    \affiliation[inst3]{
        organization={China Nuclear Data Center, China Institute of Atomic Energy},
        addressline={P.O. Box 275(41)},
        city={Beijing},
        postcode={102413},
        country={China}}

    \begin{abstract}
        Photoabsorption cross sections for 235 stable nuclei, ranging from $^{40}$Ca to $^{209}$Bi, were investigated by the quasiparticle finite amplitude method (QFAM) based on the axially deformed relativistic Hartree-Bogoliubov (RHB) approach using relativistic point-coupling interaction DD-PC1, with extensions to odd-A nuclei. GDR parameters based on the standard Lorentzian (SLO) model were extracted from QFAM results and compared with those from experimental data recommended by IAEA. Good agreement was achieved for giant dipole resonance (GDR) peak energies, while resonance widths were underestimated and hence peak cross sections were overestimated due to the lack of higher-order many-body correlations. These discrepancies were much improved in deformed nuclei.
        The effects of deformation on photoabsorption cross sections were examined systematically. The comparison of photoabsorption cross sections among QFAM results and discrepant experimental data revealed the potential of QFAM calculations in the evaluation of photonuclear data.
    \end{abstract}


\end{frontmatter}

\section{Introduction}
\label{sec:intro}

The study of photonuclear reactions is an important topic in both nuclear physics and astrophysics \cite{Zilges_2022_PPNP}. This study offers valuable insights into isovector giant dipole resonances (GDR) \cite{Bracco_2019_PPNP}, an important mode of nuclear collective vibration, and provides key inputs for astrophysical nucleosynthesis simulations \cite{Kajino_2019_PPNP,YWHao_2023_PLB}.
Photonuclear reaction cross sections also serve as essential input data for nuclear science applications \cite{Kawano_2020_NDS}, including radiation shielding \cite{Chadwick_1998_REB}, nuclear waste transmutation \cite{JGChen_2008_CPC}, production of medical radioisotopes \cite{WLuo_2016_APB,YXYang_2024_RPC}, as well as fission and fusion reactor technologies.

Extensive experimental efforts have been devoted to measuring photonuclear cross sections
using various $\gamma$-ray sources, such as bremsstrahlung beams \cite{Bogdankevich_1966_bremsstrahlung}, positron annihilation in flight \cite{Berman_1975_RMP}, and laser Compton scattering \cite{Ohgaki_2000_laser-Compton,HWWang_2022_NST}.
The collected data have been systematically compiled in the Experimental Nuclear Reaction Data (EXFOR) Library \cite{Otuka_2014_NDS}.
Among these measurements, the majority of photoneutron cross sections have been obtained using annihilation photon beams at the Lawrence Livermore National Laboratory (USA) and the CEA Saclay Nuclear Research Centre (France) \cite{Varlamov_2017_PAN}.
However, significant discrepancies have been observed among experimental cross-section data \cite{Varlamov_2019_PPN}.
To assess data reliability and meet the increasing demand for photonuclear data, the International Atomic Energy Agency (IAEA) initiated a Coordinated Research Project (CRP), leading to the development of a reference database for photon strength functions (PSF) \cite{Goriely_2019_EPJA}.

Theoretical models play an important role in studying photonuclear cross sections.
Both phenomenological and microscopic approaches have been employed to investigate photoabsorption cross sections, which are the sum of various competing photonuclear reaction channels.
Phenomenological methods, such as the Standard Lorentzian (SLO) \cite{Brink_1955_PhD,Axel_1962_PR} and Simple Modified Lorentzian (SMLO) models \cite{Plujko_2008_IJMPE}, describe photoabsorption cross sections using Lorentzian functions with phenomenological parameters fitted to experimental data \cite{YTian_2019_CPC}.  Instead of the phenomenological formulas for GDR parameters including GDR peak cross sections, energies and widths in SLO or SMLO, the GDR parameters can also be fitted to experimental data directly. 
A newly updated set of GDR parameter values, along with their corresponding uncertainties, has been extracted through least-squares fitting to experimental data and reported in Ref.~\cite{Plujko_2018_ADNDT} as part of the IAEA CRP \cite{Goriely_2019_EPJA}.
Recently, machine learning techniques have also been employed to extract GDR parameters from experimental data \cite{JHBai_2021_PLB} and to study photonuclear cross sections \cite{JSu_2021_PRC,QKSun_2025_NST}, considering the success in nuclear reaction studies \cite{KXing_2024_PLB,ZHHu_2024_PLB,WFLi_2024_PRC}.
While these models provide suitable descriptions for nuclei with experimental data, their prediction ability for nuclei without data, especially those far from the $\beta$-stability line, remains uncertain.

To provide more reliable extrapolations, microscopic methods have been developed \cite{Colo_2020_Handbook}.
Ref.~\cite{Goriely_2002_NPA} made the first attempt to predict the electric dipole (E1) strengths for more than 6000 nuclei with $8\leq Z \leq 110$ that lie between the proton and neutron drip lines with Quasiparticle Random Phase Approximation (QRPA) method based on Skyrme Hartree-Fock + Bardeen-Cooper-Schrieffer (BCS) approach.
The subsequent study~\cite{Goriely_2004_NPA} introduced a QRPA model based on Skyrme Hartree-Fock-Bogoliubov (HFB) approach, which improved descriptions of exotic nuclei, for all nuclei from $Z=8$ to $110$ stable against particle emission. In the case of odd-A and odd-odd nuclei, Ref.~\cite{Goriely_2004_NPA} considered two-quasiparticle QRPA excitations built on top of the HFB ground state using the filling approximation.
Compared to non-relativistic density functionals, such as Skyrme types, relativistic density functionals naturally include the spin-orbit interaction, and satisfy the Lorentz invariance thereby reducing the number of adjustable parameters \cite{JMeng_2016_book}. Within the continuum RPA approach based on relativistic mean-field (RMF) model, Ref.~\cite{Daoutidis_2012_PRC} performed E1 strength calculations for several even-even nuclei throughout the nuclear chart.

Spherical approximations are adopted in all the above efforts. However, many nuclei in the nuclear chart have deformations \cite{KYZhang_2022_ADNDT,PGuo_2024_ADNDT}, and deformation gives rise to a double-peak structure in the GDR strength distribution \cite{Danos_1958_NP}. To mimic the deformation effect, empirical corrections were introduced on top of the electric dipole transition strengths which were obtained from spherical Skyrme HFB + QRPA calculations~\cite{Xu_2021_PRC} for about 10000 nuclei with $8\leq Z \leq 124$ lying between the proton and the neutron drip lines. In the case of odd-A and odd-odd nuclei, the procedure developed in Refs.~\cite{Samyn_2002_NPA,Khan_2002_PRC} was adopted in the aforementioned work. However, the empirical corrections lose the advantage of good extrapolation ability inherent in microscopic models. Therefore, the inclusion of deformation degree of freedom in the microscopic models is necessary for an accurate description of nuclear photoabsorption cross sections. The axially symmetric deformed HFB + QRPA calculations based on the finite-range Gogny force were performed for nuclei with $8\leq Z \leq 94$ that lie between the proton and neutron drip lines~\cite{Martini_2016_PRC}, using an interpolation procedure to consider odd-A and odd-odd nuclei~\cite{IAEA_TECDOC_1178}.

However, including deformation complicates solving the QRPA equation since the QRPA matrix dimension increases dramatically due to the splitting of single-particle levels in deformed nuclei.
The Finite Amplitude Method (FAM)~\cite{Nakatsukasa_2007_PRC} and its extension to open shell nuclei, quasiparticle FAM (QFAM)~\cite{Avogadro_2011_PRC,Niksic_2013_PRC,XWSun_2017_PRC,Bjelcic_2020_CPC,XWSun_2022_PRC,MTMustonen_2014_PRC},
provide an efficient way for solving the QRPA model. Instead of the diagonalization of the QRPA matrix, QFAM solves the linear response formalism of the QRPA equation iteratively, thus avoiding the calculation of two-body matrix elements and matrix diagonalization. Based on relativistic energy density functionals, a QFAM model capable of studying E1 strengths in even-even nuclei was developed ~\cite{Bjelvcic_2023_CPC}. However, a systematic study accounting for deformation effects on the photoabsorption cross sections based on relativistic density functionals is still lacking. For this purpose, calculations of odd-A nuclei are indispensable, which is still missing in the present relativistic QFAM model.

In this paper, the quasiparticle finite amplitude method (QFAM) based on the axially deformed relativistic Hartree-Bogoliubov (RHB) approach with extensions to odd-A nuclei is introduced and applied to photoabsorption cross section calculations using the relativistic point-coupling interaction DD-PC1 for nuclei ranging from $^{40}$Ca to $^{209}$Bi. Sec.~\ref{sec:meth} outlines the theoretical framework, while the numerical details are presented in Sec.~\ref{sec:num}. Results and discussion are presented in Sec.~\ref{sec:res}, followed by a summary in Sec.~\ref{sec:concl}.

\section{Methods}
\label{sec:meth}
\subsection{Photoabsorption cross sections within RHB+QFAM framework}
In this work, an axially deformed RHB model \cite{Niksic_2014_dirhb} is used to determine ground-state properties as a first step.
The RHB model provides a unified description of nuclear particle-hole (ph) and particle-particle (pp) correlations on a mean-field level by the introduction of a unitary Bogoliubov transformation to quasiparticles. The RHB Hamiltonian $H$ is obtained from the variation of an energy density functional $E[R]$ with respect to the generalized density matrix $R$, which includes the density matrix and the pairing tensor. As a result, two kinds of average potentials are incorporated into the RHB equation, namely, the self-consistent nuclear mean field that encloses all the long-range ph correlations, and a pairing field that sums up the pp correlations.

The time-dependent RHB equation,
\begin{equation}\label{eq:tdfhb}
    \mathrm{i}\hbar \dot{R}(t) = [H\left[R(t)\right]+F(t),R(t)],
\end{equation}
gives the response of a nucleus under an external field. For a weak harmonic external field,
\begin{equation}
    \hat{F}(t)
    = \eta \left\{ \hat{F}(\omega) \mathrm{exp}(-\mathrm{i} \omega t)
    + \hat{F}^{\dagger} (\omega) \mathrm{exp}(\mathrm{i} \omega t) \right\},
\end{equation}
characterized by a small real parameter $\eta$ and an oscillation frequency $\omega$,
the change in the generalized density follows the form,
\begin{equation}
    \delta R(t)= \eta\left\{\delta R(\omega) \mathrm{exp}(-\mathrm{i} \omega t)+\delta R^{\dagger} (\omega) \mathrm{exp}(\mathrm{i} \omega t)\right\},
\end{equation}
and correspondingly, the RHB Hamiltonian evolves as follows,
\begin{equation}
    H(t)=H_{0}+\eta\left\{\delta H(\omega) \mathrm{exp}(-\mathrm{i} \omega t)+\delta H^{\dagger} (\omega) \mathrm{exp}(\mathrm{i} \omega t)\right\}.
\end{equation}

By linearizing the equation of motion \eqref{eq:tdfhb} with respect to $\eta$, and expressing the operator on the quasiparticle basis, the linear response equation in the frequency domain is obtained,
\begin{equation}
    \begin{aligned}
        \left(E_{\mu}+E_{\nu}-\omega\right) X_{\mu \nu}(\omega)+\delta H^{20}_{\mu \nu}(\omega)= & -F^{20}_{\mu \nu}(\omega), \\
        \left(E_{\mu}+E_{\nu}+\omega\right) Y_{\mu \nu}(\omega)+\delta H^{02}_{\mu \nu}(\omega)= & -F^{02}_{\mu \nu}(\omega).
    \end{aligned}
\end{equation}
Here, the single-particle operators are expanded on the quasiparticle basis, so $F^{20}$ ($\delta H^{20}$) and $F^{02}$ ($\delta H^{02}$) indicate matrix elements on the quasiparticle basis in front of the creation and annihilation of two quasiparticles for the operator $F$ ($\delta H$), respectively. The deviation of density matrix from ground state $\delta R$ gives the definition of $X$ and $Y$,
\begin{equation}
    \delta R(\omega)=\left(\begin{matrix} 0 & \delta R^{20}(\omega) \\ - \delta R^{02}(\omega) & 0 \end{matrix}\right)\equiv\left(\begin{matrix} 0 & X(\omega) \\ - Y(\omega) & 0 \end{matrix}\right).
\end{equation}
In practice, these equations are solved iteratively and involve only the first derivatives of $E[R]$ with respect to $R$.  This approach avoids matrix diagonalization and tedious calculations of two-body matrix elements in the matrix QRPA equation, offering a more efficient approach for large-scale studies. The details of QFAM formalism can be found in Refs. \cite{Niksic_2013_PRC,Bjelcic_2020_CPC}. 

In this work, the same nuclear Hamiltonian is employed in both RHB and QFAM calculations, thereby showing the self-consistency of the present framework.
The ph interaction of the Hamiltonian is described by the relativistic density-dependent point-coupling density functional DD-PC1 \cite{Niksic_2008_DDPC1}. The pp interaction, on the other hand, is described by a separable form taken from Ref.~\cite{YTian_2009_PLB,YTian_2009_Axial,YTian_2009_PRC}.

Since photoabsorption cross sections are primarily governed by the E1 transition, we focus on the electric dipole response of nuclei.
The E1 strength function can be obtained by,
\begin{equation}
    S(\omega) = -\frac{1}{\pi}\operatorname{Im}\left[
        \frac{1}{2}\sum_{\mu\nu} F^{20\ast}_{\mu \nu}(\omega) X_{\mu \nu}(\omega)
        + F^{02\ast}_{\mu \nu} Y_{\mu \nu}(\omega)(\omega)
        \right].
\end{equation}
By setting the frequency to be complex, $ \omega\rightarrow\omega+\mathrm{i}\gamma$, we introduce a smearing with a width of $2\gamma$.
As is widely recognized, the GDR is characterized by a large width, which arises from damping properties in the collective motion. These properties are not included in QRPA or QFAM.
Improvement can be achieved by including higher-order terms of many-body correlations, such as 2 particle-2 hole configurations \cite{Gambacurta_2011_PRL,CLBai_2010_PRL} or particle vibration coupling effects \cite{YFNiu_2015_PRL,YFNiu_2018_PLB,ZZLi_2023_PRL,ZZLi_2024_PRC,Litvinova_2019_PRC,YNZhang_2021_PRC}. However, such improvements require an extremely high computational cost.
In practical systematical studies, damping properties are usually introduced phenomenologically \cite{Martini_2016_PRC}. In QFAM, the free parameter $\gamma$ determines the width. In this study, we will not describe the damping properties in a phenomenological way by fitting $\gamma$ to the width of giant resonances. Instead, in our work this parameter is set to a fixed small value of $0.5$ MeV, which is equivalent to a Lorentz smearing with a full width at half maximum of 1 MeV.

From the E1 strength function, the photoabsorption cross section is expressed as,
\begin{equation}
    \sigma_{(\gamma,\mathrm{abs.})}(\omega)=\frac{16\pi^3}{9\hbar c} \omega S(\omega).
\end{equation}
Under the axial symmetry assumption, the E1 strength function $S(\omega)$ consists of two components,
\begin{equation}
    S(\omega)=S(K=0,\omega)+2\times S(K=1,\omega),
\end{equation}
where the projection of total angular momentum of the external field onto the symmetry axis of nucleus, $K=0$ ($K=1$), corresponds to oscillations parallel (perpendicular) to the symmetry axis.
Thus, the photoabsorption cross sections are also divided into two parts,
\begin{equation}
    \sigma_{(\gamma,\text{abs.})} (\omega)
    = \sigma_{(\gamma,\text{abs.})} (K=0,\omega)
    + 2\times\sigma_{(\gamma,\text{abs.})} (K=1,\omega).
\end{equation}
In this work, photoabsorption cross sections are calculated in the range of incident photon energy $0$ to $50$ MeV with an energy step $\Delta \omega = 0.1$ MeV.

\subsection{GDR parameters}
GDR parameters provide a convenient way to characterize the GDR properties. To quantify the GDR properties, two sets of GDR parameters, based on the SLO model, are given by fitting our QFAM results for the $K=0$ and $K=1$ components, respectively,
\begin{equation}
    \begin{aligned}
        \sigma_{\text{SLO},\iparallel}(\omega)
        = & \sigma_{\mathrm{r},\iparallel}
        \frac{\omega^2\Gamma_{\mathrm{r},\iparallel}^2}
        {(E_{\mathrm{r},\iparallel}^2-\omega^2)^2+\omega^2\Gamma_{\mathrm{r},\iparallel}^2}, \\
        \sigma_{\text{SLO},\perp}(\omega)
        = & \sigma_{\mathrm{r},\perp}
        \frac{\omega^2\Gamma_{\mathrm{r},\perp}^2}
        {(E_{\mathrm{r},\perp}^2-\omega^2)^2+\omega^2\Gamma_{\mathrm{r},\perp}^2}.
    \end{aligned}
\end{equation}
Here, $\sigma_\mathrm{r}$ is the peak cross section,
$E_\mathrm{r}$ is the resonance energy,
and $\Gamma_\mathrm{r}$ is the resonance width.
$\sigma_{\mathrm{r},\iparallel}$, $E_{\mathrm{r},\iparallel}$, and
$\Gamma_{\mathrm{r},\iparallel}$ are obtained
by fitting $\sigma_{(\gamma,\text{abs.})} (K=0,\omega)$
while $\sigma_{\mathrm{r},\perp}$, $E_{\mathrm{r},\perp}$, and
$\Gamma_{\mathrm{r},\perp}$ are obtained
by fitting $2\times\sigma_{(\gamma,\text{abs.})} (K=1,\omega)$.

However, in cases where the energy splitting between the two components is sufficiently small, i.e.,
$|E_{\mathrm{r},\iparallel}-E_{\mathrm{r},\perp}| \leq1\text{ MeV}$, we consider
the deformation effect causing the double-peak structure is negligible, hence resulting in only a single peak in GDR.
In such cases, the GDR parameters are obtained by a single SLO fit to the total photoabsorption cross sections,
\begin{equation}
    \sigma_\text{SLO}(\omega)=
    \sigma_\mathrm{r}\frac{\omega^2\Gamma_\mathrm{r}^2}{(E_\mathrm{r}^2-\omega^2)^2+\omega^2\Gamma_\mathrm{r}^2}.
\end{equation}

\subsection{Extensions to odd-A nuclei}

Usually, RHB and QFAM models are designed for the calculation of even-even nuclei with total angular momentum $J=0$. However, nuclei with an odd number of particles have nonzero angular momentum, making their treatment significantly more complex compared to the $J^\pi = 0^+$ ground-state configurations of even-even nuclei.

In RHB theory, the ground state of an odd nucleus is a one-quasiparticle excitation with respect to the quasiparticle vacuum. The unpaired valence nucleon occupies a level $k$, which is consequently blocked. For simplicity, the equal filling approximation (EFA) is often introduced. This approximation assumes that the state $k$ and its time-reversal partner state $\bar{k}$ are equally blocked, thus preserving time-reversal symmetry.
As shown in Ref.~\cite{Schunck_2010_PRC}, the EFA is equivalent to exact blocking procedure when the time-odd fields are set to zero. Furthermore, the contributions from time-odd fields to the energy of the ground state and low-lying excited states are rather small, averaging around 50 keV with variations of approximately 100$\sim$150 keV, indicating that the EFA is an excellent approximation~\cite{Schunck_2010_PRC}.

In the QFAM approach, the extension to odd-A nuclei was introduced in Ref.~\cite{Shafer_2016_pnFAModd}.
Nevertheless, this approach doubles the dimensionality of the QFAM equations, resulting in a big increase in computational costs.
Since the contribution of a single valence nucleon to the giant resonance is not significant due to the strong collectivity of giant resonances, in the QRPA method the presence of an odd valence nucleon can be introduced via occupation probabilities $v^2$ in canonical basis, providing a simplified treatment of odd-A nucleus.
Following the same spirit, in the present QFAM approach, we incorporate the canonical occupation probability $v^2$ of the valence nucleon in the special Bogoliubov transformation matrix decomposed from the general Bogoliubov transformation matrix via Bloch-Messiah's theorem \cite{Ring_2004_NMBP}.

\section{Numerical Details}
\label{sec:num}
\subsection{Test of odd-A nuclei}

\begin{figure}
    \center
    \includegraphics[scale=0.8]{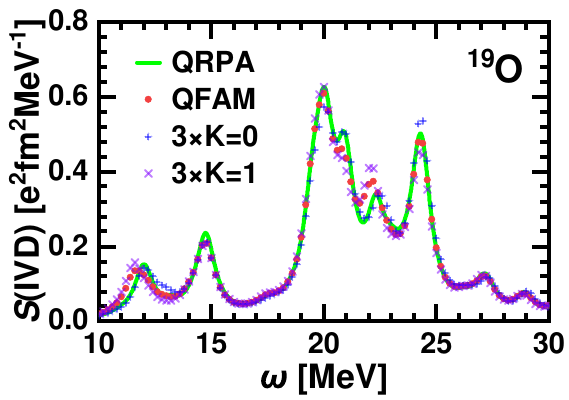}
    \caption{The isovector dipole strength functions $S(\mathrm{IVD})$ with respect to the excitation energy $\omega$ for $^{19}$O. The green line represents the QRPA result, while the red dots correspond to the QFAM results. The blue crosses and purple Xs indicate the contributions from $K=0$ and $K=1$ components in the QFAM calculation, respectively, with their magnitudes scaled by a factor of 3.}
    \label{fig1}
\end{figure}

To verify the reliability of the extensions for odd-A nuclei, numerical tests for both ground and excited states of odd-A nuclei are presented.
For ground states, taking the total energy of spherical nucleus $^{23}$N as an example,
the difference between our result ($-146.025590$ MeV) and that obtained from a spherical RHB code in Ref.~\cite{QZhao_2022_PCF-PK1} ($-146.025662$ MeV) is less than 1 keV.
For excited states, the isovector dipole strengths of $^{19}$O, calculated by our extended QFAM approach for odd-A nuclei, are compared with those obtained from a spherical QRPA code, as shown in Fig.~\ref{fig1}.
It shows that the QFAM results (red dots) are in reasonable agreement with the QRPA result (green line).
The slight differences stem from different spatial symmetry assumptions in the two codes. The QRPA approach employed in this calculation is restricted within spherical approximation, while the QFAM allows axial deformation degree of freedom.
Although $^{18}$O is a spherical nucleus, the inclusion of an extra valence neutron causes slight deformation since this neutron sits in a deformed single-particle orbital.
It also gets confirmed from the non-degeneracy of $K=0$ (blue crosses) and $K=1$ (purple Xs) components of QFAM results, since for spherical case, the $K=0$ and $K=1$ components should be exactly degenerate.

\subsection{Convergence check with respect to the basis size}
The RHB and QFAM equations are numerically solved by expanding on the axially deformed harmonic oscillator (HO) basis and the simplex-y HO basis, respectively. Correspondingly,
the basis sizes are determined by ensuring the convergence of the ground state properties, mainly the total binding energy curve as a function of quadrupole deformation, in RHB calculation, as well as the strength functions for excited states in QFAM calculation, respectively, in five randomly selected nuclei from light to heavy ones.

\begin{figure}
    \center
    \includegraphics[scale=0.8]{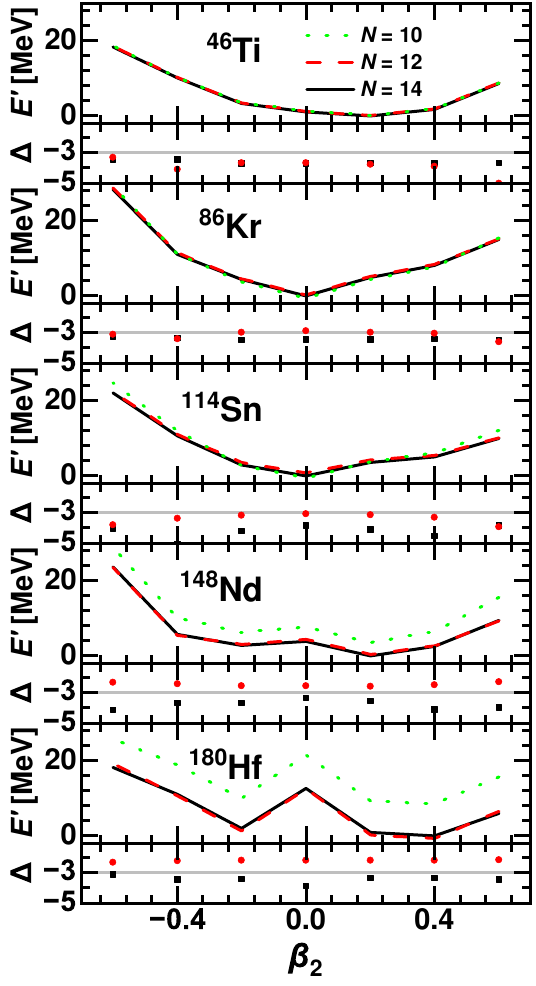}
    \caption{The normalized potential energy curves (lines) obtained with different truncations for the principal HO quantum numbers $N=10$, 12, and 14, as well as the relative changes (scatters) in total energy $\Delta=\operatorname{log}_{10}(|E_{N}/E_{N-2}-1|)$. The normalization of the PECs is performed by shifting the energy with respect to the global minimum of the PEC obtained with $N=14$, i.e., $E^\prime=E-E_{\mathrm{min.},N=14}$. The gray line indicates a relative change of $0.1\%$.}
    \label{fig2}
\end{figure}

To determine the ground states, one should firstly identify the energy minima on the potential energy curves (PECs). Therefore, we need to check the convergence of the PECs.
Fig.~\ref{fig2} illustrates normalized PECs obtained with different truncations of HO principal quantum number $N$ and the relative change in total energy $\Delta=\operatorname{log}_{10}(|E_{N}/E_{N-2}-1|)$.
The PECs are normalized by shifting the energy with respect to the global minimum of the PEC obtained with $N=14$, i.e., $E^\prime=E-E_{\mathrm{min.},N=14}$, where $E$ is the total binding energy and $E'$ is the normalized energy.
One can find that, for $^{46}$Ti and $^{86}$Kr, the PECs for $N=10,12$, and $14$ are nearly identical. With mass number increasing, for $^{114}$Sn, the PEC obtained with $N=10$ is slightly different from those with $N=12,14$. Furthermore, for even heavier nuclei $^{148}$Nd and $^{180}$Hf, the PEC obtained with $N=10$ shows a notable difference compared to the PECs for $N=12$ and $14$.
These observations suggest that larger values of $N$ are necessary to achieve convergence for nuclei with larger mass number.
This is further confirmed by the relative change $\Delta$. One can find, for $^{46}$Ti, both $\Delta_{N=12}$ and $\Delta_{N=14}$ are less than $0.1\%$, while for $^{86}$Kr, $^{114}$Sn, $^{148}$Nd, and $^{180}$Hf, $N=14$ is required to ensure $\Delta_N<0.1\%$, indicating that the convergence is achieved at a satisfactory level with $N=14$.
Thus, for PEC calculations, the truncation of HO principal quantum number is set as $N=14$.

\begin{figure}
    \center
    \includegraphics[scale=0.8]{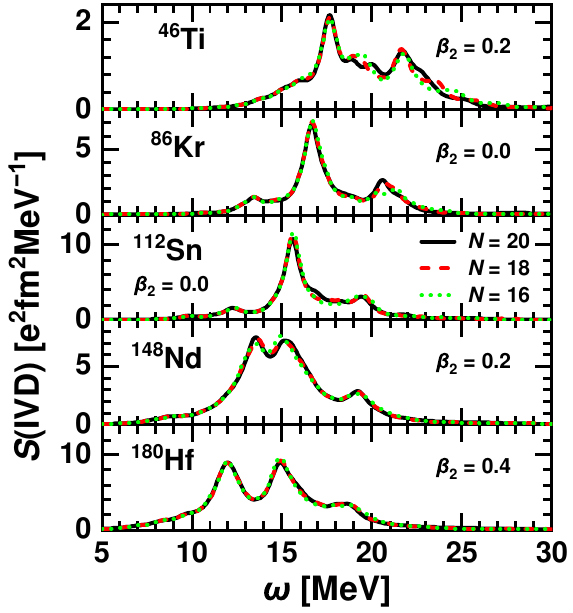}
    \caption{The isovector dipole strength functions $S(\mathrm{IVD})$
        with respect to the excitation energy $\omega$ obtained with different truncations for the principal HO quantum numbers $N=16$, 18, and 20.}
    \label{fig3}
\end{figure}

To ensure the convergence of the strength function, a convergence check with respect to the truncation of the  HO principal quantum number $N$ is shown in Fig.~\ref{fig3}.
The isovector dipole strength functions $S(\mathrm{IVD})$ as a function of excitation energy $\omega$ are shown for different values of $N$.
In the convergence check, the deformation parameters of five selected nuclei are constrained to fixed values to avoid the influences of deformation parameters on strength functions under the change of truncations.
In general, the strength functions for $N = 18$ and $N = 20$ exhibit close agreement, particularly in medium-heavy and heavy nuclei such as $^{86}$Kr, $^{112}$Sn, $^{148}$Nd, and $^{180}$Hf. However, for the light nucleus $^{46}$Ti, the strength functions at $\omega > 18$ MeV still show  slight changes from $N=18$ to $N=20$, likely due to the contributions of continuum states in these excitations.
Nevertheless, for studying GDR properties, it is enough to adopt $N = 20$ as the truncation for strength function calculations.

Based on these numerical considerations, PECs are calculated to identify their global minima, which serve as the initial deformations for an unconstrained RHB calculation. From the unconstrained RHB calculation, the true minimum of the total energy is obtained, which provides the deformation parameter of the ground state. On top of the true ground state of the nucleus under this deformation parameter, the strength functions are subsequently obtained from QFAM calculations. A detailed discussion of our results is presented in the following section.

\section{Results and Discussion}
\label{sec:res}

\begin{figure*}
    \center
    \includegraphics[scale=0.6]{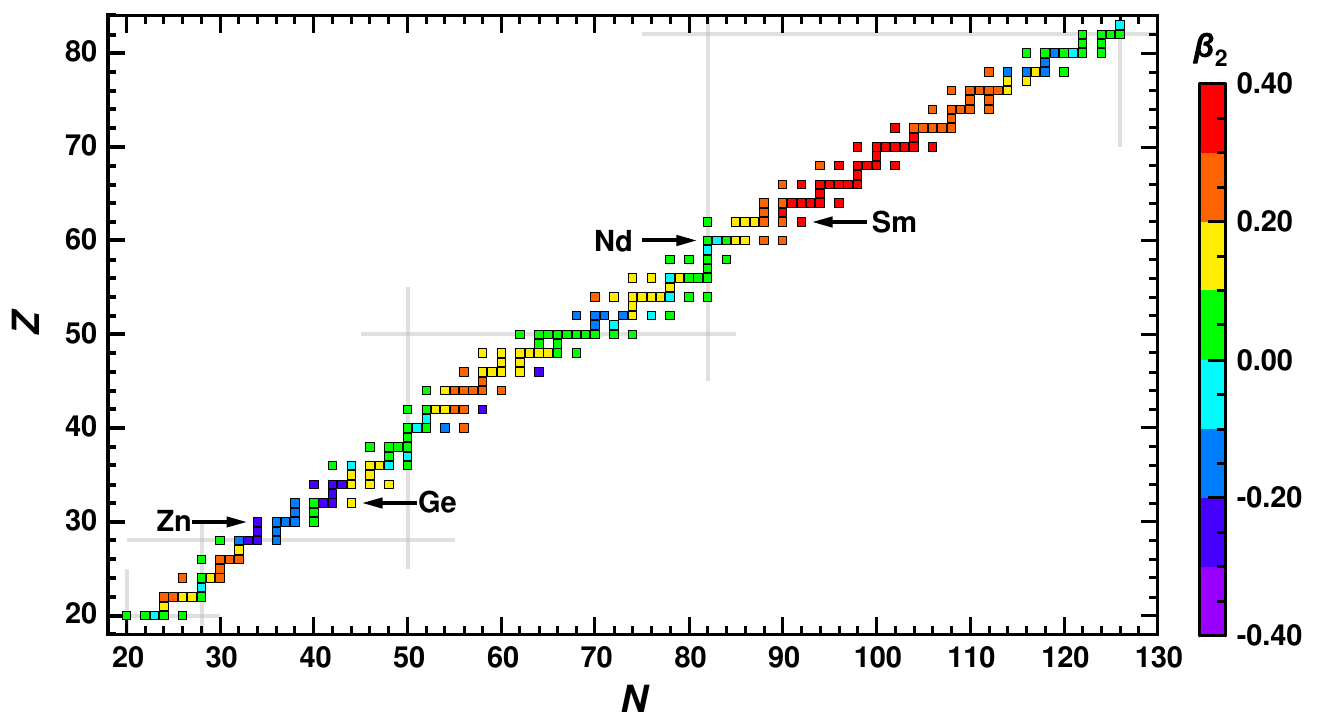}
    \caption{The ground-state quadrupole deformation parameter $\beta_2$ for 235 stable nuclei, spanning from $^{40}$Ca to $^{209}$Bi, including 146 even-even nuclei and 89 odd-A nuclei. The color of the scatters represents the value of $\beta_2$, while the gray line marks the magic numbers. }
    \label{fig4}
\end{figure*}

\begin{figure}
    \center
    \includegraphics[scale=0.7]{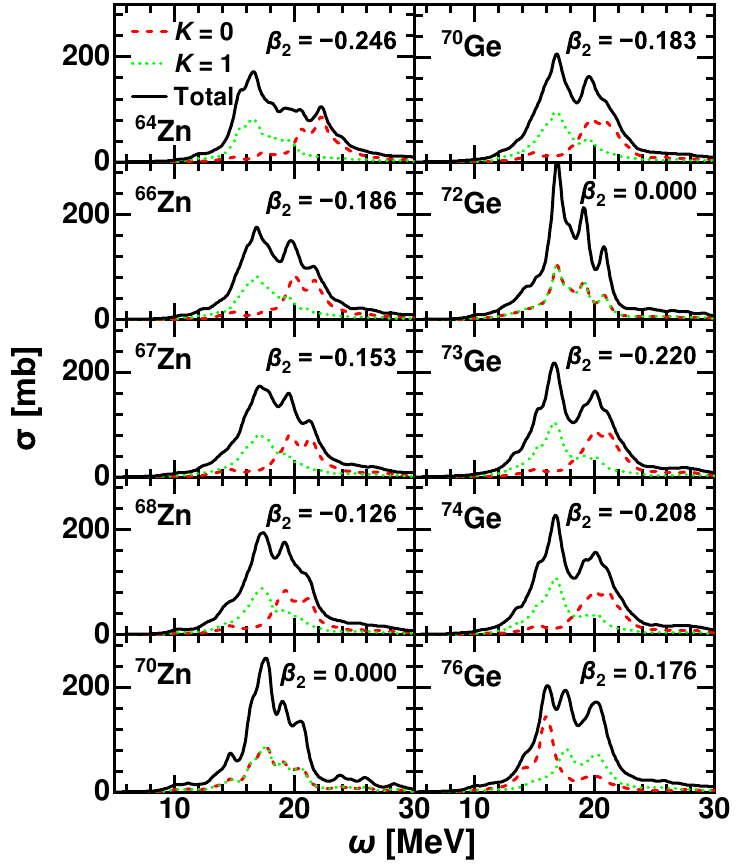}
    \caption{The photoabsorption cross sections $\sigma$ in Zn and Ge isotopes as a function of incident photon energy $\omega$ given by QFAM calculations. The green dotted lines and red dashed lines represent the contributions from $K=0$ and $K=1$ components, respectively, while the black solid lines represent the total cross sections. The $K=0$ component corresponds to the oscillations parallel to the symmetry axis of the nucleus, while the $K=1$ component corresponds to the oscillations perpendicular to the symmetry axis. The calculated ground-state quadrupole deformation parameter $\beta_2$ are also listed.}
    \label{fig5}
\end{figure}

\begin{figure}
    \center
    \includegraphics[scale=0.7]{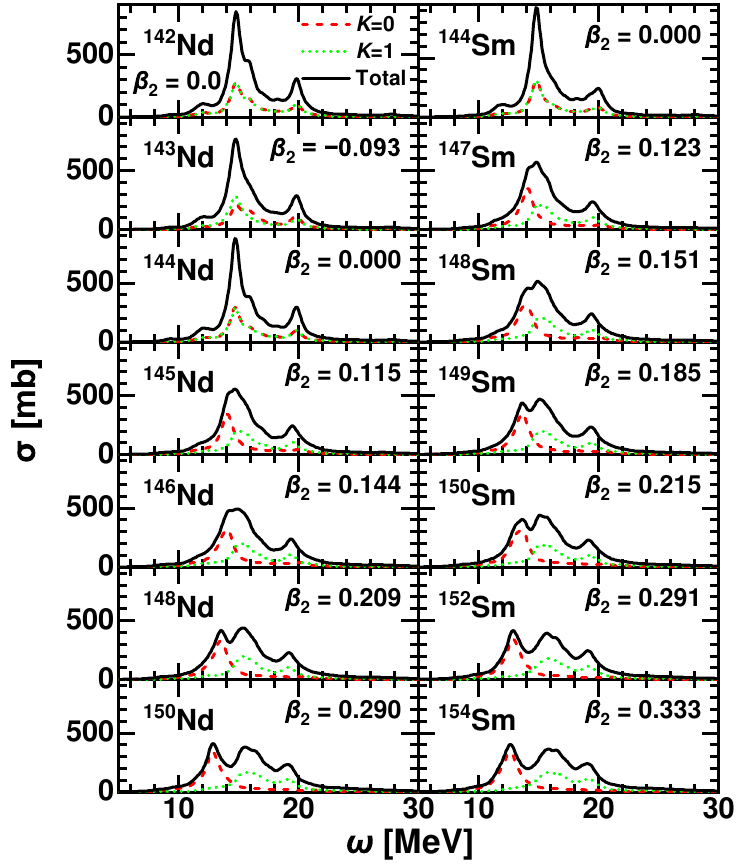}
    \caption{The same as Fig.~\ref{fig5}, but for Nd and Sm isotopes. }
    \label{fig6}
\end{figure}

In this paper, 235 stable nuclei ranging from $^{40}$Ca to $^{209}$Bi, including 146 even-even nuclei and 89 odd-A nuclei, are investigated.
The ground-state quadrupole deformation parameters $\beta_2$ for all considered nuclei are shown in Fig.~\ref{fig4}.
It can be observed that $\beta_2$ ranges from $-0.3$ to 0.4.
Nuclei near magic numbers generally exhibit nearly spherical shapes, except for the $Z=28$ isotopes, some of which display notable deformations. In contrast, open-shell nuclei are more likely to be deformed, with prolate shapes being the most prevalent.
Four typical isotopic chains,  Zn, Ge, Nd, and Sm, are pointed out in the figure, where oblate deformation dominates in the Zn and Ge isotopes, and prolate deformation is predominant in the Nd and Sm isotopes. Due to these typical features, they are used to illustrate the impact of deformation on photoabsorption cross sections in Figs.~\ref{fig5} and \ref{fig6}.

The photoabsorption cross sections $\sigma$ for Zn and Ge isotopes as a function of incident photon energy $\omega$, obtained from QFAM calculations, are presented in Fig.~\ref{fig5}. The green dotted line and red dashed line represent the contributions from the $K=0$ and $K=1$ components, respectively, while the black solid lines denote the total cross section. Furthermore, the calculated ground-state quadrupole deformation parameters $\beta_2$ are also listed. The $K=0$ component corresponds to the oscillation parallel to the symmetry axis of the nucleus, whereas the $K=1$ component corresponds to the oscillation perpendicular to the symmetry axis.

For Zn isotopes, the oblate deformation gradually decreases as the neutron number increases, eventually becoming spherical in $^{70}$Zn. For oblate-deformed $^{64,66,67,68}$Zn, the photoabsorption cross sections exhibit a double-peak structure due to the splitting of the $K=0$ and $K=1$ components, where the $K=1$ component dominates the lower-energy peak and the $K=0$ component dominates the higher-energy peak. The energy splitting between the two components decreases as the deformation diminishes. Once the deformation vanishes in $^{70}$Zn, the photoabsorption cross section becomes a single-peak structure without splitting.

Unlike Zn isotopes, the shape evolution of Ge isotopes is more complicated. $^{70}$Ge exhibits an oblate shape.
By adding two neutrons occupying $2p1/2$, $^{72}$Ge favor a spherical shape, while the oblate configuration becomes a local minimum, only hundreds of keV higher than the spherical configuration.
With one more neutron, the shape of $^{73}$Ge is slightly polarized, and the oblate configuration becomes the global minimum again. For $^{74}$Ge, the oblate configuration remains to be the global minimum, but with a softer PEC.
However, by adding two more neutrons, the prolate configuration becomes the global minimum  in $^{76}$Ge. Rich phenomena of shape coexistence and shape transition are observed in this isotopic chain. Accordingly, the strength functions of Ge isotopes evolve from a double-peak structure to a single-peak structure, and to a double-peak structure again with neutron number increasing.
It should be noted that when the shape transitions from oblate to prolate one, the positions of $K=0$ and $K=1$ components are exchanged due to the change of the symmetry axis between minor axis and major axis.

\begin{figure}
    \center
    \includegraphics[scale=0.8]{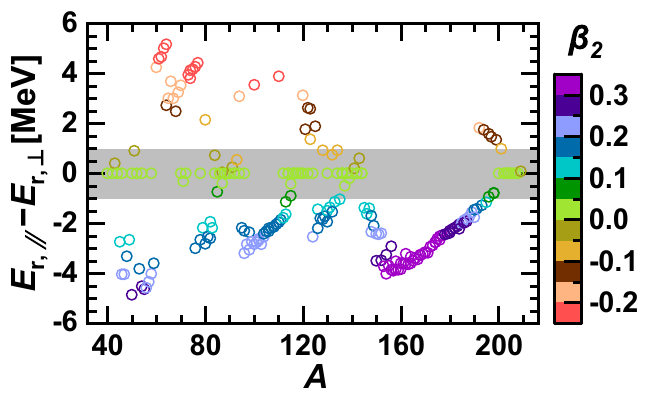}
    \caption{The energy splittings between two oscillation modes $E_{\mathrm{r},\iparallel} - E_{\mathrm{r},\perp}$ extracted from the QFAM calculations, where $E_{\mathrm{r},\iparallel}$ correspond to the oscillations parallel to the symmetry axis induced by an external field with $K=0$ and $E_{\mathrm{r},\perp}$ correspond to the oscillations perpendicular to the symmetry axis induced by an external field with $K=1$. The scatter colors indicate the deformation parameter $\beta_2$. The gray-shaded region highlights nuclei for which the energy splitting satisfies $|E_{\mathrm{r},\iparallel} - E_{\mathrm{r},\perp}| \leq 1$ MeV.}
    \label{fig7}
\end{figure}

Similarly, the photoabsorption cross sections $\sigma$ for Nd and Sm isotopes as a function of incident photon energy $\omega$, obtained from QFAM calculations, are presented in Fig.~\ref{fig6}.
The shape evolution of Nd isotopes follows a clear trend with increasing neutron number. $^{142}$Nd has a neutron number of $N=82$, resulting in a spherical configuration for this nucleus. By adding one valence neutron occupying the $K^\pi=\frac{7}{2}^-$ orbital, the polarization effect gives rise to a slight oblate deformation in  $^{143}$Nd. This polarization effect is reduced in $^{144}$Nd due to the two valence neutrons which form a pair, so the shape of this nucleus becomes spherical again.
By adding more neutrons, $^{145,146,148,150}$Nd progressively develop prolate deformation, with the deformation increasing as the neutron number moves further from $N=82$. Correspondingly, the splitting between $K=0$ and $K=1$ components increases as well. However, the splitting of GDR is only visible in the total photoabsorption cross section when the deformation parameter is larger than $0.2$.
Compared to Zn and Ge isotopes, the splitting is visible already for $\beta_2  \approx -0.13$. This is due to for lighter nuclei, the splitting of $K=0$ and $K=1$ component is larger for the same deformation parameter compared to heavy nuclei, which can be seen more clearly in Fig.~\ref{fig7}.
Similarly, for Sm isotopes, $^{144}$Sm remains spherical with a magic neutron number $N=82$. When the neutron number deviates from the magic number, the prolate deformation increases. The splitting is visible also around $\beta_2 \approx 0.2$. As mentioned earlier, for prolate nuclei, the $K=0$ component dominates the lower-energy peak and the $K=1$ component dominates the higher-energy peak, in contrast to the oblate nuclei.

The total photoabsorption cross sections and the contributions from $K=0$ and $K=1$ components given by QFAM calculations for all nuclei considered in this work are given in Figs.~\ref{fig:appendix_b} in the~\ref{sec:appendix_b}.

\begin{figure}
    \center
    \includegraphics[scale=0.8]{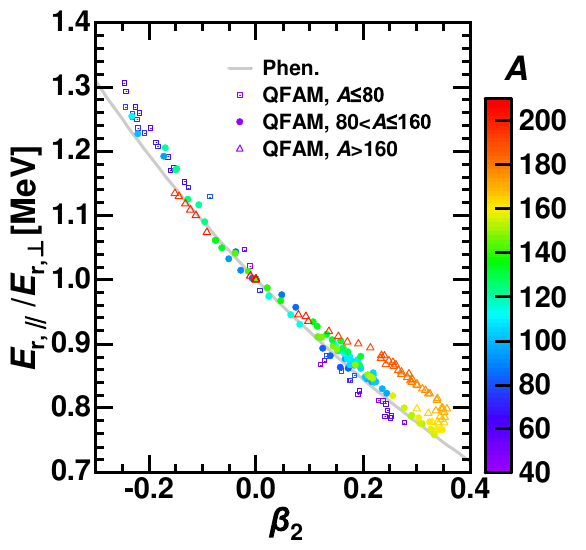}
    \caption{The splitting ratios of the GDR peak energy, $E_{\mathrm{r},\iparallel}/E_{\mathrm{r},\perp}$, extracted from QFAM results and those obtained using the phenomenological formula are shown as a function of the nuclear deformation parameter $\beta_2$.
    In the figure, open squares represent QFAM results for nuclei with $A\leq80$, solid circles correspond to those with $80<A\leq160$, and open triangles denote nuclei with $A>160$. The scatters colors indicate the mass number $A$. The gray line represents the results obtained from the phenomenological formula.}
    \label{fig8}
\end{figure}

To further investigate the deformation splitting of the GDR peak energies, the energy splittings between the two oscillation modes, i.e., $E_{\mathrm{r},\iparallel} - E_{\mathrm{r},\perp}$, are extracted from the QFAM calculations. Here, $E_{\mathrm{r},\iparallel}$ corresponds to the oscillations parallel to the symmetry axis induced by an external field with $K=0$, and $E_{\mathrm{r},\perp}$ corresponds to the oscillations perpendicular to the symmetry axis induced by an external field with $K=1$. The results are presented in Fig.~\ref{fig7}, where the color of scatters indicates the deformation parameter $\beta_2$.
The gray-shaded region highlights nuclei for which the energy splitting satisfies $|E_{\mathrm{r},\iparallel} - E_{\mathrm{r},\perp}| \leq 1$ MeV, indicating that the deformation effect is negligible in these cases. Consequently, a single-peak GDR parameterization is sufficient to describe the total photoabsorption cross sections for these nuclei. We observe these nuclei falling into the gray region in general have $|\beta_2| <0.1$ indicating that they are near-spherical nuclei.
For deformed nuclei, the splitting exhibits opposite signs depending on the deformation type, with negative values for prolate nuclei and positive values for oblate nuclei.
Furthermore, for similar mass numbers, the magnitude of the energy splitting increases with the deformation parameter.
For the same deformation parameters, the magnitude of the splitting decreases as the mass number increases as will be understood in Fig. \ref{fig8}.

Such deformation splitting has been studied in previous works, and a phenomenological formula has been proposed to describe the relationship between the GDR peak energy splitting and the nuclear deformation \cite{Danos_1958_NP,YTian_2019_CPC}.
The formula is given by
\begin{equation}\label{eq:splitting_ratio}
    E_{\mathrm{r},\iparallel}/E_{\mathrm{r},\perp}=\frac{1}{0.911a/b+0.089},
\end{equation}
where $a$ and $b$ are the major and minor axes of the nucleus, respectively, obtained from the deformation parameter $\beta_2$ following the formula in Ref. \cite{Danos_1958_NP,YTian_2019_CPC}.

Figure~\ref{fig8} presents the ratios of two GDR peak energies, $E_{\mathrm{r},\iparallel}/E_{\mathrm{r},\perp}$, extracted from QFAM calculations compared to those obtained using the phenomenological formula in Eq.~\eqref{eq:splitting_ratio}. In the figure, open squares represent nuclei with mass number $A \leq 80$; circles correspond to nuclei with $80 < A \leq 160$; and open triangles denote nuclei with $A > 160$. The scatter colors indicate the specific mass number $A$. The gray line represents the result of the phenomenological formula.

Overall, our results show good agreement with the phenomenological formula, especially for nuclei with medium mass numbers $80 < A \leq 160$. However, some discrepancies are also observed. For nuclei with $A < 80$, the splitting ratios for prolate and oblate nuclei exhibit a slight overestimation and underestimation, respectively. This indicates that the QFAM results predict more pronounced energy splittings compared to those from the phenomenological formula.

For nuclei with $A > 160$, the splitting ratios for oblate deformations agree very well with the phenomenological formula. However, for prolate nuclei, the splitting ratios are overestimated in QFAM calculations. The overestimation increases with deformation up to $\beta_2 = 0.25$, and beyond this deformation, the overestimation remains constant with further increases in deformation. This behavior indicates that for heavy nuclei, the splitting ratios also deviate from the phenomenological formula as in the case of light nuclei, but exhibit a less pronounced energy splittings in contrast to the light nuclei case.  For nuclei with deformation larger than $\beta_2 = 0.25$, the mass number is actually decreasing with the increase of deformation, which makes the energy splitting ratios coming back to the results of phenomenological formula, leading to the constant overestimation.

For the same deformation parameter, the phenomenological formula tells us the splitting ratio is fixed. The GDR energies in light nuclei are higher than those in heavy nuclei. Therefore, for a fixed splitting ratio, the energy splitting $\Delta E = |E_{\mathrm{r},\iparallel} - E_{\mathrm{r},\perp}|$ is larger in light nuclei,  which explains the behavior observed in Fig.~\ref{fig7}. In microscopic QFAM calculations, this behavior is further exaggerated, because for the same deformation parameter, the QFAM does not give the same splitting ratio for different mass numbers as the phenomenological formula does. In contrast, the QFAM calculation gives a higher splitting ratio for heavy nuclei and a smaller splitting ratio for light nuclei on the prolate side of the figure, resulting in an even larger $\Delta E$ for light nuclei.

\begin{figure*}
    \center
    \includegraphics[scale=0.9]{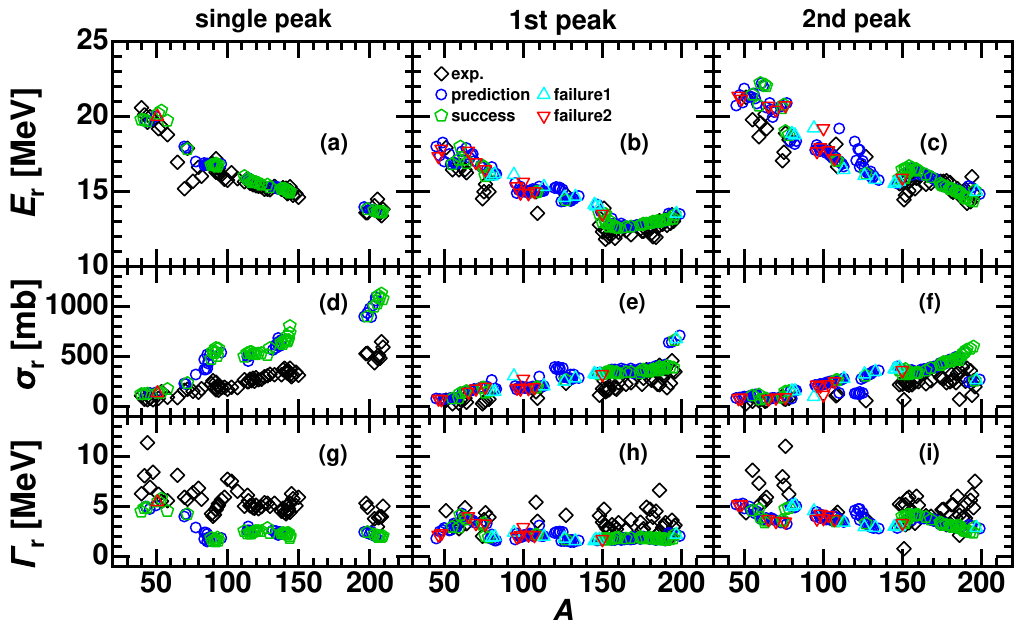}
    \caption{The comparison between GDR parameters obtained from QFAM calculations and GDR parameters extracted from experimental data recommended by IAEA~\cite{Plujko_2018_ADNDT,IAEA_GDR}. The single-peak parameters, including resonance energies $E_\mathrm{r}$, peak cross sections $\sigma_\mathrm{r}$, and resonance widths $\Gamma_\mathrm{r}$, are shown in panels (a), (d), and (g), respectively. The double-peak parameters are displayed in panels (b), (e), and (h) for the lower-energy peak (1st peak), and in panels (c), (f), and (i) for the higher-energy peak (2nd peak). Black open diamonds represent the GDR parameters recommended by IAEA. Other symbols represent the GDR parameters calculated by QFAM, among which, blue circles denote nuclei that are not included in the experimental GDR dataset; green pentagons indicate cases where QFAM results successfully reproduce the experimental classifications of single-peak or double-peak nuclei; and light blue upward triangles and red downward triangles represent cases where QFAM results fail to reproduce the experimental classifications for different reasons.}
    \label{fig9}
\end{figure*}

The GDR parameters serve as key indicators for characterizing the properties of the giant dipole resonance (GDR). The IAEA CRP~\cite{Goriely_2019_EPJA} provides a set of GDR parameters extracted from experimental data using the least-squares fitting procedure, as detailed in Ref.~\cite{Plujko_2018_ADNDT}. Additionally, a set of recommended parameters is provided in Ref.~\cite{IAEA_GDR}, selected based on the minimal least-squares deviation in comparison with other sets.

The comparison between the GDR parameters extracted from QFAM results and those recommended by IAEA is presented in Fig.~\ref{fig9}.
The experimental GDR parameters represented by black diamonds are categorized into two types: single-peak and double-peak parameters.
The single-peak parameters $E^\mathrm{exp.}_\mathrm{r}$, $\sigma^\mathrm{exp.}_\mathrm{r}$, and $\Gamma^\mathrm{exp.}_\mathrm{r}$ are shown in panels (a), (d), and (g), respectively. The double-peak parameters are displayed in panels (b), (e), and (h) as $E_{\mathrm{r},1}$, $\sigma_{\mathrm{r},1}$, and $\Gamma_{\mathrm{r},1}$ for the lower-energy peak, and in panels (c), (f), and (i) as $E_{\mathrm{r},2}$, $\sigma_{\mathrm{r},2}$, and $\Gamma_{\mathrm{r},2}$ for the higher-energy peak.
Correspondingly, the GDR parameters extracted from QFAM results are also categorized into two types, single-peak and double-peak parameters, based on the energy splitting of the GDR peaks $|E_{\mathrm{r},\iparallel} - E_{\mathrm{r},\perp}|$. For the single-peak case, where $|E_{\mathrm{r},\iparallel} - E_{\mathrm{r},\perp}| \leq 1$ MeV, the GDR parameters $E_\mathrm{r}$, $\sigma_\mathrm{r}$, and $\Gamma_\mathrm{r}$ are extracted from the total photoabsorption cross sections and displayed in panels (a), (d), and (g), respectively.
For the double-peak case, where $|E_{\mathrm{r},\iparallel} - E_{\mathrm{r},\perp}| > 1$ MeV, the GDR parameters for the lower-energy peak are displayed in panels (b), (e), and (h). For prolate nuclei, the GDR parameters of this lower-energy peak correspond to $E_{\mathrm{r},\iparallel}$, $\sigma_{\mathrm{r},\iparallel}$, and $\Gamma_{\mathrm{r},\iparallel}$, while for oblate nuclei, they correspond to $E_{\mathrm{r},\perp}$, $\sigma_{\mathrm{r},\perp}$, and $\Gamma_{\mathrm{r},\perp}$, due to the opposite ordering of these components in prolate and oblate nuclei.
Conversely, the GDR parameters for the higher-energy peak are shown in panels (c), (f), and (i), where $E_{\mathrm{r},\perp}$, $\sigma_{\mathrm{r},\perp}$, and $\Gamma_{\mathrm{r},\perp}$ are GDR parameters of this higher-energy peak in prolate nuclei, and $E_{\mathrm{r},\iparallel}$, $\sigma_{\mathrm{r},\iparallel}$, and $\Gamma_{\mathrm{r},\iparallel}$ are those in oblate nuclei.

The shapes and colors of scatters indicate different statuses regarding the reproduction of experimentally classified single- and double-peak structures: green pentagons represent cases where QFAM results successfully reproduce the experimental classifications; light blue upward triangles and red downward triangles represent cases where QFAM results fail to replicate these structures, with the color and pattern indicating the reason for the failure; and blue circles indicate nuclei that are not included in the GDR parameter set recommended by IAEA.

For the classification-reproduced QFAM results, the GDR peak energies reach reasonable agreement with experimental results for both single- and double-peak structures. The energy differences between our QFAM results and experimental data are quite small, while an energy shift of $\sim$ 2 MeV is required to reproduce experimental data in the deformed QRPA calculation with the Gogny interaction \cite{Martini_2016_PRC}.
However, for GDR peak cross sections and resonance widths, QFAM tends to overestimate and underestimate the results, respectively. This situation is much improved in nuclei exhibiting double-peak structures. The discrepancy arises from the small width obtained in QFAM calculations, where only the Landau width is included, and here we use a small smearing parameter of $\gamma =0.5$ MeV to mimic other damping effects. To improve these results, incorporating higher-order terms in the many-body correlation, such as 2 particle-2 hole configurations\cite{Gambacurta_2011_PRL,CLBai_2010_PRL} or particle vibration coupling effects \cite{YFNiu_2015_PRL,YFNiu_2018_PLB,ZZLi_2023_PRL,ZZLi_2024_PRC,Litvinova_2019_PRC,YNZhang_2021_PRC}, is necessary. For double-peak nuclei, the presence of deformation leads to more fragmented strengths and induces larger widths compared to spherical nuclei.

For QFAM results that fail to reproduce the experimentally classified single- or double-peak structures, there are two situations. The first situation in panels (a), (d), and (g) shows that almost all QFAM results exhibiting single-peak structures align with experimental classifications, with only one exception: $^{51}$V. For $^{51}$V, which is experimentally classified as having a double-peak structure, the QFAM results categorize the cross section as a single-peak case, consistent with its nearly spherical shape.

The second situation in the rest panels shows that there are 21 nuclei experimentally classified as single-peak structures while exhibiting double-peak structures in the QFAM results, which will be discussed in the following.
For the double-peak parameters shown in panels (b), (e), and (h), as well as (c), (f), and (i), the QFAM results exhibit good agreement with experimental data, particularly for heavy nuclei.
However, in the QFAM results, 21 nuclei experimentally classified as single-peak structures are categorized as the double-peak types based on our criterion.
For $^{46,48}$Ti,  $^{65}$Cu, $^{70}$Ge, $^{76}$Se, $^{96,98,100}$Mo, $^{103}$Rh, $^{107}$Ag, and $^{150}$Sm shown by red downward triangles, the total cross sections given by QFAM calculations do exhibit double-peak structures, while for $^{80,82}$Se, $^{94}$Zr, $^{113}$In, $^{126}$Te, $^{133}$Cs, $^{145,146}$Nd, $^{148}$Sm, and $^{197}$Au shown by light blue upward triangles, the total cross sections obtained from QFAM calculations actually exhibit single-peak structures, which indicates our criterion for classification is not good enough to reproduce experimental classifications.
In detail, for $^{80,82}$Se, the higher-energy component exhibits two peaks, but with only one remaining in the GDR region. However, the Lorentzian fitting is influenced by the peak outside the GDR region, leading to a more pronounced energy splitting than it should be.
For $^{94}$Zr and $^{197}$Au, the peak cross sections of $K=0$ and $K=1$ component differ by more than a factor of two, which merges these two peaks into a single peak. However, the criterion we used here $|E_{\mathrm{r},\iparallel} - E_{\mathrm{r},\perp}| \leq 1$ MeV cannot reflect the information of peak cross sections.
Finally, for $^{113}$In, $^{126}$Te, $^{133}$Cs, $^{145,146}$Nd, and $^{148}$Sm, the energy splitting between $K=0$ and $K=1$ components is not enough to separate them and exhibit a double-peak structure. If the criterion were changed to $|E_{\mathrm{r},\iparallel} - E_{\mathrm{r},\perp}| \leq 1.7$ MeV, these nuclei would be classified as single-peak nuclei.

In summary, the GDR parameters extracted from QFAM calculations are consistent with the general trends observed in the experimental data, particularly for the resonance energies and the nuclear region of $A\geq150$, suggesting that our results are reliable. For nuclei not included in the experimental GDR dataset, our results are helpful to provide a reference for future experimental studies. The GDR parameters extracted from QFAM calculations are also listed in Tables~\ref{table:appendix_a1} for double-peak nuclei and \ref{table:appendix_a2} for single-peak nuclei in the \ref{sec:appendix_a}.

Following a systematic comparison of GDR parameters, we further present a comparison of photoabsorption cross sections.
Since experimental photoabsorption cross sections are limited, the comparison is made between QFAM results and  experimental data of the following three types, taken from EXFOR  \cite{Otuka_2014_NDS}.
(1). (G,ABS): data for photoabsorption cross sections, used directly for comparison;
(2). equivalent photoabsorption cross sections derived from inelastic proton scattering experiments, taken from (G,TOT) in EXFOR;
(3). total photoneutron cross sections expressed as
\begin{equation}
    \sigma(\gamma,s\mathrm{n})=\sigma(\gamma,\mathrm{n})+\sigma(\gamma,\mathrm{np}) + \sigma(\gamma,2\mathrm{n})+\cdots.
\end{equation}
Some of the data used here are taken from (G,X) in EXFOR directly, and some of the data are given by summing up all the branching cross sections given in EXFOR according to the above formula.
For heavy nuclei, the contribution of the photoproton cross section $\sigma(\gamma,\mathrm{p})$ to the photoabsorption cross section $\sigma(\gamma,\mathrm{abs.})$ is negligible, allowing a direct comparison between $\sigma(\gamma,s\mathrm{n})$ and $\sigma(\gamma,\mathrm{abs.})$.
However, for lighter nuclei, this assumption does not hold.
In this work, comparisons between calculated $\sigma(\gamma,\mathrm{abs.})$ and experimental $\sigma(\gamma,s\mathrm{n})$ are made for both medium-heavy and heavy nuclei due to the lack of sufficient data. 
The results of the comparison between QFAM results and experimental data for all nuclei considered in this work, with available experimental data, are shown in Figs.~\ref{fig:appendix_c} in \ref{sec:appendix_c}.
In addition, photoabsorption cross sections form IAEA Evaluated Photonuclear Data Library (IAEA/PD-2019.2) \cite{Kawano_2020_NDS} 
and GDR components of photoabsorption cross sections calculated with GDR parameters extracted from experimental data recommended by IAEA \cite{Plujko_2018_ADNDT}
are also plotted in Figs.~\ref{fig:appendix_c} for comparisons.

\begin{figure*}
    \center
    \includegraphics[scale=0.9]{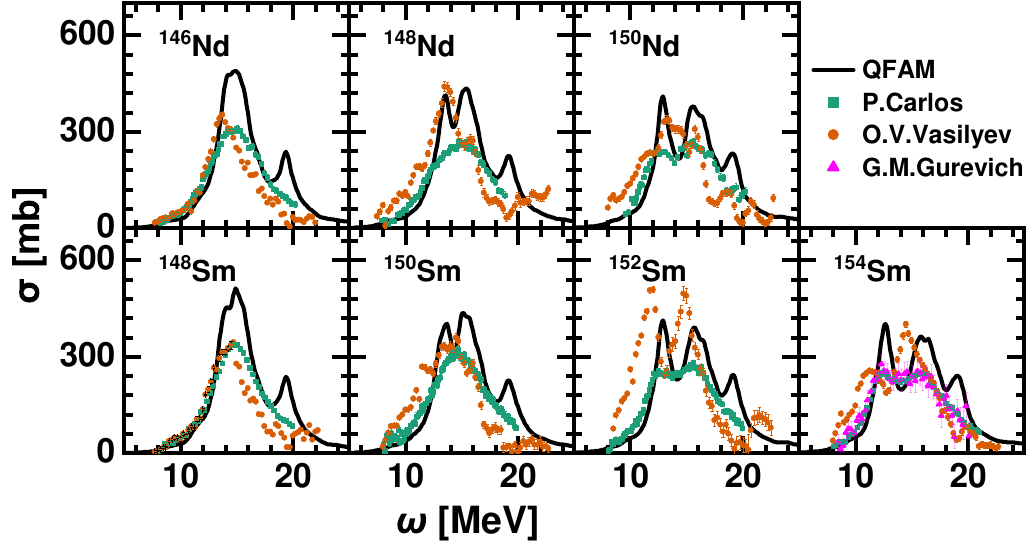}
    \caption{The comparisons between photoabsorption cross sections given by QFAM calculations and experimental data from EXFOR in Nd and Sm isotopes.
        The black lines represent our QFAM results, while the colored scatter points represent experimental data.
        Green squares correspond to total photoneutron data from P. Carlos\cite{Carlos_1971_NPA,Carlos_1974_NPA}, while orange circles correspond to total photoneutron data from O. V. Vasilyev~\cite{VasileV_1969_PLB}.
        Additionally, the photoabsorption data for $^{154}$Sm from G. M. Gurevich~\cite{Gurevich_1981_NPA} are represented by magenta triangles.}
    \label{fig10}
\end{figure*}

Here we show some examples of this comparison between photoabsorption cross sections given by QFAM calculations and experimental data from EXFOR in Nd and Sm isotopes in Fig. \ref{fig10}, to illustrate the role of our calculation in photoabsorption data evaluation. The black lines represent our QFAM results, while the colored scatters represent experimental data.
Green squares correspond to total photoneutron data from P. Carlos~\cite{Carlos_1971_NPA,Carlos_1974_NPA}, while orange circles represent total photoneutron data from O. V. Vasilyev~\cite{VasileV_1969_PLB}.
The photoabsorption data for $^{154}$Sm from G. M. Gurevich~\cite{Gurevich_1981_NPA} are represented by magenta triangles.

In $^{146}$Nd and $^{148}$Sm, both sets of experimental data exhibit single-peak structures. However, P. Carlos' data show a higher GDR energy compared to O. V. Vasilyev's data.
Our results also show a single-peak structure, with peak energies more consistent with P. Carlos' data. As we have discussed earlier, we are able to provide a good description of GDR energies, which can also be seen here from the good agreement with P. Carlos' data, however, our results yield a smaller resonance width and hence a much higher peak cross section, which is due to the lack of higher-order many-body correlations of 2p-2h type in QFAM approach.
For $^{148}$Nd and $^{150}$Sm, both sets of experimental data present overall single peaks, with O. V. Vasilyev's data exhibiting more complicated fine structures. Again, the GDR energies of P. Carlos' data are higher than those of O. V. Vasilyev's data.
Although our results already show a double-peak structure with a small energy splitting in these two deformed nuclei, it can still be seen that the overall structures of GDR are more consistent with P. Carlos' data.
For $^{150}$Nd and $^{152}$Sm, both sets of experimental data exhibit double-peak structures. Systematically, the GDR energies of P. Carlos' data are still higher than those of O. V. Vasilyev's data for these two nuclei, as in all previous cases. Additionally, for O. V. Vasilyev's data, the two peaks in $^{150}$Nd are very asymmetric, and the cross sections in $^{152}$Sm are much higher. Our results also exhibit double-peak structures, and show much better agreement with P. Carlos' data, like all the previous cases.
Finally, for $^{154}$Sm, which has three sets of experimental data, a clear consistency is observed between P. Carlos' and G. M. Gurevich's data, which is also supported by our results.

Systematic discrepancies are observed in the experimental photoneutron cross sections for Nd and Sm isotopes, with O. V. Vasilyev's data consistently showing lower peak energies compared to P. Carlos' data. Our results all exhibit better agreement with P. Carlos' data, which is further supported by G. M. Gurevich's data for $^{154}$Sm. In this regard, our calculations may be useful in photonuclear data evaluation.

\section{Conclusions}
\label{sec:concl}
In this study, photoabsorption cross sections for 235 stable nuclei, including 146 even-even nuclei and 89 odd-A nuclei, ranging from $^{40}$Ca to $^{209}$Bi, were investigated using the quasiparticle finite amplitude method (QFAM) based on the relativistic axially deformed Hartree-Bogoliubov (RHB) model. The extension to odd-A nuclei was introduced in this work and checked with spherical QRPA results. GDR parameters based on the standard Lorentzian (SLO) model were extracted from QFAM results.

The deformation effects on photoabsorption cross sections were studied systematically and illustrated in detail by the oblate-dominated isotopes such as Zn and Ge, as well as prolate-dominated isotopes including Nd and Sm. In deformed nuclei, the giant dipole resonance (GDR) peaks split into two components, corresponding to oscillations along the symmetry axis and perpendicular to it. The ordering of these two components is reversed in prolate and oblate nuclei.
Systematic trends were also observed between these two components, with the energy splitting increasing as deformation increases and being larger for smaller mass numbers with the same deformation parameter. The splitting ratios of peak energies for these two components show good agreement with the phenomenological formula.

Further comparisons with GDR parameters extracted from experimental data show good agreement for GDR peak energies, though peak cross sections are overestimated and resonance widths are underestimated due to the damping mechanisms not considered in the QFAM approach. Furthermore, these discrepancies in peak cross sections and resonance widths are less pronounced in deformed nuclei due to the deformation effect.

A comparison of photoabsorption cross sections from QFAM calculations with experimental data was performed systematically. For illustration, we selected Nd and Sm isotopes, which exhibit systematic discrepancies in experimental data, as examples to show the role of our calculations in photonuclear data evaluation. Our results show better agreement with P. Carlos' total photoneutron cross-section data, which are also supported by G. M. Gurevich's photoabsorption data. These findings highlight the potential of our calculations in distinguishing discrepant data.


\section{Acknowledgements}
The authors gratefully acknowledge Profs. P. Dimitriou, R. Capote, 
and A. Koning at the IAEA for their insightful discussions.
This work was supported by 
the Lingchuang Research Project of China National Nuclear Corporation under Grant No.CNNC-LCKY-2024-082, 
the National Key Research and Development Program under Grants No.2022YFA1602403 and No.2021YFA1601500, 
the National Natural Science Foundation of China under Grants No.12075104 and No.12447106, 
and the Fundamental Research Funds for the Central Universities under Grant No. lzujbky-2023-stlt01.

\bibliographystyle{elsarticle-num}

\appendix

\section{GDR parameters extracted from QFAM calculations}
\label{sec:appendix_a}

Tables~\ref{table:appendix_a1} present the GDR parameters extracted from QFAM calculations 
for nuclei categorized as the double-peak type according to our criterion.
The table is organized as follows:
Column 1 lists the corresponding nuclides. 
For nuclides with odd mass numbers $A$, 
Column 2 specifies the projected total angular momentum and parity $K^\pi$.
Column 3 provides the deformation parameter $\beta_2$.
The GDR parameters for oscillations parallel to the symmetry axis 
induced by an external dipole field with $K=0$ are detailed in Columns 4-6. 
These columns present the peak cross sections $\sigma_{\mathrm{r},\iparallel}$, 
resonance energies $E_{\mathrm{r},\iparallel}$, 
and resonance widths $\Gamma_{\mathrm{r},\iparallel}$.
Similarly, Columns 7-9 describe the GDR parameters for the oscillations perpendicular to the symmetry axis,  
induced by an external dipole field with $K=1$. 
These columns list the peak cross sections $\sigma_{\mathrm{r},\perp}$, 
resonance energies $E_{\mathrm{r},\perp}$, 
and resonance widths $\Gamma_{\mathrm{r},\perp}$.

Tables~\ref{table:appendix_a2} present the GDR parameters extracted from QFAM calculations 
for nuclei categorized as the single-peak type according to our criterion.
Columns 1-3 represent the same information as Columns 1-3 in Table~\ref{table:appendix_a1}.
Columns 4-6 show the GDR parameters for the total photoabsorption cross sections, 
the resonance energies $E_{\mathrm{r}}$, peak cross sections $\sigma_{\mathrm{r}}$,
and resonance widths $\Gamma_{\mathrm{r}}$.

\begin{table*}
    \centering
    \caption{}\label{table:appendix_a1}
    \begin{tabular*}{0.9\textwidth}{@{\extracolsep{\fill}}ccccccccc}
        \hline\hline
        Nuclide    & $K^\pi$  & $\beta_2$ & $\sigma_{\mathrm{r},\iparallel}$ [mb] & $E_{\mathrm{r},\iparallel}$ [MeV]& $\Gamma_{\mathrm{r},\iparallel}$ [MeV]& $\sigma_{\mathrm{r},\perp}$ [mb]& $E_{\mathrm{r},\perp}$ [MeV]& $\Gamma_{\mathrm{r},\perp}$ [MeV]\\
        \hline
$^{ 45}$Sc & $ 1/2^-$ & $ 0.121$ & $ 76.906$ & $17.964$ & $1.747$ & $ 80.209$ & $20.708$ & $5.158$ \\
$^{ 46}$Ti & $      $ & $ 0.238$ & $ 76.118$ & $17.324$ & $2.144$ & $ 80.205$ & $21.369$ & $5.202$ \\
$^{ 47}$Ti & $ 5/2^-$ & $ 0.229$ & $ 70.358$ & $17.366$ & $2.441$ & $ 82.652$ & $21.414$ & $5.130$ \\
$^{ 48}$Ti & $      $ & $ 0.177$ & $ 72.319$ & $17.833$ & $2.290$ & $ 84.597$ & $21.159$ & $5.204$ \\
$^{ 49}$Ti & $ 7/2^-$ & $ 0.128$ & $ 65.952$ & $18.243$ & $2.609$ & $ 85.464$ & $20.895$ & $5.238$ \\
$^{ 50}$Cr & $      $ & $ 0.277$ & $ 70.994$ & $16.977$ & $2.792$ & $ 92.842$ & $21.843$ & $4.751$ \\
$^{ 53}$Cr & $ 1/2^-$ & $ 0.191$ & $ 76.434$ & $17.466$ & $2.563$ & $ 96.693$ & $21.279$ & $4.944$ \\
$^{ 54}$Cr & $      $ & $ 0.253$ & $ 82.189$ & $16.801$ & $2.579$ & $103.523$ & $21.315$ & $4.586$ \\
$^{ 55}$Mn & $ 5/2^-$ & $ 0.251$ & $ 76.936$ & $16.798$ & $2.887$ & $105.869$ & $21.432$ & $4.570$ \\
$^{ 56}$Fe & $      $ & $ 0.241$ & $ 71.886$ & $16.911$ & $3.269$ & $108.286$ & $21.498$ & $4.617$ \\
$^{ 57}$Fe & $ 1/2^-$ & $ 0.246$ & $ 77.810$ & $16.843$ & $2.953$ & $116.785$ & $21.174$ & $4.255$ \\
$^{ 58}$Fe & $      $ & $ 0.242$ & $ 79.131$ & $16.788$ & $2.918$ & $115.519$ & $20.817$ & $4.437$ \\
$^{ 59}$Co & $ 7/2^-$ & $ 0.194$ & $ 81.714$ & $17.107$ & $2.829$ & $113.427$ & $20.709$ & $4.667$ \\
$^{ 60}$Ni & $      $ & $-0.196$ & $ 65.802$ & $22.207$ & $3.829$ & $127.870$ & $17.963$ & $4.088$ \\
$^{ 61}$Ni & $ 1/2^-$ & $-0.218$ & $ 68.836$ & $22.204$ & $3.661$ & $126.595$ & $17.633$ & $4.245$ \\
$^{ 62}$Ni & $      $ & $-0.225$ & $ 70.316$ & $21.997$ & $3.645$ & $137.440$ & $17.343$ & $3.921$ \\
$^{ 64}$Ni & $      $ & $-0.133$ & $ 70.758$ & $20.564$ & $3.806$ & $142.929$ & $17.854$ & $3.778$ \\
$^{ 63}$Cu & $ 5/2^-$ & $-0.242$ & $ 75.976$ & $22.087$ & $3.367$ & $142.467$ & $17.084$ & $3.842$ \\
$^{ 65}$Cu & $ 3/2^-$ & $-0.159$ & $ 74.430$ & $20.715$ & $3.696$ & $143.291$ & $17.708$ & $3.985$ \\
$^{ 64}$Zn & $      $ & $-0.246$ & $ 74.243$ & $22.003$ & $3.567$ & $150.380$ & $16.851$ & $3.662$ \\
$^{ 66}$Zn & $      $ & $-0.186$ & $ 75.204$ & $20.874$ & $3.721$ & $153.829$ & $17.203$ & $3.664$ \\
$^{ 67}$Zn & $ 3/2^-$ & $-0.153$ & $ 74.299$ & $20.418$ & $3.794$ & $156.577$ & $17.410$ & $3.622$ \\
$^{ 68}$Zn & $      $ & $-0.126$ & $ 77.665$ & $19.944$ & $3.710$ & $160.057$ & $17.459$ & $3.548$ \\
$^{ 69}$Ga & $ 3/2^-$ & $-0.164$ & $ 80.486$ & $20.253$ & $3.604$ & $167.924$ & $17.006$ & $3.334$ \\
$^{ 70}$Ge & $      $ & $-0.183$ & $ 83.899$ & $20.449$ & $3.517$ & $176.100$ & $16.936$ & $3.306$ \\
$^{ 73}$Ge & $ 7/2^+$ & $-0.220$ & $ 90.006$ & $20.625$ & $3.346$ & $182.629$ & $16.691$ & $3.213$ \\
$^{ 74}$Ge & $      $ & $-0.208$ & $ 86.257$ & $20.523$ & $3.580$ & $187.119$ & $16.716$ & $3.154$ \\
$^{ 76}$Ge & $      $ & $ 0.176$ & $132.698$ & $16.004$ & $1.989$ & $144.976$ & $19.004$ & $4.709$ \\
$^{ 75}$As & $ 1/2^-$ & $-0.230$ & $ 90.720$ & $20.763$ & $3.436$ & $185.703$ & $16.595$ & $3.314$ \\
$^{ 74}$Se & $      $ & $-0.222$ & $ 90.426$ & $20.696$ & $3.461$ & $190.888$ & $16.567$ & $3.273$ \\
$^{ 76}$Se & $      $ & $-0.231$ & $ 92.440$ & $20.739$ & $3.455$ & $194.459$ & $16.480$ & $3.254$ \\
$^{ 77}$Se & $ 5/2^+$ & $-0.243$ & $ 98.491$ & $20.854$ & $3.190$ & $195.258$ & $16.438$ & $3.244$ \\
$^{ 78}$Se & $      $ & $ 0.159$ & $145.255$ & $16.055$ & $1.866$ & $149.946$ & $18.718$ & $4.812$ \\
$^{ 80}$Se & $      $ & $ 0.184$ & $146.283$ & $16.008$ & $1.864$ & $149.862$ & $18.821$ & $4.933$ \\
$^{ 82}$Se & $      $ & $ 0.172$ & $144.919$ & $16.139$ & $1.898$ & $147.644$ & $18.733$ & $5.117$ \\
$^{ 79}$Br & $ 5/2^-$ & $ 0.131$ & $151.912$ & $16.111$ & $1.773$ & $146.600$ & $18.298$ & $5.018$ \\
$^{ 81}$Br & $ 5/2^-$ & $ 0.157$ & $159.301$ & $16.070$ & $1.692$ & $145.468$ & $18.606$ & $5.251$ \\
$^{ 80}$Kr & $      $ & $-0.086$ & $ 76.691$ & $18.736$ & $4.930$ & $260.000$ & $16.600$ & $2.175$ \\
$^{ 82}$Kr & $      $ & $ 0.125$ & $173.279$ & $16.191$ & $1.539$ & $157.502$ & $18.136$ & $4.831$ \\
$^{ 83}$Kr & $ 7/2^+$ & $ 0.139$ & $172.002$ & $16.146$ & $1.555$ & $154.706$ & $18.326$ & $5.016$ \\
$^{ 94}$Zr & $      $ & $-0.172$ & $ 95.281$ & $19.190$ & $4.481$ & $298.239$ & $16.100$ & $2.349$ \\
$^{ 96}$Zr & $      $ & $ 0.244$ & $165.173$ & $14.868$ & $2.233$ & $192.051$ & $18.072$ & $4.421$ \\
$^{ 95}$Mo & $ 1/2^+$ & $ 0.167$ & $201.021$ & $15.540$ & $1.651$ & $208.924$ & $17.721$ & $3.945$ \\
$^{ 96}$Mo & $      $ & $ 0.186$ & $184.904$ & $15.350$ & $1.939$ & $215.193$ & $17.659$ & $3.766$ \\
$^{ 97}$Mo & $ 3/2^+$ & $ 0.216$ & $182.912$ & $15.058$ & $1.988$ & $206.127$ & $17.907$ & $4.143$ \\
$^{ 98}$Mo & $      $ & $ 0.236$ & $176.707$ & $14.895$ & $2.109$ & $204.085$ & $17.948$ & $4.248$ \\
$^{100}$Mo & $      $ & $-0.221$ & $113.771$ & $19.194$ & $3.778$ & $275.855$ & $15.650$ & $2.915$ \\
$^{ 98}$Ru & $      $ & $ 0.183$ & $188.433$ & $15.302$ & $1.956$ & $220.711$ & $17.666$ & $3.818$ \\
$^{ 99}$Ru & $ 3/2^+$ & $ 0.207$ & $194.078$ & $15.080$ & $1.876$ & $214.865$ & $17.829$ & $4.087$ \\
$^{100}$Ru & $      $ & $ 0.224$ & $184.601$ & $14.960$ & $2.054$ & $216.662$ & $17.801$ & $4.064$ \\
$^{101}$Ru & $ 3/2^+$ & $ 0.220$ & $181.291$ & $14.958$ & $2.098$ & $225.994$ & $17.713$ & $3.902$ \\
$^{102}$Ru & $      $ & $ 0.219$ & $178.932$ & $14.919$ & $2.169$ & $234.605$ & $17.536$ & $3.665$ \\
$^{104}$Ru & $      $ & $ 0.219$ & $175.819$ & $14.865$ & $2.233$ & $238.822$ & $17.395$ & $3.649$ \\
        \hline\hline
    \end{tabular*}
\end{table*}
\begin{table*}\ContinuedFloat
    \centering
    \begin{tabular*}{0.9\textwidth}{@{\extracolsep{\fill}}ccccccccc}
        \hline\hline
        Nuclide    & $K^\pi$  & $\beta_2$ & $\sigma_{\mathrm{r},\iparallel}$ [mb] & $E_{\mathrm{r},\iparallel}$ [MeV]& $\Gamma_{\mathrm{r},\iparallel}$ [MeV]& $\sigma_{\mathrm{r},\perp}$ [mb]& $E_{\mathrm{r},\perp}$ [MeV]& $\Gamma_{\mathrm{r},\perp}$ [MeV]\\
        \hline
$^{103}$Rh & $ 1/2^-$ & $ 0.221$ & $180.565$ & $14.881$ & $2.167$ & $222.374$ & $17.723$ & $4.085$ \\
$^{102}$Pd & $      $ & $ 0.204$ & $189.969$ & $15.019$ & $2.034$ & $225.738$ & $17.693$ & $4.010$ \\
$^{104}$Pd & $      $ & $ 0.196$ & $185.581$ & $15.006$ & $2.122$ & $242.740$ & $17.416$ & $3.664$ \\
$^{105}$Pd & $ 5/2^+$ & $ 0.193$ & $181.336$ & $15.000$ & $2.165$ & $246.803$ & $17.314$ & $3.615$ \\
$^{106}$Pd & $      $ & $ 0.189$ & $182.906$ & $15.009$ & $2.167$ & $251.261$ & $17.215$ & $3.556$ \\
$^{108}$Pd & $      $ & $ 0.190$ & $183.755$ & $14.920$ & $2.167$ & $258.537$ & $17.051$ & $3.467$ \\
$^{110}$Pd & $      $ & $-0.232$ & $129.630$ & $19.167$ & $3.634$ & $293.249$ & $15.284$ & $3.067$ \\
$^{107}$Ag & $ 7/2^+$ & $ 0.179$ & $188.067$ & $15.016$ & $2.084$ & $248.791$ & $17.193$ & $3.696$ \\
$^{109}$Ag & $ 7/2^+$ & $ 0.175$ & $189.240$ & $14.948$ & $2.078$ & $261.105$ & $16.988$ & $3.501$ \\
$^{106}$Cd & $      $ & $ 0.175$ & $193.761$ & $15.117$ & $2.010$ & $234.977$ & $17.429$ & $4.011$ \\
$^{108}$Cd & $      $ & $ 0.170$ & $191.766$ & $15.111$ & $2.047$ & $243.295$ & $17.242$ & $3.919$ \\
$^{110}$Cd & $      $ & $ 0.165$ & $193.271$ & $15.049$ & $2.037$ & $256.469$ & $17.007$ & $3.687$ \\
$^{111}$Cd & $ 1/2^+$ & $ 0.157$ & $190.950$ & $15.001$ & $2.067$ & $261.895$ & $16.862$ & $3.565$ \\
$^{112}$Cd & $      $ & $ 0.150$ & $197.318$ & $15.014$ & $1.995$ & $268.797$ & $16.789$ & $3.490$ \\
$^{113}$Cd & $ 5/2^+$ & $ 0.145$ & $197.756$ & $14.986$ & $1.989$ & $272.586$ & $16.631$ & $3.405$ \\
$^{113}$In & $ 9/2^+$ & $ 0.082$ & $198.540$ & $15.288$ & $1.964$ & $274.879$ & $16.438$ & $3.369$ \\
$^{121}$Sb & $ 7/2^+$ & $-0.107$ & $126.925$ & $17.028$ & $4.221$ & $369.368$ & $15.259$ & $2.383$ \\
$^{123}$Sb & $ 7/2^+$ & $-0.096$ & $139.842$ & $16.589$ & $3.588$ & $362.574$ & $15.223$ & $2.457$ \\
$^{120}$Te & $      $ & $-0.169$ & $123.505$ & $18.311$ & $4.534$ & $391.044$ & $15.195$ & $2.304$ \\
$^{122}$Te & $      $ & $-0.148$ & $121.279$ & $17.803$ & $4.695$ & $390.778$ & $15.184$ & $2.312$ \\
$^{123}$Te & $ 7/2^-$ & $-0.150$ & $120.949$ & $17.719$ & $4.719$ & $379.485$ & $15.133$ & $2.404$ \\
$^{125}$Te & $ 7/2^-$ & $-0.127$ & $128.413$ & $16.999$ & $4.277$ & $379.310$ & $15.115$ & $2.401$ \\
$^{126}$Te & $      $ & $ 0.119$ & $244.635$ & $14.582$ & $1.740$ & $321.382$ & $16.022$ & $2.917$ \\
$^{127}$I  & $ 3/2^+$ & $ 0.153$ & $261.241$ & $14.388$ & $1.648$ & $304.009$ & $16.183$ & $3.227$ \\
$^{124}$Xe & $      $ & $ 0.216$ & $292.062$ & $14.295$ & $1.479$ & $272.289$ & $16.841$ & $3.954$ \\
$^{126}$Xe & $      $ & $ 0.193$ & $292.415$ & $14.347$ & $1.478$ & $285.600$ & $16.542$ & $3.679$ \\
$^{128}$Xe & $      $ & $ 0.171$ & $286.834$ & $14.404$ & $1.512$ & $312.094$ & $16.235$ & $3.174$ \\
$^{129}$Xe & $ 3/2^+$ & $ 0.163$ & $282.250$ & $14.416$ & $1.540$ & $314.987$ & $16.126$ & $3.120$ \\
$^{130}$Xe & $      $ & $ 0.142$ & $276.216$ & $14.489$ & $1.579$ & $332.352$ & $15.996$ & $2.899$ \\
$^{131}$Xe & $ 1/2^+$ & $ 0.132$ & $269.625$ & $14.525$ & $1.620$ & $334.640$ & $15.883$ & $2.861$ \\
$^{133}$Cs & $ 5/2^+$ & $ 0.114$ & $281.392$ & $14.617$ & $1.523$ & $342.984$ & $15.763$ & $2.829$ \\
$^{130}$Ba & $      $ & $ 0.196$ & $317.292$ & $14.345$ & $1.381$ & $317.893$ & $16.293$ & $3.254$ \\
$^{132}$Ba & $      $ & $ 0.158$ & $313.397$ & $14.461$ & $1.393$ & $348.667$ & $15.992$ & $2.815$ \\
$^{135}$Ba & $ 1/2^+$ & $ 0.107$ & $281.389$ & $14.679$ & $1.568$ & $356.619$ & $15.719$ & $2.753$ \\
$^{145}$Nd & $ 3/2^-$ & $ 0.115$ & $322.894$ & $14.115$ & $1.511$ & $371.556$ & $15.496$ & $2.942$ \\
$^{146}$Nd & $      $ & $ 0.144$ & $312.955$ & $13.987$ & $1.631$ & $372.373$ & $15.605$ & $2.973$ \\
$^{148}$Nd & $      $ & $ 0.209$ & $318.377$ & $13.503$ & $1.664$ & $357.365$ & $15.849$ & $3.225$ \\
$^{150}$Nd & $      $ & $ 0.290$ & $321.731$ & $12.927$ & $1.685$ & $317.911$ & $16.422$ & $3.970$ \\
$^{147}$Sm & $ 3/2^-$ & $ 0.123$ & $327.579$ & $14.091$ & $1.552$ & $393.095$ & $15.486$ & $2.803$ \\
$^{148}$Sm & $      $ & $ 0.151$ & $317.849$ & $13.923$ & $1.671$ & $385.914$ & $15.623$ & $2.923$ \\
$^{149}$Sm & $ 1/2^-$ & $ 0.185$ & $334.947$ & $13.648$ & $1.584$ & $369.461$ & $15.741$ & $3.128$ \\
$^{150}$Sm & $      $ & $ 0.215$ & $322.105$ & $13.484$ & $1.716$ & $361.258$ & $15.900$ & $3.301$ \\
$^{152}$Sm & $      $ & $ 0.291$ & $335.505$ & $12.943$ & $1.664$ & $327.868$ & $16.419$ & $3.922$ \\
$^{154}$Sm & $      $ & $ 0.333$ & $333.954$ & $12.646$ & $1.714$ & $319.006$ & $16.675$ & $4.163$ \\
$^{151}$Eu & $ 7/2^+$ & $ 0.218$ & $325.095$ & $13.433$ & $1.719$ & $352.775$ & $15.876$ & $3.418$ \\
$^{153}$Eu & $ 5/2^-$ & $ 0.310$ & $334.078$ & $12.836$ & $1.699$ & $322.130$ & $16.547$ & $4.083$ \\
$^{152}$Gd & $      $ & $ 0.213$ & $334.620$ & $13.474$ & $1.682$ & $369.979$ & $15.876$ & $3.301$ \\
$^{154}$Gd & $      $ & $ 0.277$ & $347.642$ & $13.036$ & $1.635$ & $341.623$ & $16.287$ & $3.809$ \\
$^{155}$Gd & $ 3/2^+$ & $ 0.306$ & $344.858$ & $12.827$ & $1.668$ & $330.933$ & $16.525$ & $4.043$ \\
$^{156}$Gd & $      $ & $ 0.326$ & $343.490$ & $12.702$ & $1.708$ & $329.671$ & $16.566$ & $4.083$ \\
$^{157}$Gd & $ 5/2^-$ & $ 0.332$ & $340.224$ & $12.635$ & $1.748$ & $324.630$ & $16.533$ & $4.177$ \\
$^{158}$Gd & $      $ & $ 0.338$ & $333.626$ & $12.607$ & $1.786$ & $332.009$ & $16.455$ & $4.062$ \\
$^{160}$Gd & $      $ & $ 0.348$ & $325.264$ & $12.535$ & $1.846$ & $332.599$ & $16.368$ & $4.081$ \\
\hline\hline
    \end{tabular*}
\end{table*}

\begin{table*}\ContinuedFloat
    \centering
    \caption{}
    \begin{tabular*}{0.9\textwidth}{@{\extracolsep{\fill}}ccccccccc}
        \hline\hline
        Nuclide    & $K^\pi$  & $\beta_2$ & $\sigma_{\mathrm{r},\iparallel}$ [mb] & $E_{\mathrm{r},\iparallel}$ [MeV]& $\Gamma_{\mathrm{r},\iparallel}$ [MeV]& $\sigma_{\mathrm{r},\perp}$ [mb]& $E_{\mathrm{r},\perp}$ [MeV]& $\Gamma_{\mathrm{r},\perp}$ [MeV]\\
        \hline
$^{159}$Tb & $ 3/2^+$ & $ 0.339$ & $339.155$ & $12.607$ & $1.749$ & $332.168$ & $16.476$ & $4.146$ \\
$^{156}$Dy & $      $ & $ 0.255$ & $362.894$ & $13.151$ & $1.577$ & $360.355$ & $16.063$ & $3.618$ \\
$^{158}$Dy & $      $ & $ 0.304$ & $363.602$ & $12.834$ & $1.616$ & $344.555$ & $16.348$ & $3.947$ \\
$^{160}$Dy & $      $ & $ 0.324$ & $354.132$ & $12.699$ & $1.690$ & $346.500$ & $16.304$ & $3.954$ \\
$^{161}$Dy & $ 3/2^-$ & $ 0.329$ & $347.919$ & $12.637$ & $1.729$ & $346.160$ & $16.255$ & $3.964$ \\
$^{162}$Dy & $      $ & $ 0.338$ & $343.230$ & $12.623$ & $1.761$ & $349.874$ & $16.235$ & $3.937$ \\
$^{163}$Dy & $ 5/2^+$ & $ 0.341$ & $338.635$ & $12.572$ & $1.784$ & $350.149$ & $16.211$ & $3.952$ \\
$^{164}$Dy & $      $ & $ 0.351$ & $334.693$ & $12.559$ & $1.816$ & $352.883$ & $16.166$ & $3.914$ \\
$^{165}$Ho & $ 7/2^-$ & $ 0.348$ & $338.451$ & $12.587$ & $1.766$ & $354.561$ & $16.094$ & $3.910$ \\
$^{162}$Er & $      $ & $ 0.300$ & $375.812$ & $12.828$ & $1.593$ & $366.712$ & $16.057$ & $3.732$ \\
$^{164}$Er & $      $ & $ 0.322$ & $366.893$ & $12.716$ & $1.647$ & $367.757$ & $16.058$ & $3.766$ \\
$^{166}$Er & $      $ & $ 0.344$ & $355.732$ & $12.630$ & $1.712$ & $370.320$ & $16.039$ & $3.757$ \\
$^{167}$Er & $ 1/2^-$ & $ 0.352$ & $354.960$ & $12.588$ & $1.709$ & $370.356$ & $16.029$ & $3.770$ \\
$^{168}$Er & $      $ & $ 0.351$ & $341.330$ & $12.630$ & $1.791$ & $378.391$ & $15.917$ & $3.659$ \\
$^{170}$Er & $      $ & $ 0.349$ & $323.496$ & $12.645$ & $1.924$ & $378.982$ & $15.809$ & $3.663$ \\
$^{169}$Tm & $ 1/2^+$ & $ 0.347$ & $344.044$ & $12.641$ & $1.763$ & $377.881$ & $15.856$ & $3.694$ \\
$^{168}$Yb & $      $ & $ 0.336$ & $376.463$ & $12.665$ & $1.636$ & $388.574$ & $15.915$ & $3.585$ \\
$^{170}$Yb & $      $ & $ 0.347$ & $360.739$ & $12.666$ & $1.712$ & $396.275$ & $15.821$ & $3.509$ \\
$^{171}$Yb & $ 7/2^+$ & $ 0.356$ & $352.600$ & $12.652$ & $1.751$ & $391.803$ & $15.839$ & $3.582$ \\
$^{172}$Yb & $      $ & $ 0.337$ & $342.671$ & $12.727$ & $1.825$ & $402.070$ & $15.671$ & $3.448$ \\
$^{173}$Yb & $ 7/2^-$ & $ 0.338$ & $335.480$ & $12.719$ & $1.875$ & $400.150$ & $15.606$ & $3.459$ \\
$^{174}$Yb & $      $ & $ 0.324$ & $331.414$ & $12.744$ & $1.920$ & $409.713$ & $15.503$ & $3.372$ \\
$^{176}$Yb & $      $ & $ 0.316$ & $329.821$ & $12.745$ & $1.932$ & $413.051$ & $15.398$ & $3.368$ \\
$^{175}$Lu & $ 7/2^+$ & $ 0.310$ & $352.766$ & $12.776$ & $1.778$ & $420.768$ & $15.402$ & $3.259$ \\
$^{174}$Hf & $      $ & $ 0.322$ & $373.059$ & $12.758$ & $1.682$ & $430.712$ & $15.525$ & $3.174$ \\
$^{176}$Hf & $      $ & $ 0.300$ & $362.725$ & $12.830$ & $1.743$ & $440.549$ & $15.355$ & $3.087$ \\
$^{177}$Hf & $ 5/2^-$ & $ 0.300$ & $357.165$ & $12.818$ & $1.784$ & $443.420$ & $15.303$ & $3.047$ \\
$^{178}$Hf & $      $ & $ 0.291$ & $352.366$ & $12.843$ & $1.820$ & $440.710$ & $15.285$ & $3.121$ \\
$^{179}$Hf & $ 9/2^+$ & $ 0.288$ & $346.484$ & $12.855$ & $1.863$ & $437.266$ & $15.268$ & $3.161$ \\
$^{180}$Hf & $      $ & $ 0.282$ & $344.302$ & $12.868$ & $1.874$ & $445.942$ & $15.173$ & $3.070$ \\
$^{181}$Ta & $ 9/2^-$ & $ 0.271$ & $360.595$ & $12.907$ & $1.767$ & $450.775$ & $15.146$ & $3.035$ \\
$^{180}$W  & $      $ & $ 0.292$ & $379.908$ & $12.869$ & $1.685$ & $465.600$ & $15.239$ & $2.931$ \\
$^{182}$W  & $      $ & $ 0.271$ & $367.131$ & $12.956$ & $1.746$ & $467.049$ & $15.126$ & $2.923$ \\
$^{183}$W  & $ 3/2^-$ & $ 0.266$ & $363.993$ & $12.964$ & $1.759$ & $459.386$ & $15.091$ & $3.001$ \\
$^{184}$W  & $      $ & $ 0.257$ & $358.670$ & $12.977$ & $1.796$ & $461.480$ & $15.010$ & $2.998$ \\
$^{186}$W  & $      $ & $ 0.248$ & $353.816$ & $12.975$ & $1.825$ & $458.905$ & $14.856$ & $3.037$ \\
$^{185}$Re & $ 5/2^+$ & $ 0.242$ & $375.291$ & $12.994$ & $1.701$ & $474.549$ & $14.973$ & $2.943$ \\
$^{187}$Re & $ 5/2^+$ & $ 0.233$ & $366.818$ & $13.000$ & $1.755$ & $476.865$ & $14.823$ & $2.945$ \\
$^{184}$Os & $      $ & $ 0.285$ & $401.209$ & $12.906$ & $1.603$ & $490.905$ & $15.121$ & $2.780$ \\
$^{186}$Os & $      $ & $ 0.261$ & $390.552$ & $12.982$ & $1.642$ & $496.049$ & $14.971$ & $2.782$ \\
$^{187}$Os & $ 1/2^-$ & $ 0.255$ & $388.113$ & $12.996$ & $1.645$ & $490.989$ & $14.918$ & $2.843$ \\
$^{188}$Os & $      $ & $ 0.234$ & $379.748$ & $13.032$ & $1.695$ & $509.270$ & $14.779$ & $2.718$ \\
$^{189}$Os & $ 9/2^-$ & $ 0.213$ & $379.520$ & $13.051$ & $1.700$ & $533.171$ & $14.601$ & $2.537$ \\
$^{190}$Os & $      $ & $ 0.191$ & $377.426$ & $13.081$ & $1.725$ & $543.474$ & $14.540$ & $2.499$ \\
$^{191}$Ir & $ 3/2^+$ & $ 0.170$ & $398.619$ & $13.101$ & $1.613$ & $564.693$ & $14.512$ & $2.444$ \\
$^{193}$Ir & $ 3/2^+$ & $ 0.154$ & $388.698$ & $13.127$ & $1.683$ & $574.046$ & $14.399$ & $2.408$ \\
$^{190}$Pt & $      $ & $ 0.239$ & $415.072$ & $13.008$ & $1.550$ & $536.375$ & $14.763$ & $2.593$ \\
$^{192}$Pt & $      $ & $-0.152$ & $242.076$ & $15.327$ & $3.097$ & $634.620$ & $13.513$ & $2.280$ \\
$^{194}$Pt & $      $ & $-0.144$ & $247.542$ & $15.217$ & $3.007$ & $642.378$ & $13.475$ & $2.251$ \\
$^{195}$Pt & $11/2^+$ & $ 0.136$ & $382.933$ & $13.161$ & $1.745$ & $592.794$ & $14.313$ & $2.349$ \\
$^{196}$Pt & $      $ & $-0.132$ & $253.232$ & $15.026$ & $2.913$ & $660.946$ & $13.442$ & $2.166$ \\
$^{197}$Au & $ 1/2^-$ & $-0.123$ & $259.121$ & $14.952$ & $2.864$ & $670.601$ & $13.486$ & $2.132$ \\
$^{199}$Hg & $ 1/2^+$ & $-0.112$ & $265.727$ & $14.793$ & $2.766$ & $706.701$ & $13.454$ & $2.001$ \\
\hline\hline
    \end{tabular*}
\end{table*}

\begin{table*}
    \centering
    \caption{}\label{table:appendix_a2}
    \begin{tabular*}{0.9\textwidth}{@{\extracolsep{\fill}}cccccc}
        \hline\hline
        Nuclide    & $K^\pi$  & $\beta_2$ & $\sigma_{\mathrm{r}}$ [mb] & $E_{\mathrm{r}}$ [MeV] & $\Gamma_{\mathrm{r}}$ [MeV] \\
        \hline
$^{ 40}$Ca & $      $ & $ 0.000$ & $ 119.632$ & $19.775$ & $4.441$\\
$^{ 42}$Ca & $      $ & $ 0.000$ & $ 119.192$ & $19.746$ & $4.805$\\
$^{ 43}$Ca & $ 7/2^-$ & $-0.011$ & $ 117.661$ & $19.623$ & $4.946$\\
$^{ 44}$Ca & $      $ & $ 0.000$ & $ 123.158$ & $19.665$ & $4.841$\\
$^{ 46}$Ca & $      $ & $ 0.000$ & $ 129.572$ & $19.650$ & $4.680$\\
$^{ 50}$Ti & $      $ & $ 0.000$ & $ 134.205$ & $19.954$ & $5.062$\\
$^{ 51}$V  & $ 7/2^-$ & $-0.021$ & $ 129.582$ & $19.981$ & $5.406$\\
$^{ 52}$Cr & $      $ & $ 0.000$ & $ 134.116$ & $20.153$ & $5.470$\\
$^{ 54}$Fe & $      $ & $ 0.000$ & $ 135.706$ & $20.375$ & $5.726$\\
$^{ 58}$Ni & $      $ & $ 0.000$ & $ 168.171$ & $19.696$ & $4.512$\\
$^{ 70}$Zn & $      $ & $ 0.000$ & $ 226.368$ & $17.961$ & $3.958$\\
$^{ 71}$Ga & $ 1/2^-$ & $ 0.008$ & $ 221.458$ & $17.852$ & $4.196$\\
$^{ 72}$Ge & $      $ & $ 0.000$ & $ 228.933$ & $17.810$ & $4.192$\\
$^{ 78}$Kr & $      $ & $ 0.000$ & $ 313.951$ & $16.980$ & $2.865$\\
$^{ 84}$Kr & $      $ & $-0.038$ & $ 364.415$ & $16.851$ & $2.332$\\
$^{ 86}$Kr & $      $ & $ 0.000$ & $ 376.517$ & $16.829$ & $2.208$\\
$^{ 85}$Rb & $ 3/2^-$ & $ 0.074$ & $ 365.175$ & $16.772$ & $2.360$\\
$^{ 87}$Rb & $ 1/2^-$ & $-0.007$ & $ 423.468$ & $16.744$ & $1.868$\\
$^{ 84}$Sr & $      $ & $ 0.000$ & $ 462.961$ & $16.688$ & $1.695$\\
$^{ 86}$Sr & $      $ & $ 0.000$ & $ 515.138$ & $16.724$ & $1.458$\\
$^{ 87}$Sr & $ 9/2^+$ & $ 0.048$ & $ 446.622$ & $16.757$ & $1.802$\\
$^{ 88}$Sr & $      $ & $ 0.000$ & $ 507.550$ & $16.767$ & $1.511$\\
$^{ 89}$Y  & $ 1/2^-$ & $ 0.000$ & $ 471.661$ & $16.721$ & $1.792$\\
$^{ 90}$Zr & $      $ & $ 0.000$ & $ 546.350$ & $16.778$ & $1.501$\\
$^{ 91}$Zr & $ 5/2^+$ & $-0.028$ & $ 528.020$ & $16.773$ & $1.643$\\
$^{ 92}$Zr & $      $ & $ 0.000$ & $ 581.884$ & $16.694$ & $1.457$\\
$^{ 93}$Nb & $ 9/2^+$ & $-0.051$ & $ 501.035$ & $16.639$ & $1.836$\\
$^{ 92}$Mo & $      $ & $ 0.000$ & $ 538.791$ & $16.832$ & $1.634$\\
$^{ 94}$Mo & $      $ & $ 0.000$ & $ 575.301$ & $16.752$ & $1.568$\\
$^{ 96}$Ru & $      $ & $ 0.000$ & $ 536.466$ & $16.791$ & $1.804$\\
$^{114}$Cd & $      $ & $ 0.000$ & $ 496.495$ & $15.773$ & $2.538$\\
$^{116}$Cd & $      $ & $ 0.000$ & $ 504.684$ & $15.674$ & $2.528$\\
$^{115}$In & $ 9/2^+$ & $ 0.065$ & $ 450.177$ & $15.871$ & $2.959$\\
$^{112}$Sn & $      $ & $ 0.000$ & $ 493.140$ & $15.984$ & $2.493$\\
$^{114}$Sn & $      $ & $ 0.000$ & $ 530.742$ & $15.802$ & $2.237$\\
$^{115}$Sn & $ 1/2^+$ & $ 0.024$ & $ 494.652$ & $15.763$ & $2.543$\\
$^{116}$Sn & $      $ & $ 0.000$ & $ 538.329$ & $15.724$ & $2.252$\\
$^{117}$Sn & $ 1/2^+$ & $ 0.000$ & $ 540.537$ & $15.616$ & $2.248$\\
$^{118}$Sn & $      $ & $ 0.000$ & $ 527.279$ & $15.676$ & $2.402$\\
$^{119}$Sn & $ 1/2^+$ & $ 0.000$ & $ 523.372$ & $15.602$ & $2.438$\\
$^{120}$Sn & $      $ & $ 0.000$ & $ 521.566$ & $15.608$ & $2.495$\\
$^{122}$Sn & $      $ & $ 0.000$ & $ 523.101$ & $15.523$ & $2.511$\\
$^{124}$Sn & $      $ & $ 0.000$ & $ 527.975$ & $15.432$ & $2.490$\\
$^{128}$Te & $      $ & $-0.076$ & $ 507.607$ & $15.424$ & $2.781$\\
$^{130}$Te & $      $ & $ 0.000$ & $ 578.041$ & $15.253$ & $2.307$\\
$^{132}$Xe & $      $ & $-0.064$ & $ 559.261$ & $15.315$ & $2.541$\\
$^{134}$Xe & $      $ & $ 0.000$ & $ 612.053$ & $15.200$ & $2.235$\\
$^{136}$Xe & $      $ & $ 0.000$ & $ 611.991$ & $15.140$ & $2.233$\\
\hline\hline
    \end{tabular*}
\end{table*}

\begin{table*}\ContinuedFloat
    \centering
    \caption{}
    \begin{tabular*}{0.9\textwidth}{@{\extracolsep{\fill}}cccccc}
        \hline\hline
        Nuclide    & $K^\pi$ & $\beta_2$ & $\sigma_{\mathrm{r}}$ [mb] & $E_{\mathrm{r}}$ [MeV] & $\Gamma_{\mathrm{r}}$ [MeV]\\
        \hline
$^{134}$Ba & $      $ & $-0.075$ & $ 581.504$ & $15.291$ & $2.494$\\
$^{136}$Ba & $      $ & $ 0.000$ & $ 639.862$ & $15.209$ & $2.195$\\
$^{137}$Ba & $11/2^-$ & $ 0.046$ & $ 598.219$ & $15.229$ & $2.401$\\
$^{138}$Ba & $      $ & $ 0.000$ & $ 633.212$ & $15.163$ & $2.231$\\
$^{139}$La & $ 7/2^+$ & $ 0.021$ & $ 621.245$ & $15.136$ & $2.305$\\
$^{136}$Ce & $      $ & $ 0.000$ & $ 680.097$ & $15.221$ & $2.082$\\
$^{138}$Ce & $      $ & $ 0.000$ & $ 662.358$ & $15.188$ & $2.182$\\
$^{140}$Ce & $      $ & $ 0.000$ & $ 647.740$ & $15.158$ & $2.265$\\
$^{142}$Ce & $      $ & $ 0.000$ & $ 644.994$ & $15.007$ & $2.290$\\
$^{141}$Pr & $ 5/2^+$ & $-0.014$ & $ 661.258$ & $15.094$ & $2.227$\\
$^{142}$Nd & $      $ & $ 0.000$ & $ 697.755$ & $15.056$ & $2.128$\\
$^{143}$Nd & $ 9/2^-$ & $-0.039$ & $ 659.074$ & $15.016$ & $2.305$\\
$^{144}$Nd & $      $ & $ 0.000$ & $ 728.050$ & $14.911$ & $1.996$\\
$^{144}$Sm & $      $ & $ 0.000$ & $ 799.827$ & $14.969$ & $1.792$\\
$^{198}$Pt & $      $ & $ 0.093$ & $ 896.527$ & $13.796$ & $2.378$\\
$^{196}$Hg & $      $ & $ 0.098$ & $ 894.807$ & $13.936$ & $2.424$\\
$^{198}$Hg & $      $ & $ 0.079$ & $ 932.180$ & $13.851$ & $2.305$\\
$^{200}$Hg & $      $ & $ 0.000$ & $ 973.265$ & $13.768$ & $2.201$\\
$^{201}$Hg & $ 3/2^-$ & $-0.092$ & $ 889.227$ & $13.748$ & $2.450$\\
$^{202}$Hg & $      $ & $ 0.000$ & $1004.102$ & $13.691$ & $2.114$\\
$^{204}$Hg & $      $ & $ 0.000$ & $1054.882$ & $13.617$ & $1.981$\\
$^{203}$Tl & $ 1/2^+$ & $ 0.000$ & $ 993.166$ & $13.704$ & $2.162$\\
$^{205}$Tl & $ 1/2^+$ & $ 0.000$ & $1026.260$ & $13.640$ & $2.067$\\
$^{204}$Pb & $      $ & $ 0.000$ & $1091.800$ & $13.673$ & $1.954$\\
$^{206}$Pb & $      $ & $ 0.000$ & $1106.677$ & $13.626$ & $1.924$\\
$^{207}$Pb & $ 1/2^-$ & $ 0.000$ & $1080.516$ & $13.608$ & $1.981$\\
$^{208}$Pb & $      $ & $ 0.000$ & $1123.169$ & $13.552$ & $1.891$\\
$^{209}$Bi & $ 9/2^-$ & $-0.011$ & $1070.519$ & $13.626$ & $2.021$\\
        \hline\hline
    \end{tabular*}
\end{table*}

\section{Photoabsorption cross sections obtained from QFAM calculations}
\label{sec:appendix_b}
Figs.~\ref{fig:appendix_b} present the total photoabsorption cross sections 
and the contributions from $K=0$ and $K=1$ components, given by QFAM calculations, 
for all nuclei considered in this work.
In the figures, the cross sections given by QFAM calculations are shown by solid lines, 
while the fitted results are shown by dashed lines.
The red lines represent oscillations parallel to the symmetry axis 
induced by an external dipole field with $K=0$, 
while the blue lines represent oscillations perpendicular to the symmetry axis 
induced by an external dipole field with $K=1$.
The black lines indicate the total cross section.
Additionally, the corresponding GDR parameters are provided for both oscillation modes:
peak cross sections ($\sigma_{\mathrm{r},\iparallel}$ and $\sigma_{\mathrm{r},\perp}$), 
resonance energies ($E_{\mathrm{r},\iparallel}$ and $E_{\mathrm{r},\perp}$), 
and resonance widths ($\Gamma_{\mathrm{r},\iparallel}$ and $\Gamma_{\mathrm{r},\perp}$).
The GDR parameters for total cross sections are also included, showing 
peak cross sections $\sigma_{\mathrm{r}}$, resonance energies $E_{\mathrm{r}}$, 
and resonance widths $\Gamma_{\mathrm{r}}$.
The deformation $\beta_2$ and the total energies $E_\mathrm{tot}$ are also listed in the figures for reference.

\begin{figure*}
    \centering
    \includegraphics[width=0.4\textwidth]{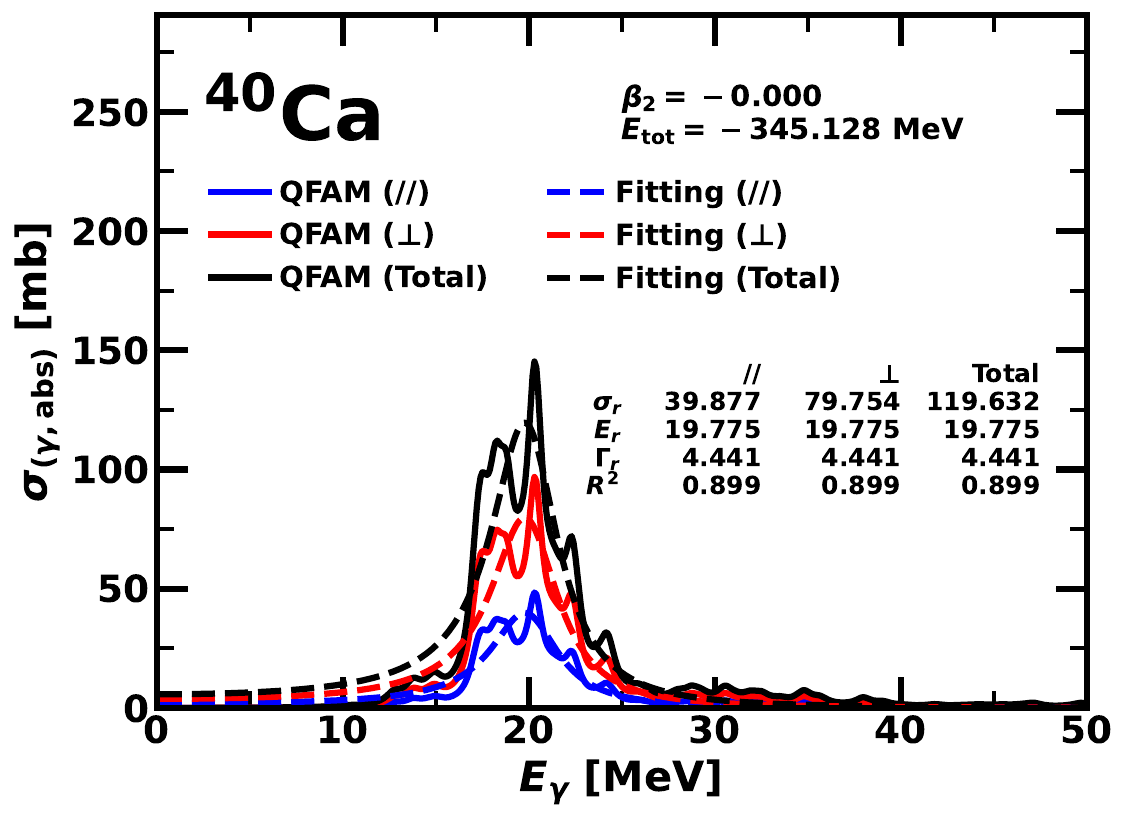}
    \includegraphics[width=0.4\textwidth]{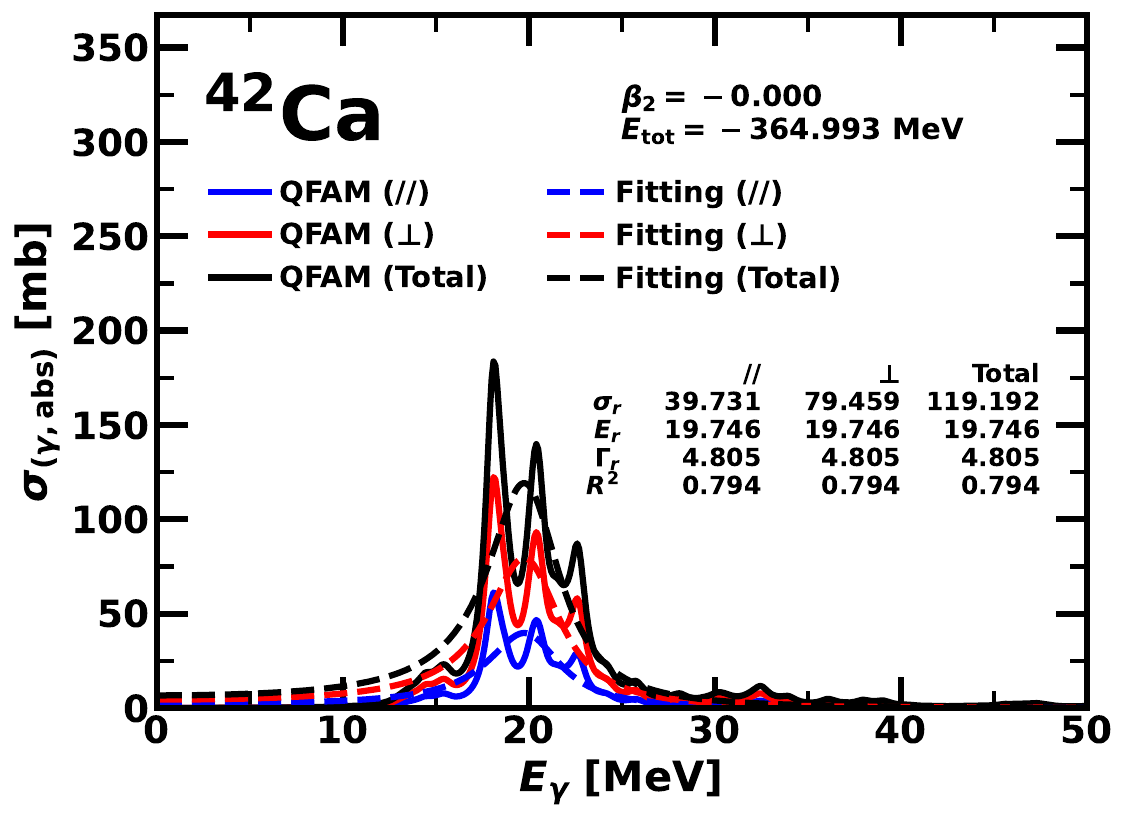}
    \includegraphics[width=0.4\textwidth]{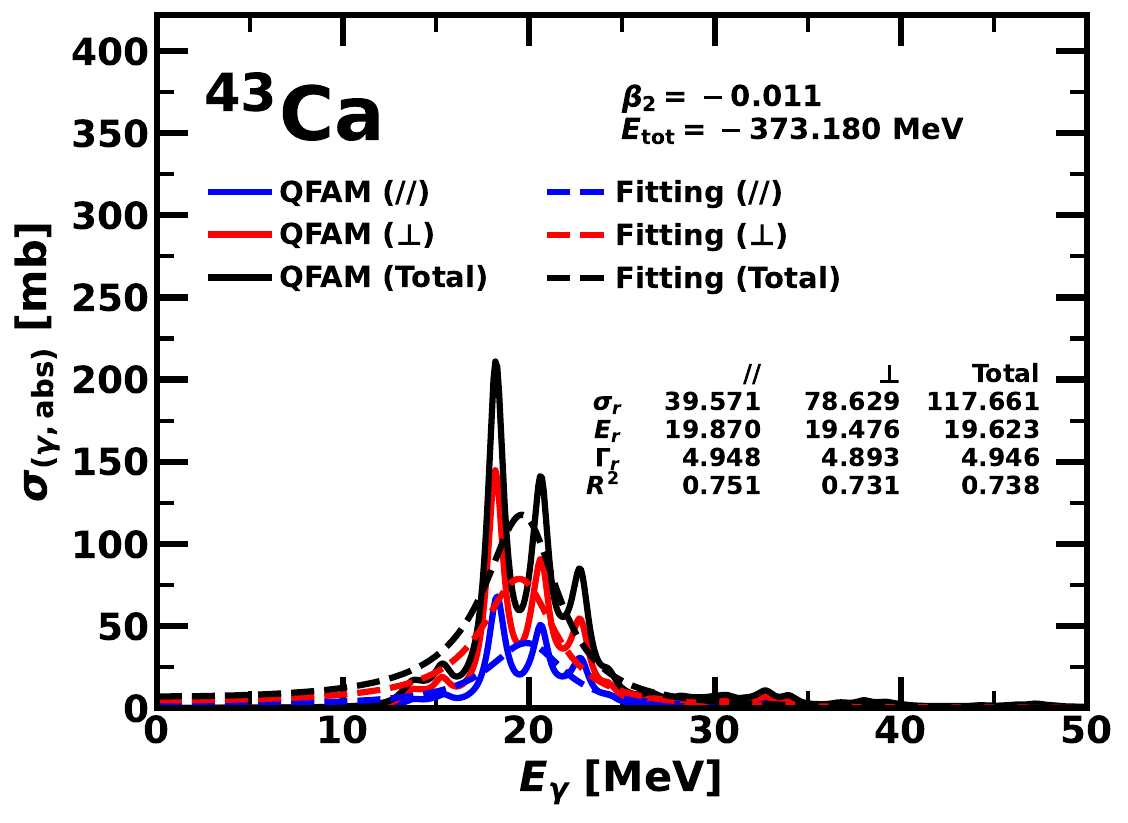}
    \includegraphics[width=0.4\textwidth]{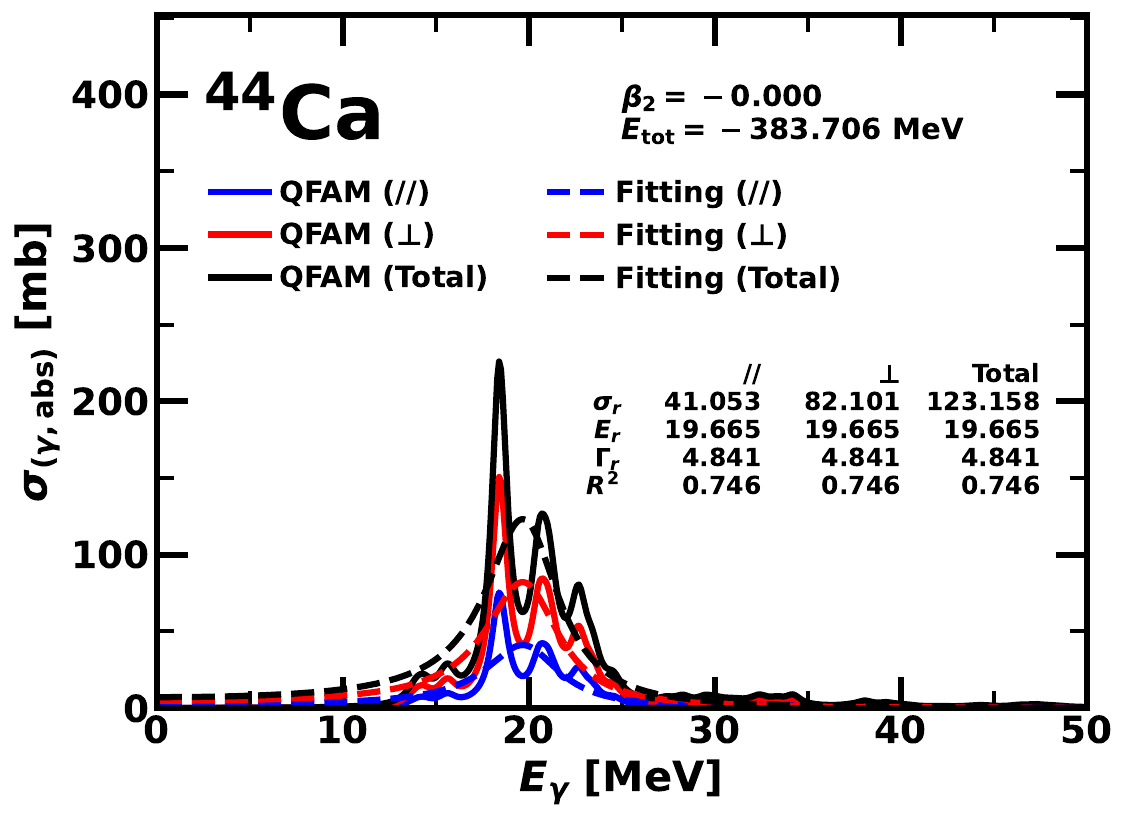}
    \includegraphics[width=0.4\textwidth]{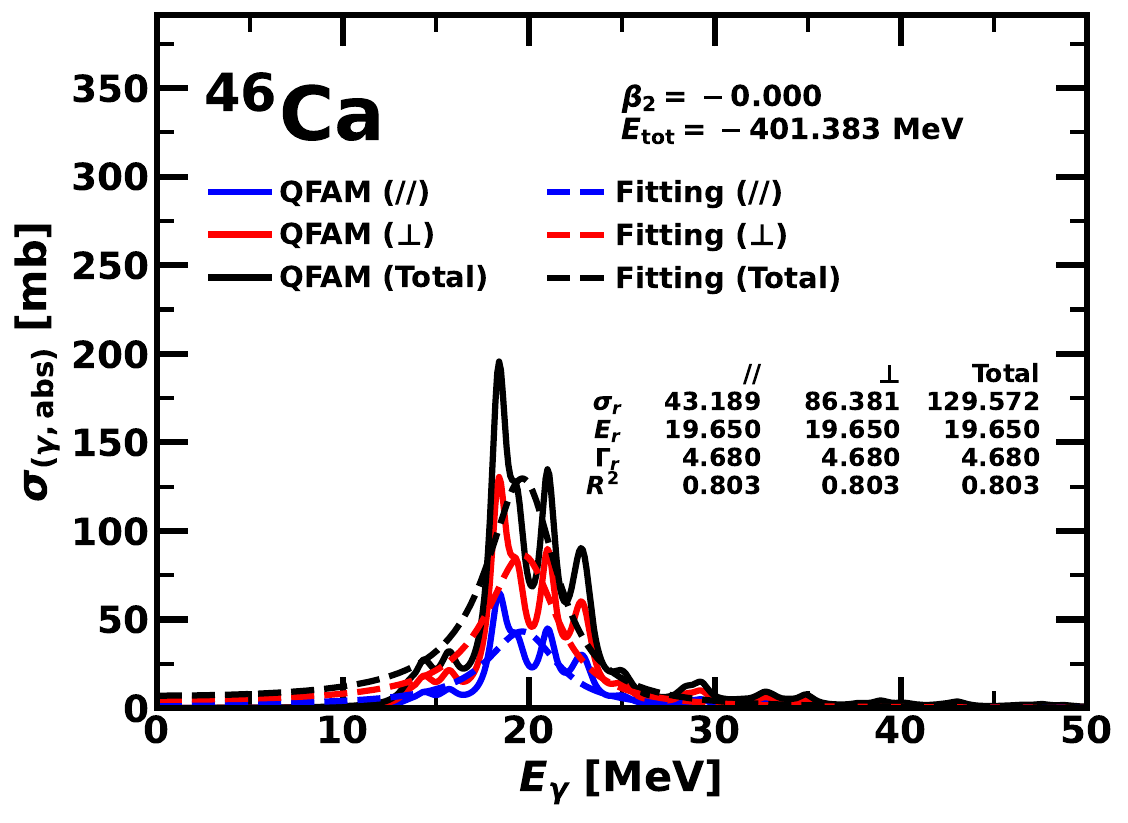}
    \includegraphics[width=0.4\textwidth]{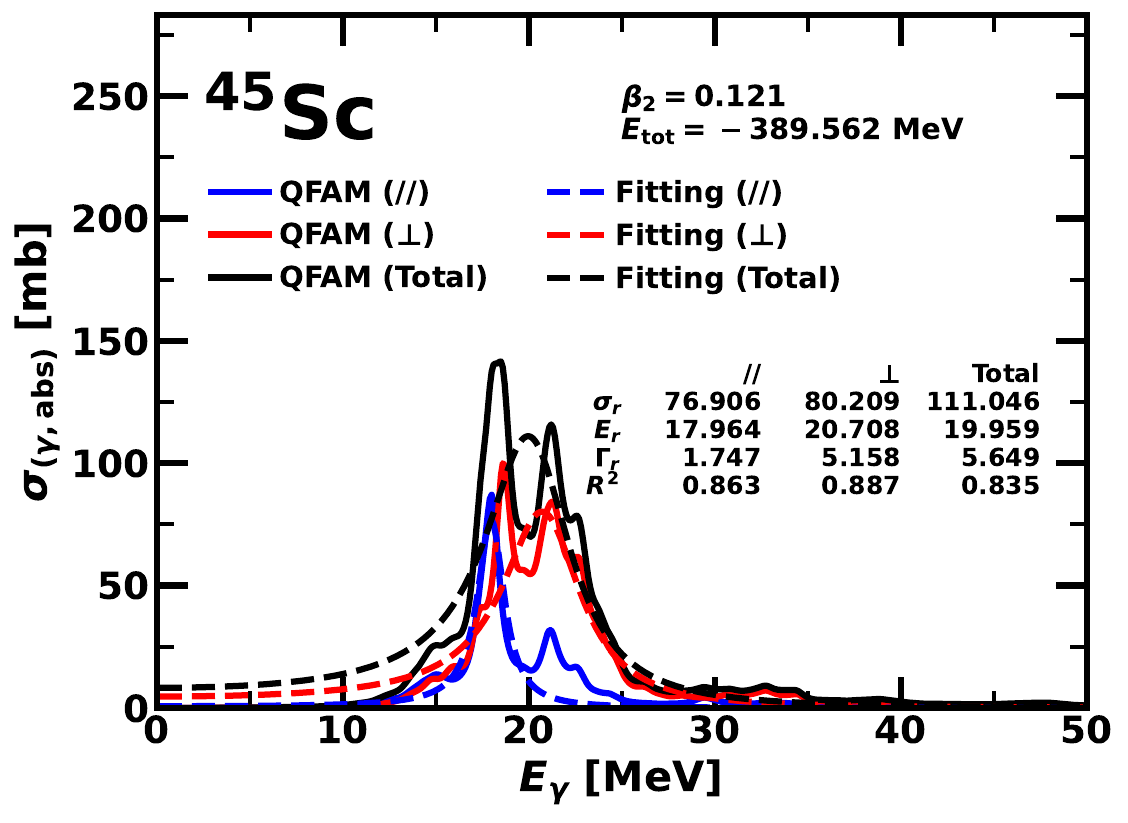}
    \includegraphics[width=0.4\textwidth]{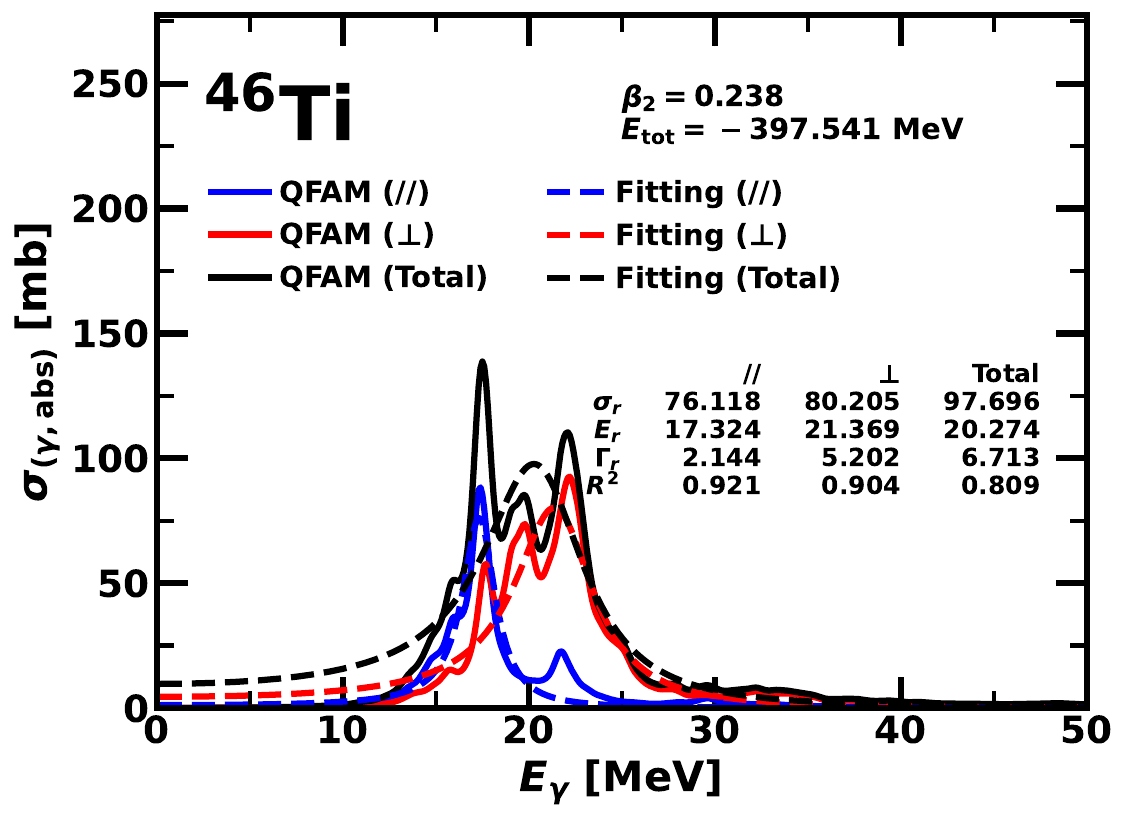}
    \includegraphics[width=0.4\textwidth]{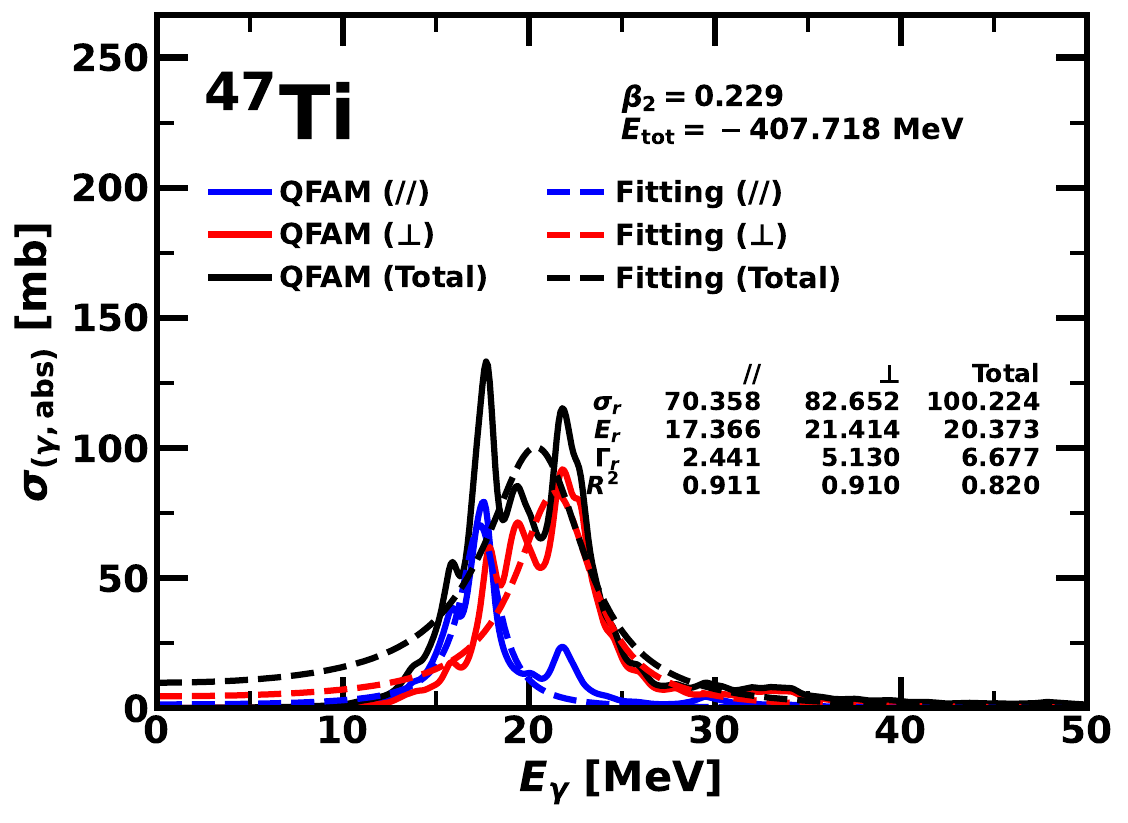}
    \caption{}
    \label{fig:appendix_b}
\end{figure*}
\begin{figure*}\ContinuedFloat
    \centering
    \includegraphics[width=0.4\textwidth]{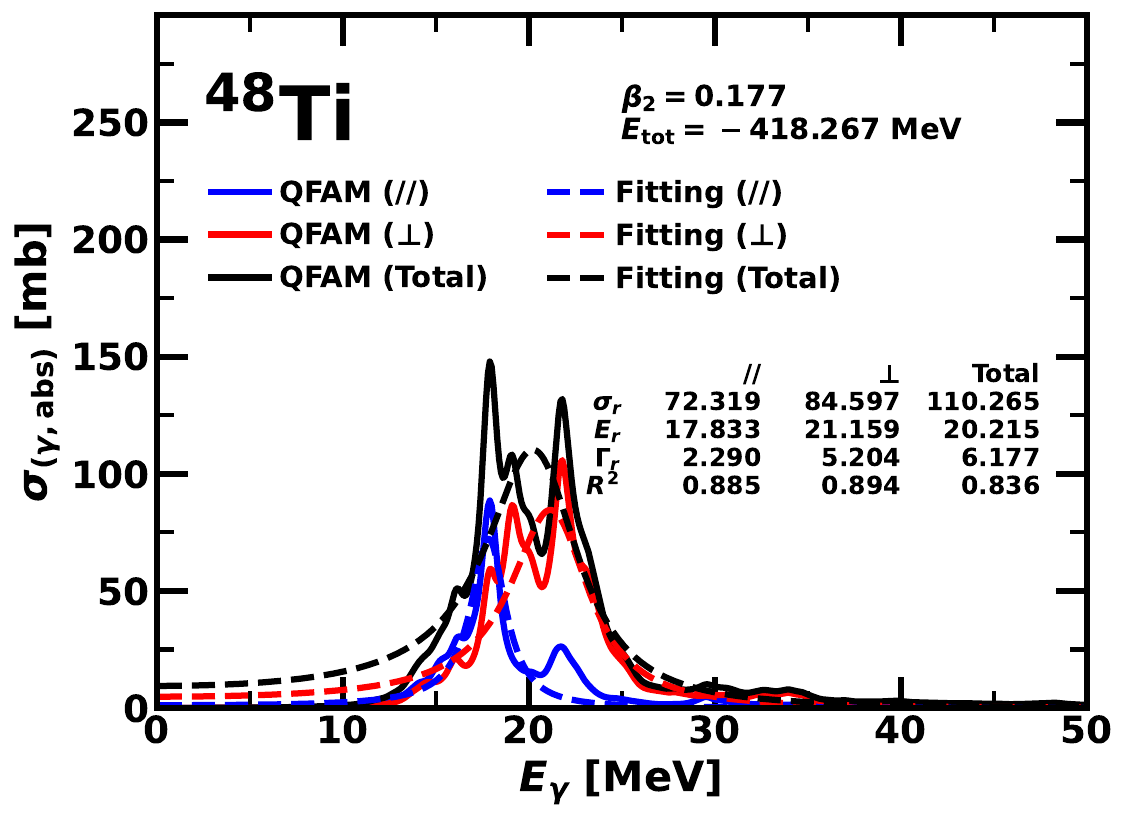}
    \includegraphics[width=0.4\textwidth]{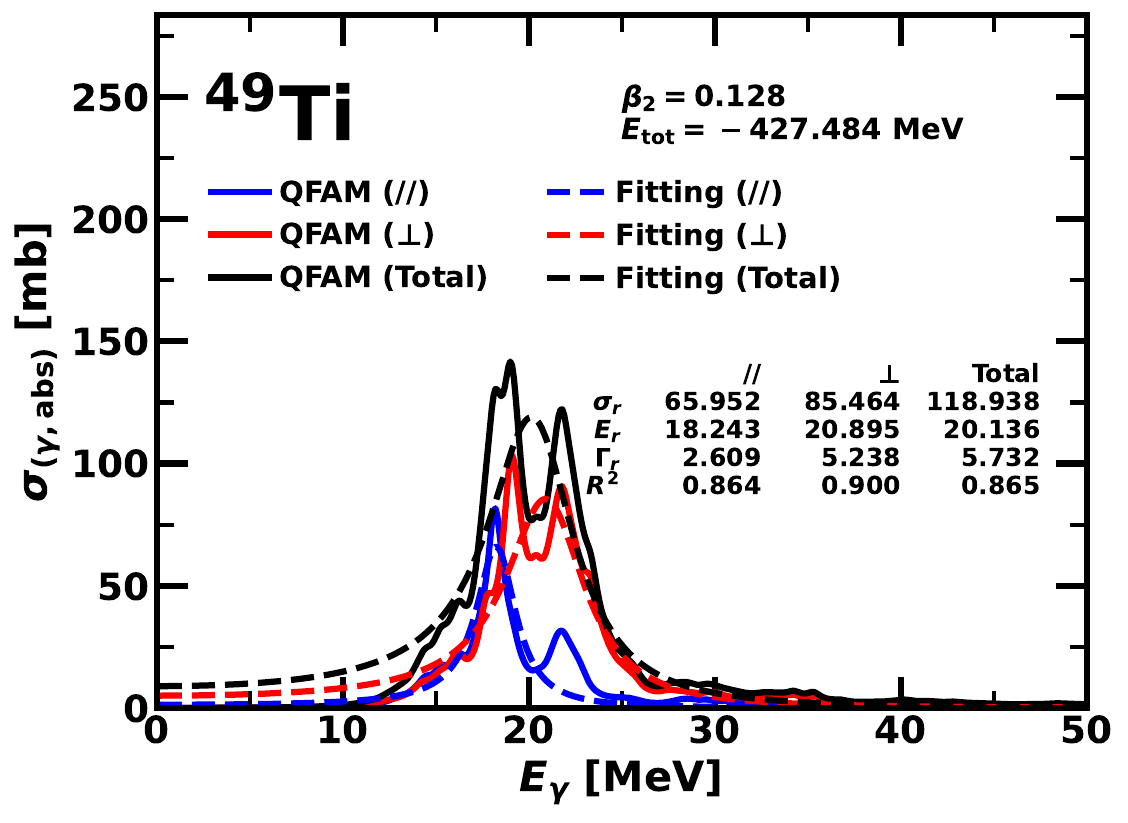}
    \includegraphics[width=0.4\textwidth]{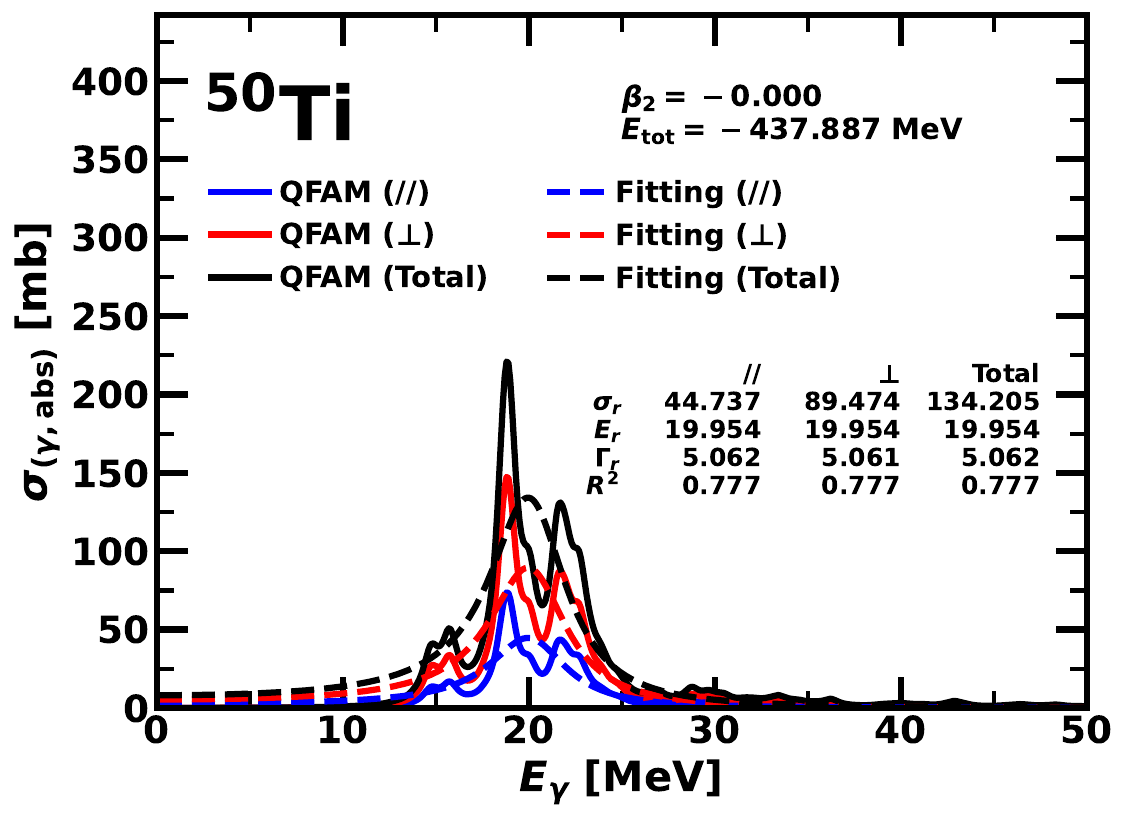}
    \includegraphics[width=0.4\textwidth]{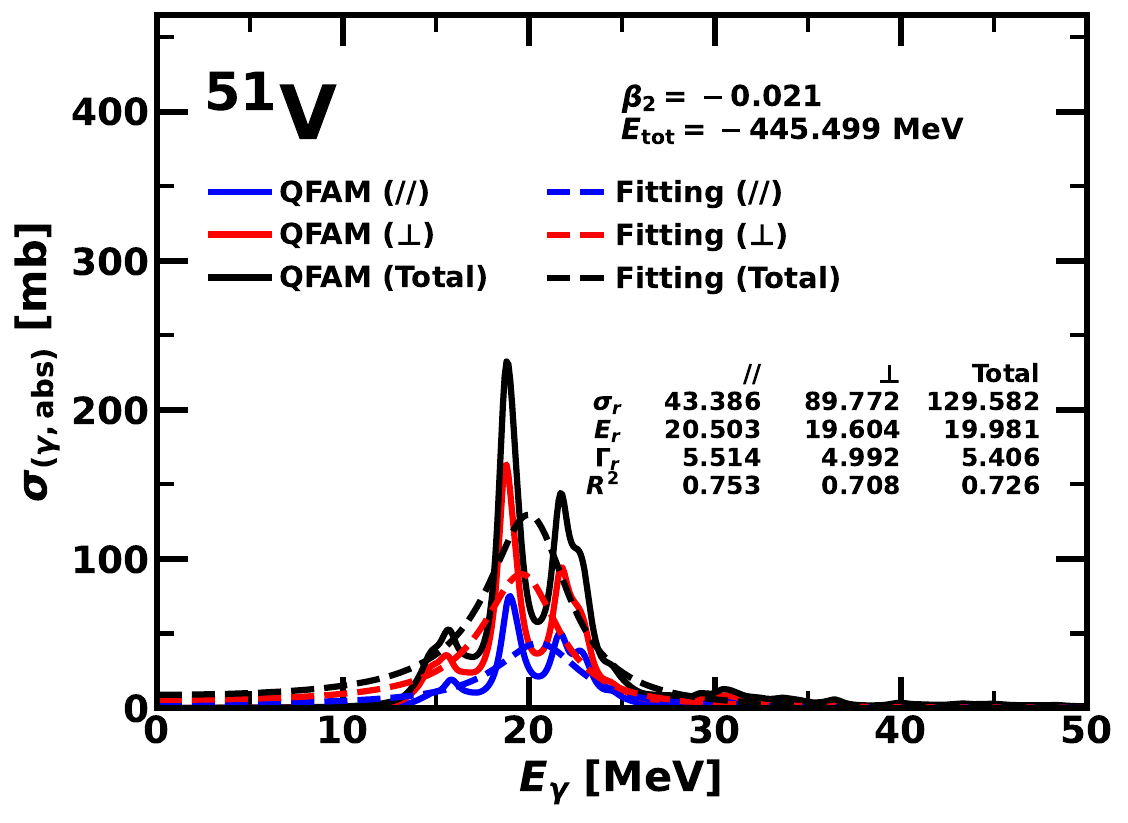}
    \includegraphics[width=0.4\textwidth]{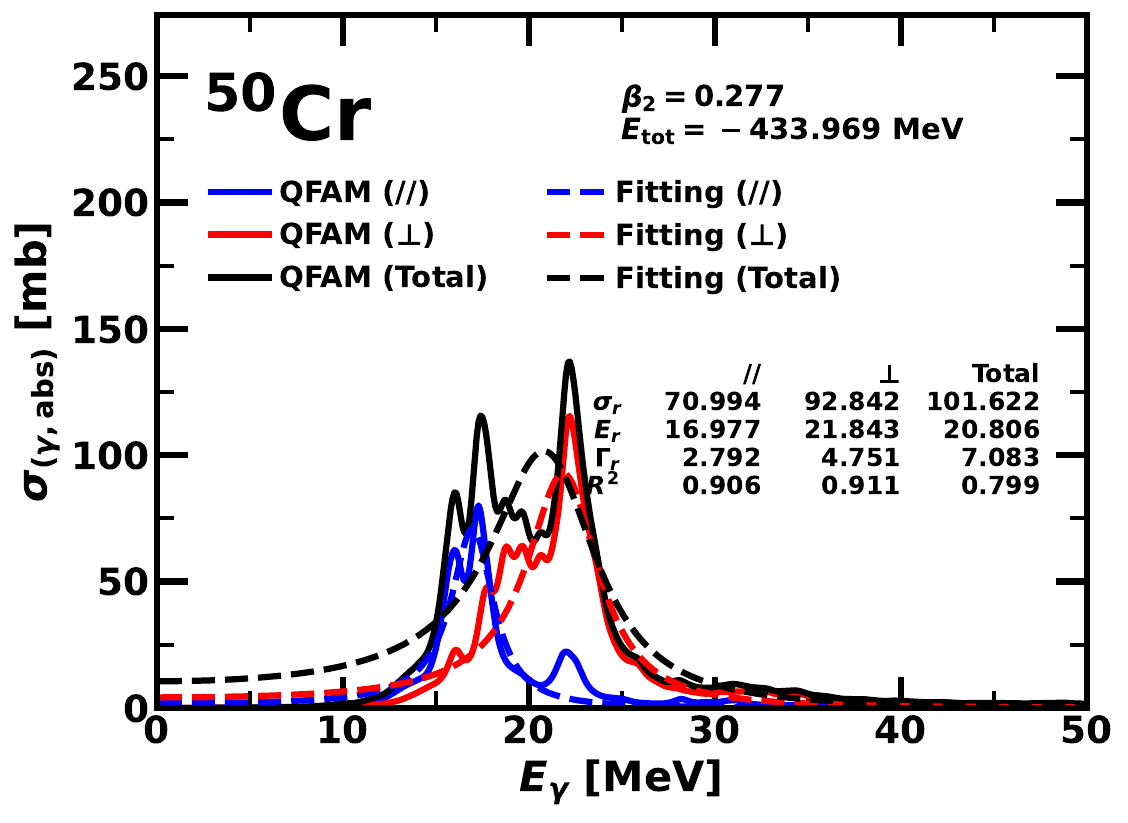}
    \includegraphics[width=0.4\textwidth]{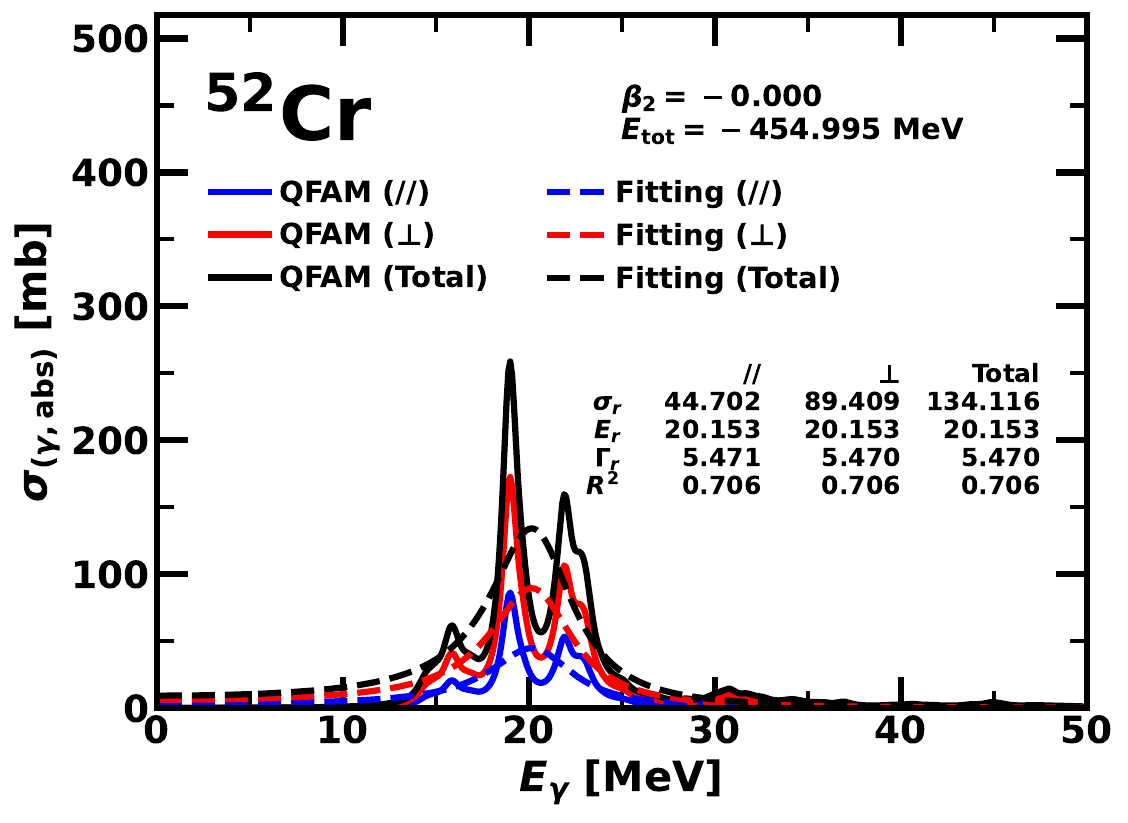}
    \includegraphics[width=0.4\textwidth]{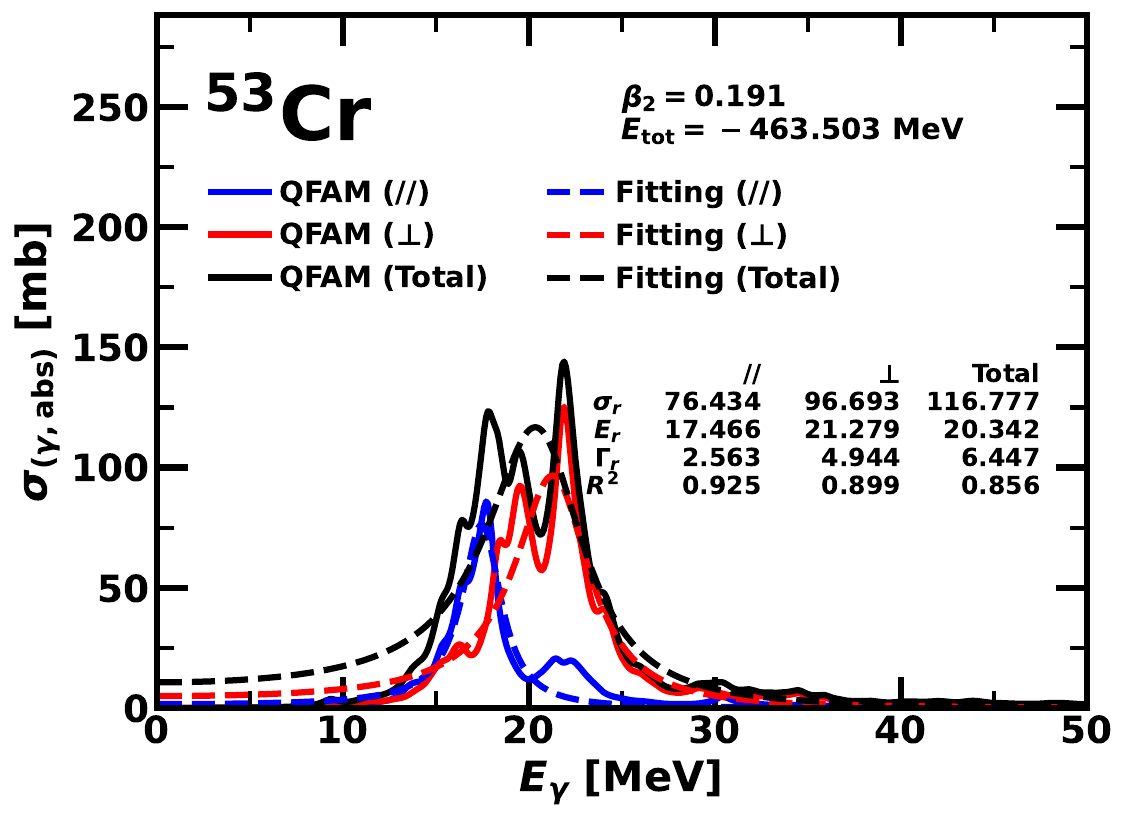}
    \includegraphics[width=0.4\textwidth]{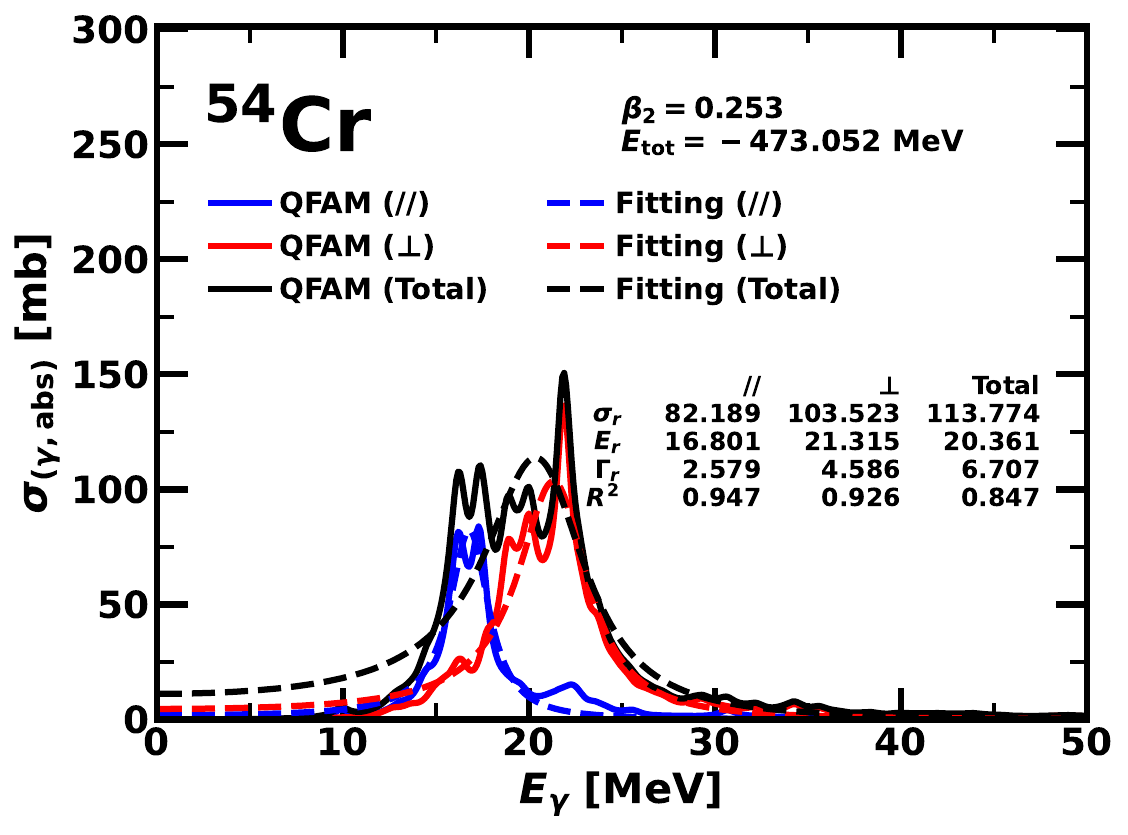}
\end{figure*}
\begin{figure*}\ContinuedFloat
    \centering
    \includegraphics[width=0.4\textwidth]{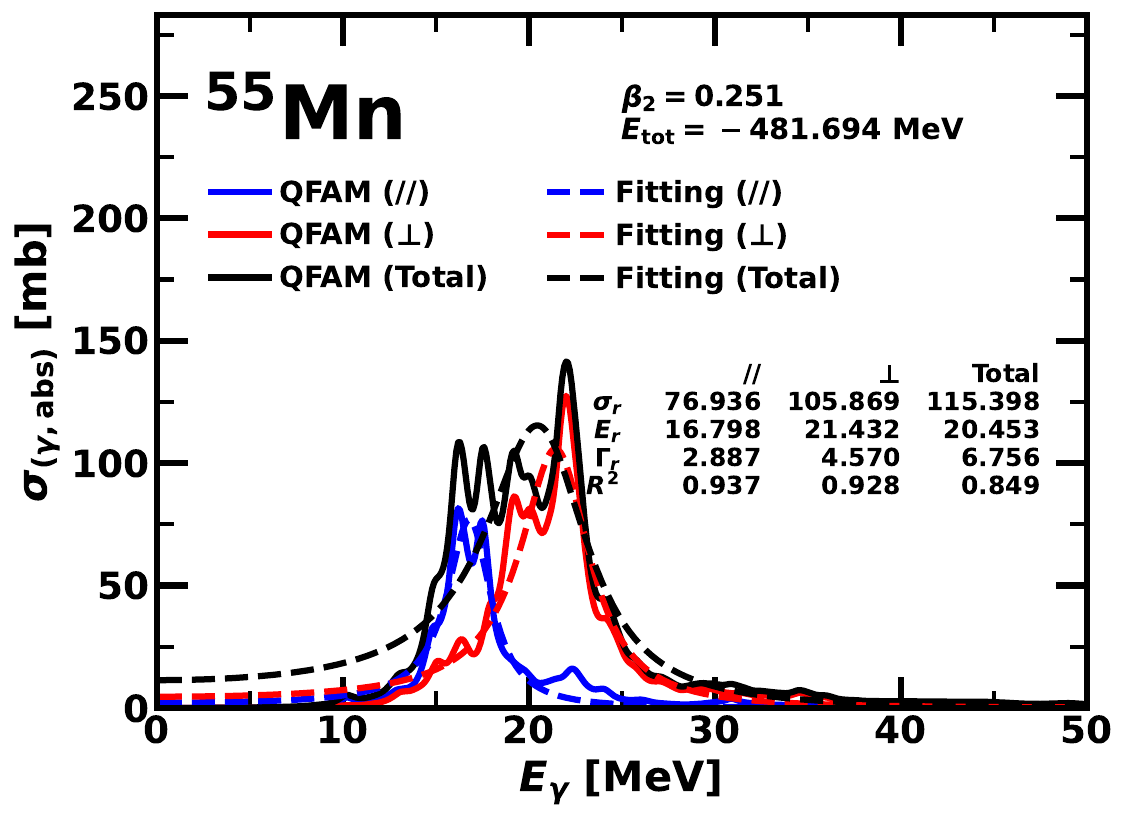}
    \includegraphics[width=0.4\textwidth]{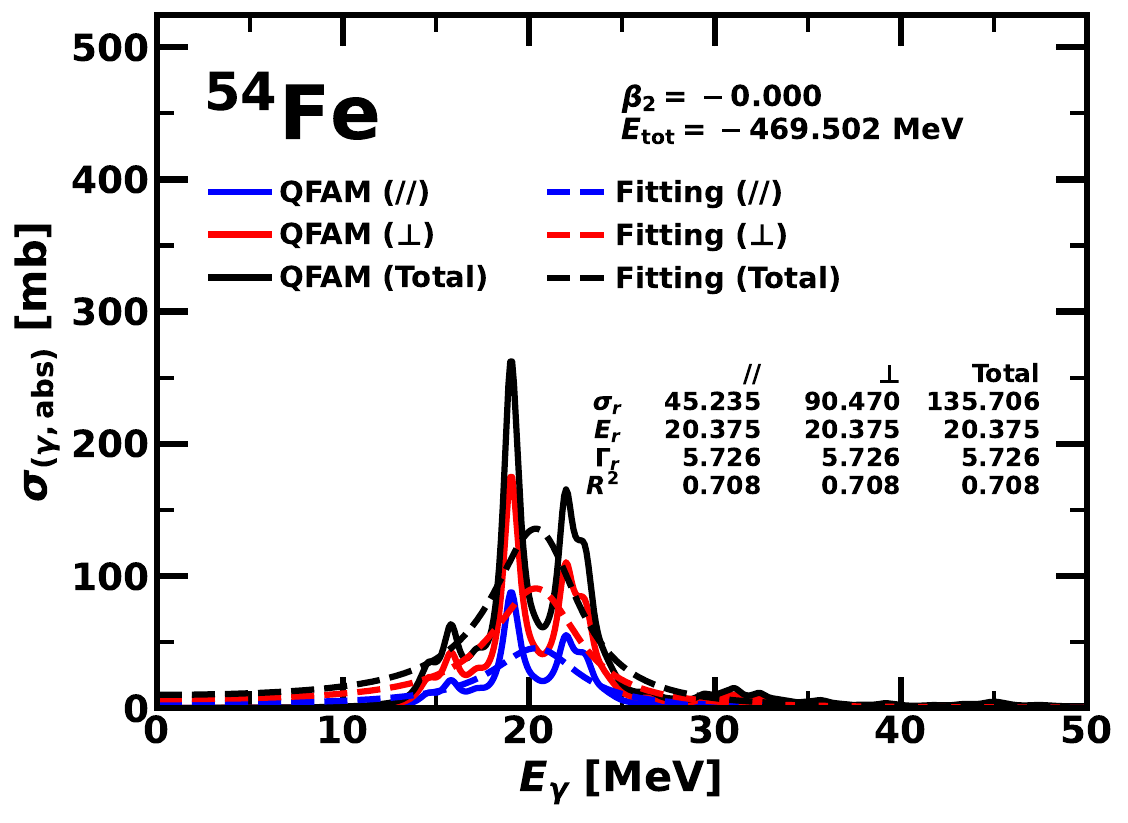}
    \includegraphics[width=0.4\textwidth]{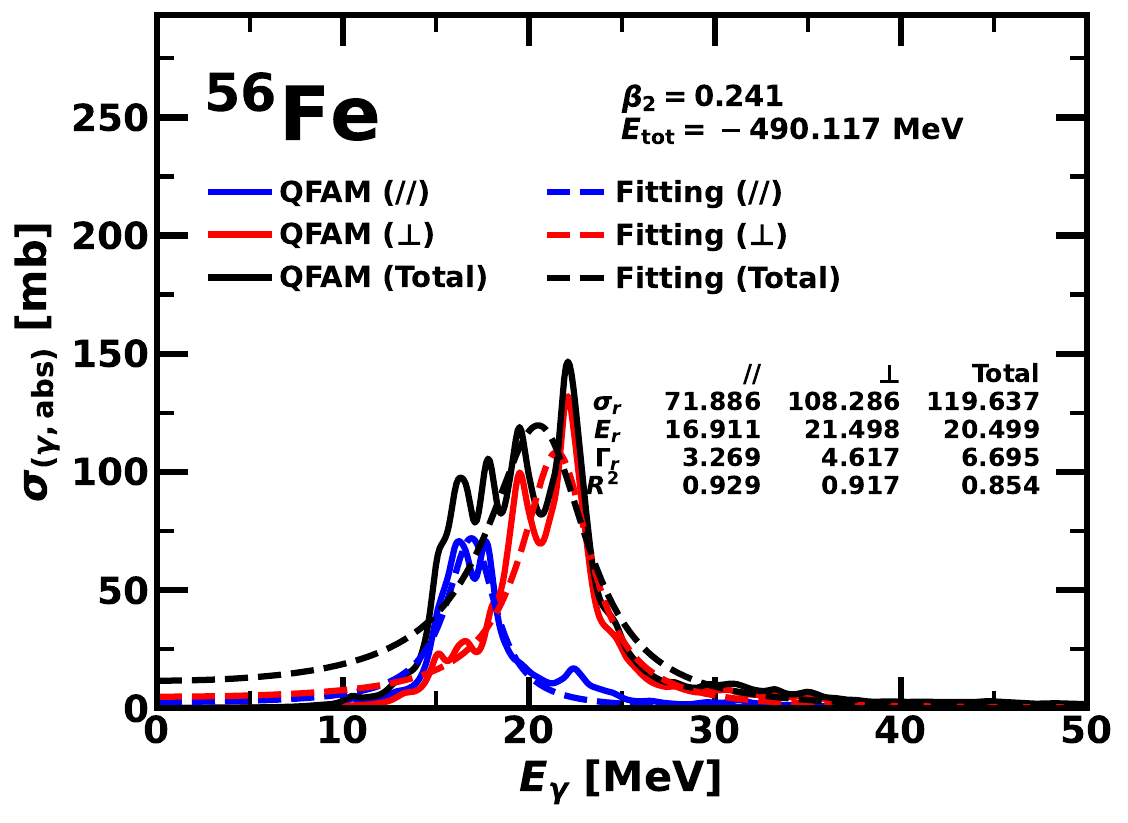}
    \includegraphics[width=0.4\textwidth]{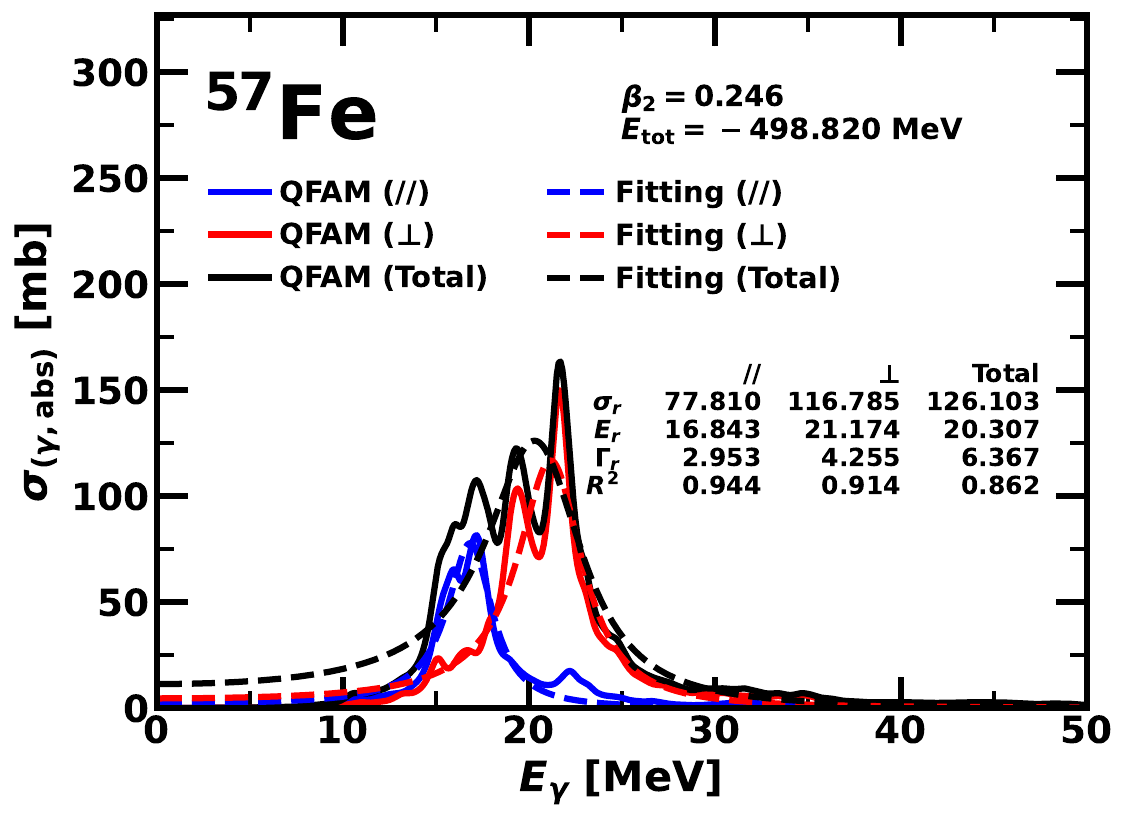}
    \includegraphics[width=0.4\textwidth]{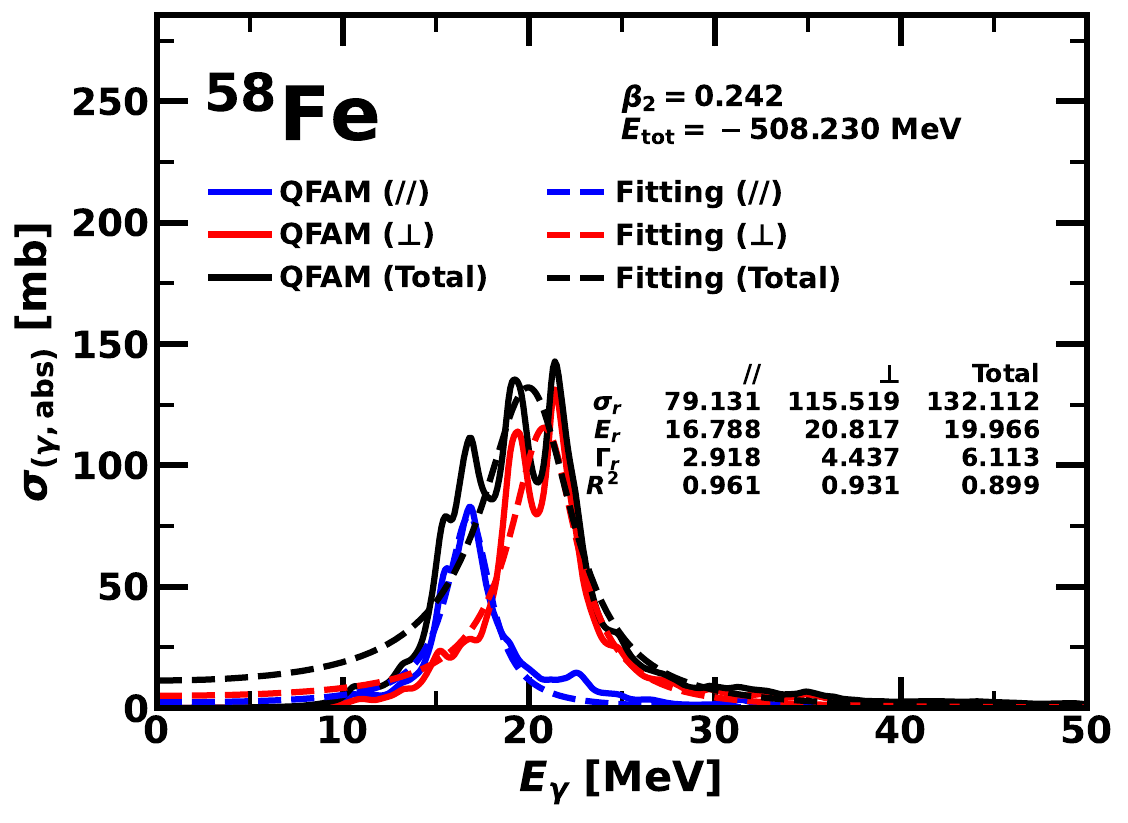}
    \includegraphics[width=0.4\textwidth]{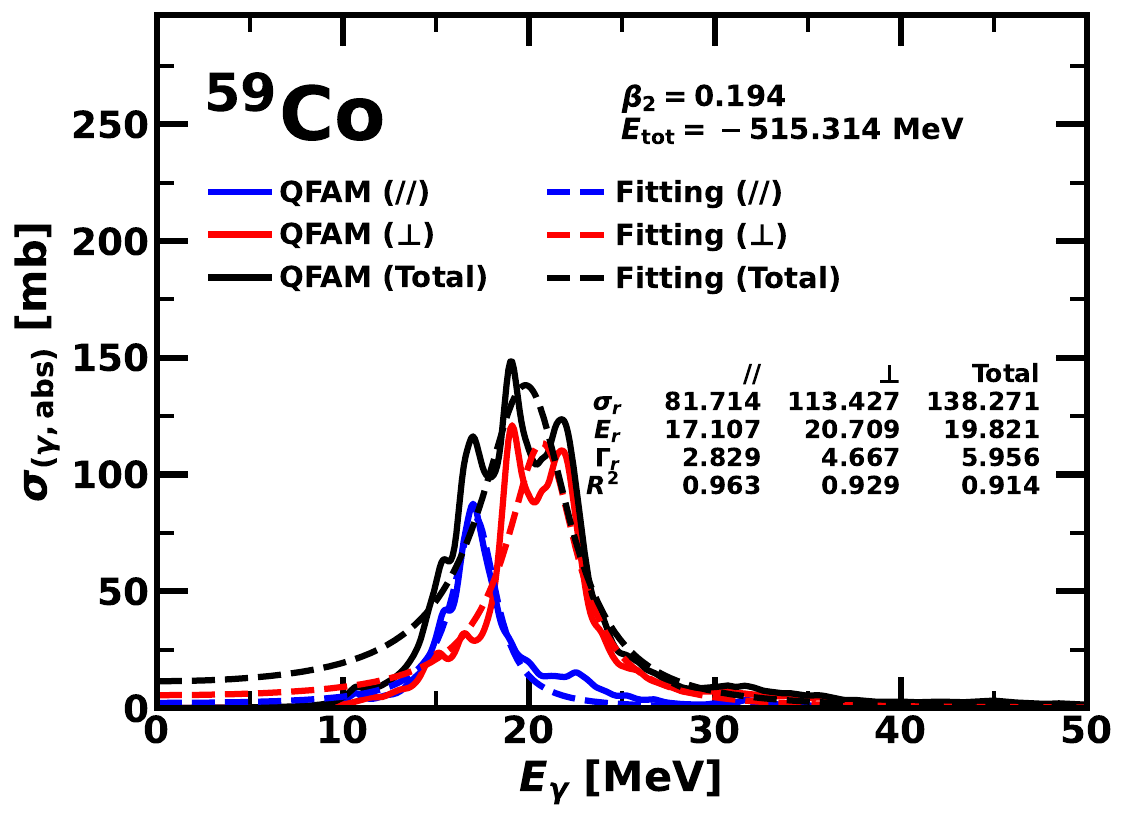}
    \includegraphics[width=0.4\textwidth]{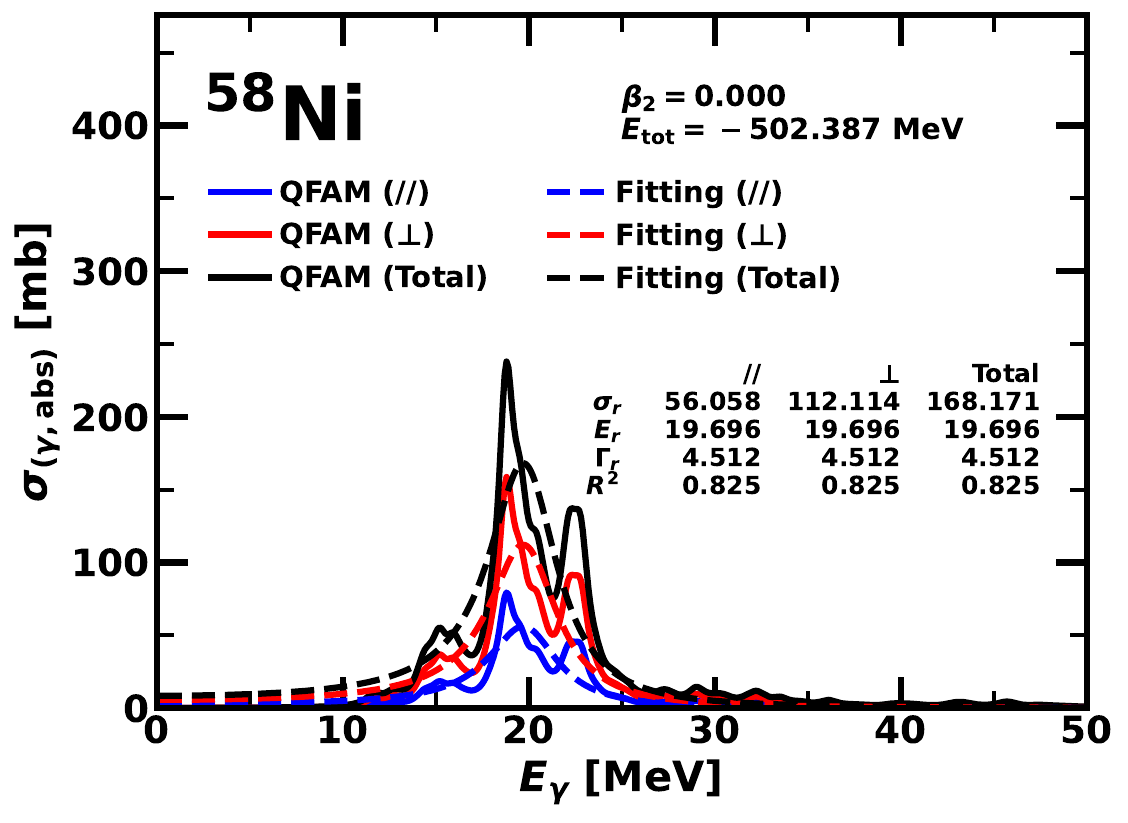}
    \includegraphics[width=0.4\textwidth]{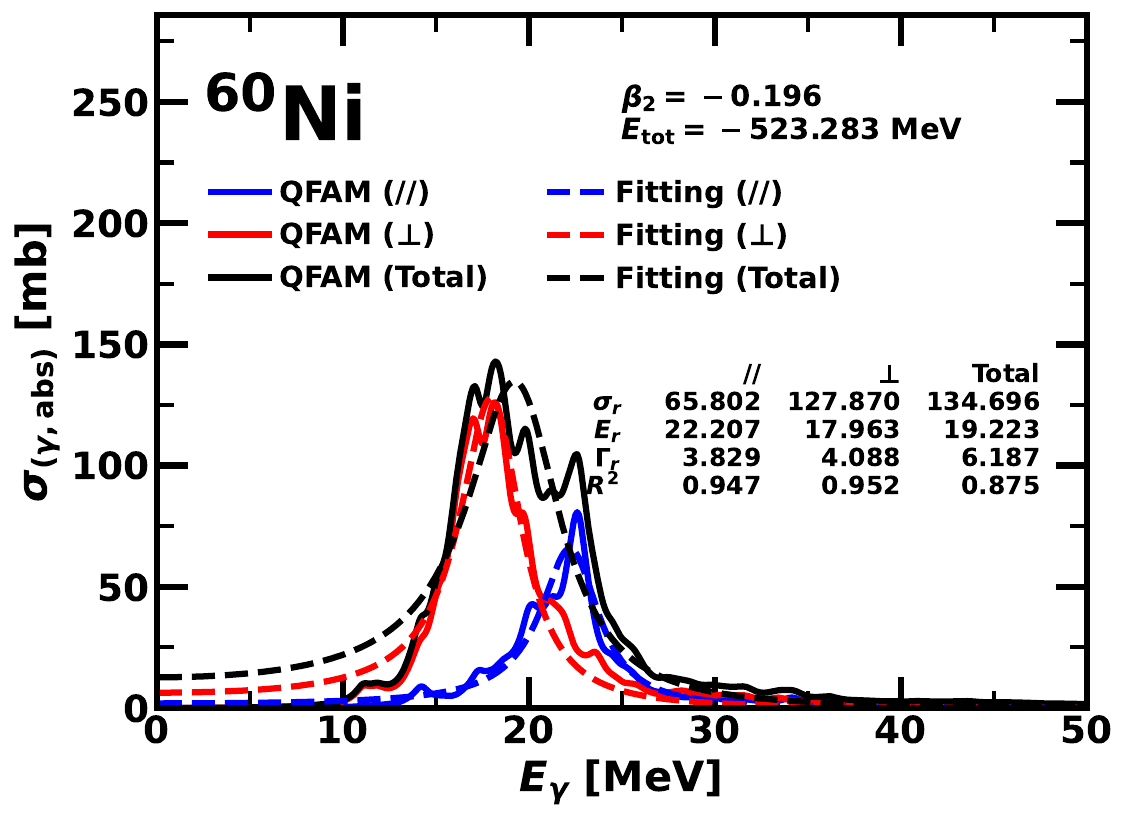}
\end{figure*}
\begin{figure*}\ContinuedFloat
    \centering
    \includegraphics[width=0.4\textwidth]{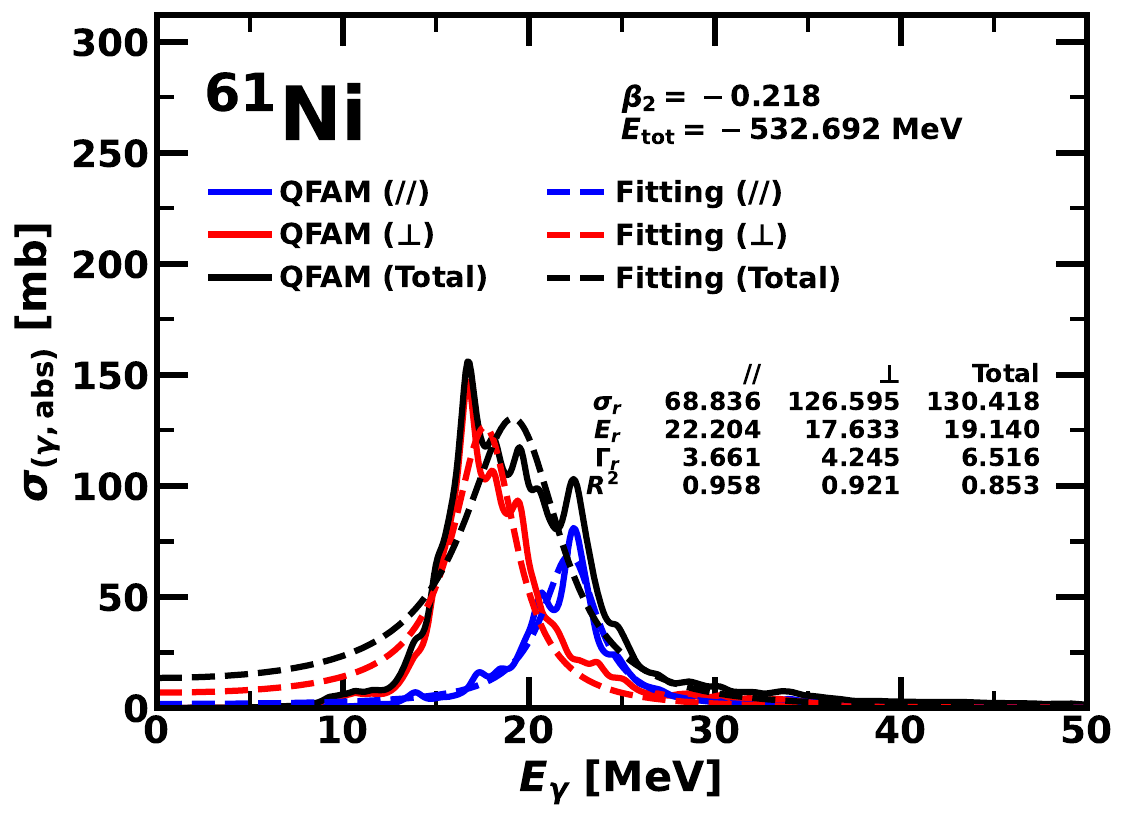}
    \includegraphics[width=0.4\textwidth]{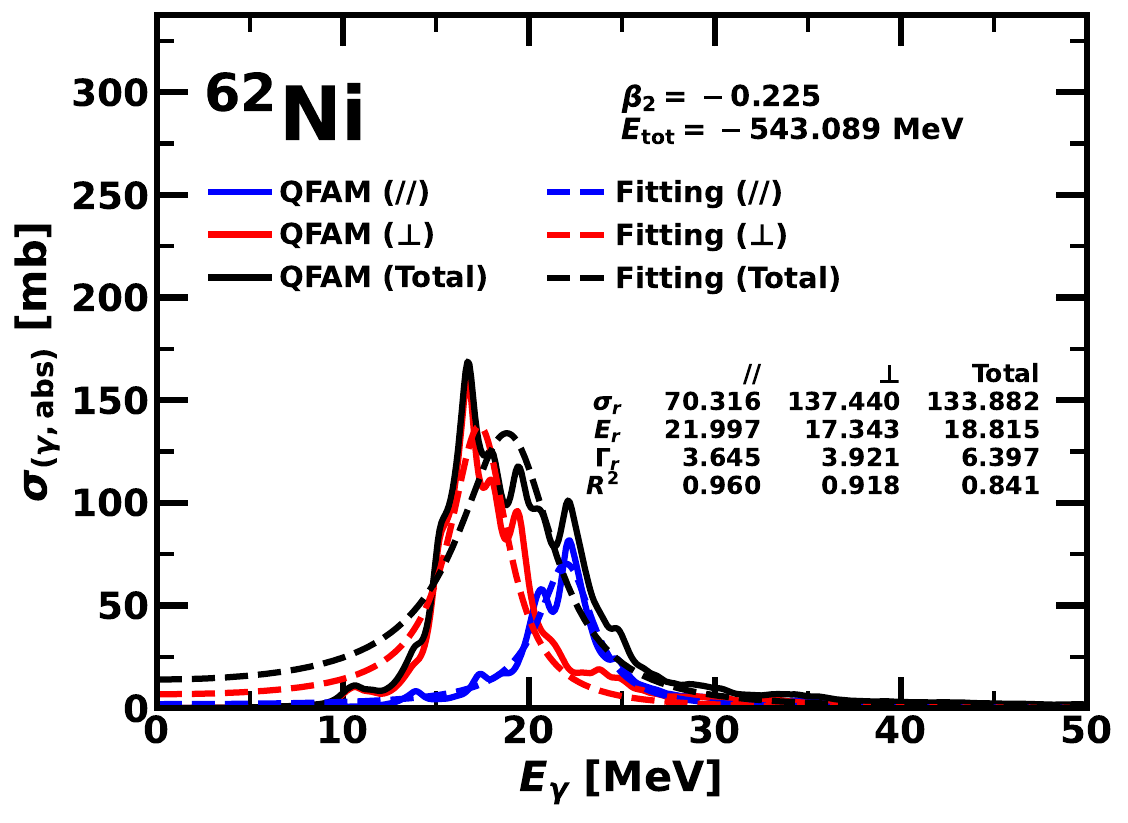}
    \includegraphics[width=0.4\textwidth]{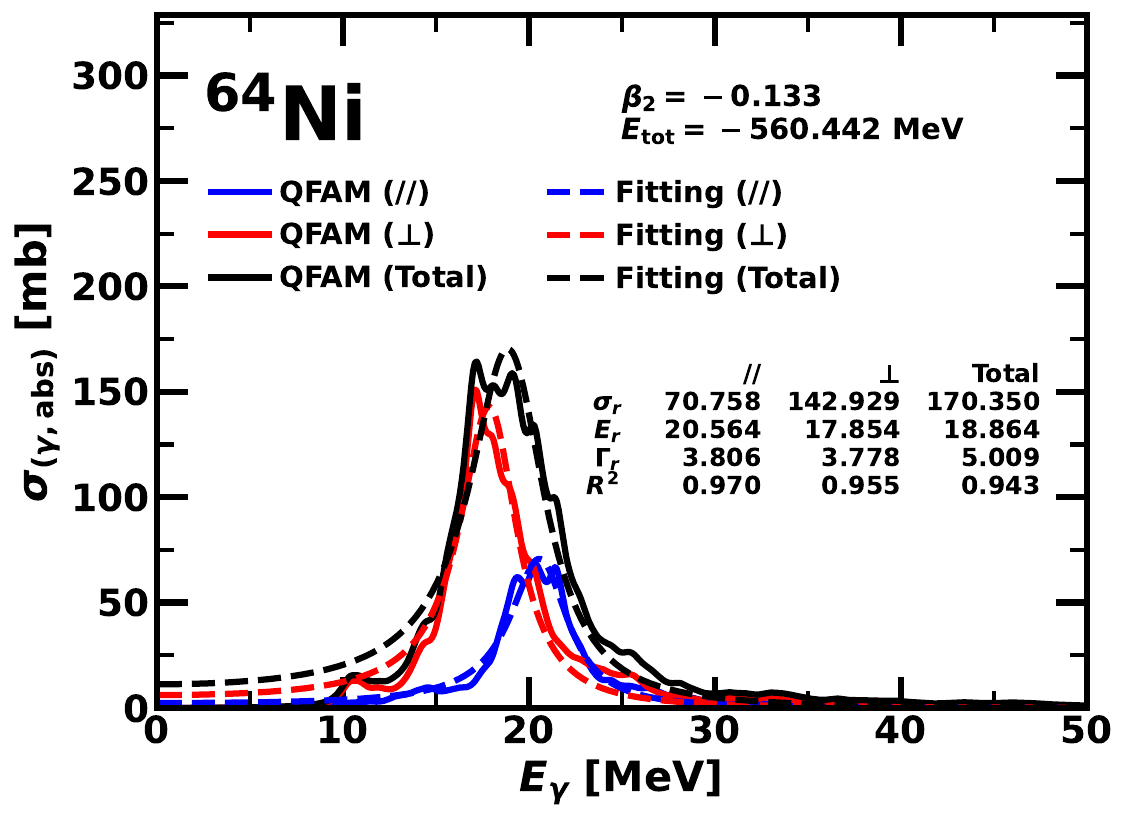}
    \includegraphics[width=0.4\textwidth]{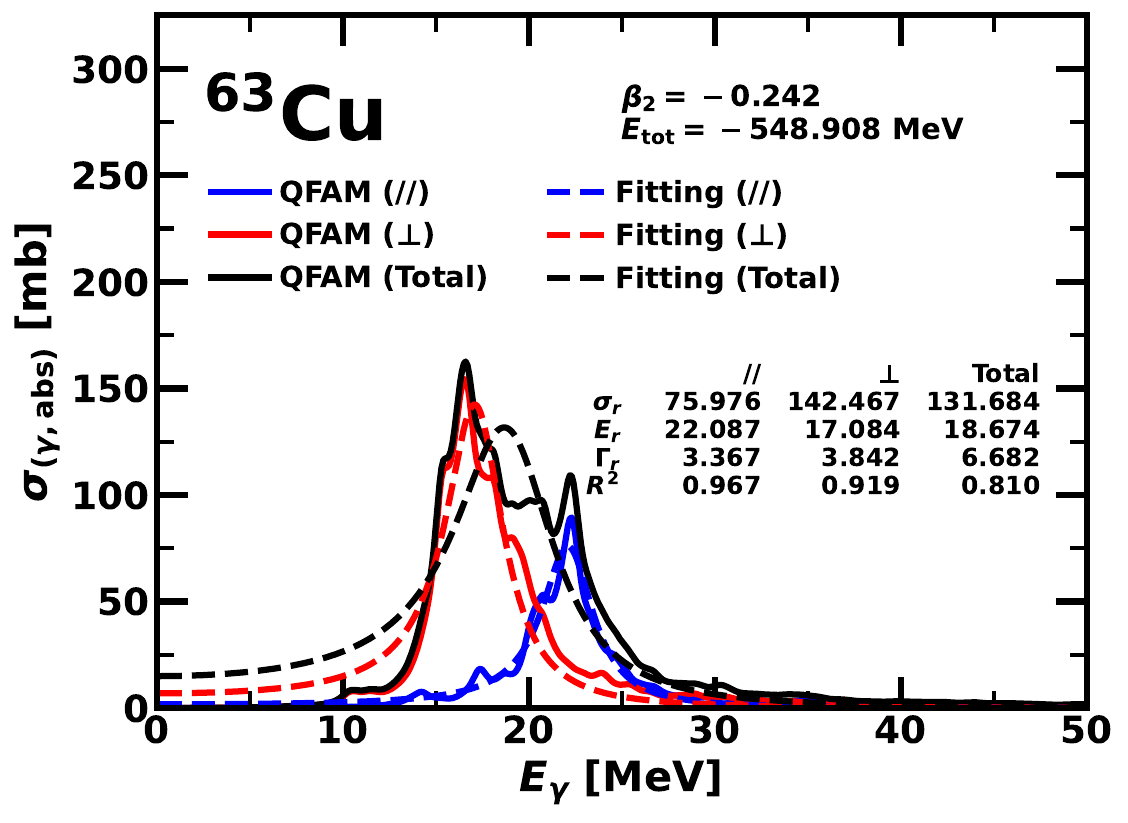}
    \includegraphics[width=0.4\textwidth]{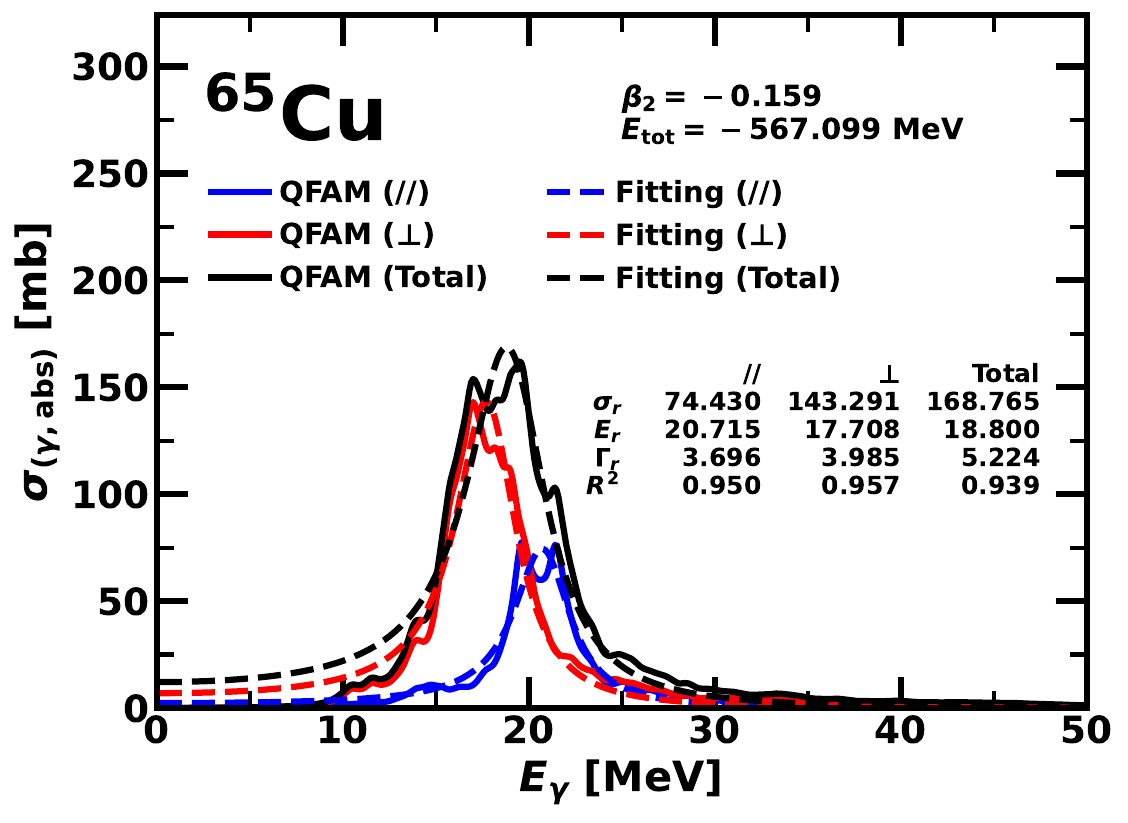}
    \includegraphics[width=0.4\textwidth]{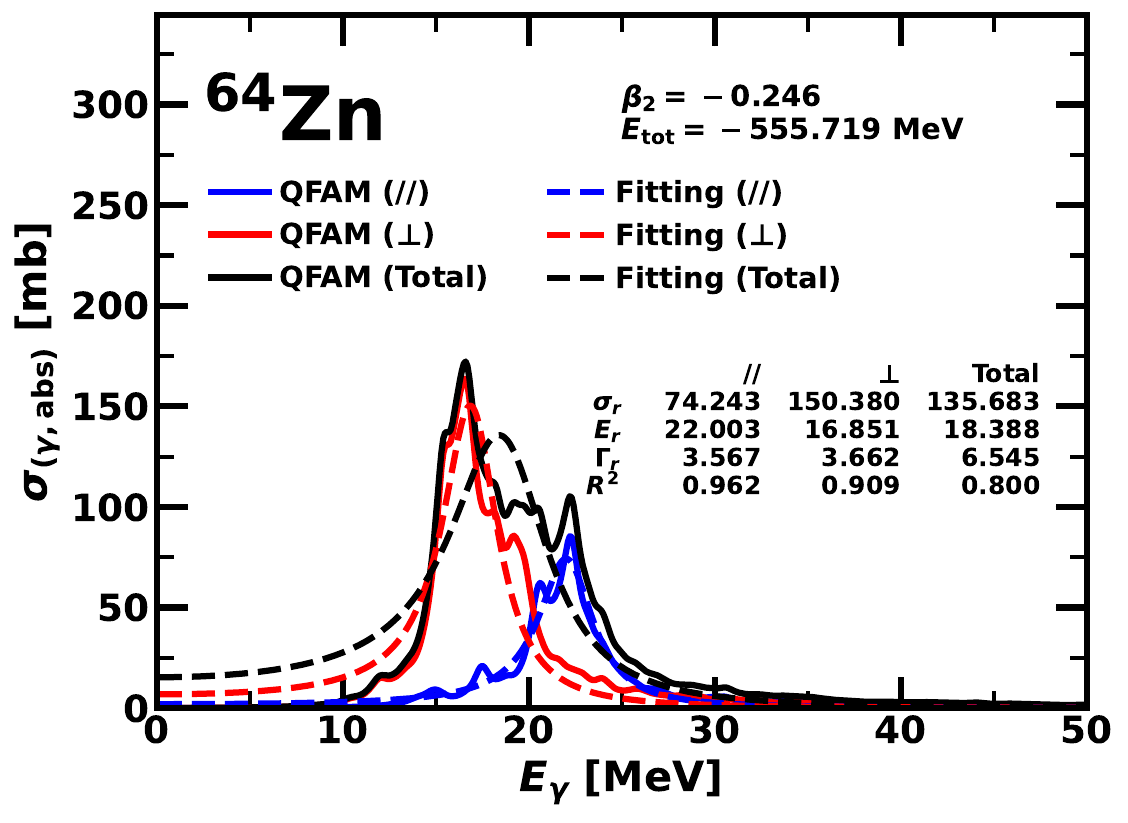}
    \includegraphics[width=0.4\textwidth]{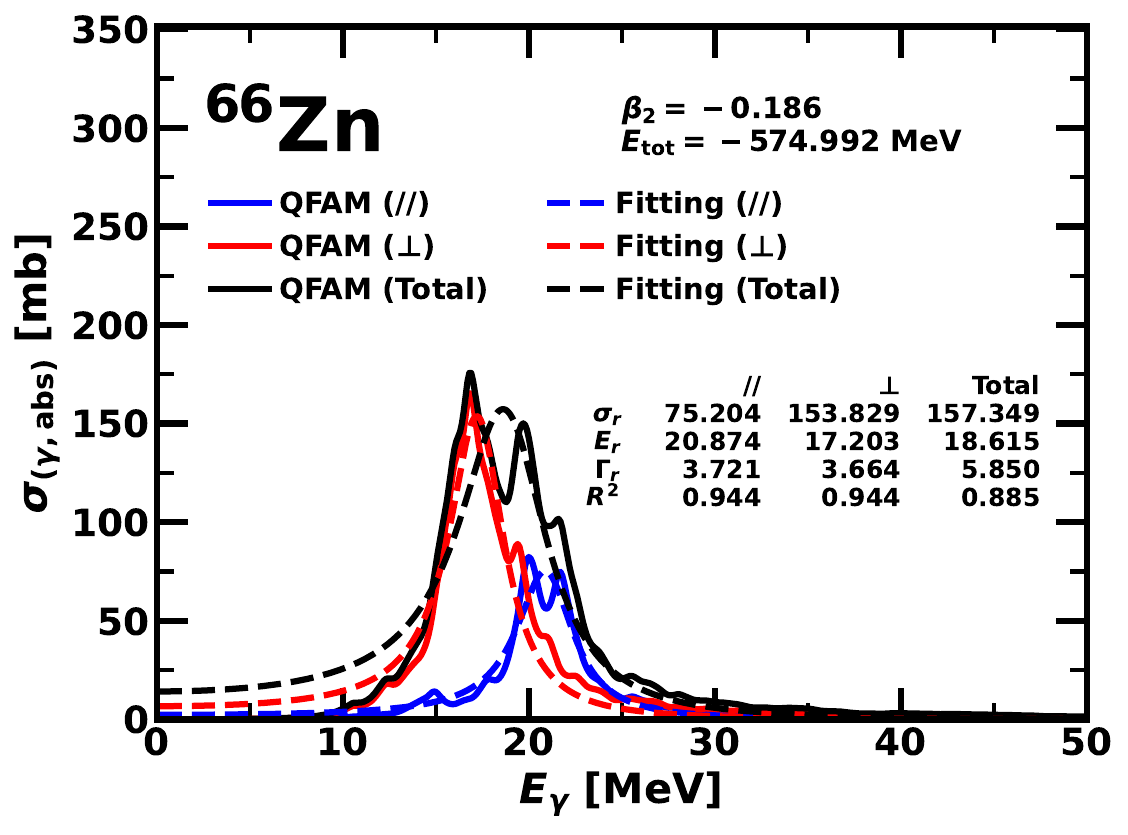}
    \includegraphics[width=0.4\textwidth]{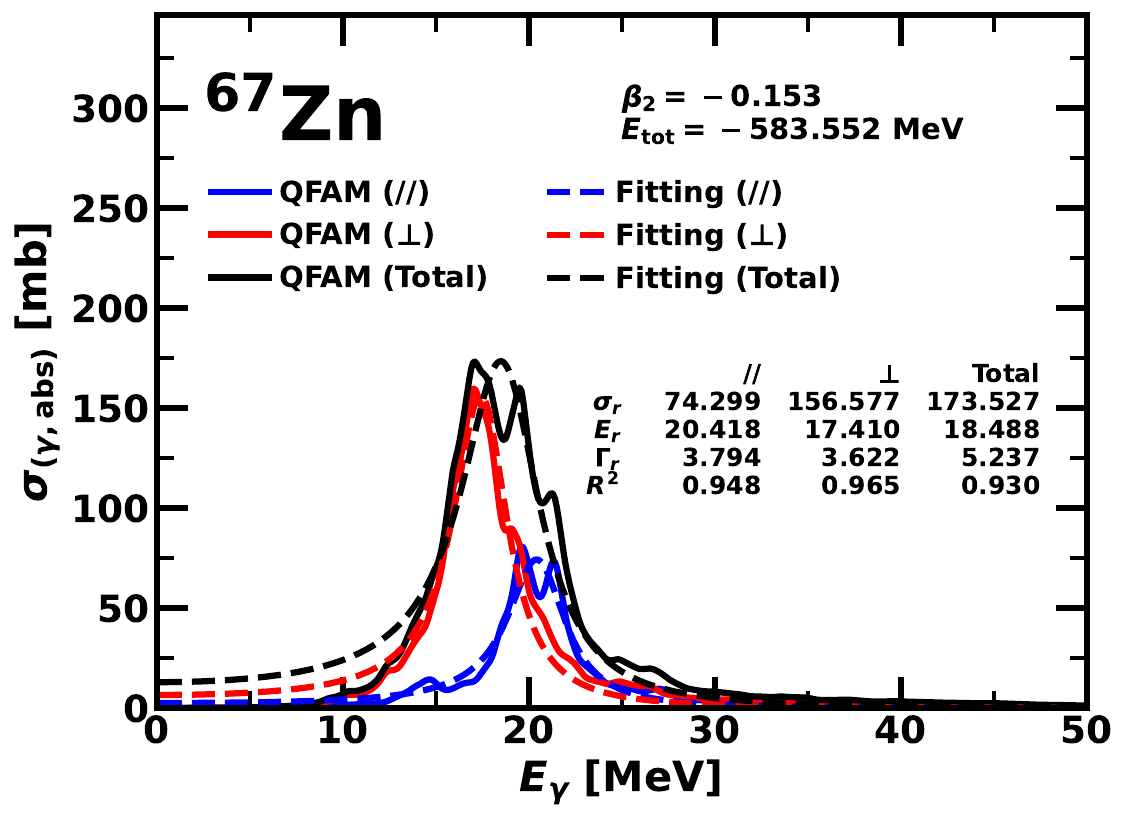}
\end{figure*}
\begin{figure*}\ContinuedFloat
    \centering
    \includegraphics[width=0.4\textwidth]{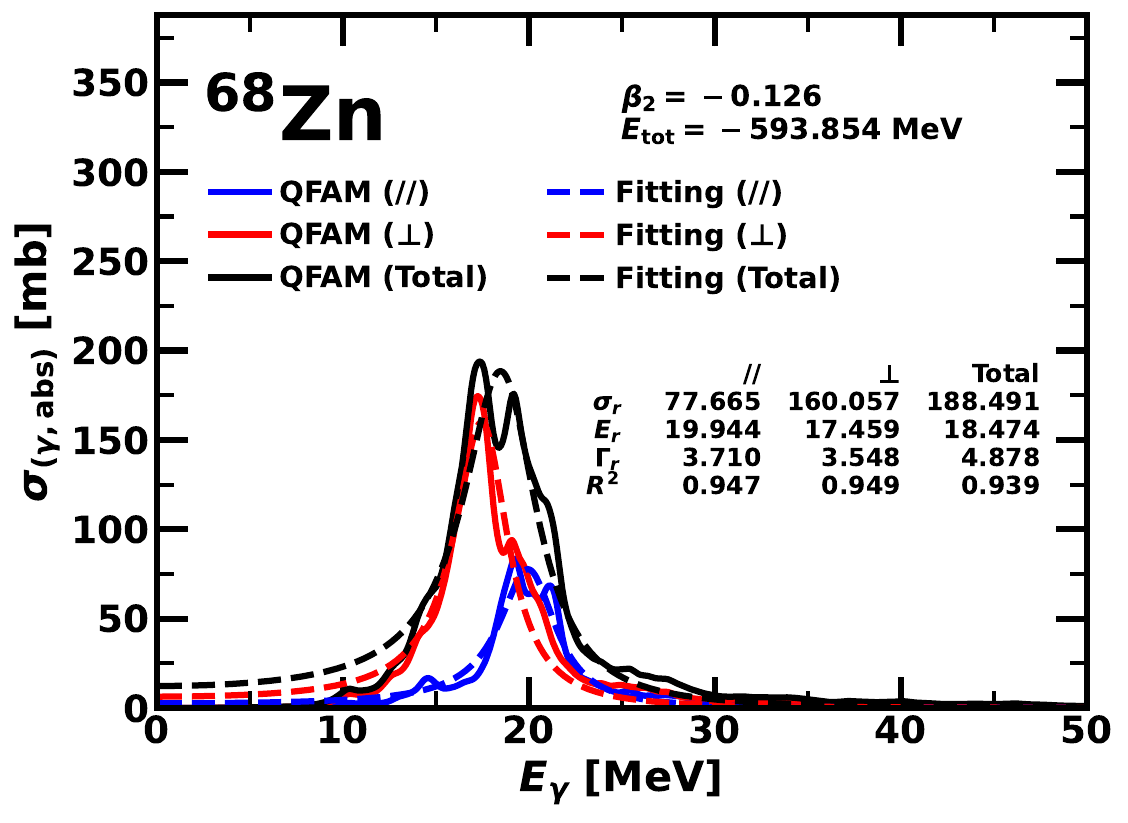}
    \includegraphics[width=0.4\textwidth]{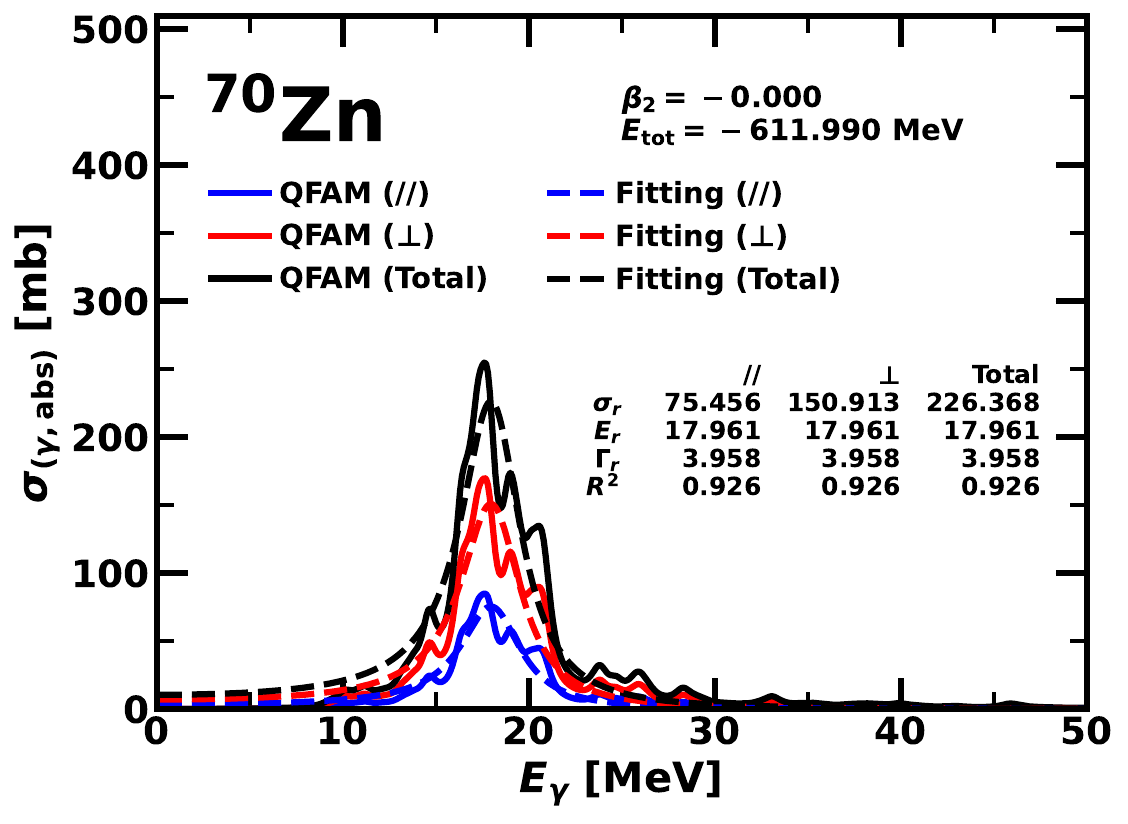}
    \includegraphics[width=0.4\textwidth]{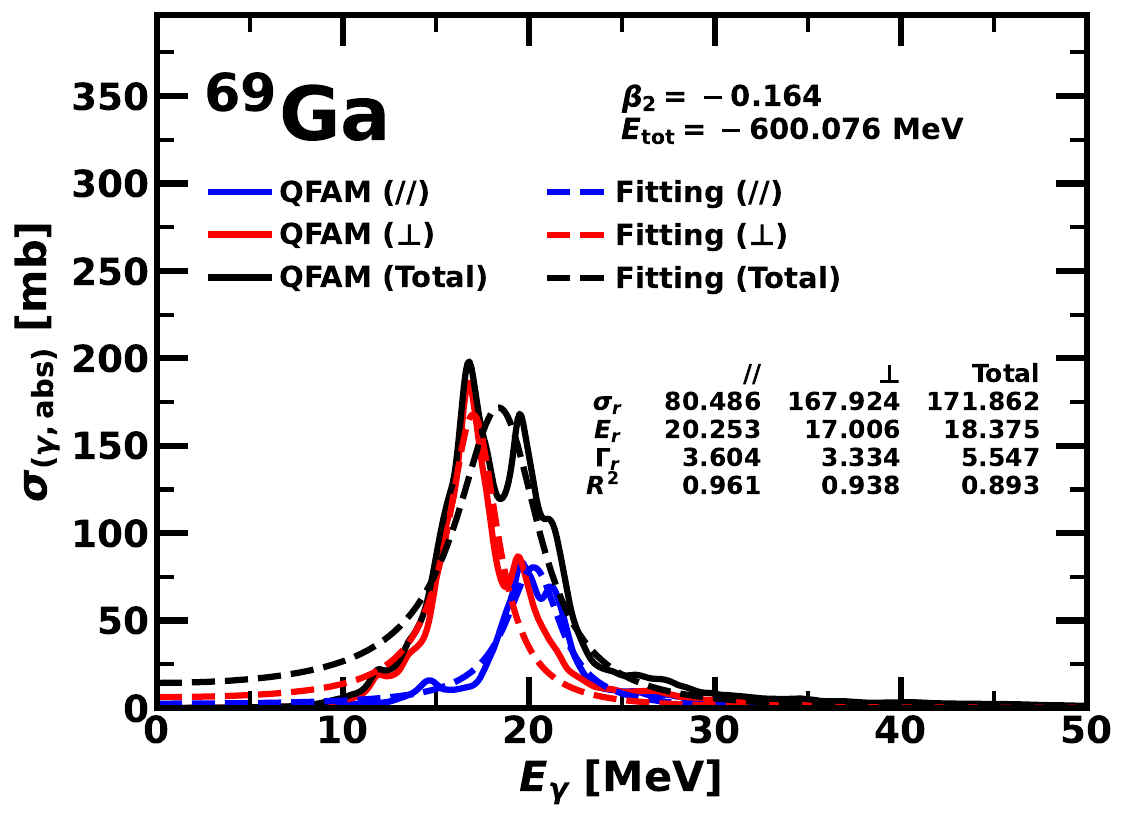}
    \includegraphics[width=0.4\textwidth]{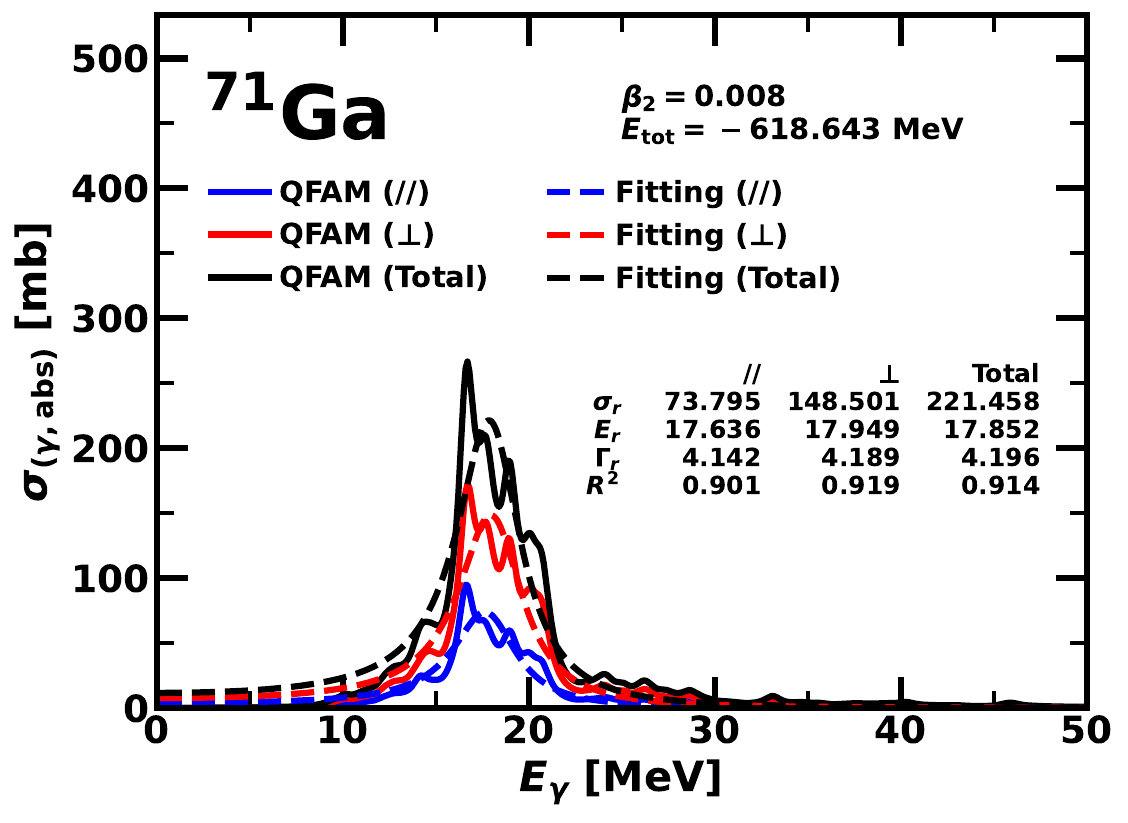}
    \includegraphics[width=0.4\textwidth]{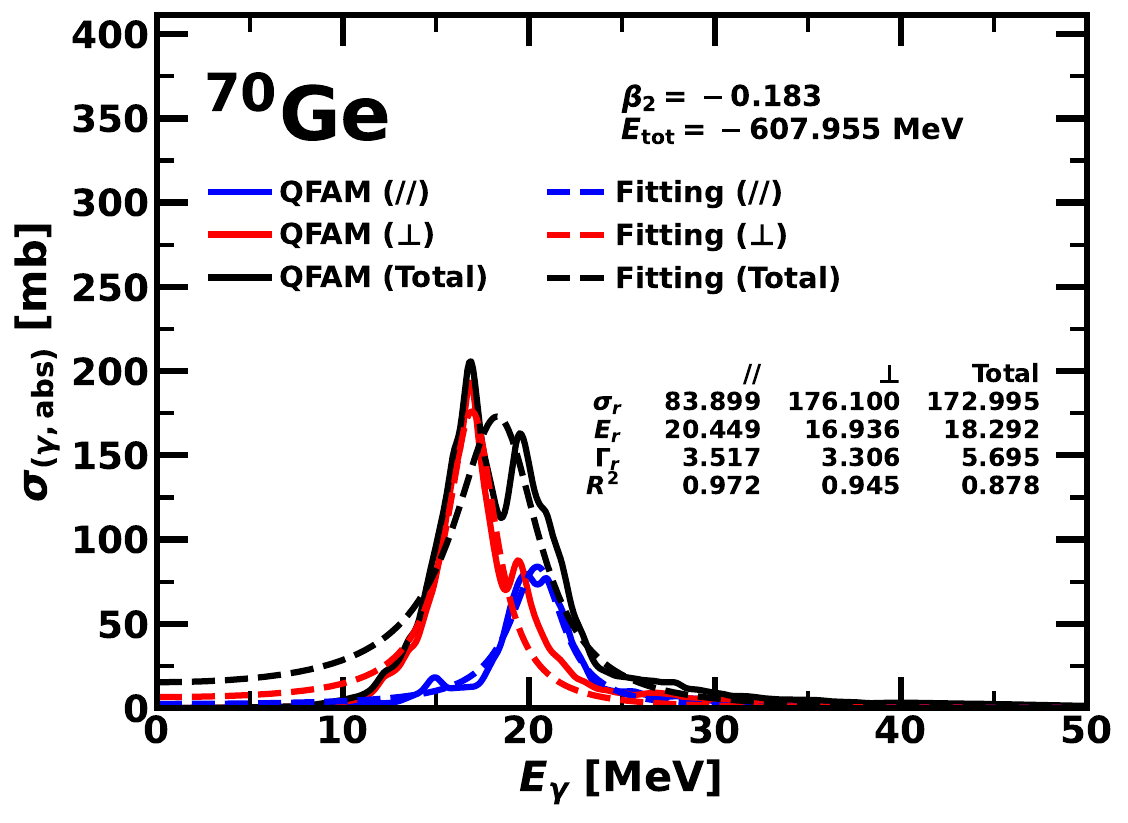}
    \includegraphics[width=0.4\textwidth]{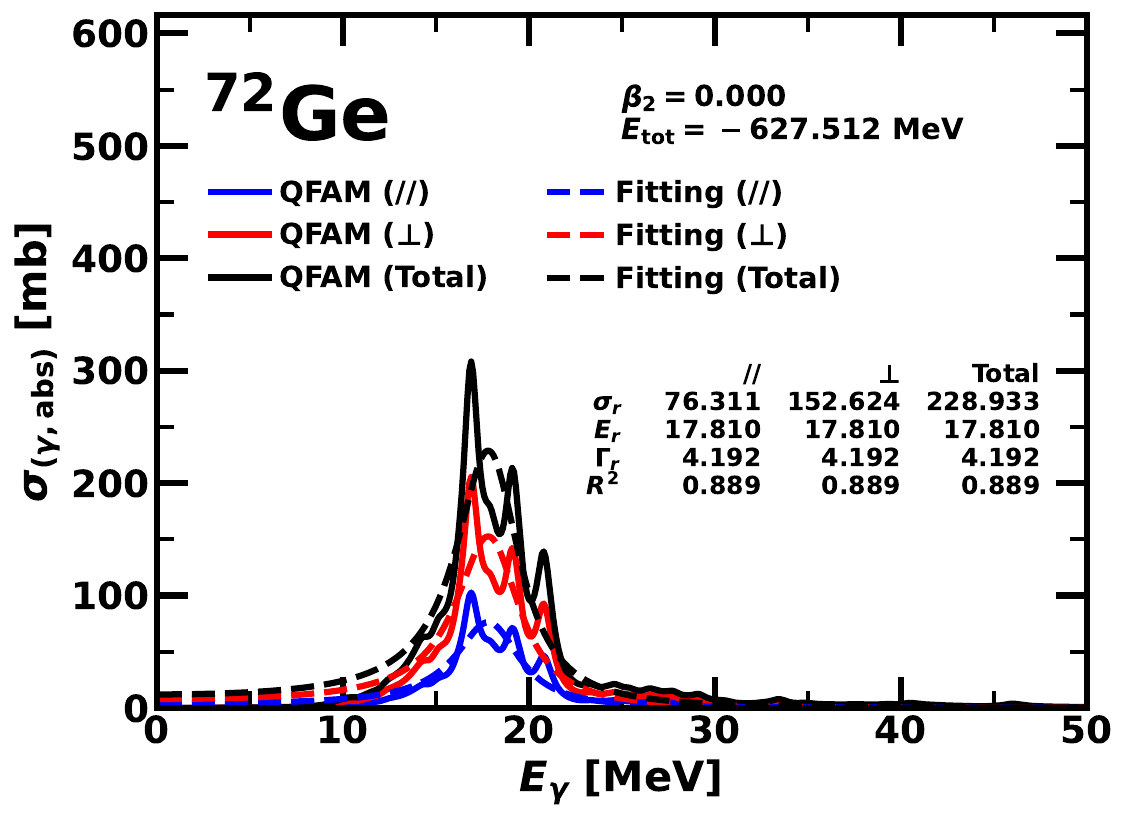}
    \includegraphics[width=0.4\textwidth]{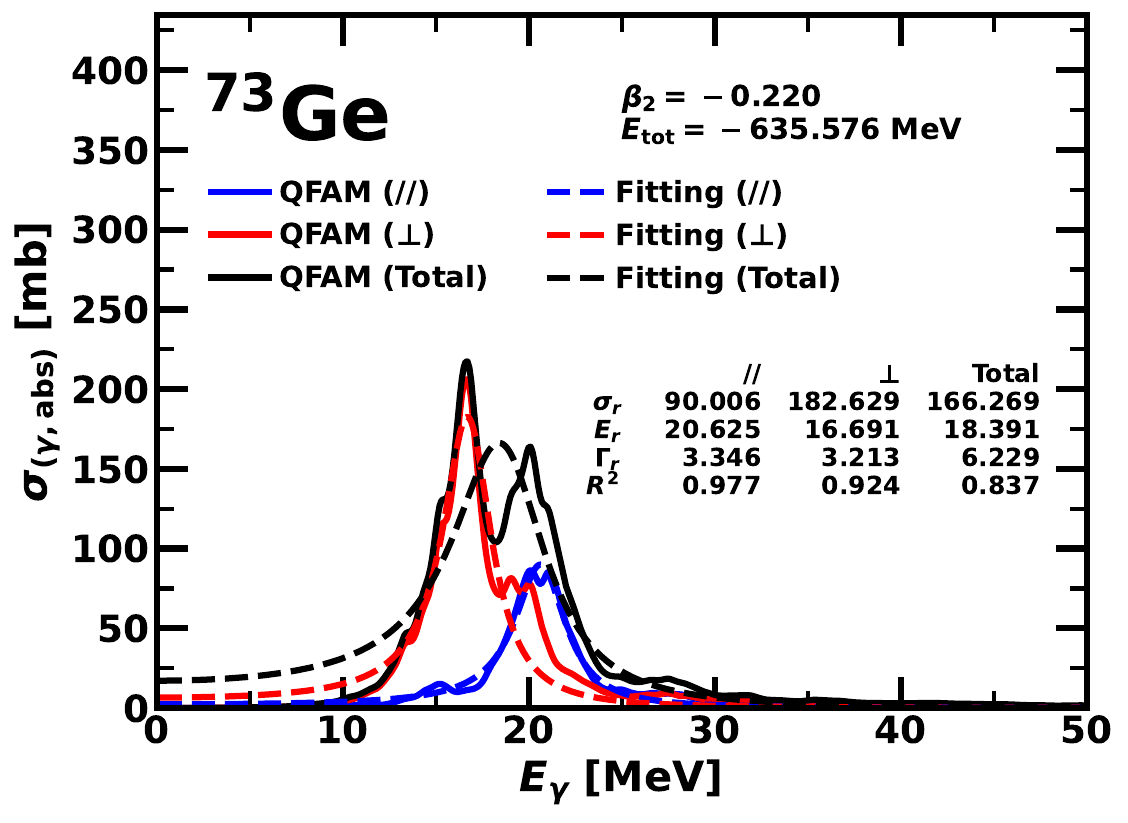}
    \includegraphics[width=0.4\textwidth]{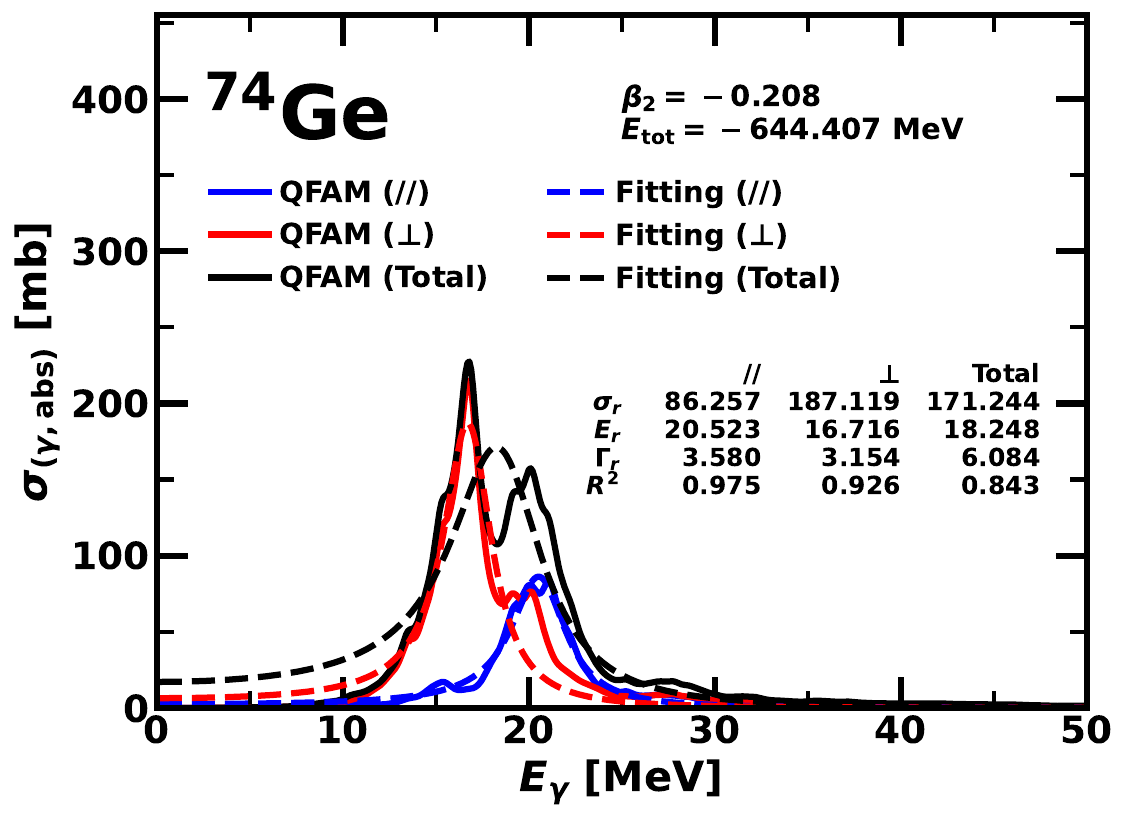}
\end{figure*}
\begin{figure*}\ContinuedFloat
    \centering
    \includegraphics[width=0.4\textwidth]{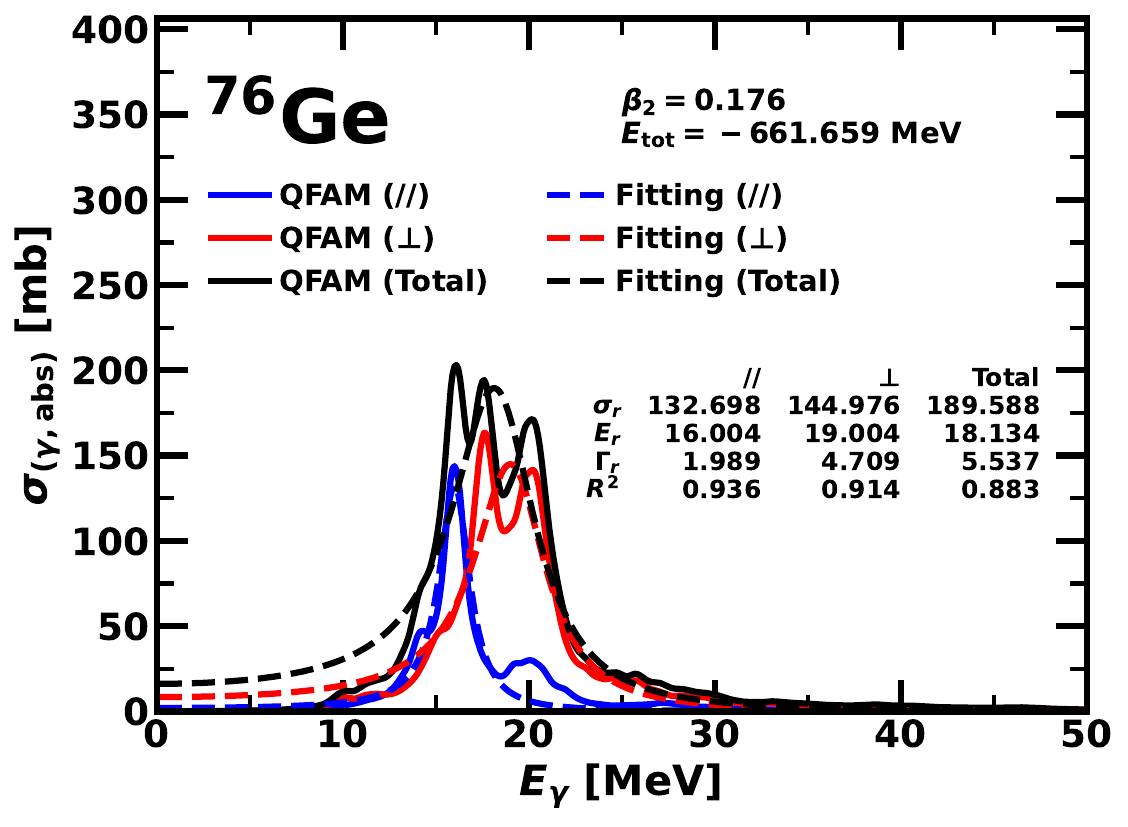}
    \includegraphics[width=0.4\textwidth]{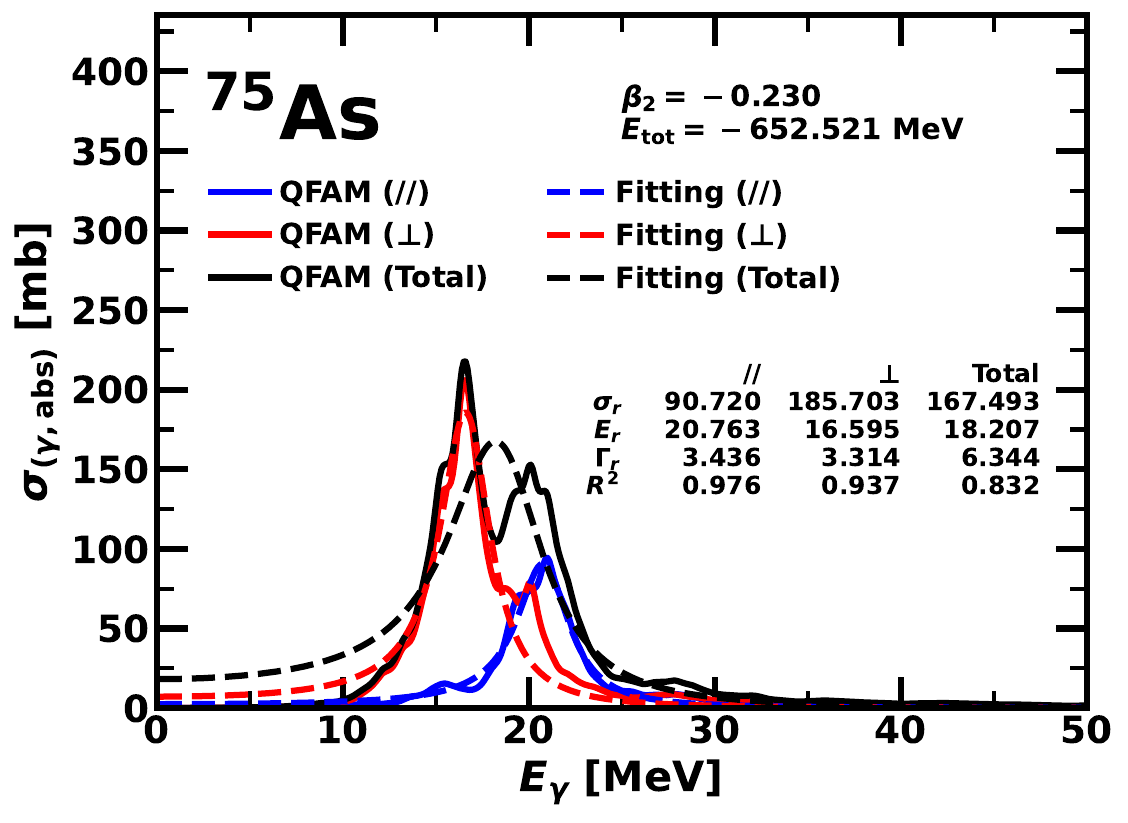}
    \includegraphics[width=0.4\textwidth]{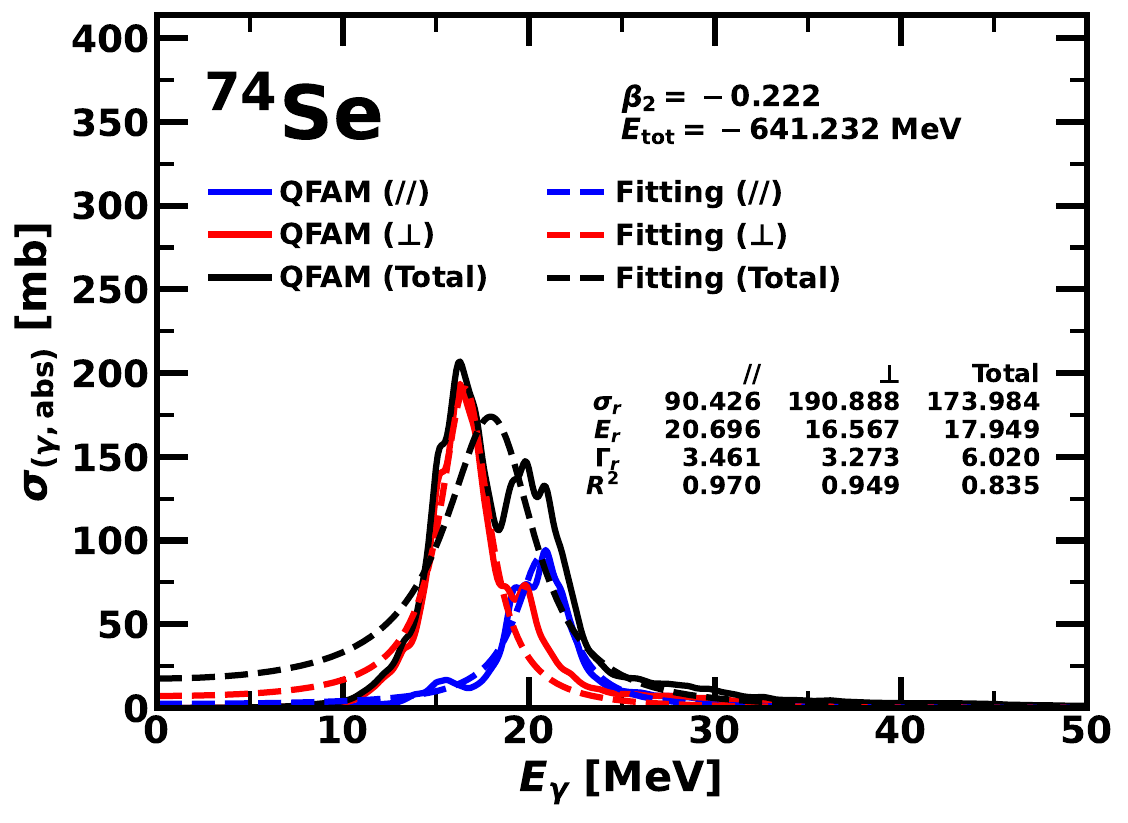}
    \includegraphics[width=0.4\textwidth]{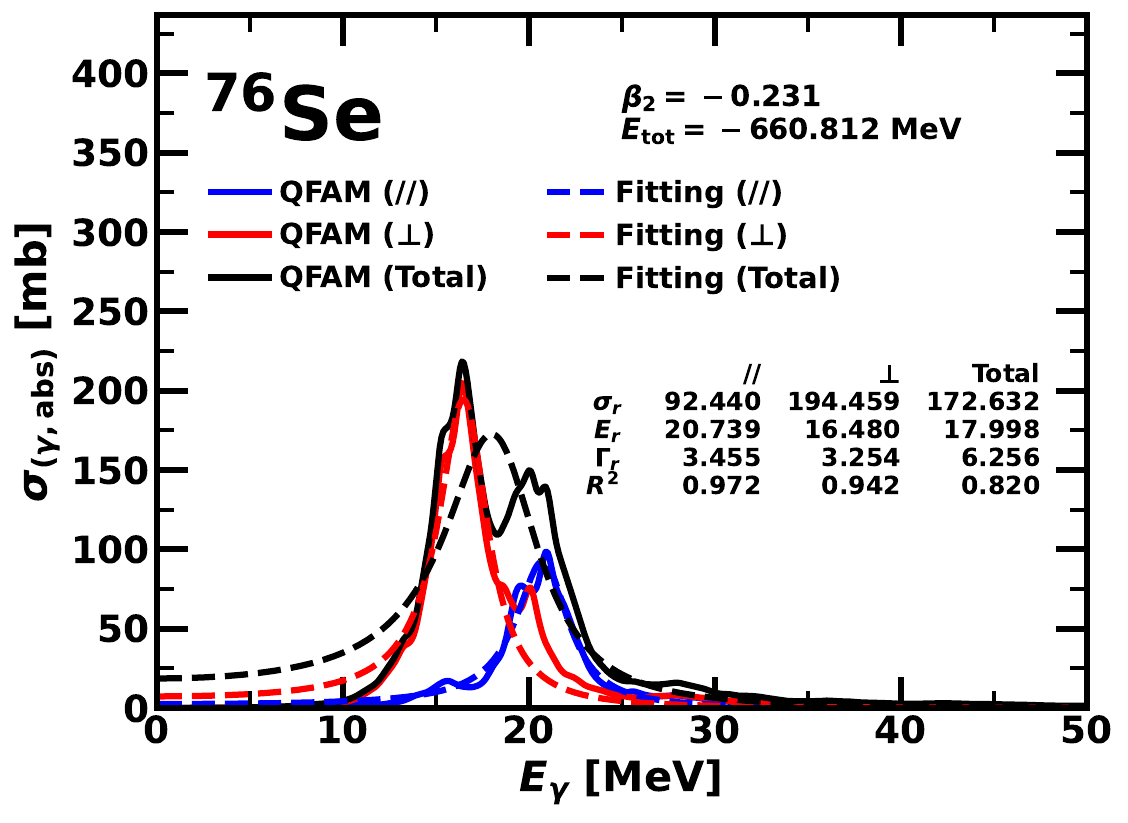}
    \includegraphics[width=0.4\textwidth]{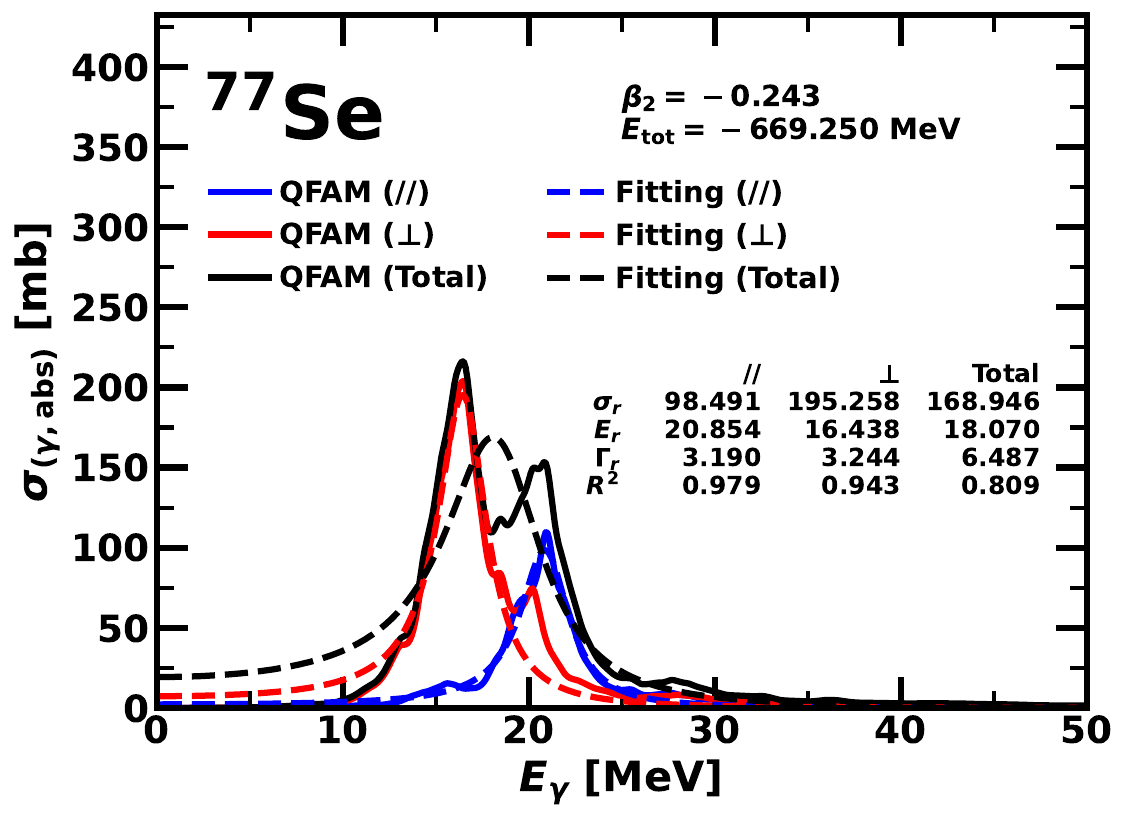}
    \includegraphics[width=0.4\textwidth]{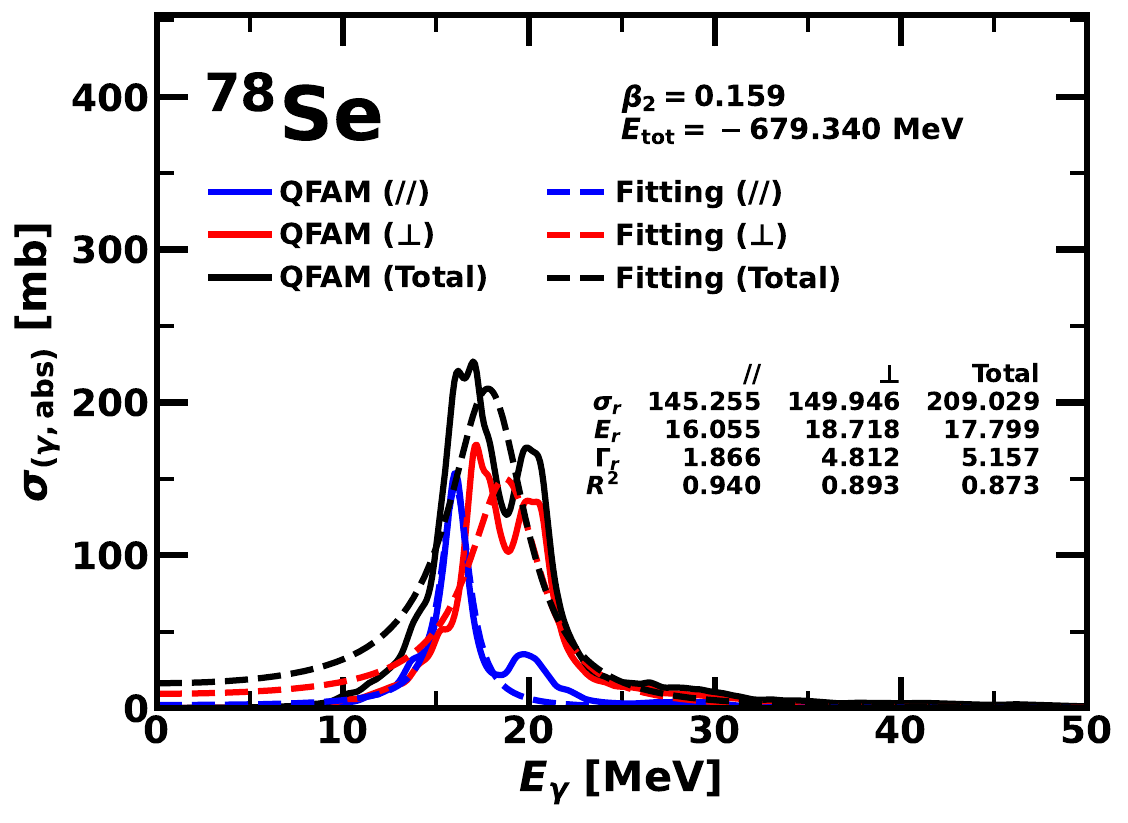}
    \includegraphics[width=0.4\textwidth]{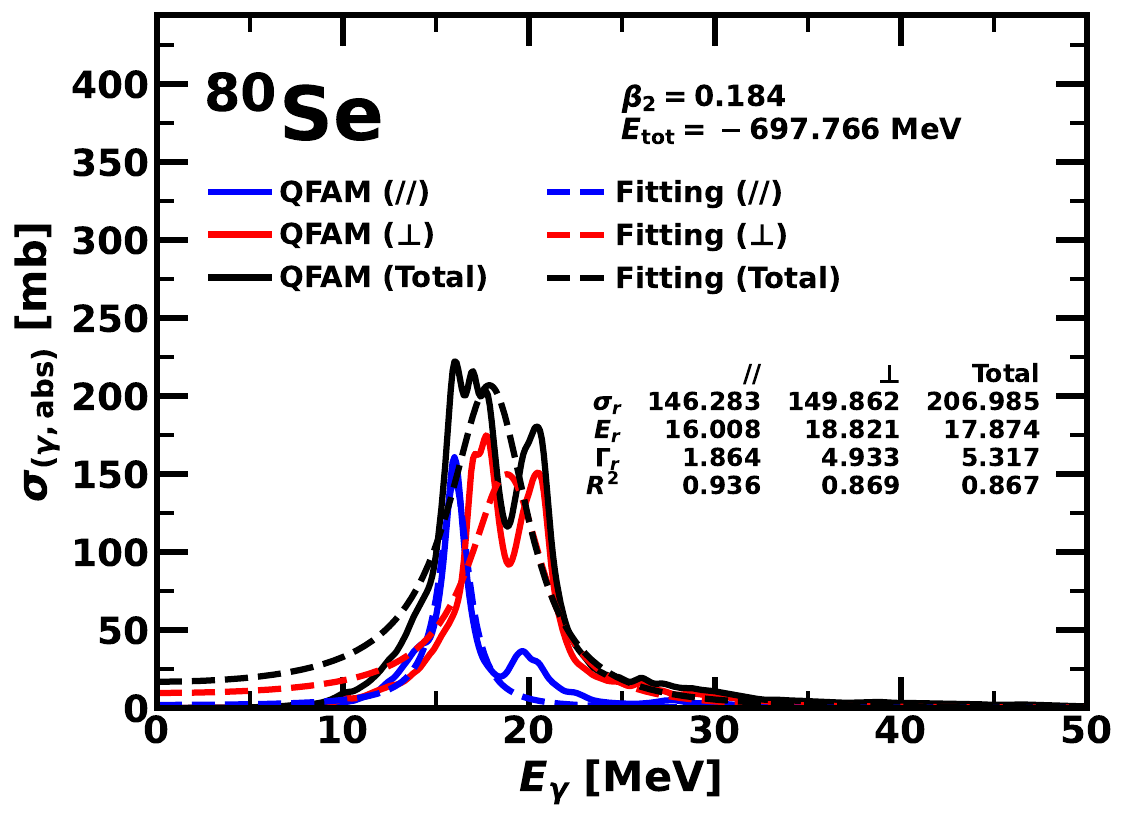}
    \includegraphics[width=0.4\textwidth]{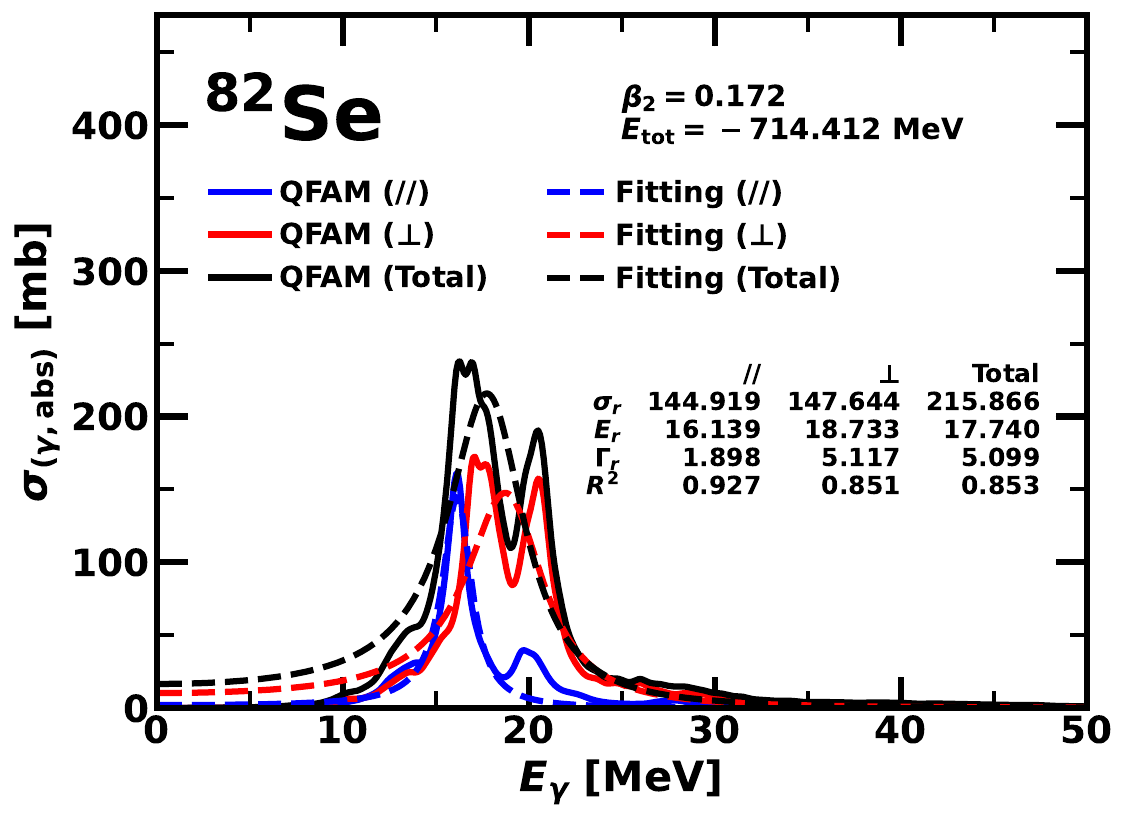}
\end{figure*}
\begin{figure*}\ContinuedFloat
    \centering
    \includegraphics[width=0.4\textwidth]{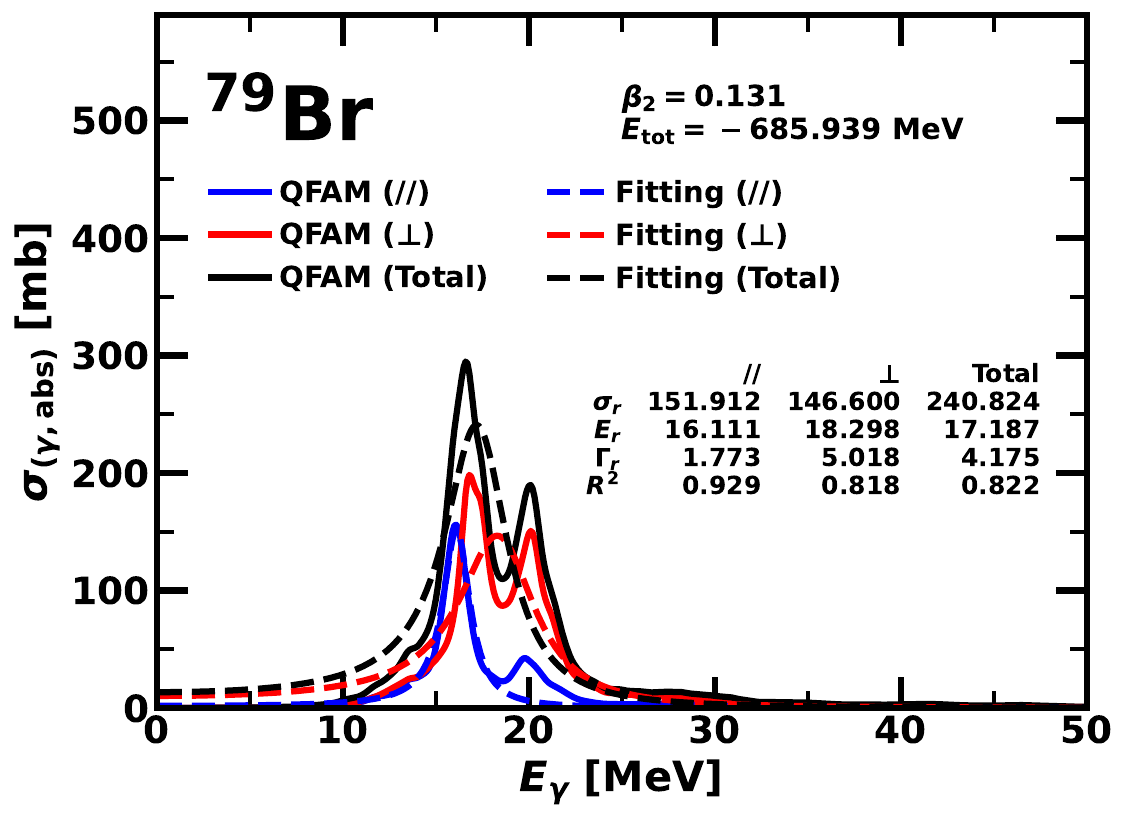}
    \includegraphics[width=0.4\textwidth]{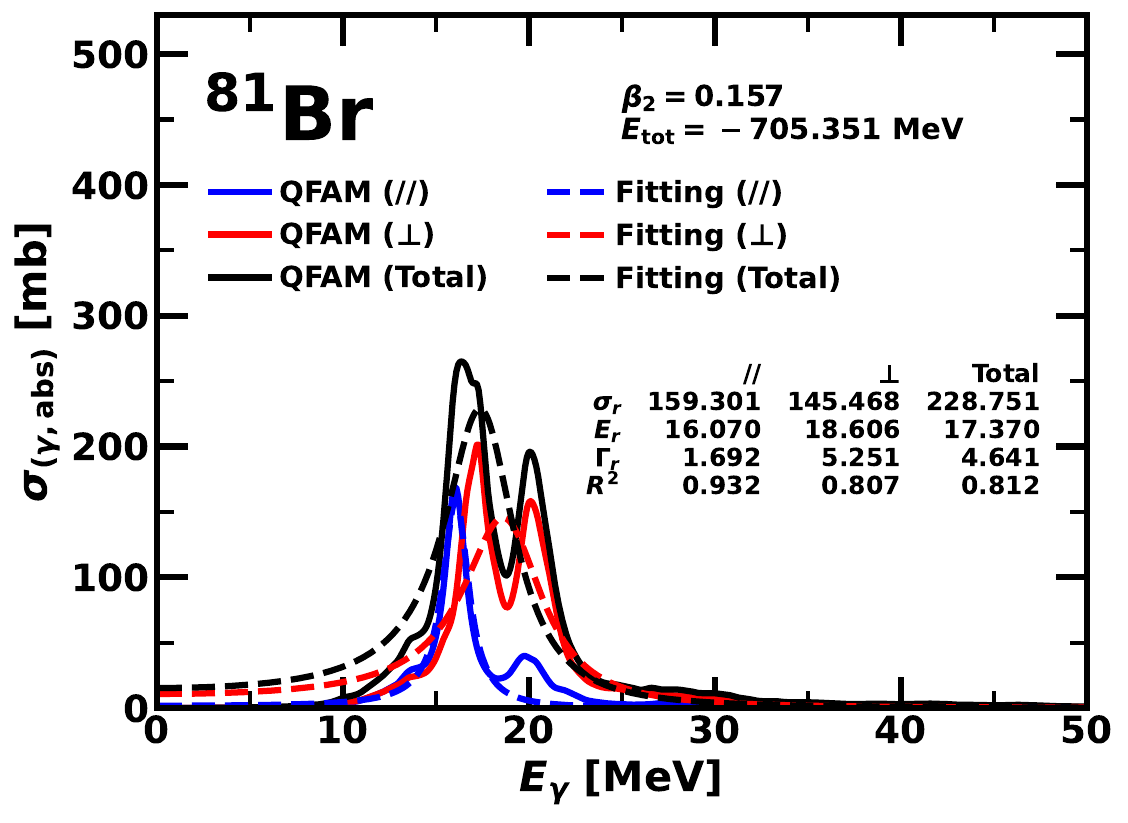}
    \includegraphics[width=0.4\textwidth]{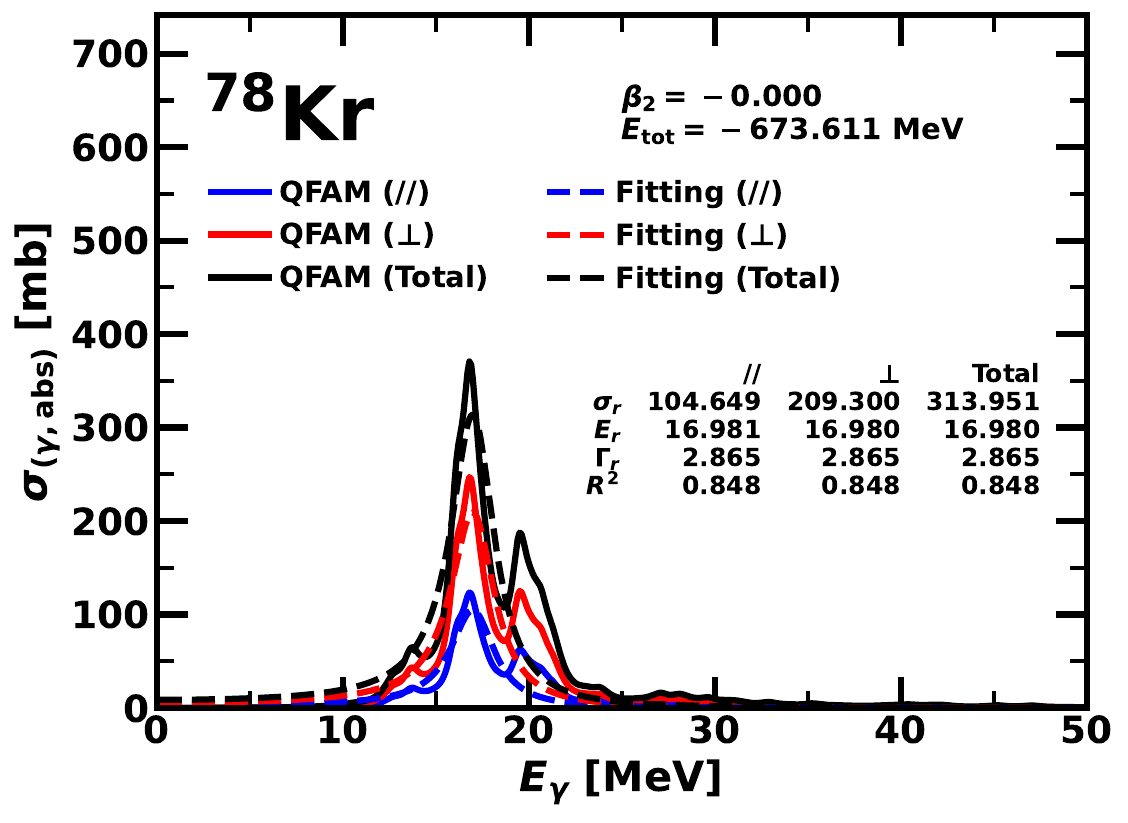}
    \includegraphics[width=0.4\textwidth]{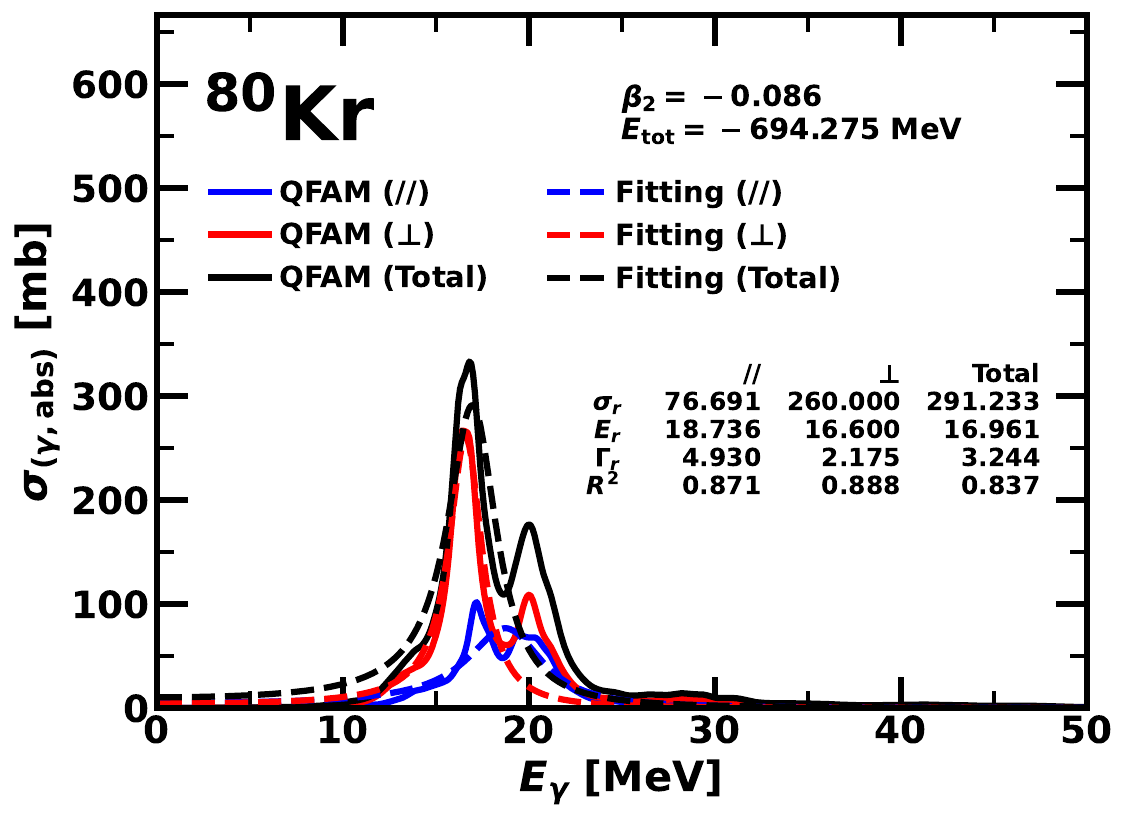}
    \includegraphics[width=0.4\textwidth]{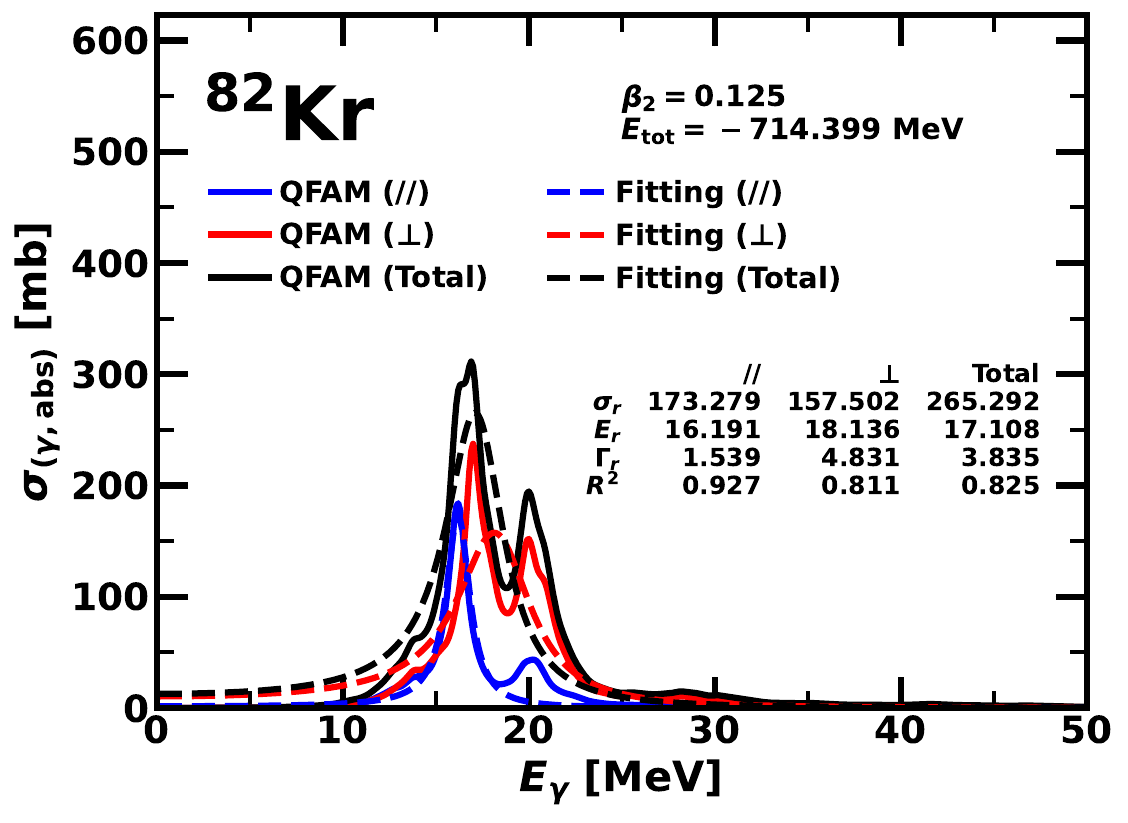}
    \includegraphics[width=0.4\textwidth]{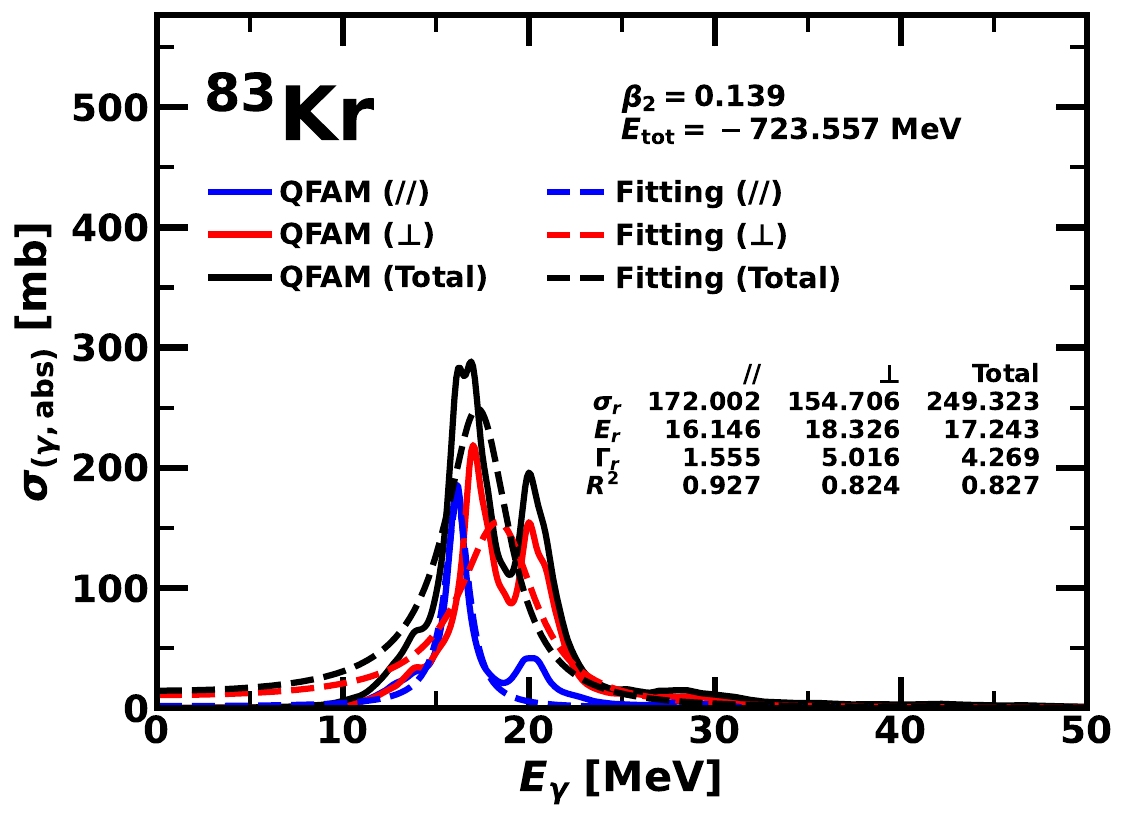}
    \includegraphics[width=0.4\textwidth]{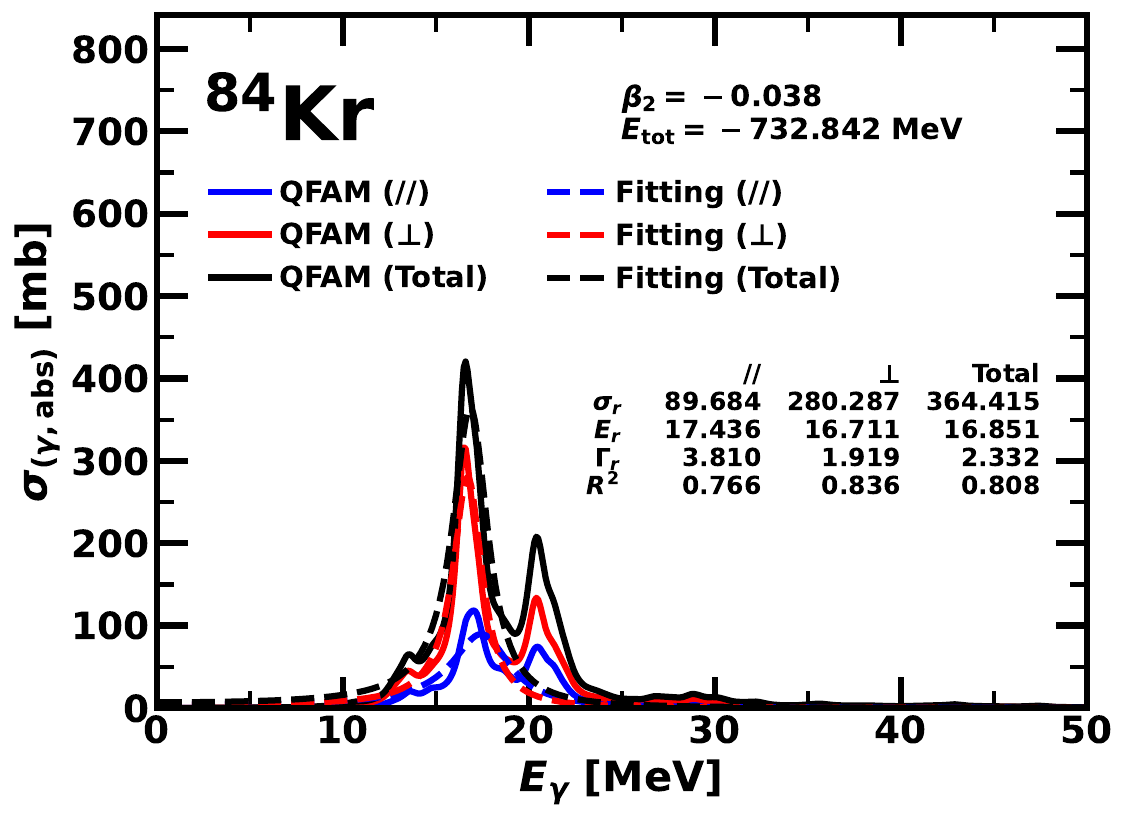}
    \includegraphics[width=0.4\textwidth]{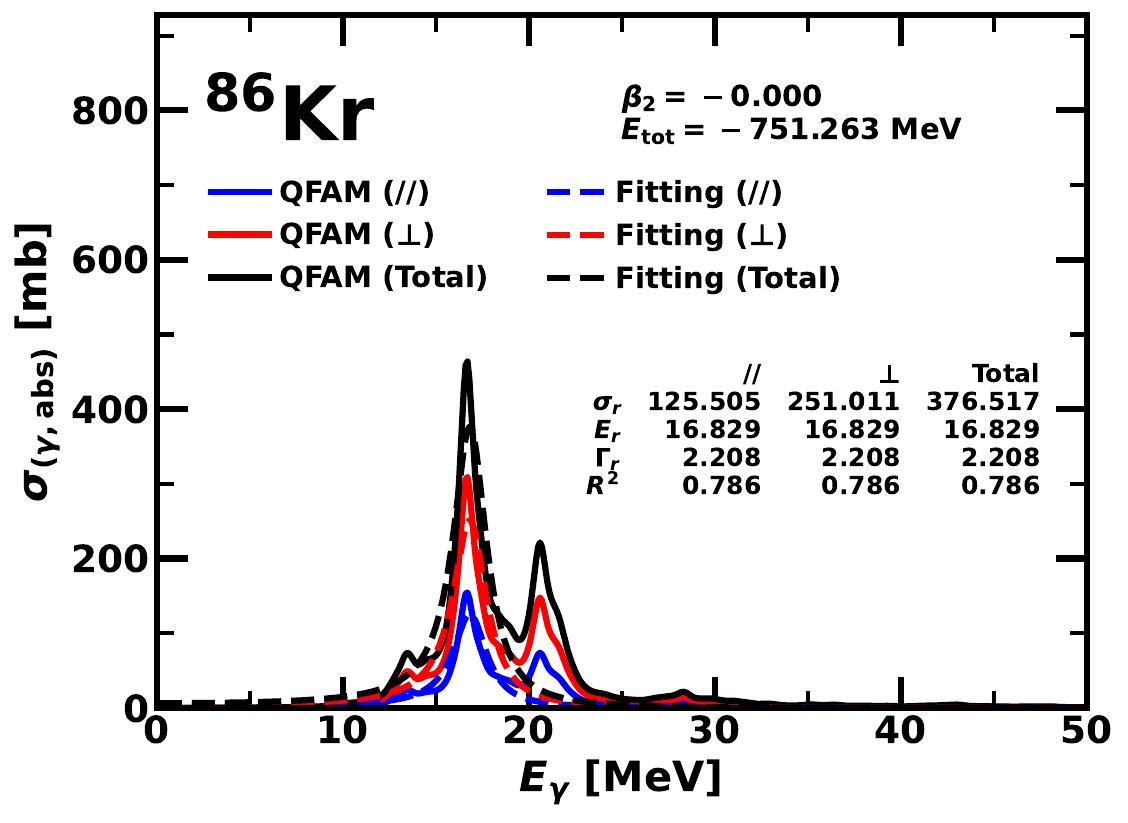}
\end{figure*}
\begin{figure*}\ContinuedFloat
    \centering
    \includegraphics[width=0.4\textwidth]{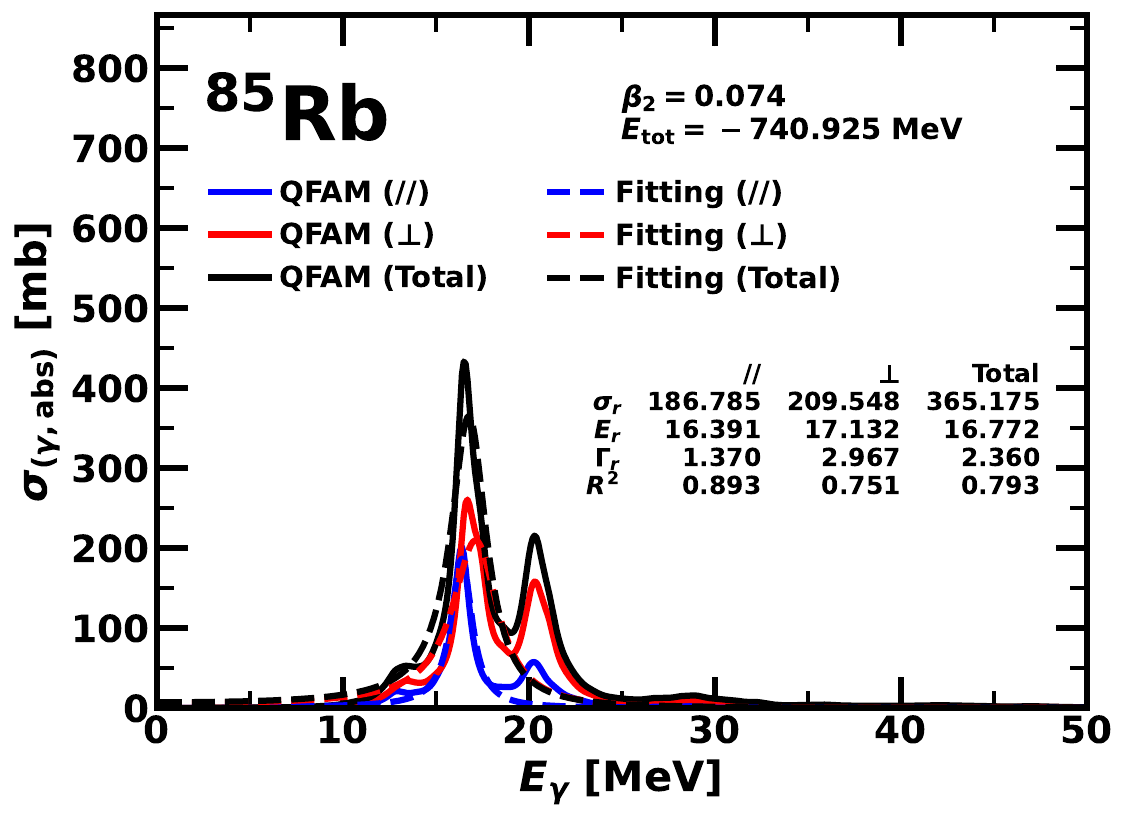}
    \includegraphics[width=0.4\textwidth]{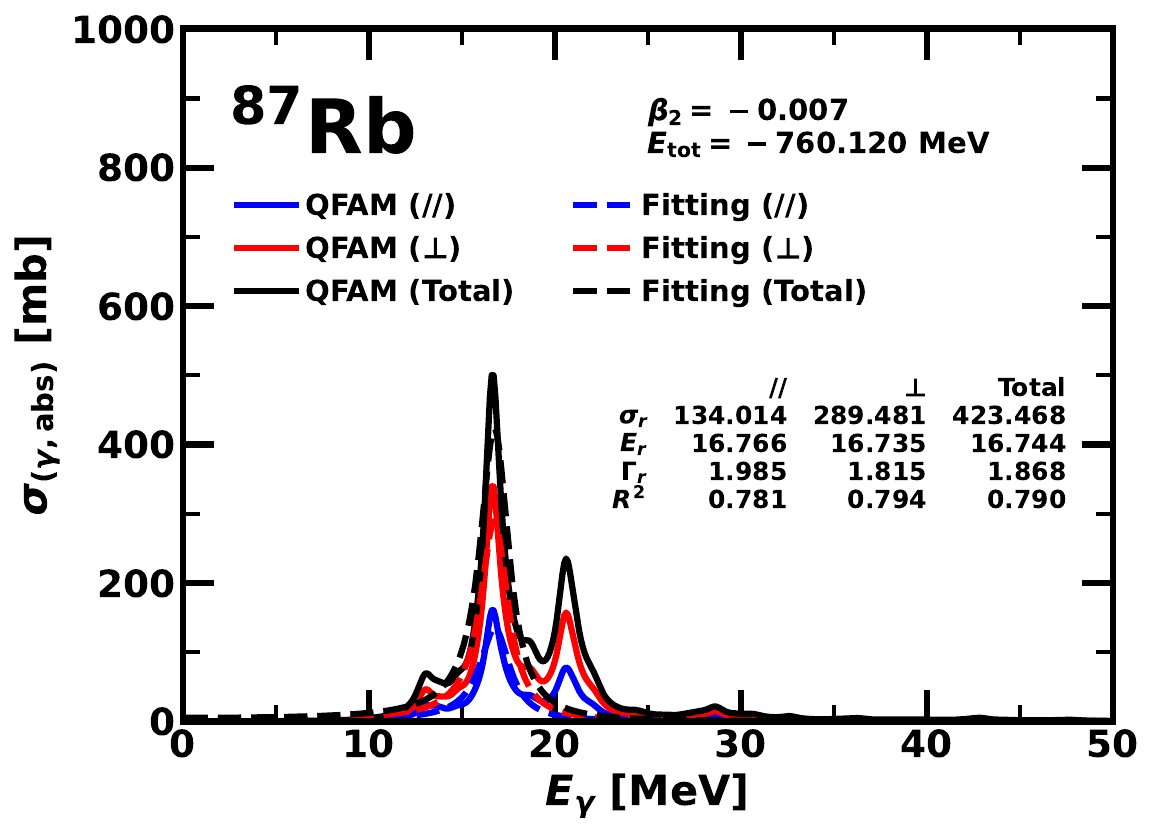}
    \includegraphics[width=0.4\textwidth]{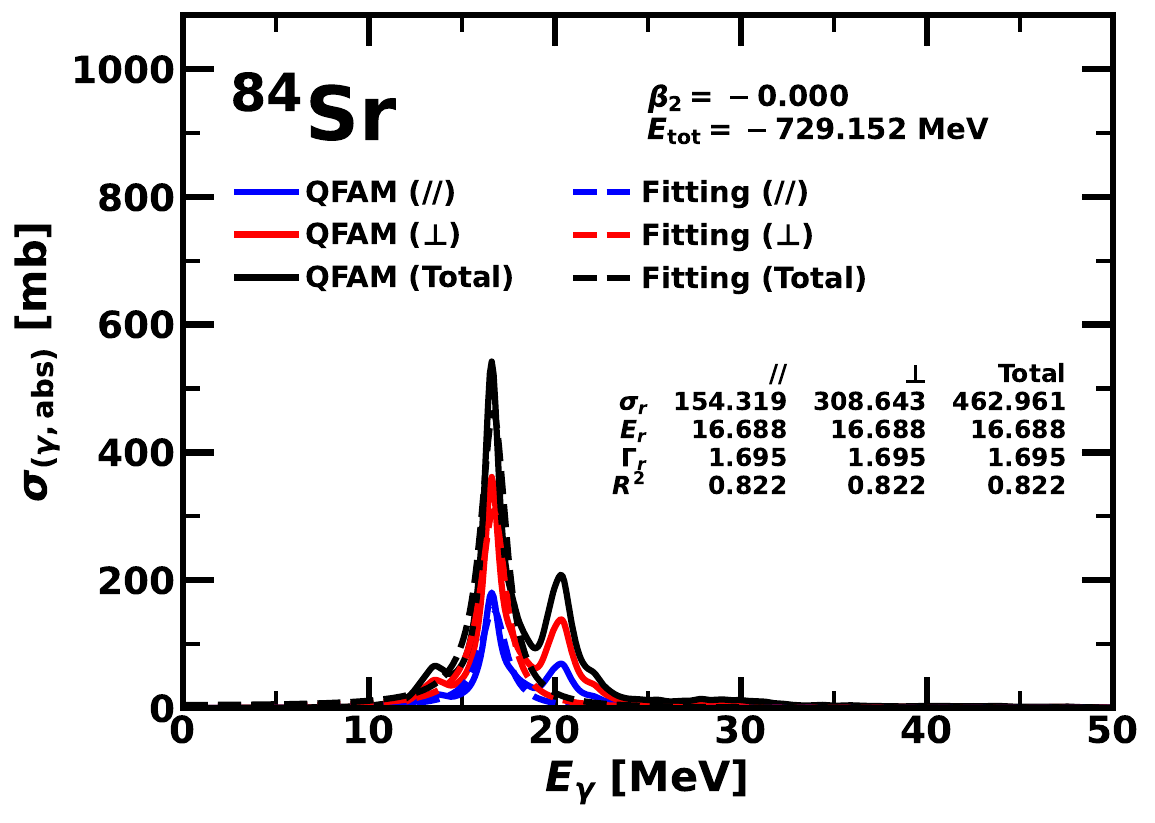}
    \includegraphics[width=0.4\textwidth]{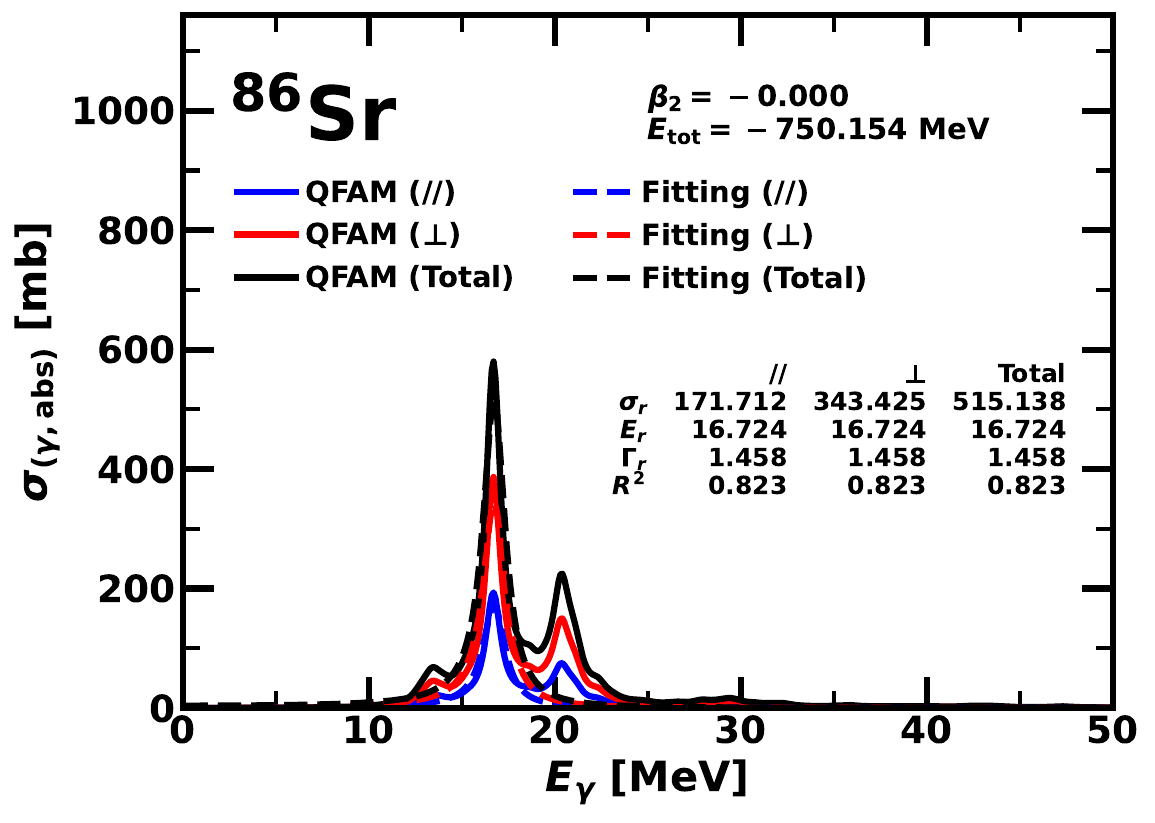}
    \includegraphics[width=0.4\textwidth]{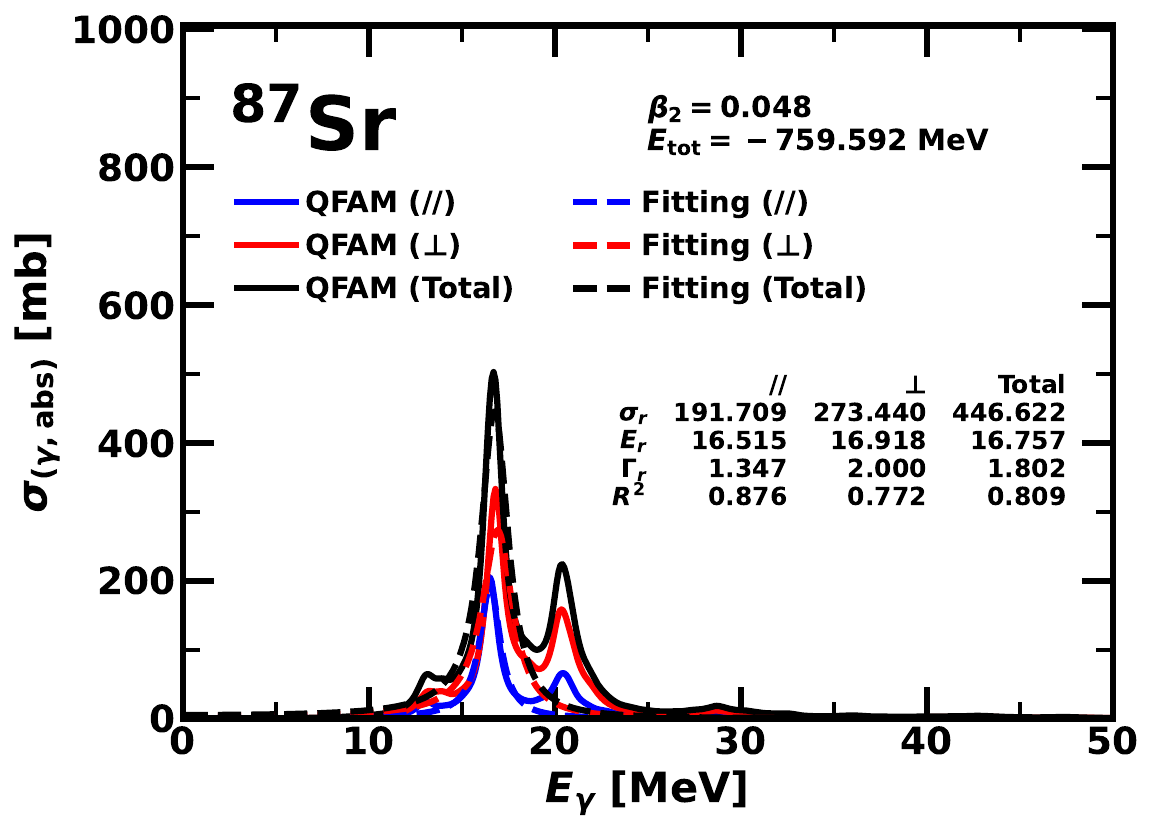}
    \includegraphics[width=0.4\textwidth]{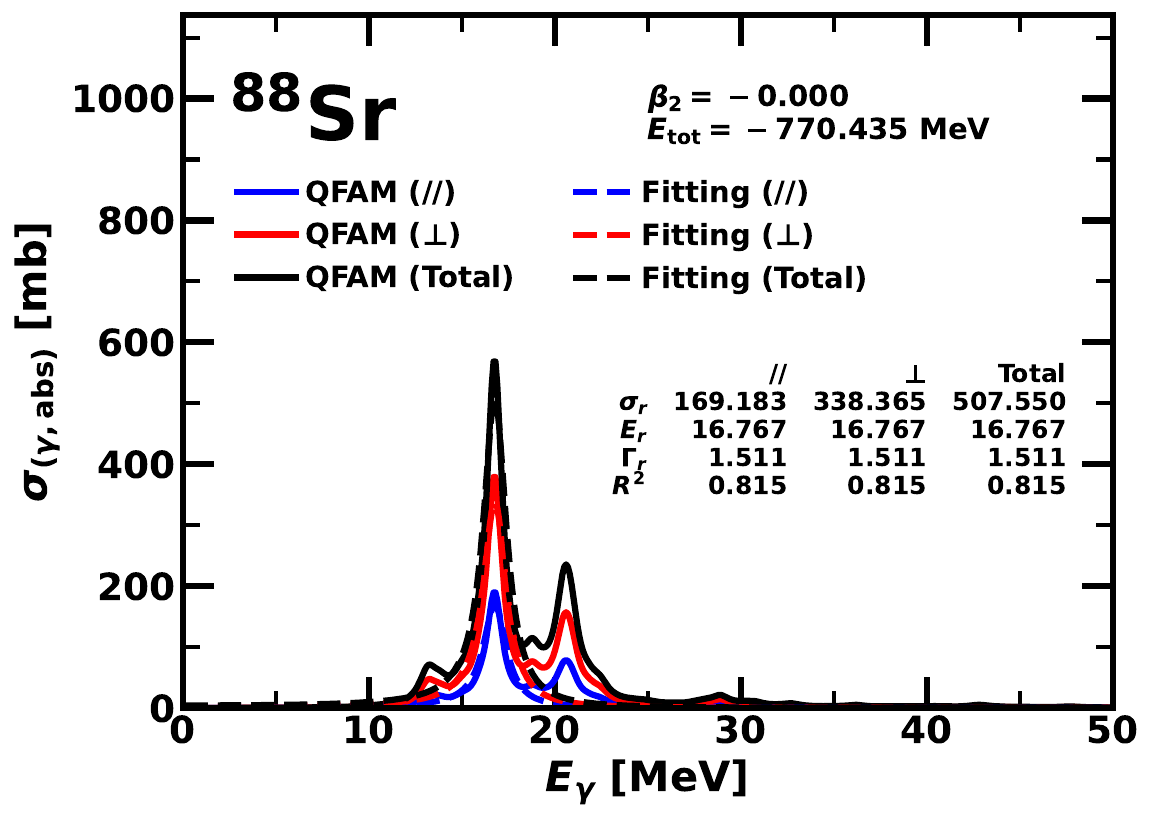}
    \includegraphics[width=0.4\textwidth]{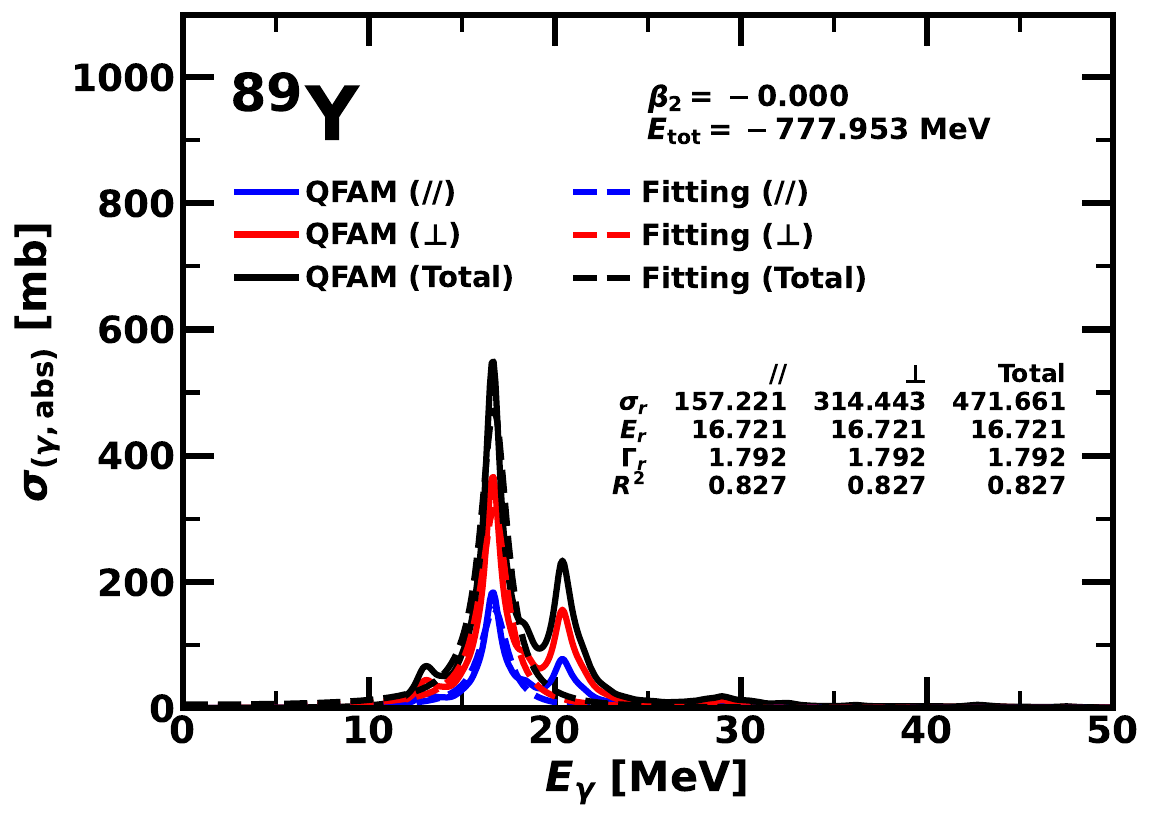}
    \includegraphics[width=0.4\textwidth]{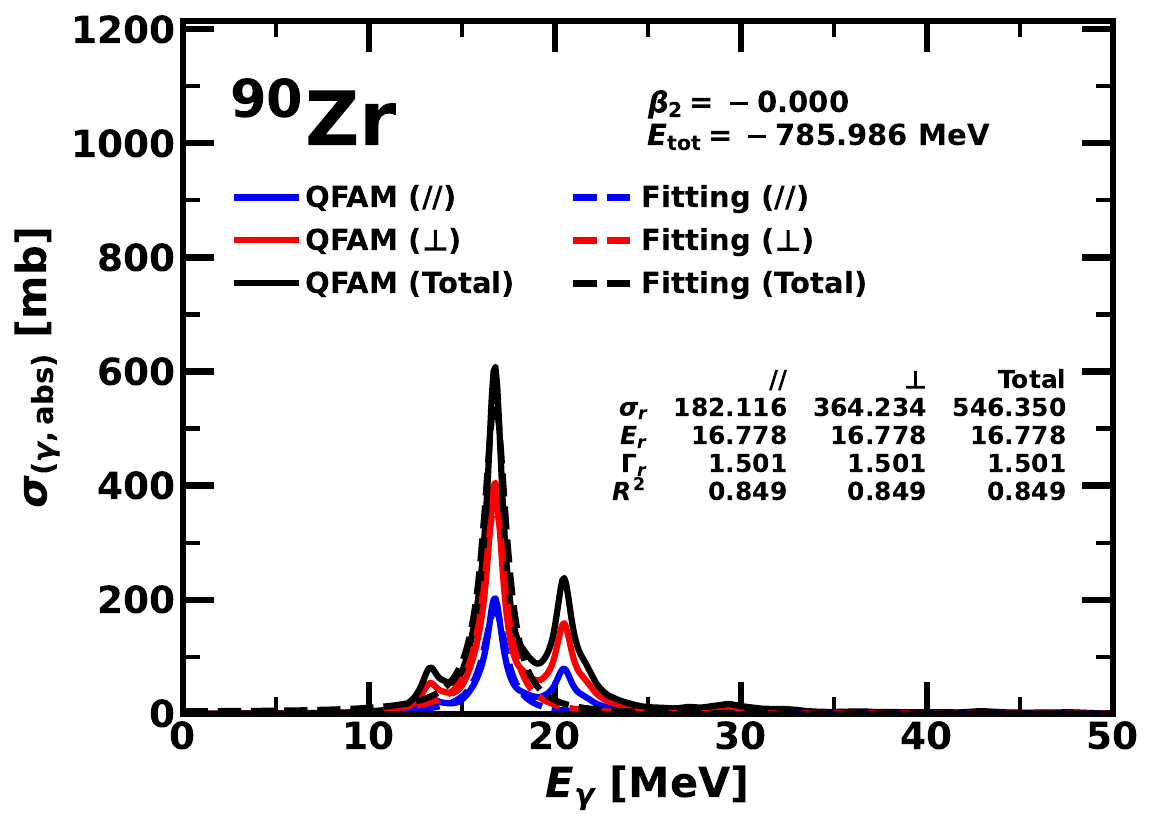}
\end{figure*}
\begin{figure*}\ContinuedFloat
    \centering
    \includegraphics[width=0.4\textwidth]{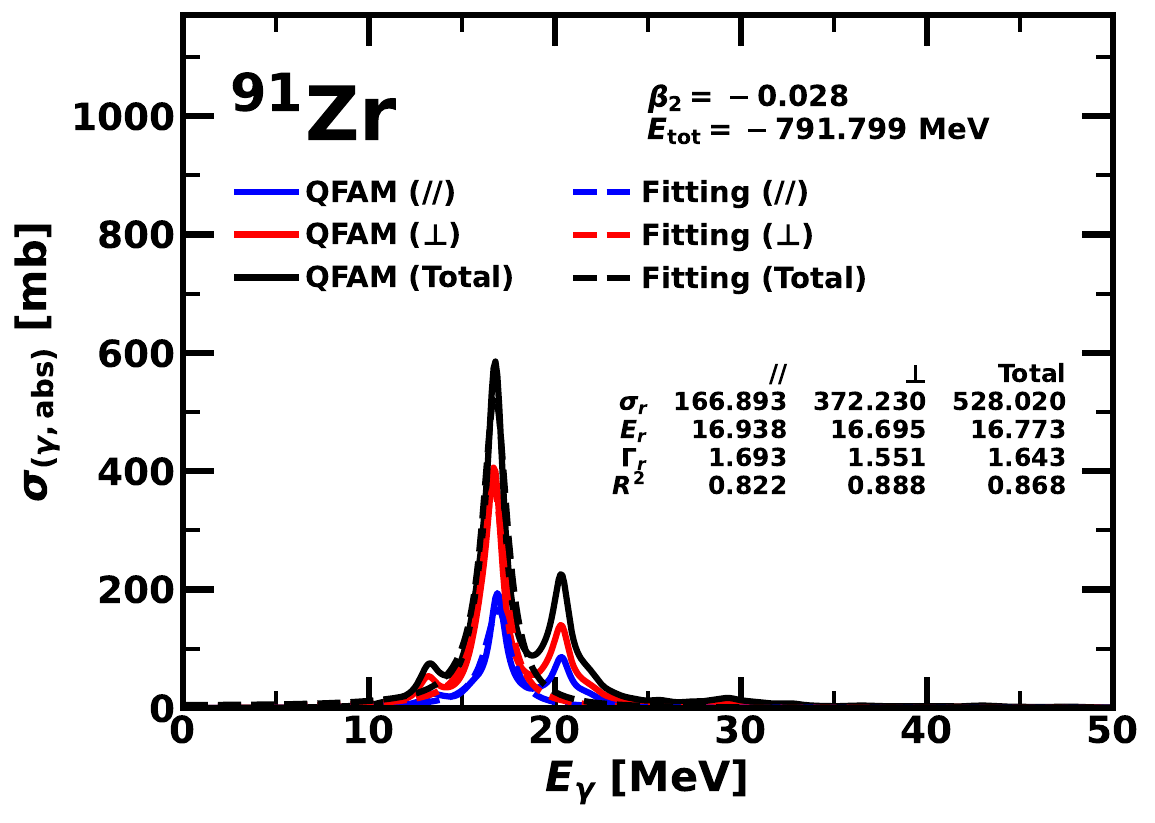}
    \includegraphics[width=0.4\textwidth]{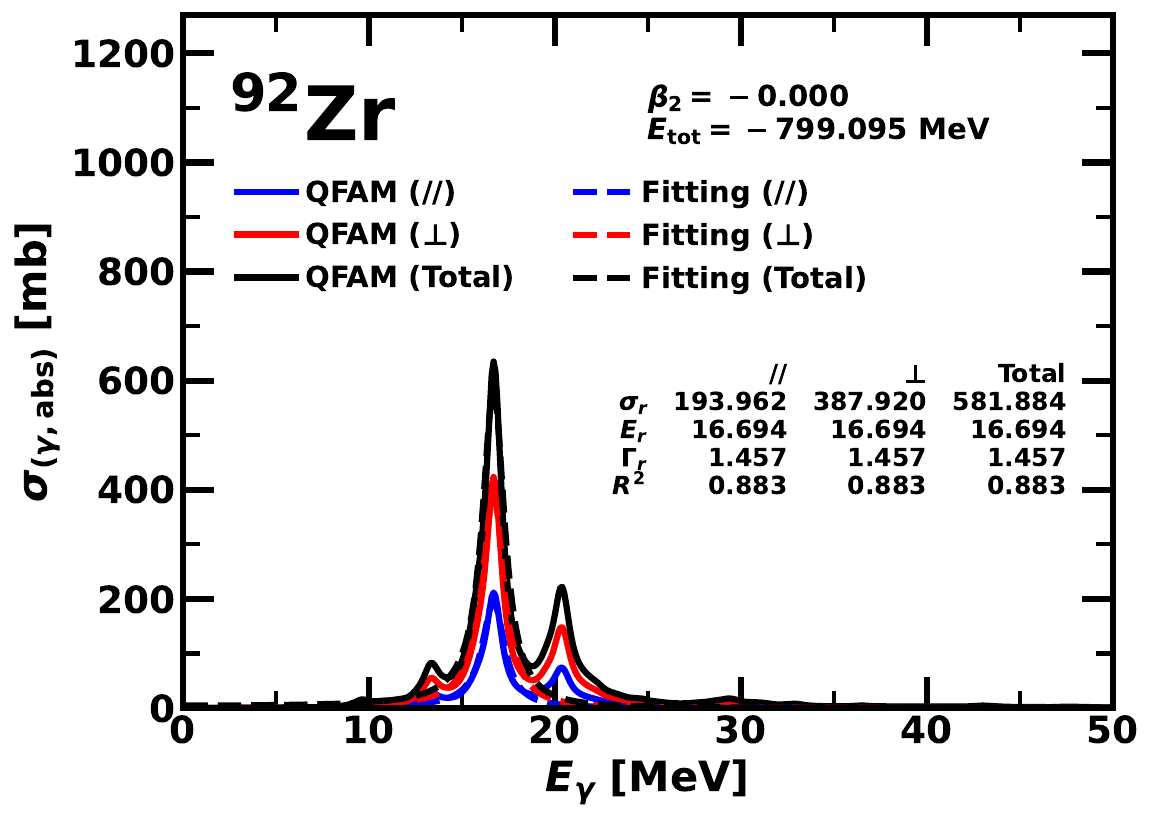}
    \includegraphics[width=0.4\textwidth]{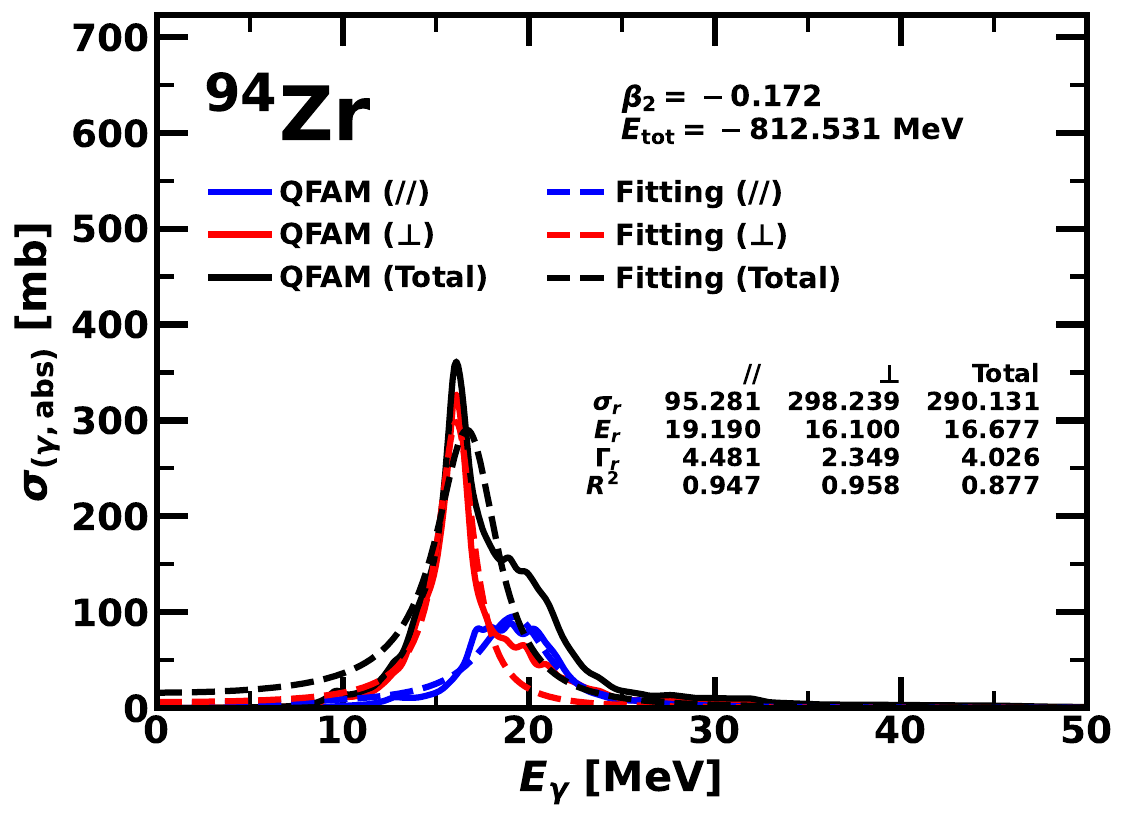}
    \includegraphics[width=0.4\textwidth]{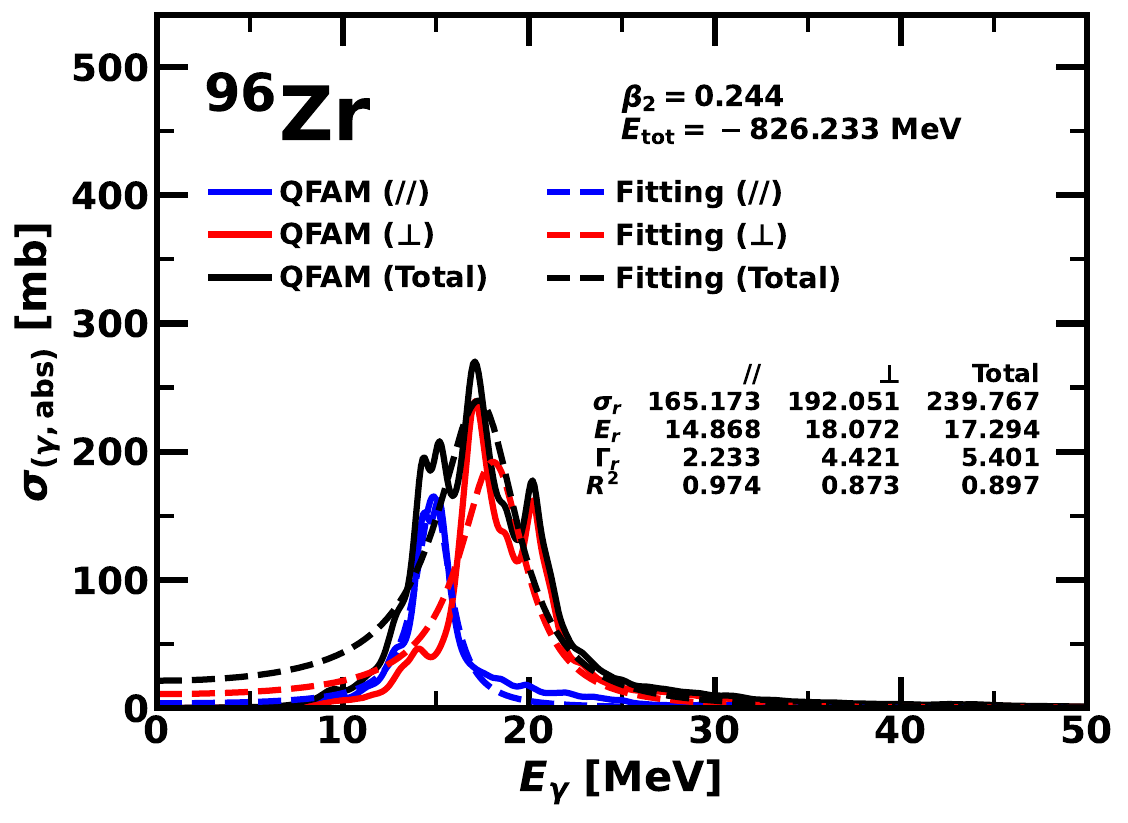}
    \includegraphics[width=0.4\textwidth]{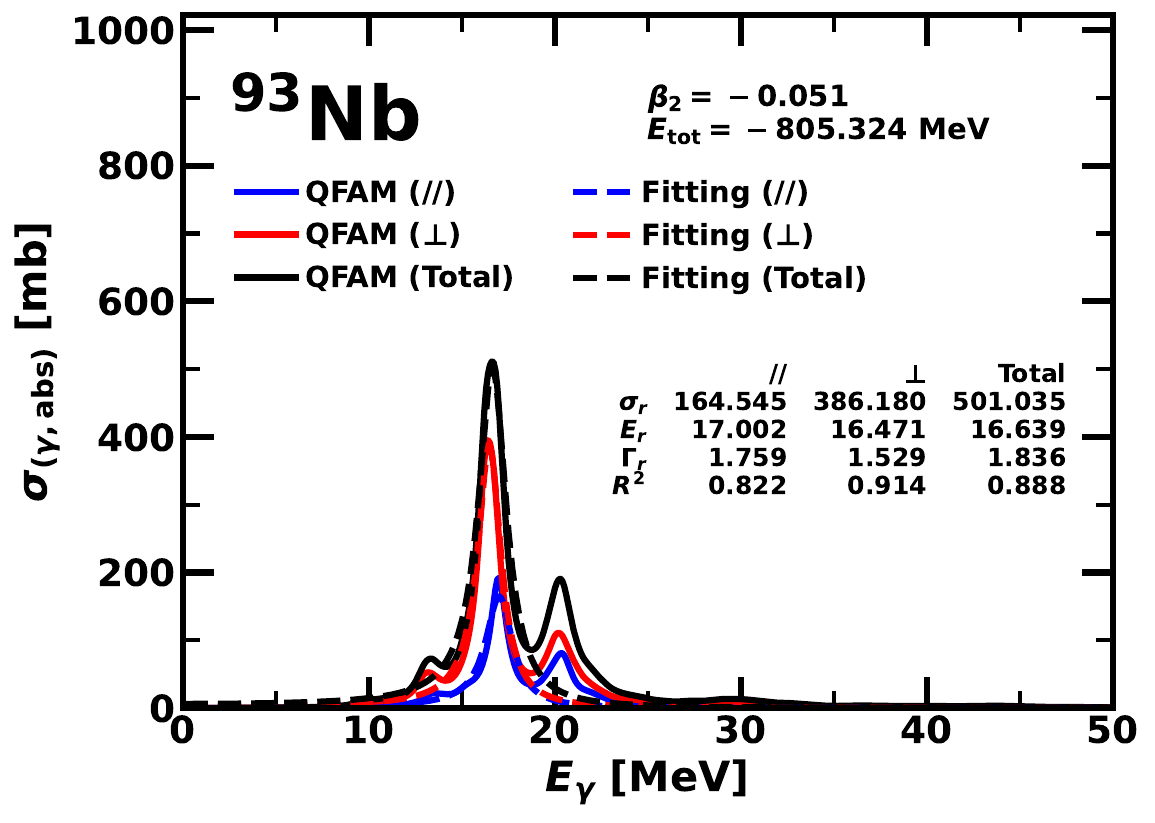}
    \includegraphics[width=0.4\textwidth]{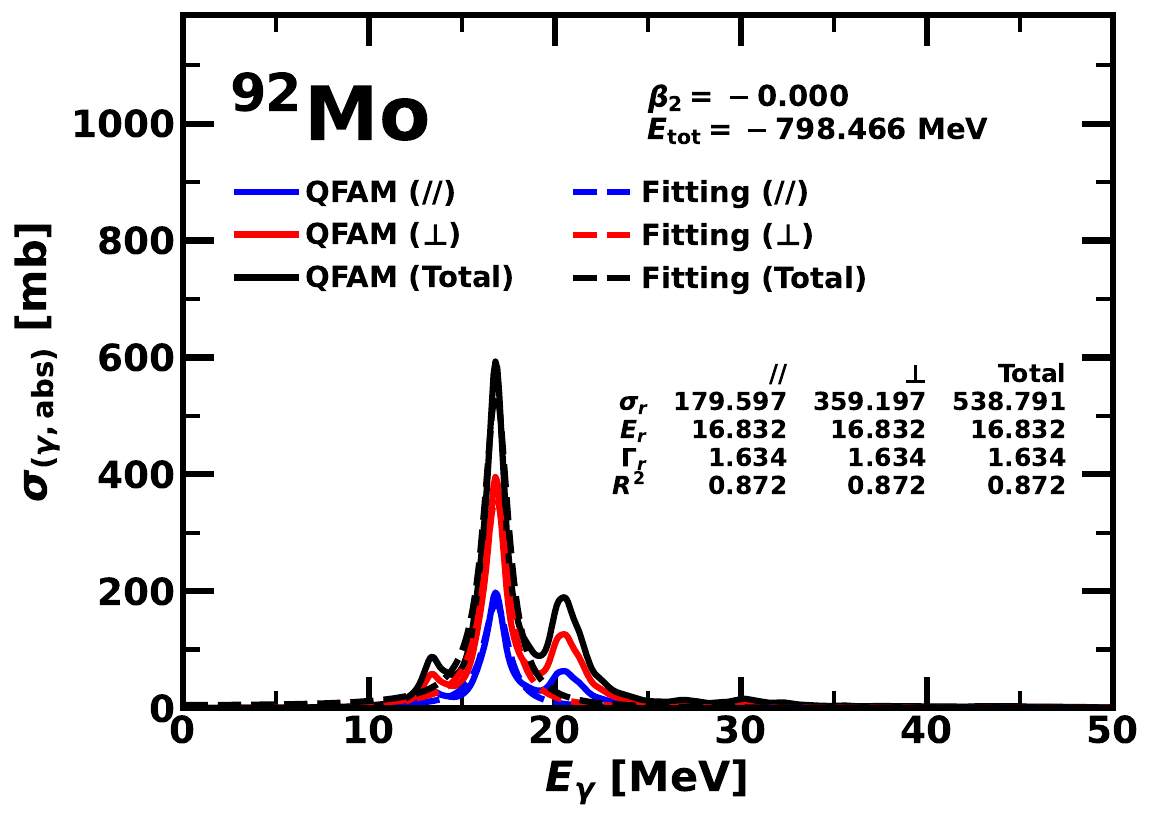}
    \includegraphics[width=0.4\textwidth]{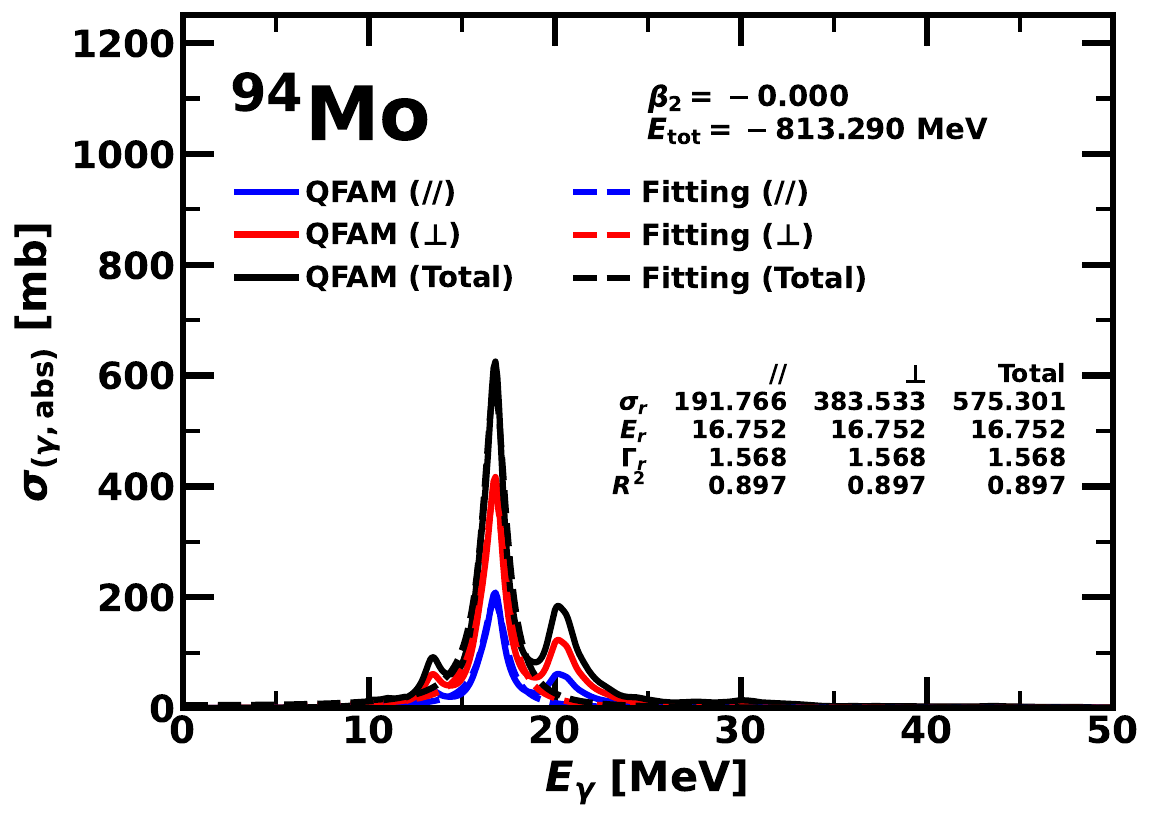}
    \includegraphics[width=0.4\textwidth]{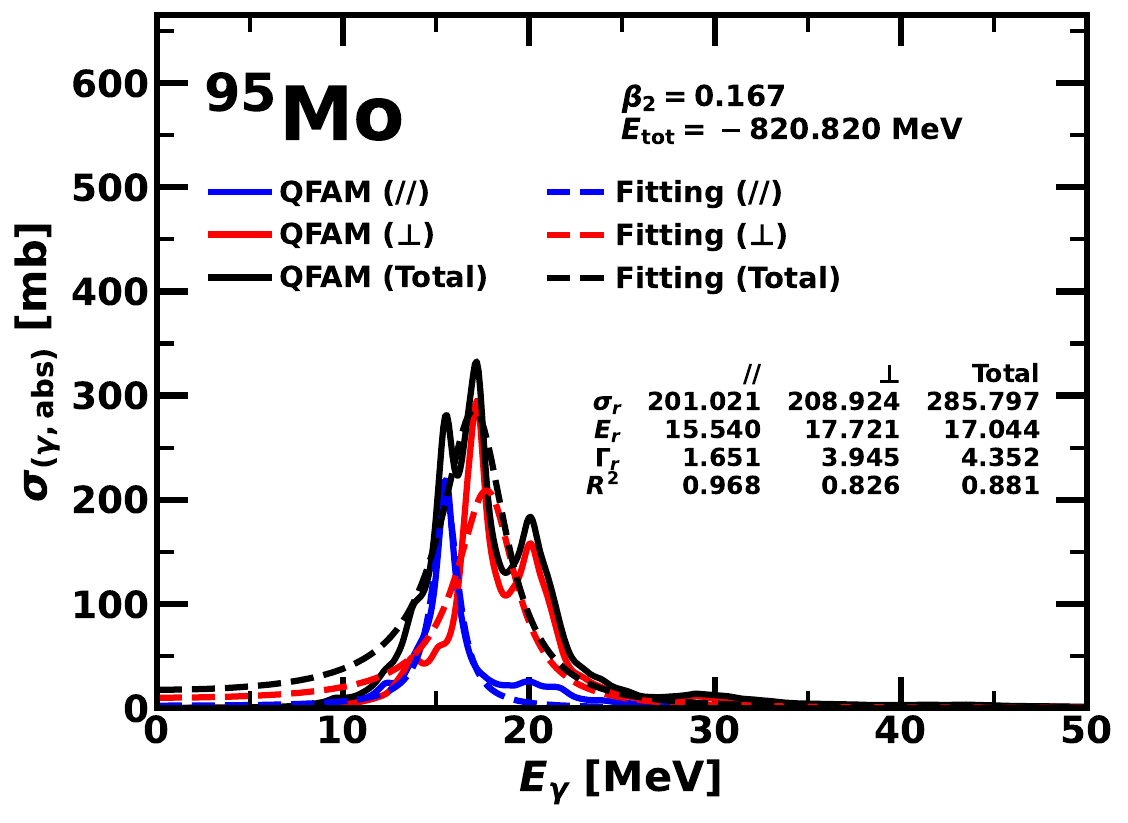}
\end{figure*}
\begin{figure*}\ContinuedFloat
    \centering
    \includegraphics[width=0.4\textwidth]{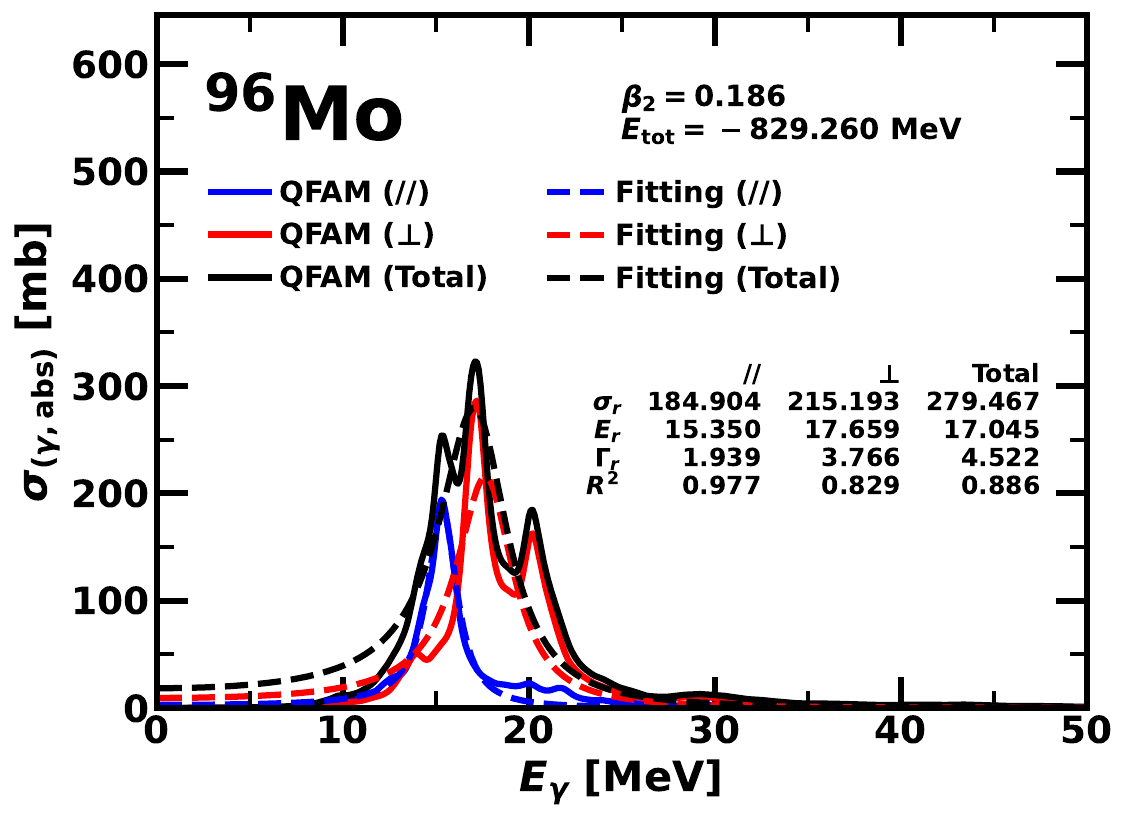}
    \includegraphics[width=0.4\textwidth]{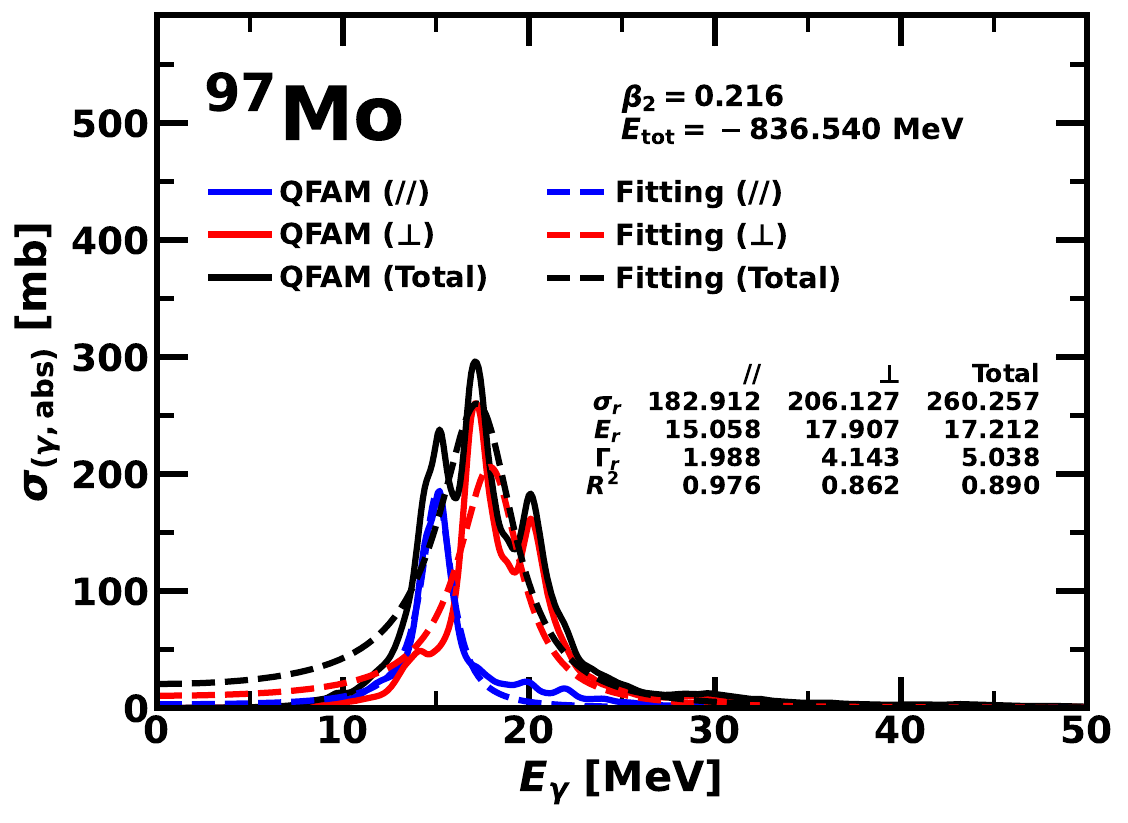}
    \includegraphics[width=0.4\textwidth]{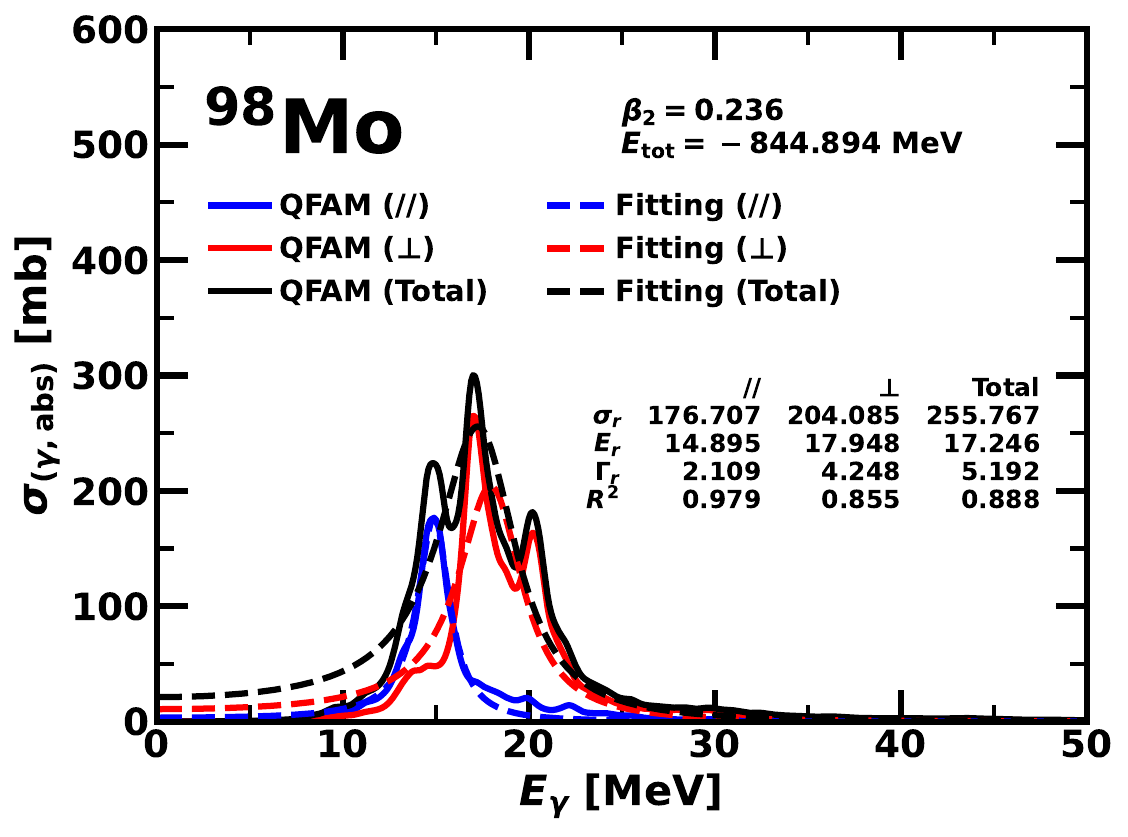}
    \includegraphics[width=0.4\textwidth]{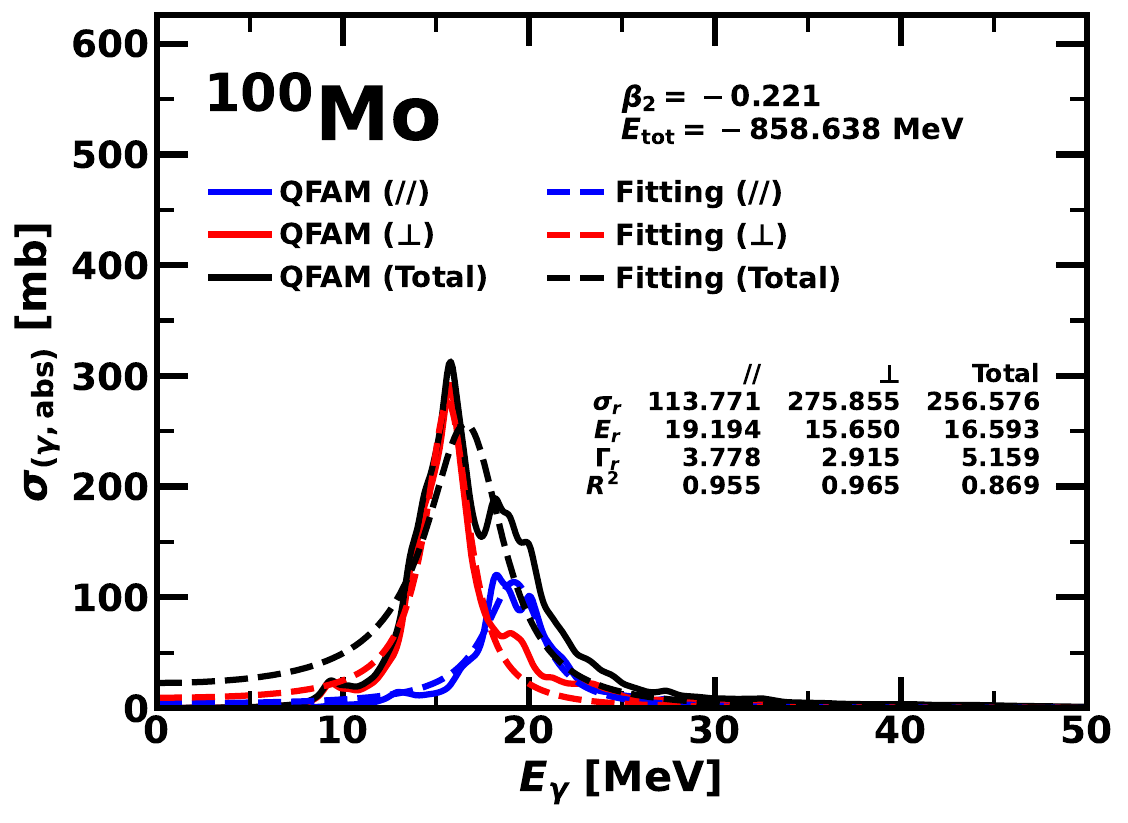}
    \includegraphics[width=0.4\textwidth]{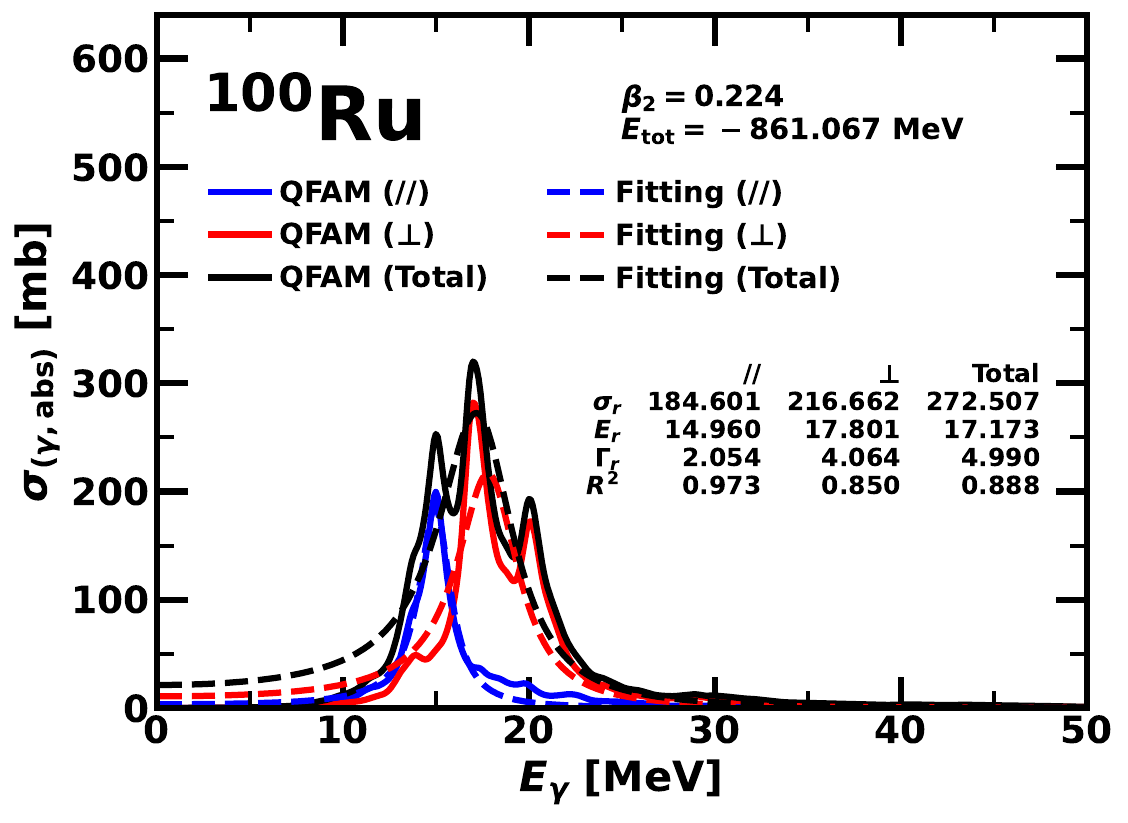}
    \includegraphics[width=0.4\textwidth]{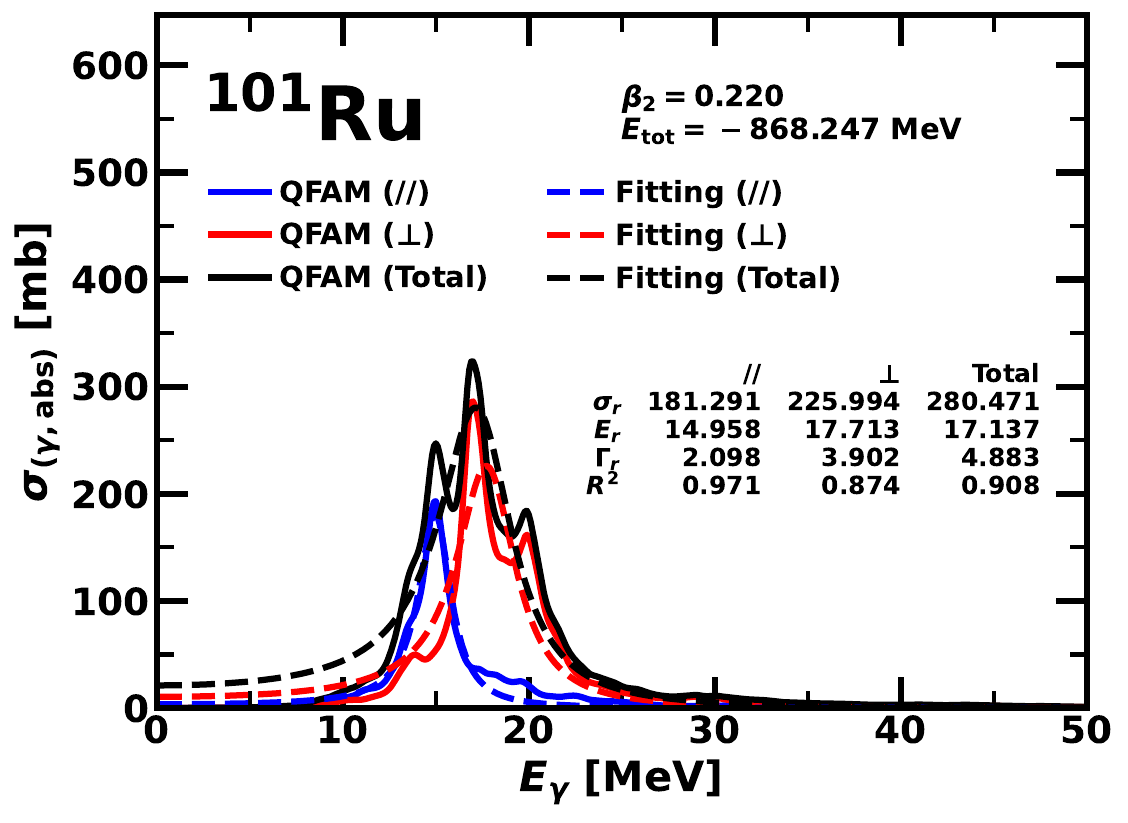}
    \includegraphics[width=0.4\textwidth]{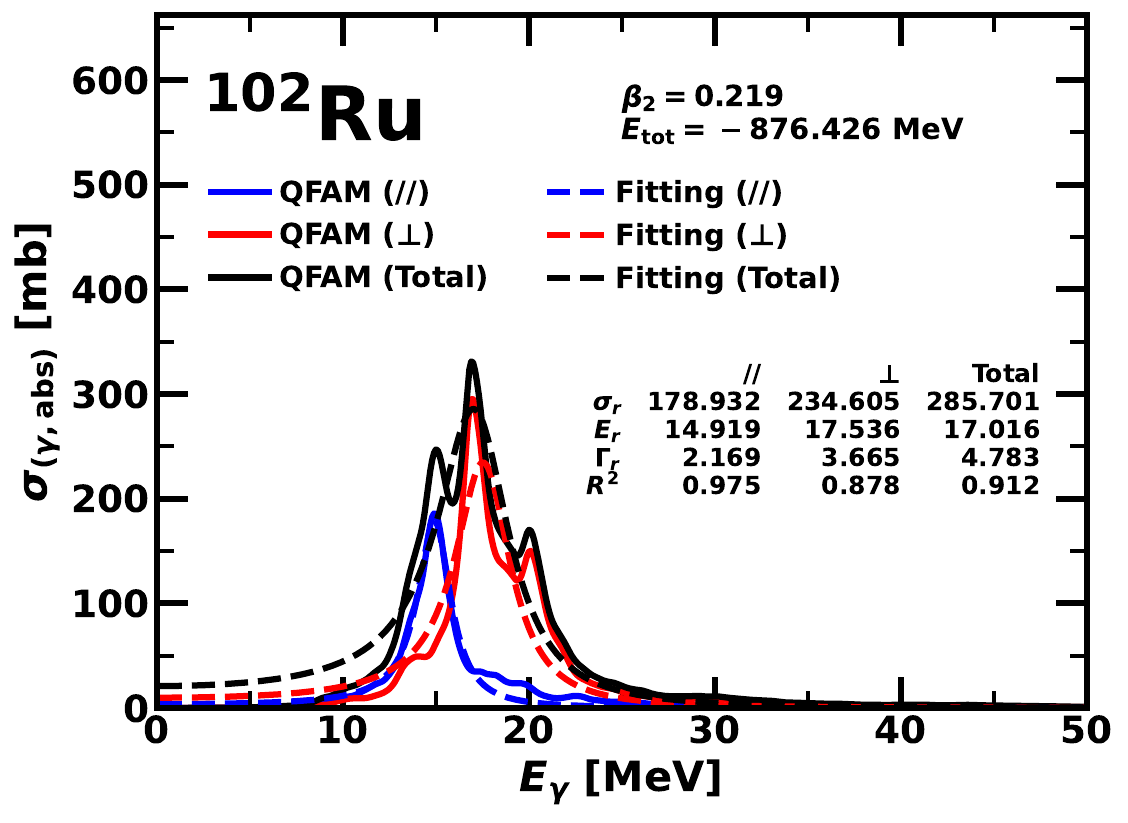}
    \includegraphics[width=0.4\textwidth]{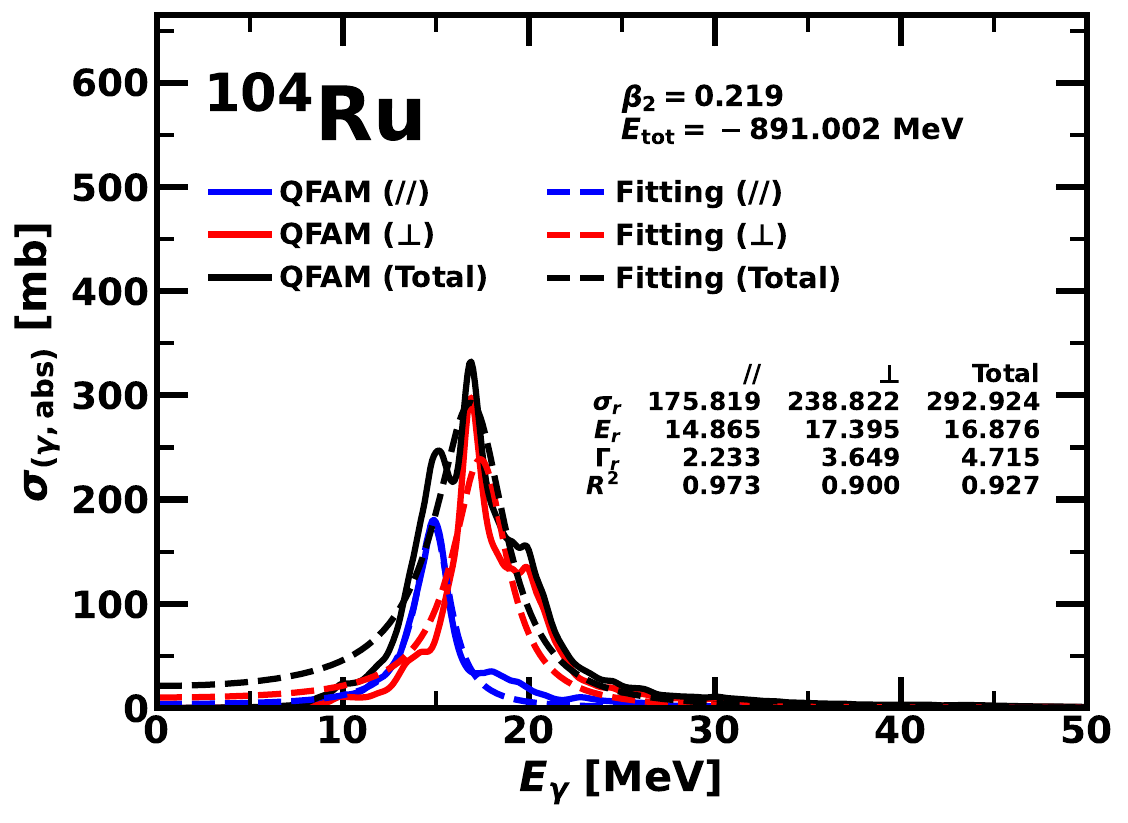}
\end{figure*}
\begin{figure*}\ContinuedFloat
    \centering
    \includegraphics[width=0.4\textwidth]{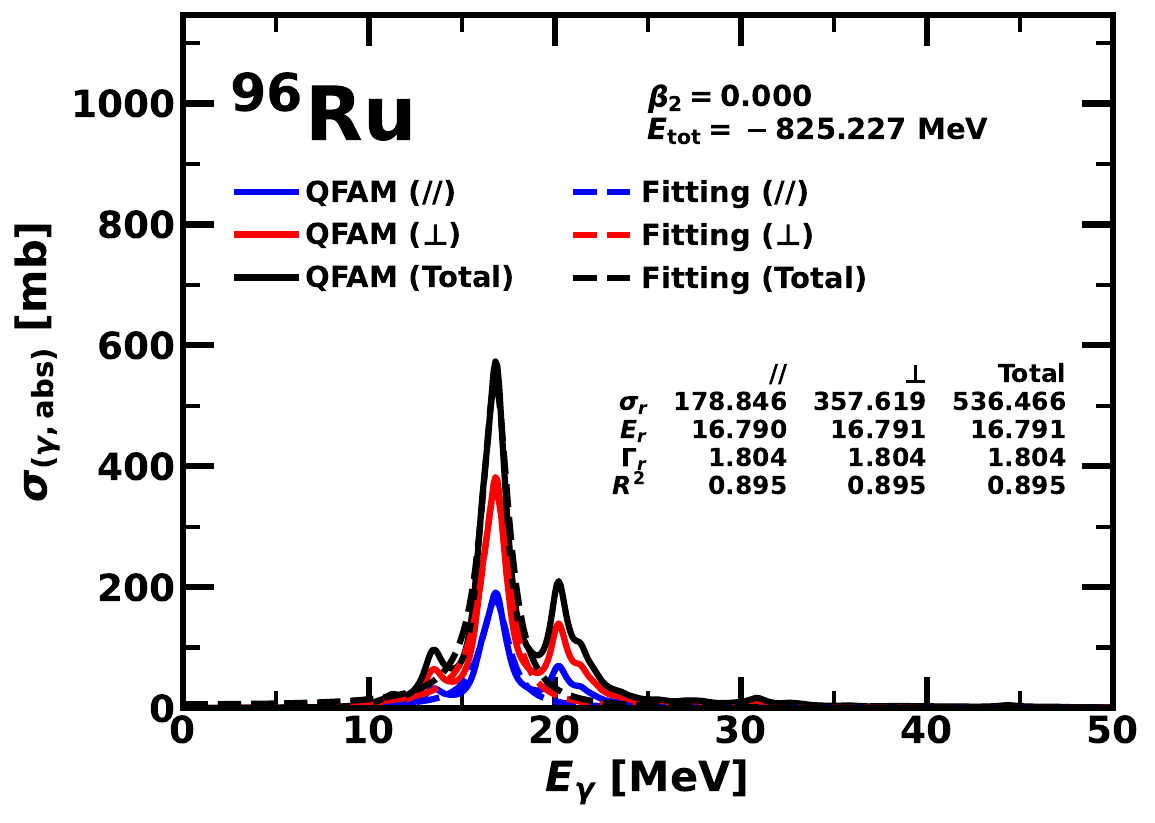}
    \includegraphics[width=0.4\textwidth]{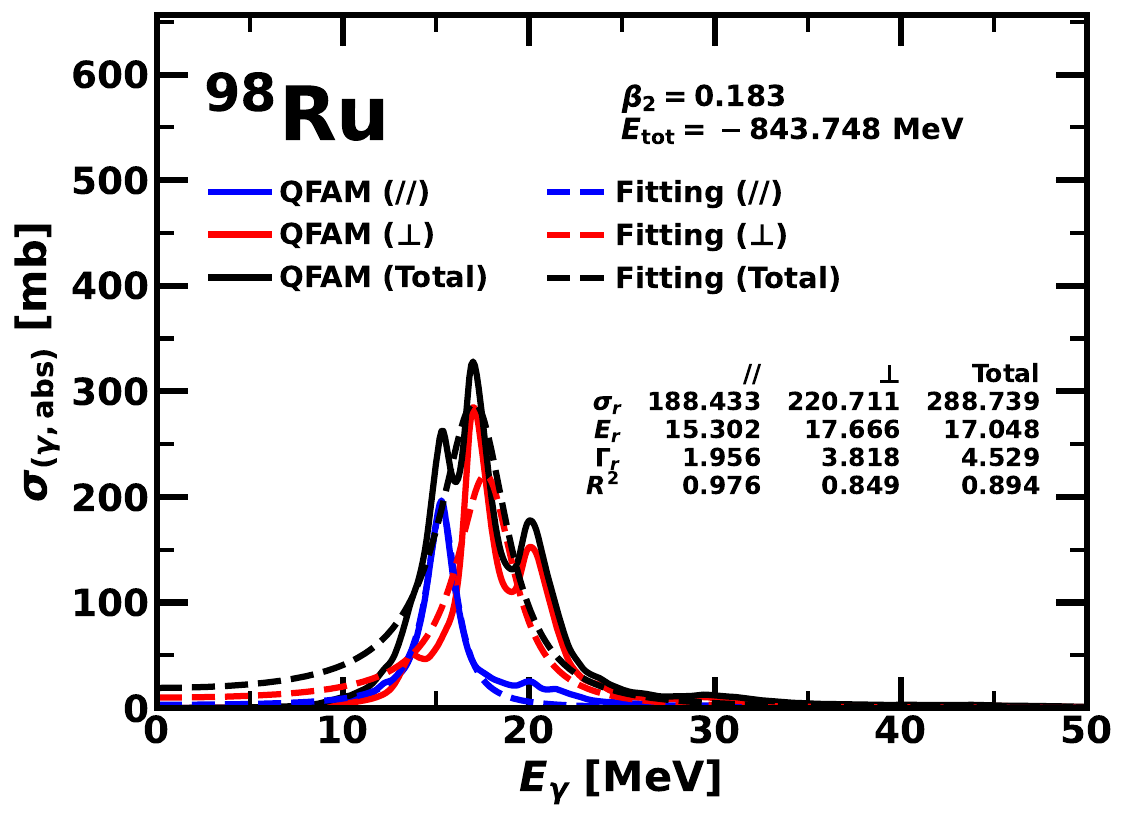}
    \includegraphics[width=0.4\textwidth]{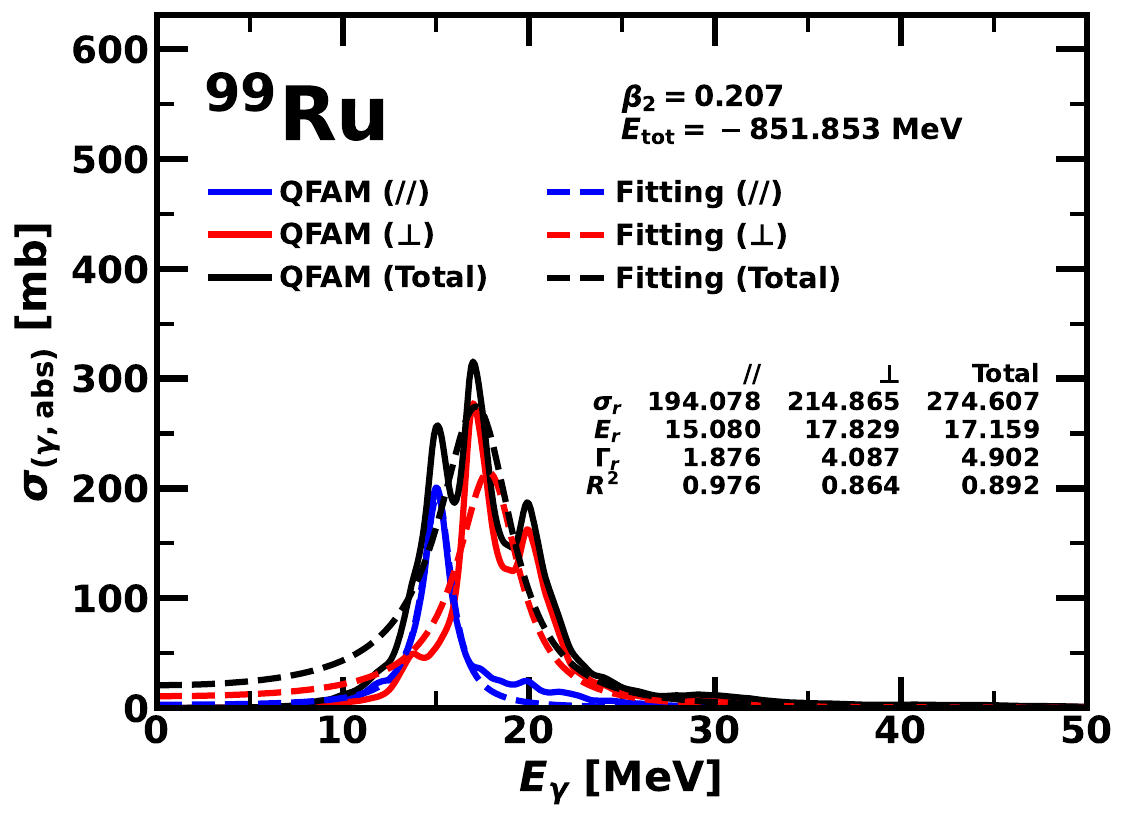}
    \includegraphics[width=0.4\textwidth]{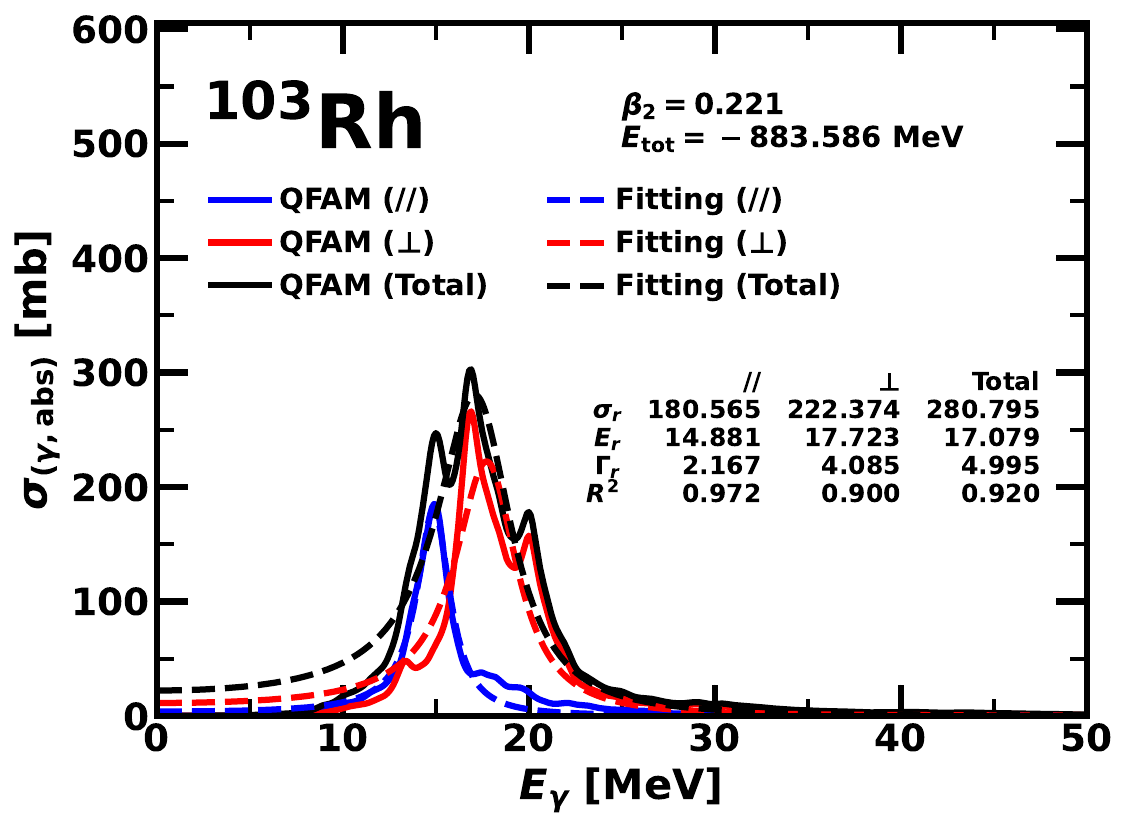}
    \includegraphics[width=0.4\textwidth]{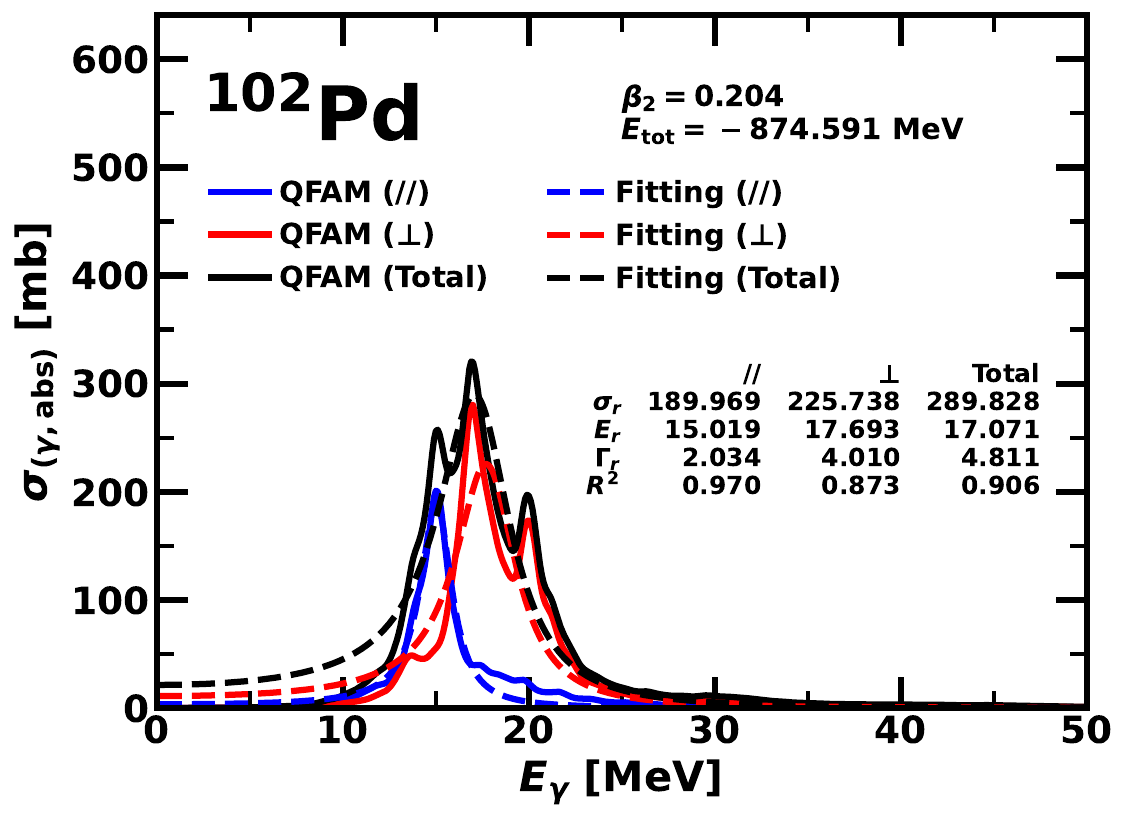}
    \includegraphics[width=0.4\textwidth]{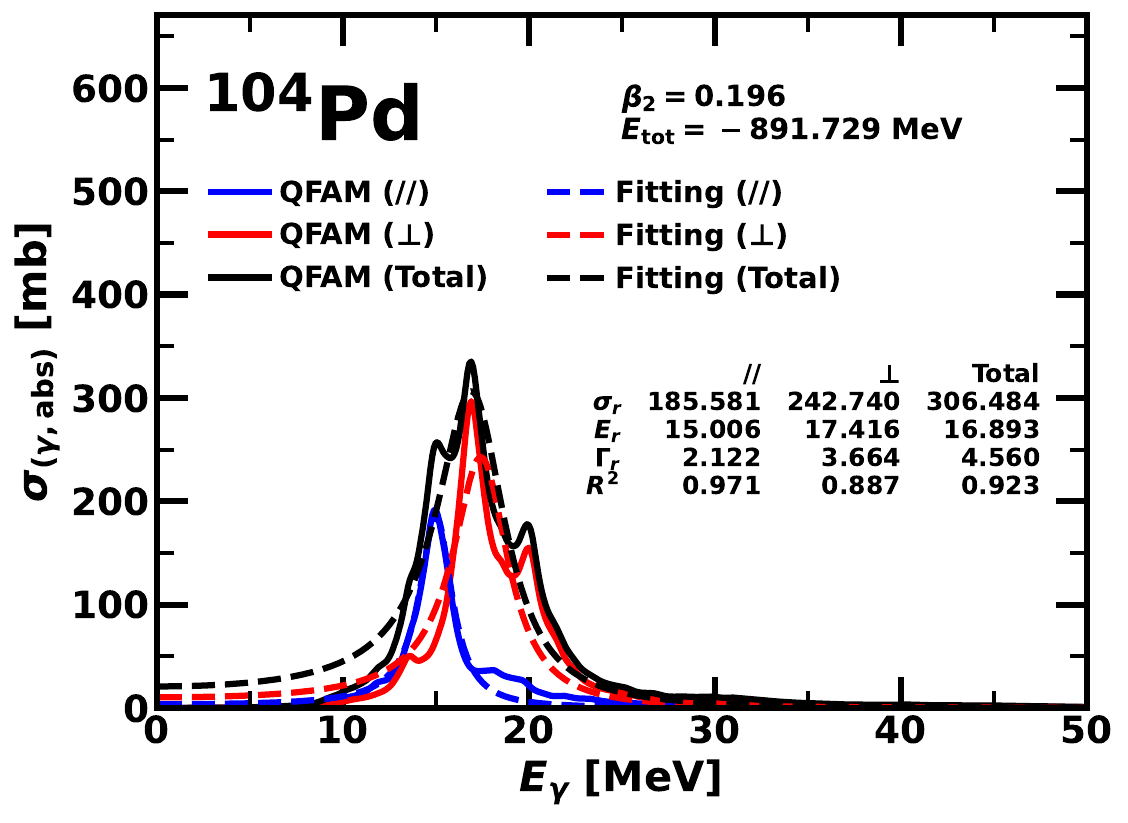}
    \includegraphics[width=0.4\textwidth]{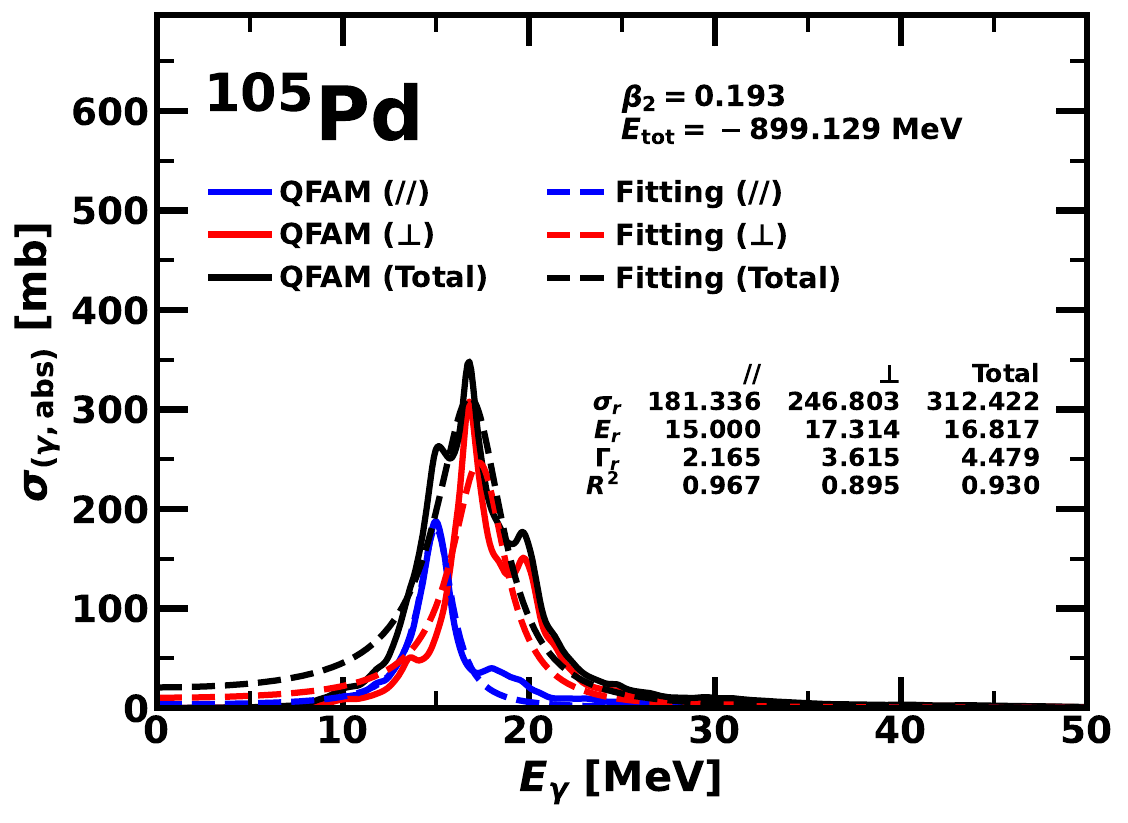}
    \includegraphics[width=0.4\textwidth]{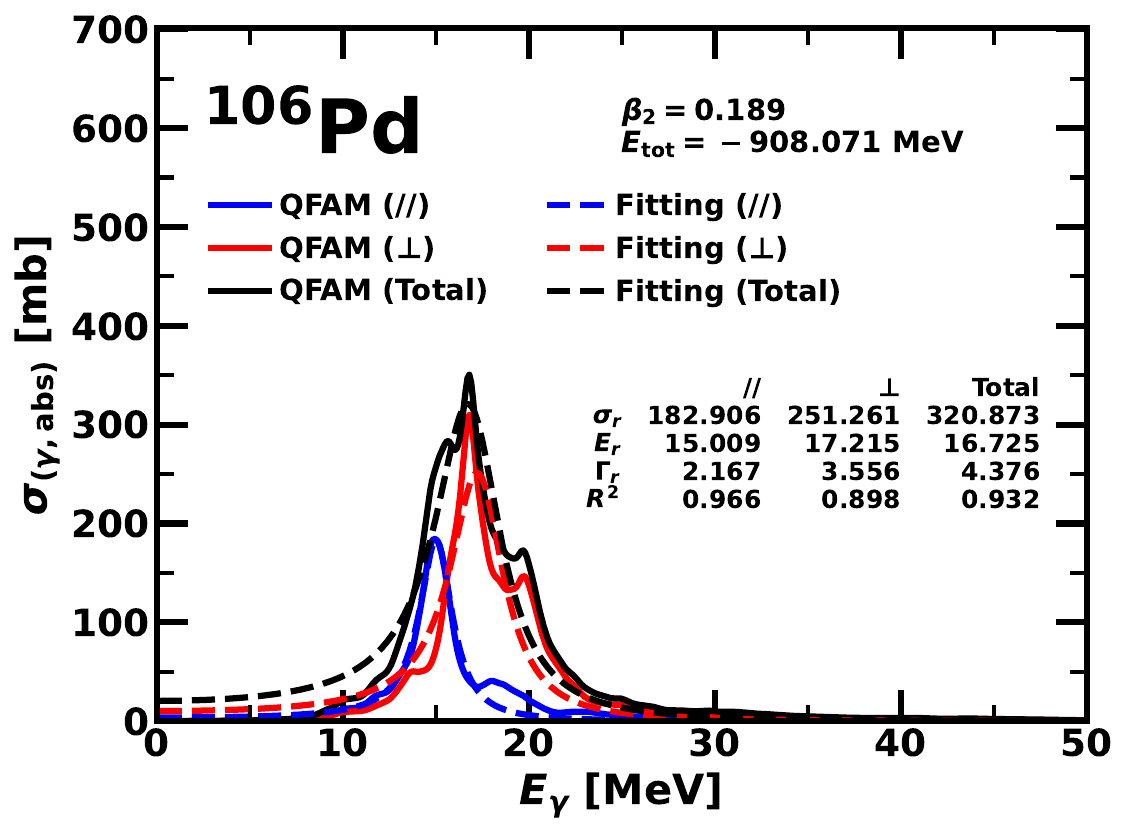}
\end{figure*}
\begin{figure*}\ContinuedFloat
    \centering
    \includegraphics[width=0.4\textwidth]{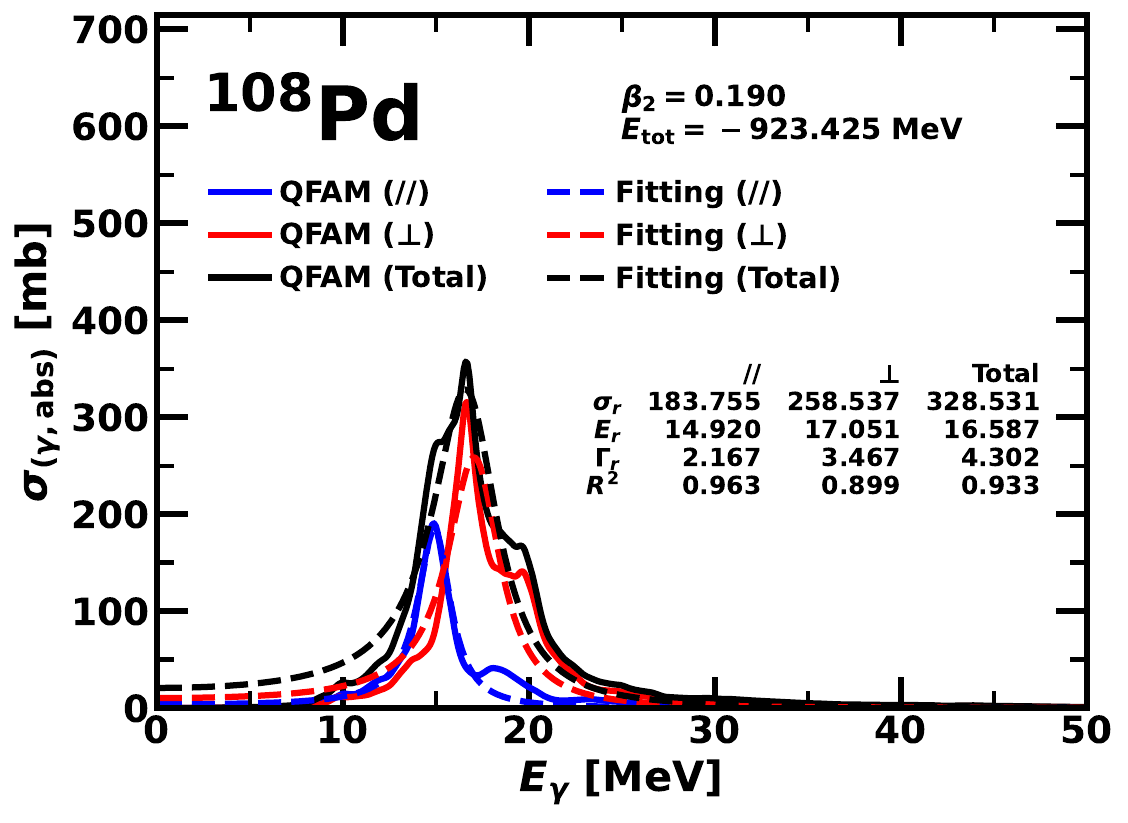}
    \includegraphics[width=0.4\textwidth]{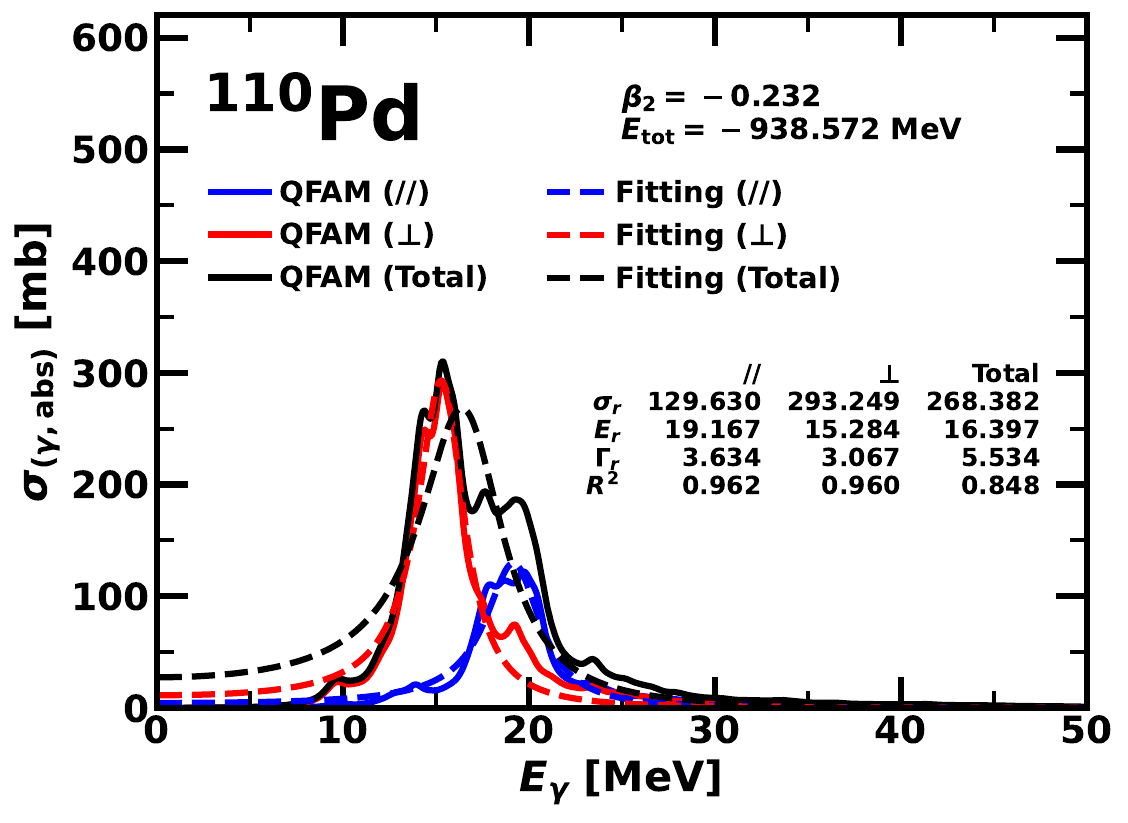}
    \includegraphics[width=0.4\textwidth]{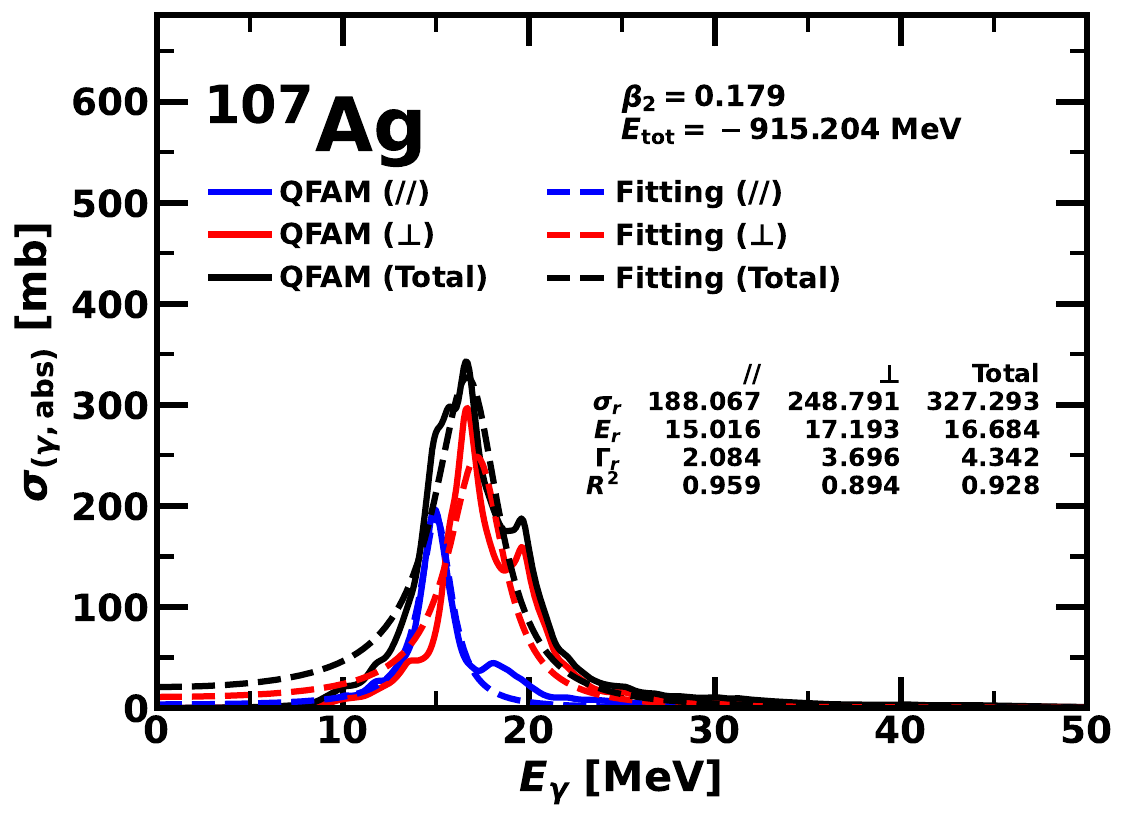}
    \includegraphics[width=0.4\textwidth]{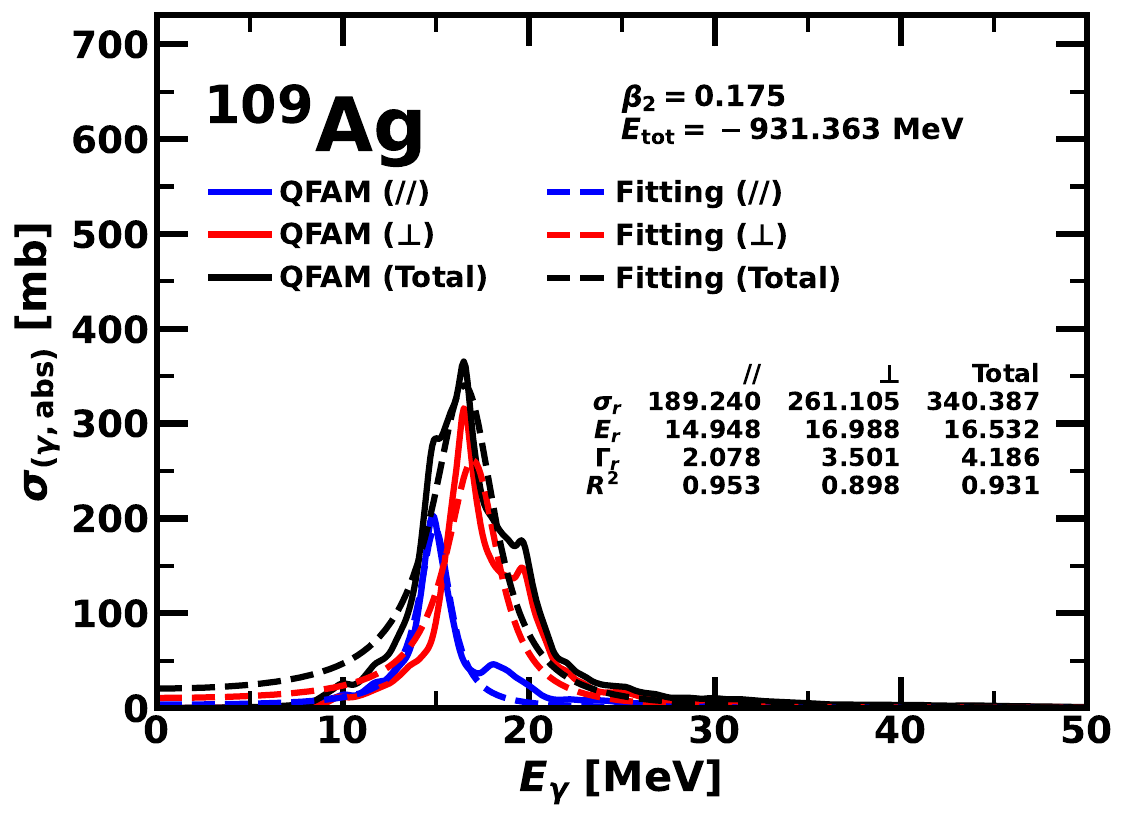}
    \includegraphics[width=0.4\textwidth]{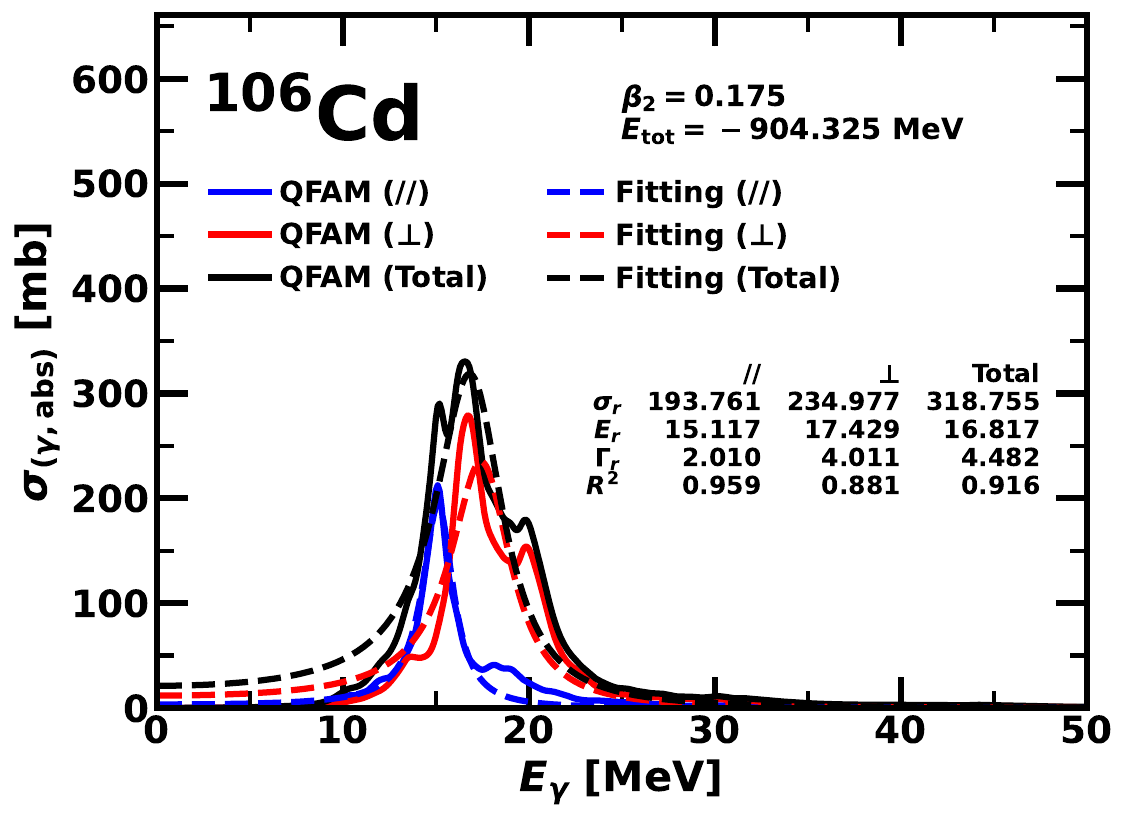}
    \includegraphics[width=0.4\textwidth]{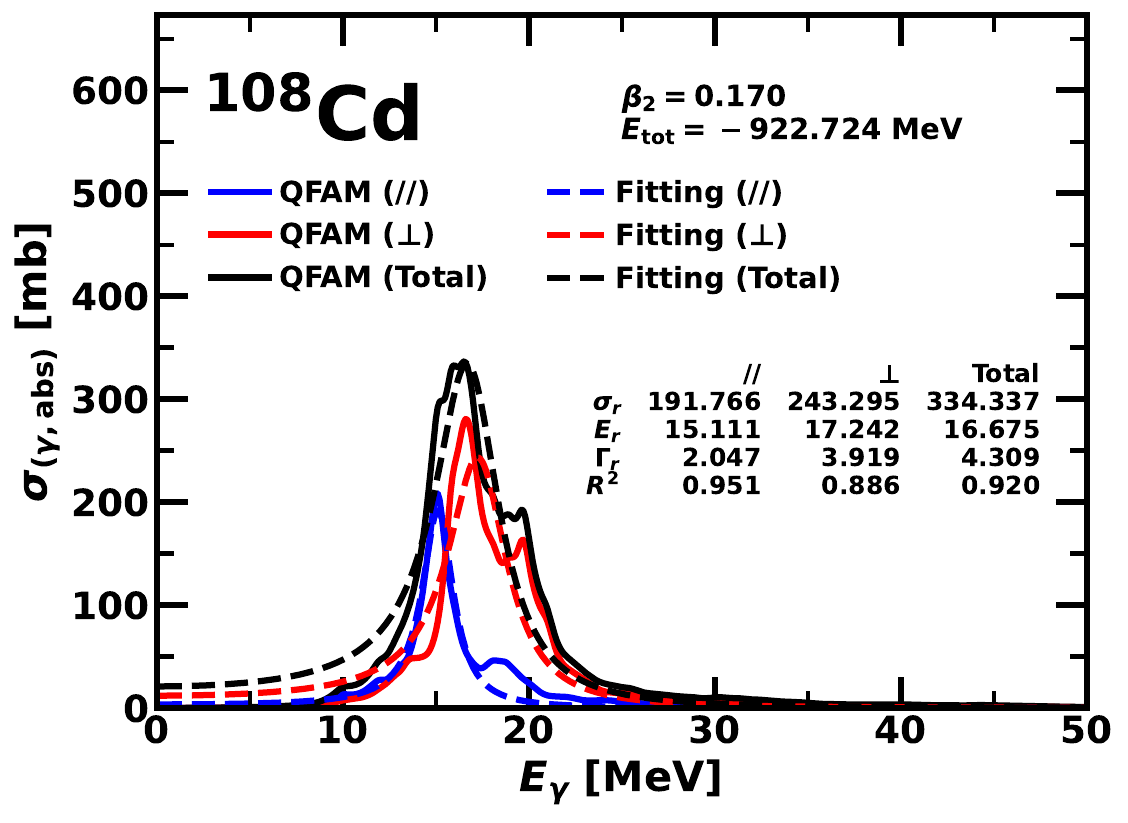}
    \includegraphics[width=0.4\textwidth]{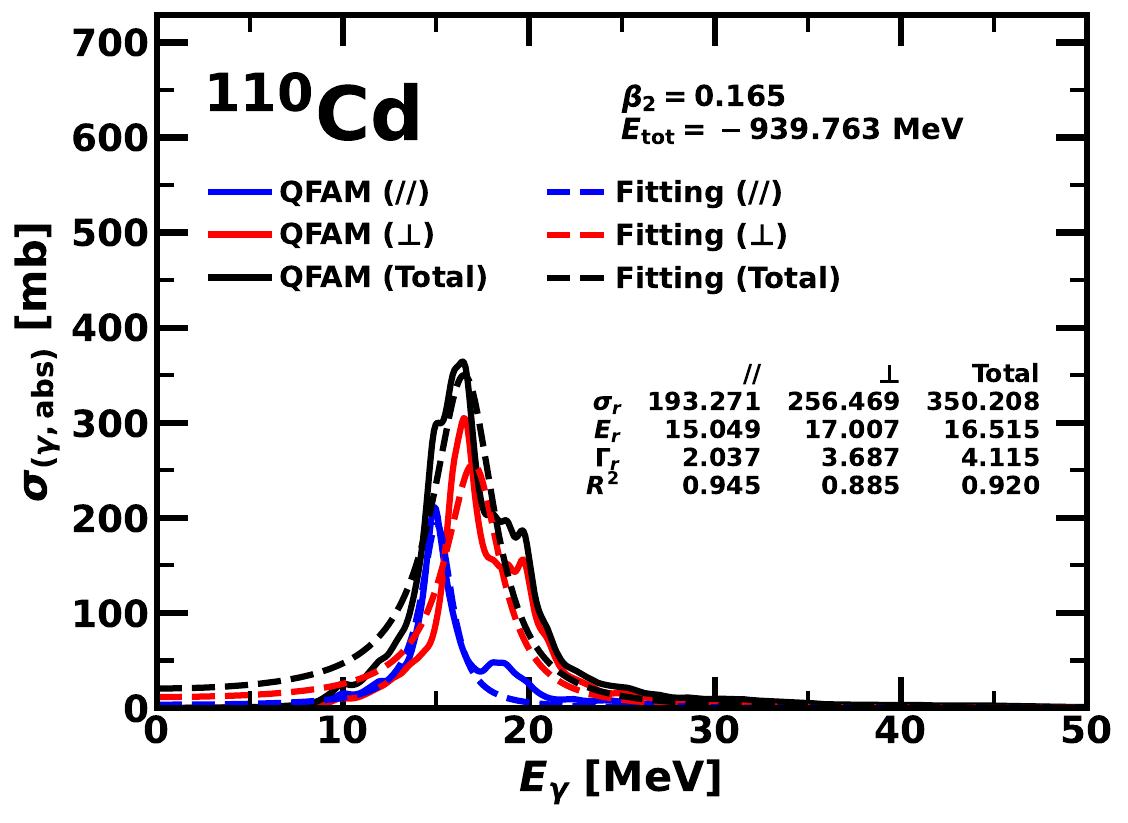}
    \includegraphics[width=0.4\textwidth]{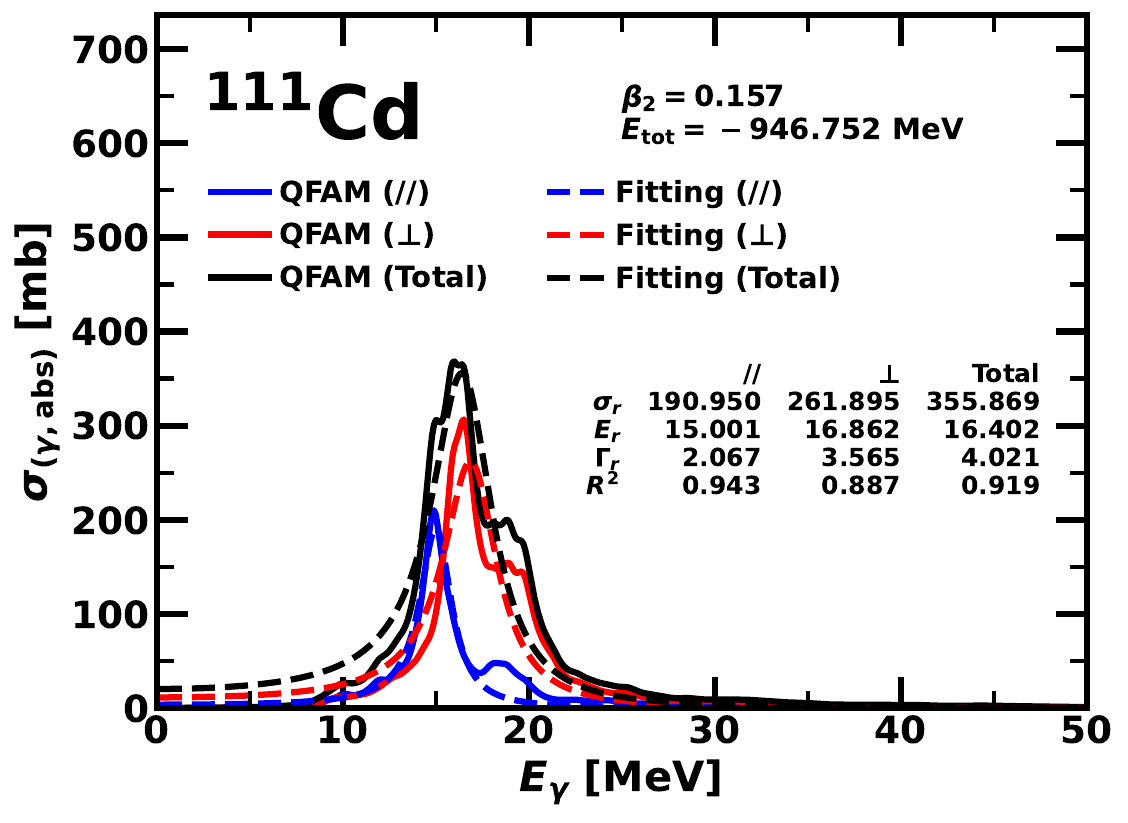}
\end{figure*}
\begin{figure*}\ContinuedFloat
    \centering
    \includegraphics[width=0.4\textwidth]{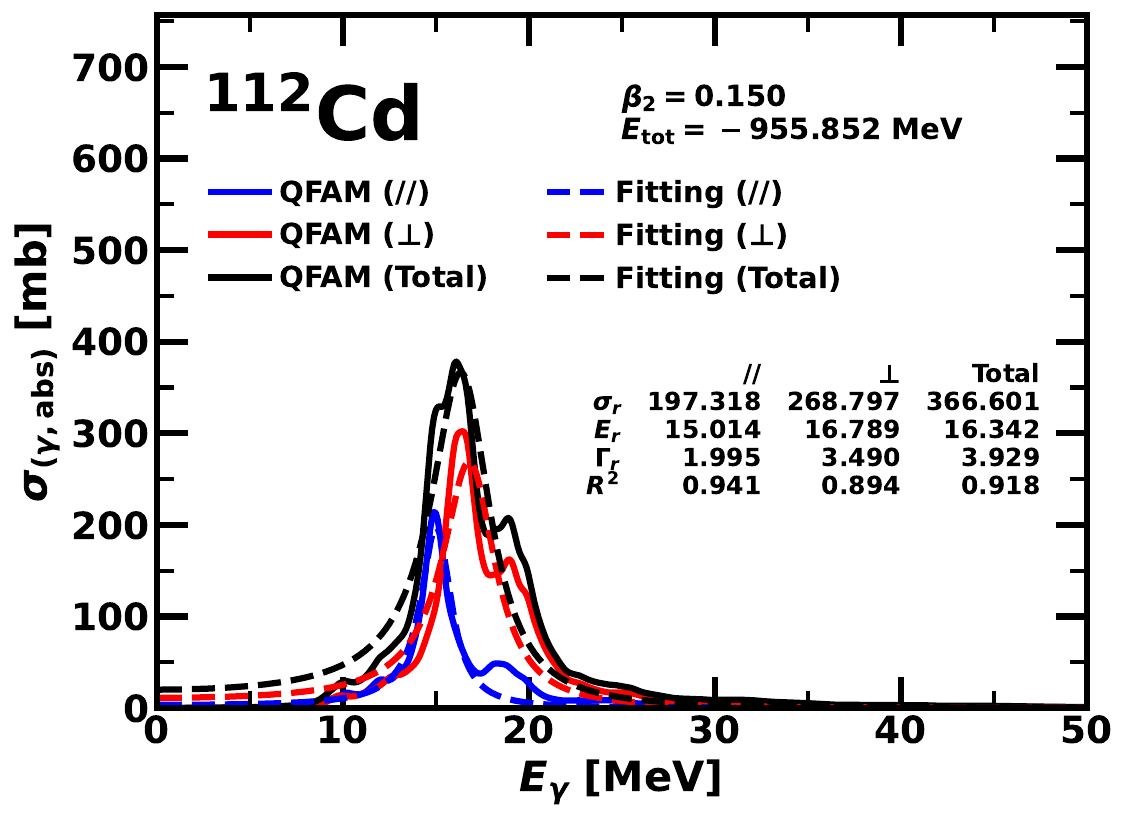}
    \includegraphics[width=0.4\textwidth]{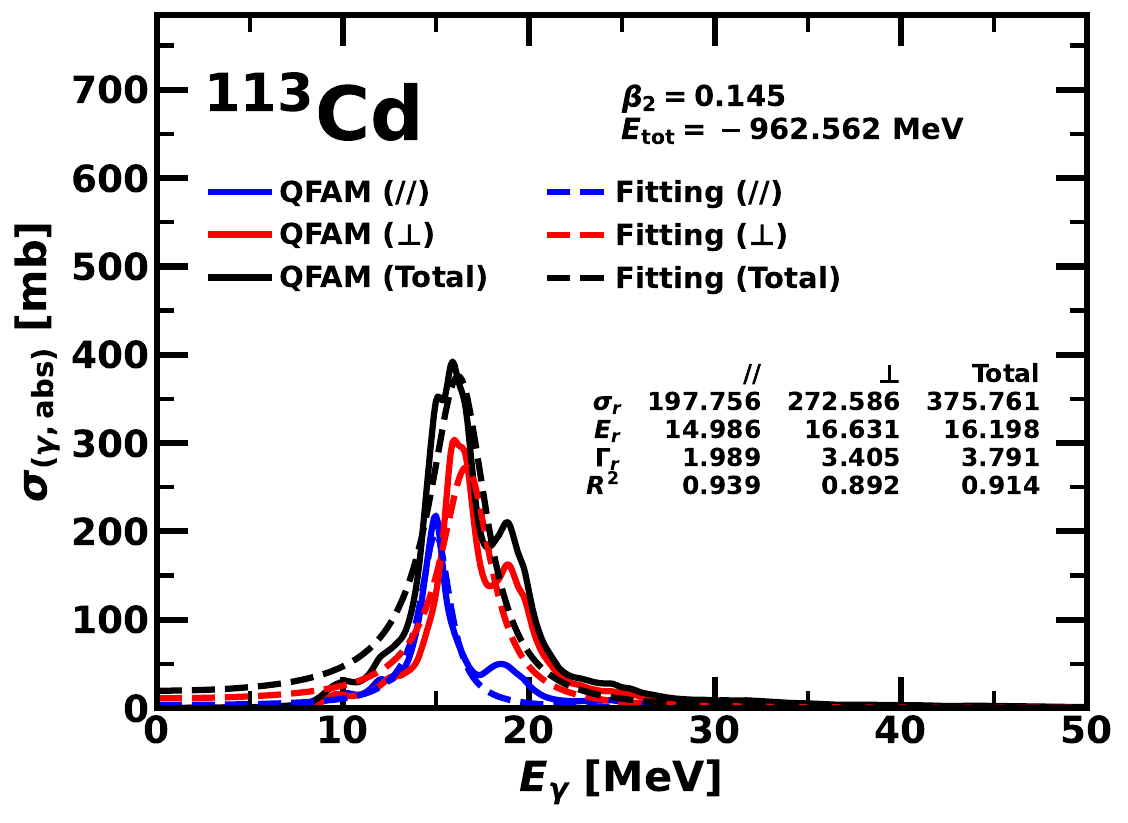}
    \includegraphics[width=0.4\textwidth]{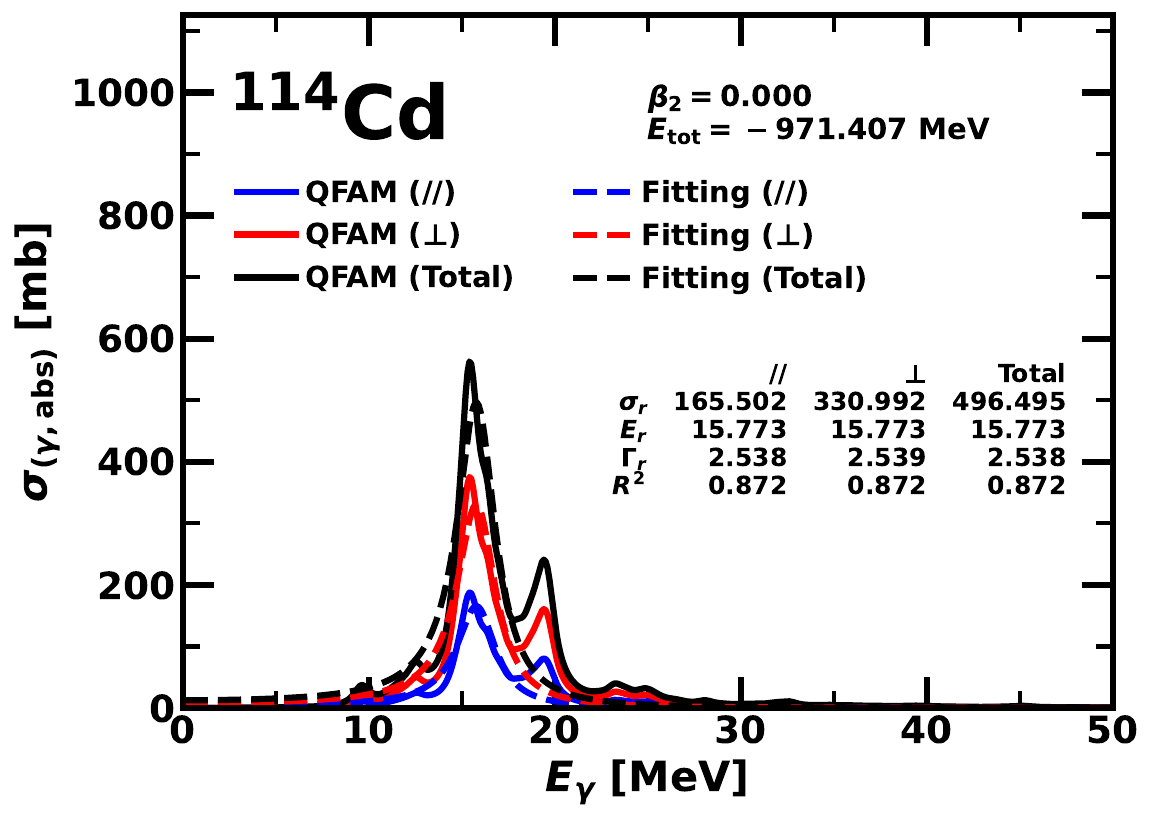}
    \includegraphics[width=0.4\textwidth]{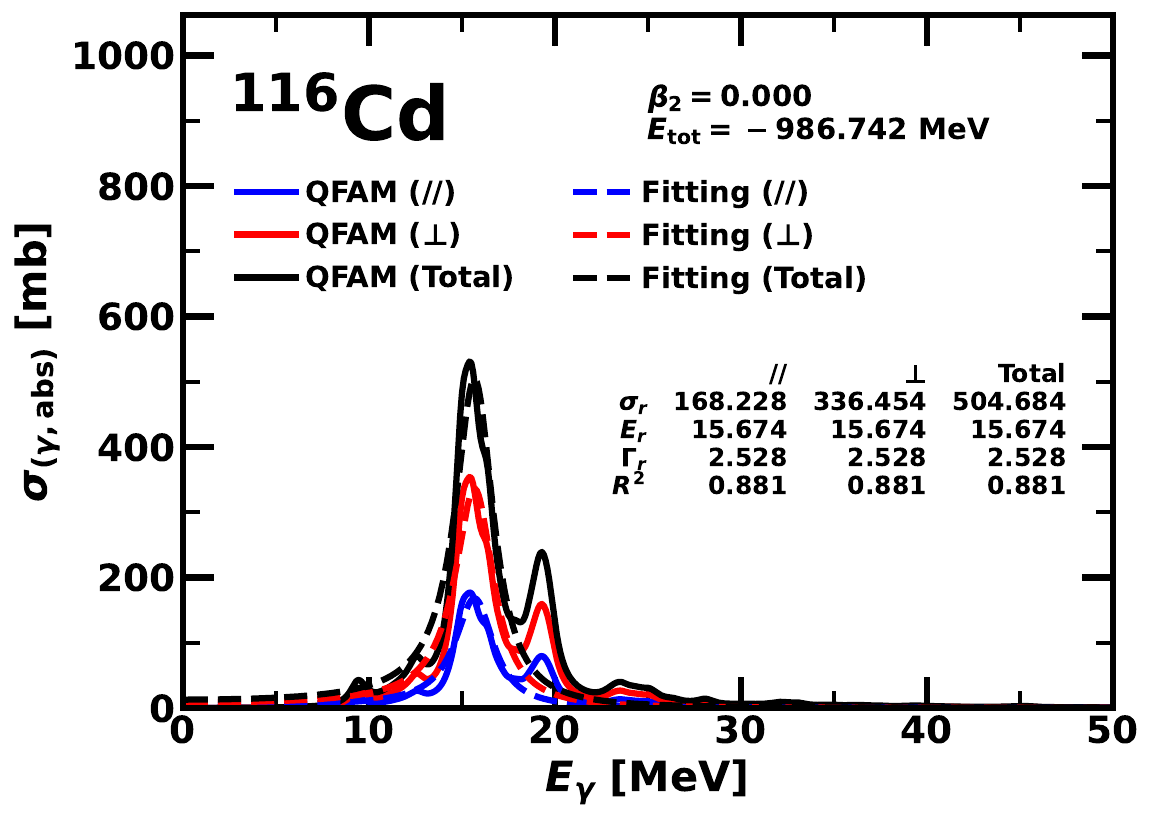}
    \includegraphics[width=0.4\textwidth]{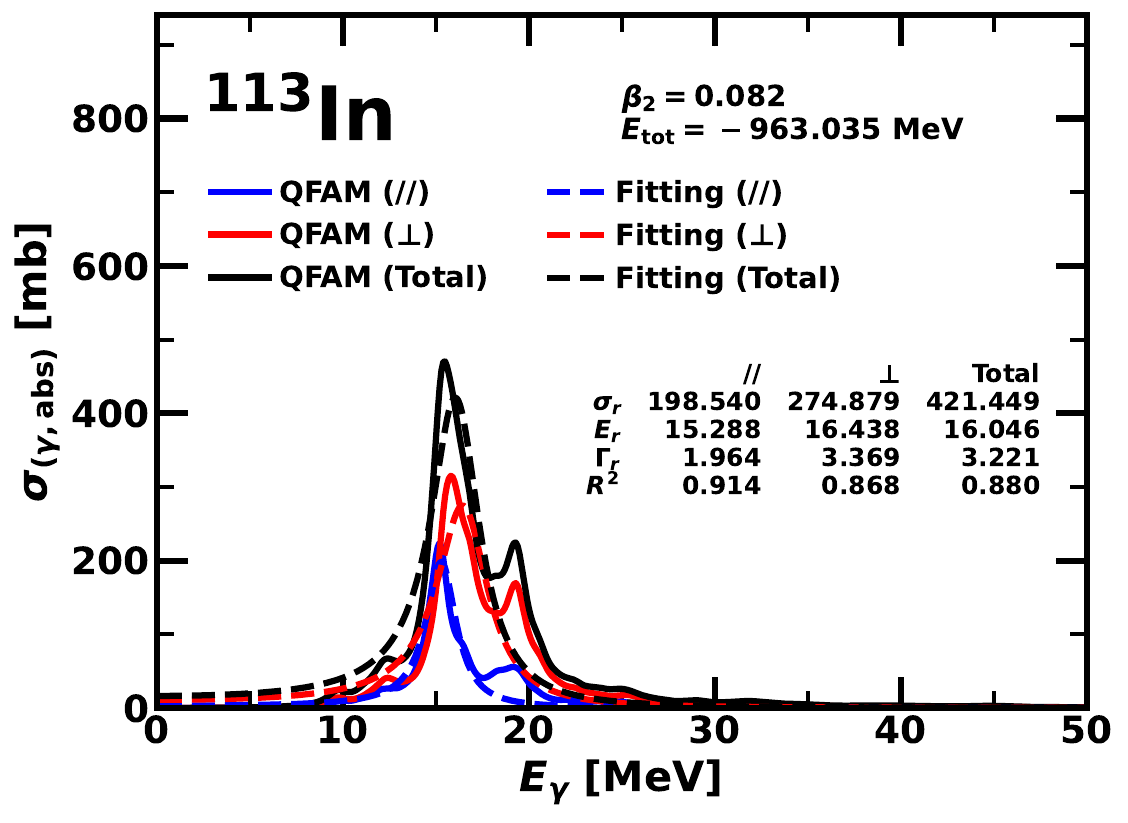}
    \includegraphics[width=0.4\textwidth]{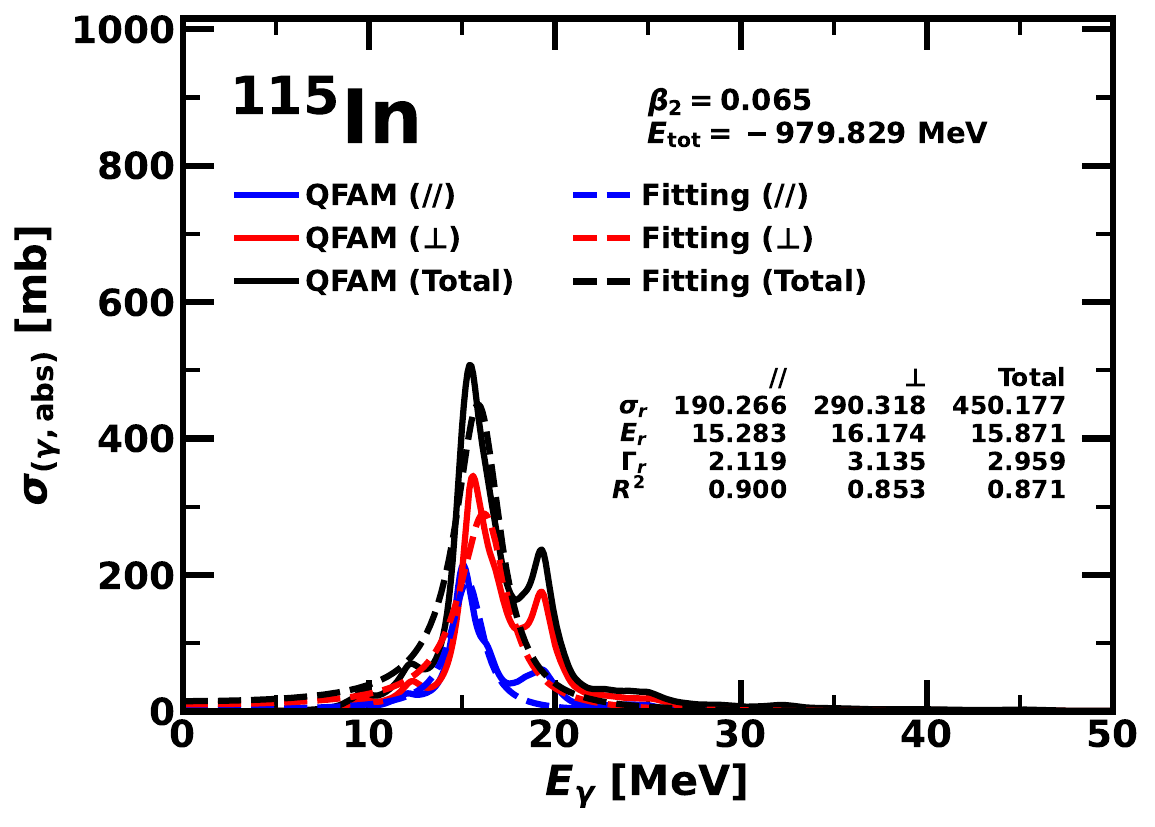}
    \includegraphics[width=0.4\textwidth]{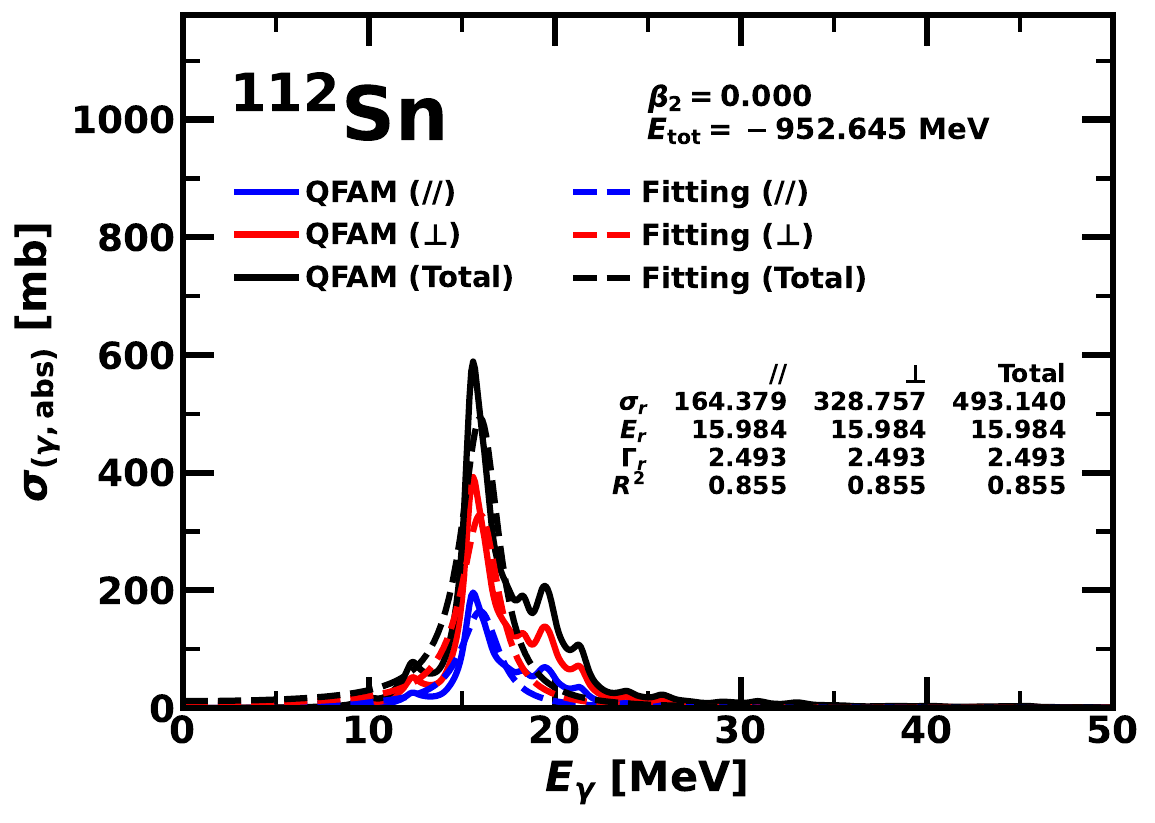}
    \includegraphics[width=0.4\textwidth]{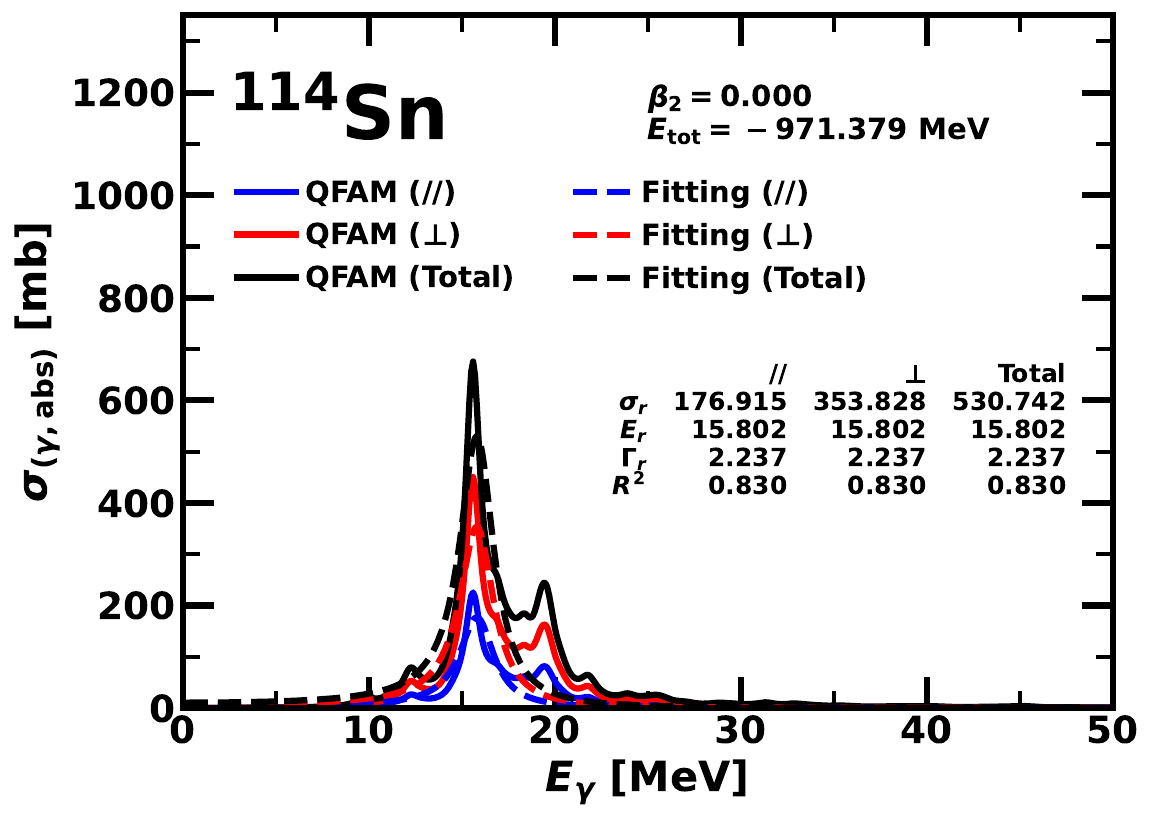}
\end{figure*}
\begin{figure*}\ContinuedFloat
    \centering
    \includegraphics[width=0.4\textwidth]{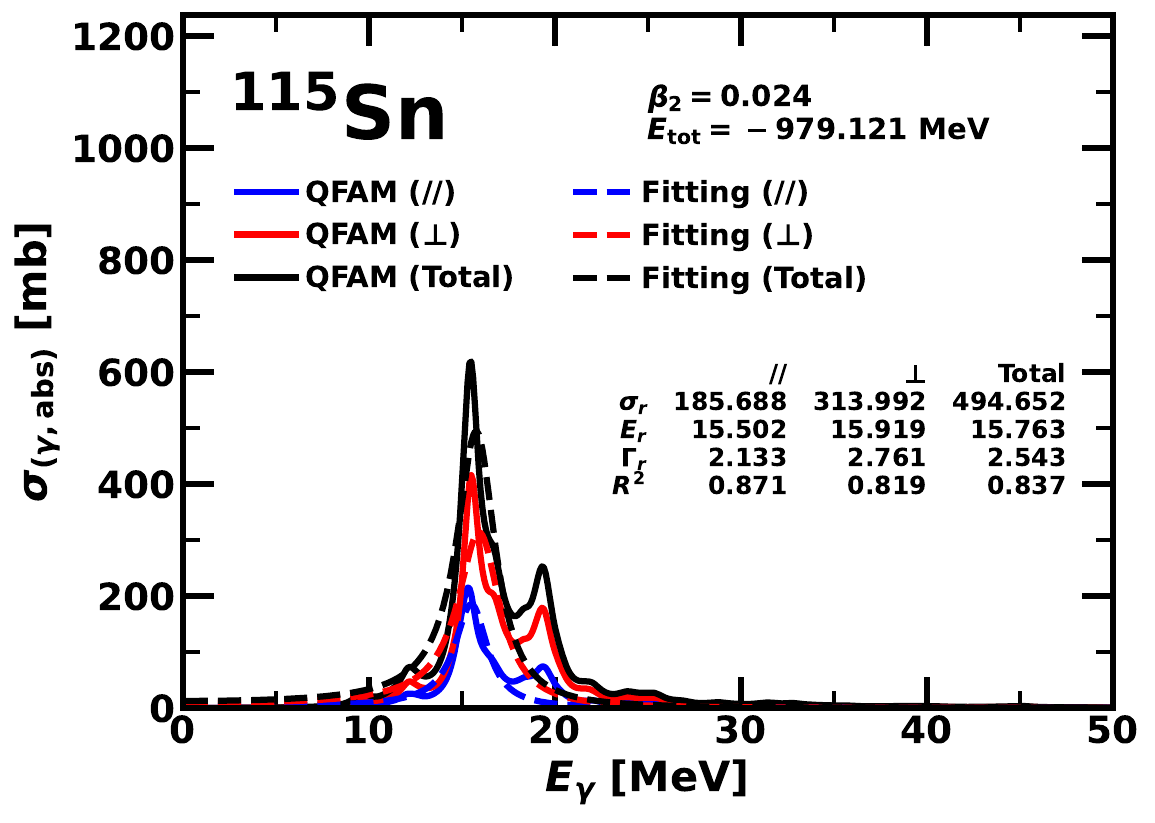}
    \includegraphics[width=0.4\textwidth]{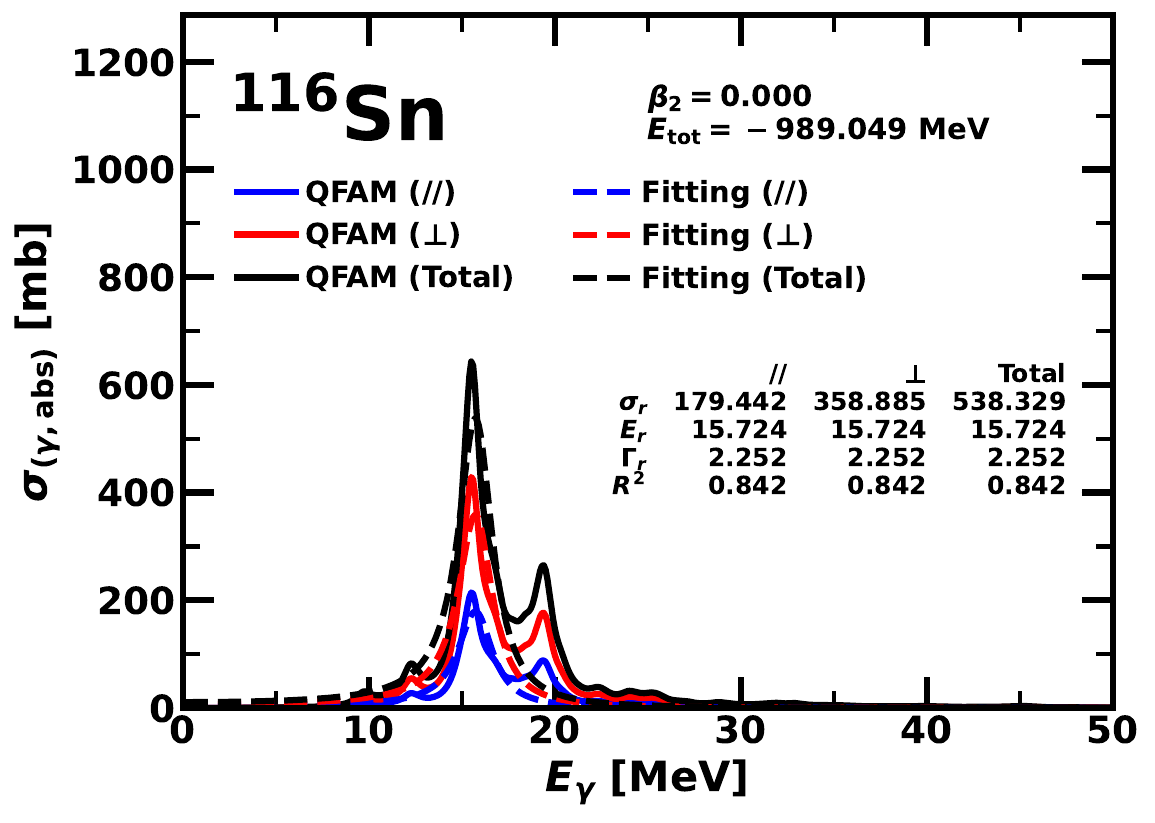}
    \includegraphics[width=0.4\textwidth]{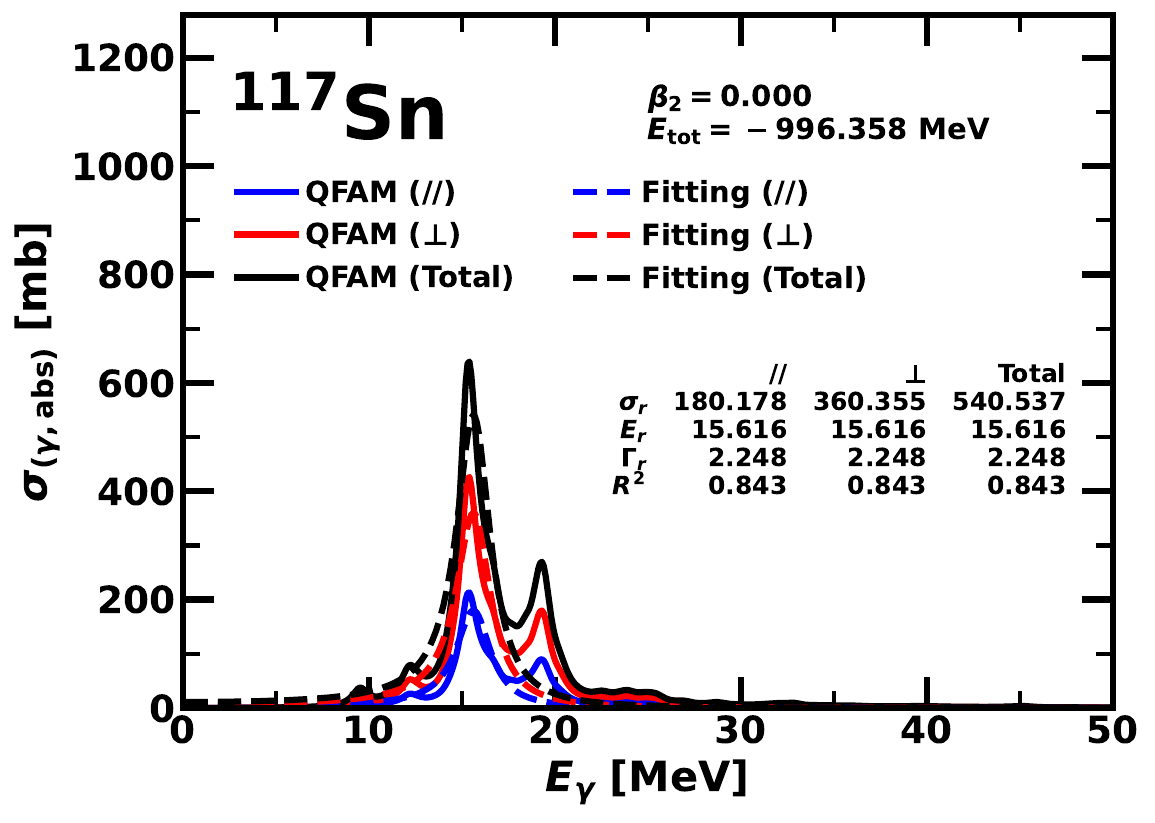}
    \includegraphics[width=0.4\textwidth]{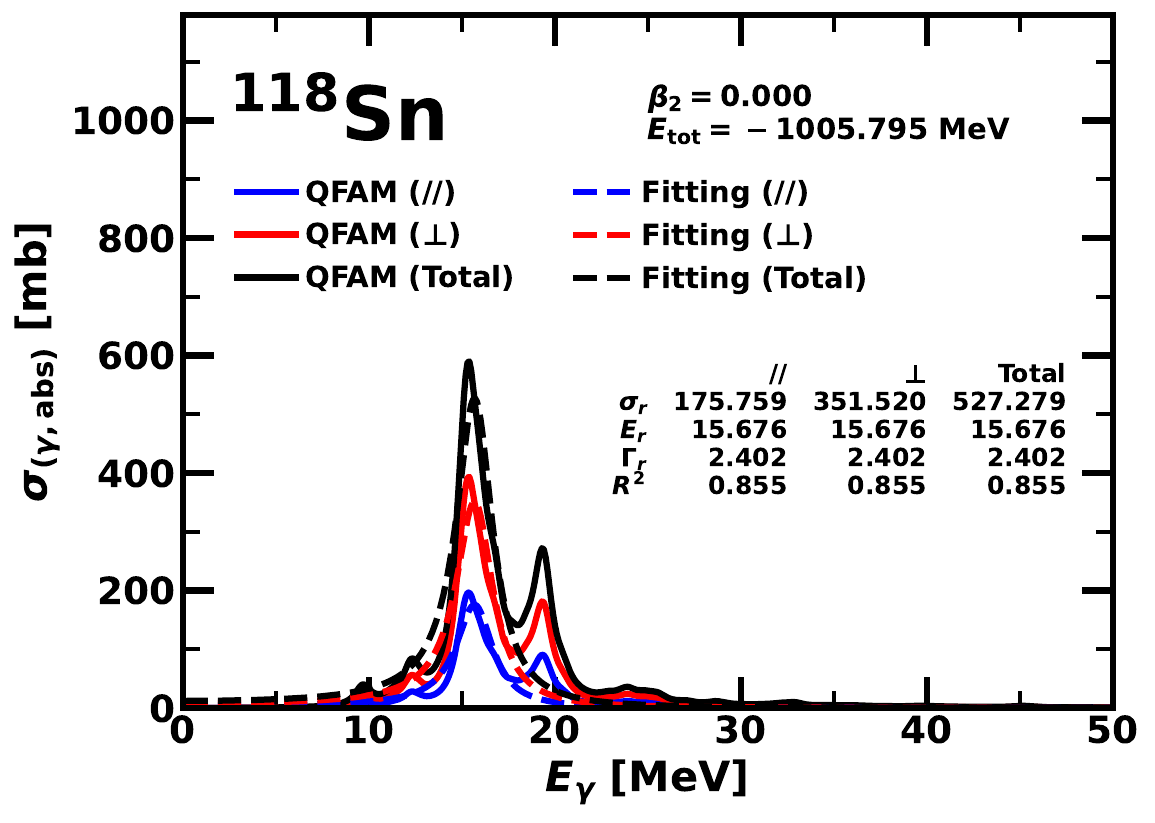}
    \includegraphics[width=0.4\textwidth]{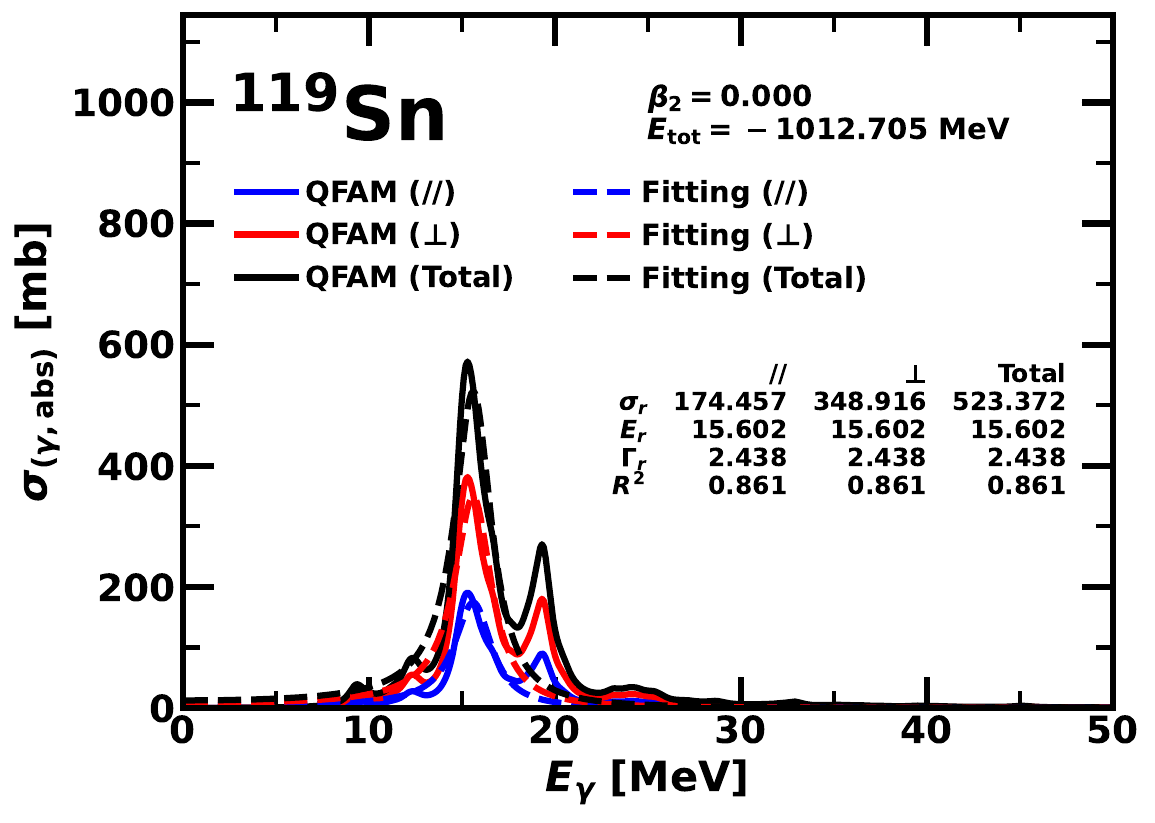}
    \includegraphics[width=0.4\textwidth]{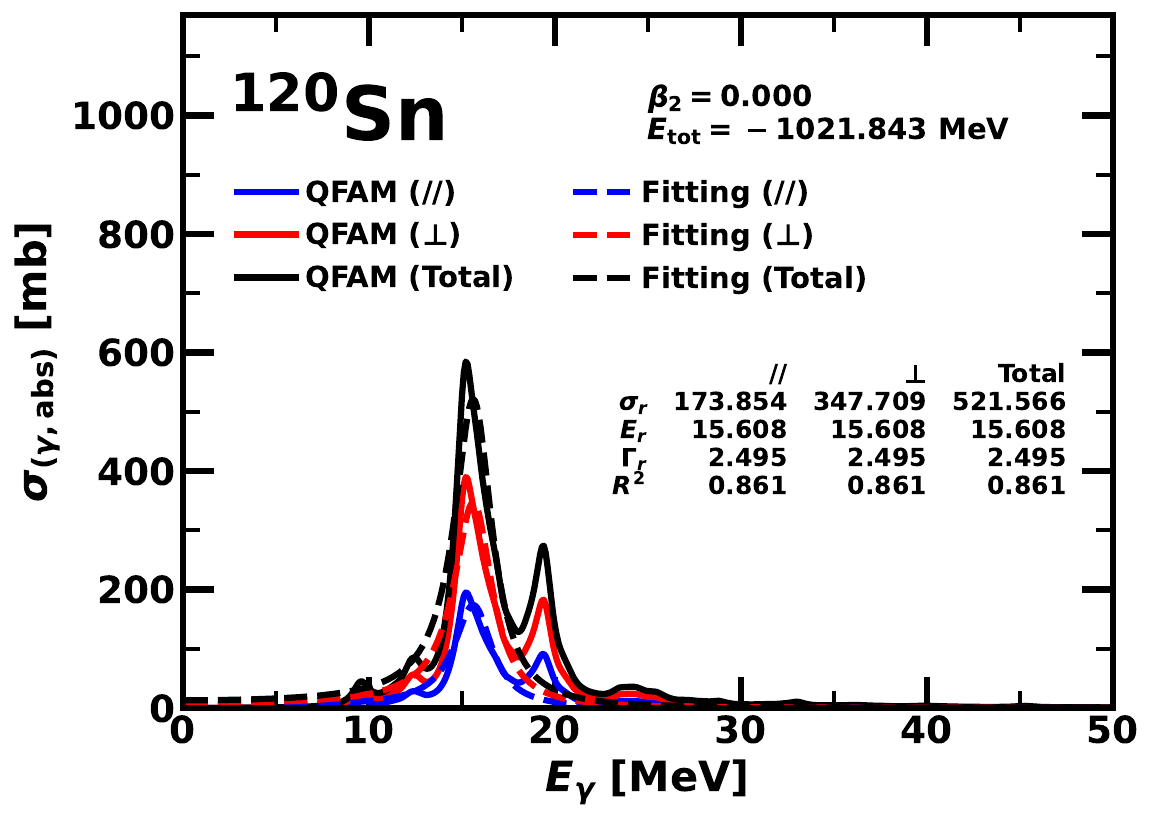}
    \includegraphics[width=0.4\textwidth]{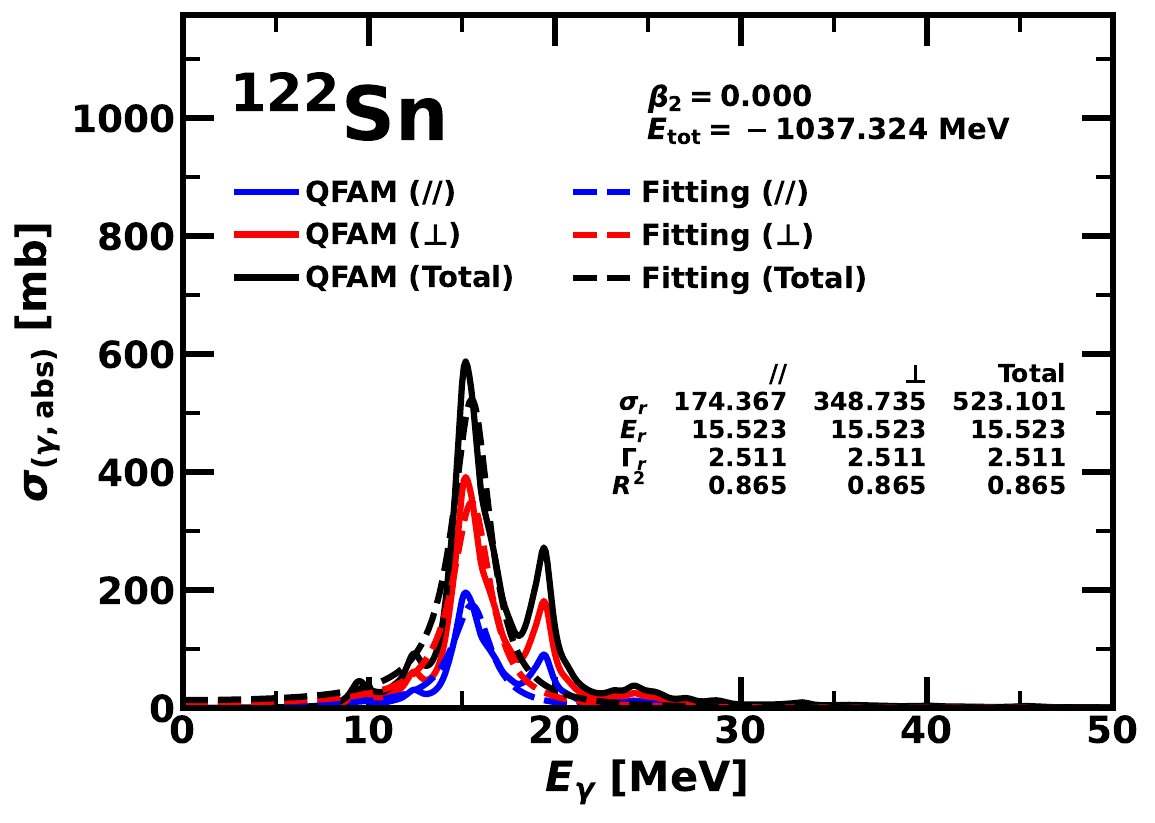}
    \includegraphics[width=0.4\textwidth]{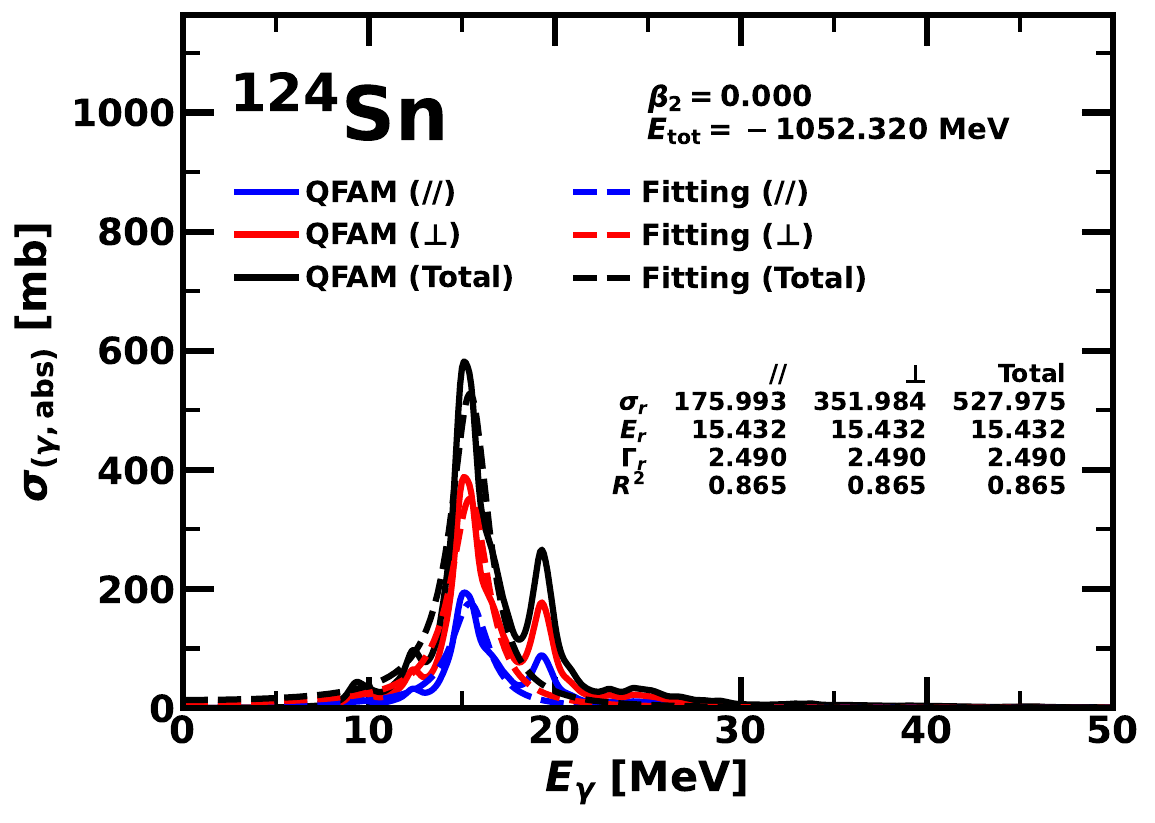}
\end{figure*}
\begin{figure*}\ContinuedFloat
    \centering
    \includegraphics[width=0.4\textwidth]{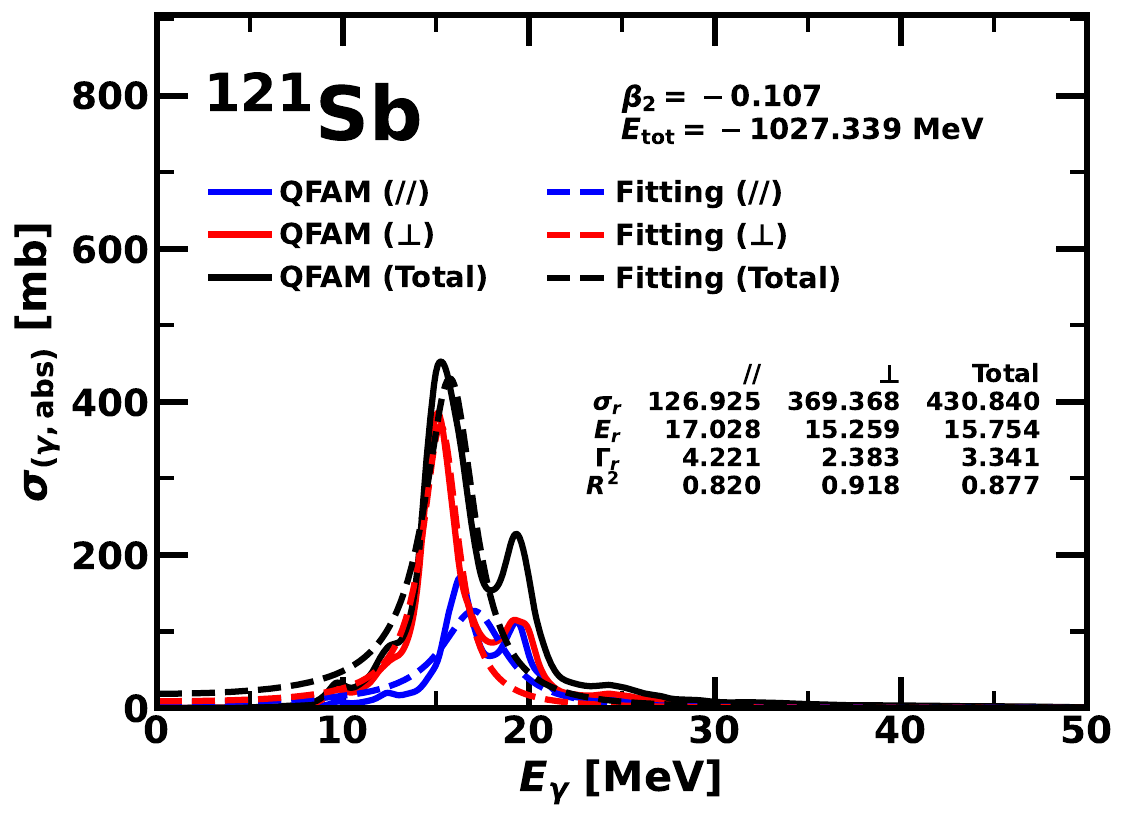}
    \includegraphics[width=0.4\textwidth]{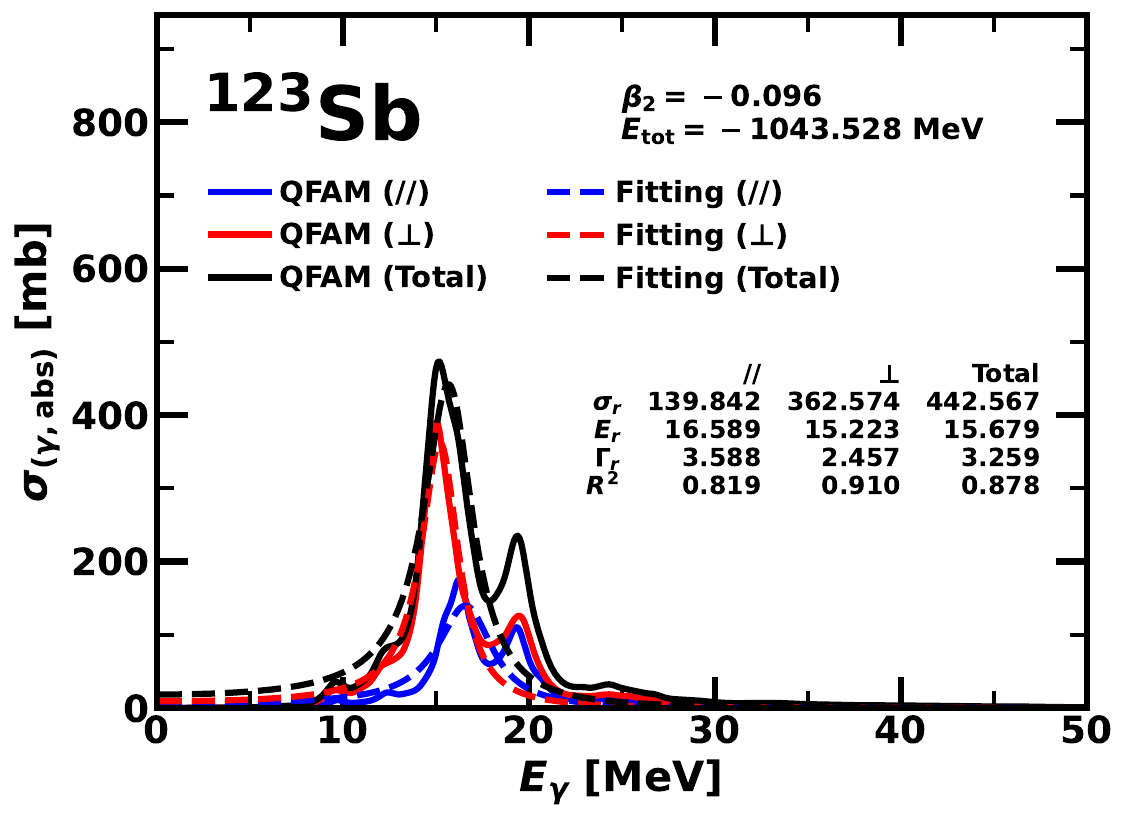}
    \includegraphics[width=0.4\textwidth]{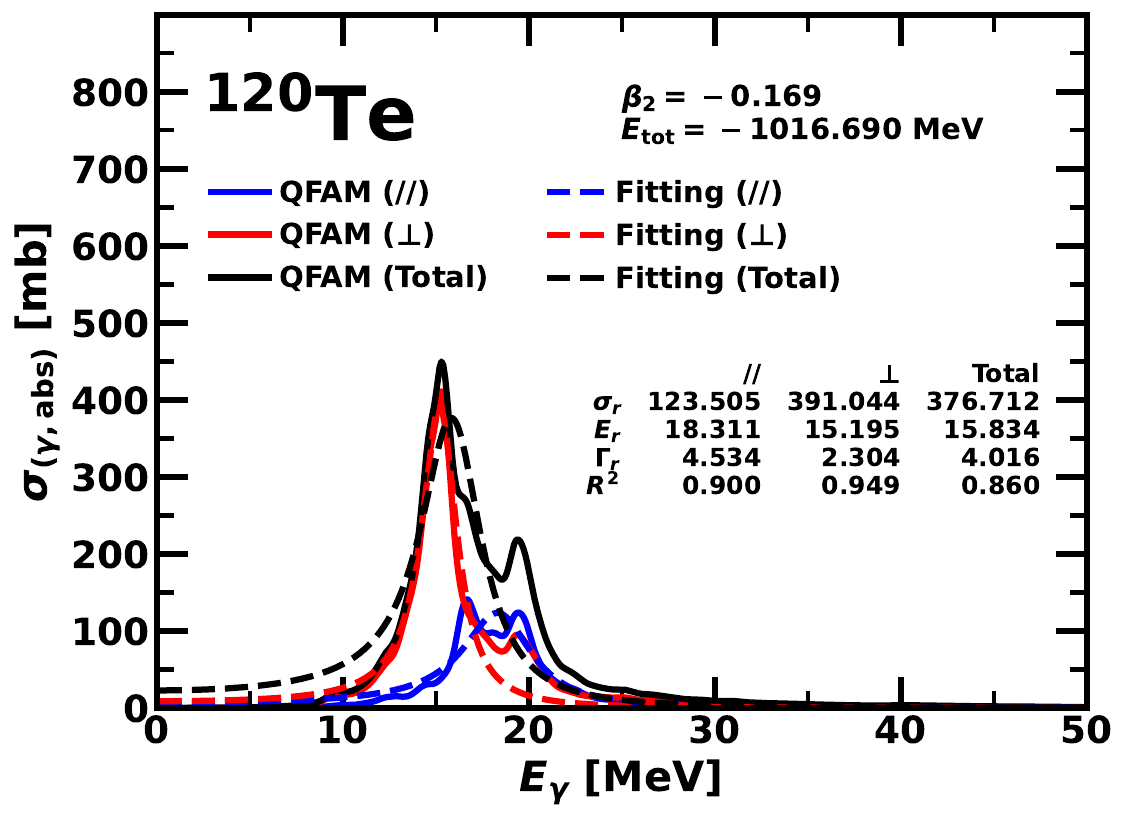}
    \includegraphics[width=0.4\textwidth]{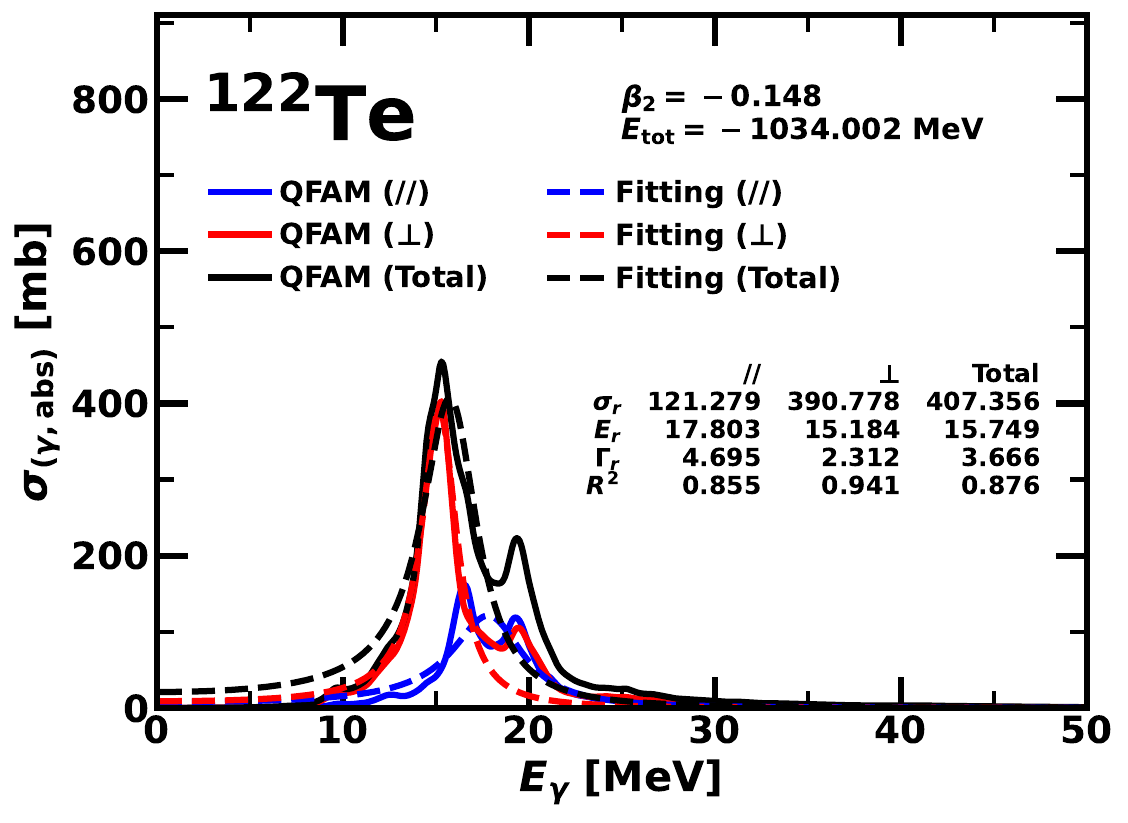}
    \includegraphics[width=0.4\textwidth]{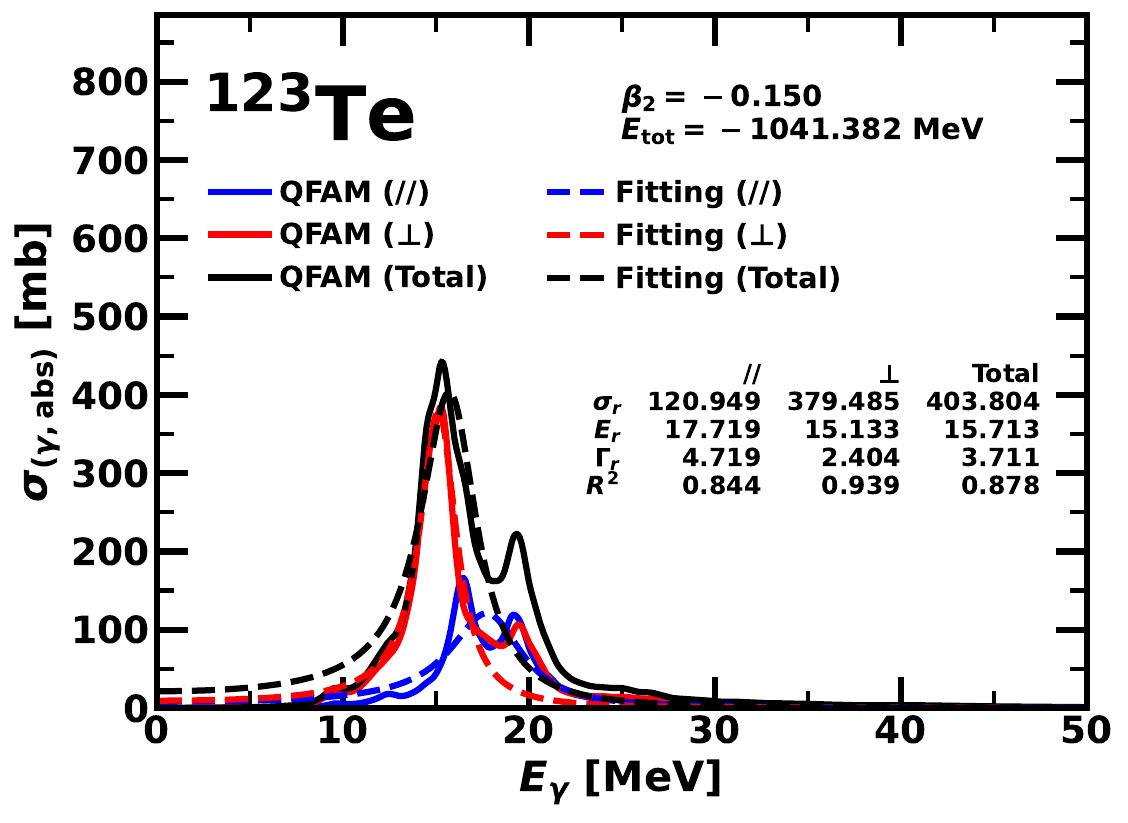}
    \includegraphics[width=0.4\textwidth]{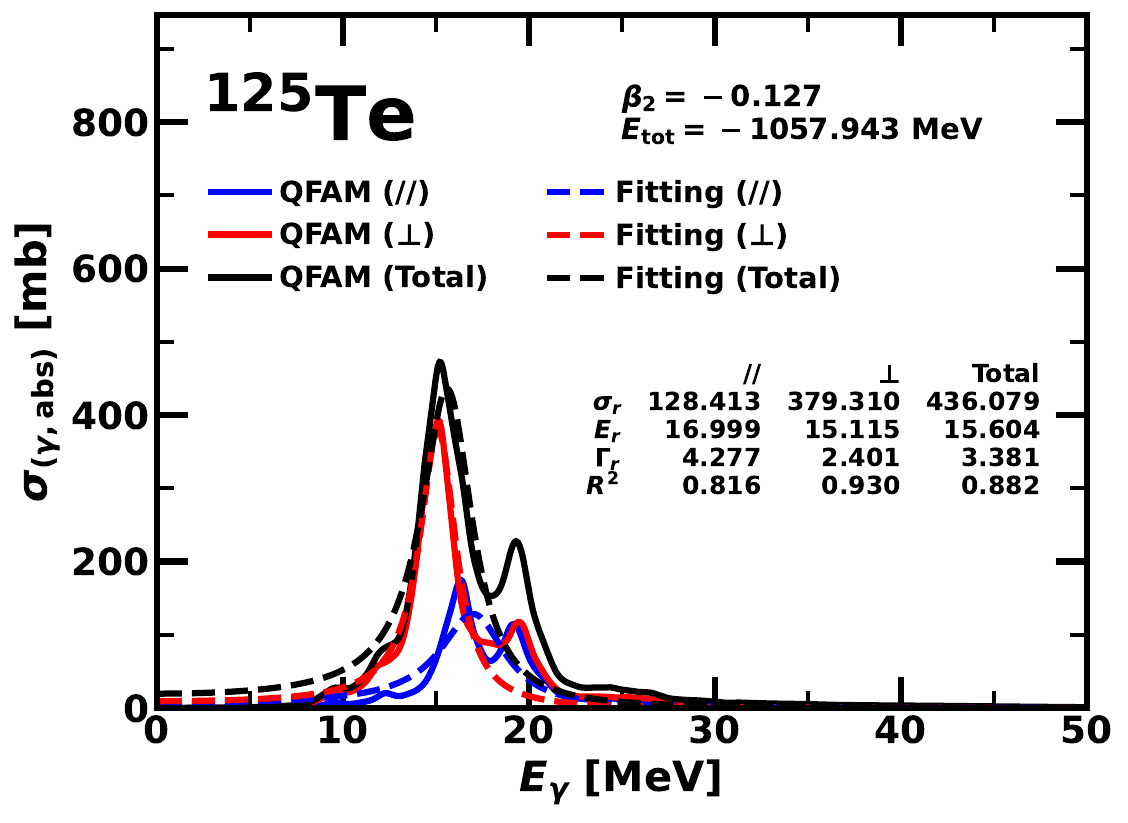}
    \includegraphics[width=0.4\textwidth]{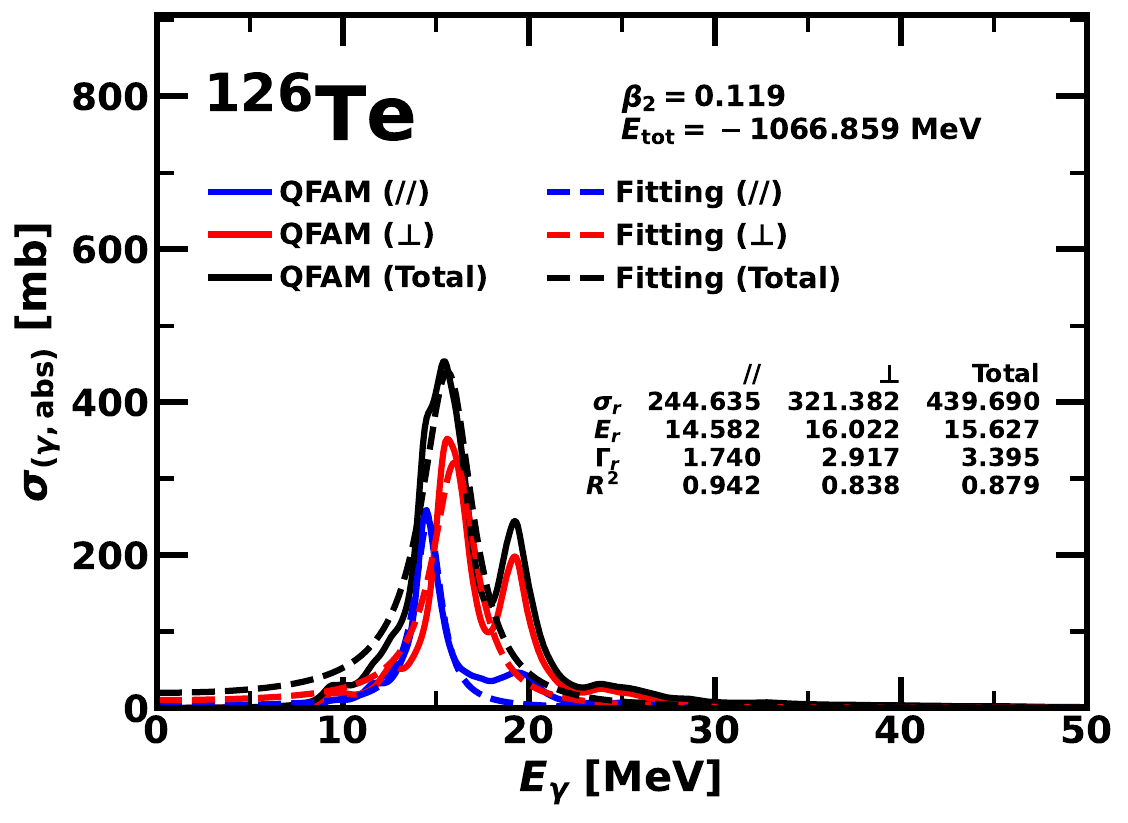}
    \includegraphics[width=0.4\textwidth]{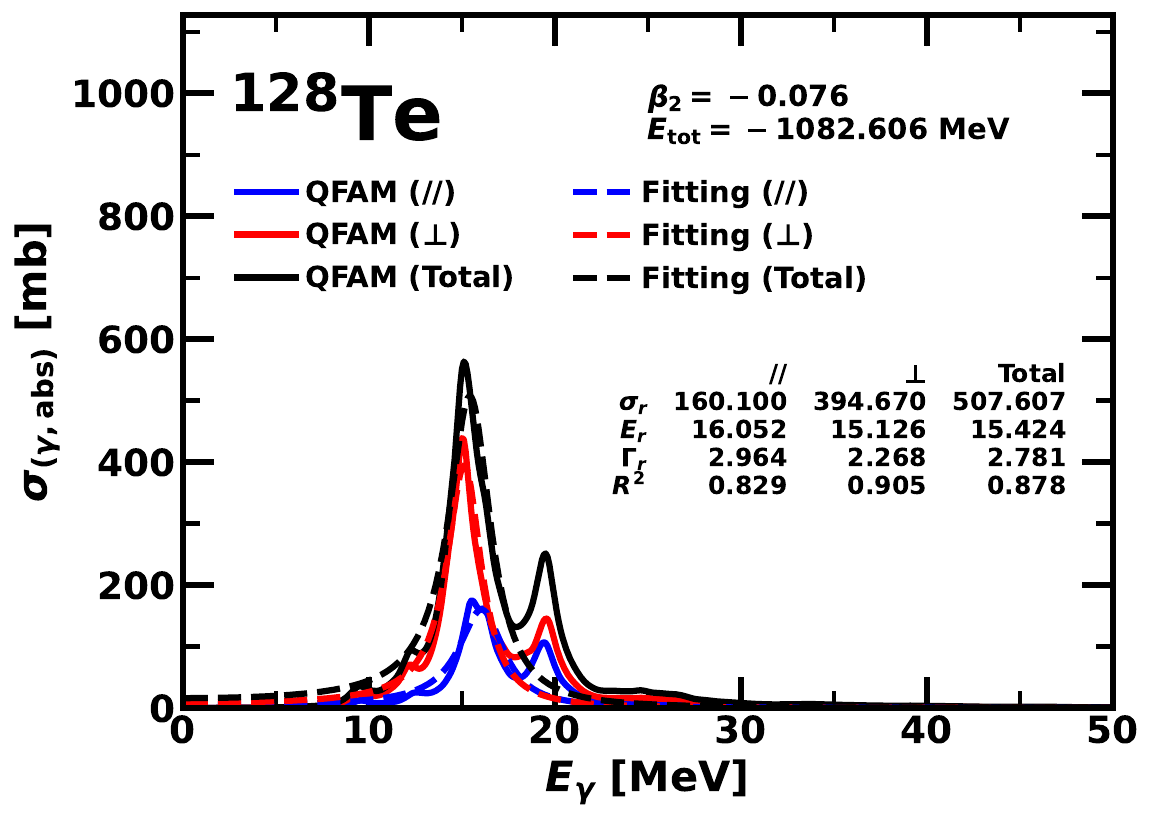}
\end{figure*}
\begin{figure*}\ContinuedFloat
    \centering
    \includegraphics[width=0.4\textwidth]{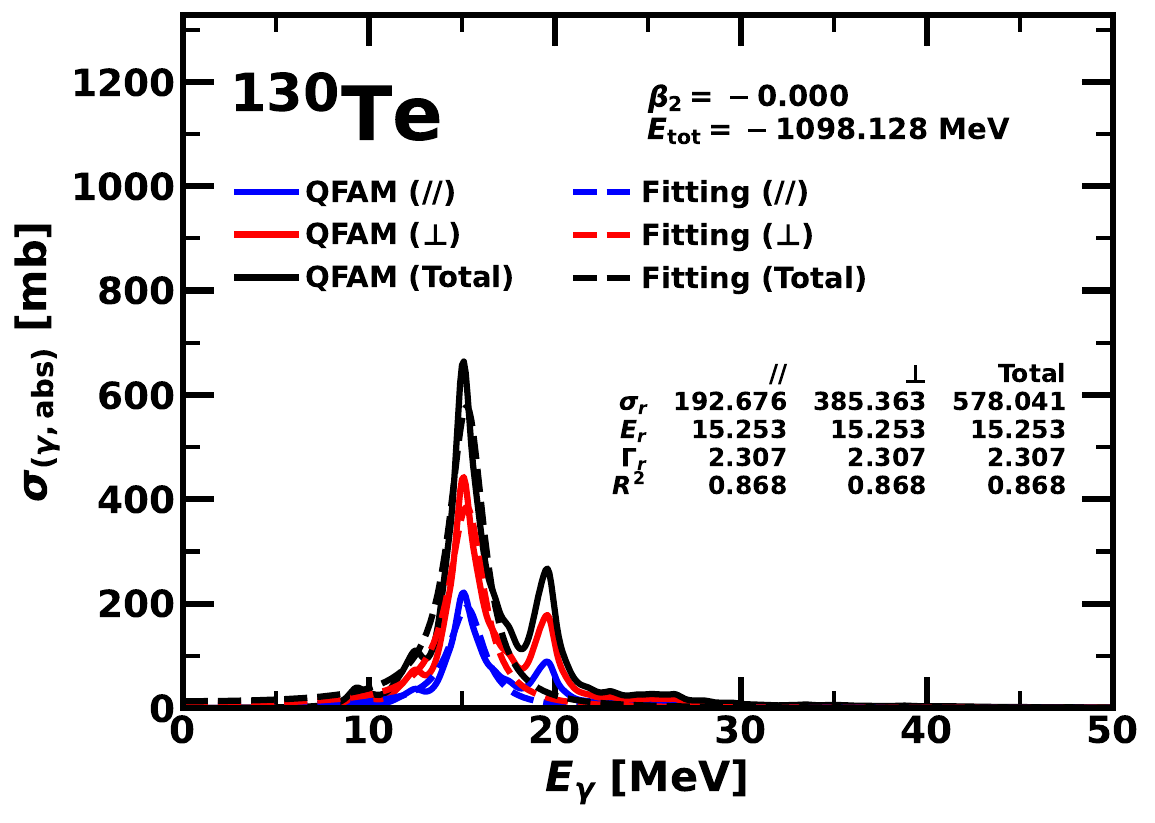}
    \includegraphics[width=0.4\textwidth]{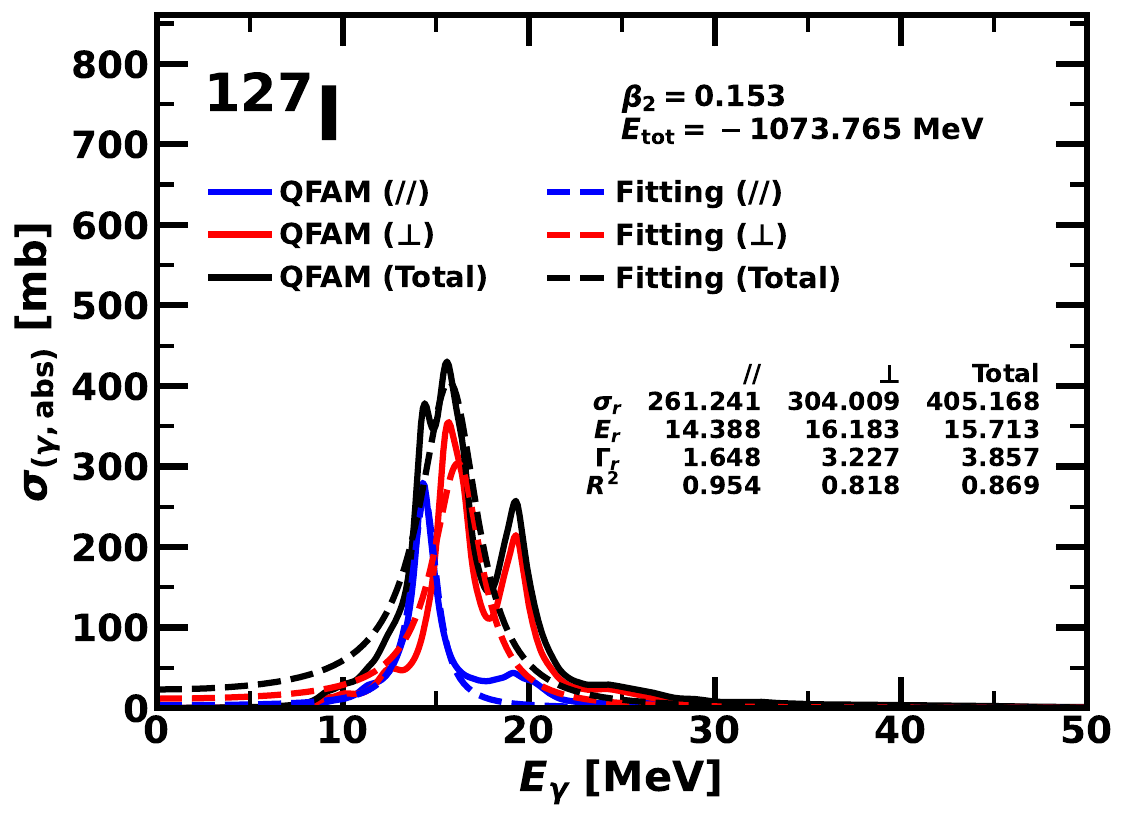}
    \includegraphics[width=0.4\textwidth]{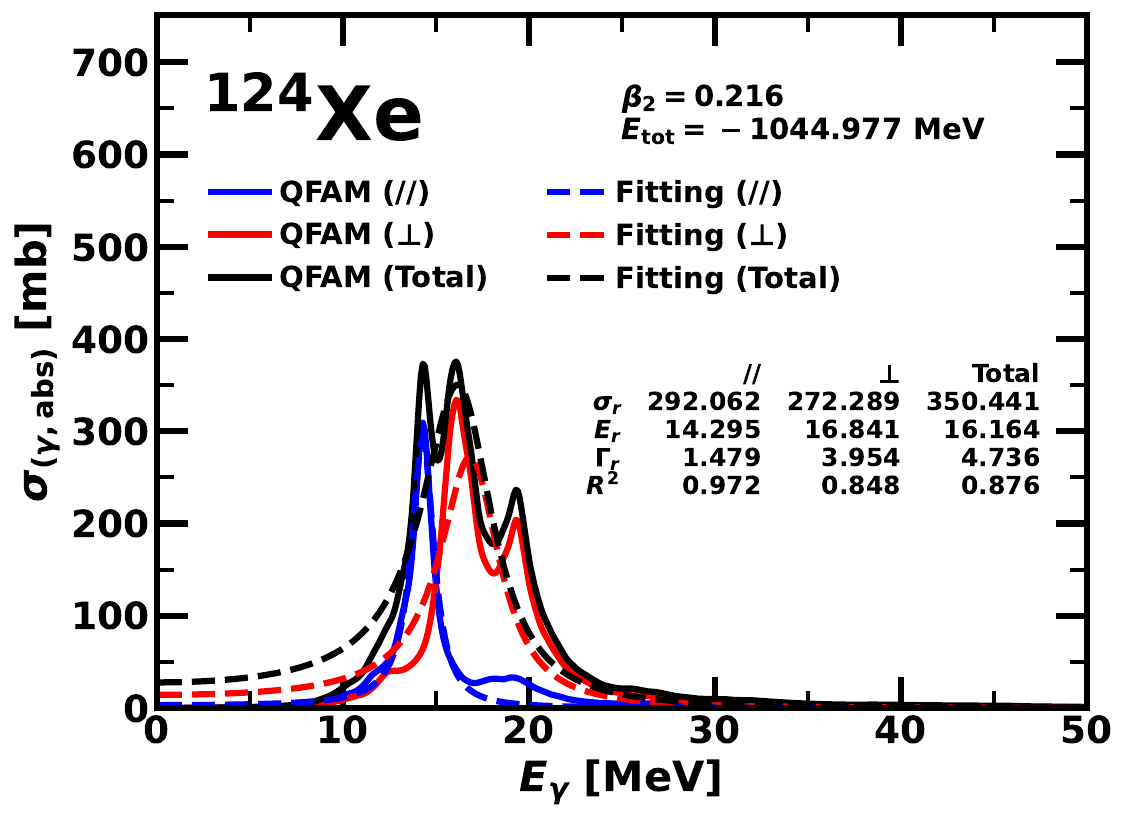}
    \includegraphics[width=0.4\textwidth]{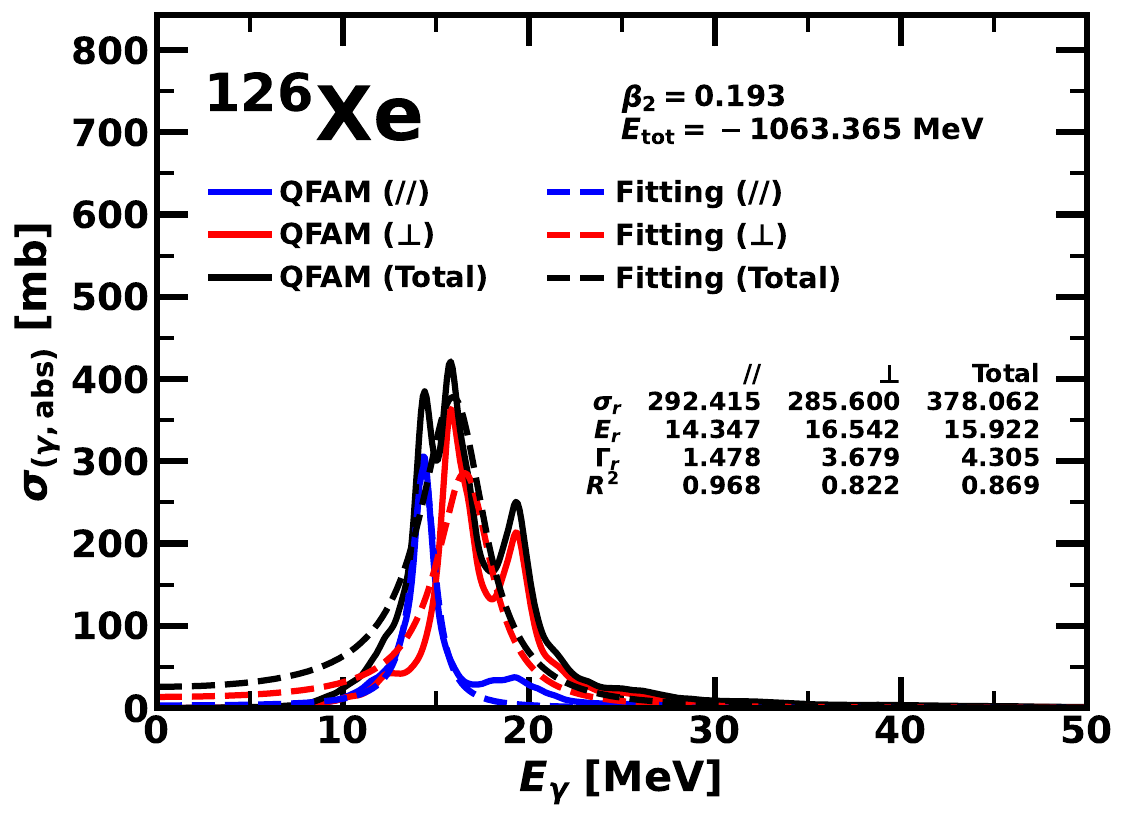}
    \includegraphics[width=0.4\textwidth]{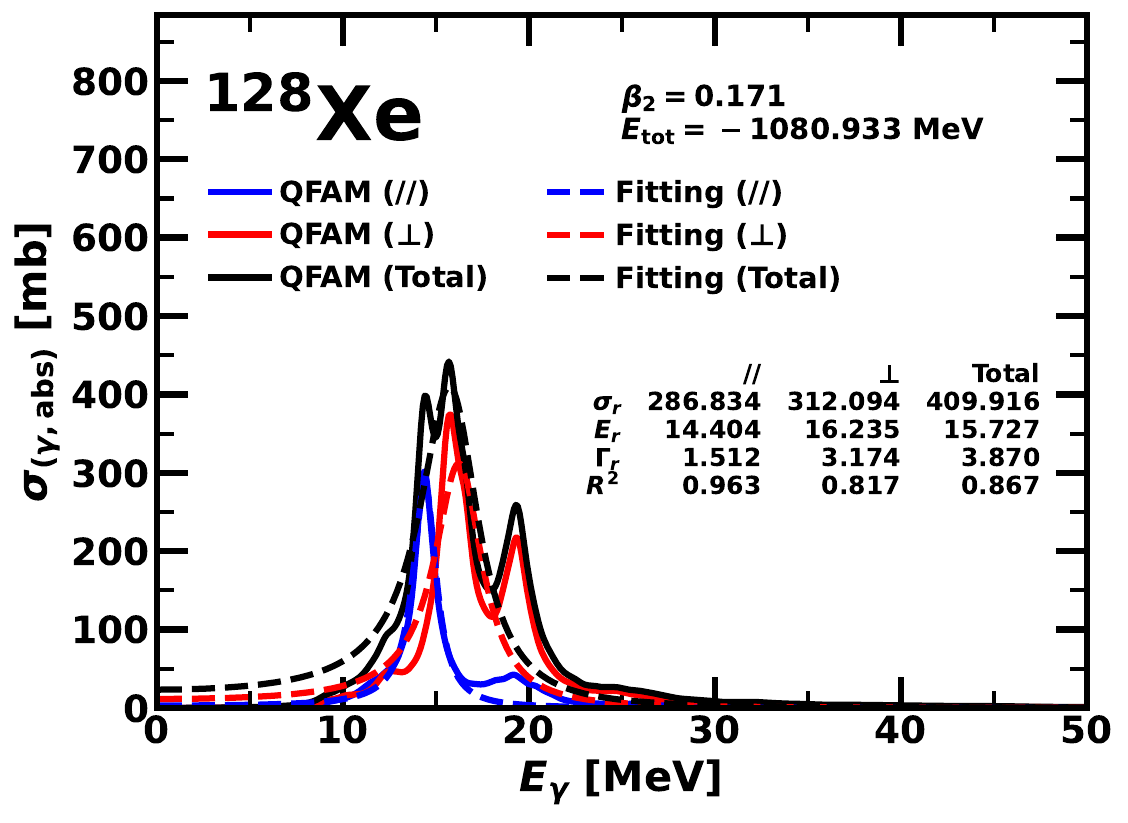}
    \includegraphics[width=0.4\textwidth]{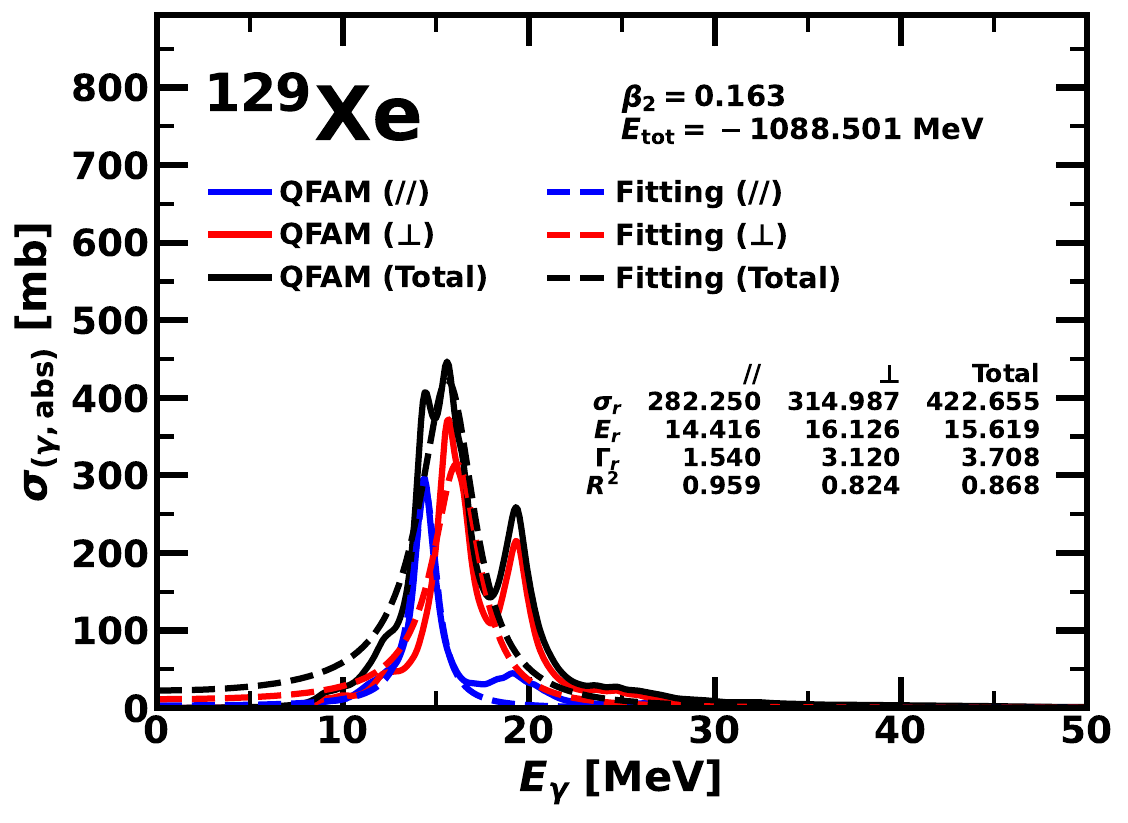}
    \includegraphics[width=0.4\textwidth]{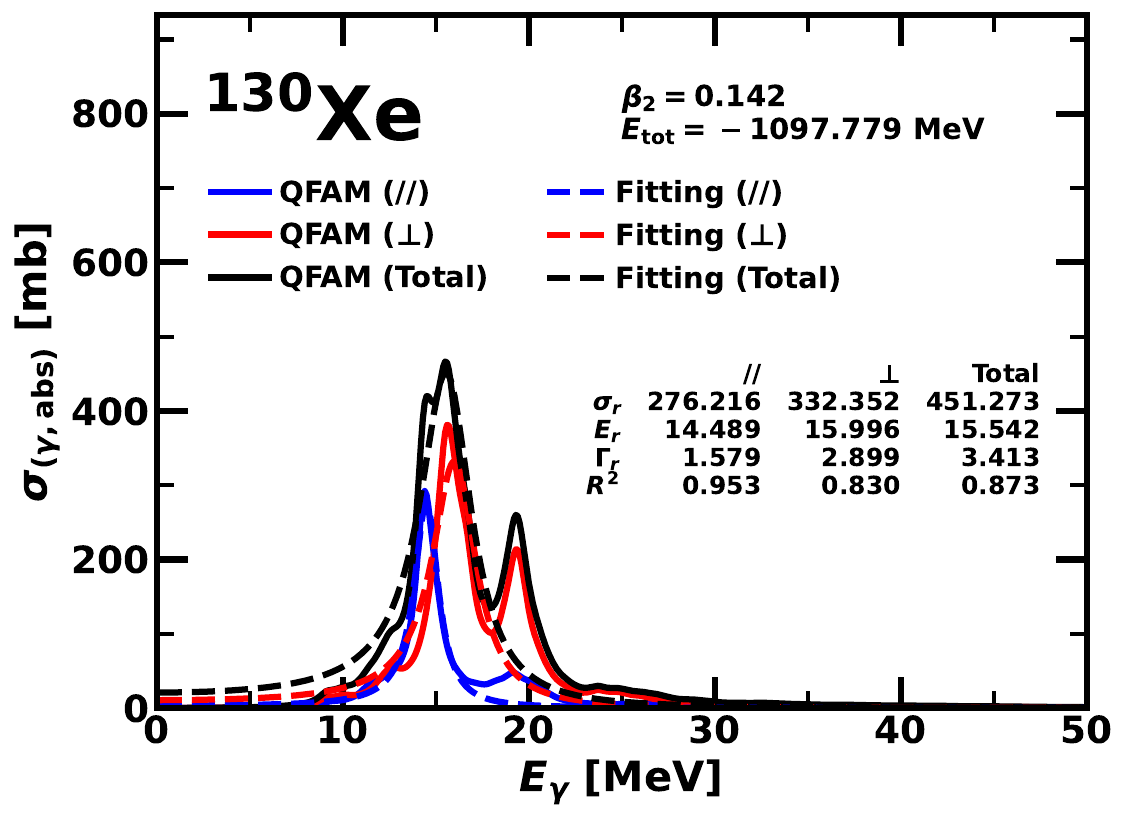}
    \includegraphics[width=0.4\textwidth]{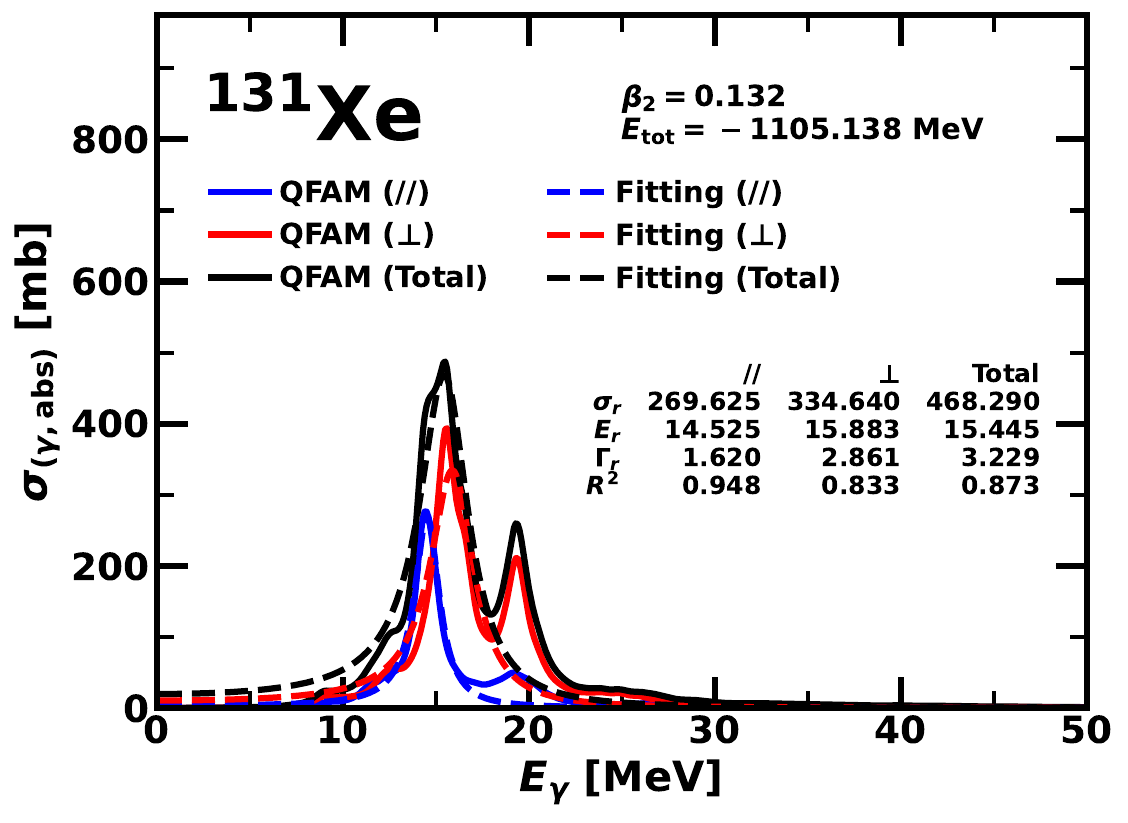}
\end{figure*}
\begin{figure*}\ContinuedFloat
    \centering
    \includegraphics[width=0.4\textwidth]{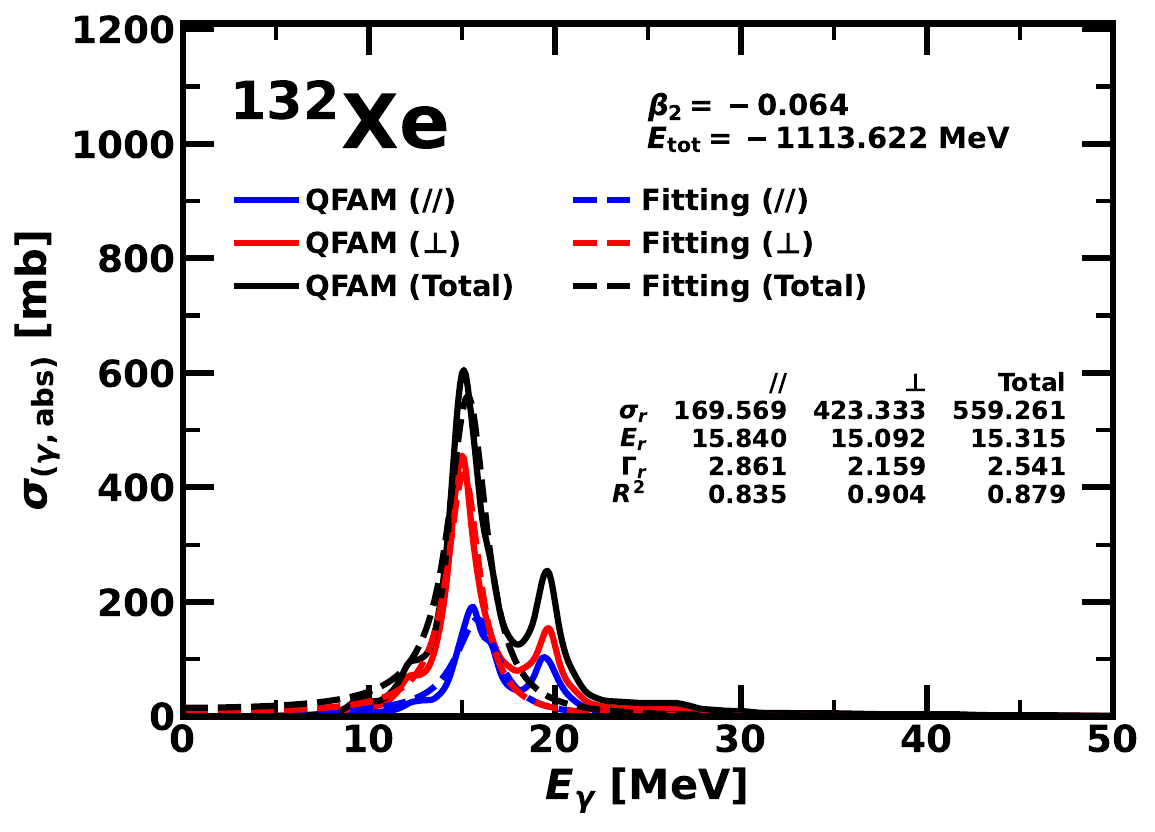}
    \includegraphics[width=0.4\textwidth]{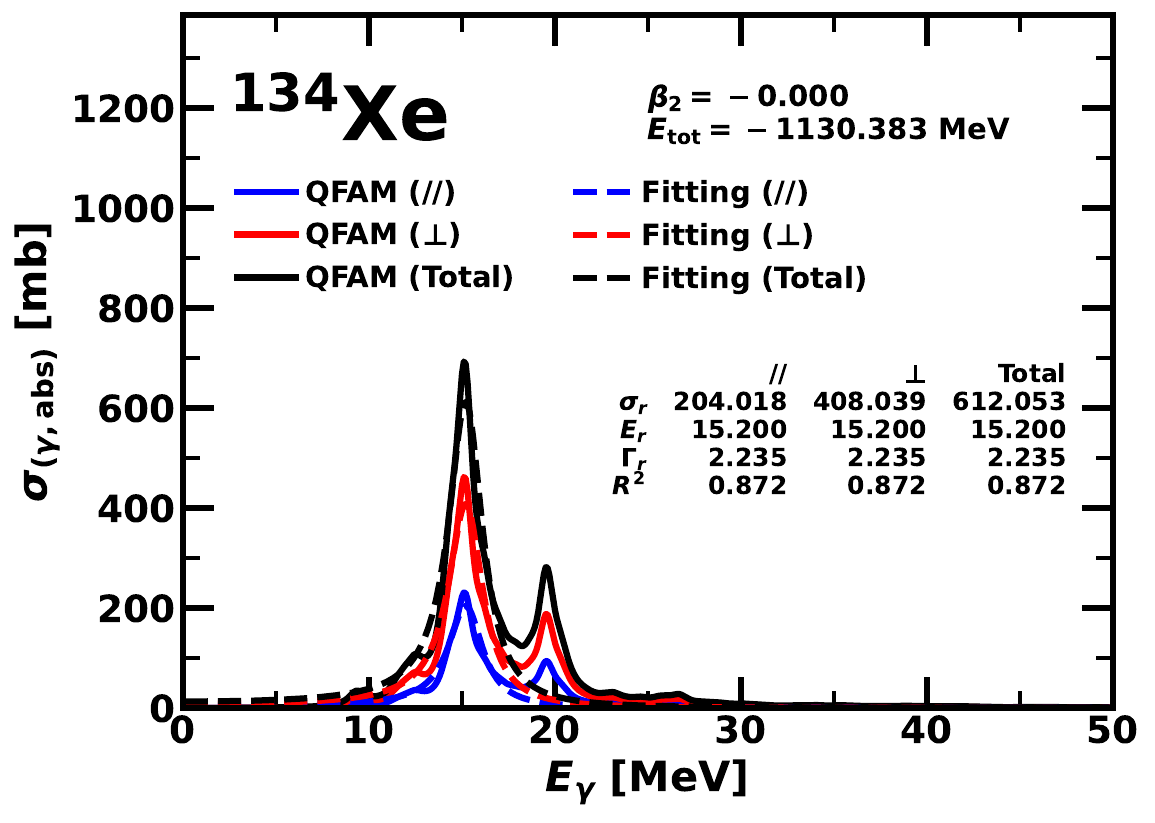}
    \includegraphics[width=0.4\textwidth]{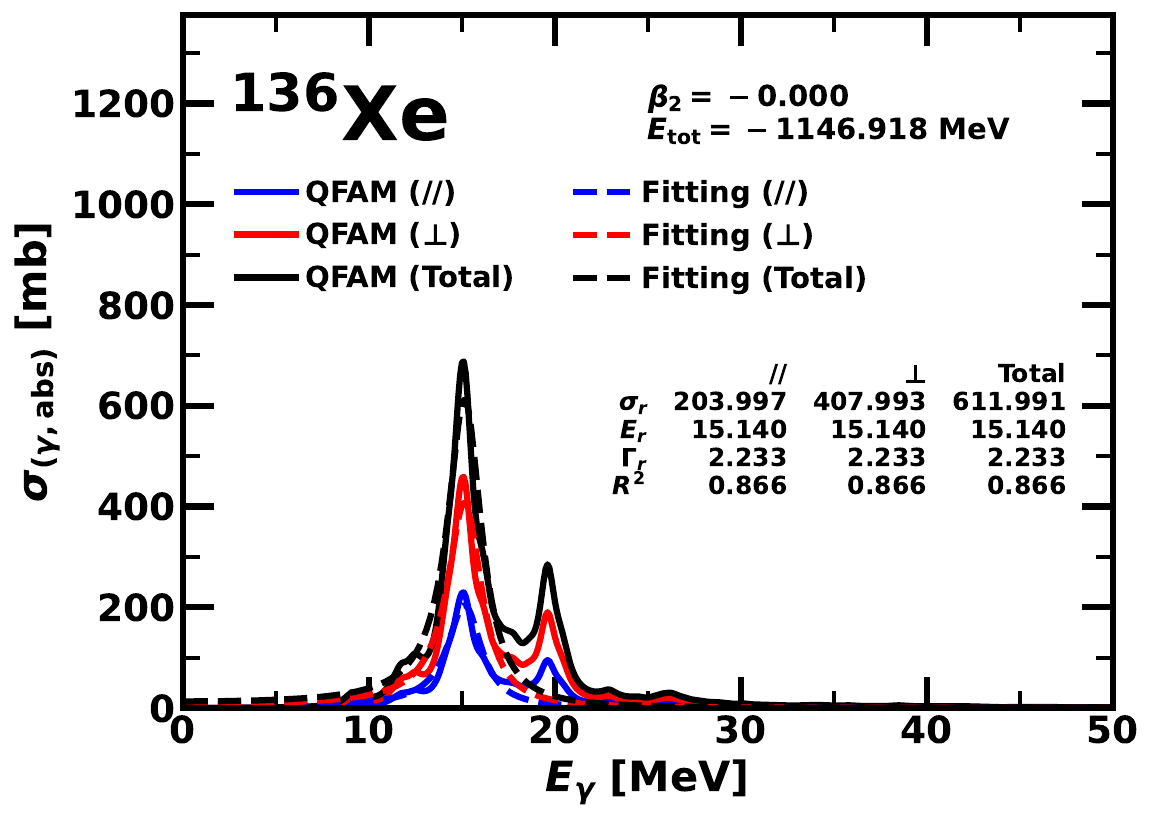}
    \includegraphics[width=0.4\textwidth]{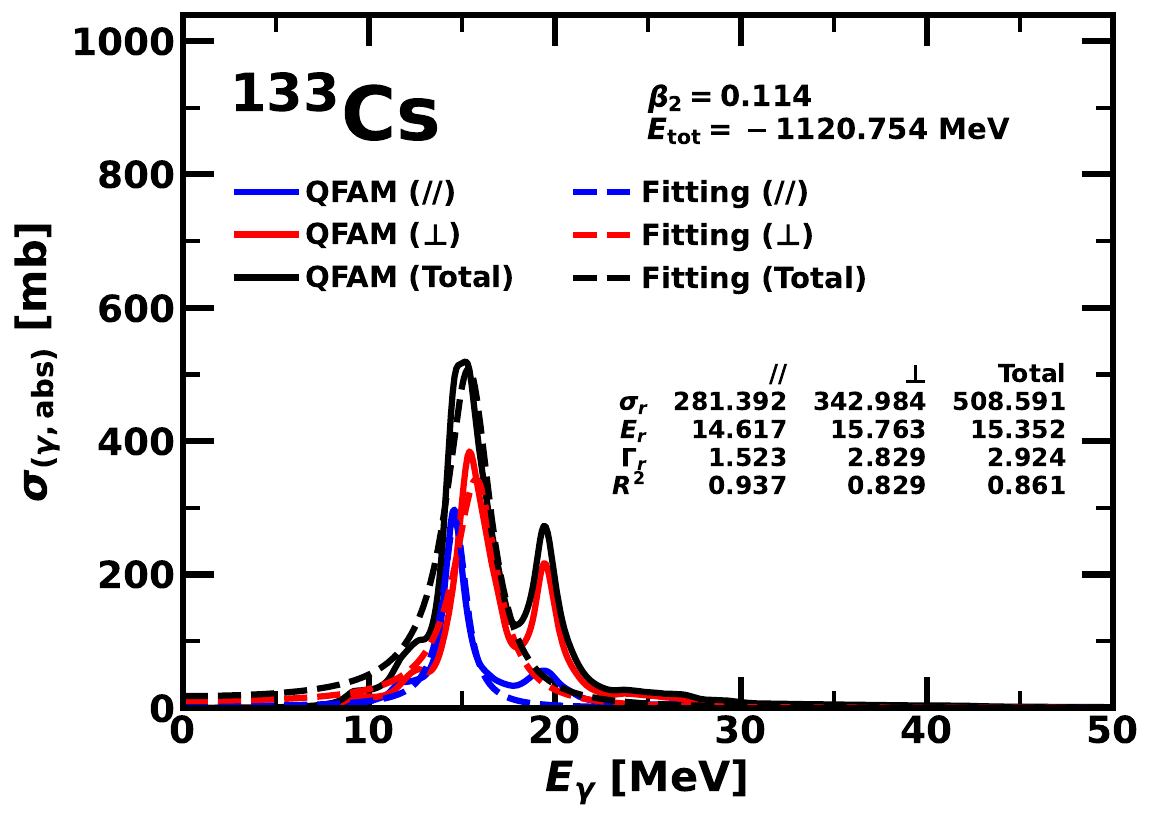}
    \includegraphics[width=0.4\textwidth]{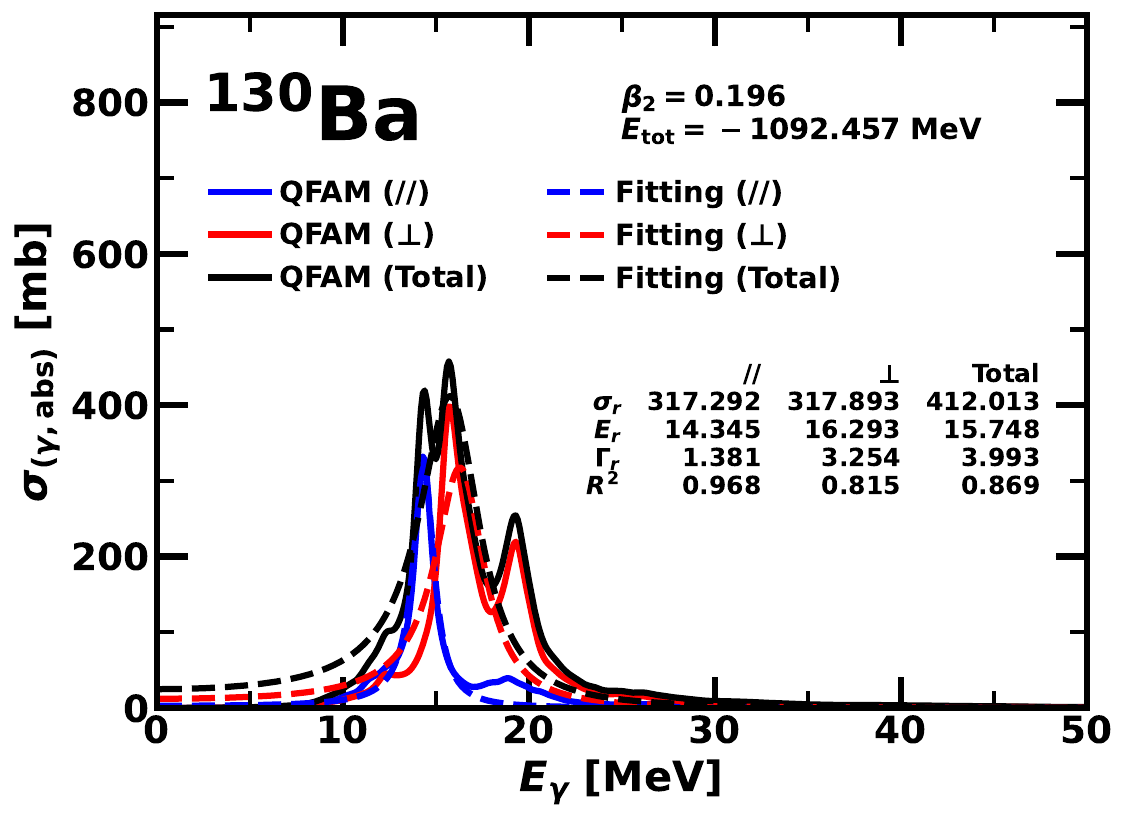}
    \includegraphics[width=0.4\textwidth]{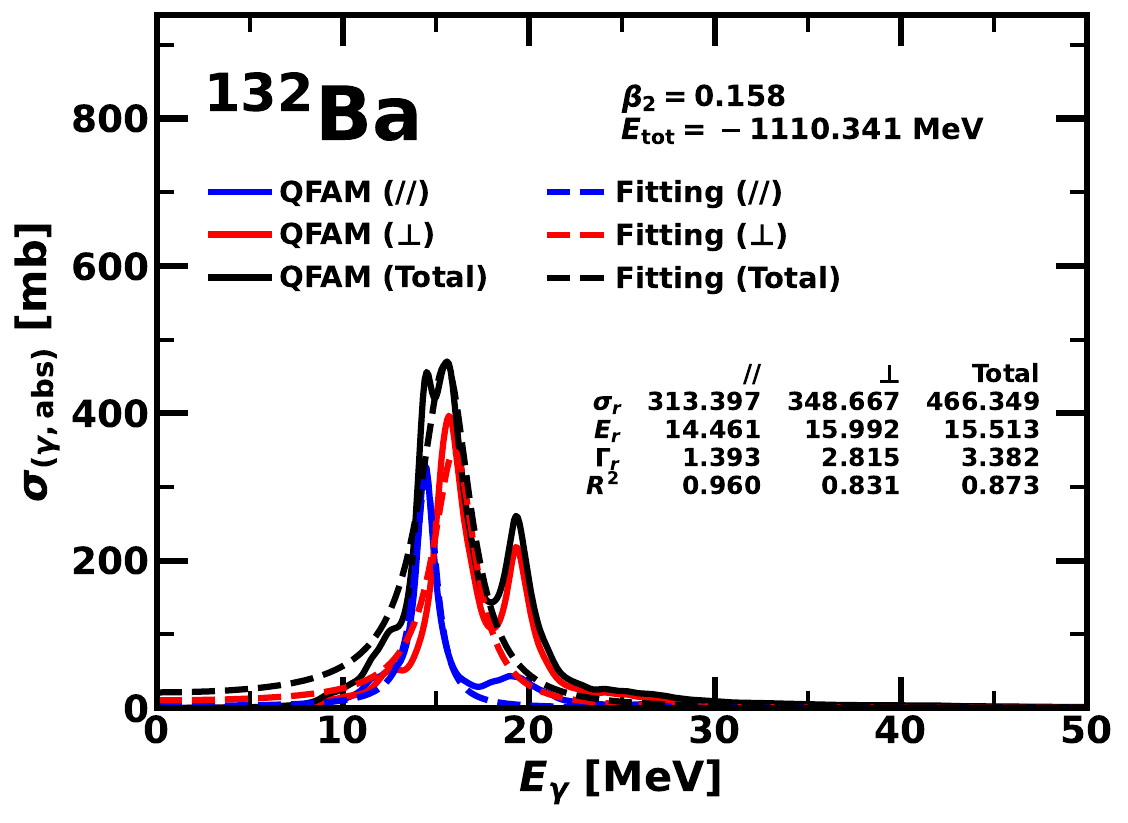}
    \includegraphics[width=0.4\textwidth]{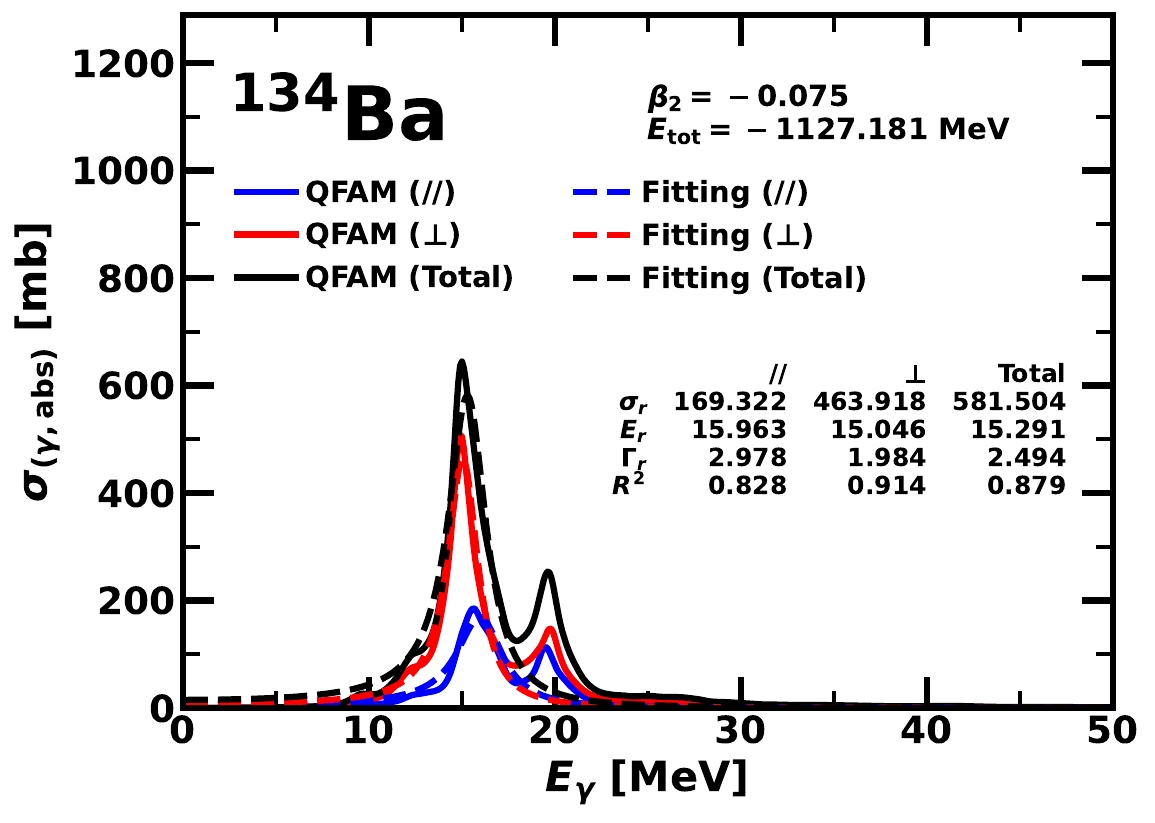}
    \includegraphics[width=0.4\textwidth]{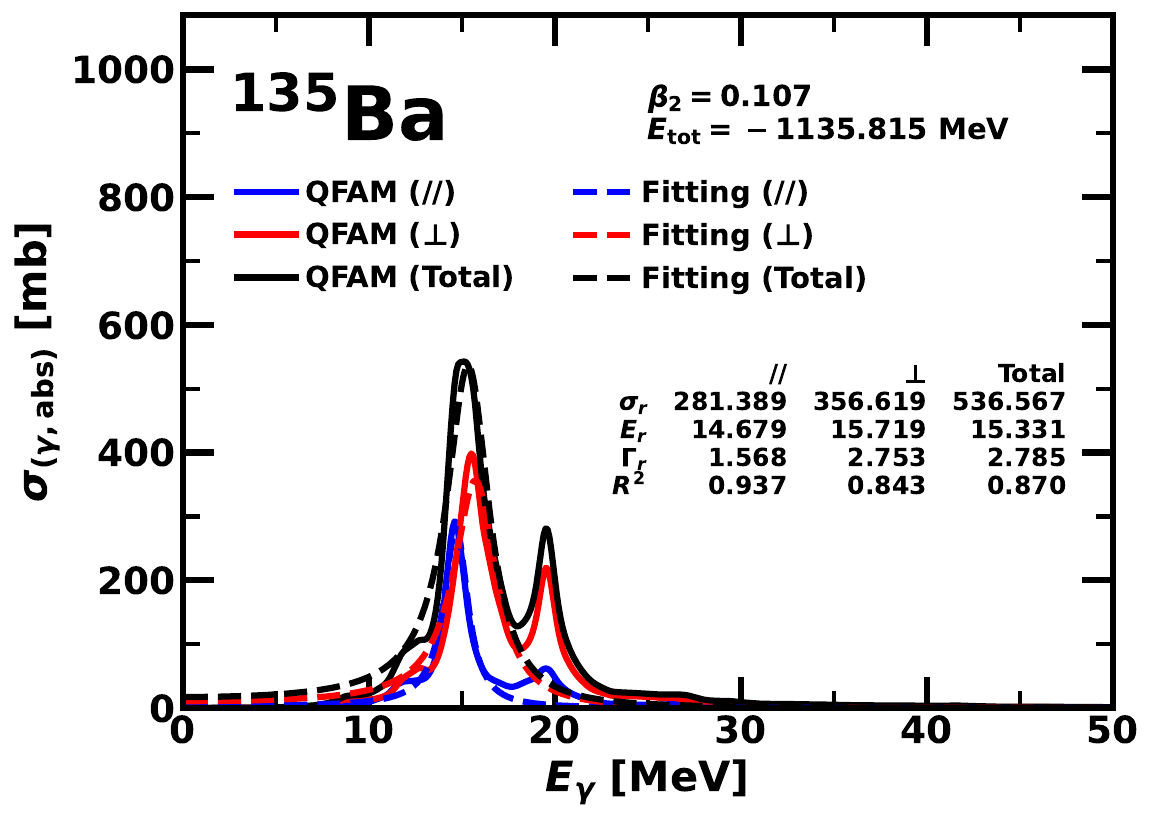}
\end{figure*}
\begin{figure*}\ContinuedFloat
    \centering
    \includegraphics[width=0.4\textwidth]{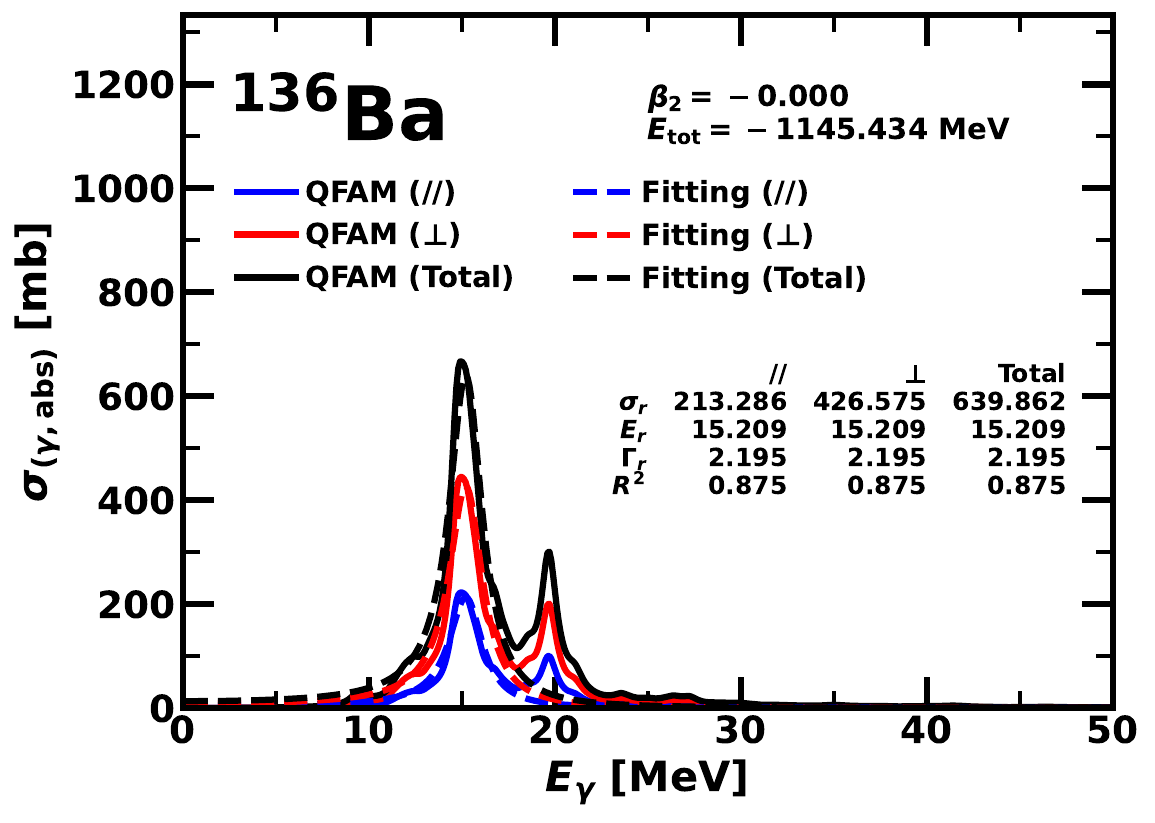}
    \includegraphics[width=0.4\textwidth]{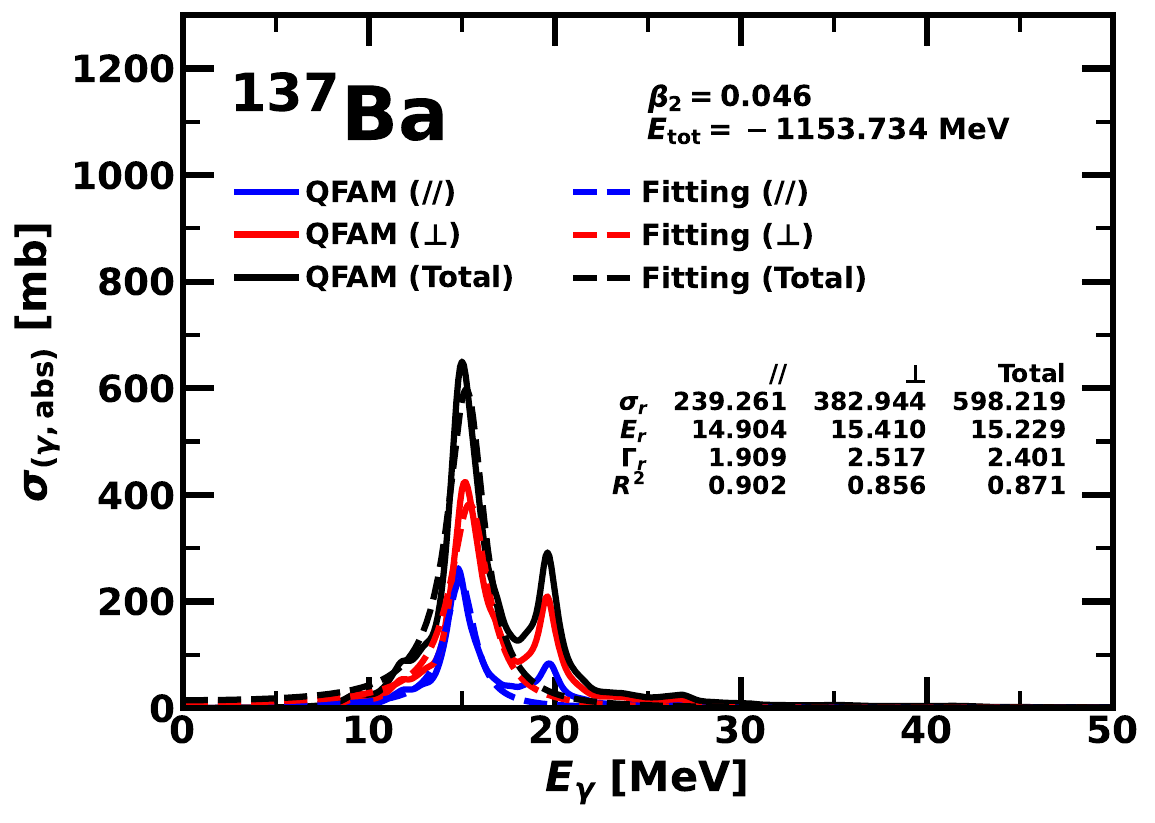}
    \includegraphics[width=0.4\textwidth]{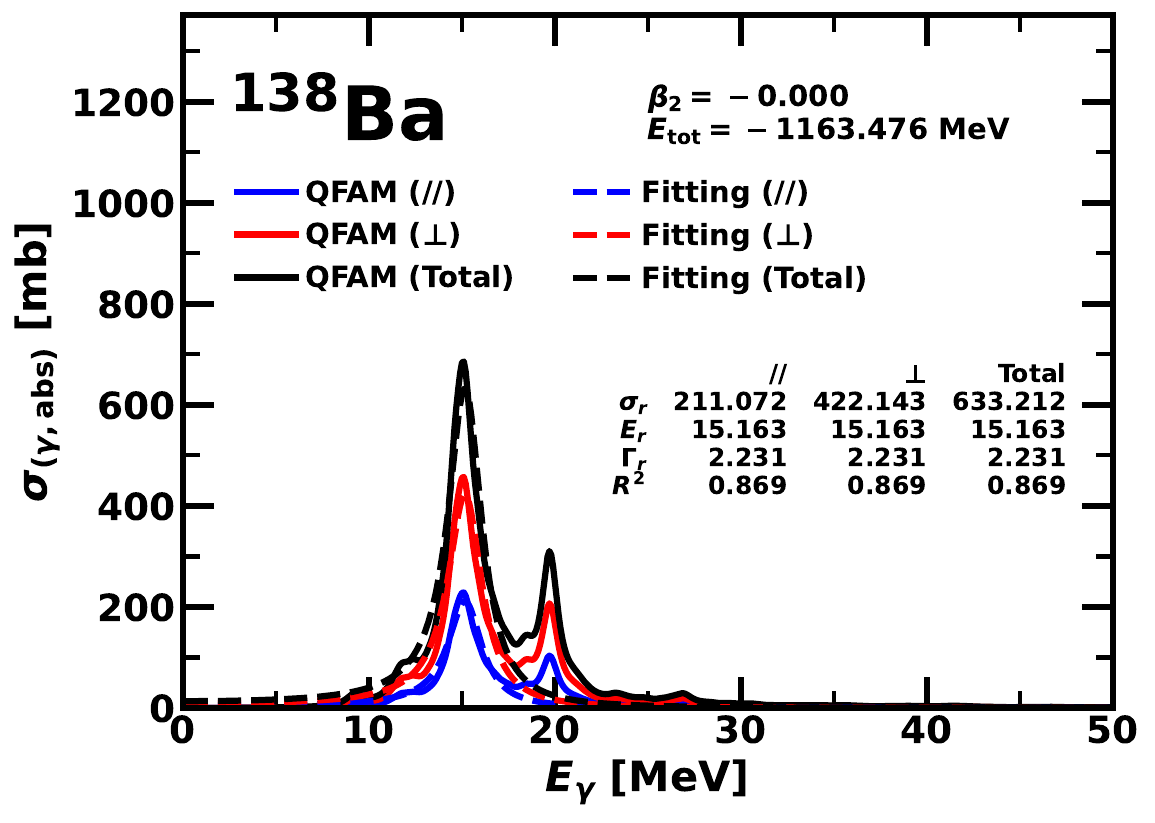}
    \includegraphics[width=0.4\textwidth]{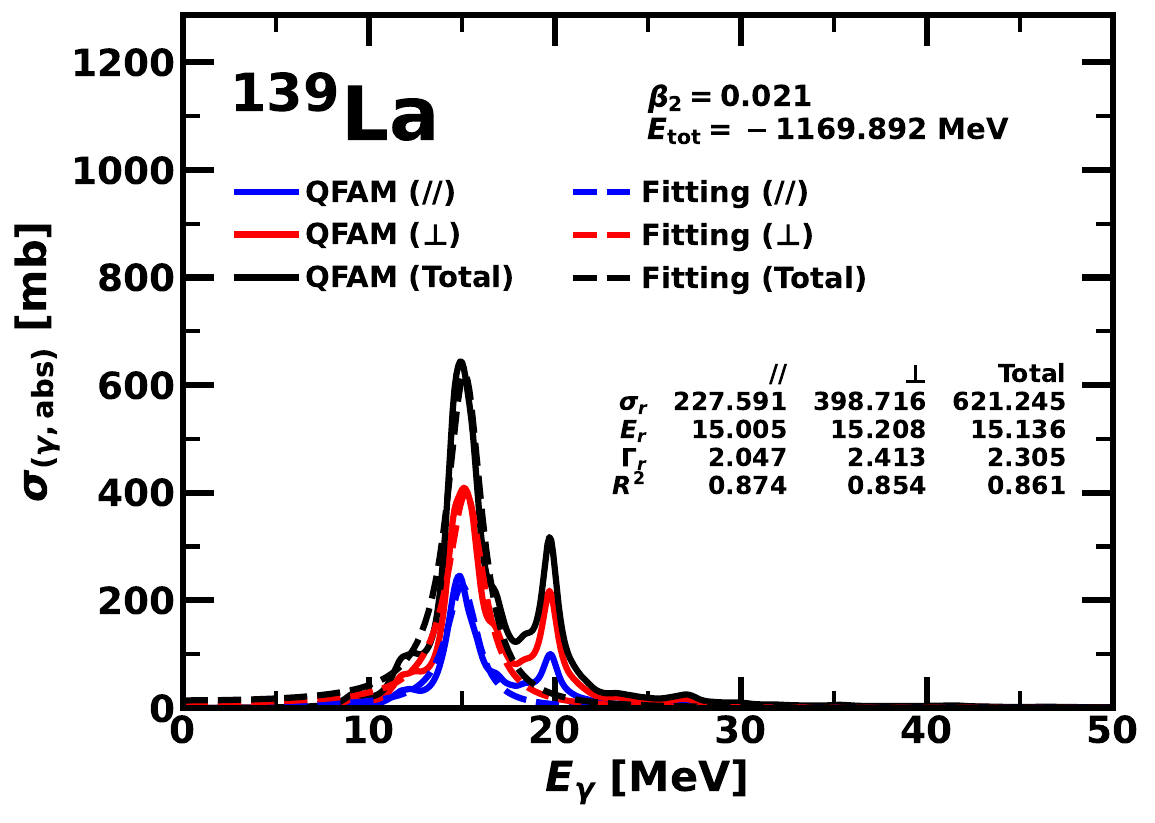}
    \includegraphics[width=0.4\textwidth]{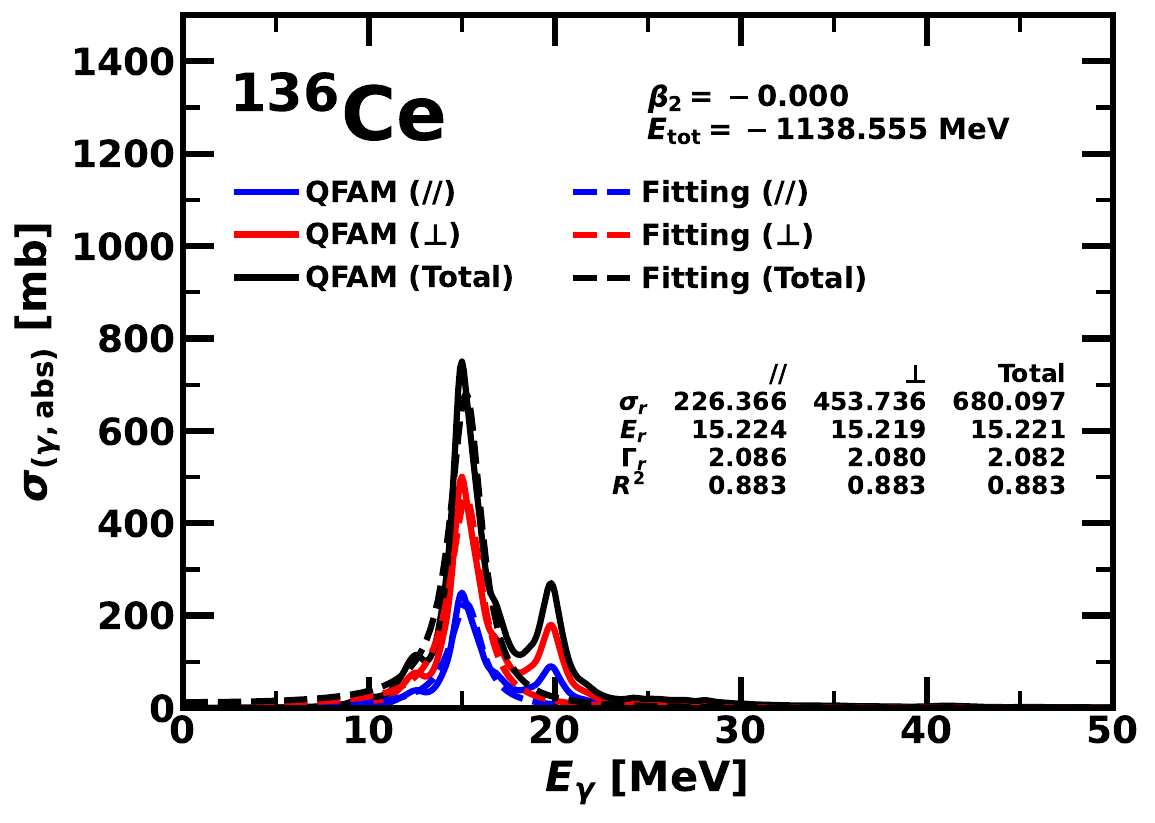}
    \includegraphics[width=0.4\textwidth]{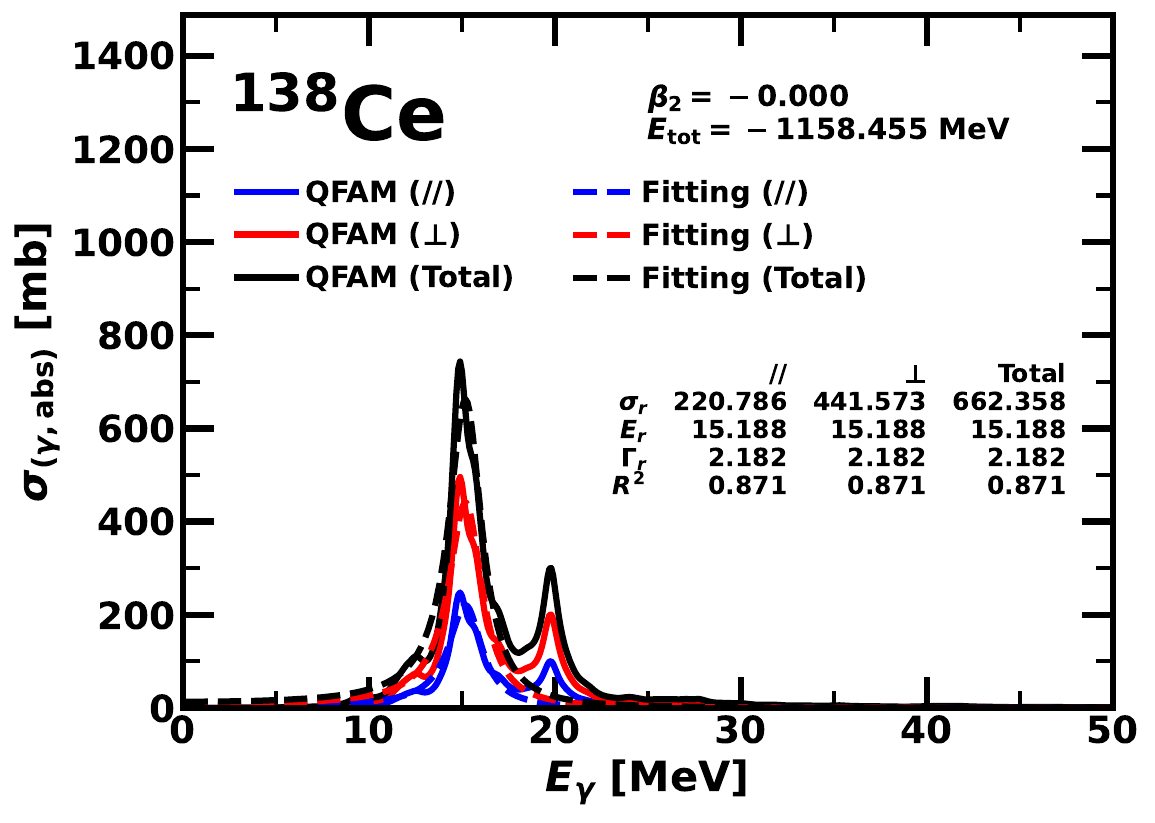}
    \includegraphics[width=0.4\textwidth]{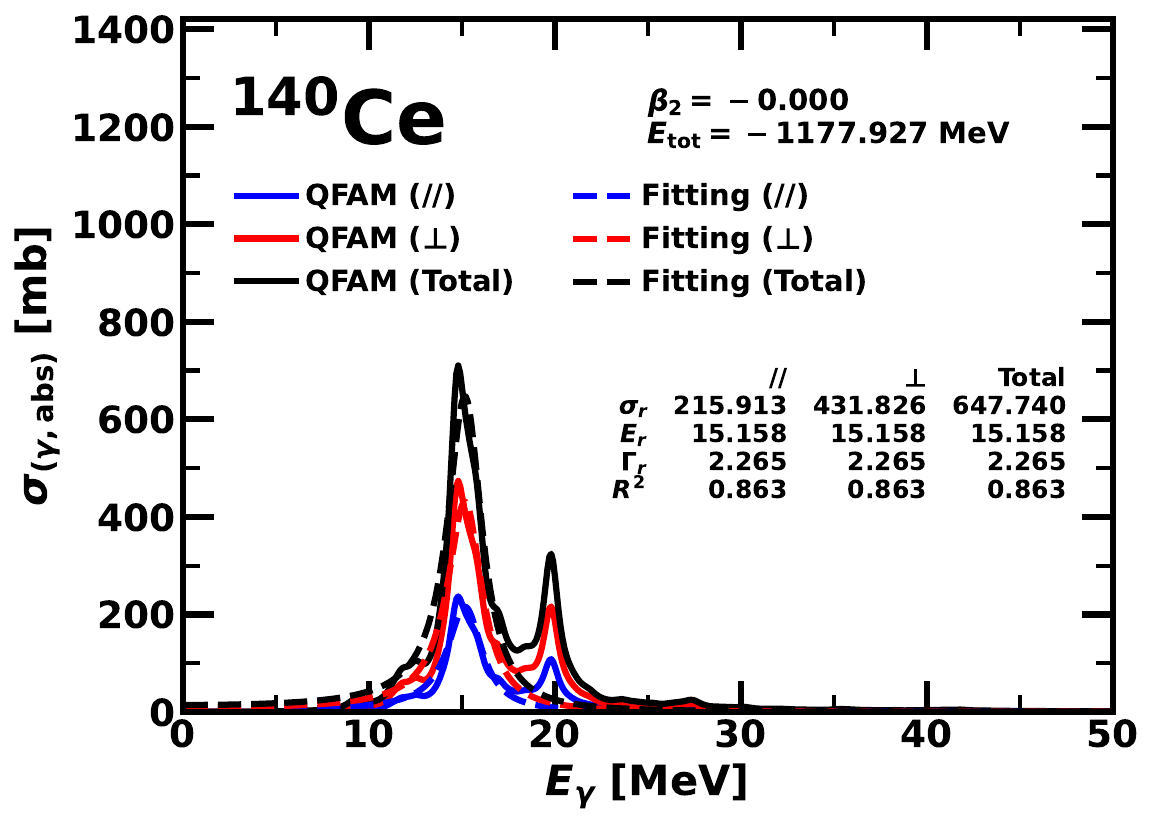}
    \includegraphics[width=0.4\textwidth]{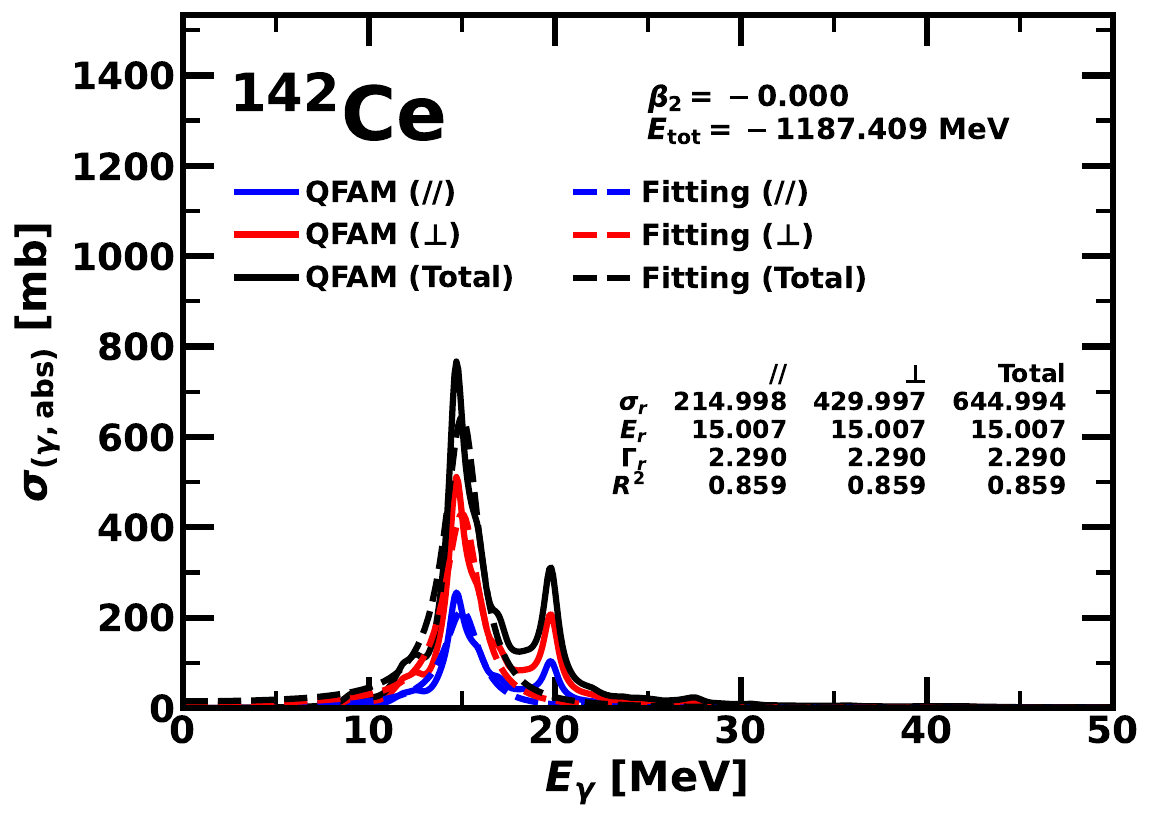}
\end{figure*}
\begin{figure*}\ContinuedFloat
    \centering
    \includegraphics[width=0.4\textwidth]{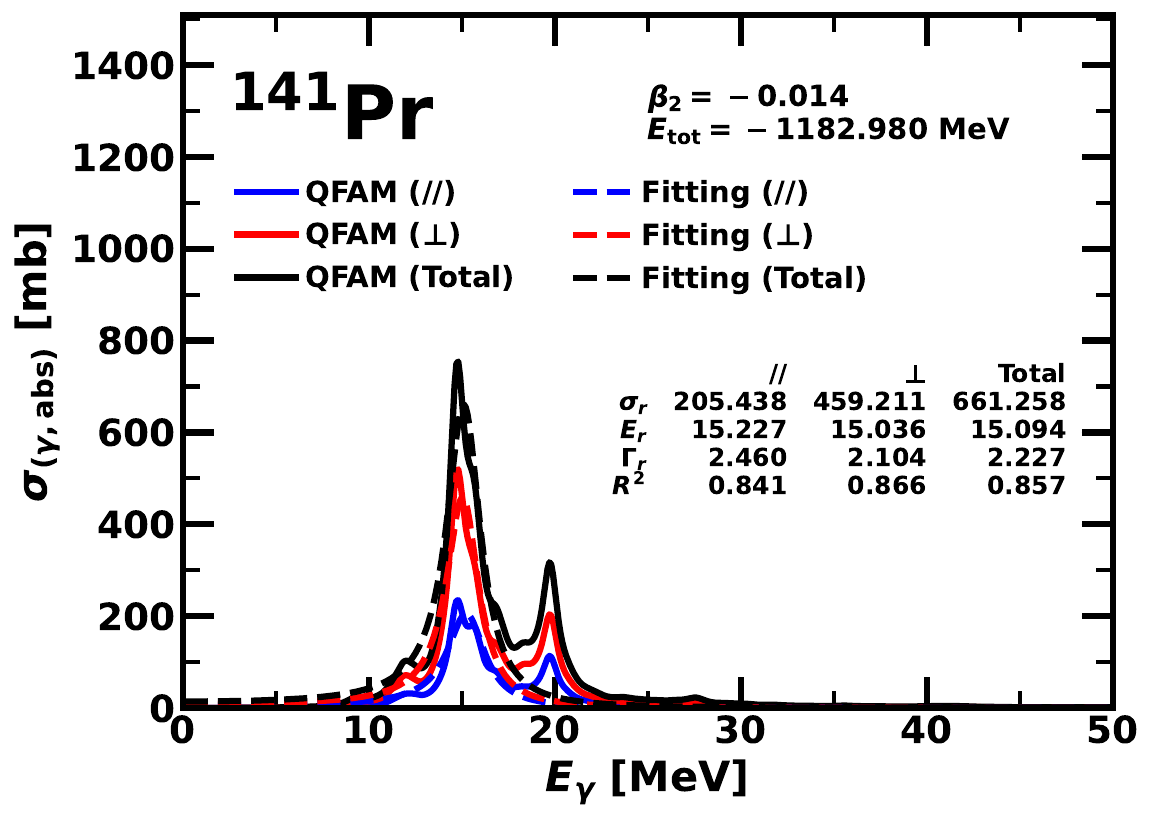}
    \includegraphics[width=0.4\textwidth]{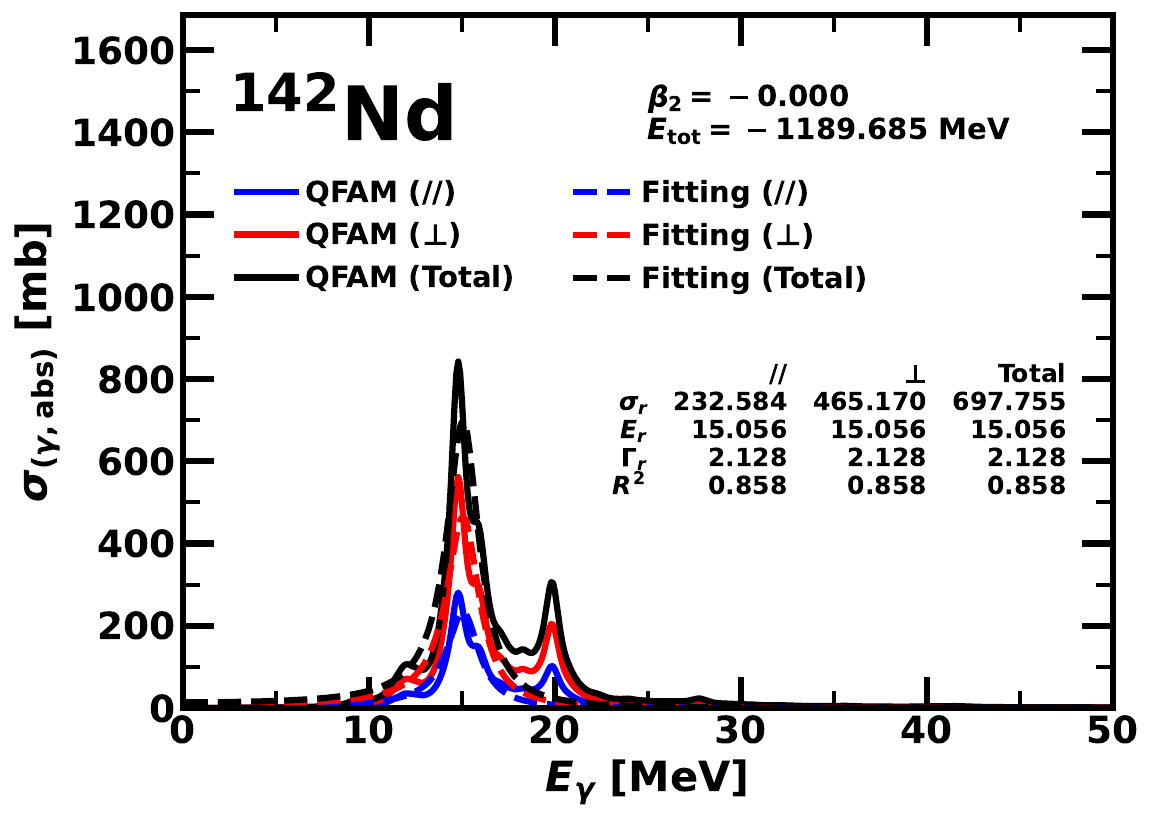}
    \includegraphics[width=0.4\textwidth]{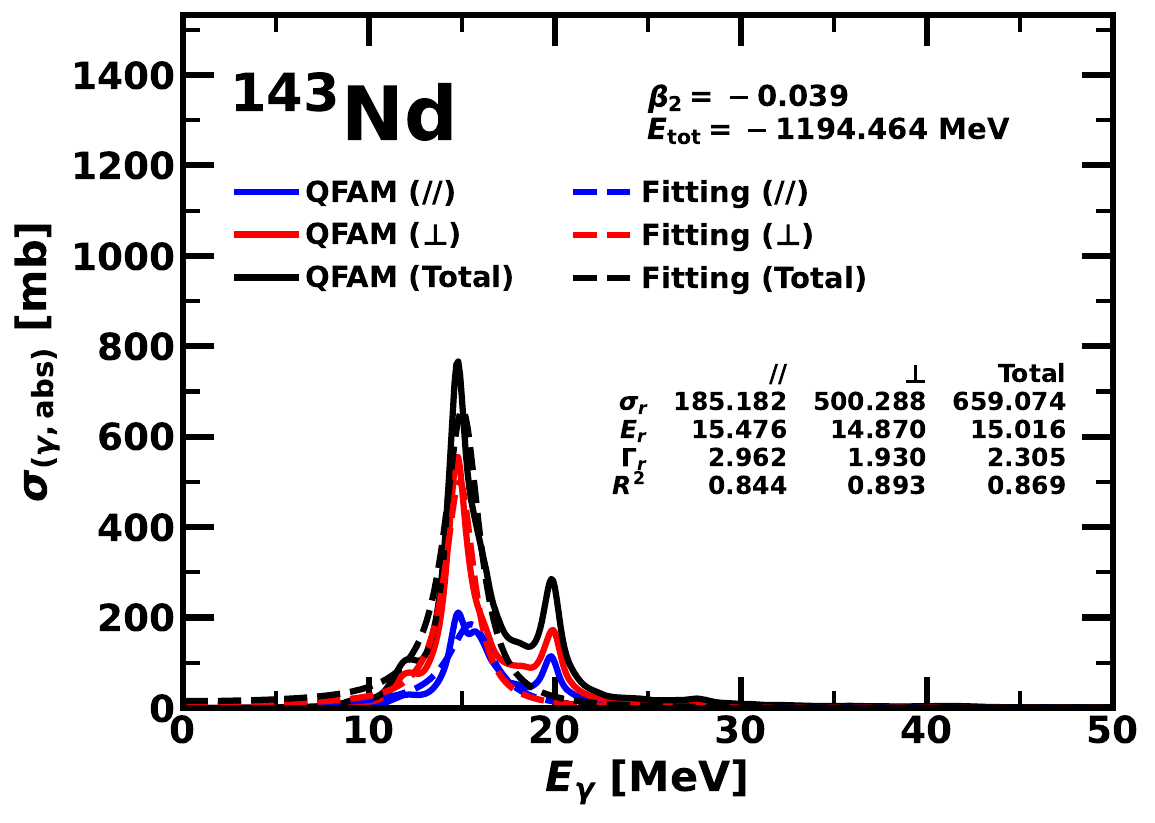}
    \includegraphics[width=0.4\textwidth]{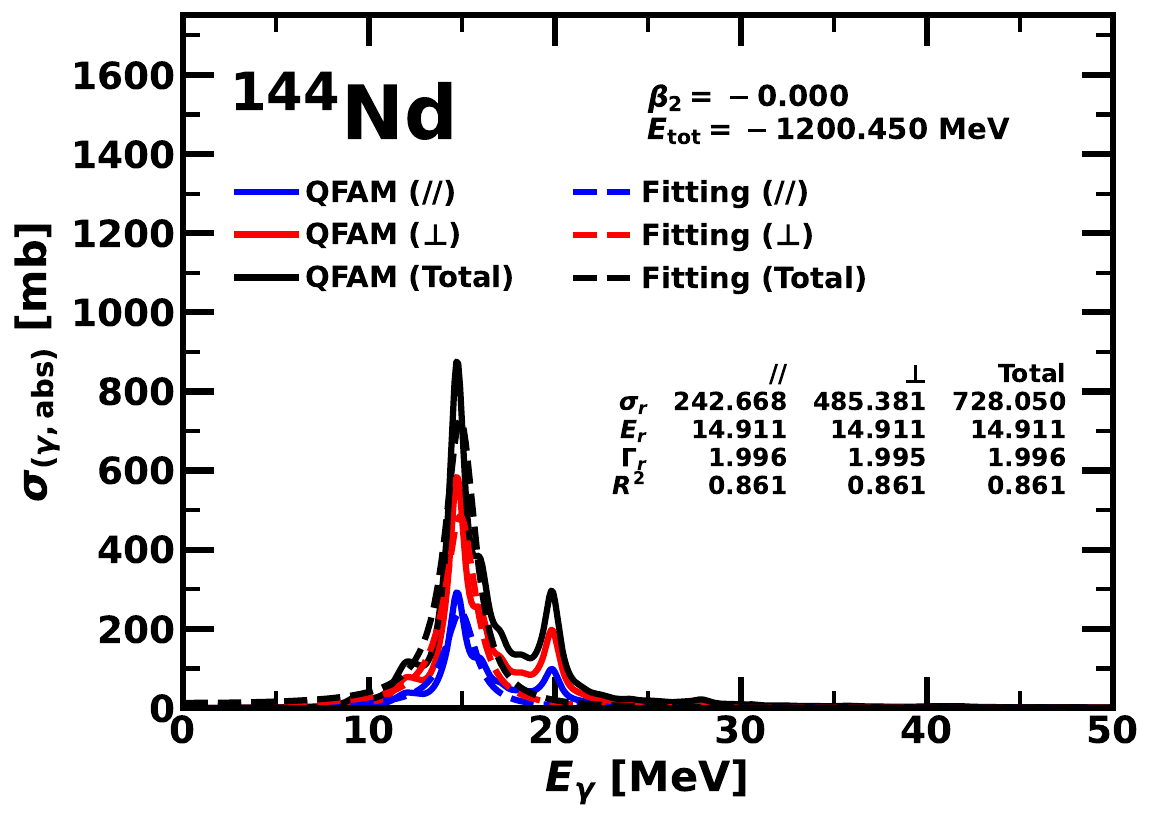}
    \includegraphics[width=0.4\textwidth]{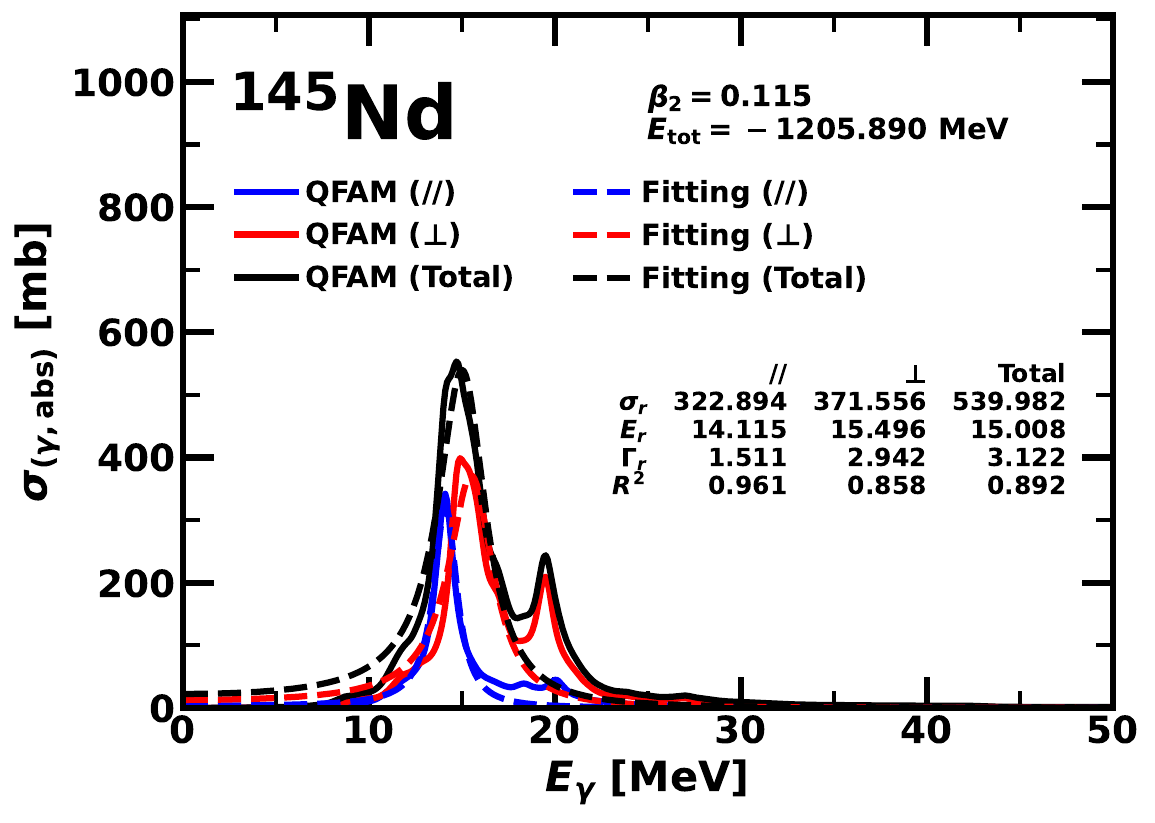}
    \includegraphics[width=0.4\textwidth]{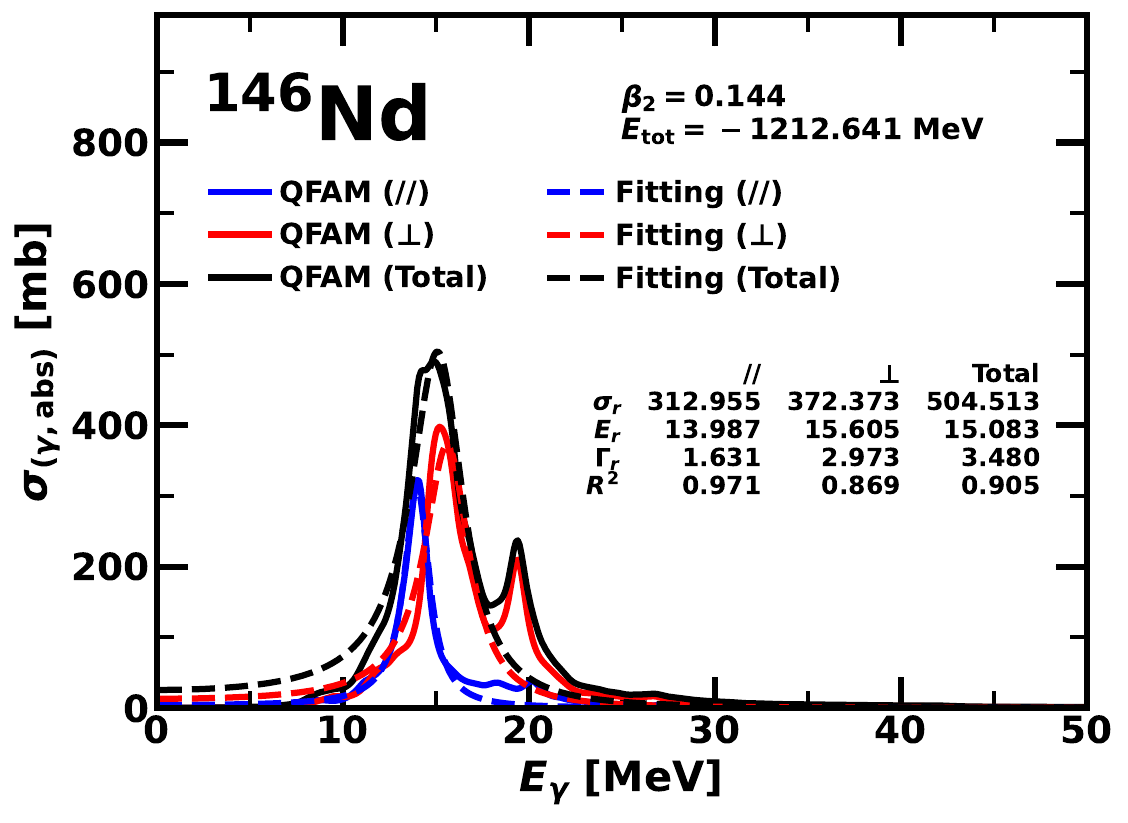}
    \includegraphics[width=0.4\textwidth]{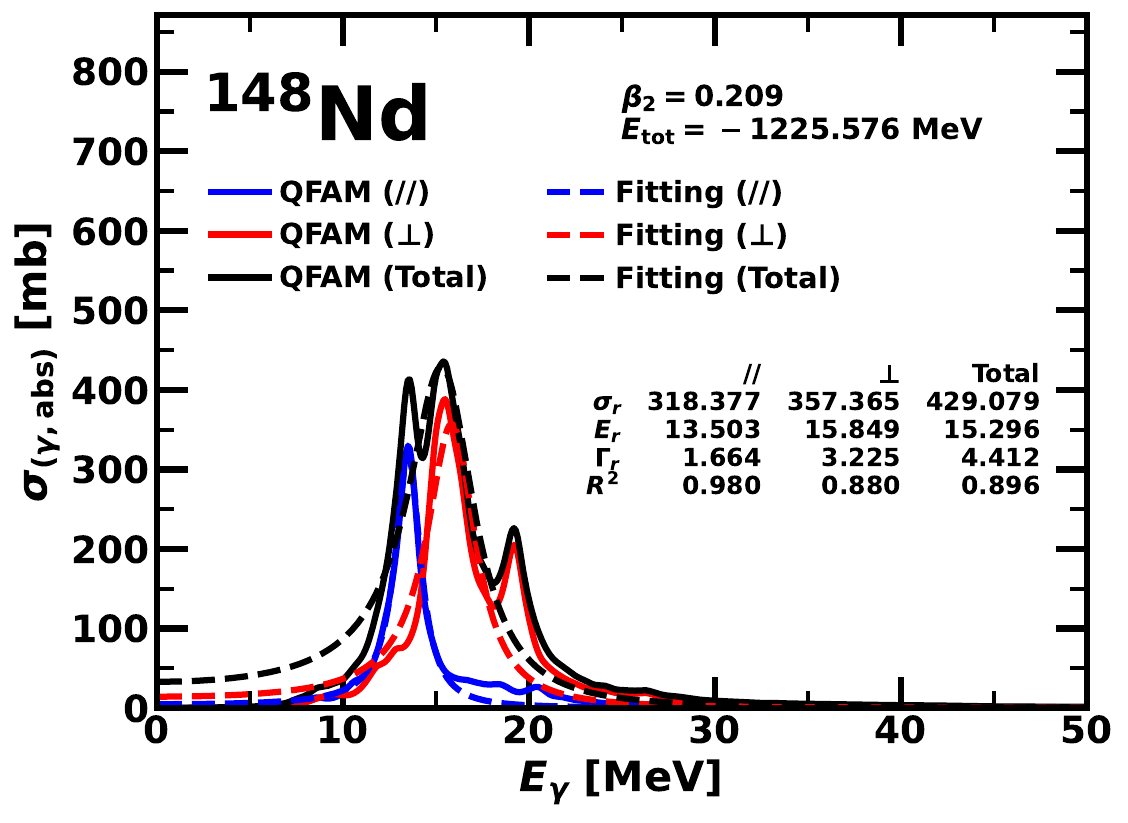}
    \includegraphics[width=0.4\textwidth]{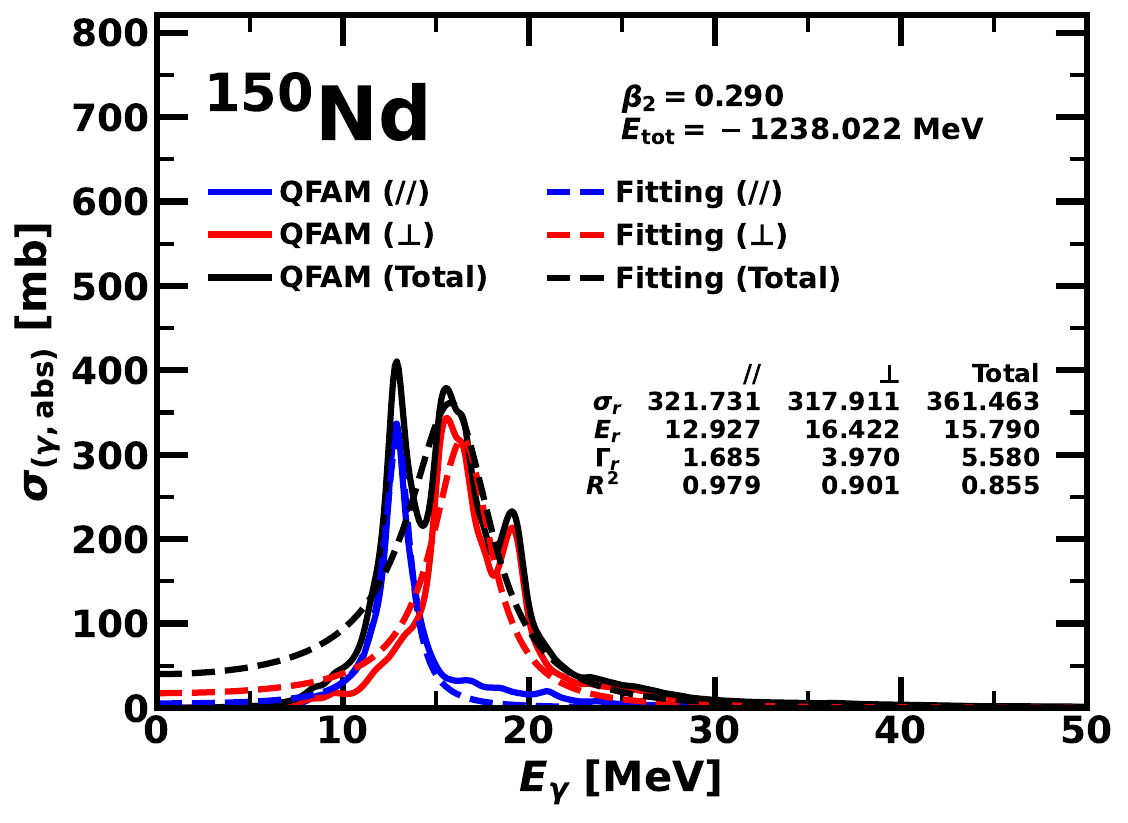}
\end{figure*}
\begin{figure*}\ContinuedFloat
    \centering
    \includegraphics[width=0.4\textwidth]{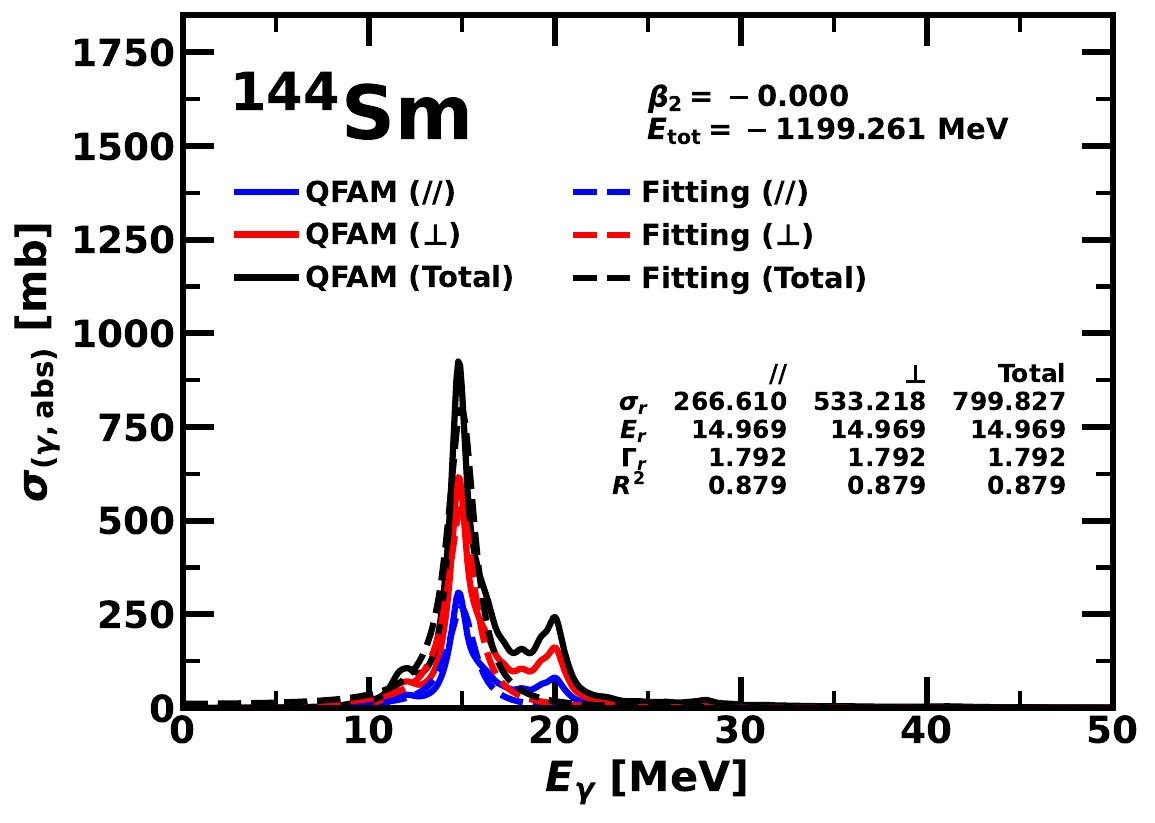}
    \includegraphics[width=0.4\textwidth]{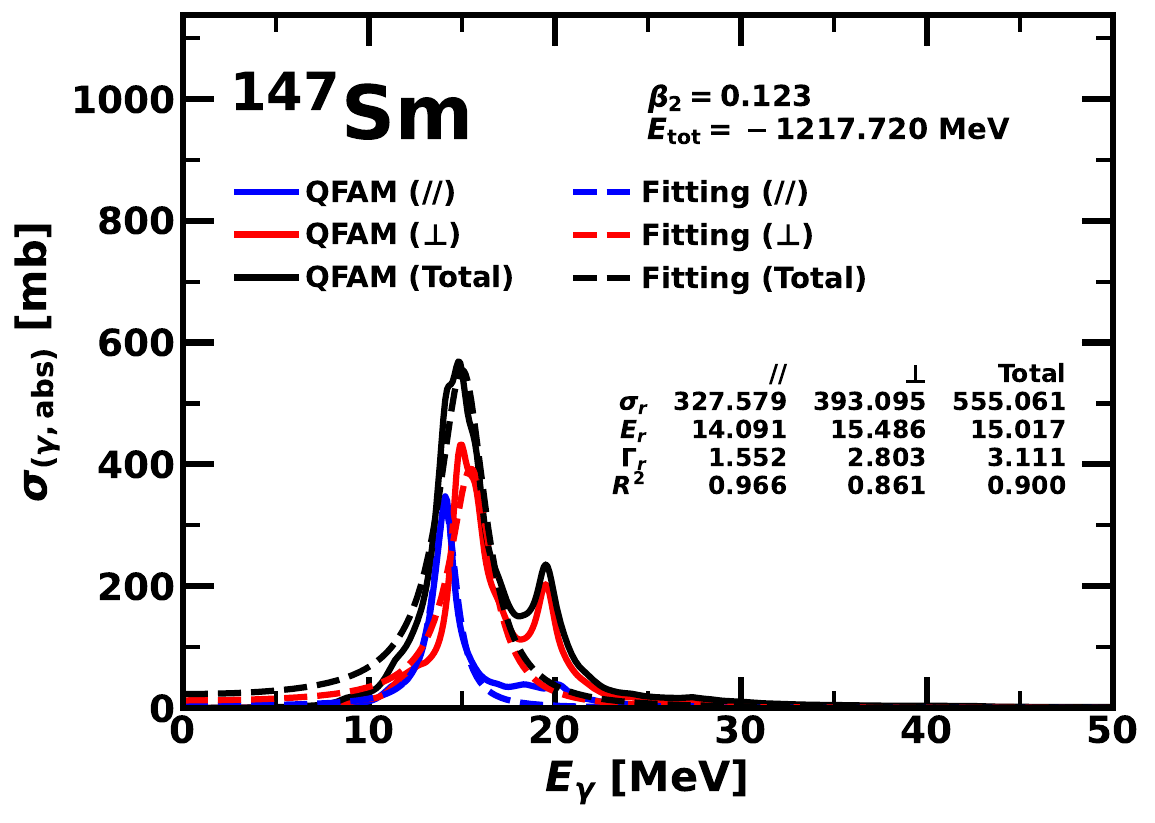}
    \includegraphics[width=0.4\textwidth]{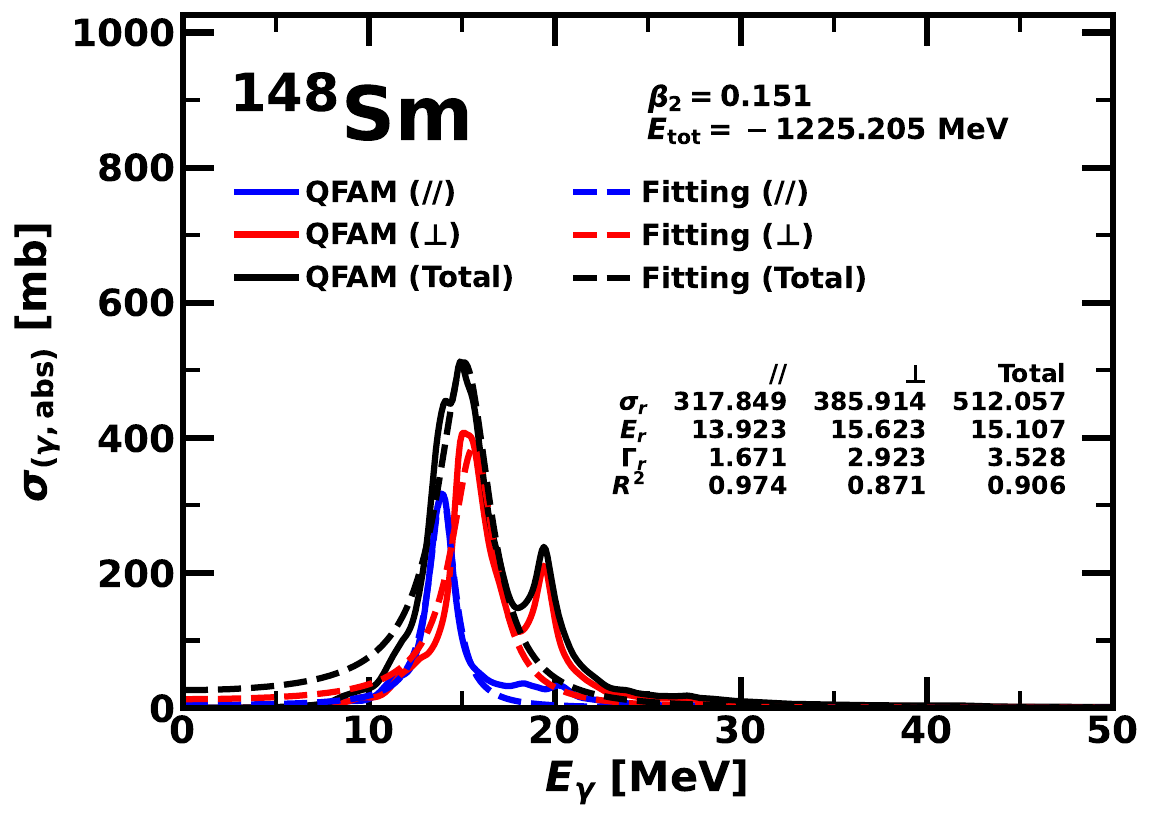}
    \includegraphics[width=0.4\textwidth]{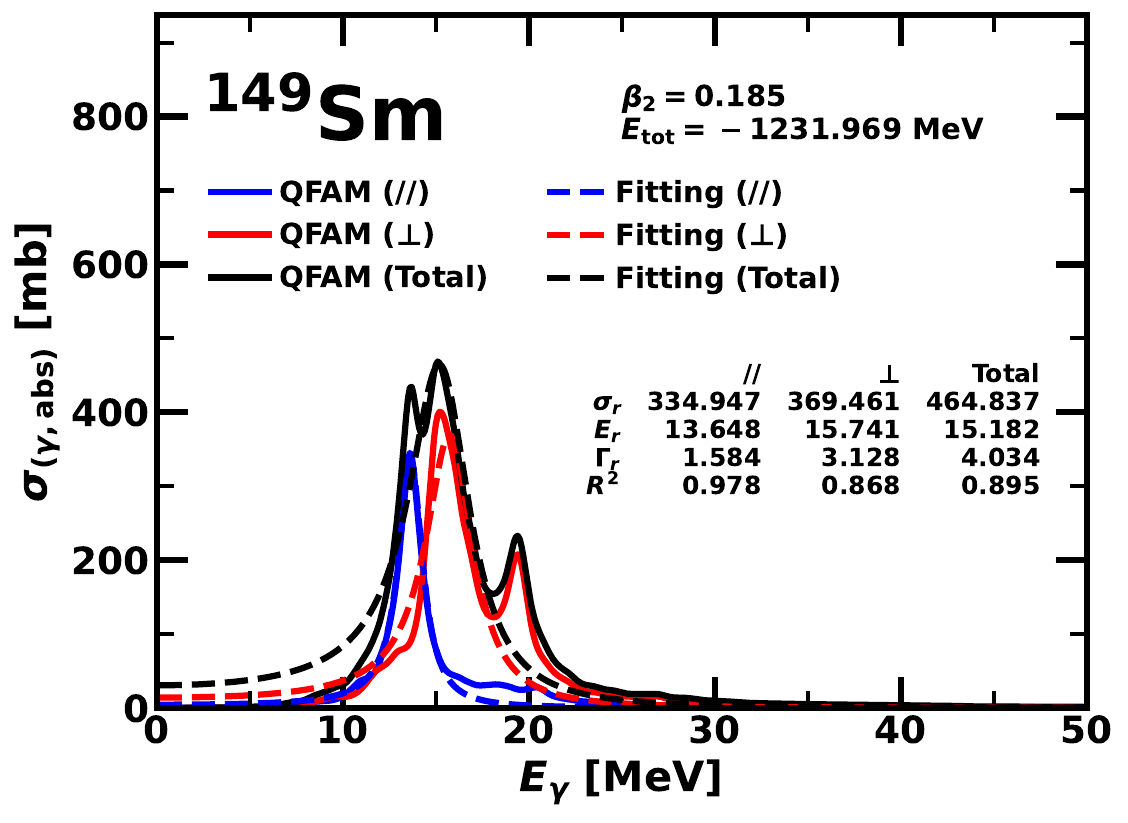}
    \includegraphics[width=0.4\textwidth]{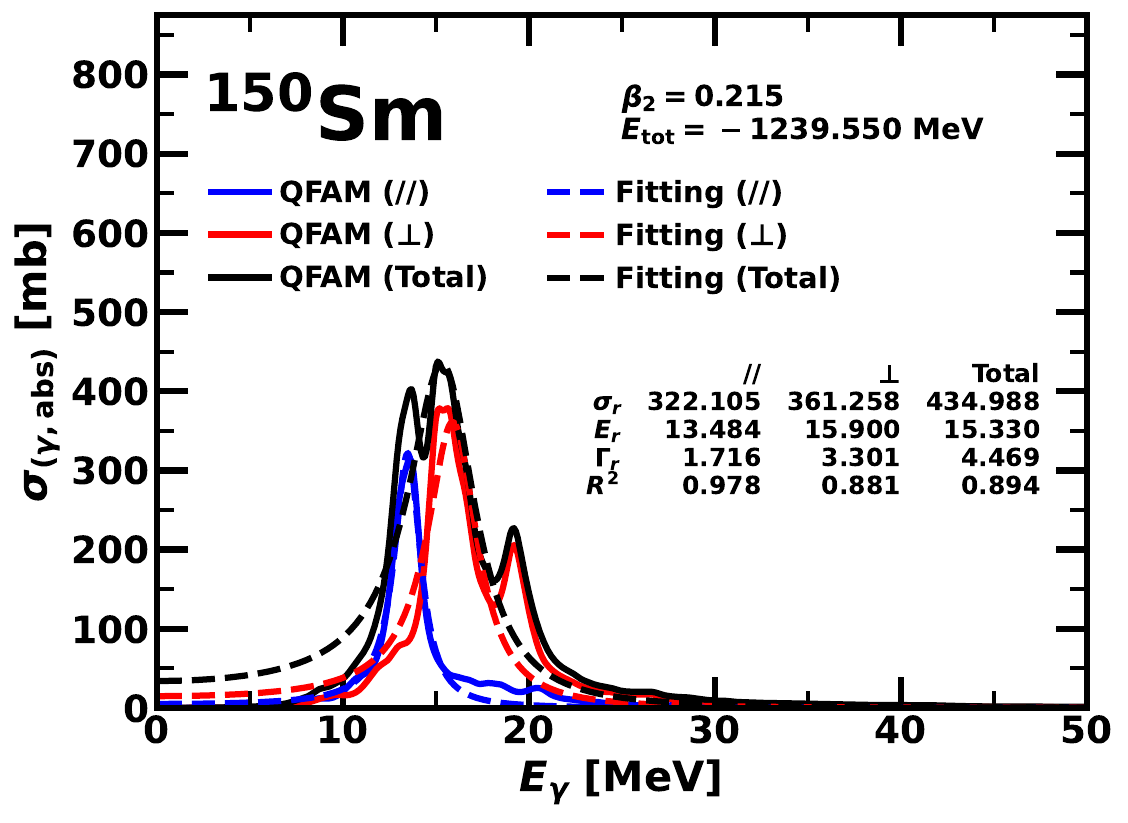}
    \includegraphics[width=0.4\textwidth]{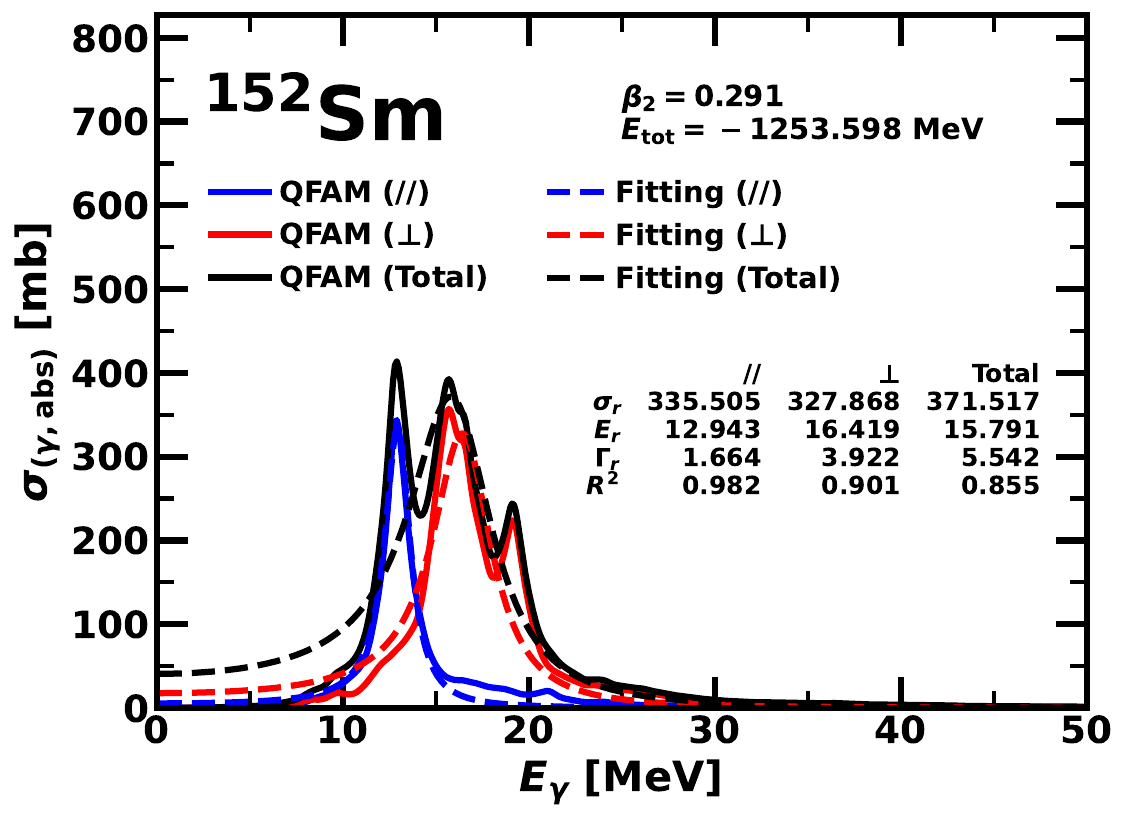}
    \includegraphics[width=0.4\textwidth]{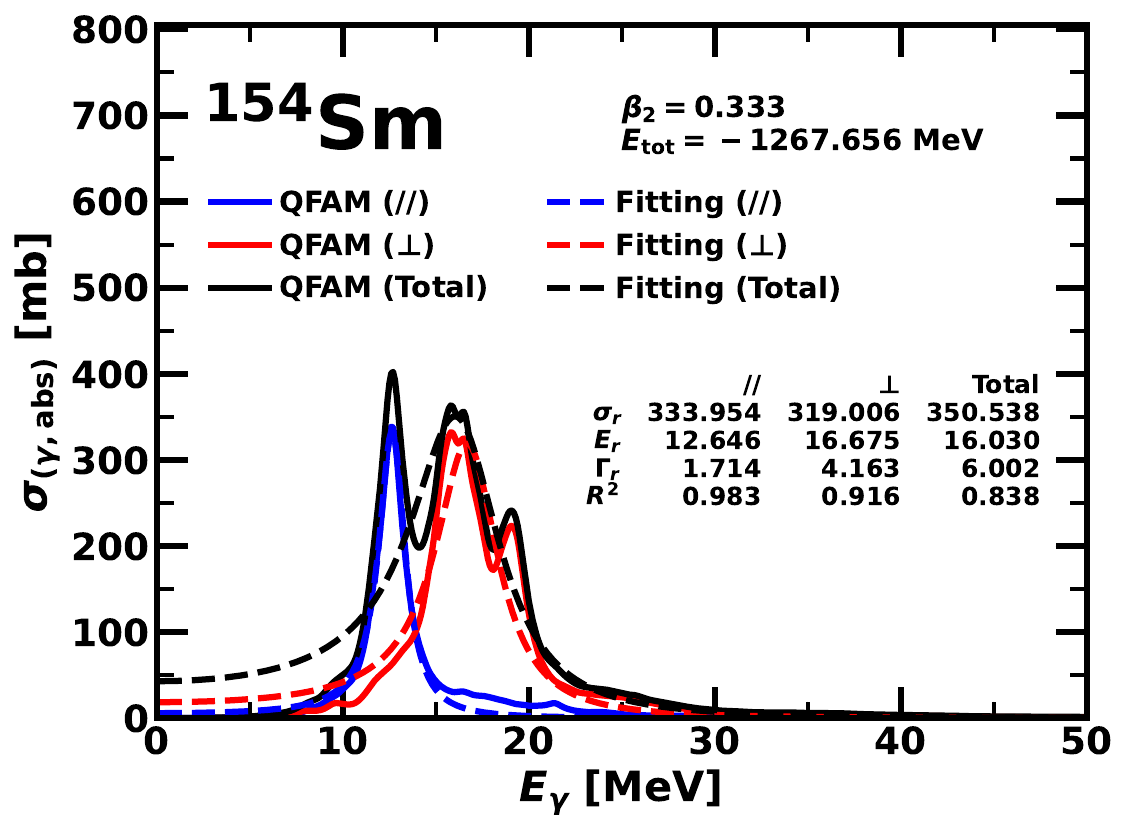}
    \includegraphics[width=0.4\textwidth]{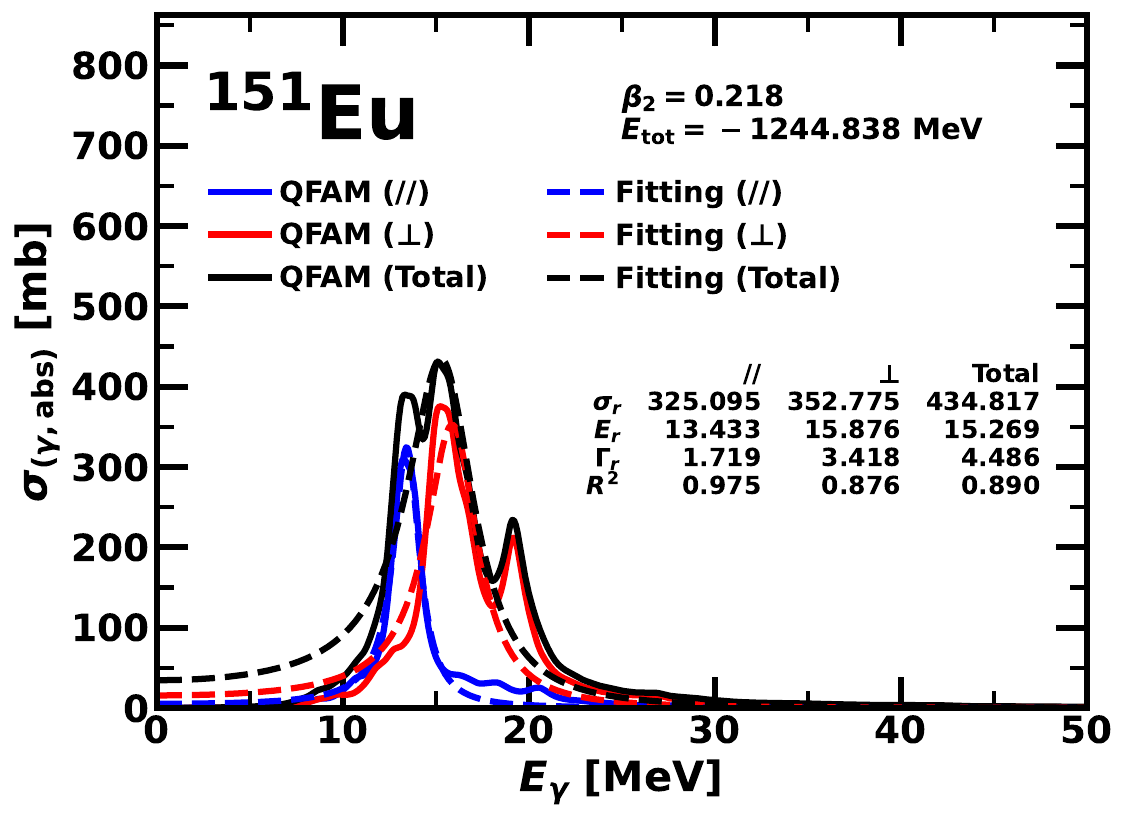}
\end{figure*}
\begin{figure*}\ContinuedFloat
    \centering
    \includegraphics[width=0.4\textwidth]{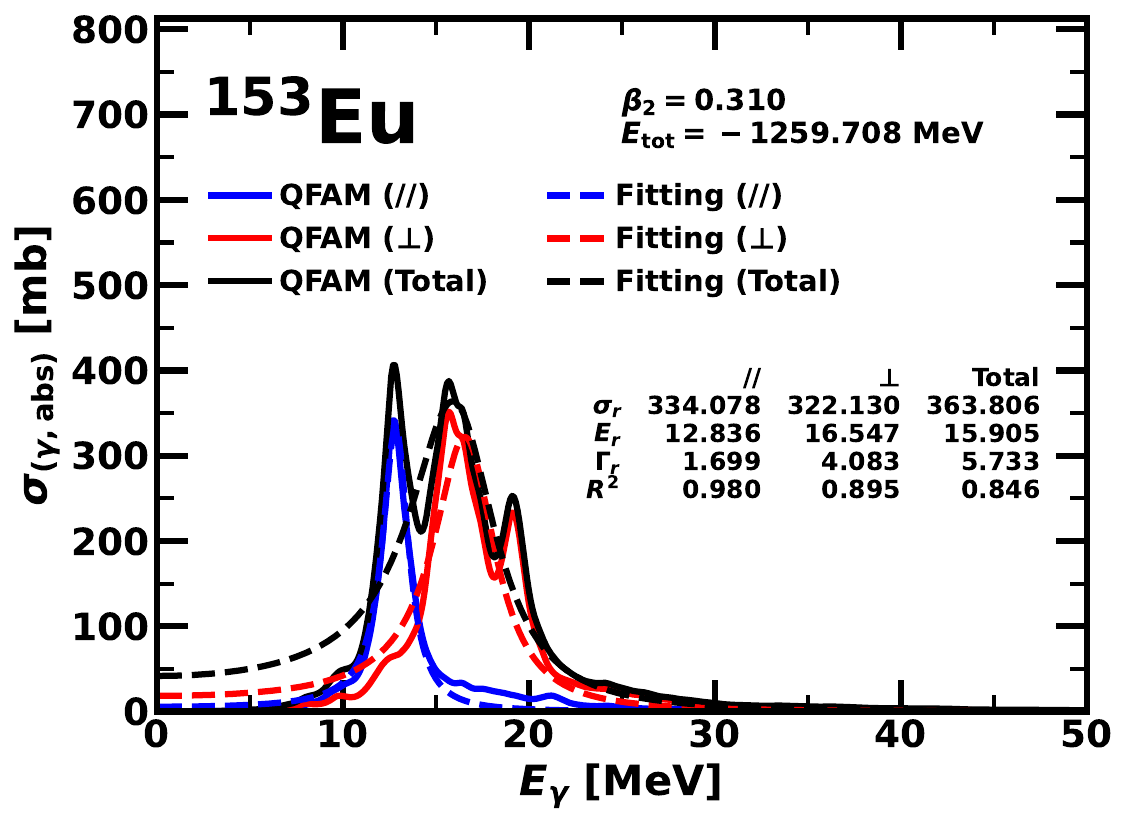}
    \includegraphics[width=0.4\textwidth]{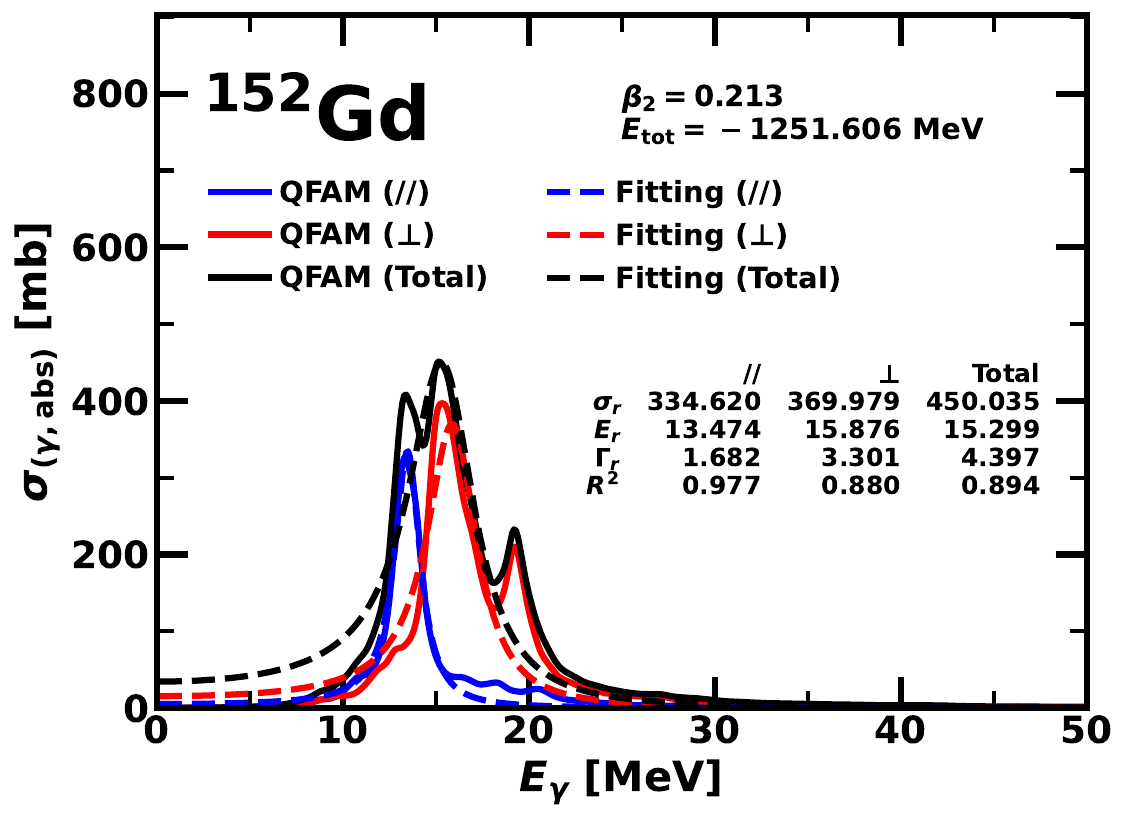}
    \includegraphics[width=0.4\textwidth]{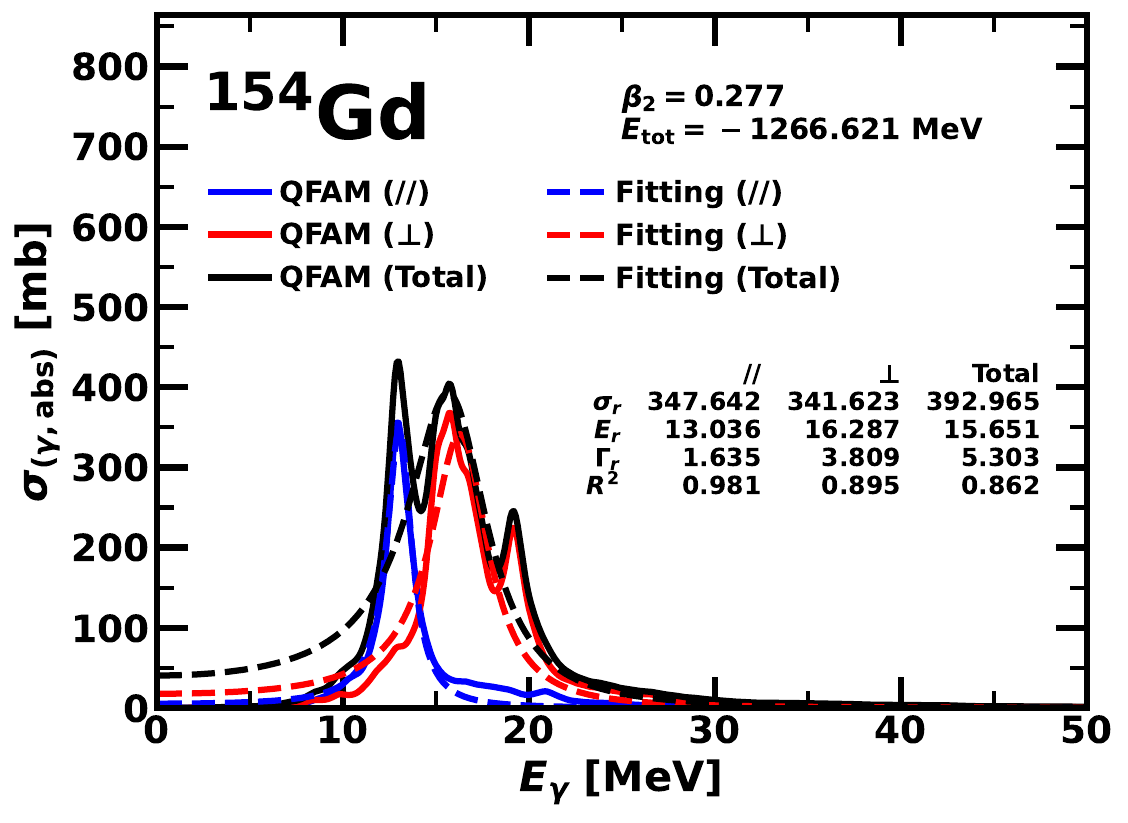}
    \includegraphics[width=0.4\textwidth]{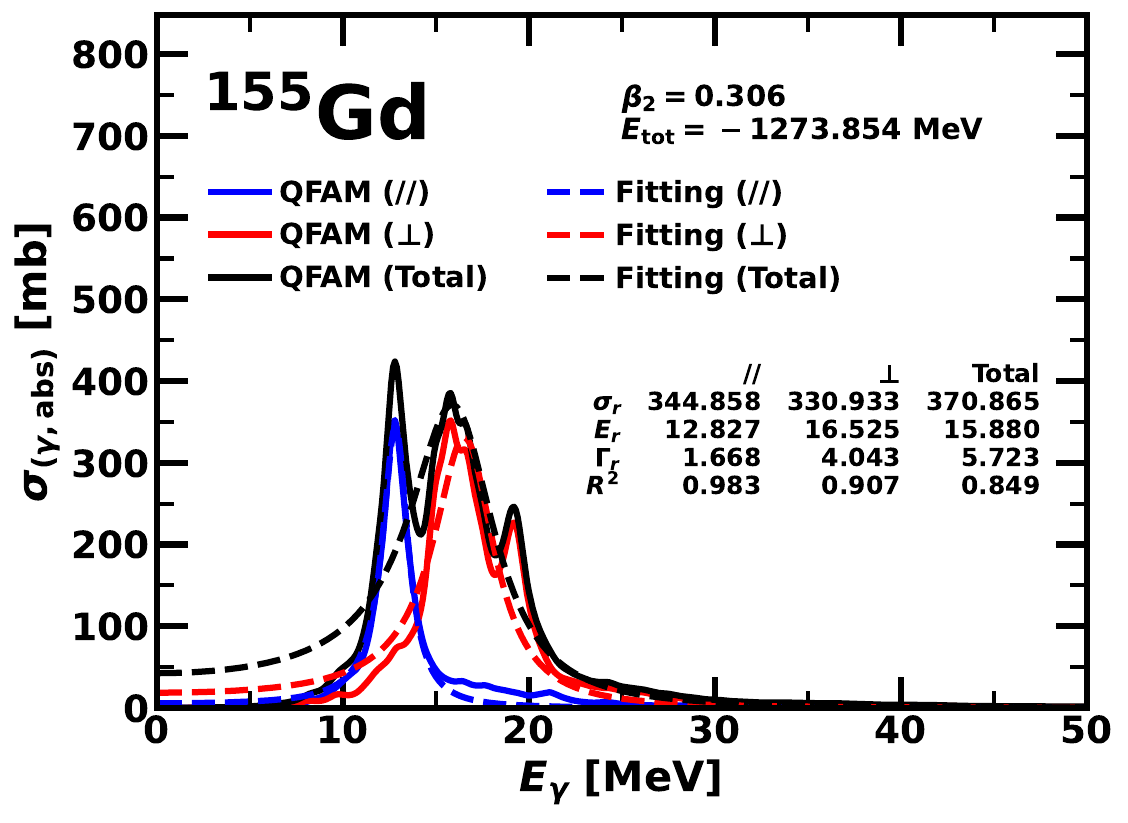}
    \includegraphics[width=0.4\textwidth]{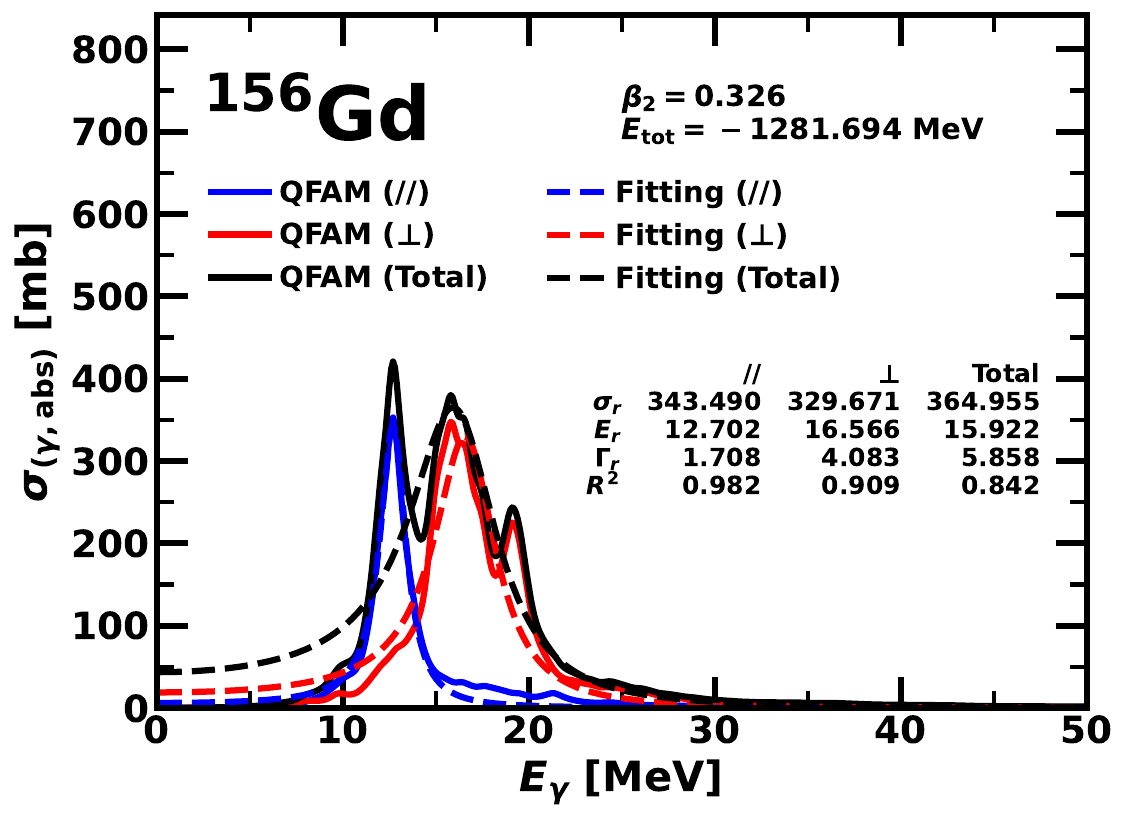}
    \includegraphics[width=0.4\textwidth]{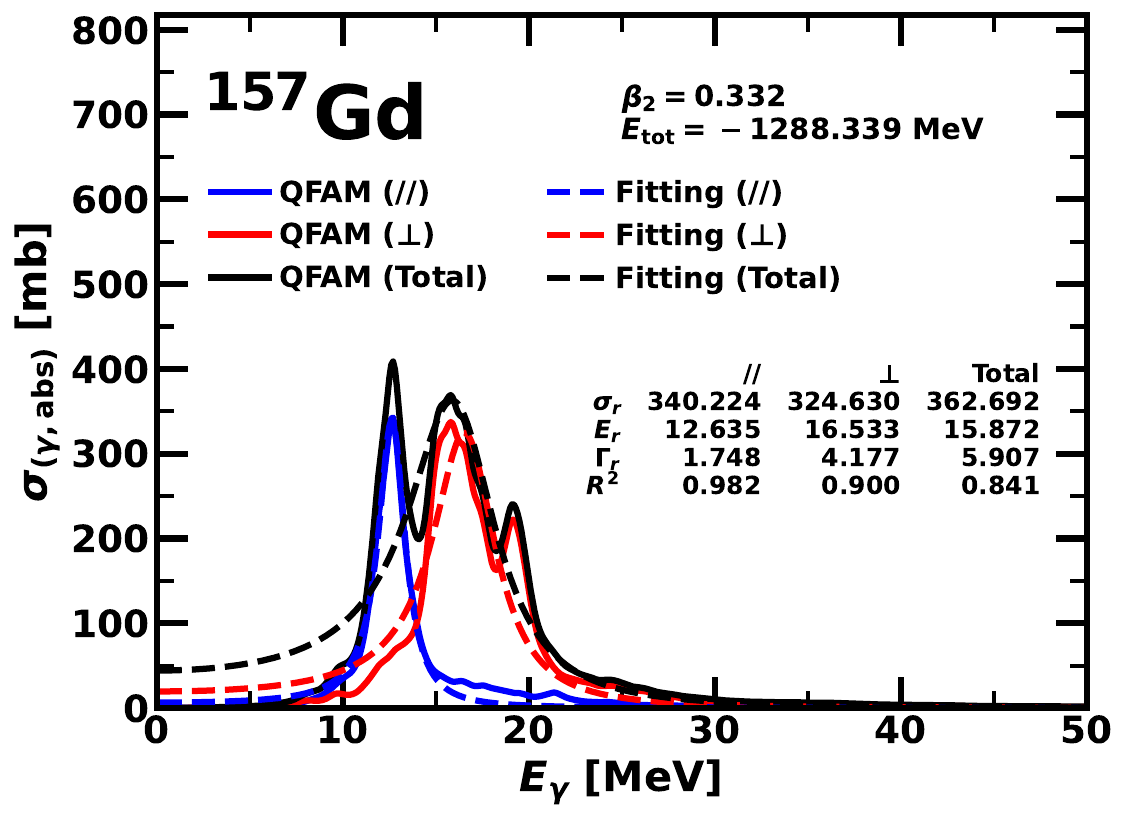}
    \includegraphics[width=0.4\textwidth]{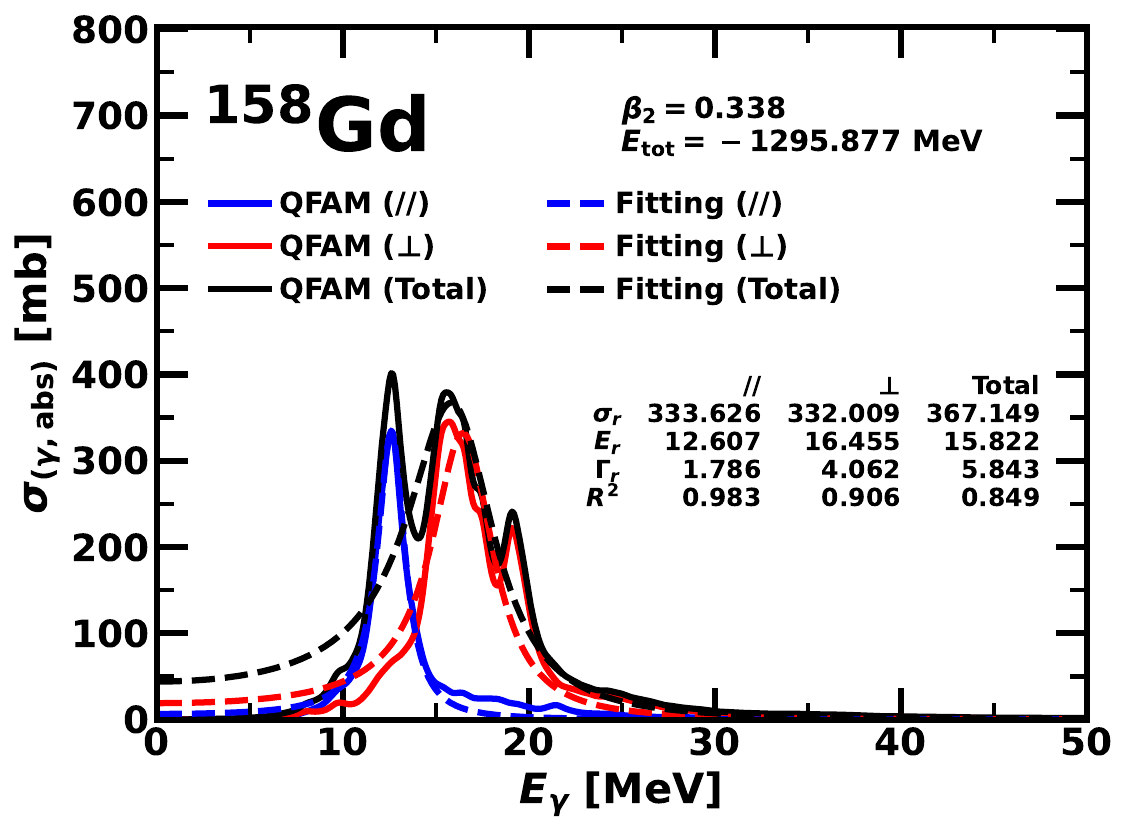}
    \includegraphics[width=0.4\textwidth]{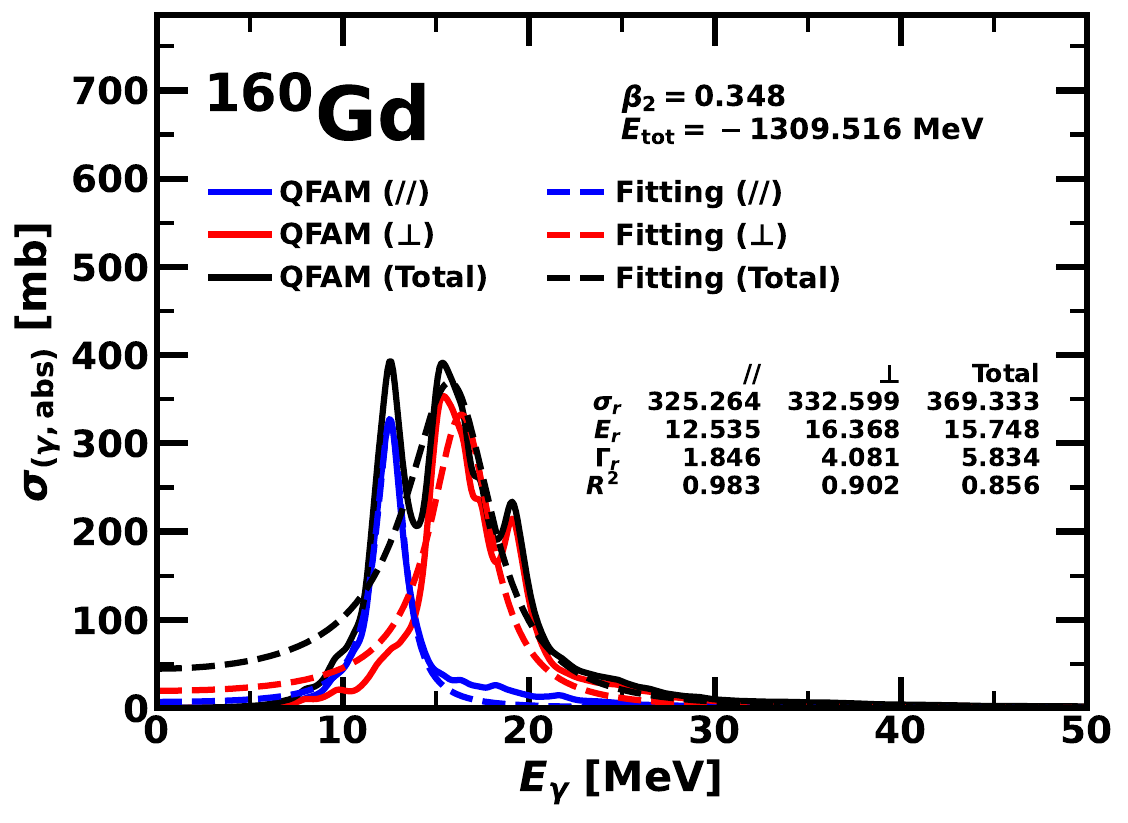}
\end{figure*}
\begin{figure*}\ContinuedFloat
    \centering
    \includegraphics[width=0.4\textwidth]{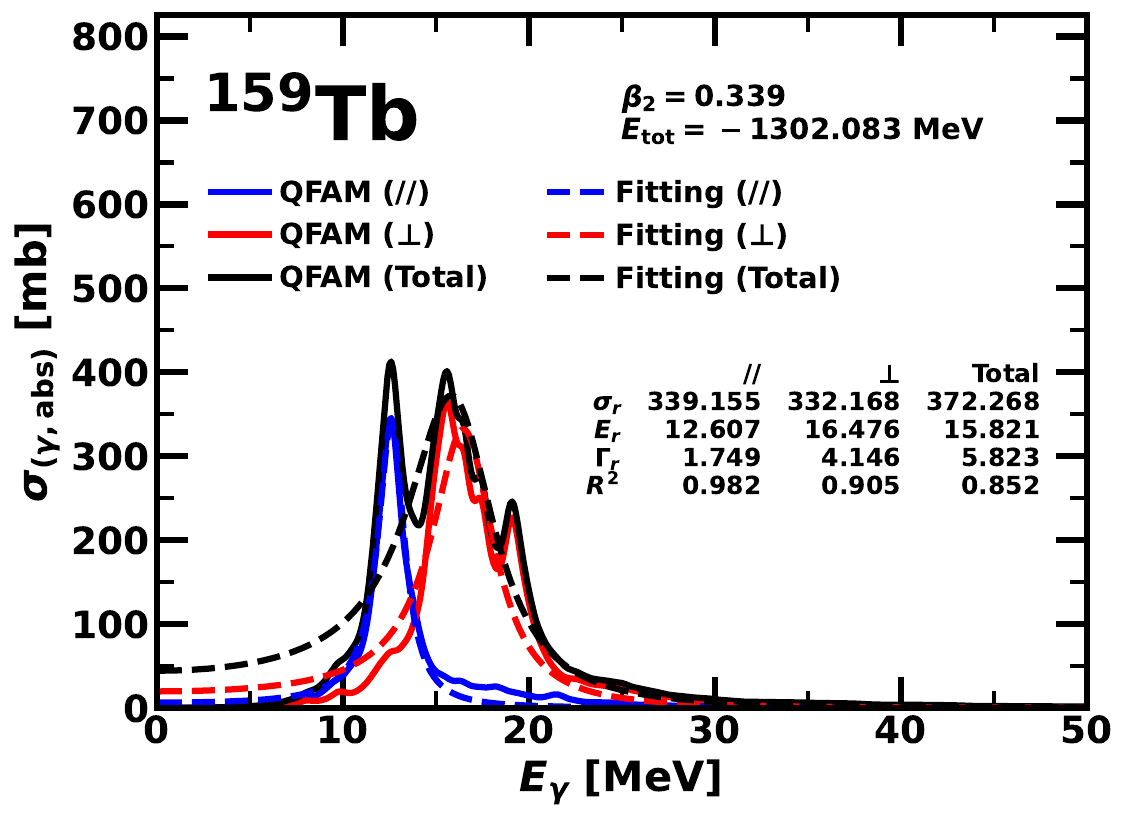}
    \includegraphics[width=0.4\textwidth]{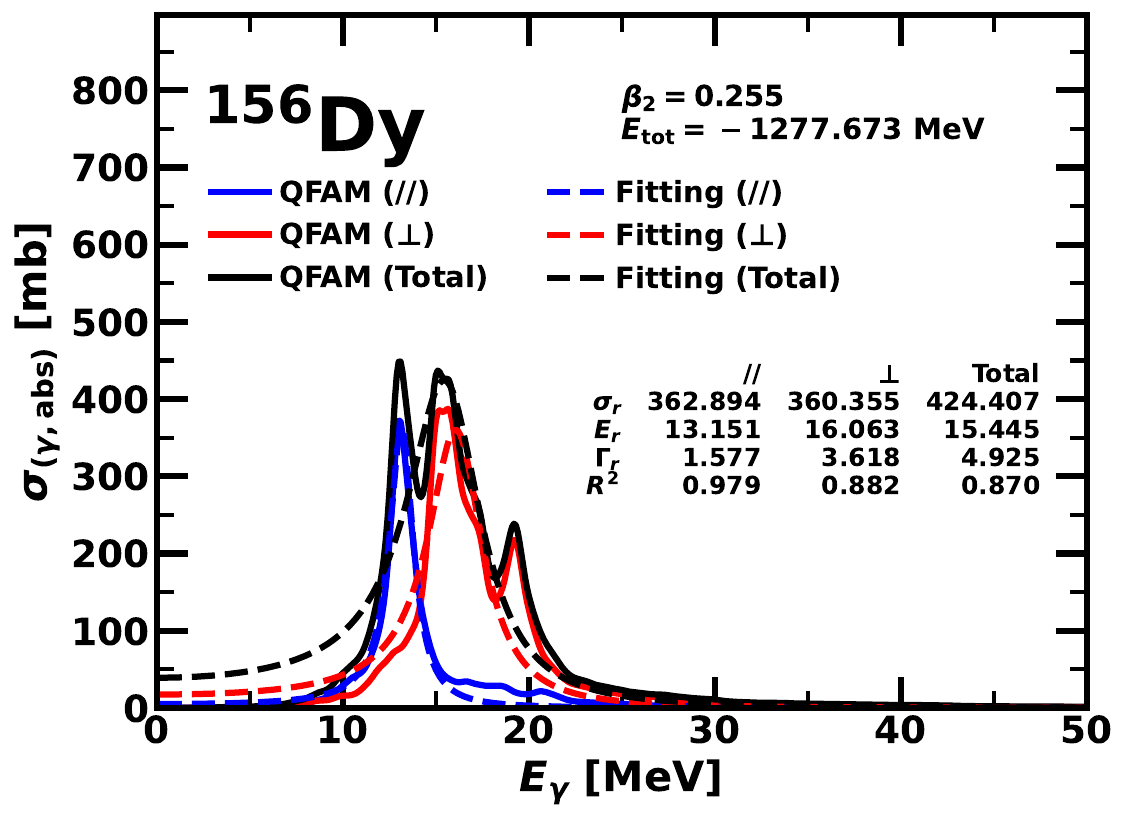}
    \includegraphics[width=0.4\textwidth]{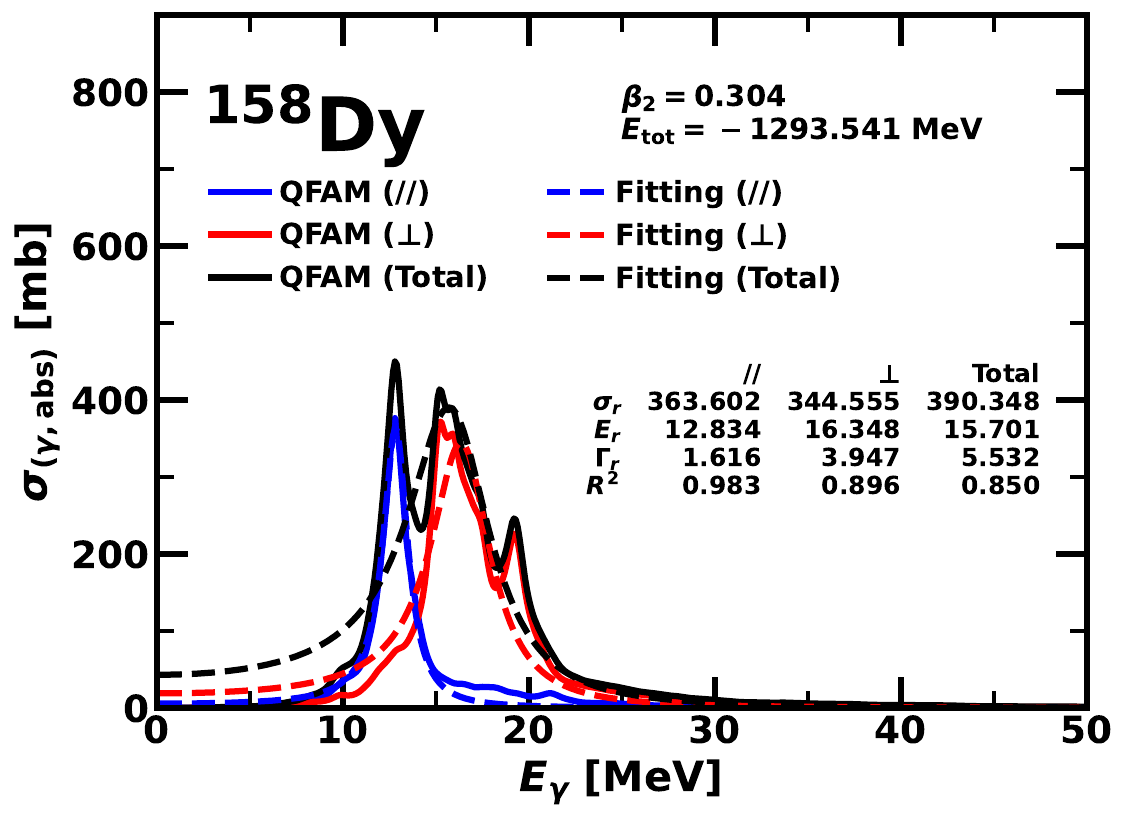}
    \includegraphics[width=0.4\textwidth]{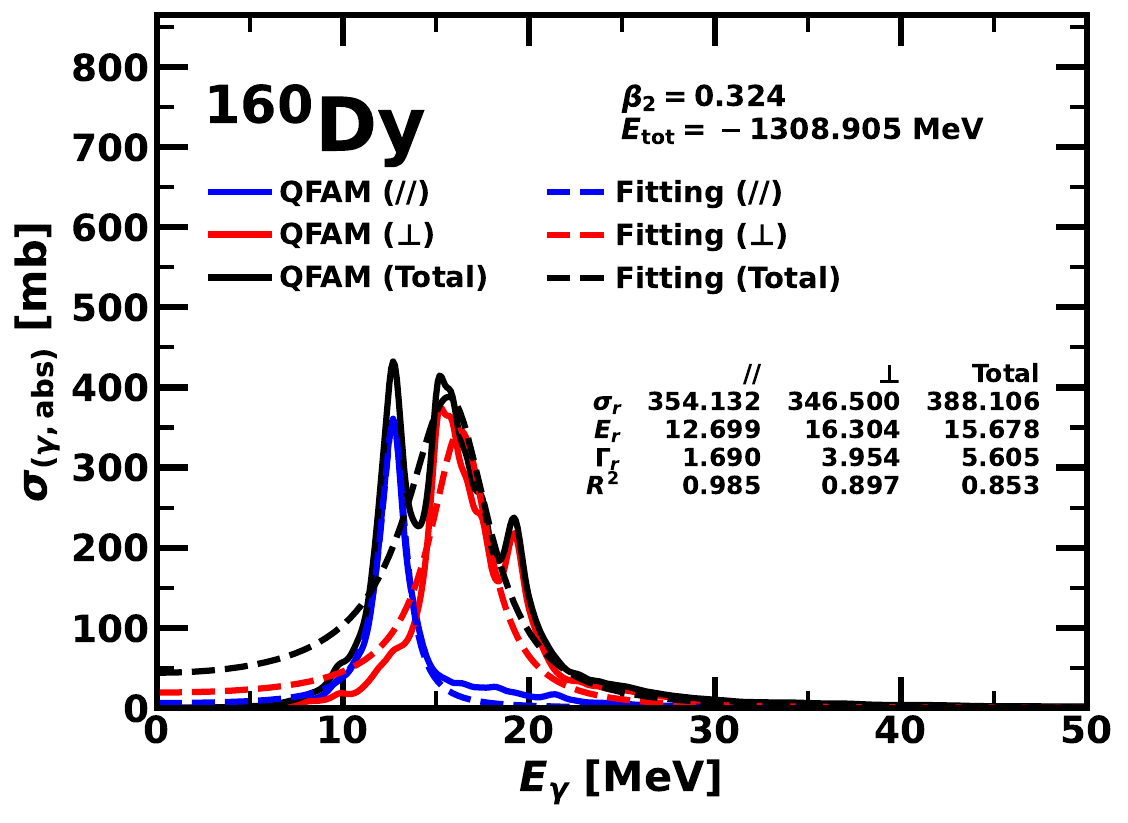}
    \includegraphics[width=0.4\textwidth]{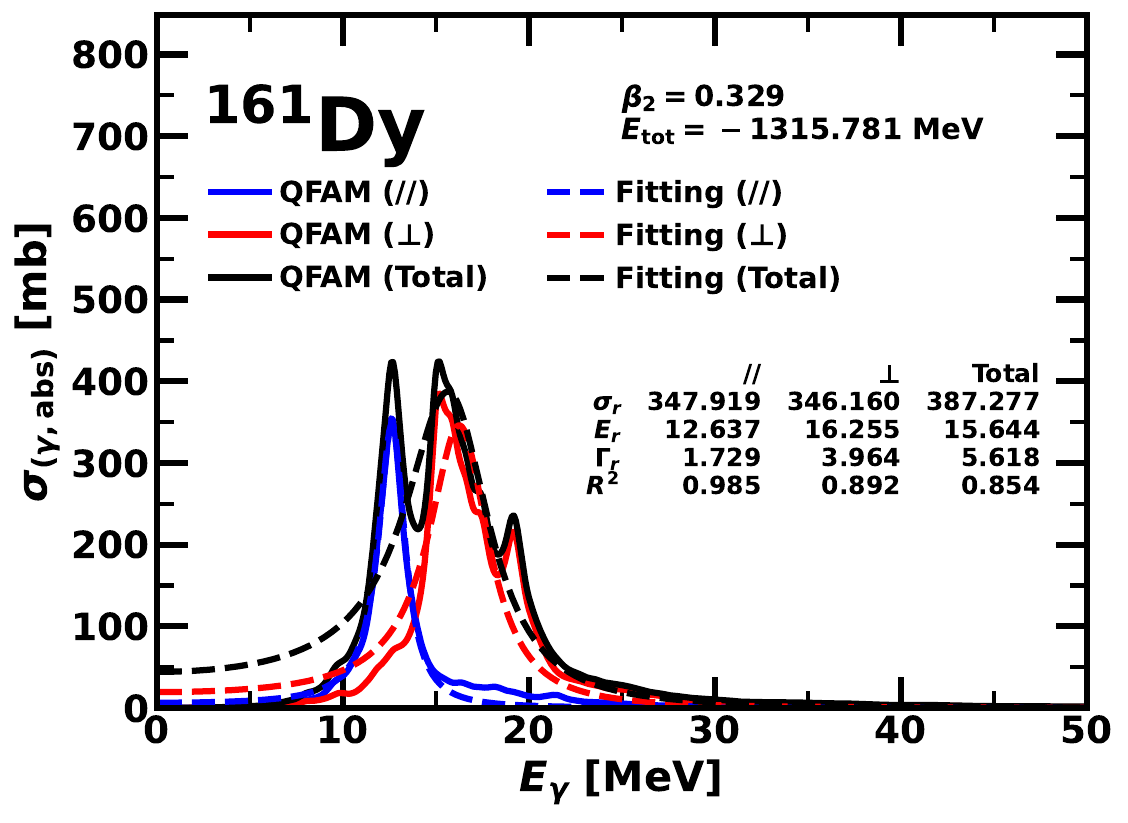}
    \includegraphics[width=0.4\textwidth]{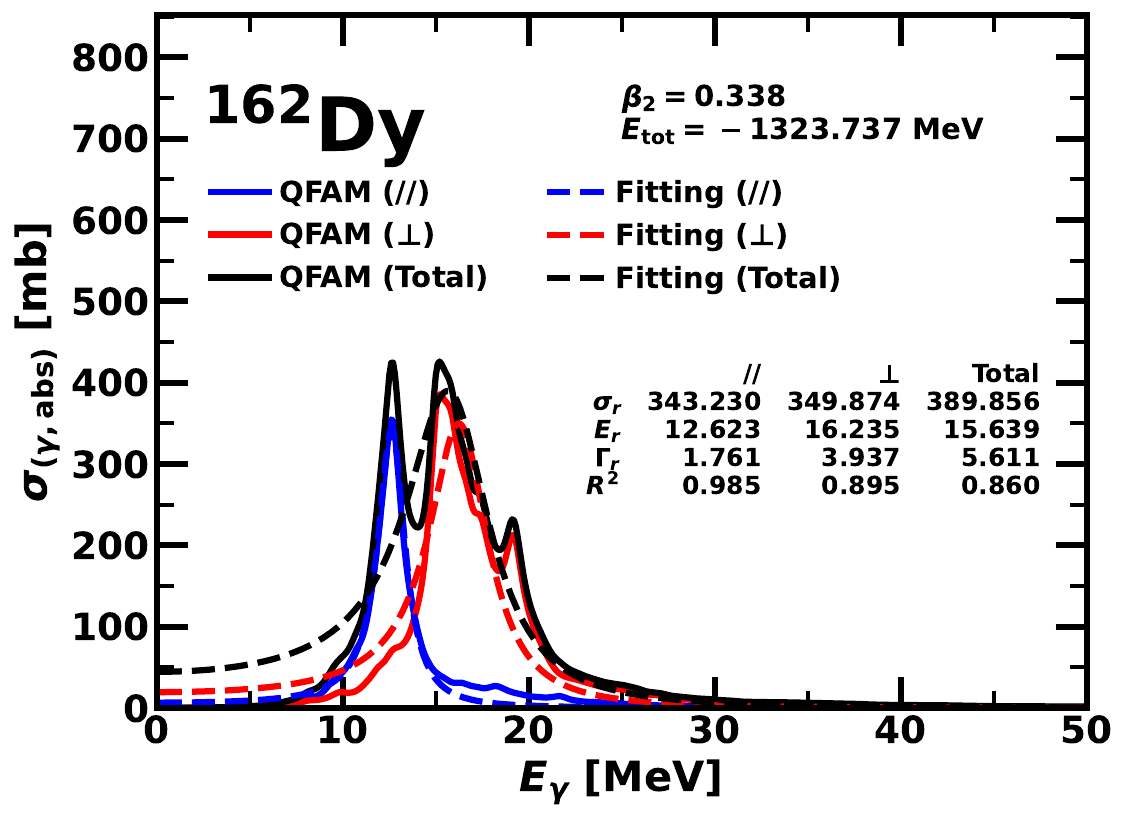}
    \includegraphics[width=0.4\textwidth]{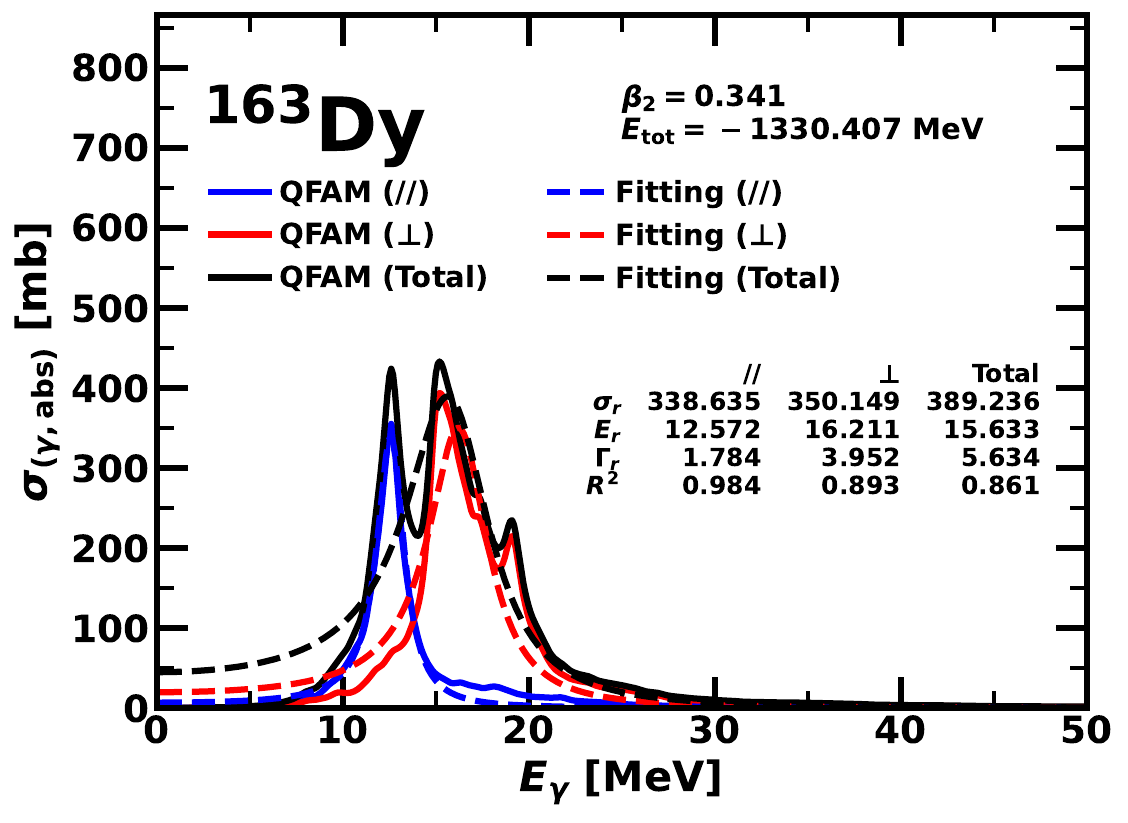}
    \includegraphics[width=0.4\textwidth]{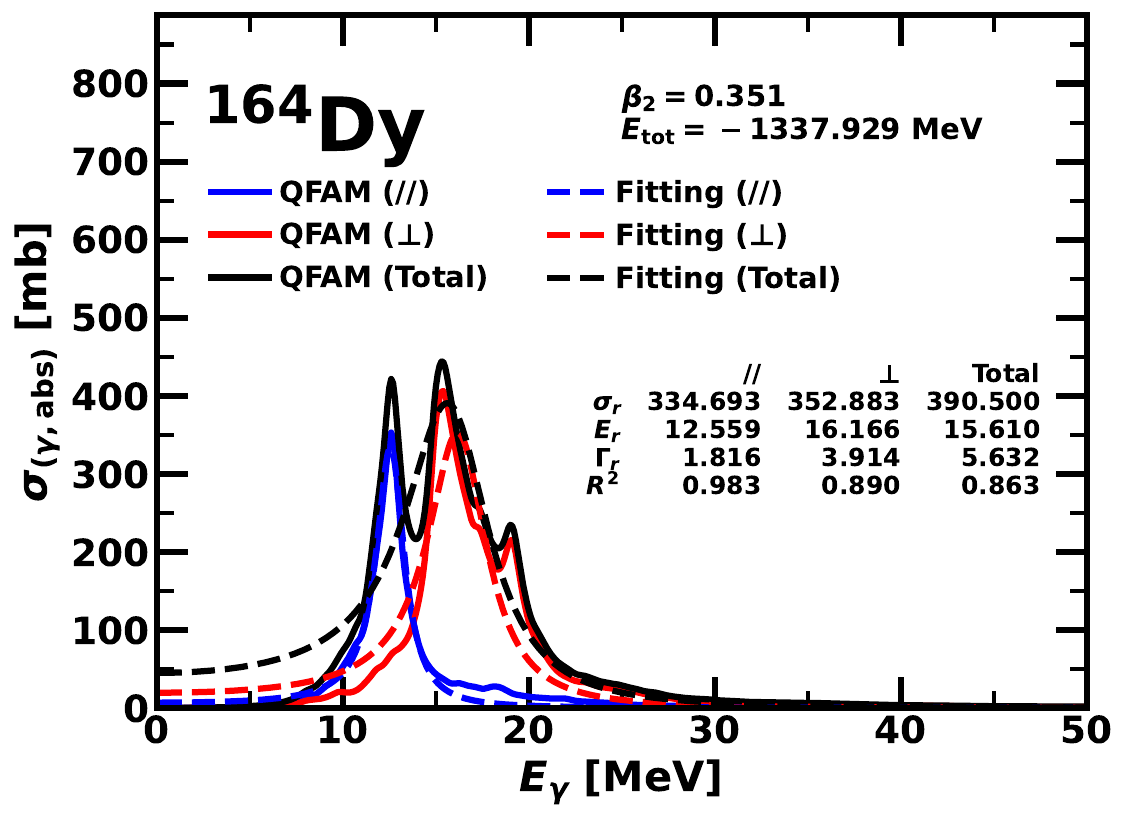}
\end{figure*}
\begin{figure*}\ContinuedFloat
    \centering
    \includegraphics[width=0.4\textwidth]{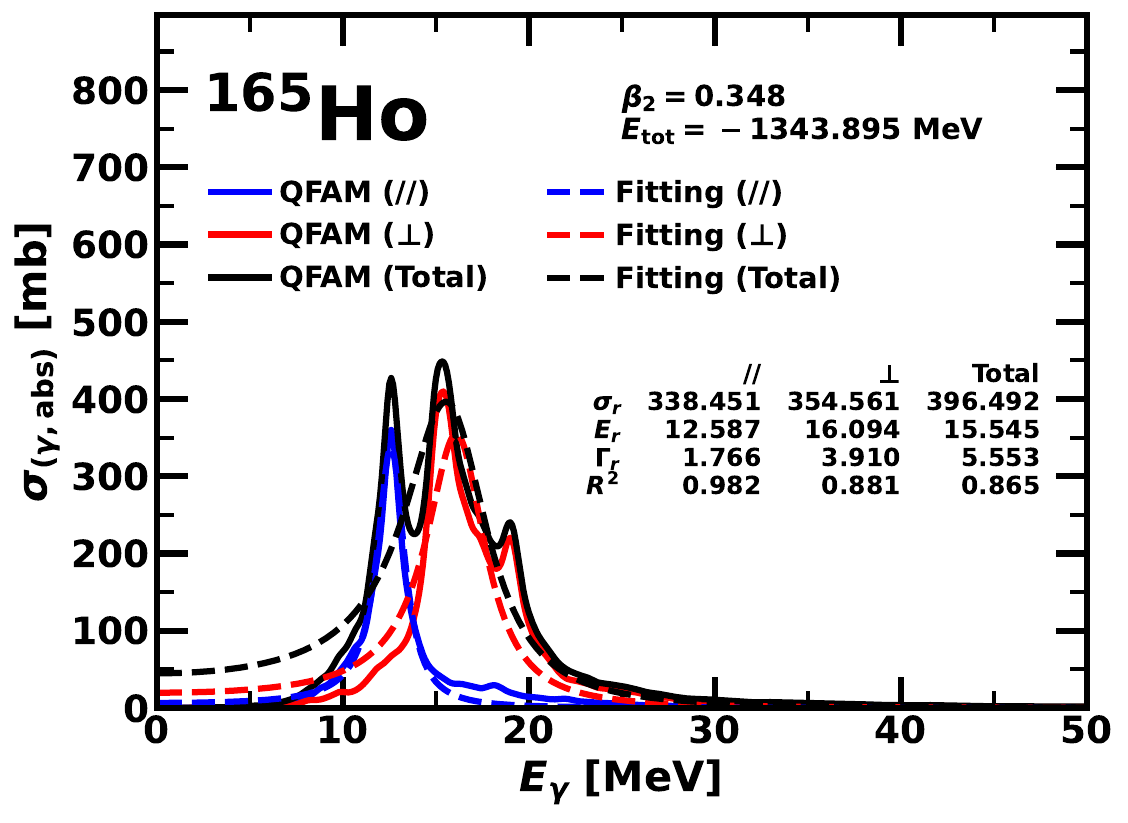}
    \includegraphics[width=0.4\textwidth]{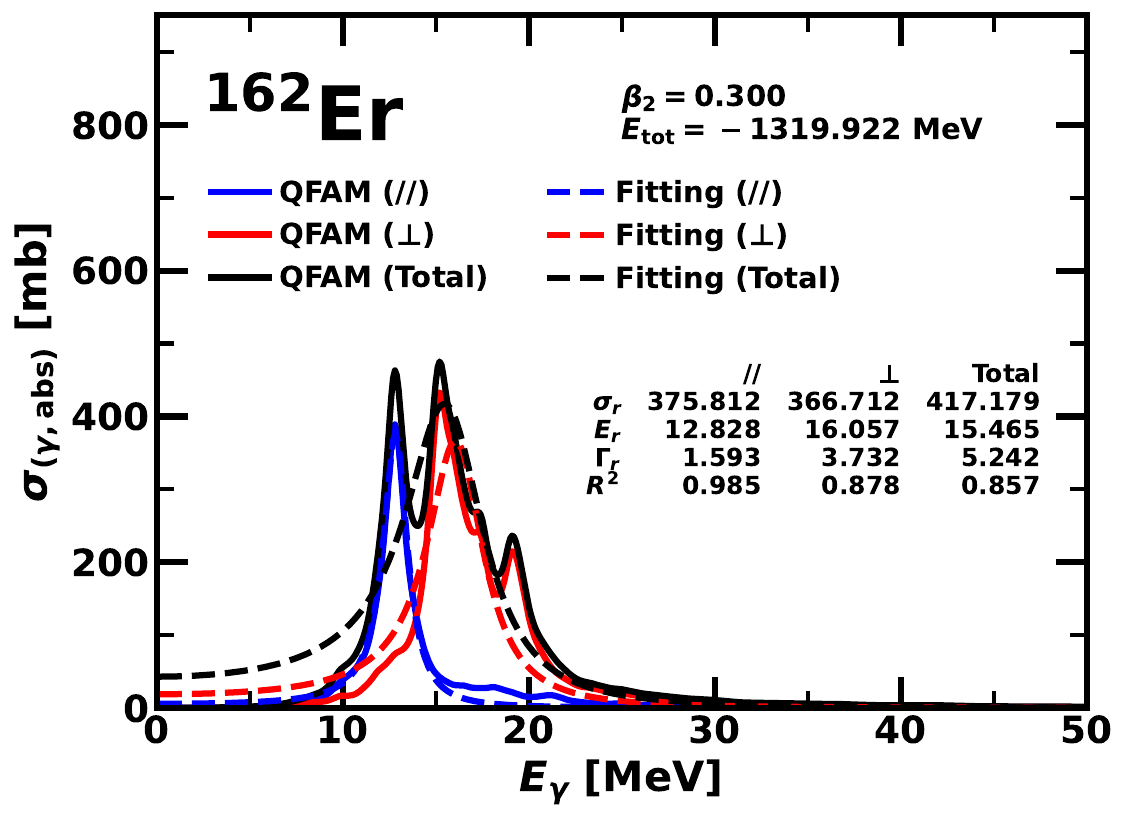}
    \includegraphics[width=0.4\textwidth]{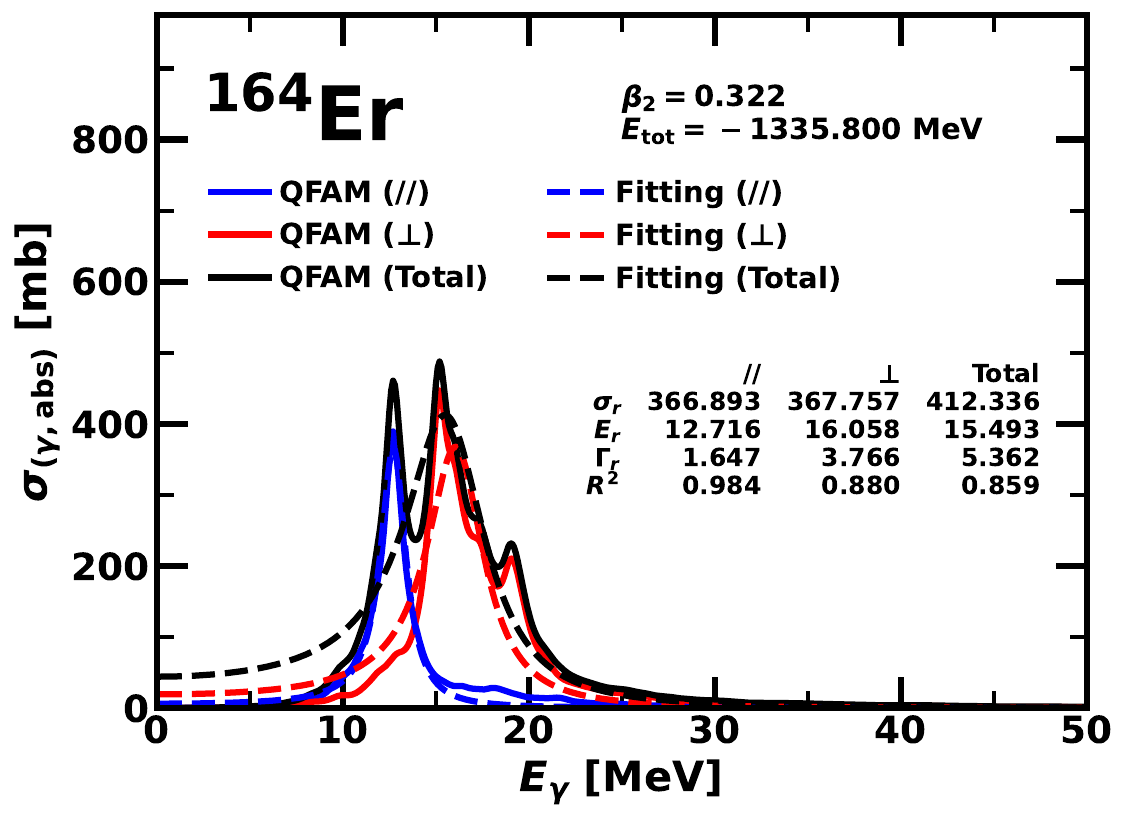}
    \includegraphics[width=0.4\textwidth]{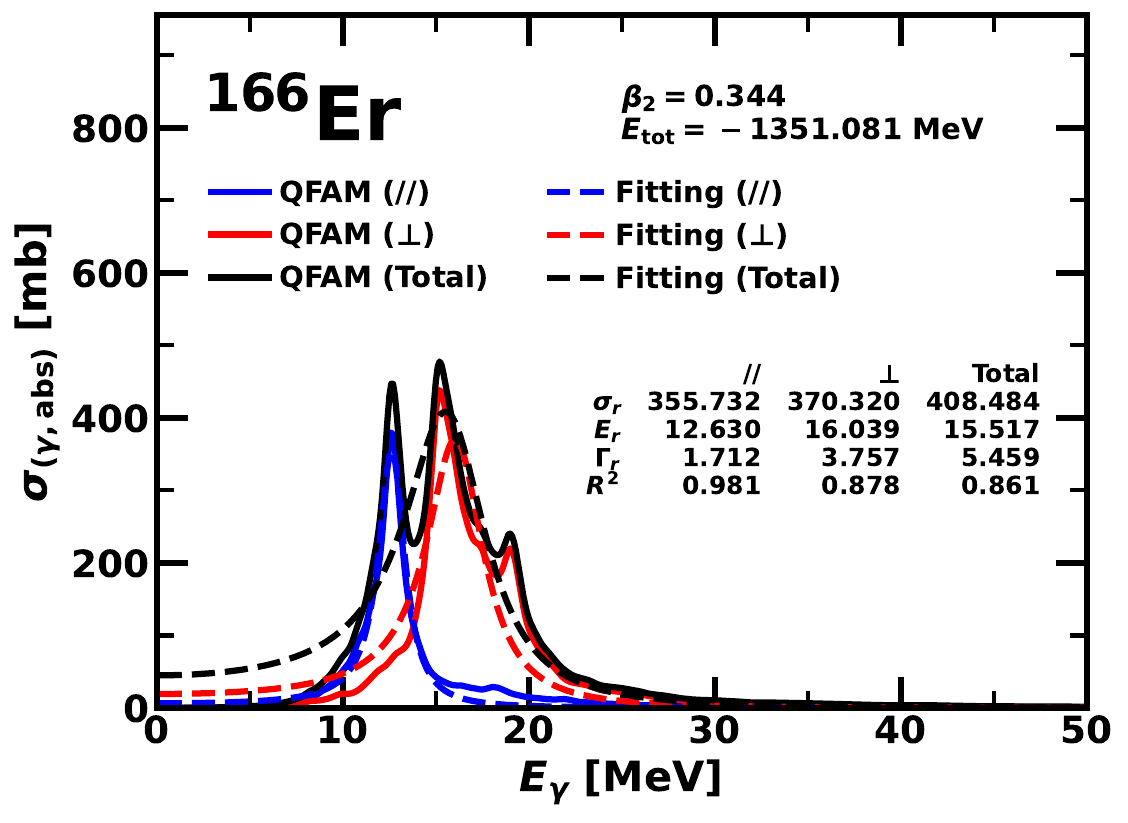}
    \includegraphics[width=0.4\textwidth]{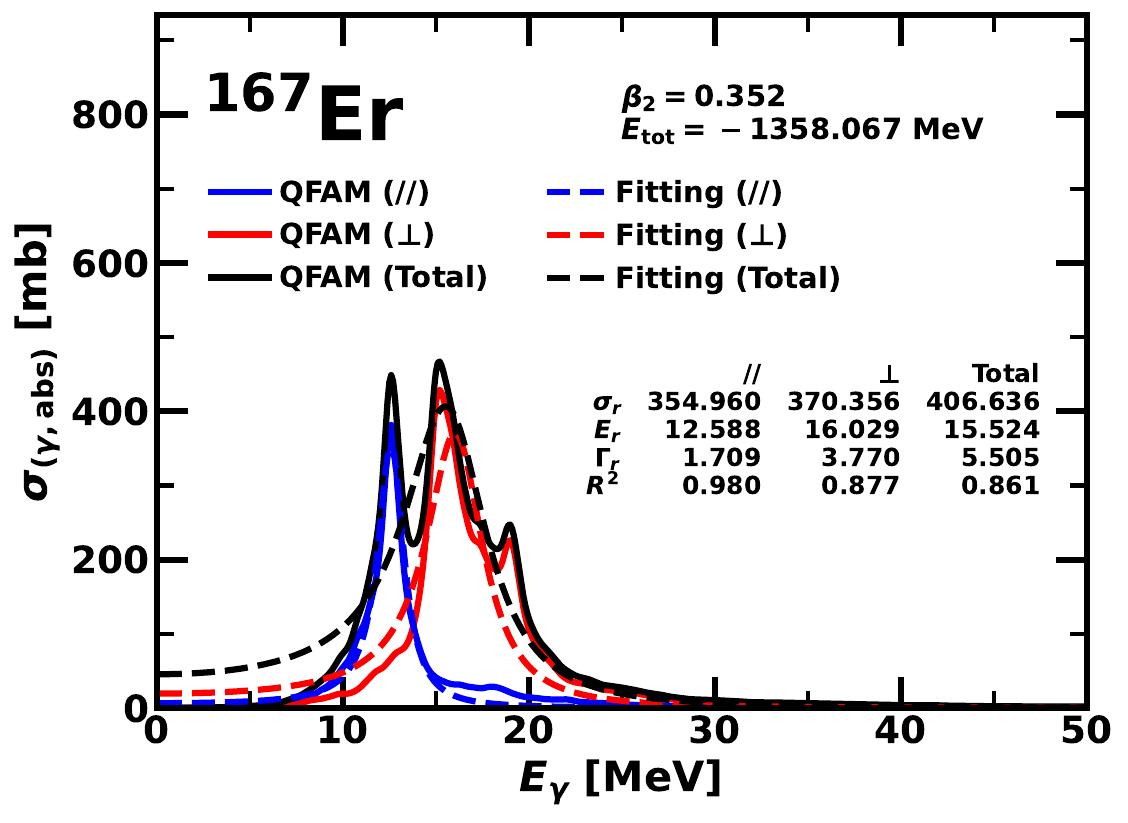}
    \includegraphics[width=0.4\textwidth]{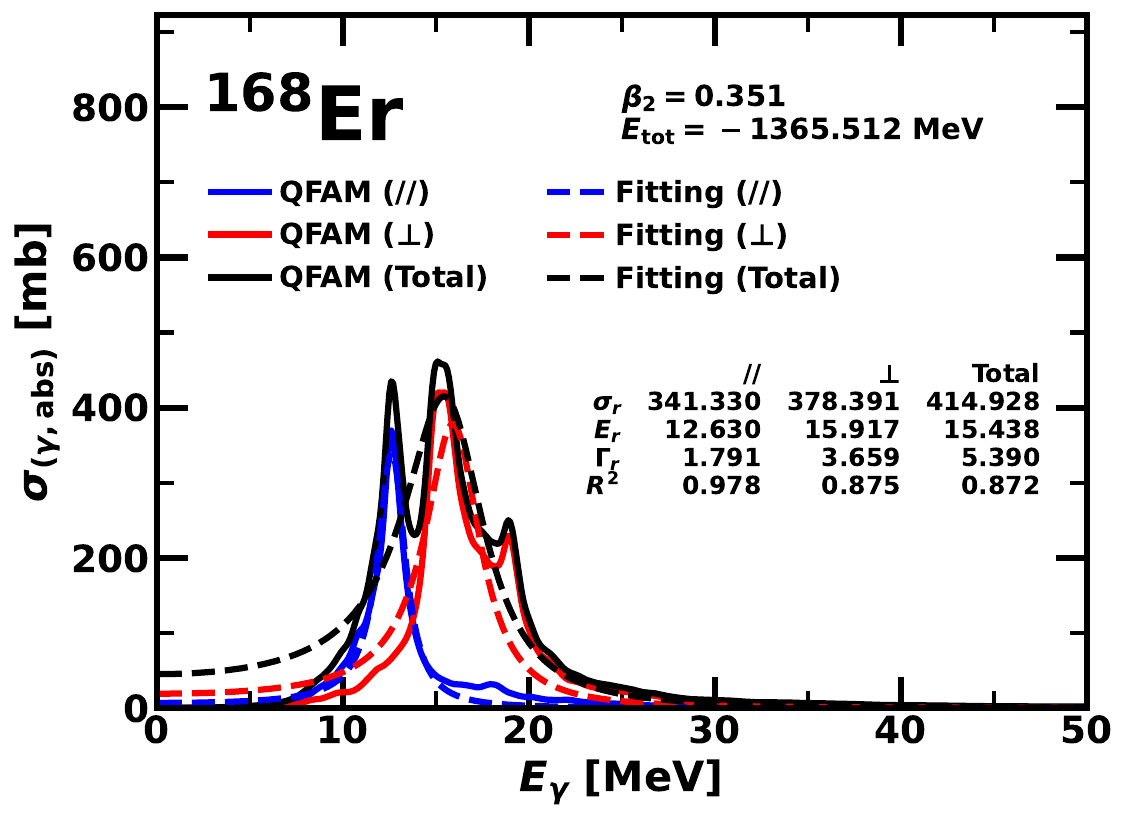}
    \includegraphics[width=0.4\textwidth]{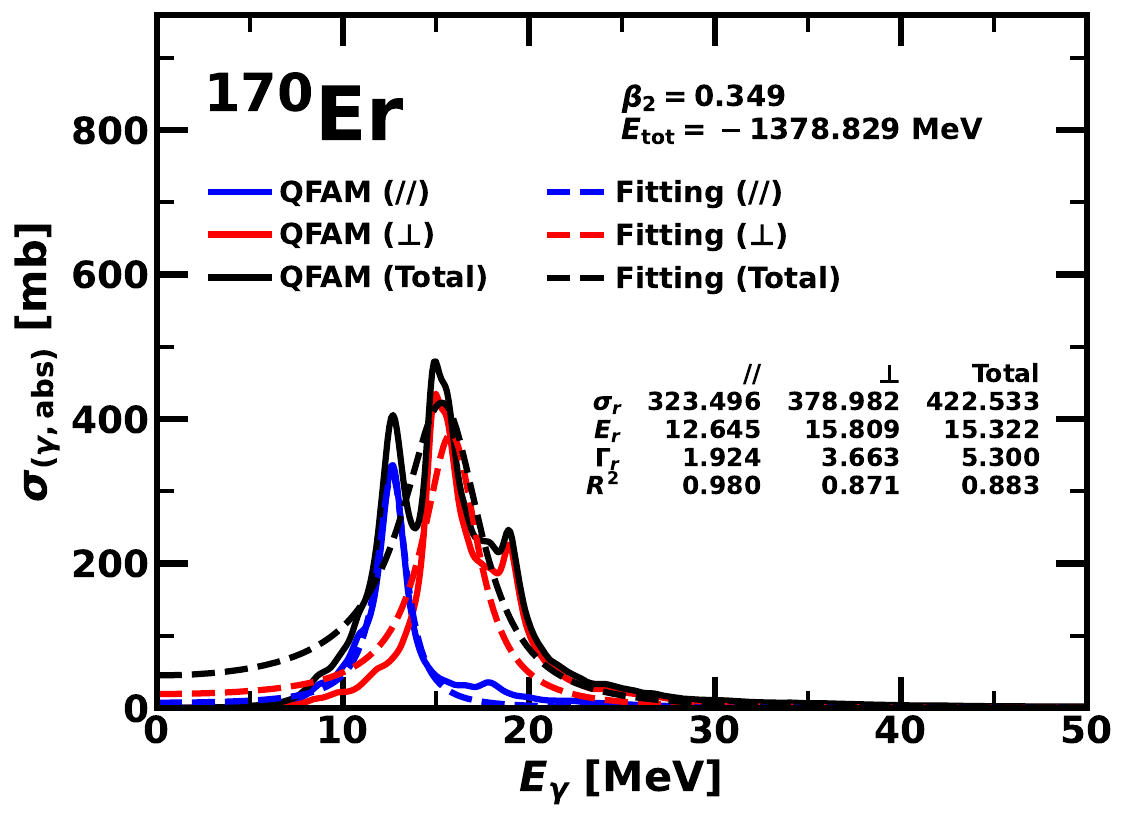}
    \includegraphics[width=0.4\textwidth]{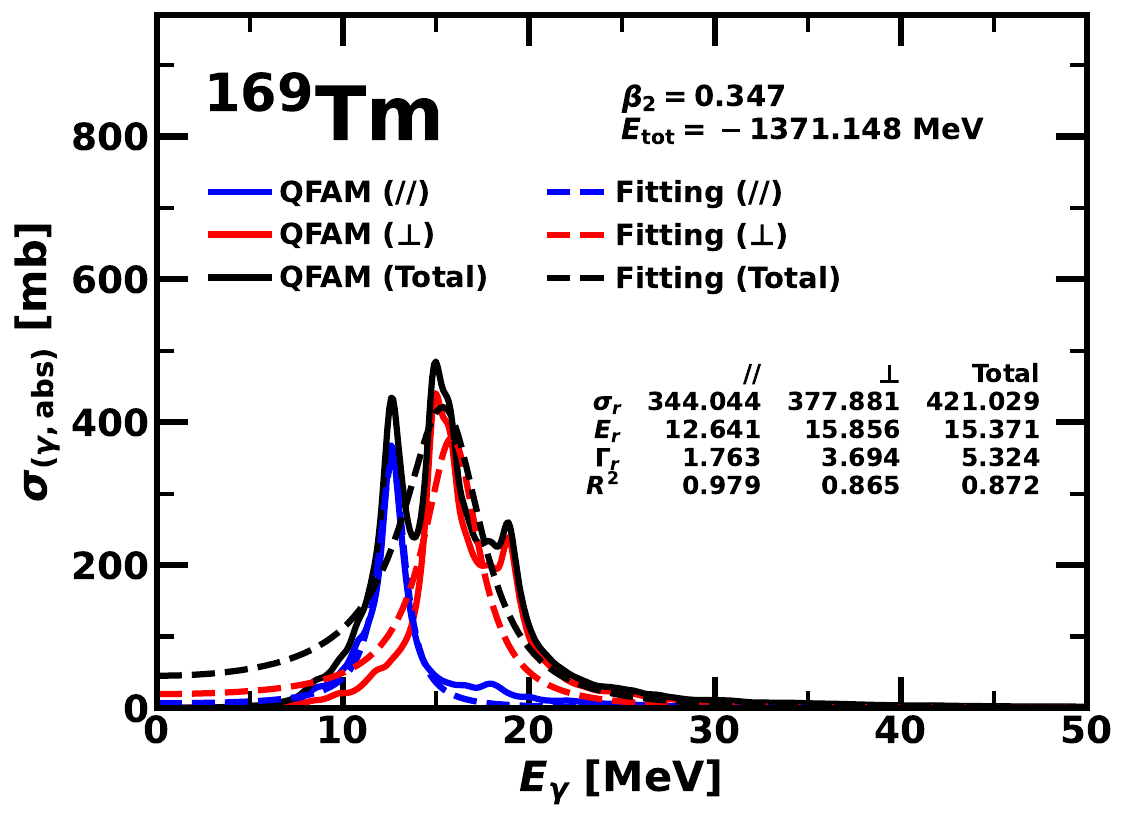}
\end{figure*}
\begin{figure*}\ContinuedFloat
    \centering
    \includegraphics[width=0.4\textwidth]{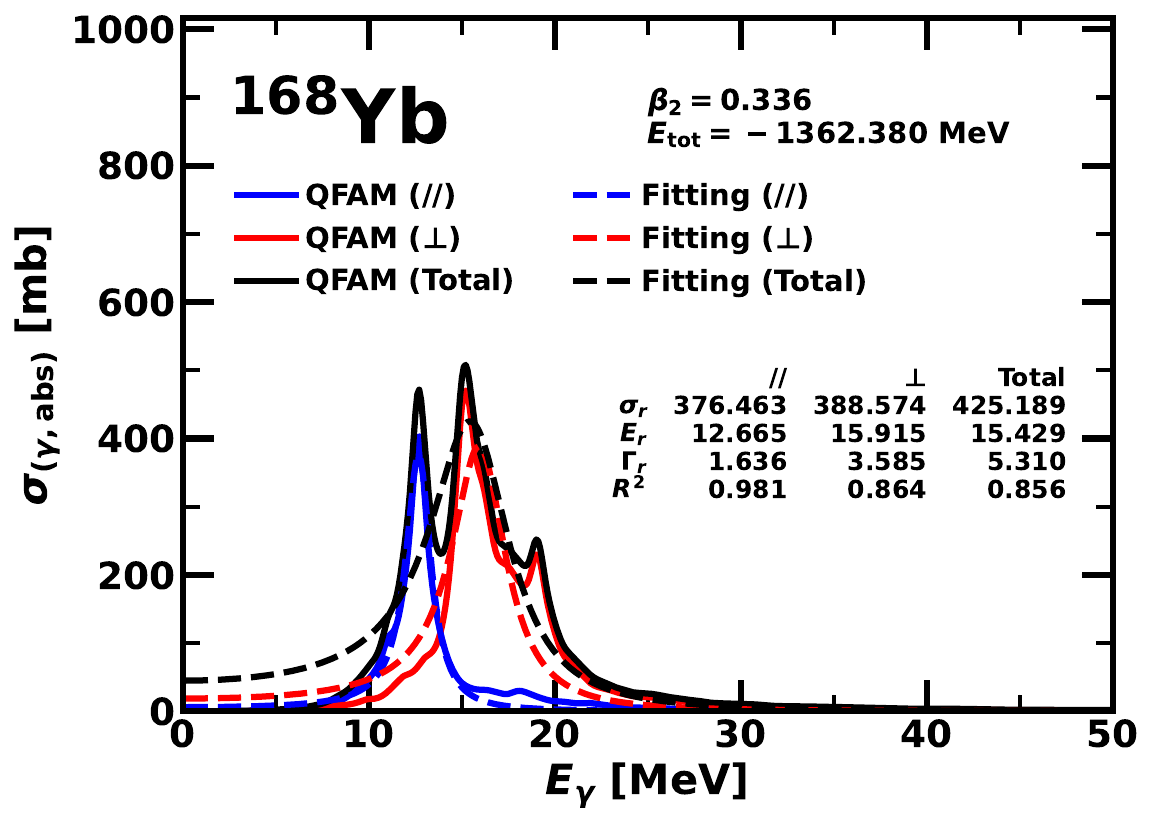}
    \includegraphics[width=0.4\textwidth]{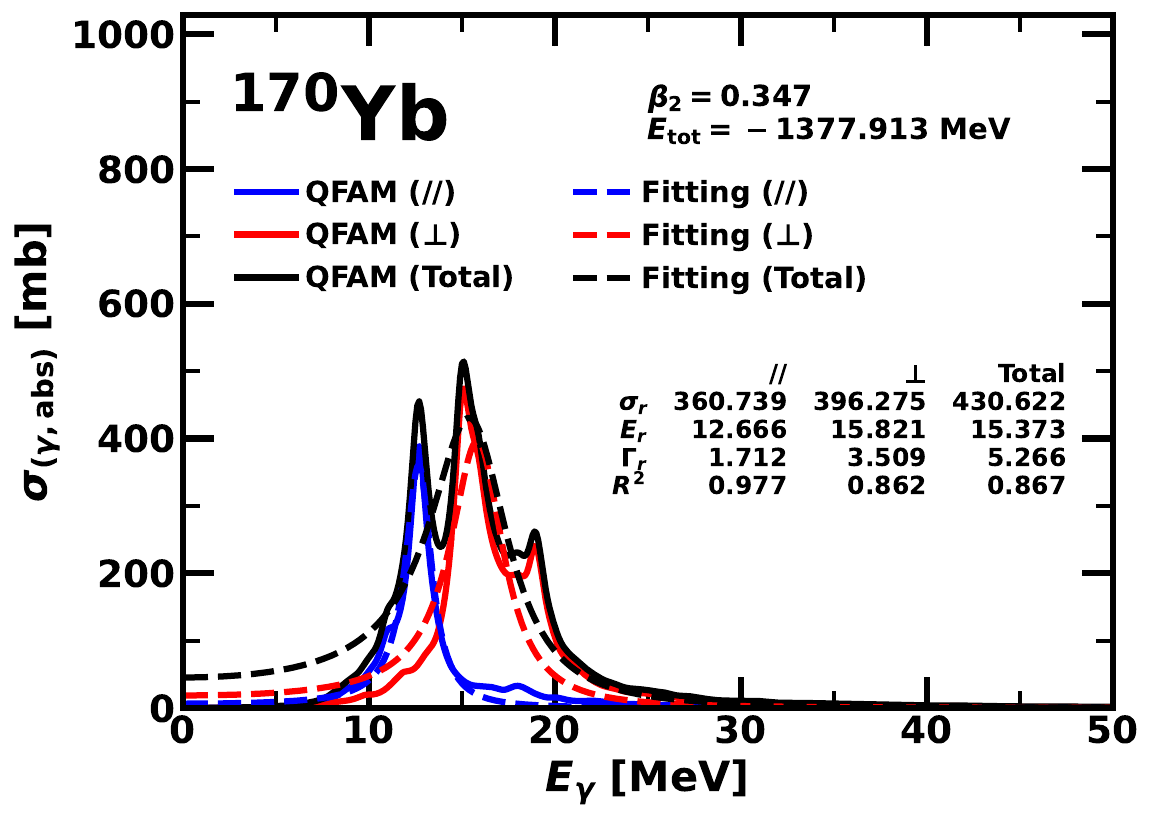}
    \includegraphics[width=0.4\textwidth]{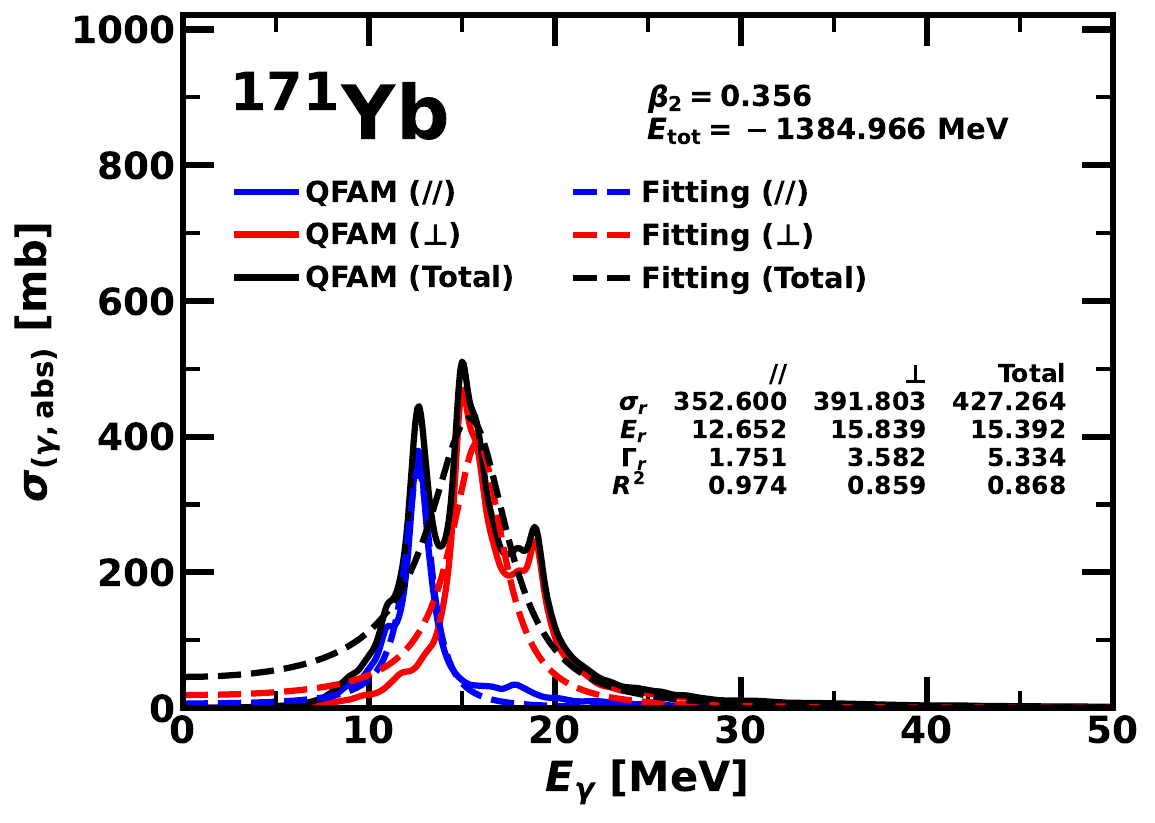}
    \includegraphics[width=0.4\textwidth]{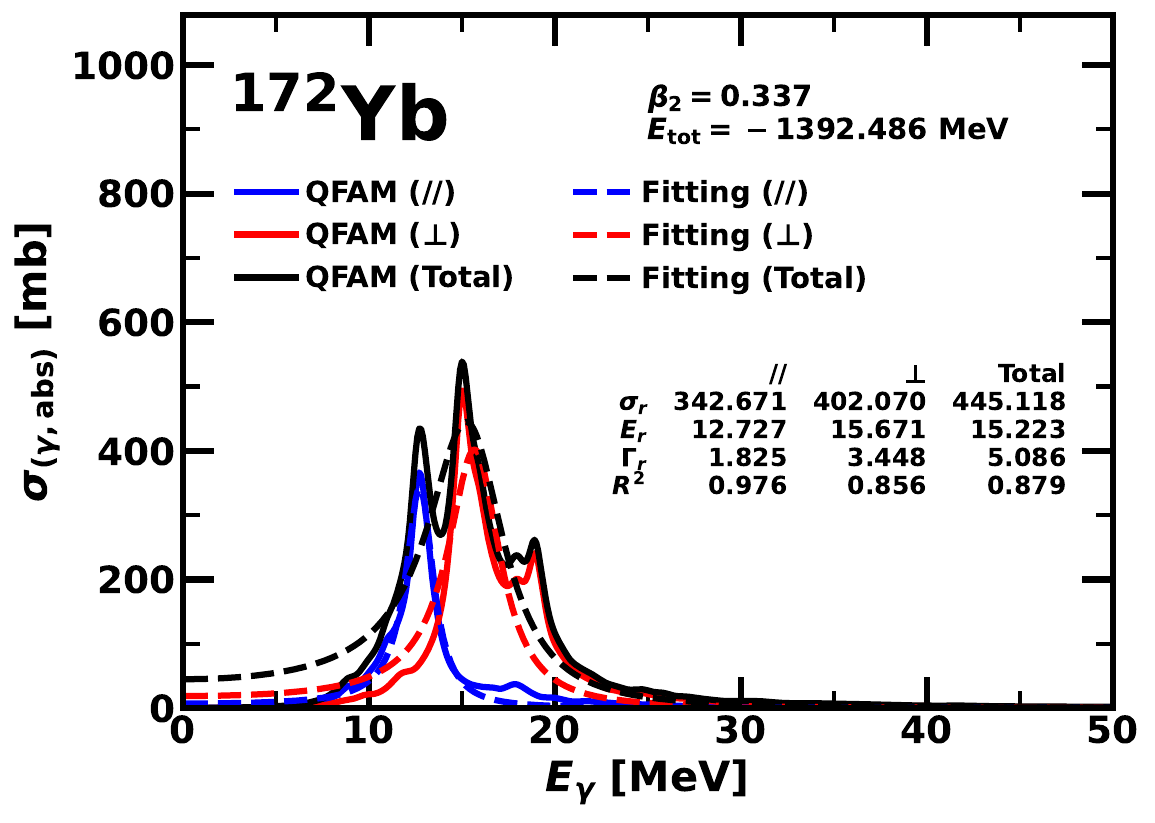}
    \includegraphics[width=0.4\textwidth]{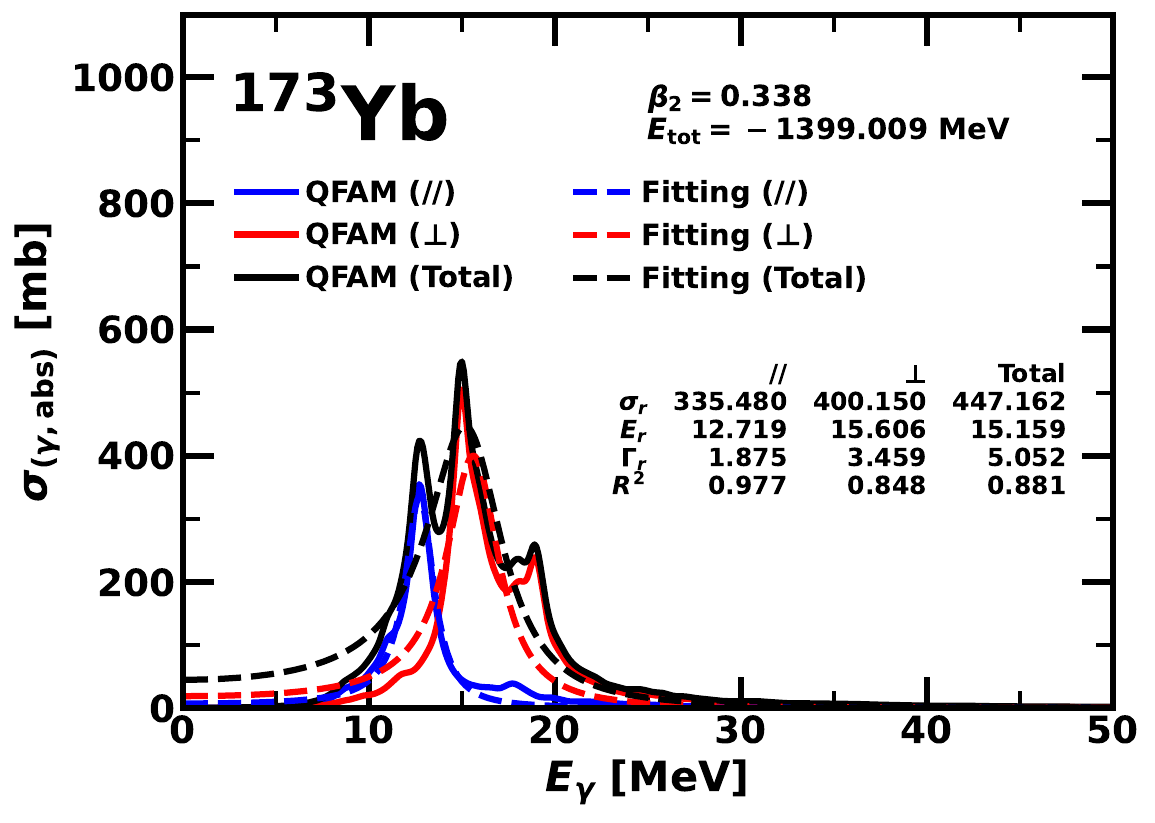}
    \includegraphics[width=0.4\textwidth]{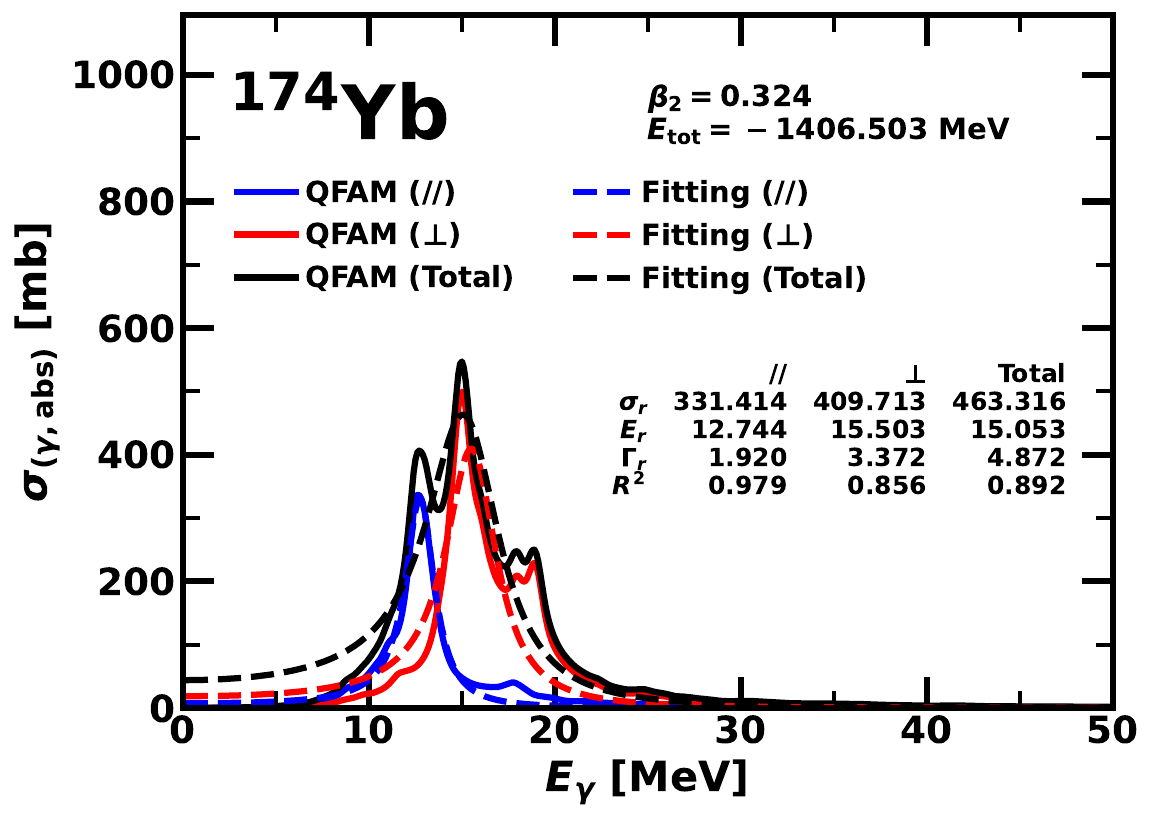}
    \includegraphics[width=0.4\textwidth]{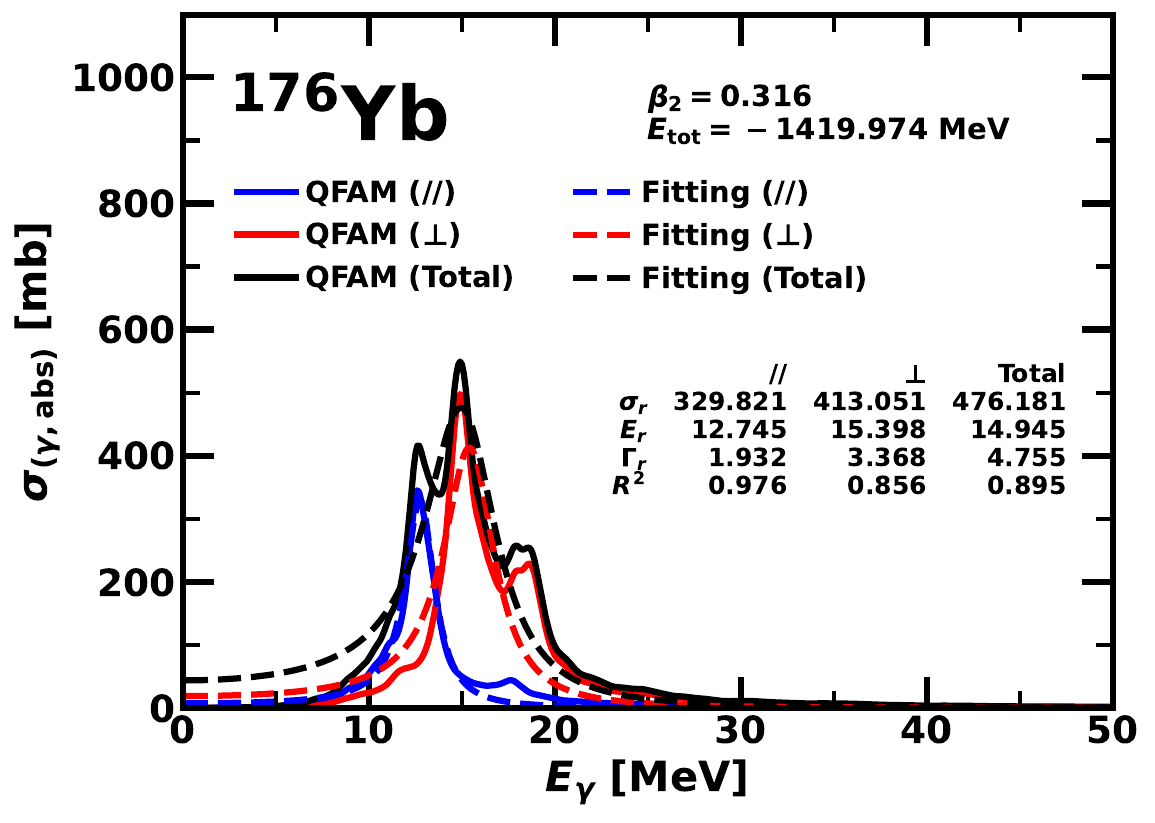}
    \includegraphics[width=0.4\textwidth]{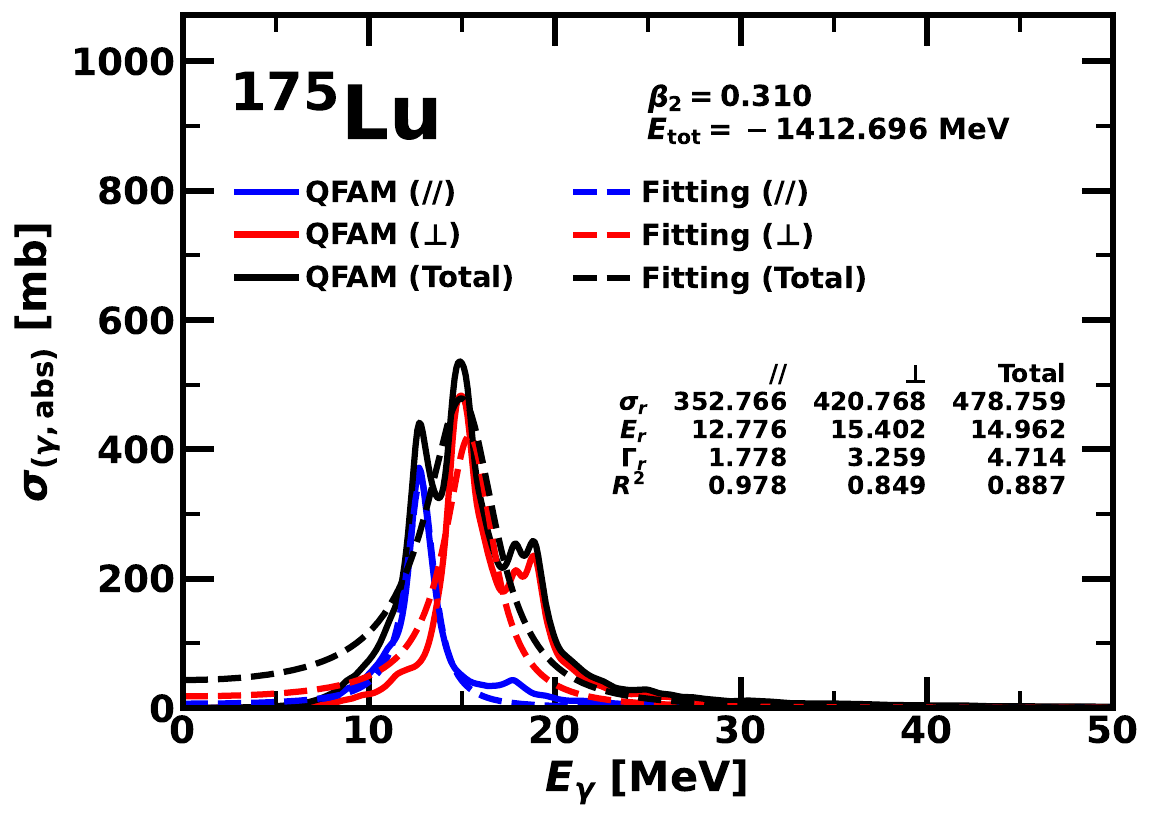}
\end{figure*}
\begin{figure*}\ContinuedFloat
    \centering
    \includegraphics[width=0.4\textwidth]{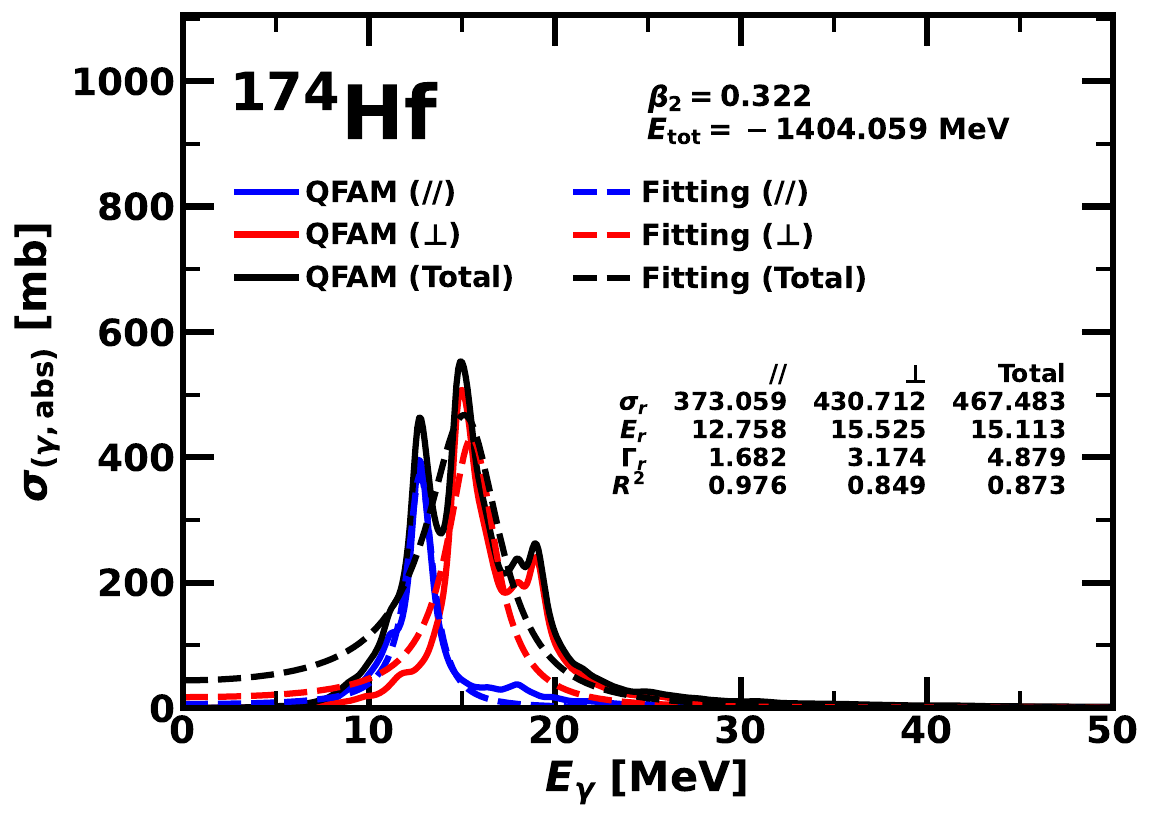}
    \includegraphics[width=0.4\textwidth]{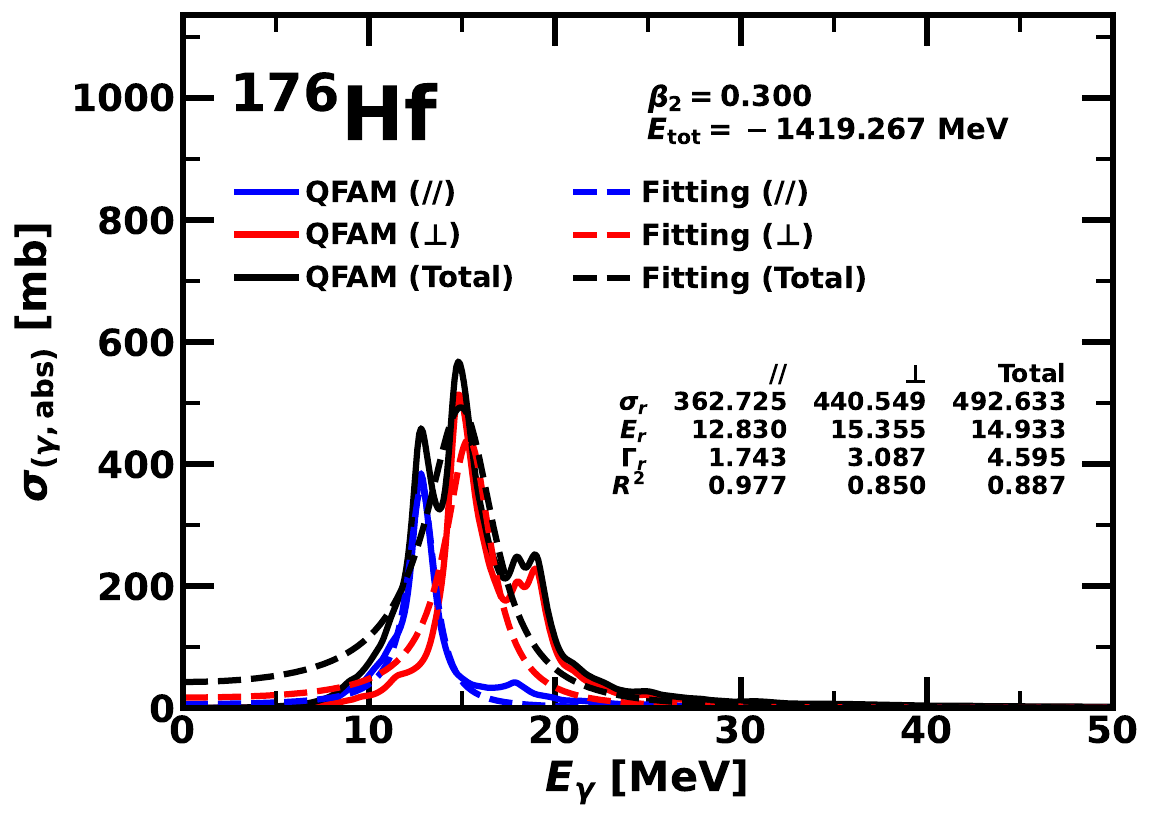}
    \includegraphics[width=0.4\textwidth]{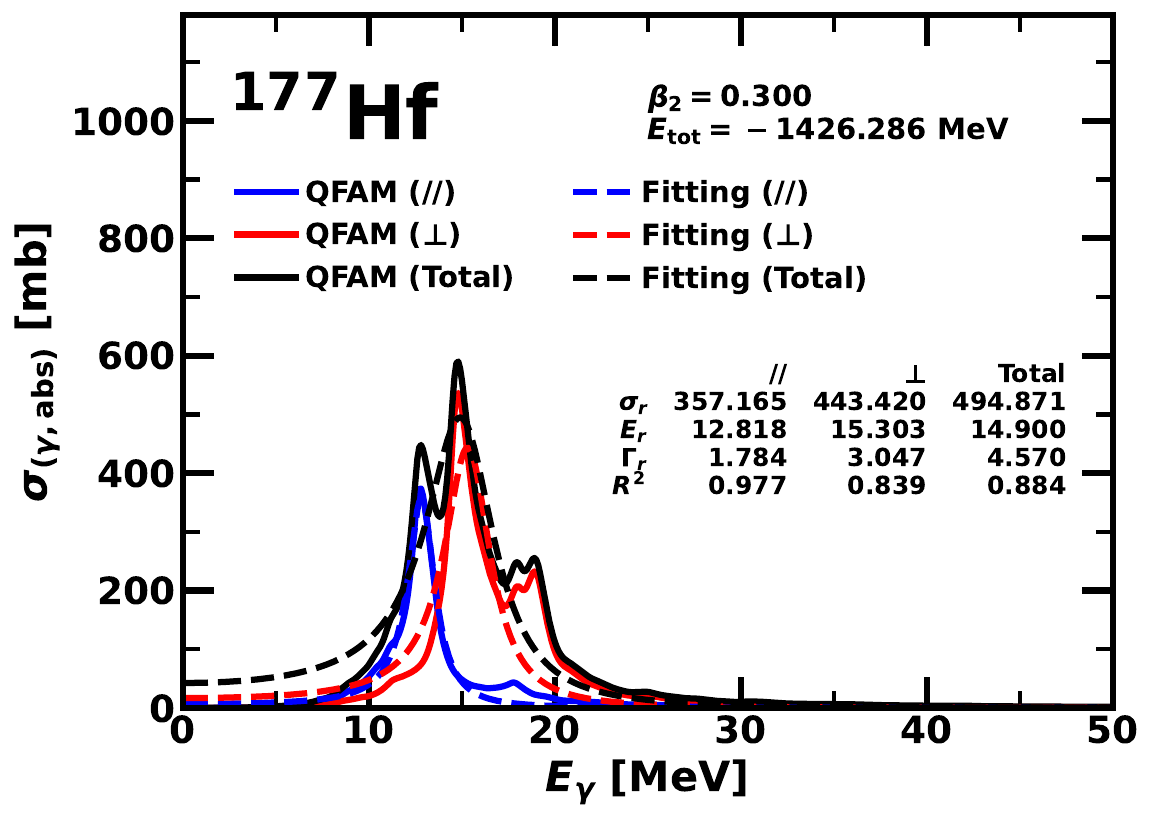}
    \includegraphics[width=0.4\textwidth]{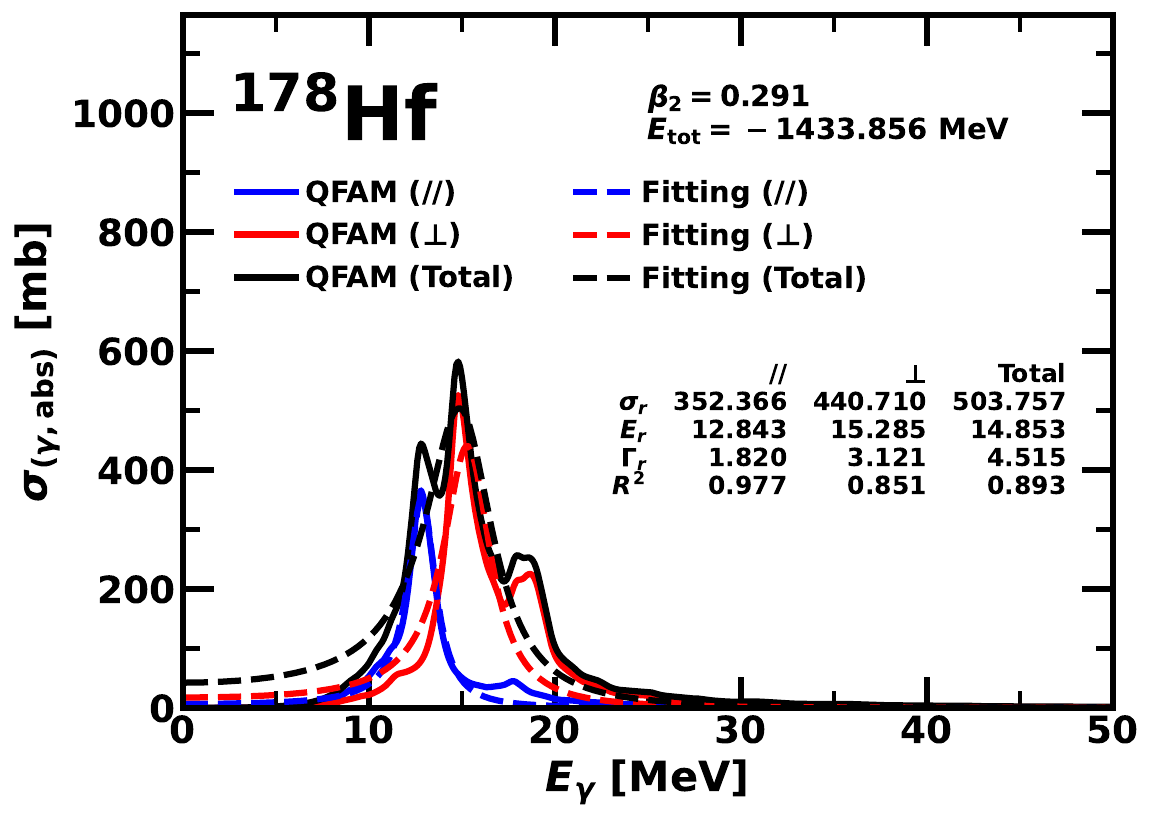}
    \includegraphics[width=0.4\textwidth]{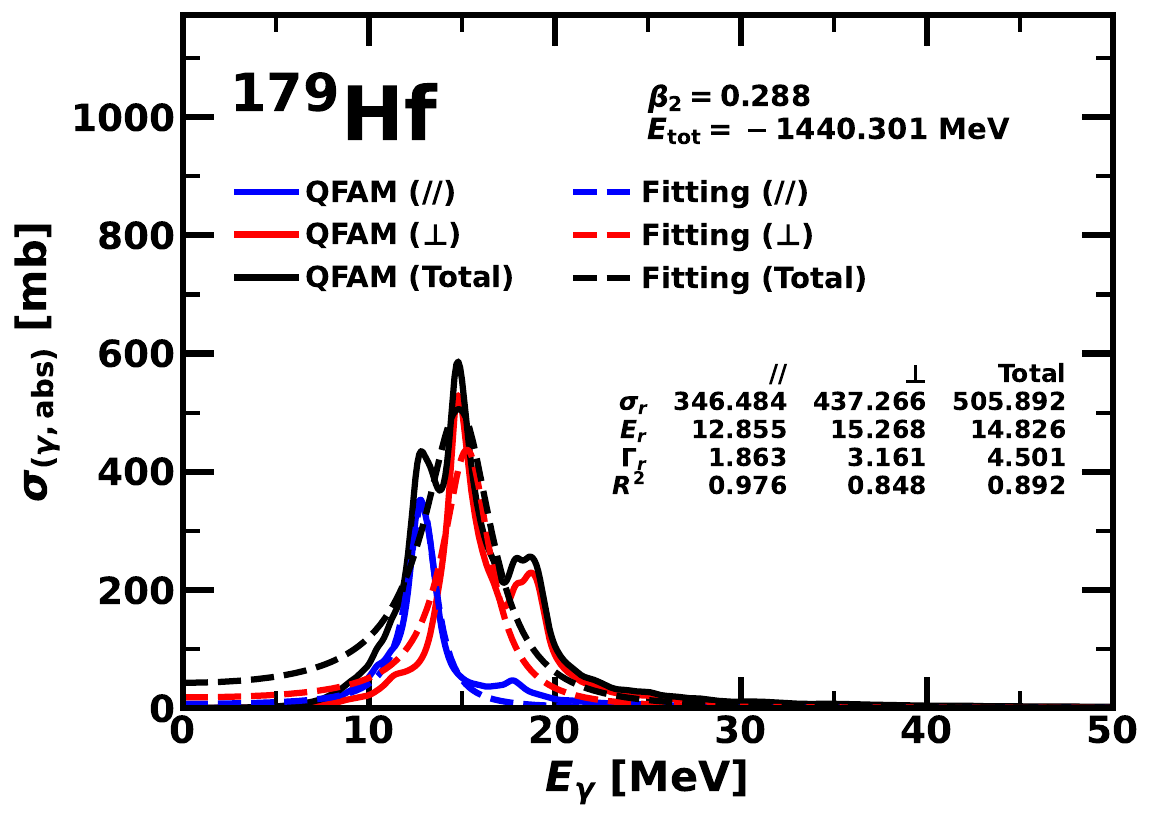}
    \includegraphics[width=0.4\textwidth]{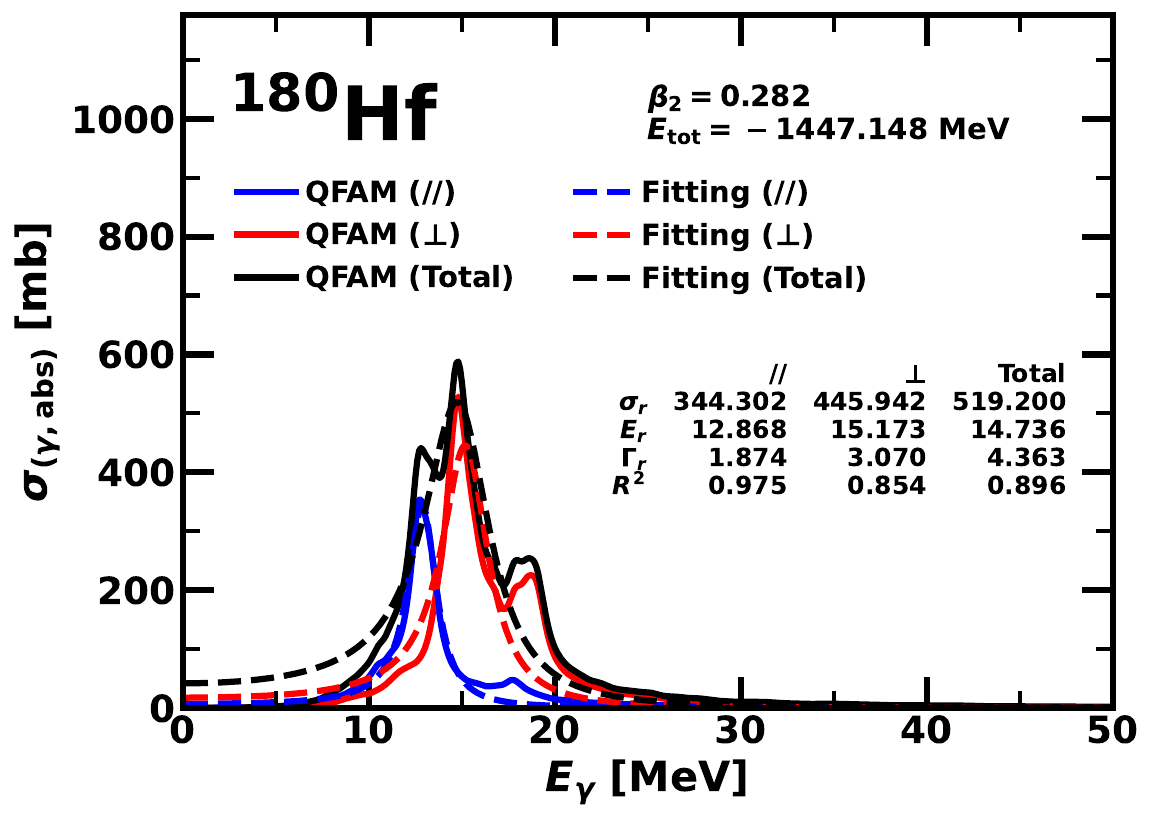}
    \includegraphics[width=0.4\textwidth]{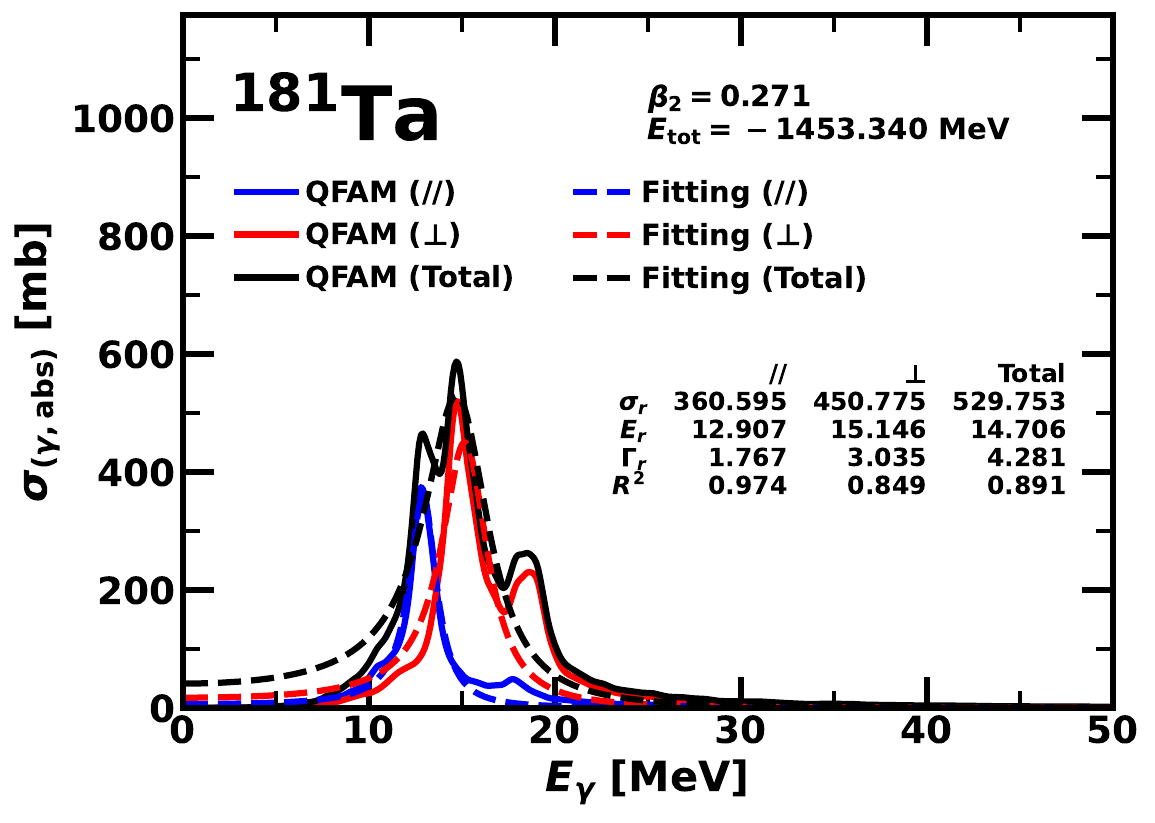}
    \includegraphics[width=0.4\textwidth]{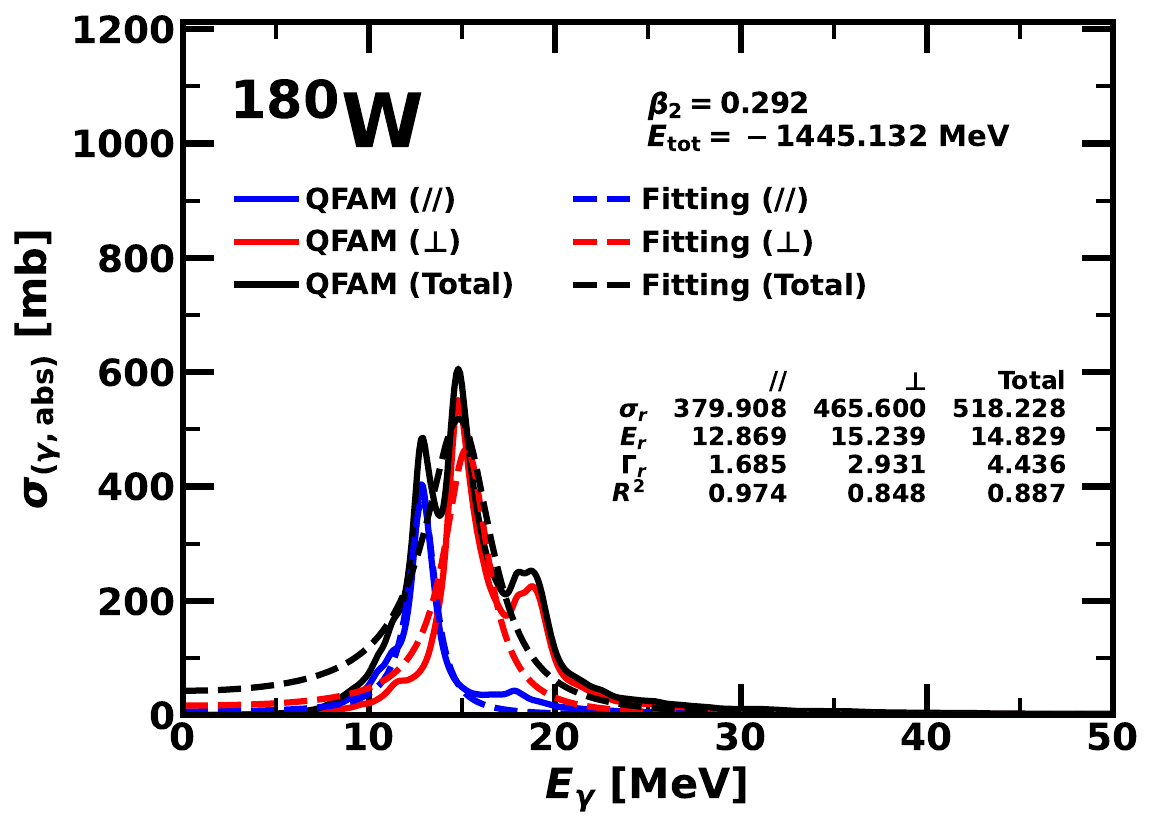}
\end{figure*}
\begin{figure*}\ContinuedFloat
    \centering
    \includegraphics[width=0.4\textwidth]{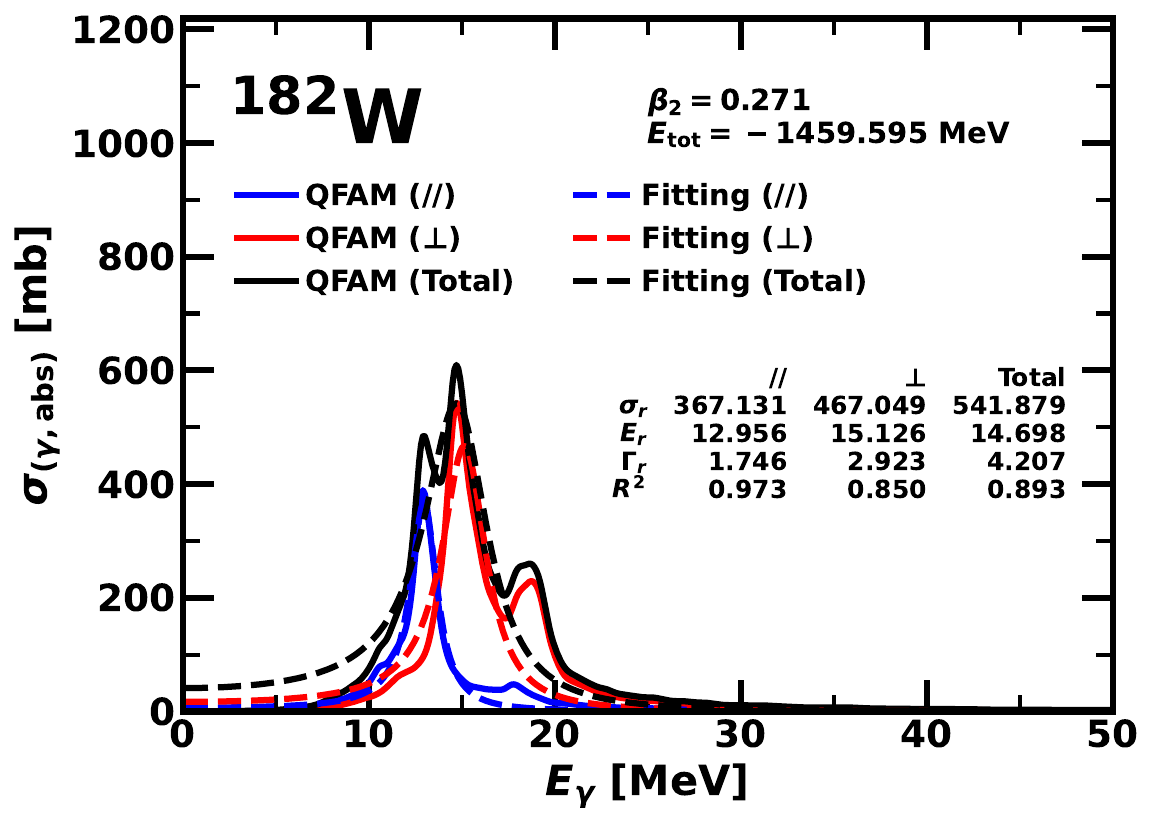}
    \includegraphics[width=0.4\textwidth]{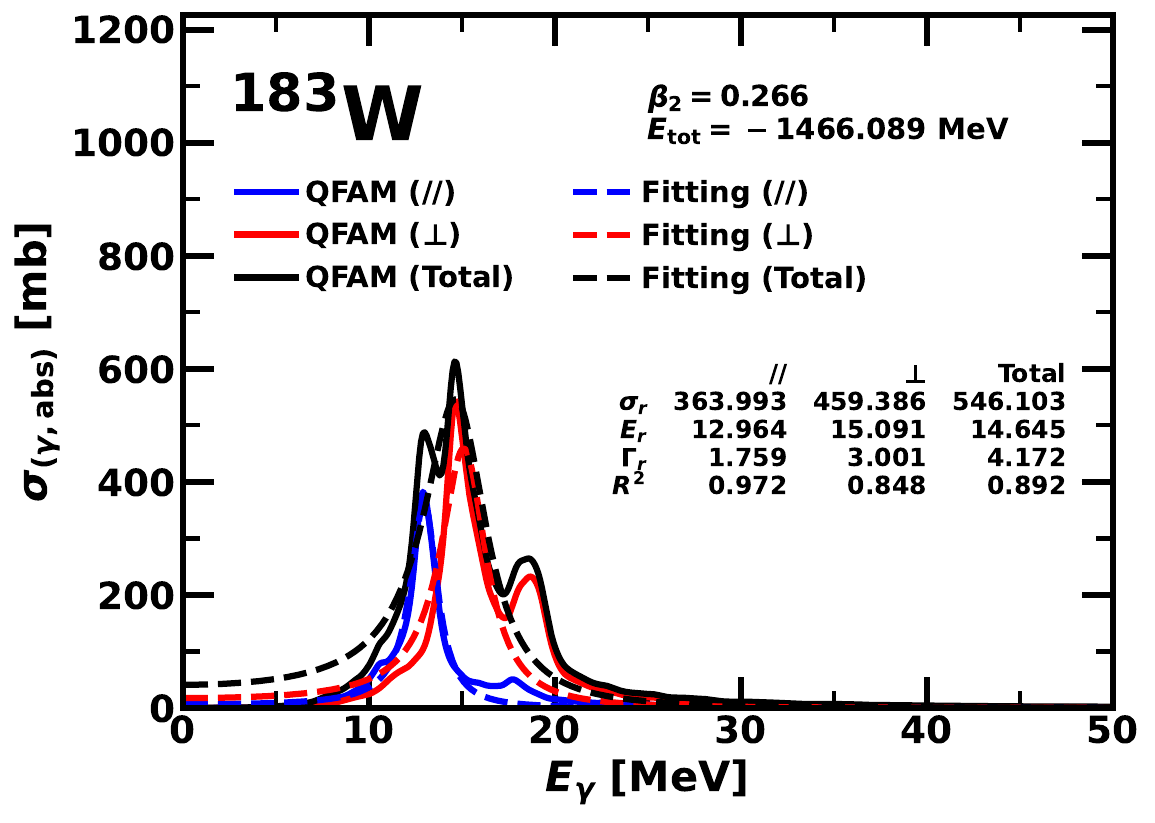}
    \includegraphics[width=0.4\textwidth]{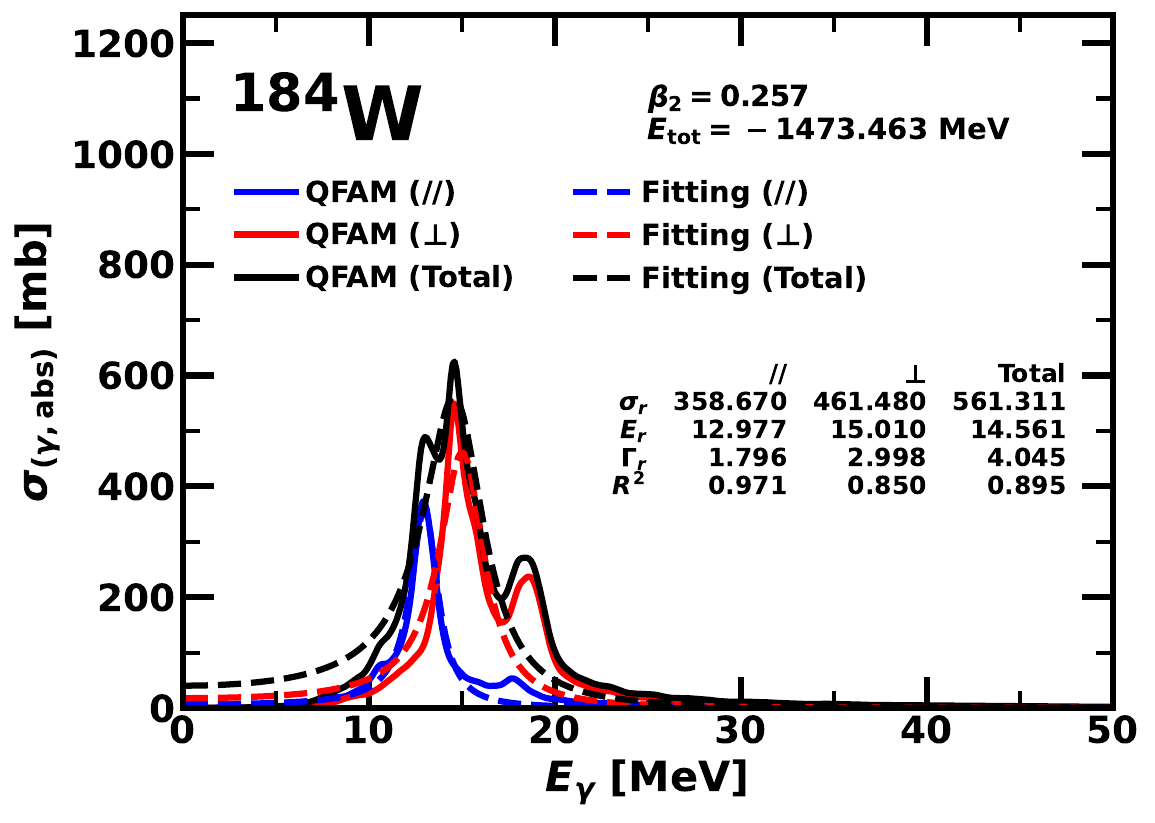}
    \includegraphics[width=0.4\textwidth]{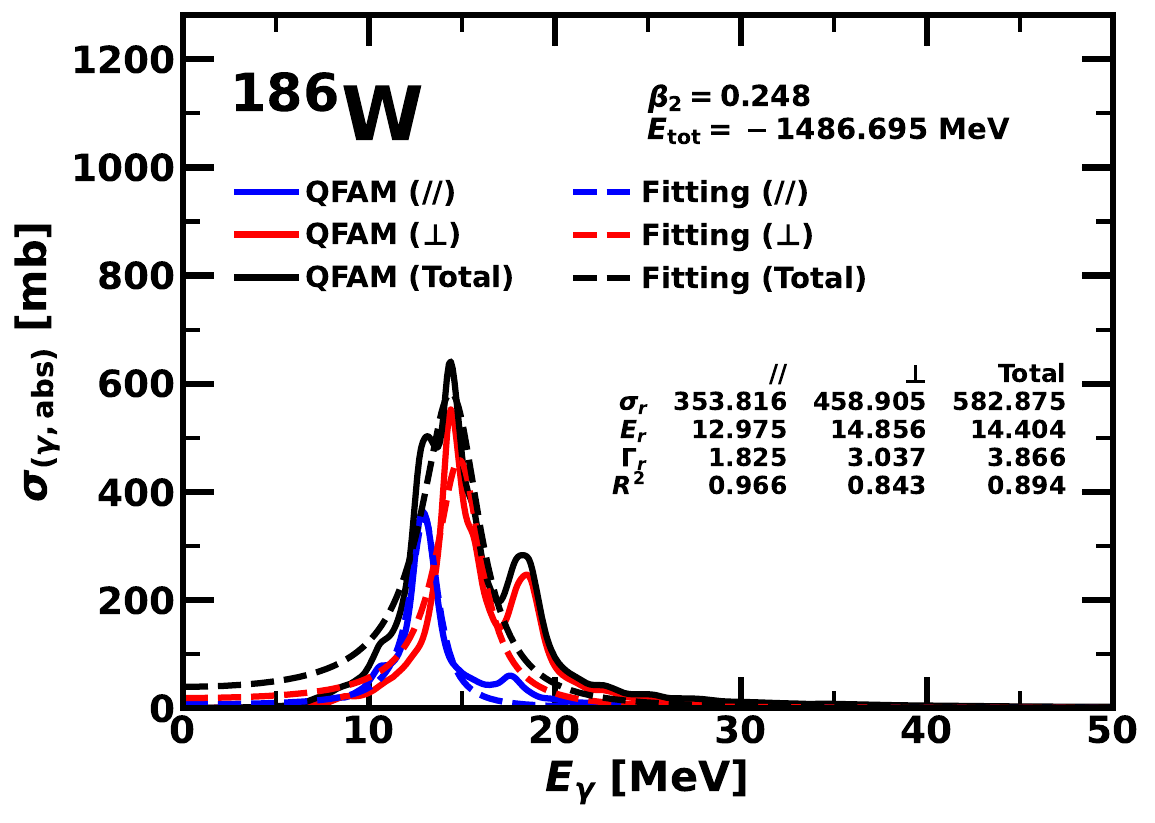}
    \includegraphics[width=0.4\textwidth]{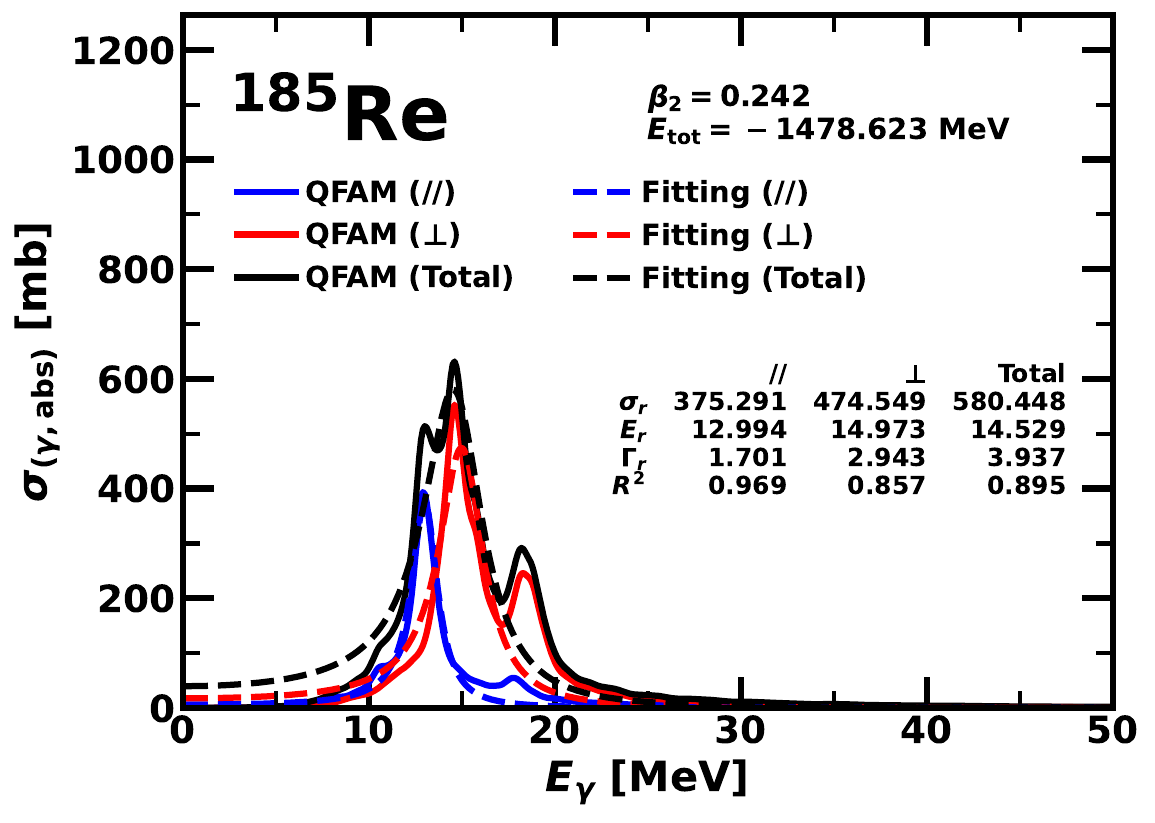}
    \includegraphics[width=0.4\textwidth]{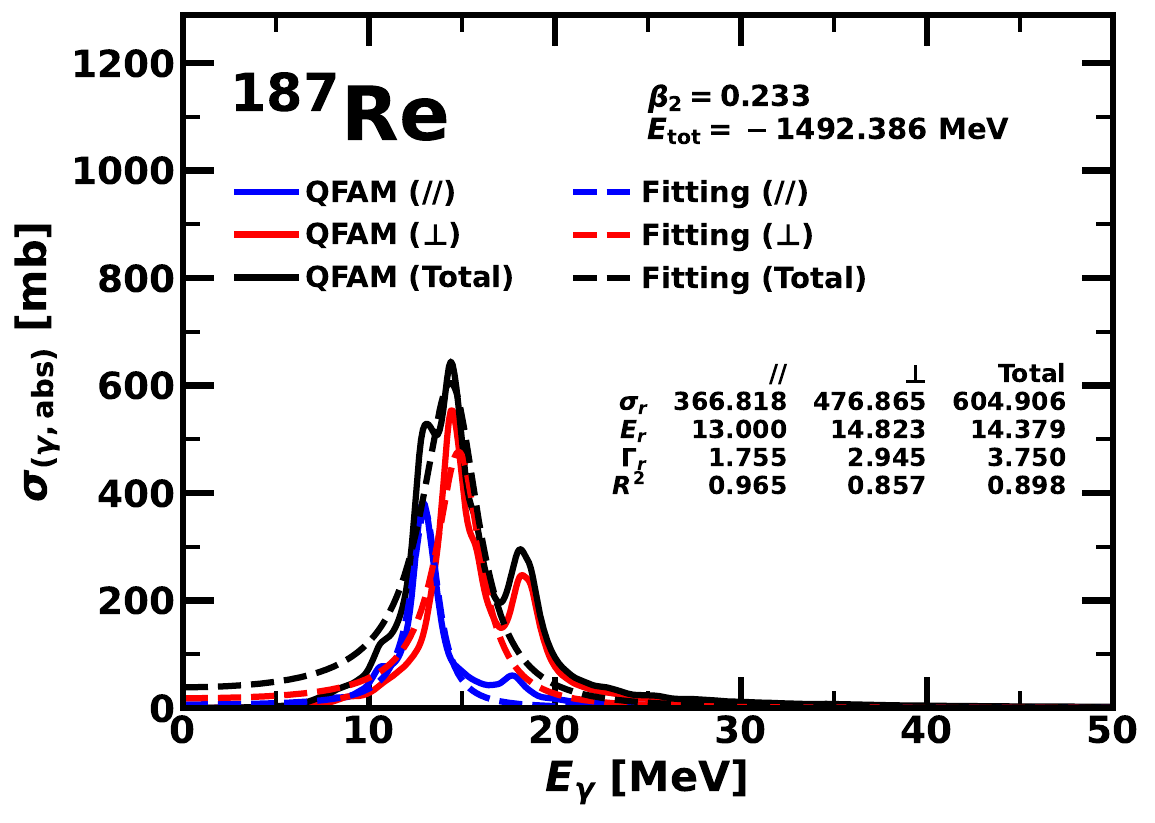}
    \includegraphics[width=0.4\textwidth]{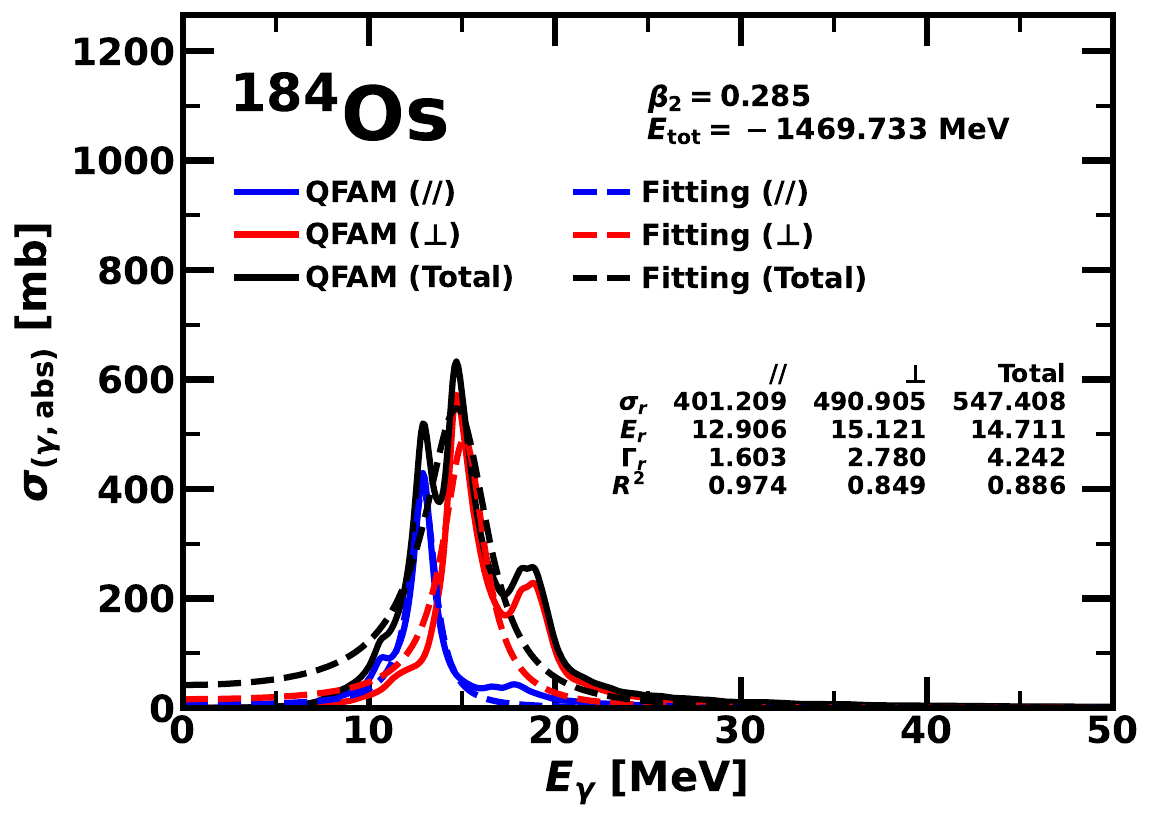}
    \includegraphics[width=0.4\textwidth]{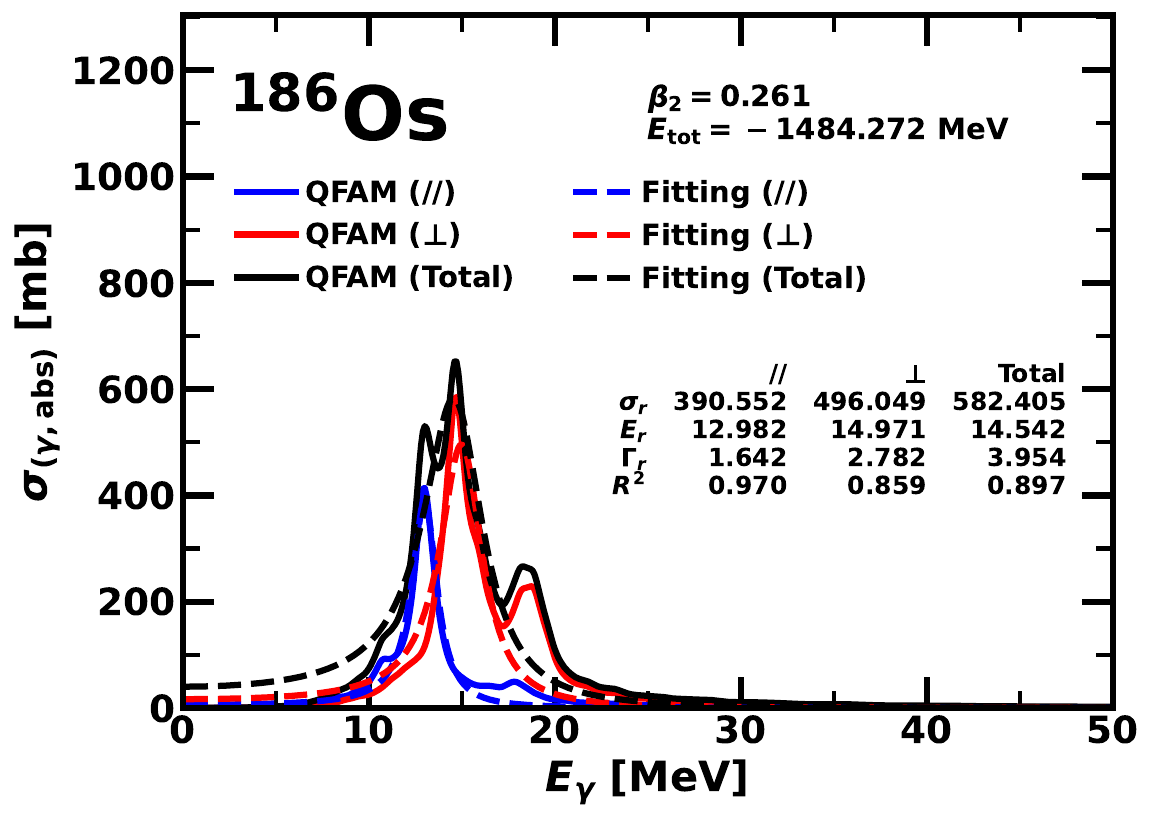}
\end{figure*}
\begin{figure*}\ContinuedFloat
    \centering
    \includegraphics[width=0.4\textwidth]{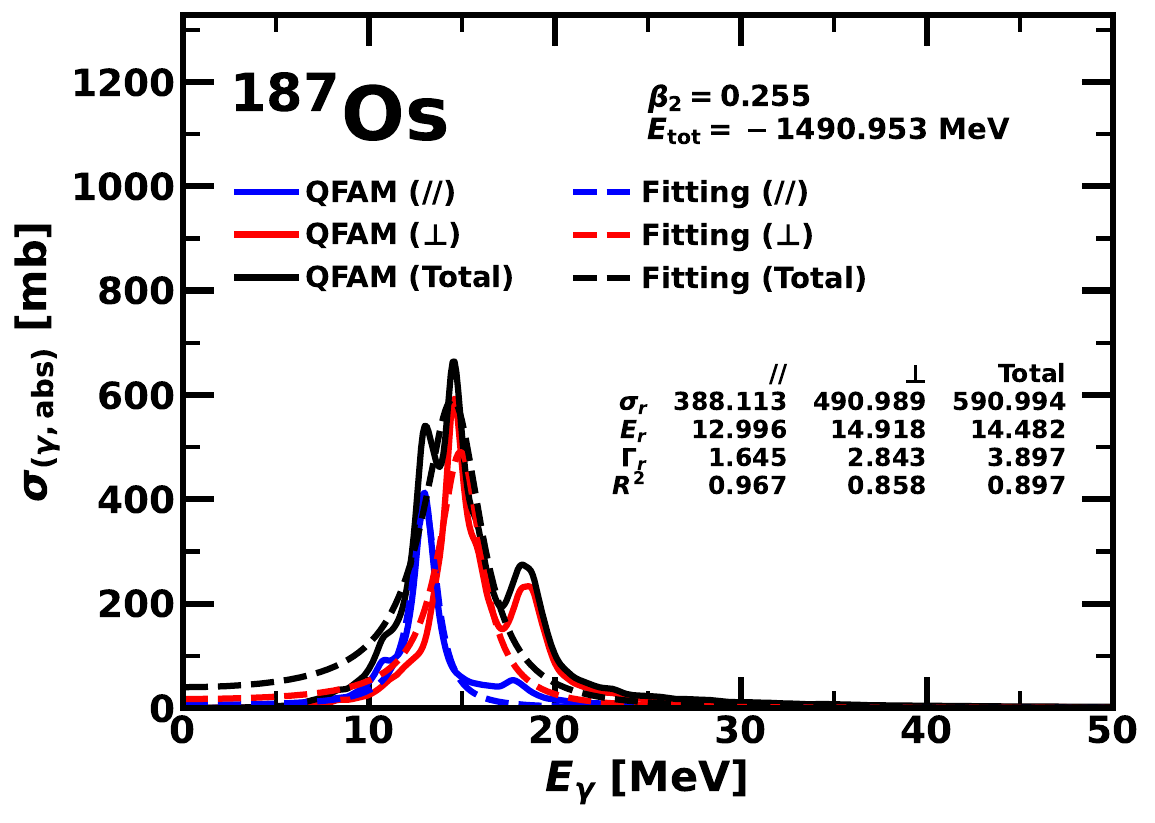}
    \includegraphics[width=0.4\textwidth]{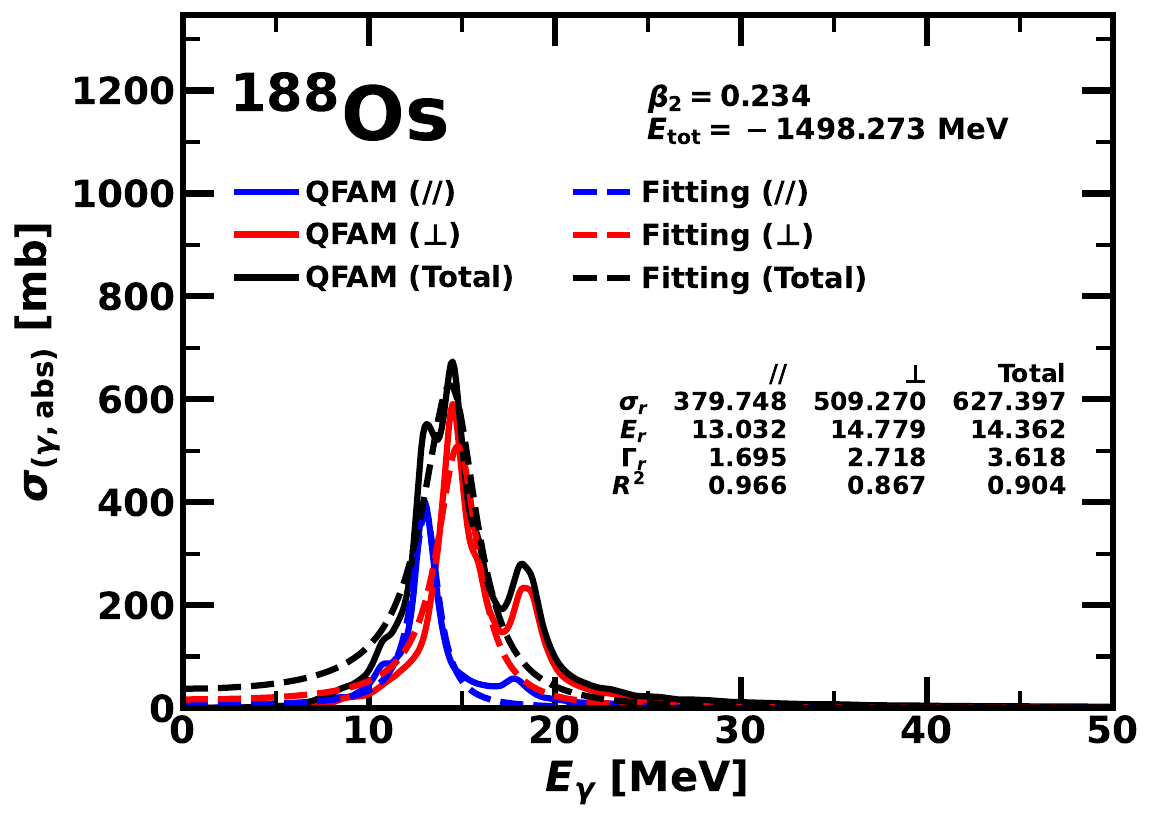}
    \includegraphics[width=0.4\textwidth]{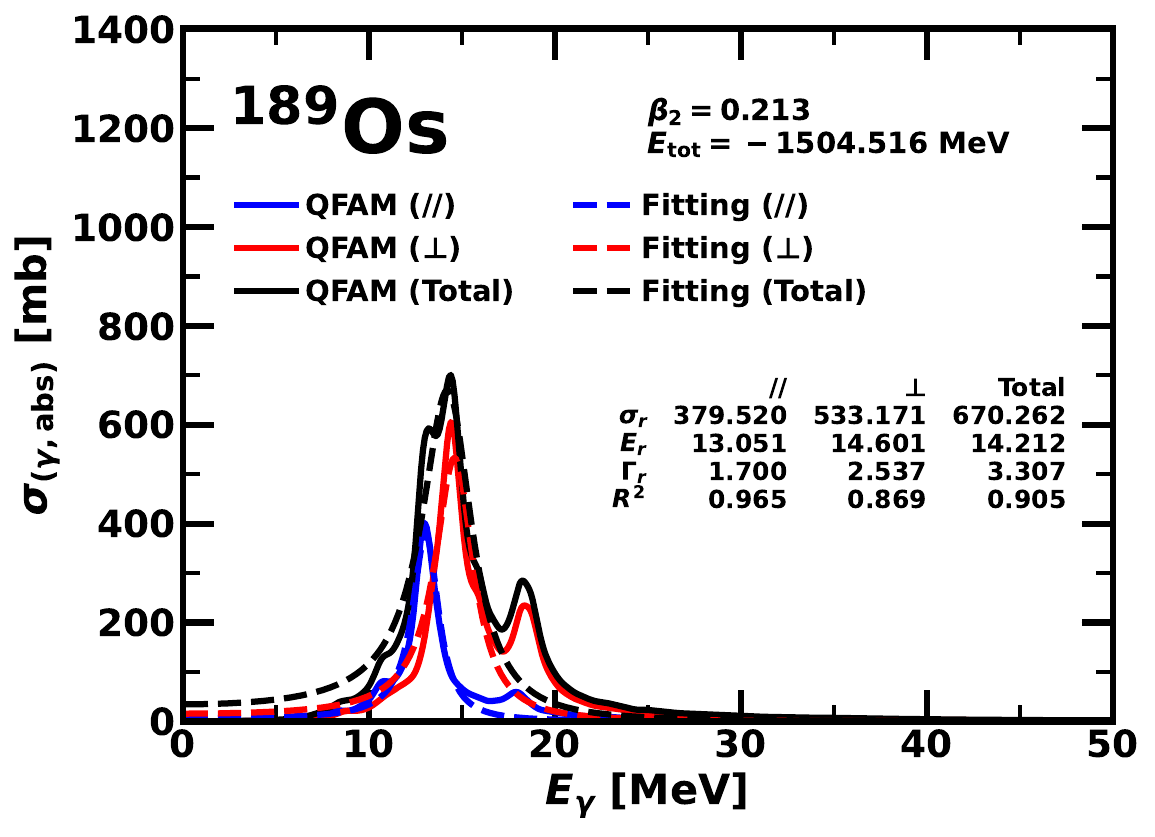}
    \includegraphics[width=0.4\textwidth]{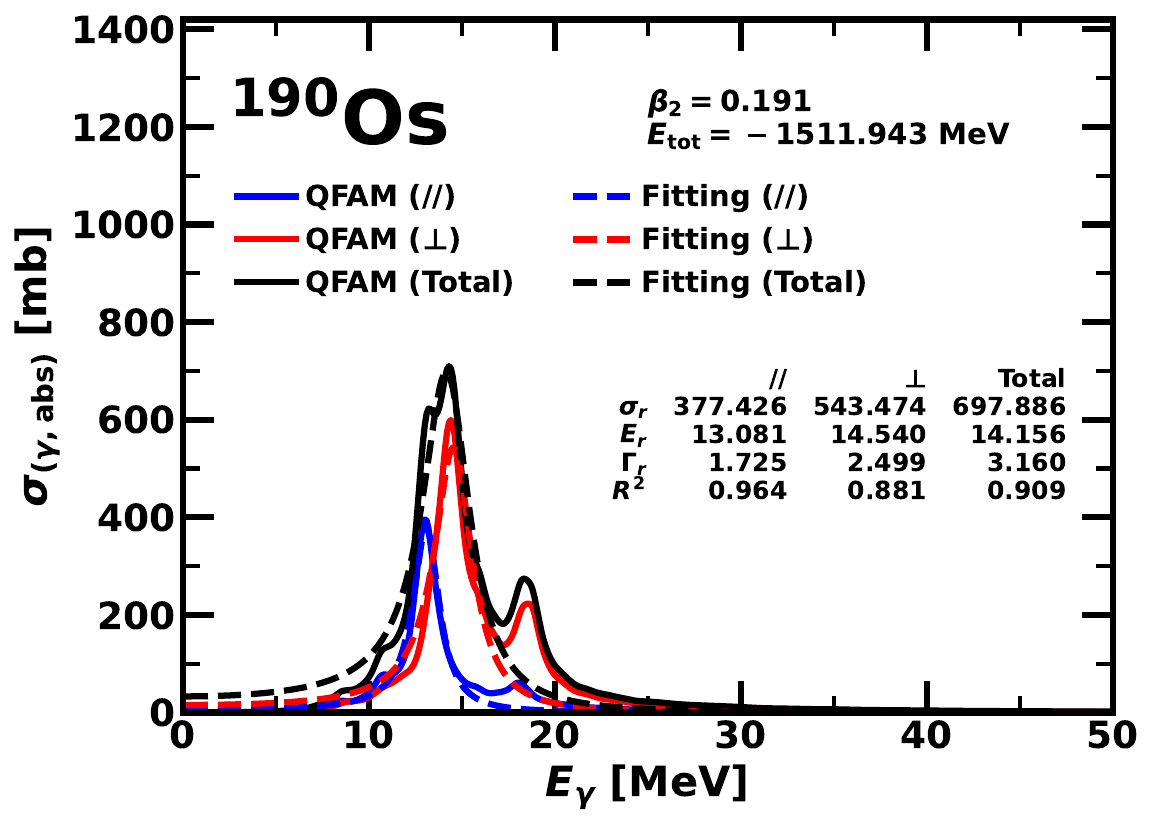}
    \includegraphics[width=0.4\textwidth]{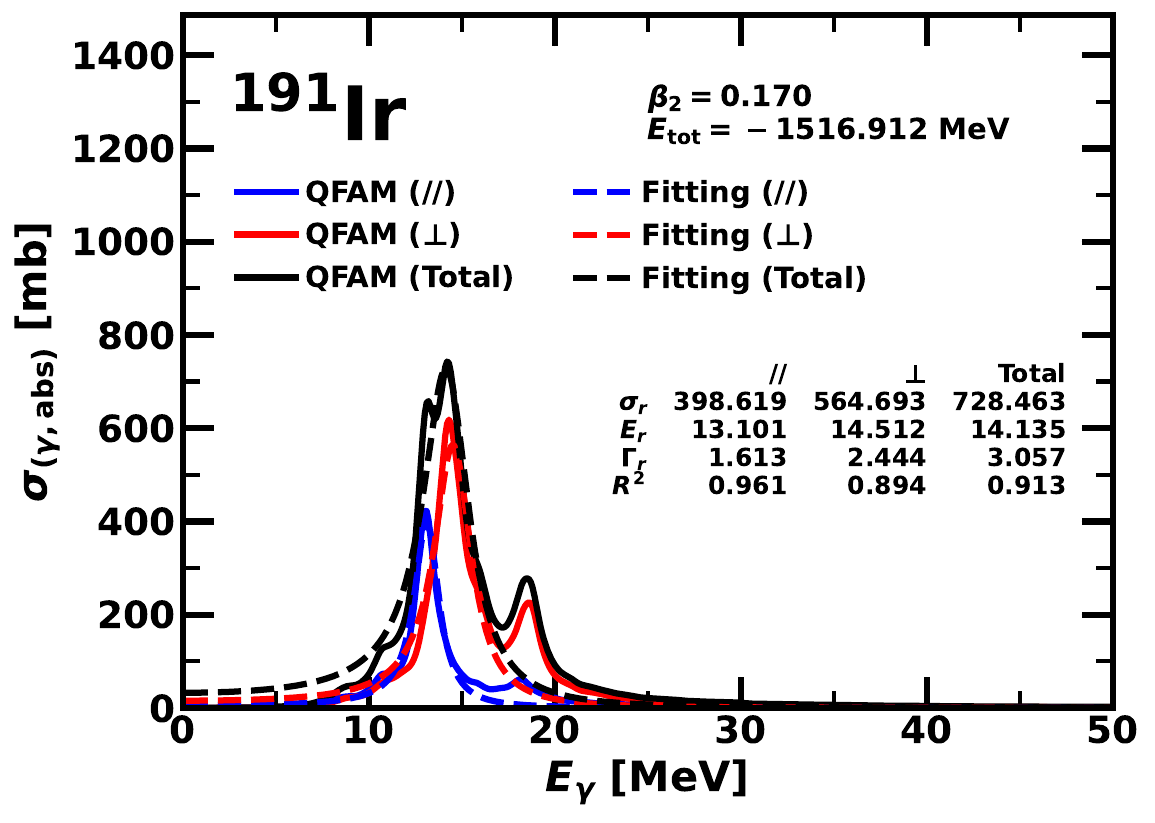}
    \includegraphics[width=0.4\textwidth]{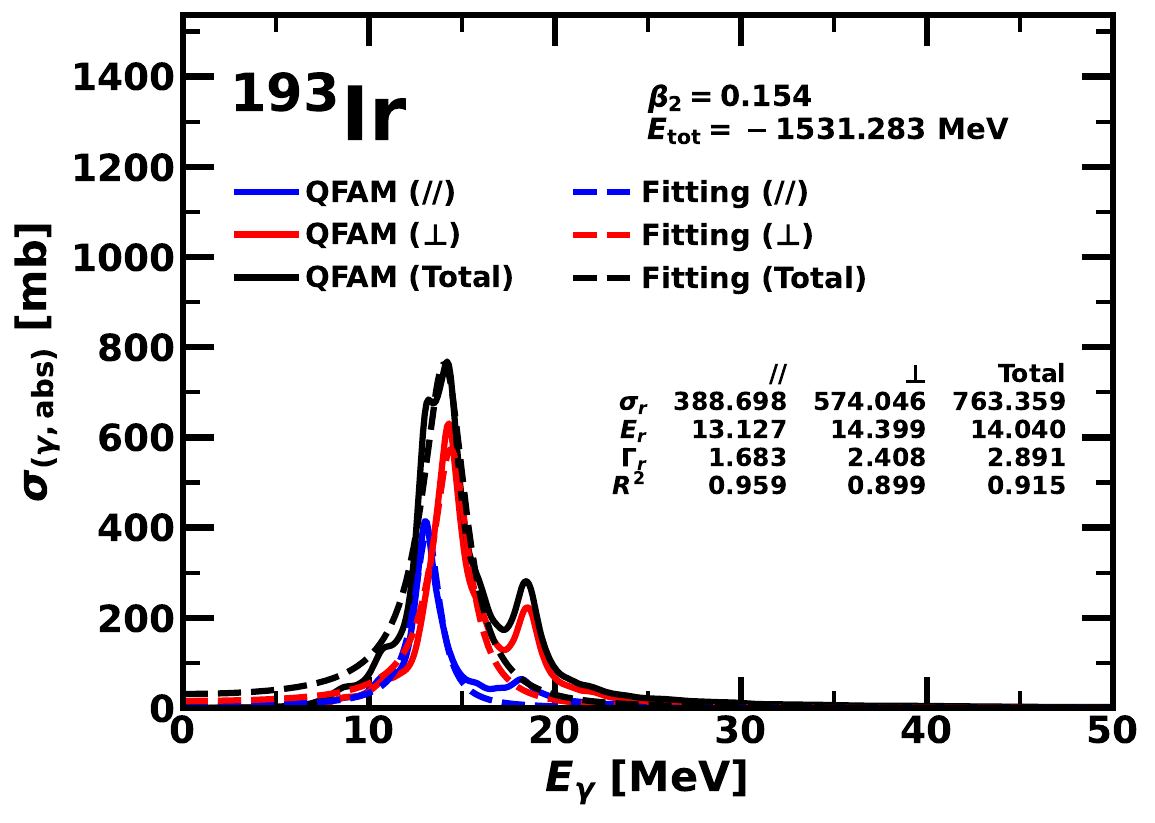}
    \includegraphics[width=0.4\textwidth]{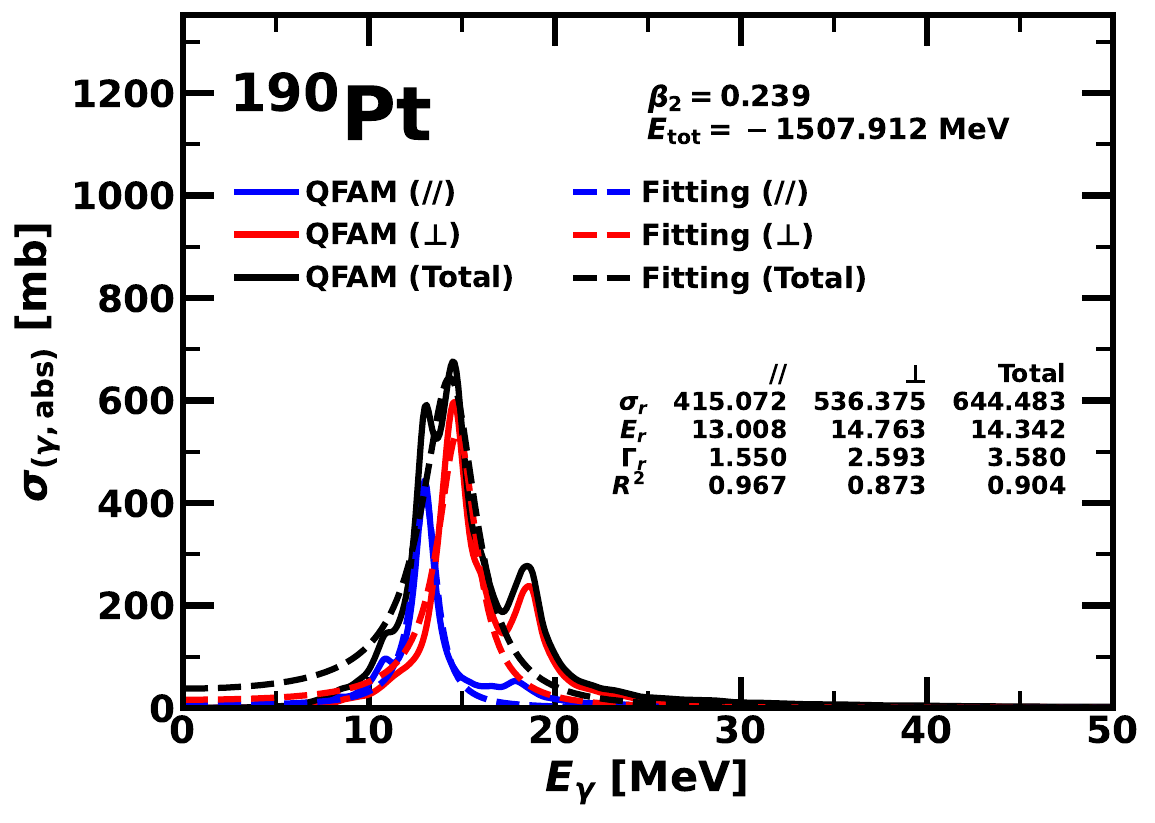}
    \includegraphics[width=0.4\textwidth]{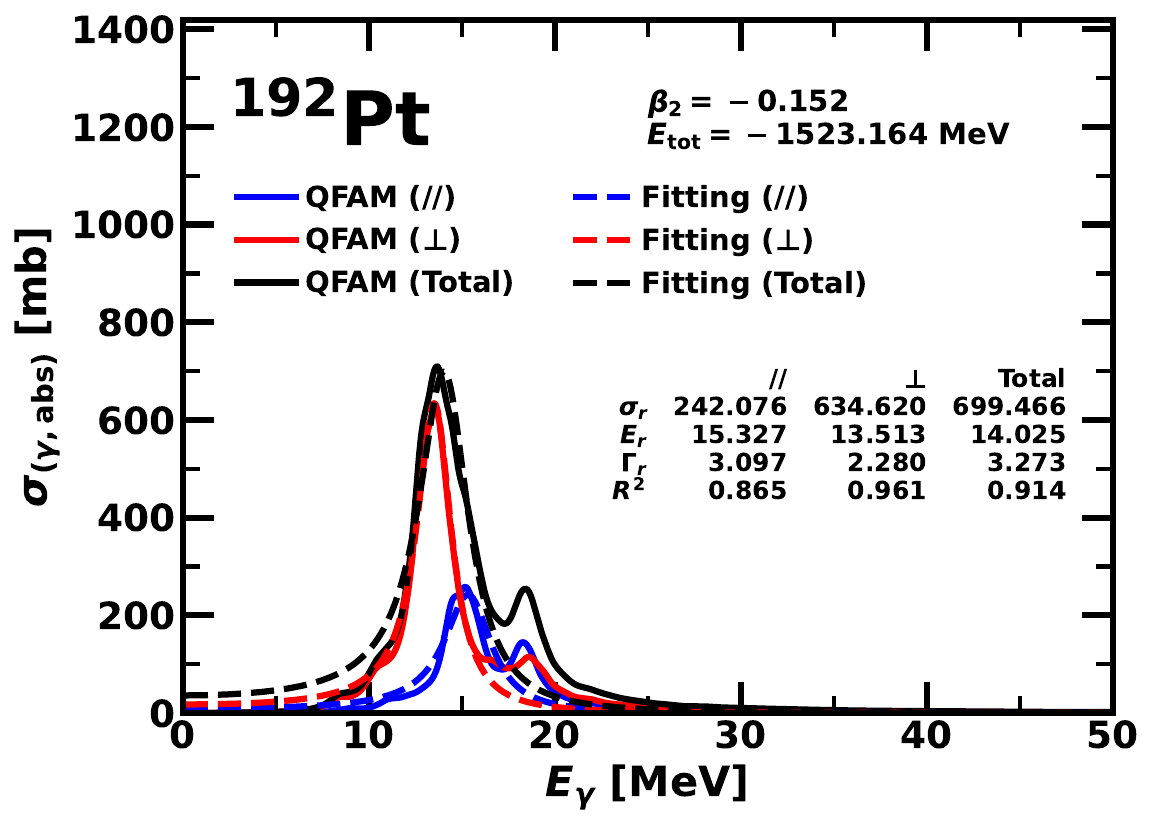}
\end{figure*}
\begin{figure*}\ContinuedFloat
    \centering
    \includegraphics[width=0.4\textwidth]{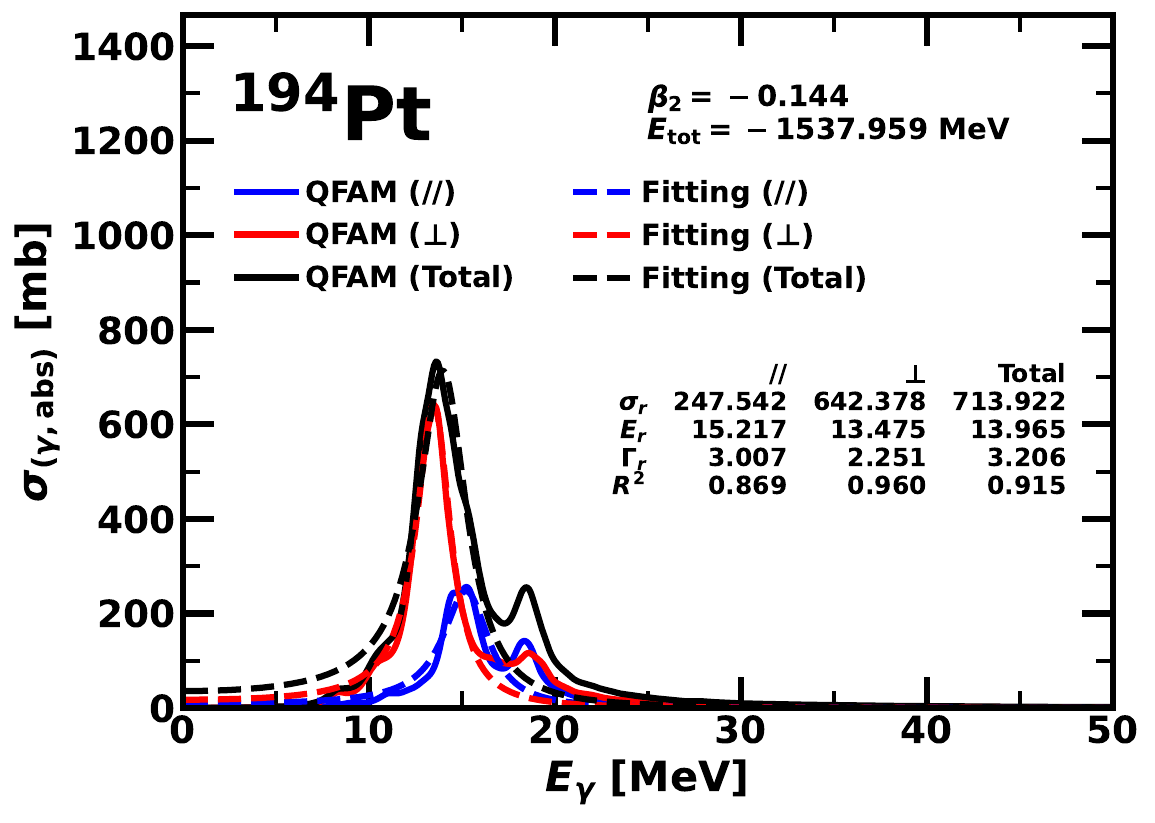}
    \includegraphics[width=0.4\textwidth]{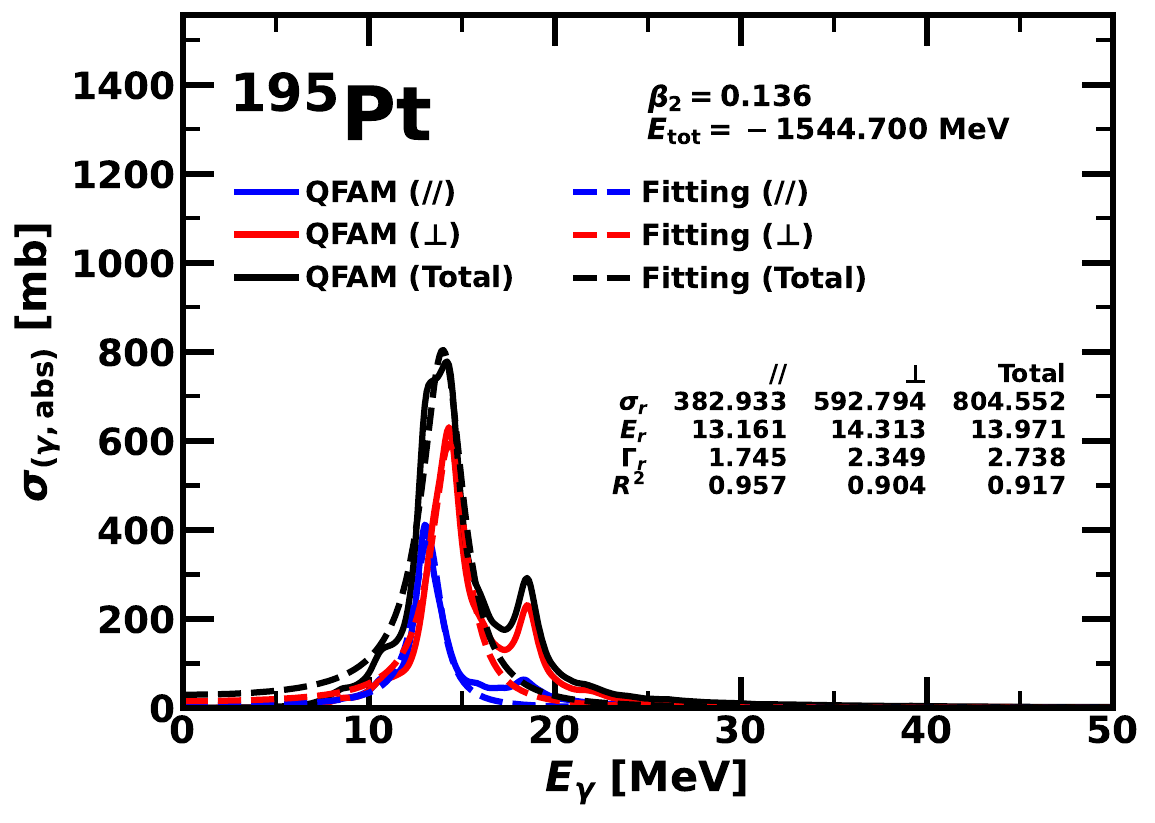}
    \includegraphics[width=0.4\textwidth]{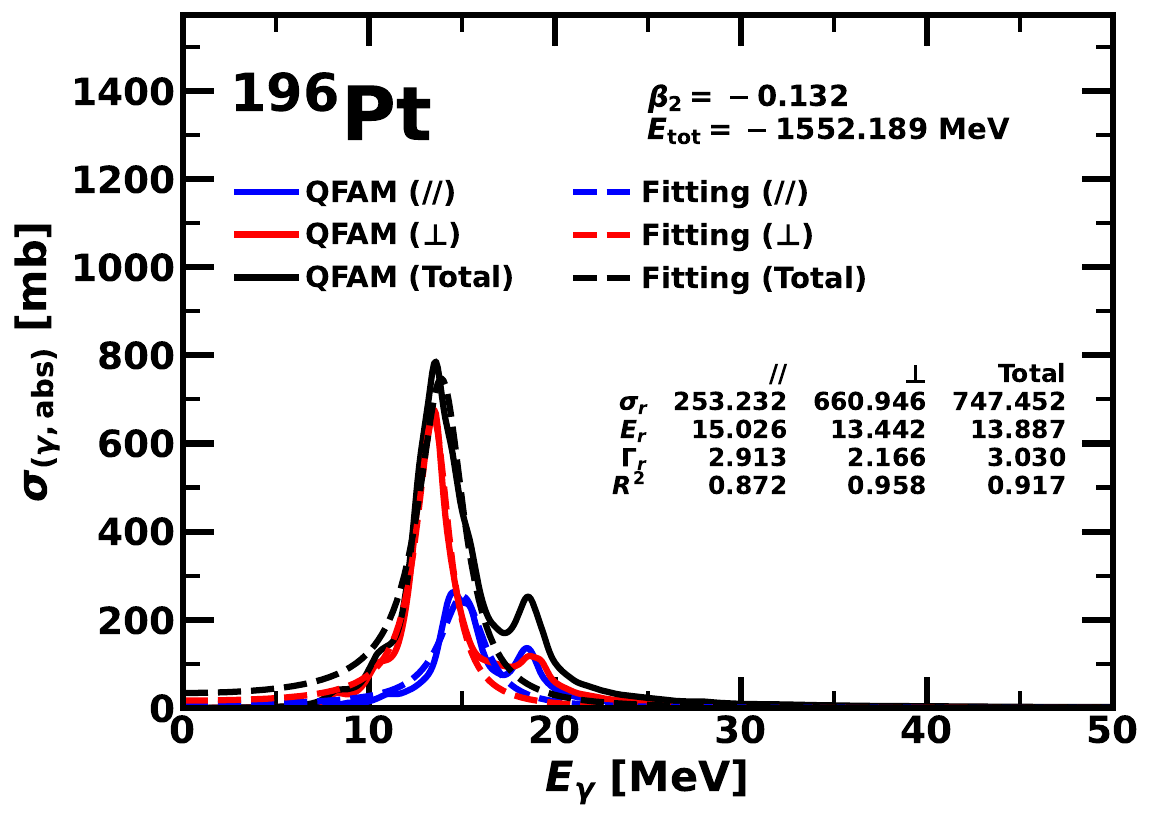}
    \includegraphics[width=0.4\textwidth]{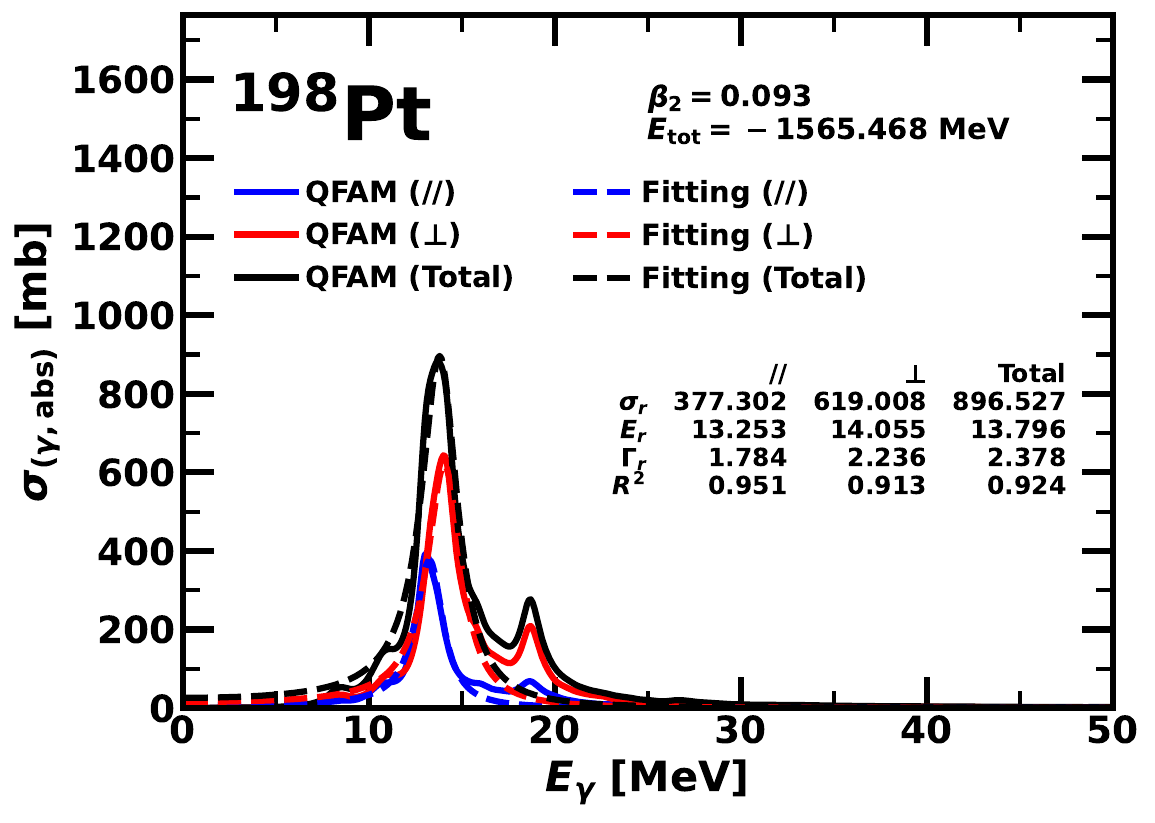}
    \includegraphics[width=0.4\textwidth]{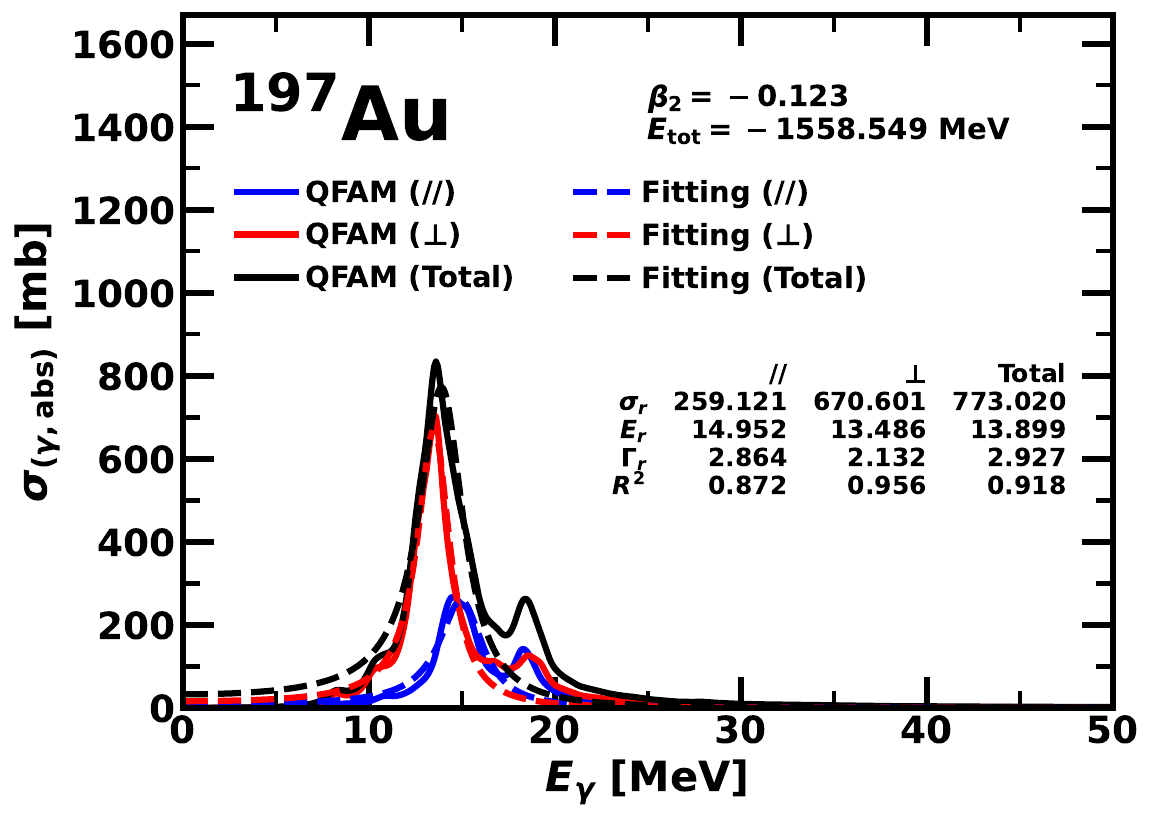}
    \includegraphics[width=0.4\textwidth]{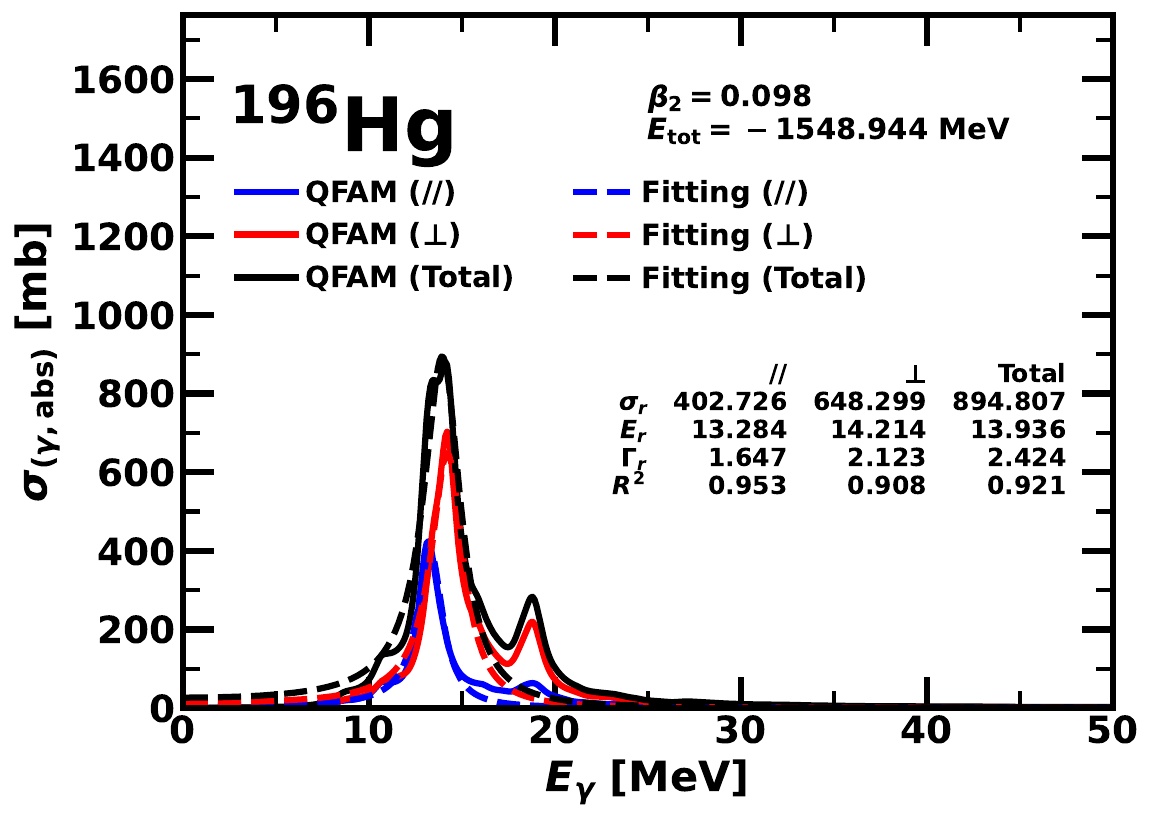}
    \includegraphics[width=0.4\textwidth]{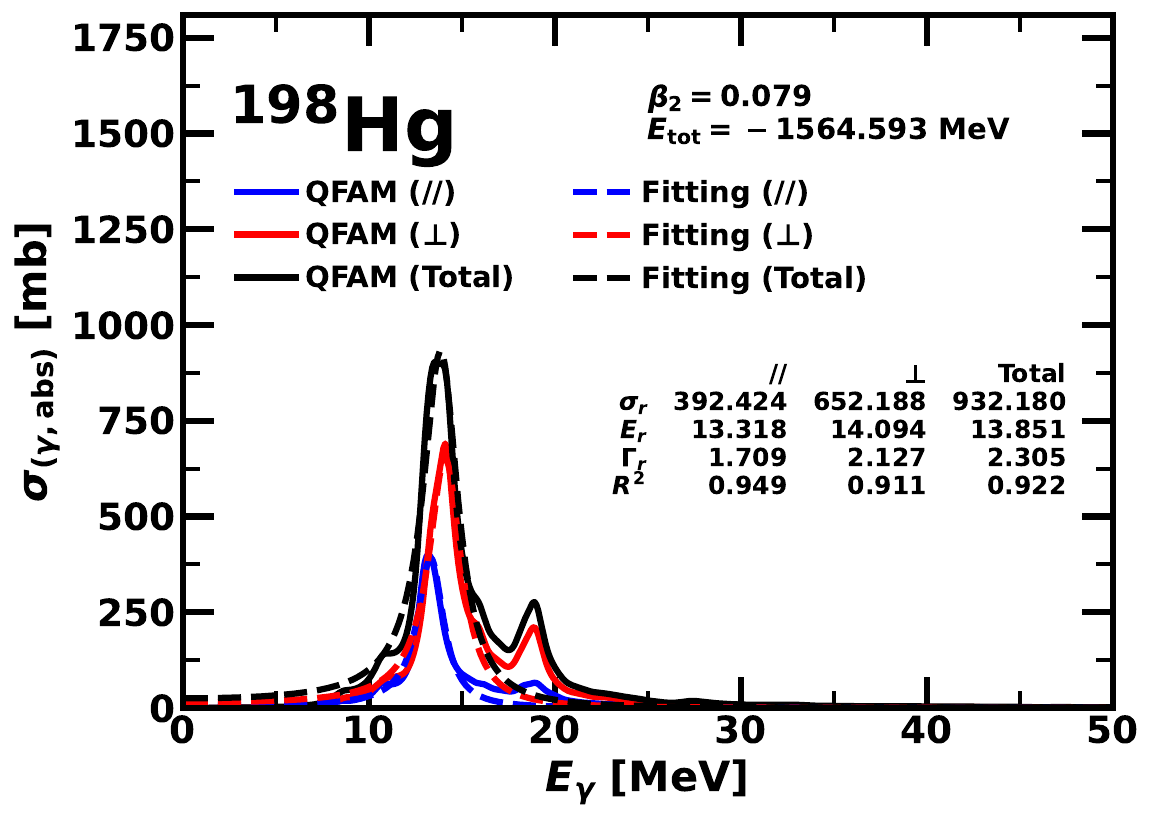}
    \includegraphics[width=0.4\textwidth]{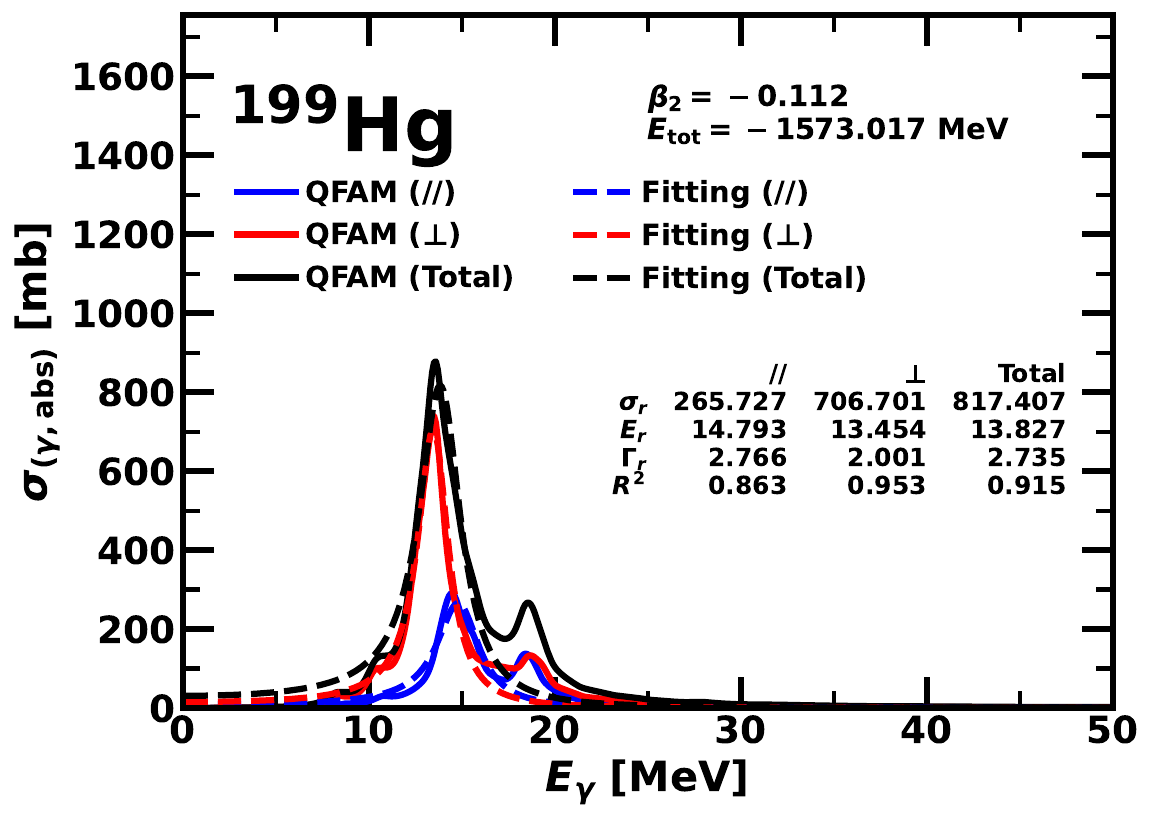}
\end{figure*}
\begin{figure*}\ContinuedFloat
    \centering
    \includegraphics[width=0.4\textwidth]{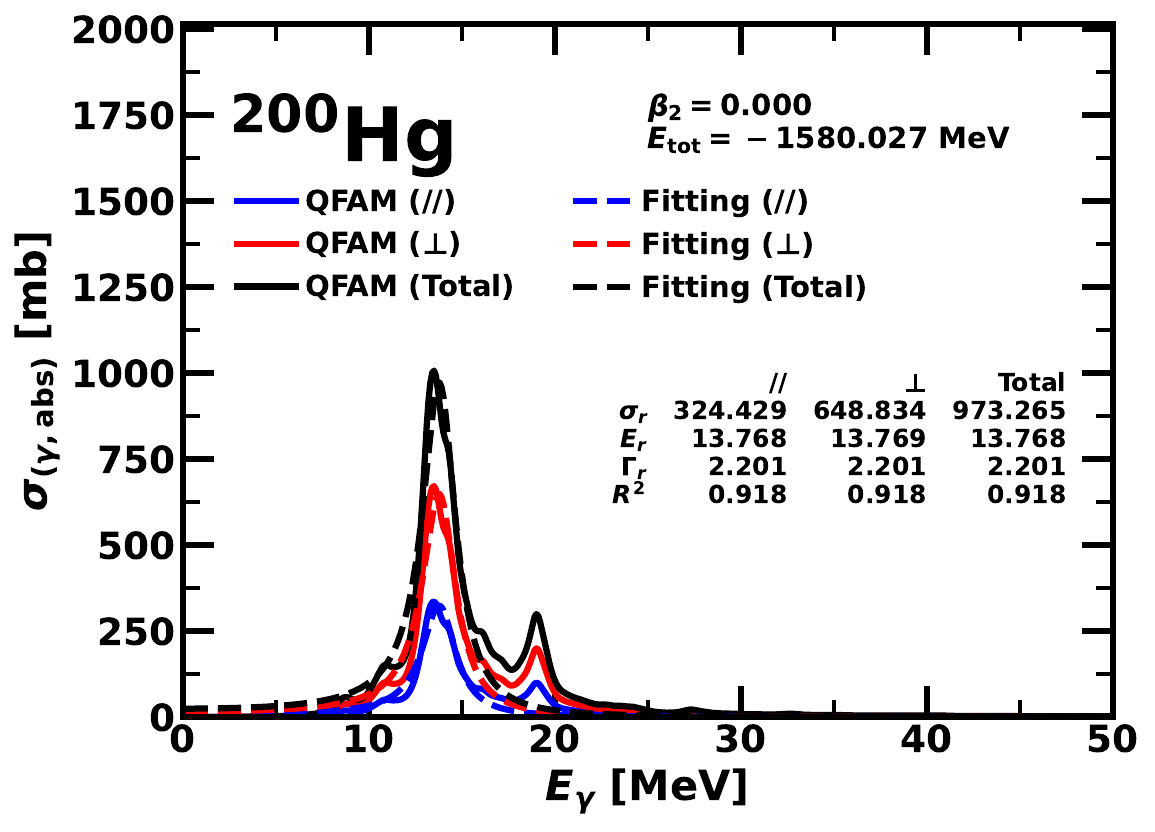}
    \includegraphics[width=0.4\textwidth]{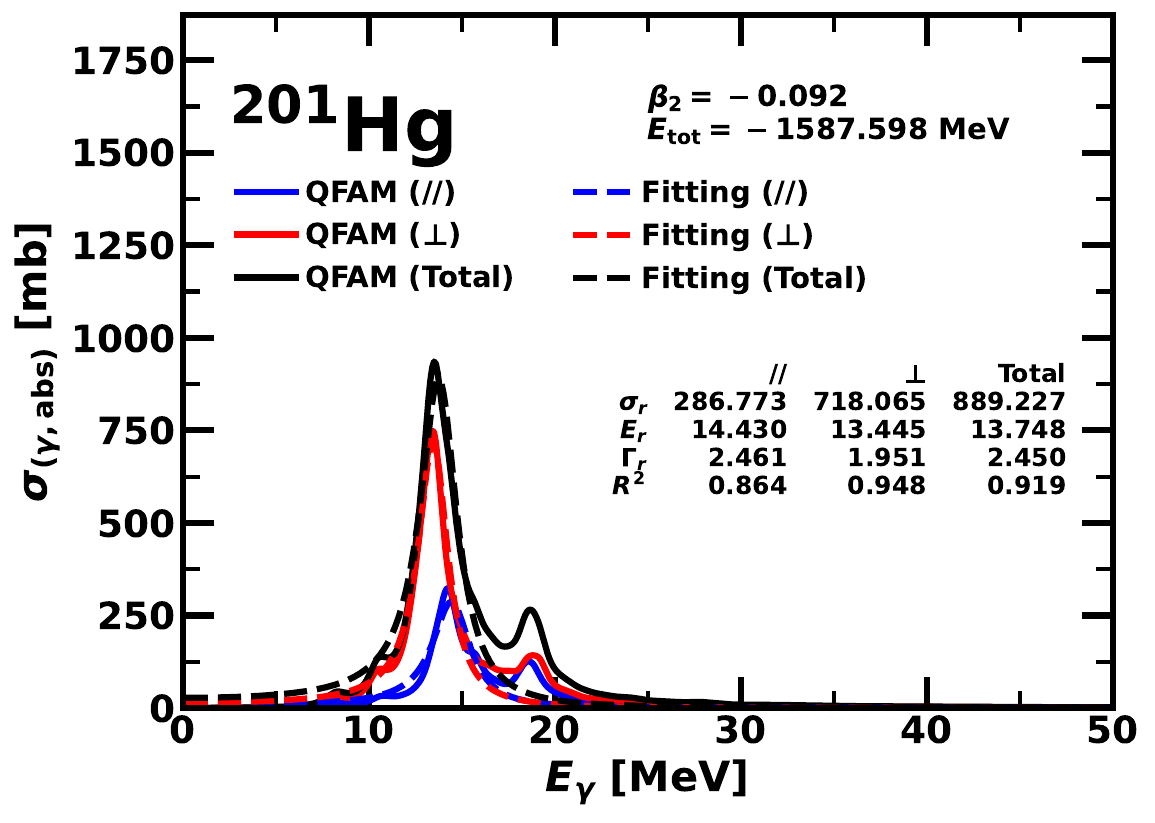}
    \includegraphics[width=0.4\textwidth]{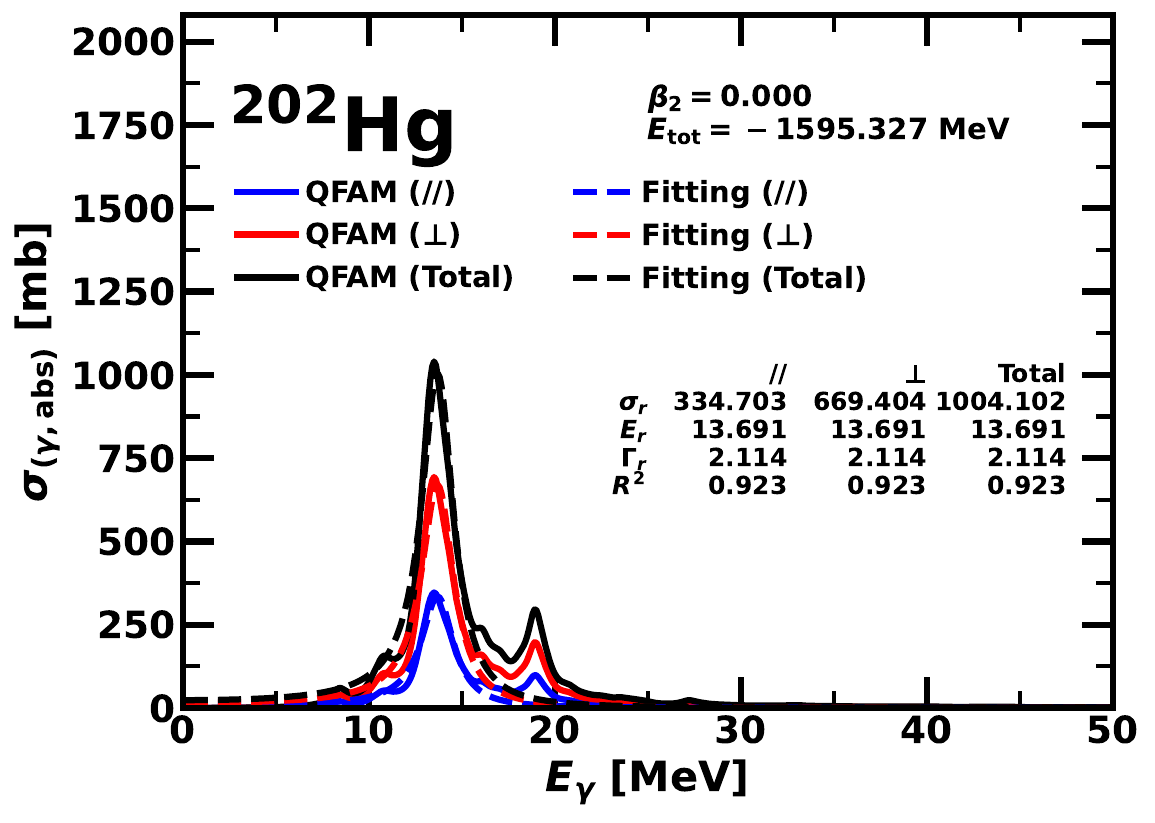}
    \includegraphics[width=0.4\textwidth]{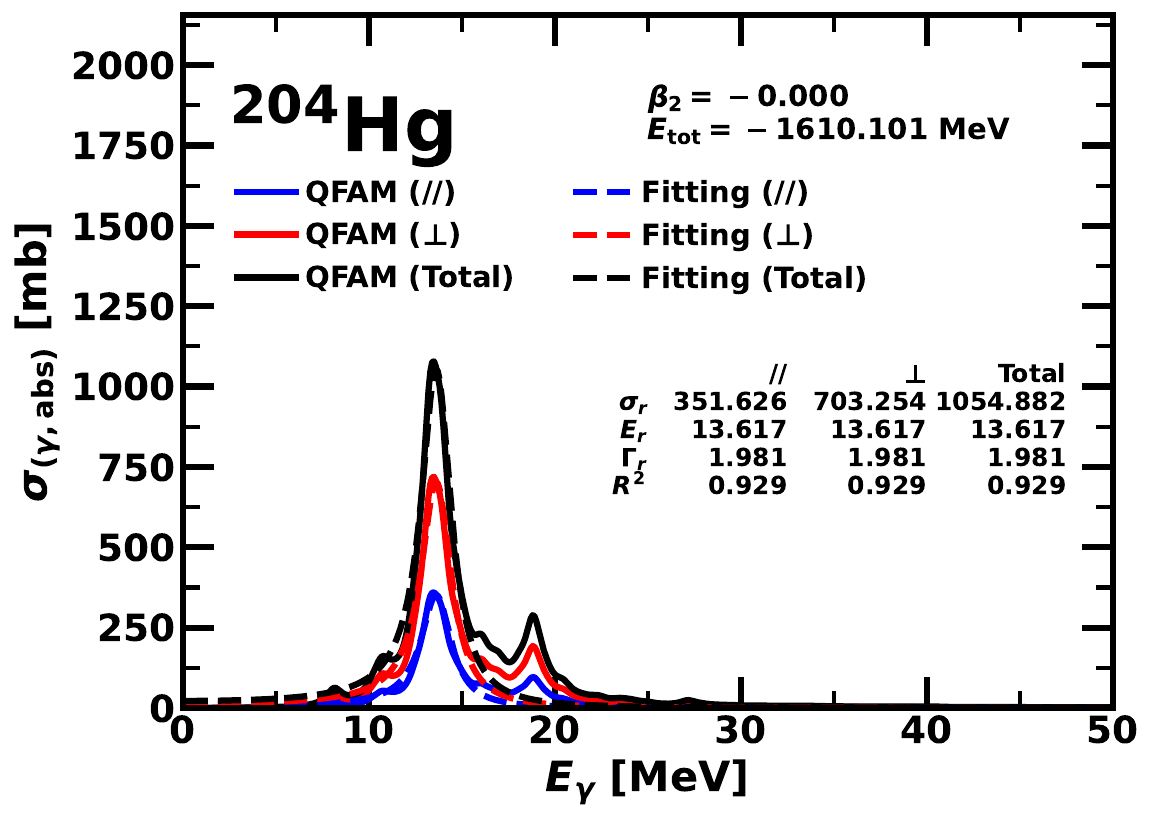}
    \includegraphics[width=0.4\textwidth]{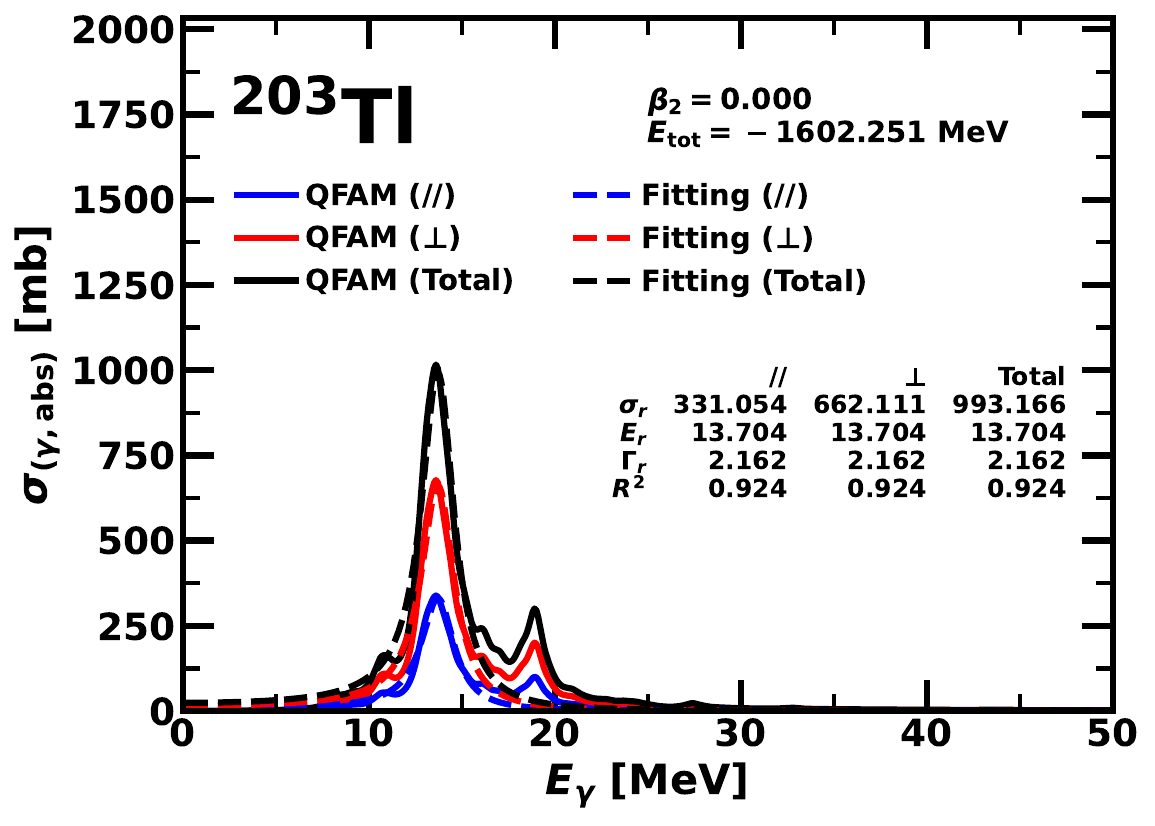}
    \includegraphics[width=0.4\textwidth]{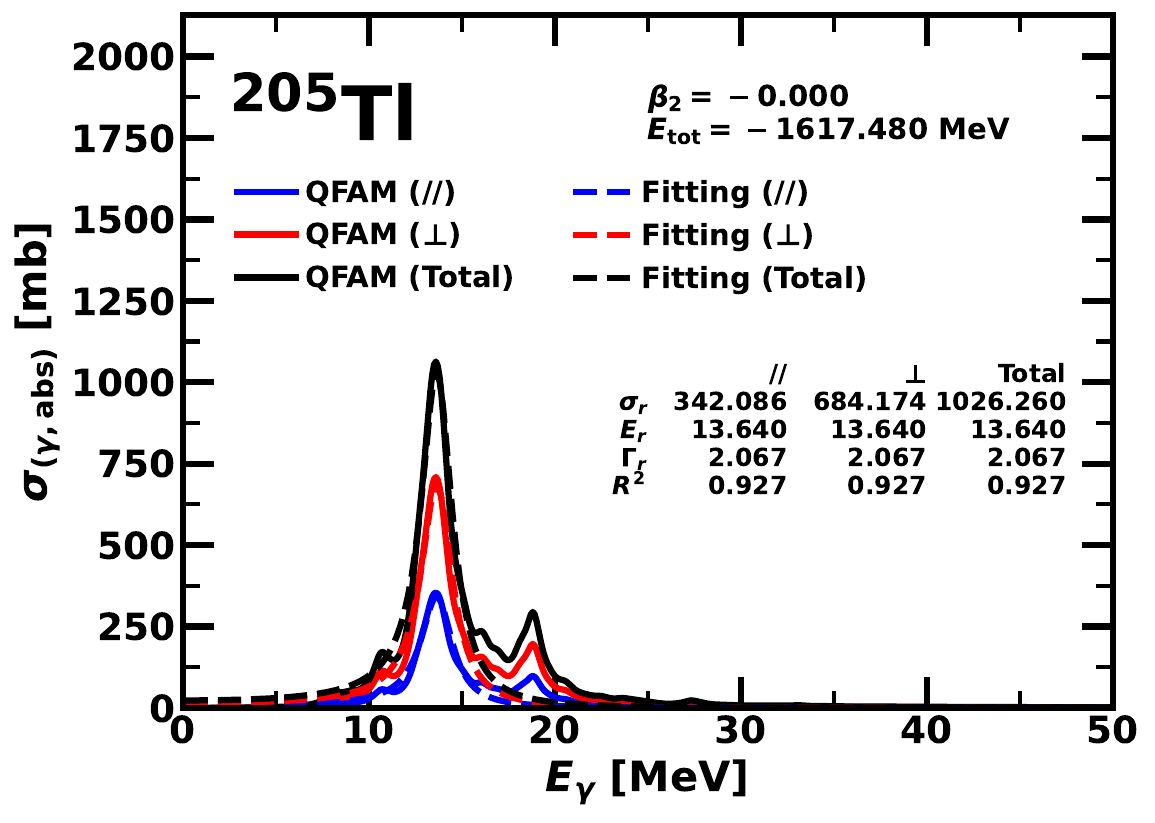}
    \includegraphics[width=0.4\textwidth]{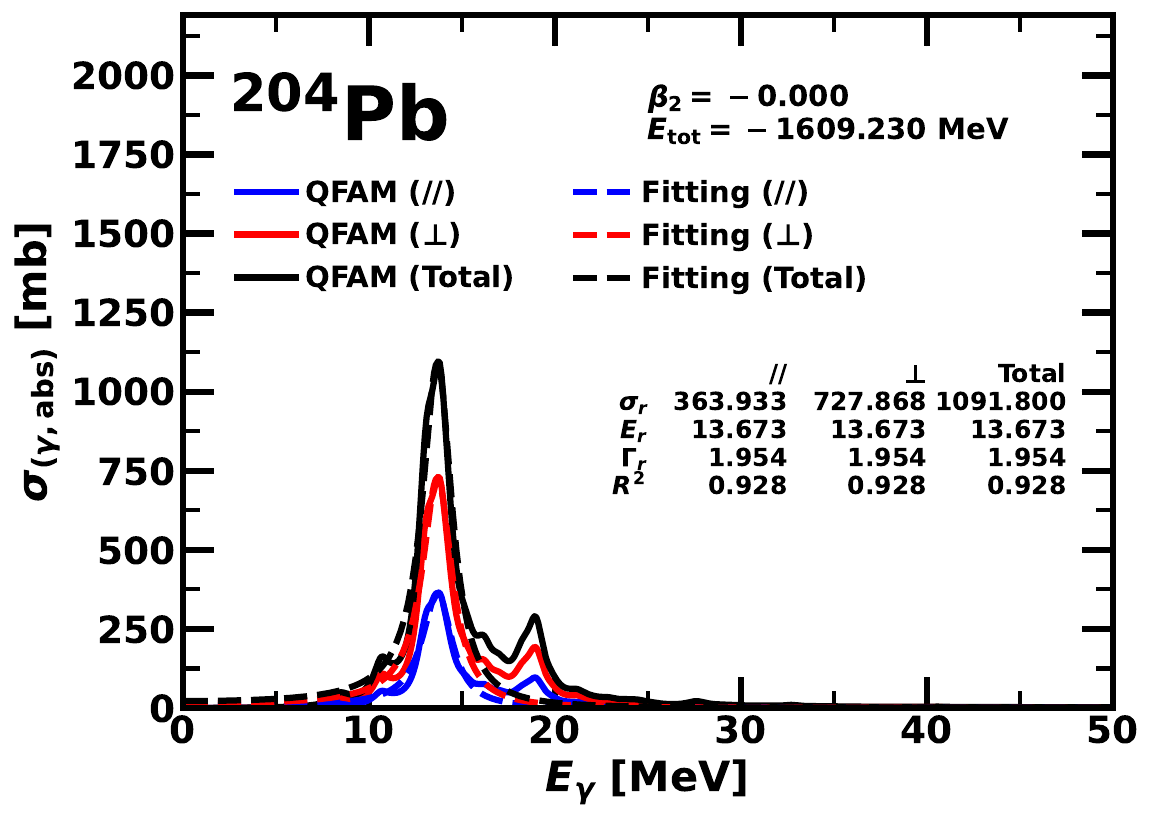}
    \includegraphics[width=0.4\textwidth]{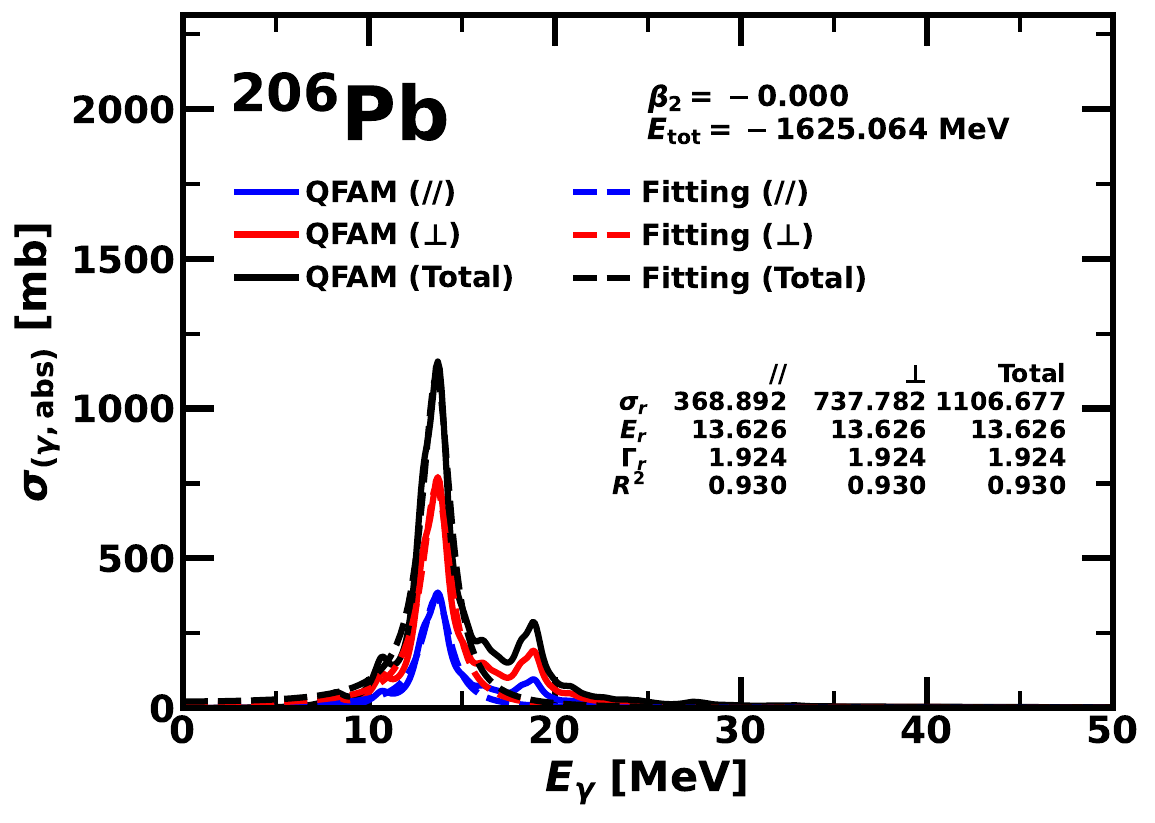}
\end{figure*}
\begin{figure*}\ContinuedFloat
    \centering
    \includegraphics[width=0.4\textwidth]{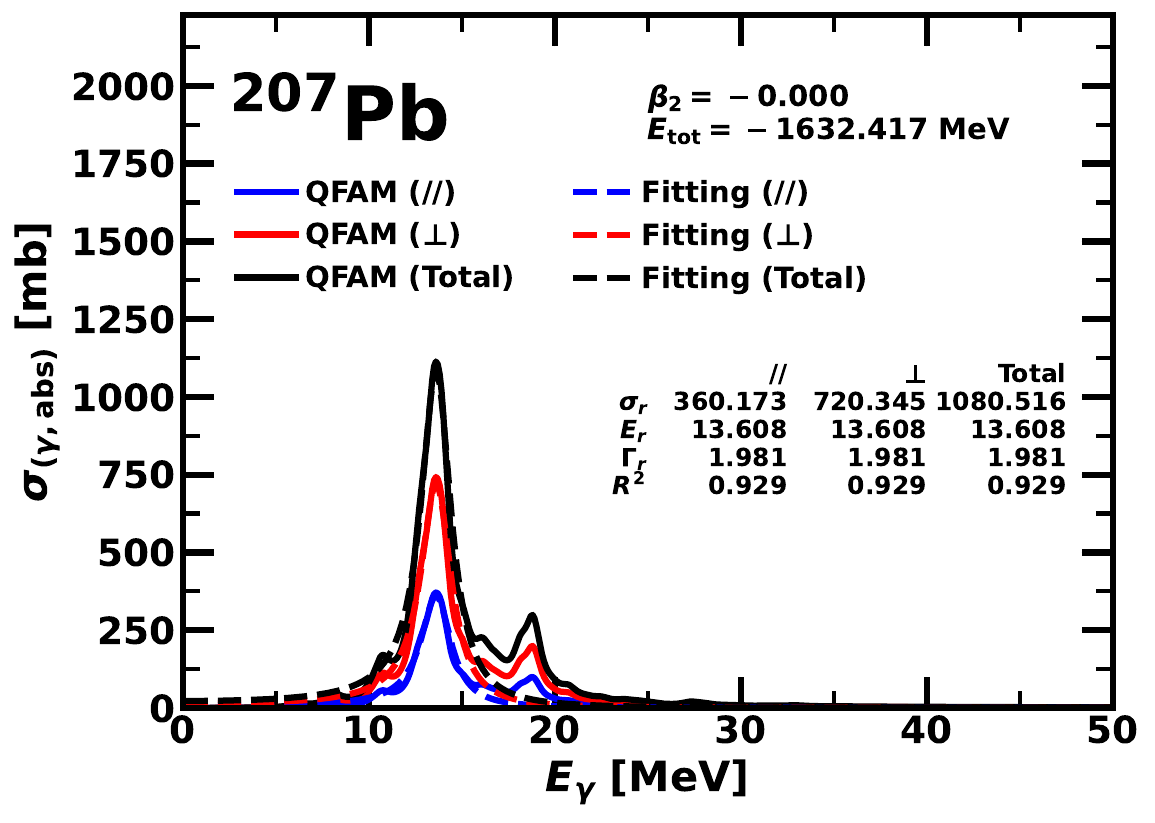}
    \includegraphics[width=0.4\textwidth]{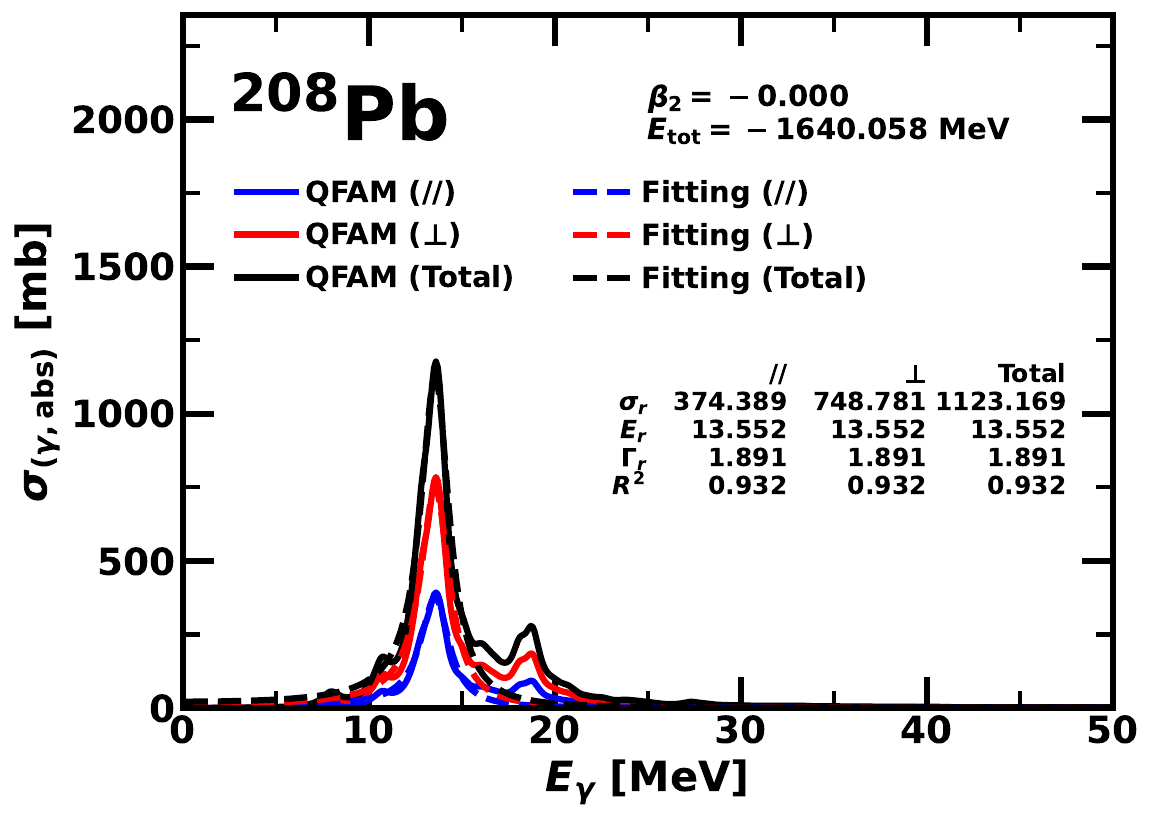}
    \includegraphics[width=0.4\textwidth]{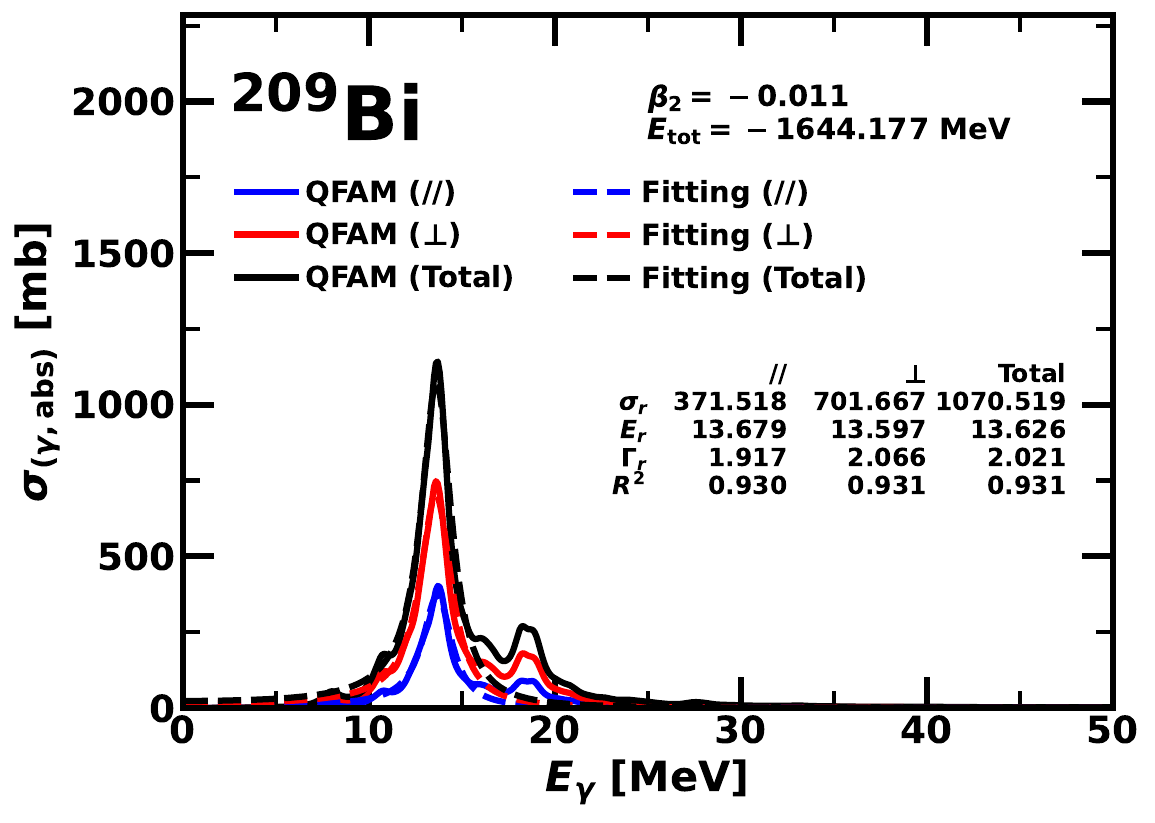}
\end{figure*}

\section{Comparisons between QFAM results and experimental photonuclear cross section}
\label{sec:appendix_c}
Figs.~\ref{fig:appendix_c} show comparisons among photoabsorption cross sections from QFAM calculations, 
experimental photonuclear cross section data from EXFOR database, 
photoabsorption cross sections form IAEA Evaluated Photonuclear Data Library (IAEA/PD-2019.2) \cite{Kawano_2020_NDS} 
and GDR components of photoabsorption cross sections calculated with GDR parameters extracted from experimental data recommended by IAEA \cite{Plujko_2018_ADNDT}.
In these figures, the QFAM results are shown as black solid lines, 
those from IAEA/PD-2019.2 are shown as blue dashed lines, 
those calculated with IAEA-recommended GDR parameters are shown as purple dashed dotted lines,
and the experimental data are shown as colored scatters with error bars.
The experimental data include the photoabsorption cross section $\sigma(\gamma,\text{abs.})$ 
and total photoneutron cross section $\sigma(\gamma,sn)$, as described in Section~\ref{sec:res}.
For each nucleus, the type of experimental data 
and the corresponding EXFOR subentry numbers are listed in the figures.
Additionally, 
the deformation parameters $\beta_2$ obtained from RHB calculations are provided for all nuclei. 
For even-even nuclei, the absolute values of experimental deformation parameter $|\beta_2|$ are provided. 
For odd-A nuclei, the projection of angular momentum and parity $K^\pi$, determined by RHB calculations, 
as well as the experimental angular momentum and parity $J^\pi_\mathrm{exp.}$ are given.

\begin{figure*}
    \centering
    \includegraphics[width=0.35\textwidth]{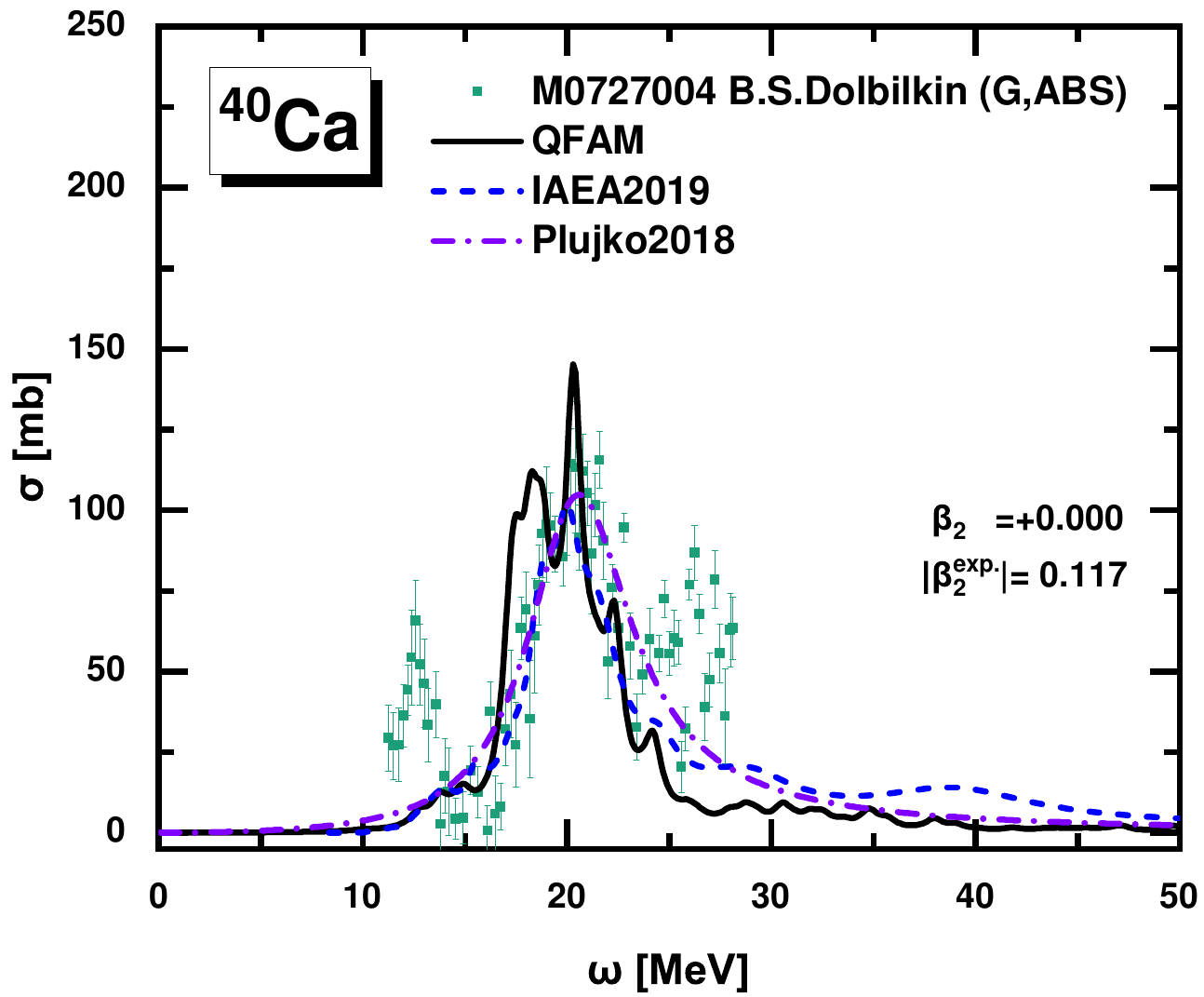}
    \includegraphics[width=0.35\textwidth]{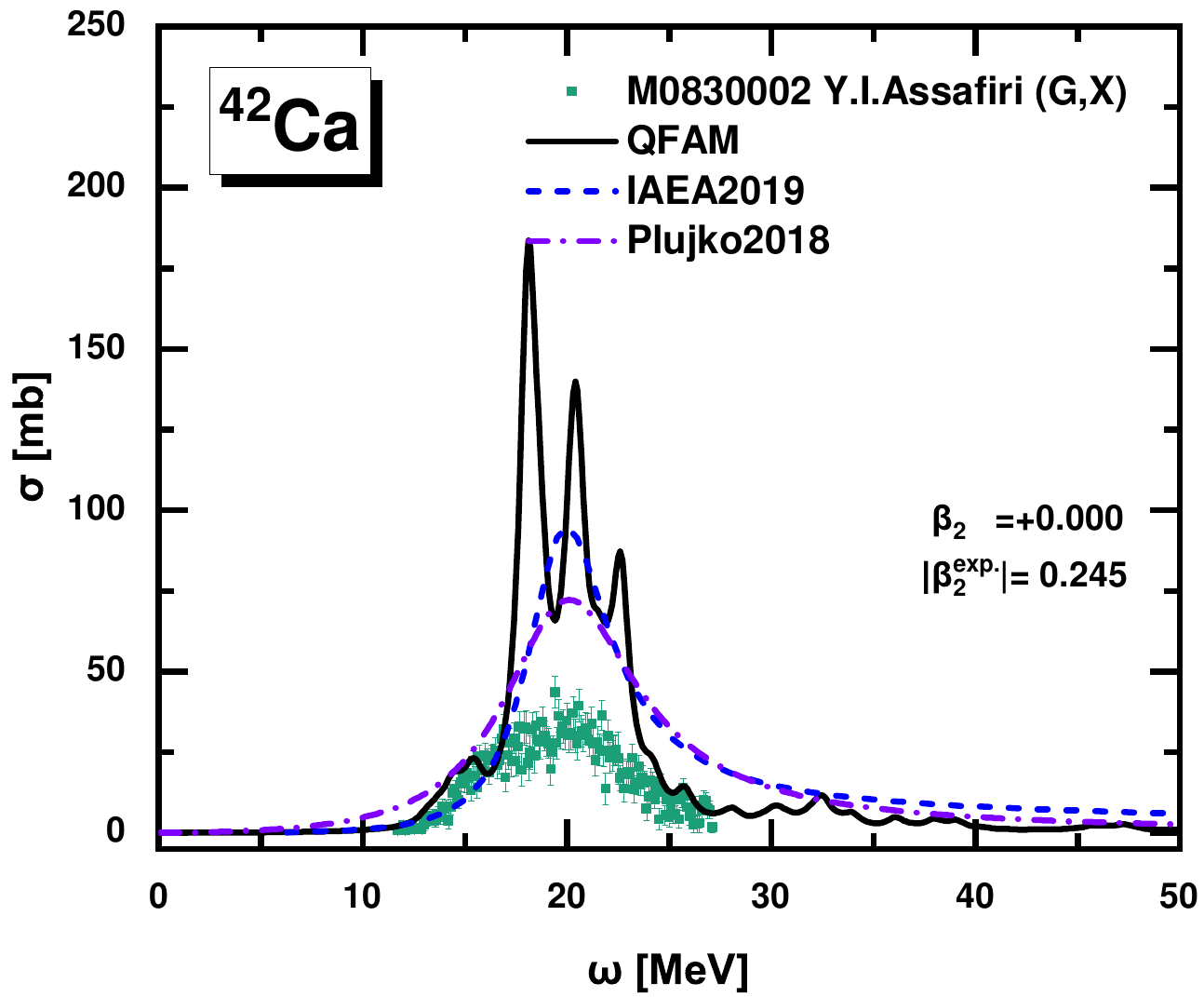}
    \includegraphics[width=0.35\textwidth]{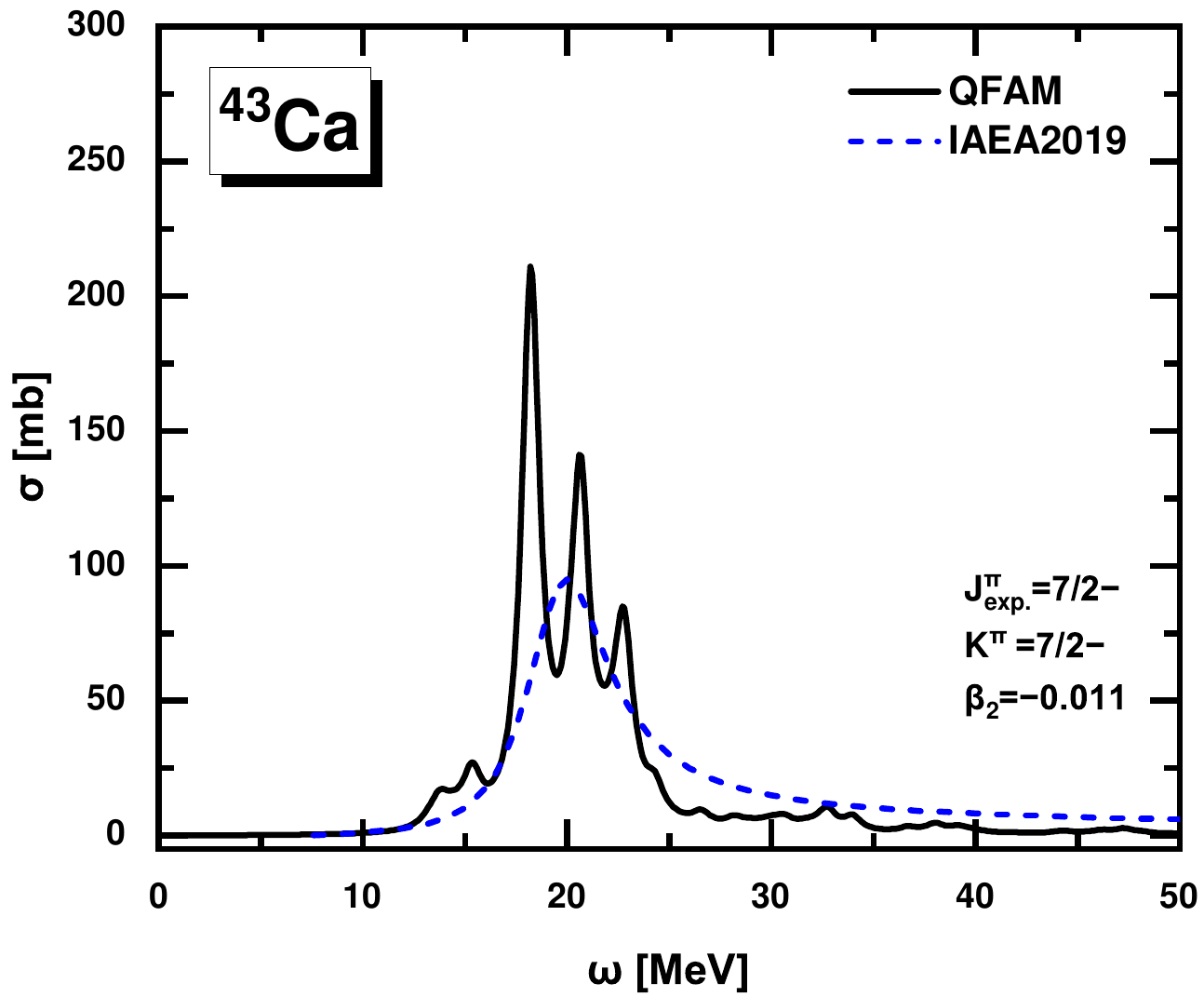}
    \includegraphics[width=0.35\textwidth]{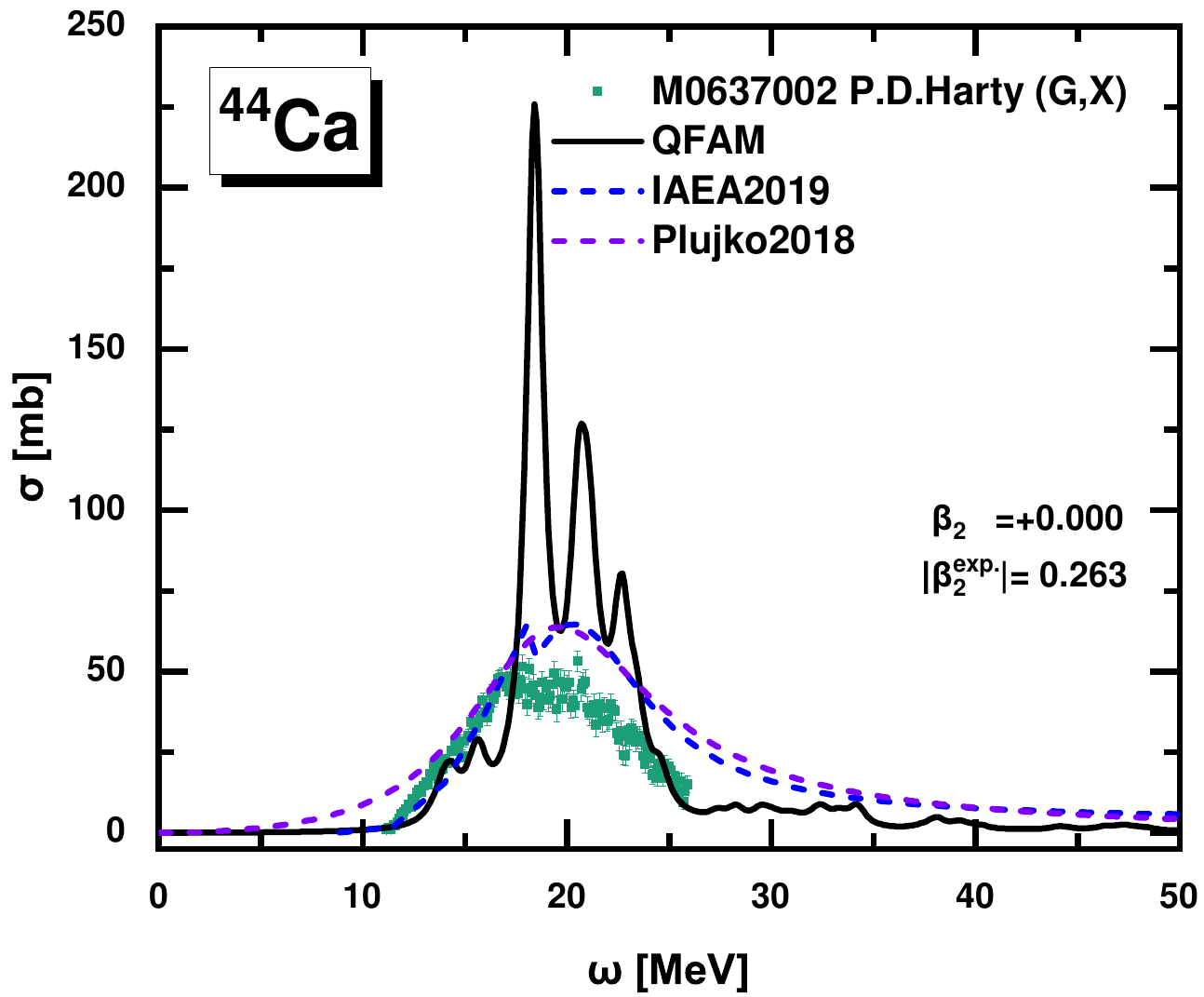}
    \includegraphics[width=0.35\textwidth]{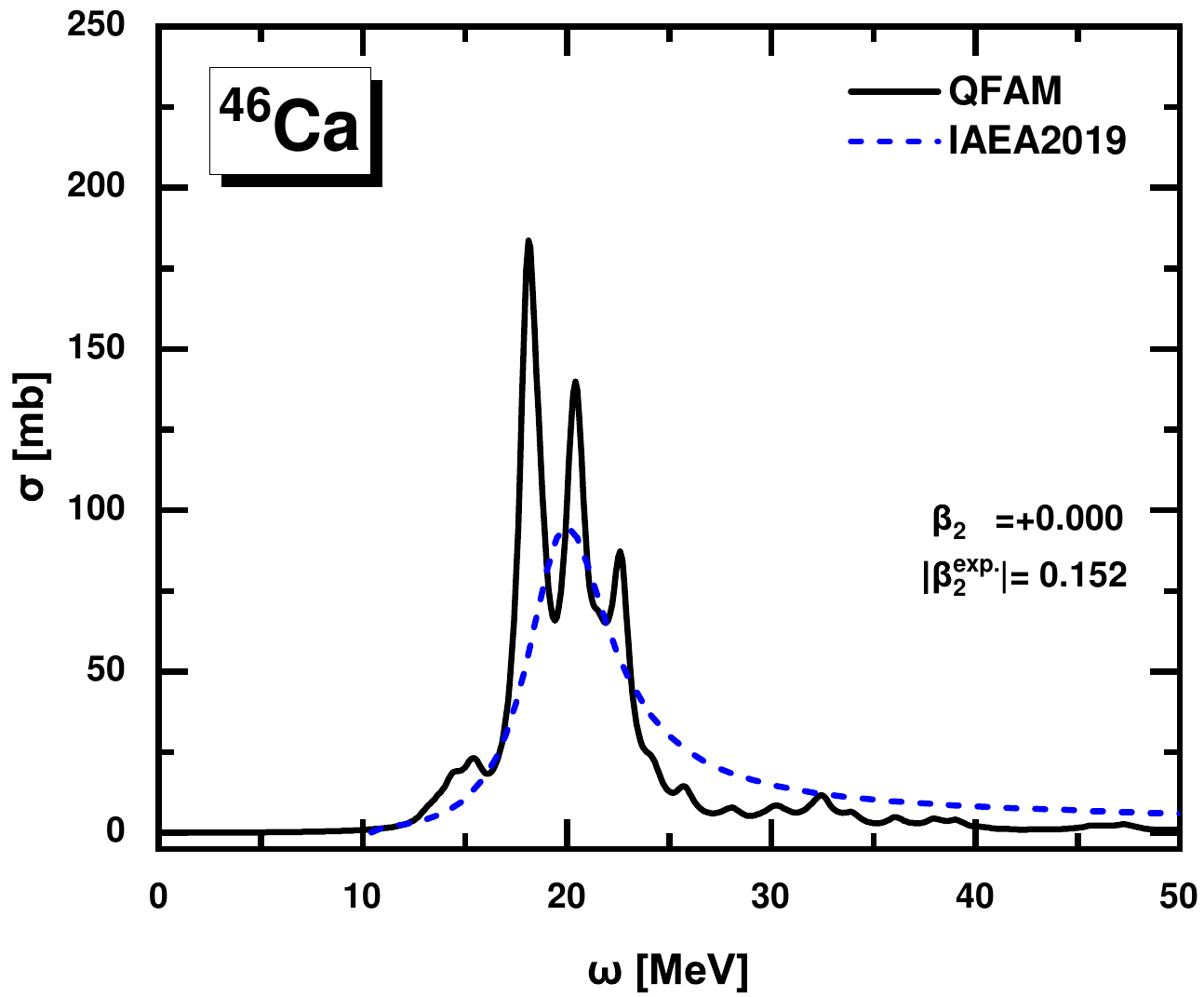}
    \includegraphics[width=0.35\textwidth]{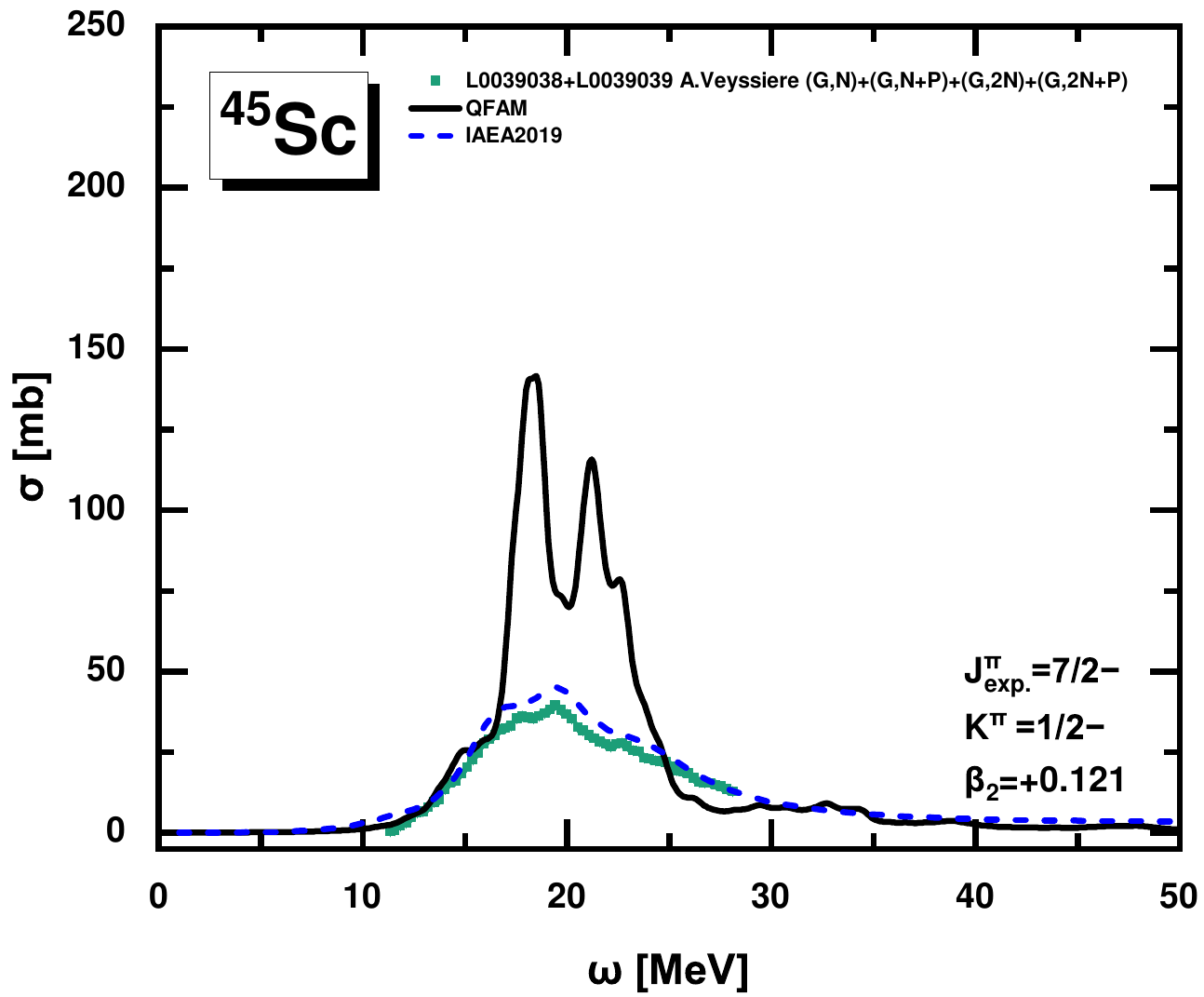}
    \includegraphics[width=0.35\textwidth]{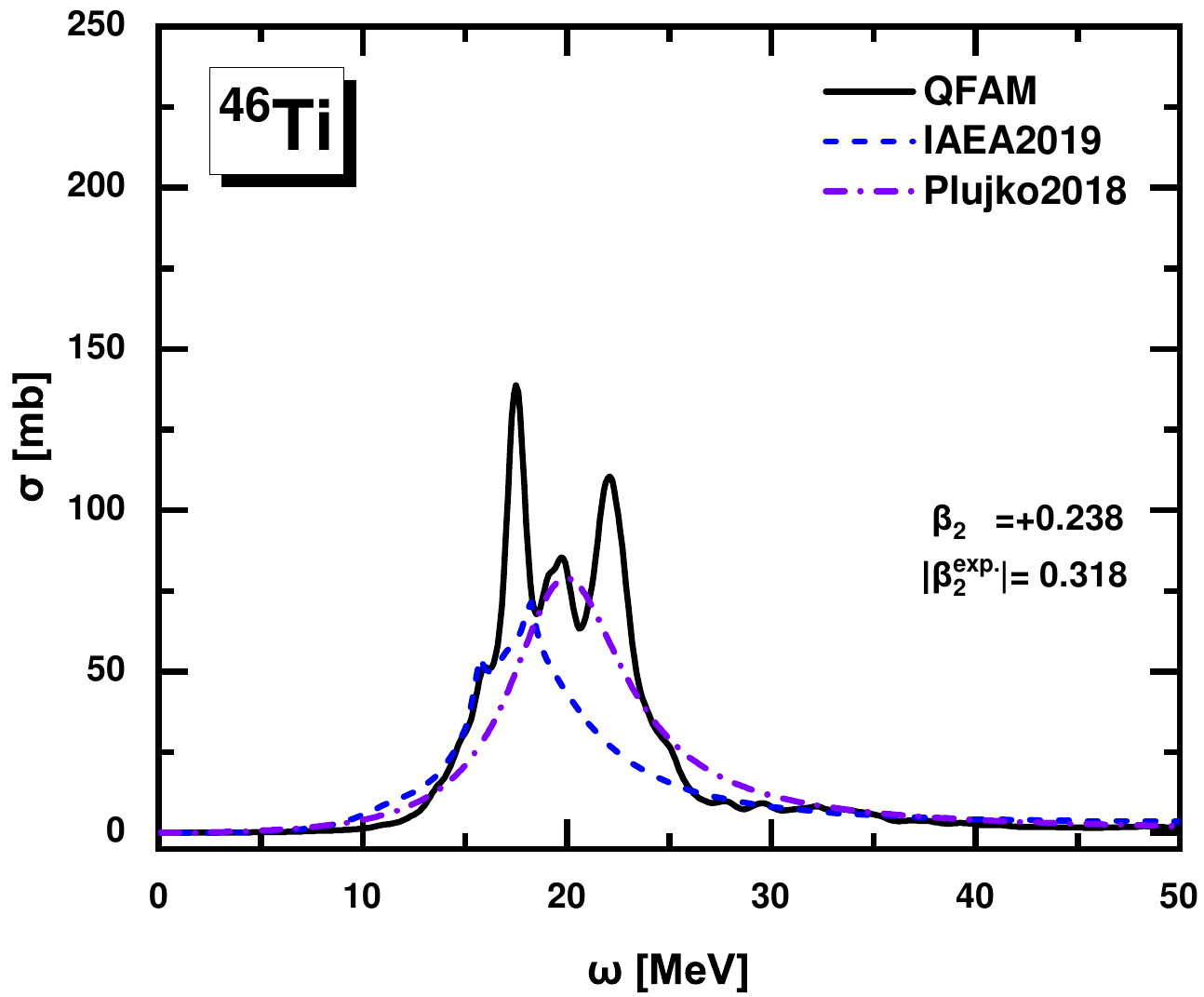}
    \includegraphics[width=0.35\textwidth]{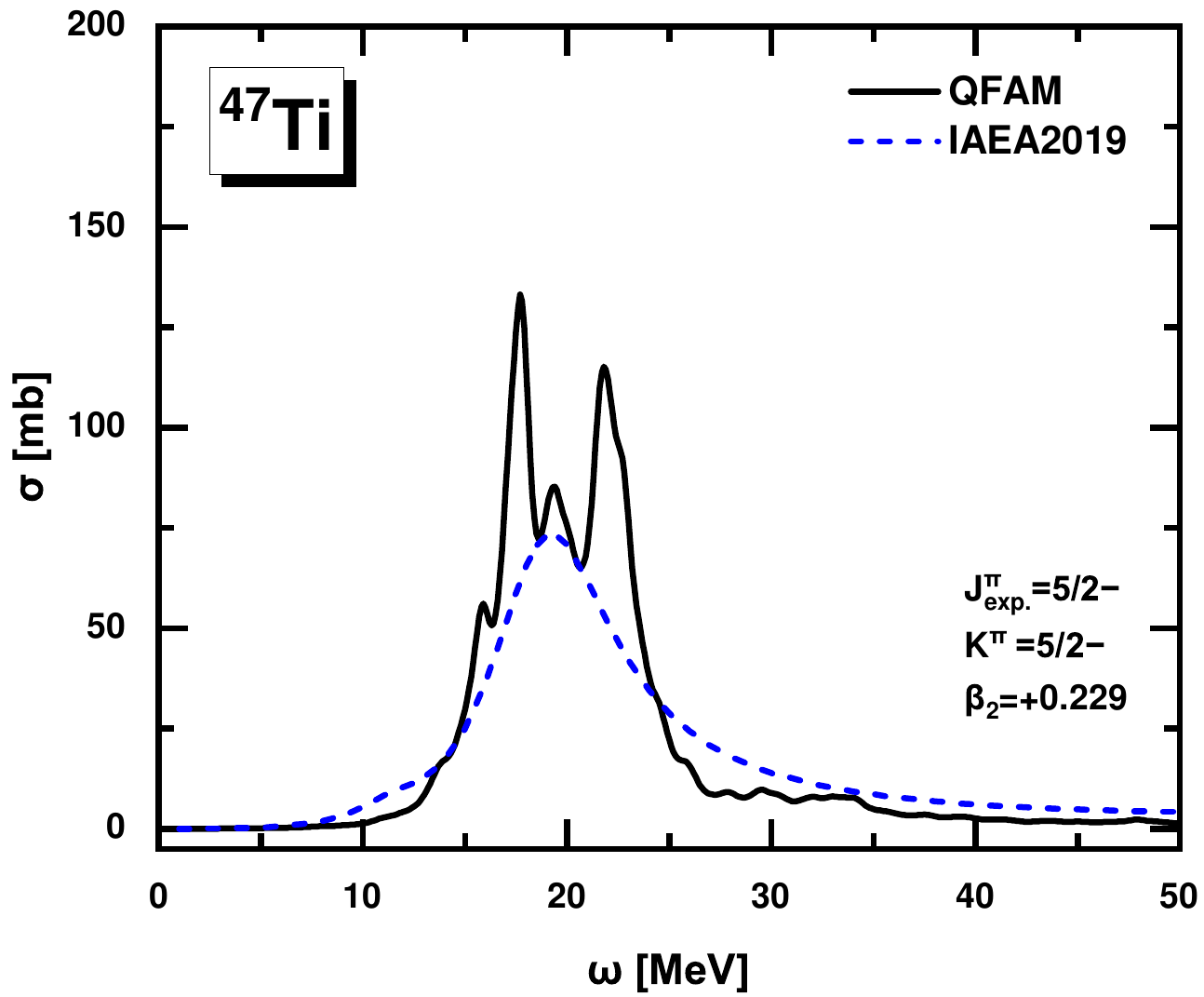}
    \caption{}\label{fig:appendix_c}
\end{figure*}
\begin{figure*}\ContinuedFloat
    \centering
    \includegraphics[width=0.35\textwidth]{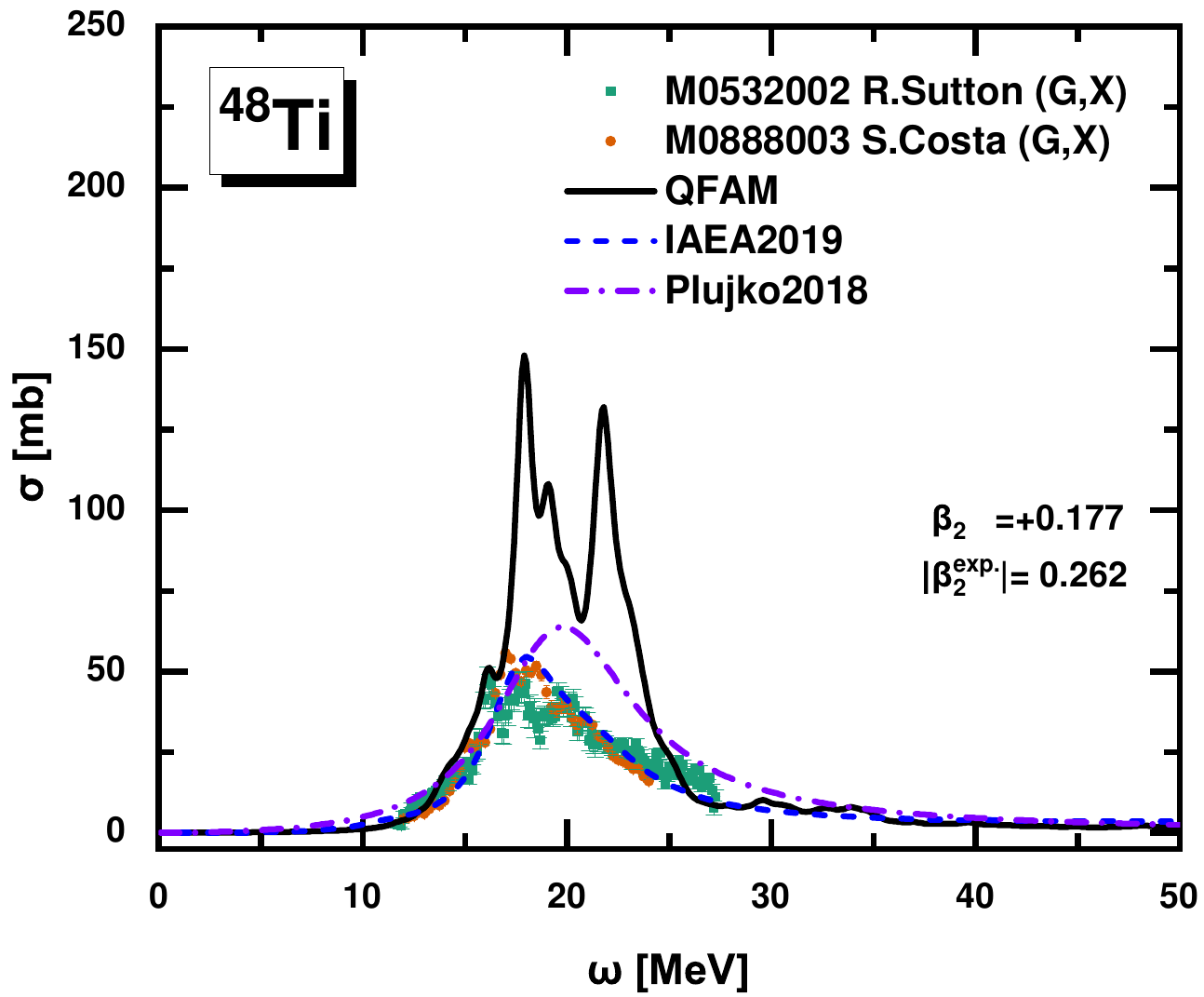}
    \includegraphics[width=0.35\textwidth]{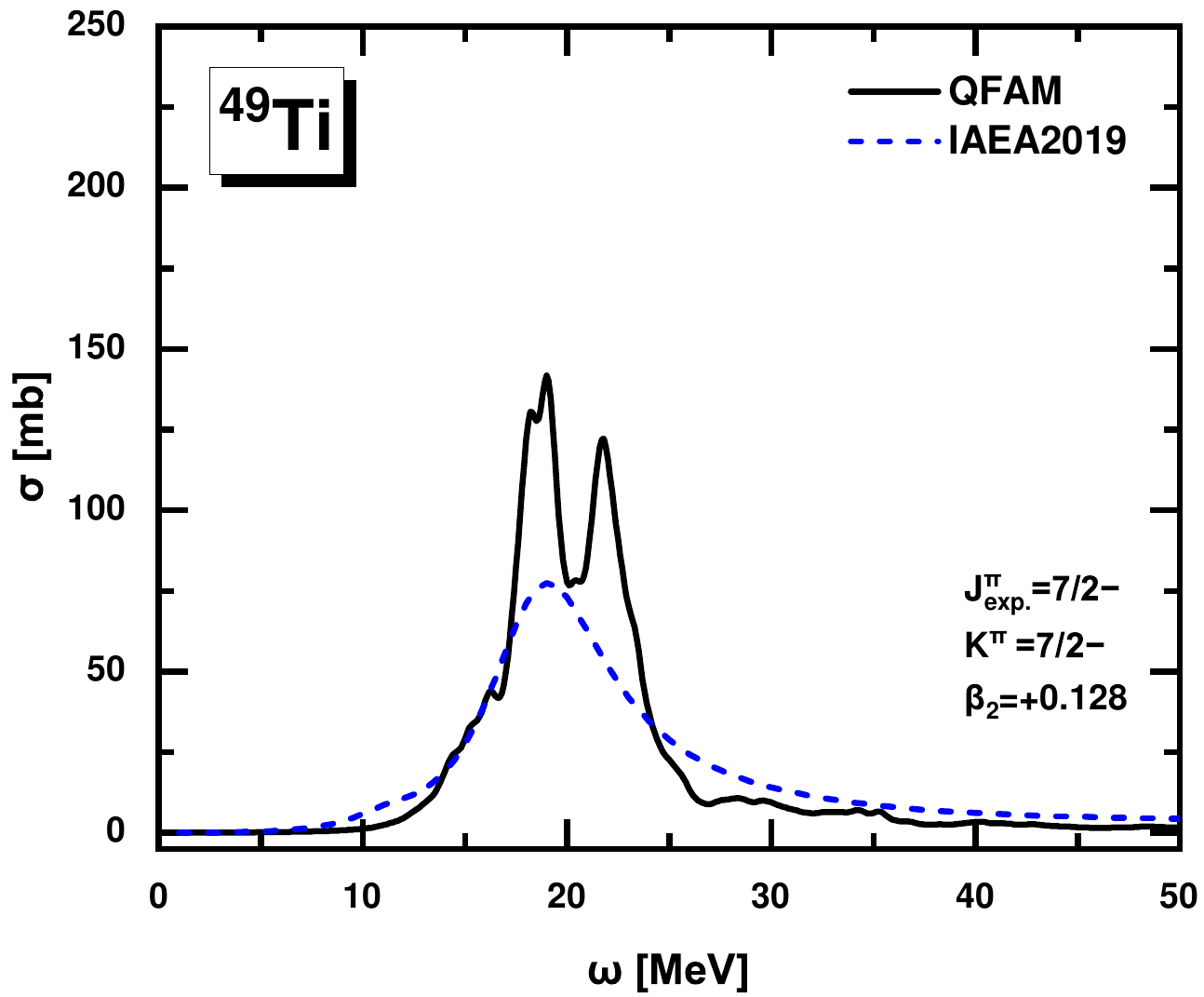}
    \includegraphics[width=0.35\textwidth]{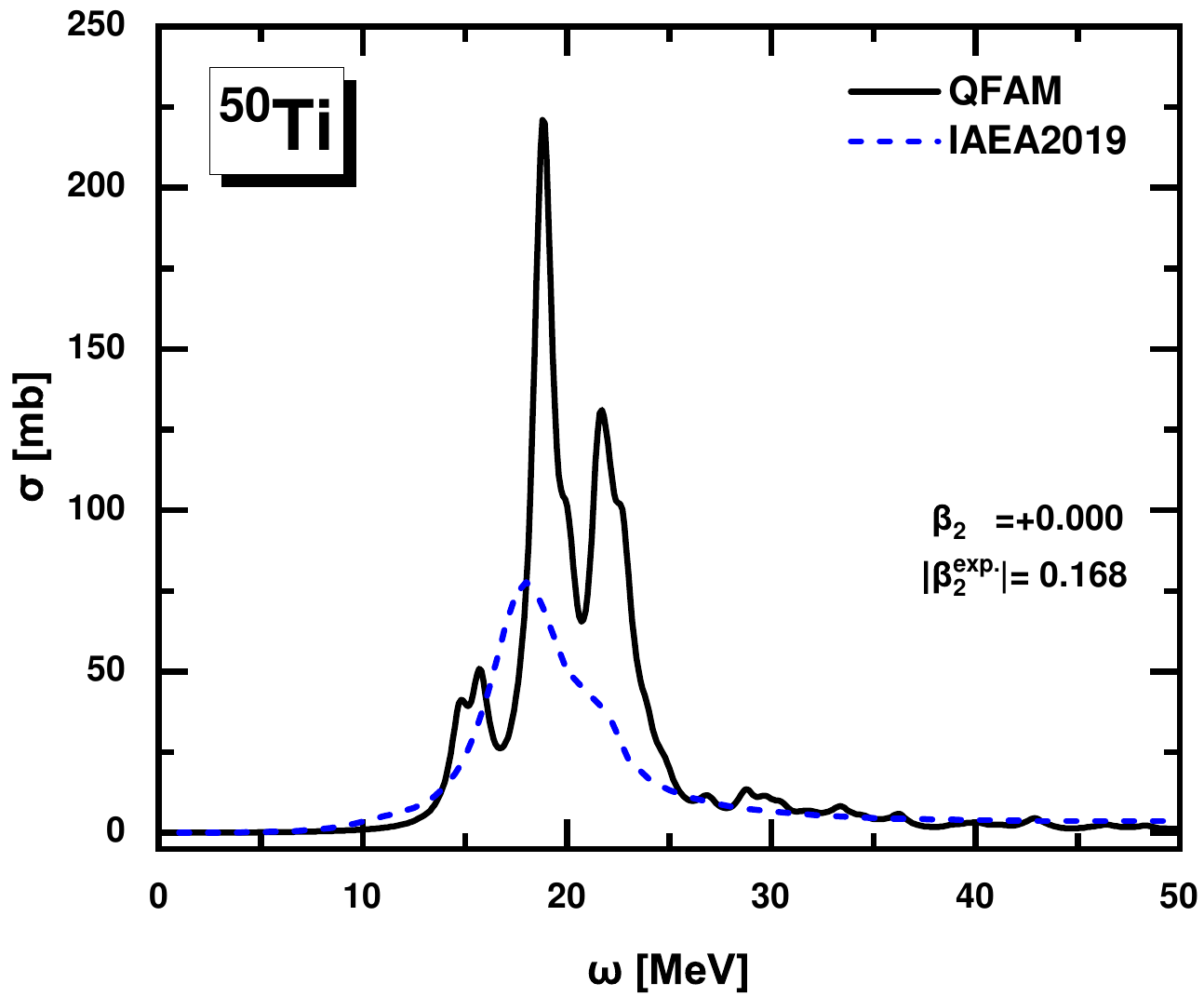}
    \includegraphics[width=0.35\textwidth]{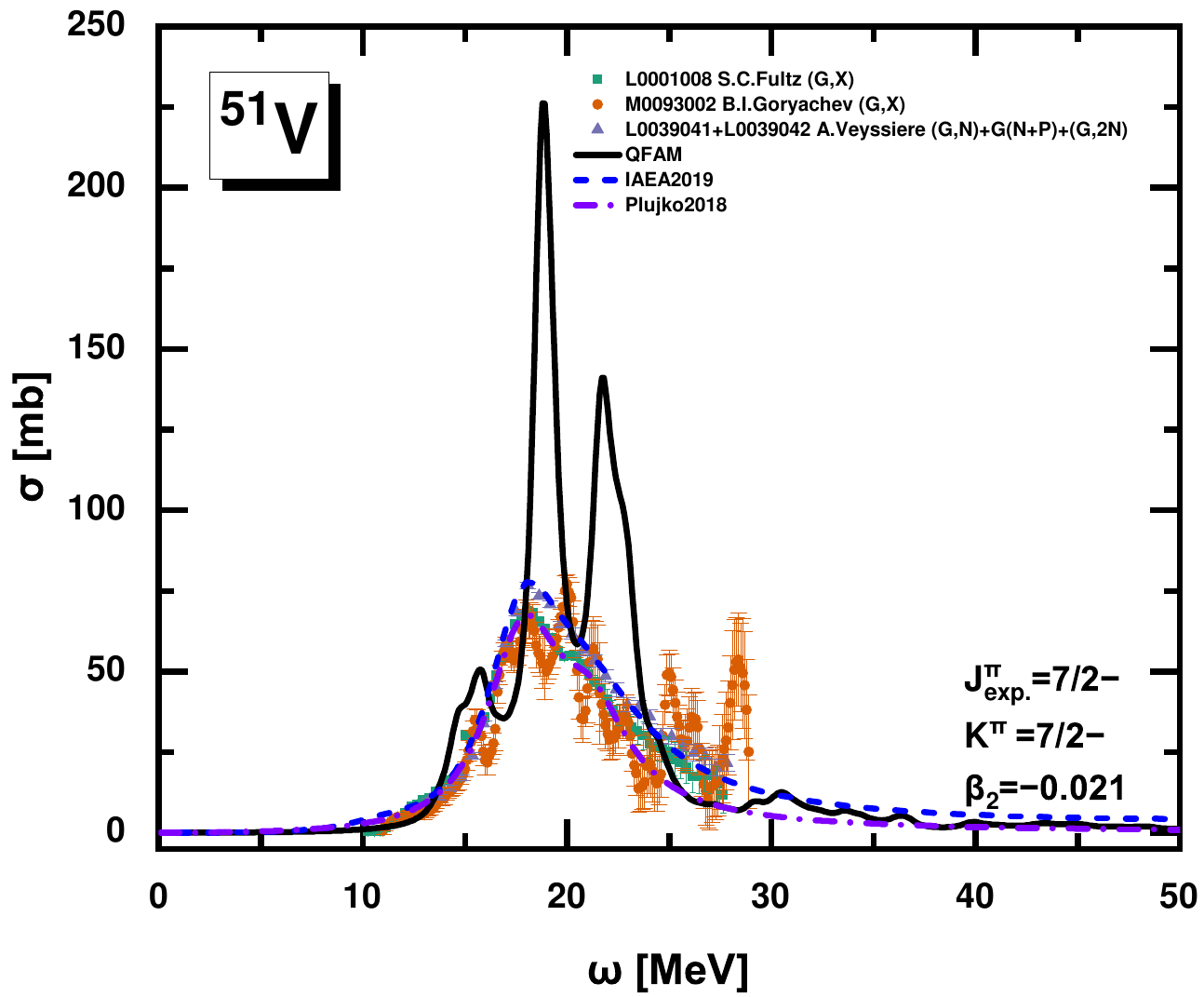}
    \includegraphics[width=0.35\textwidth]{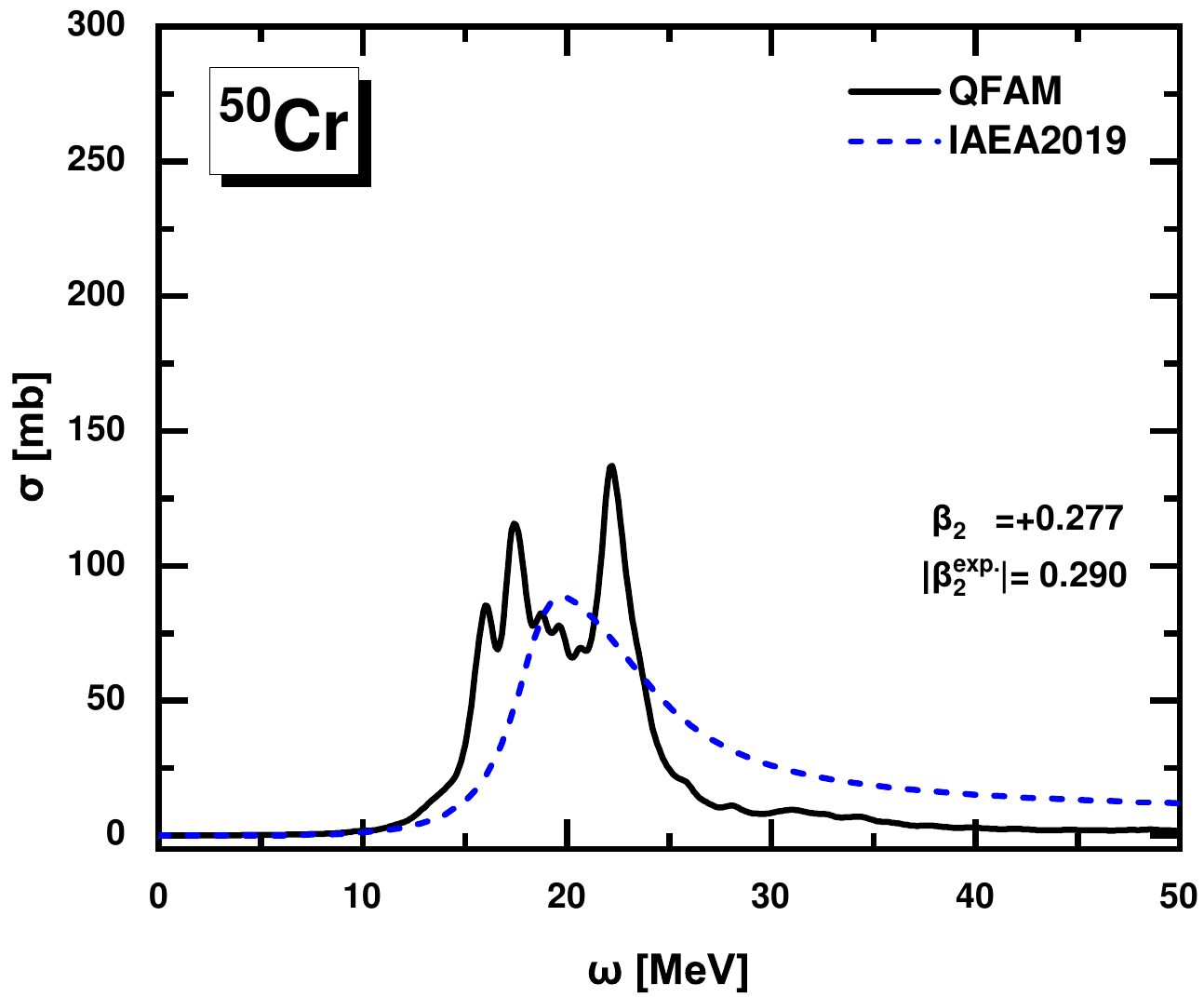}
    \includegraphics[width=0.35\textwidth]{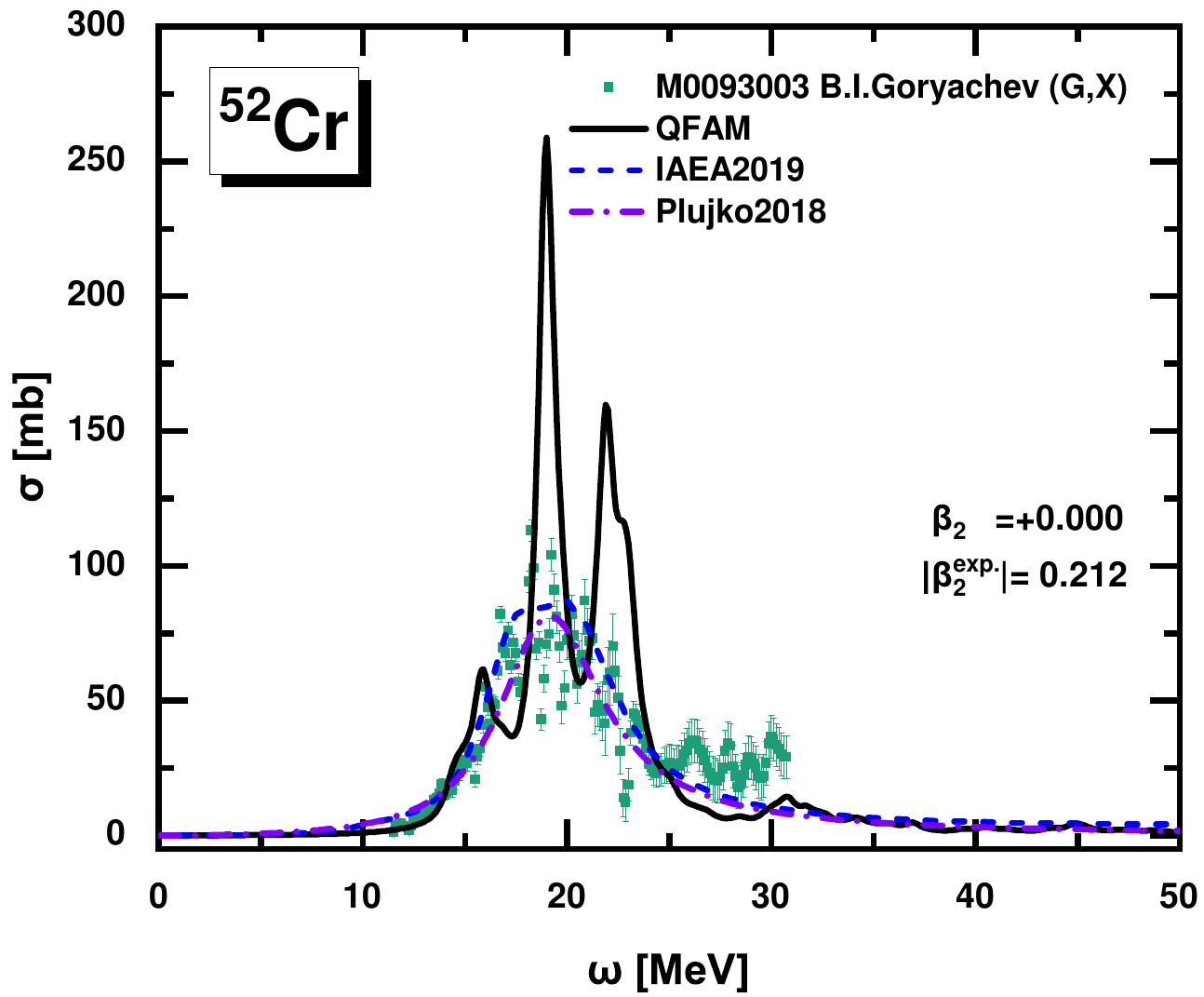}
    \includegraphics[width=0.35\textwidth]{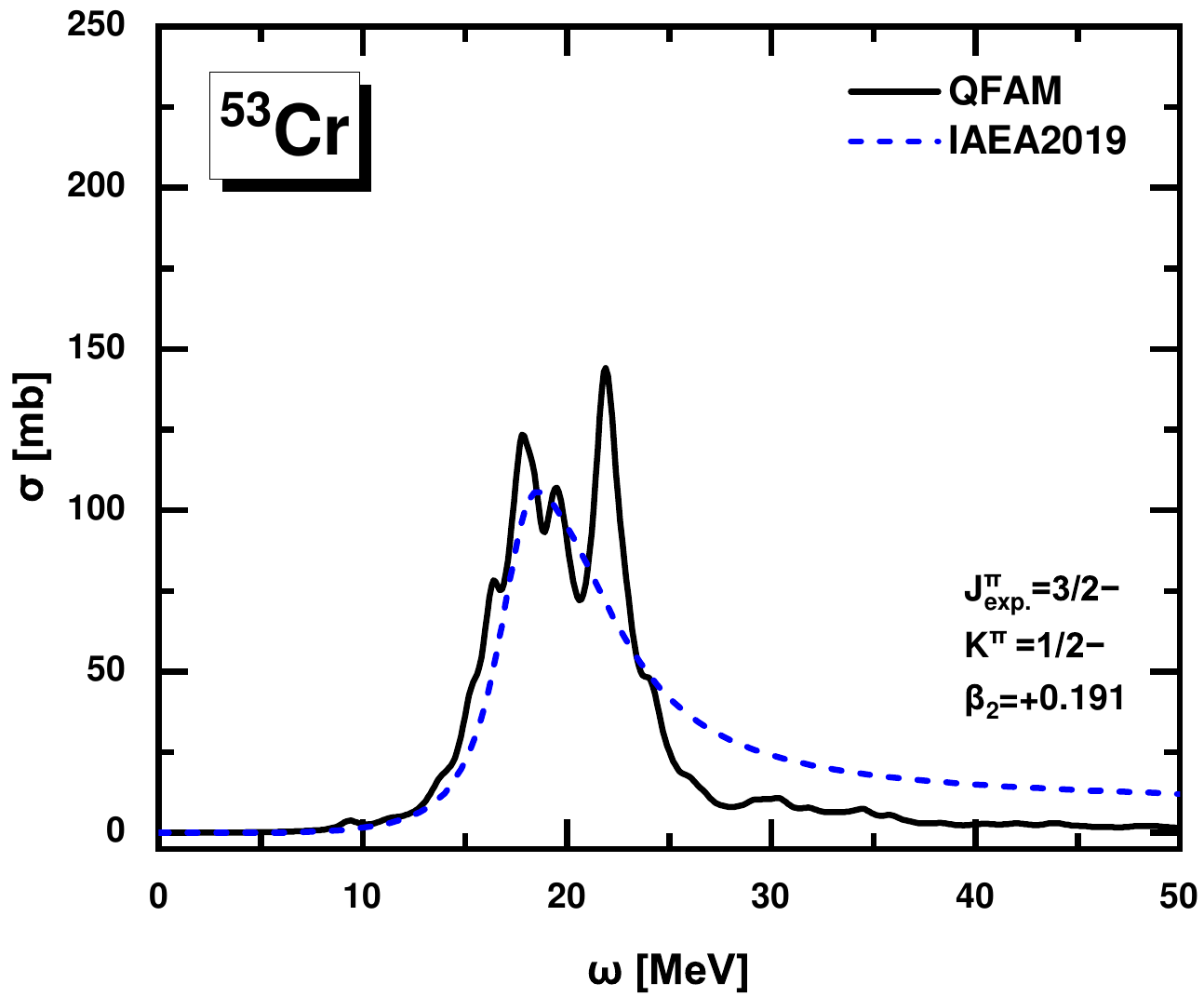}
    \includegraphics[width=0.35\textwidth]{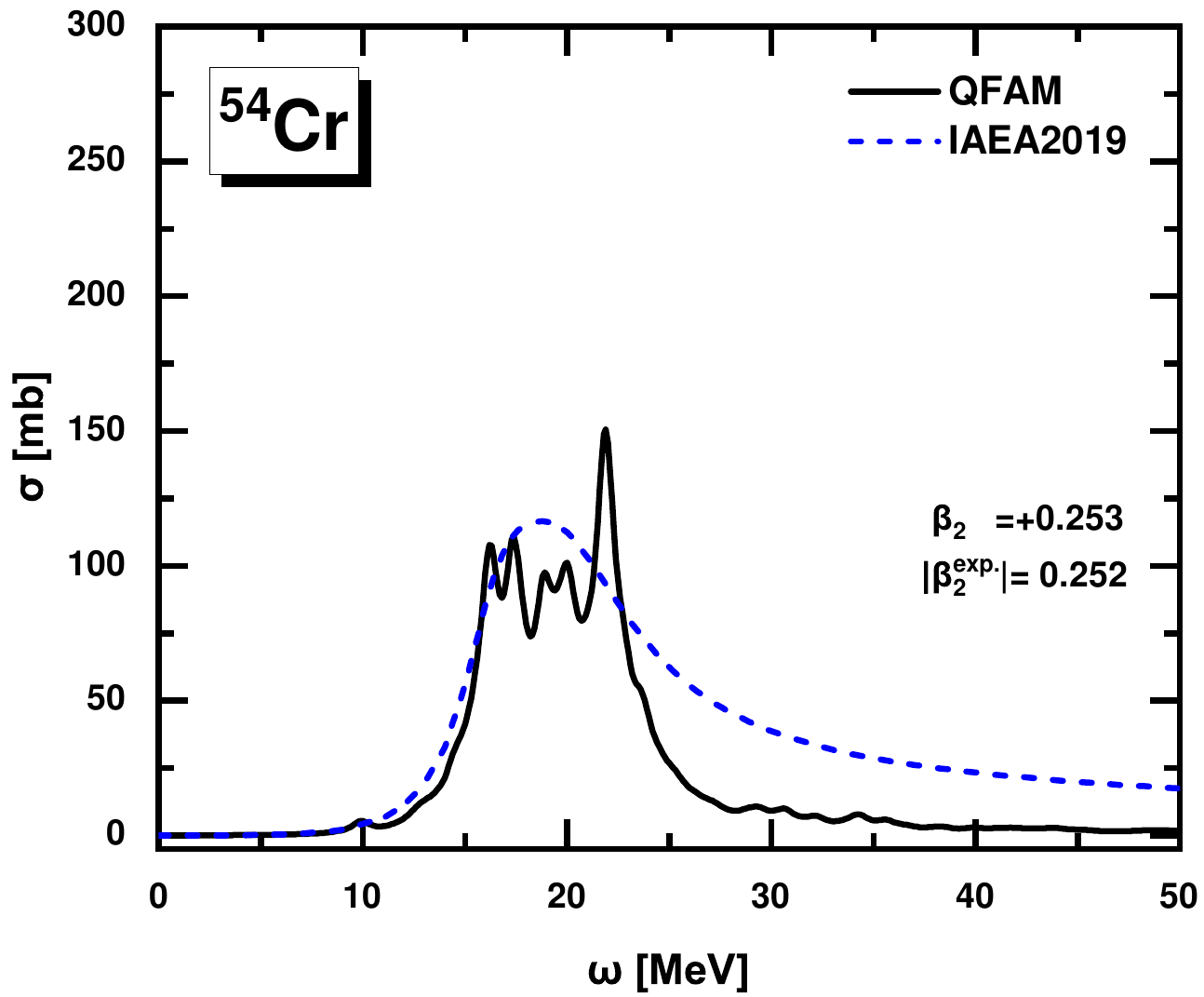}
\end{figure*}
\begin{figure*}\ContinuedFloat
    \centering
    \includegraphics[width=0.35\textwidth]{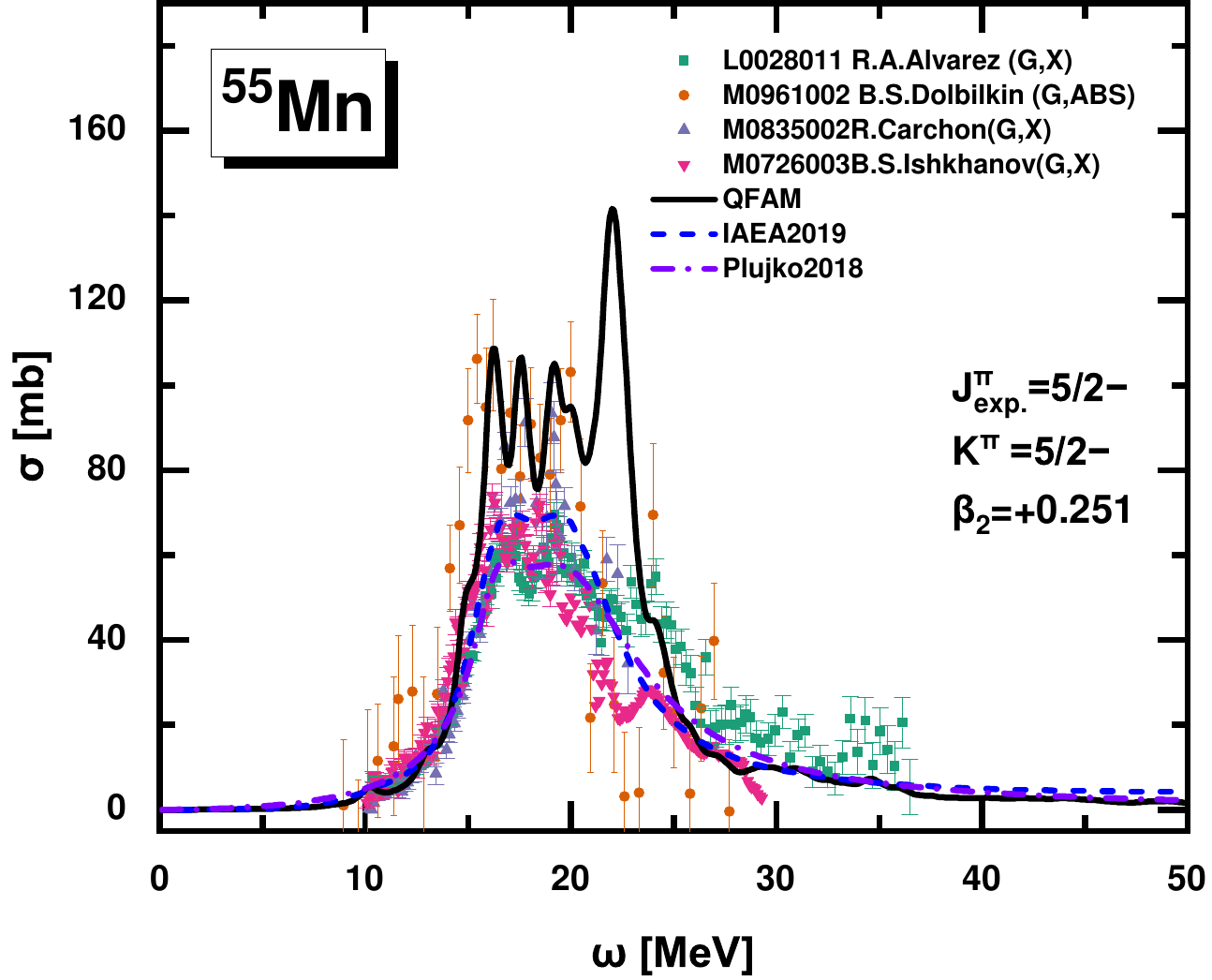}
    \includegraphics[width=0.35\textwidth]{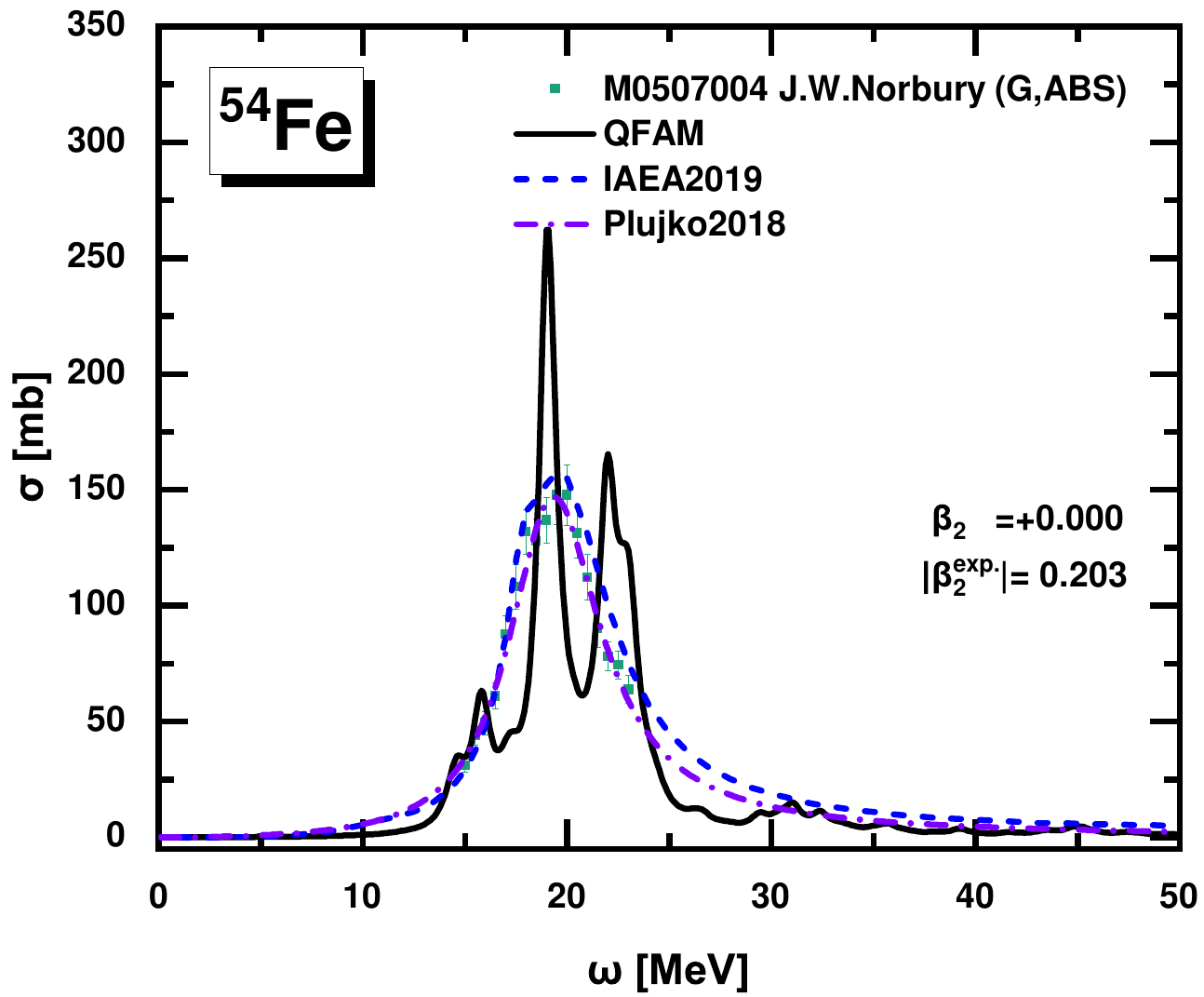}
    \includegraphics[width=0.35\textwidth]{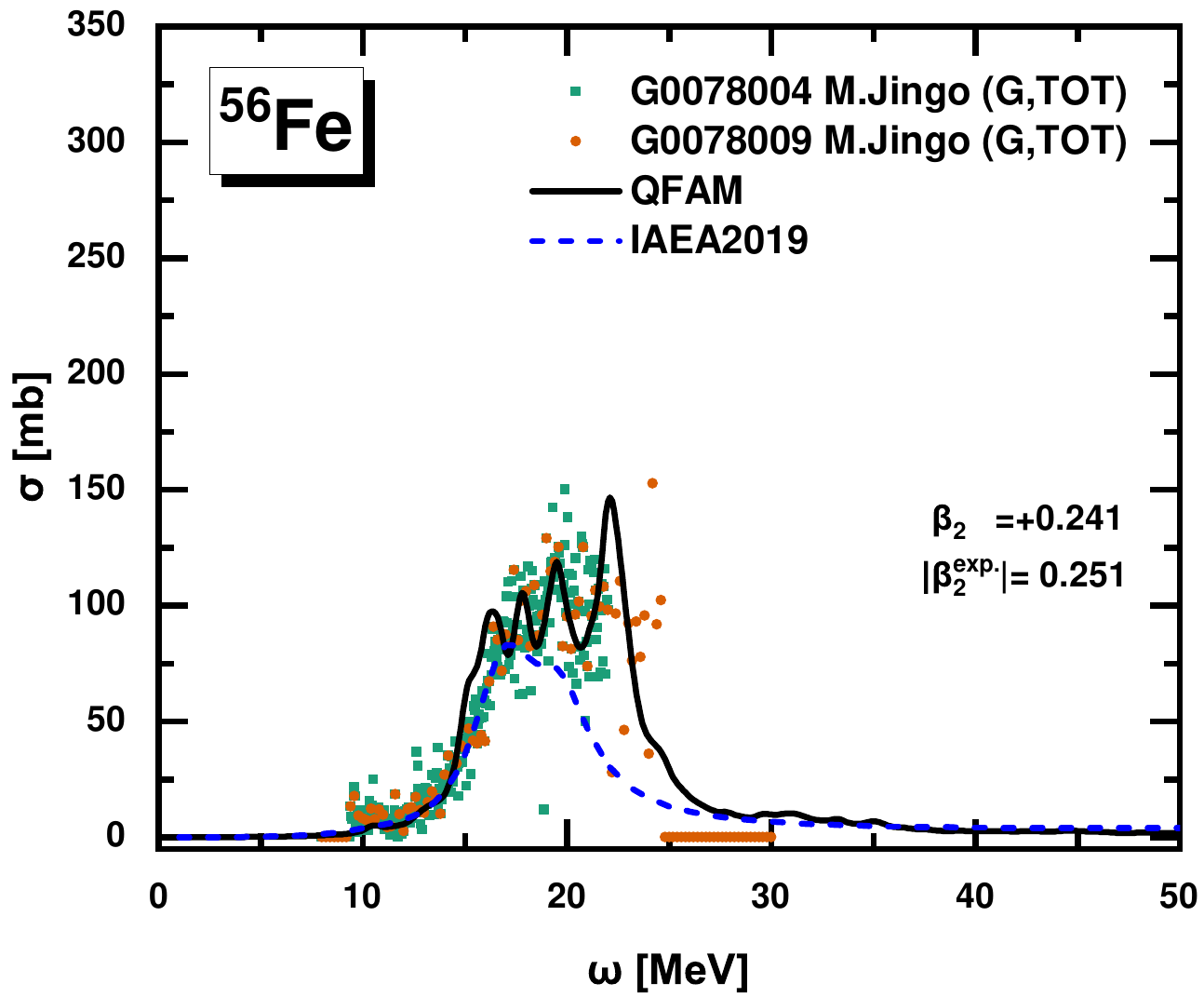}
    \includegraphics[width=0.35\textwidth]{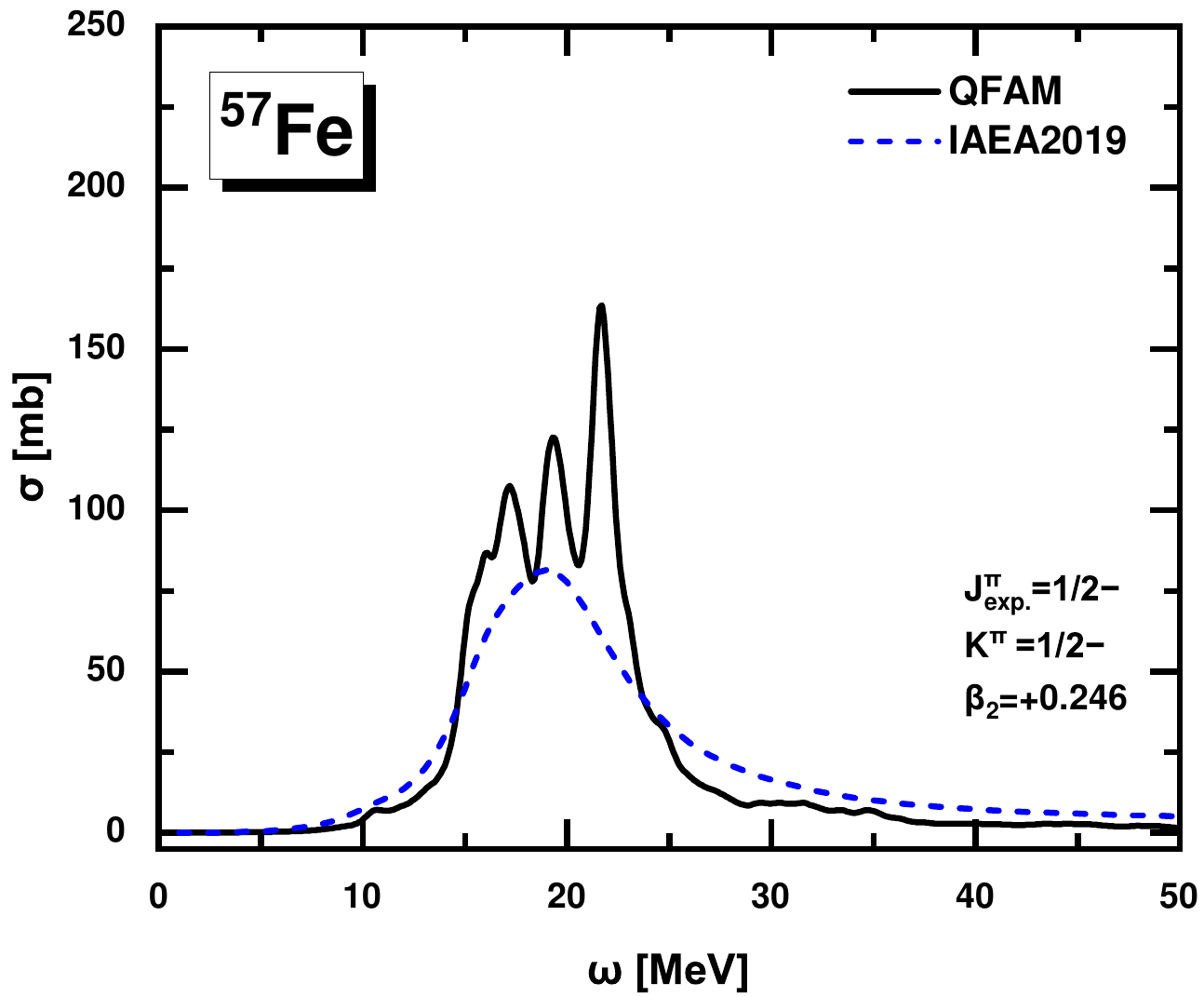}
    \includegraphics[width=0.35\textwidth]{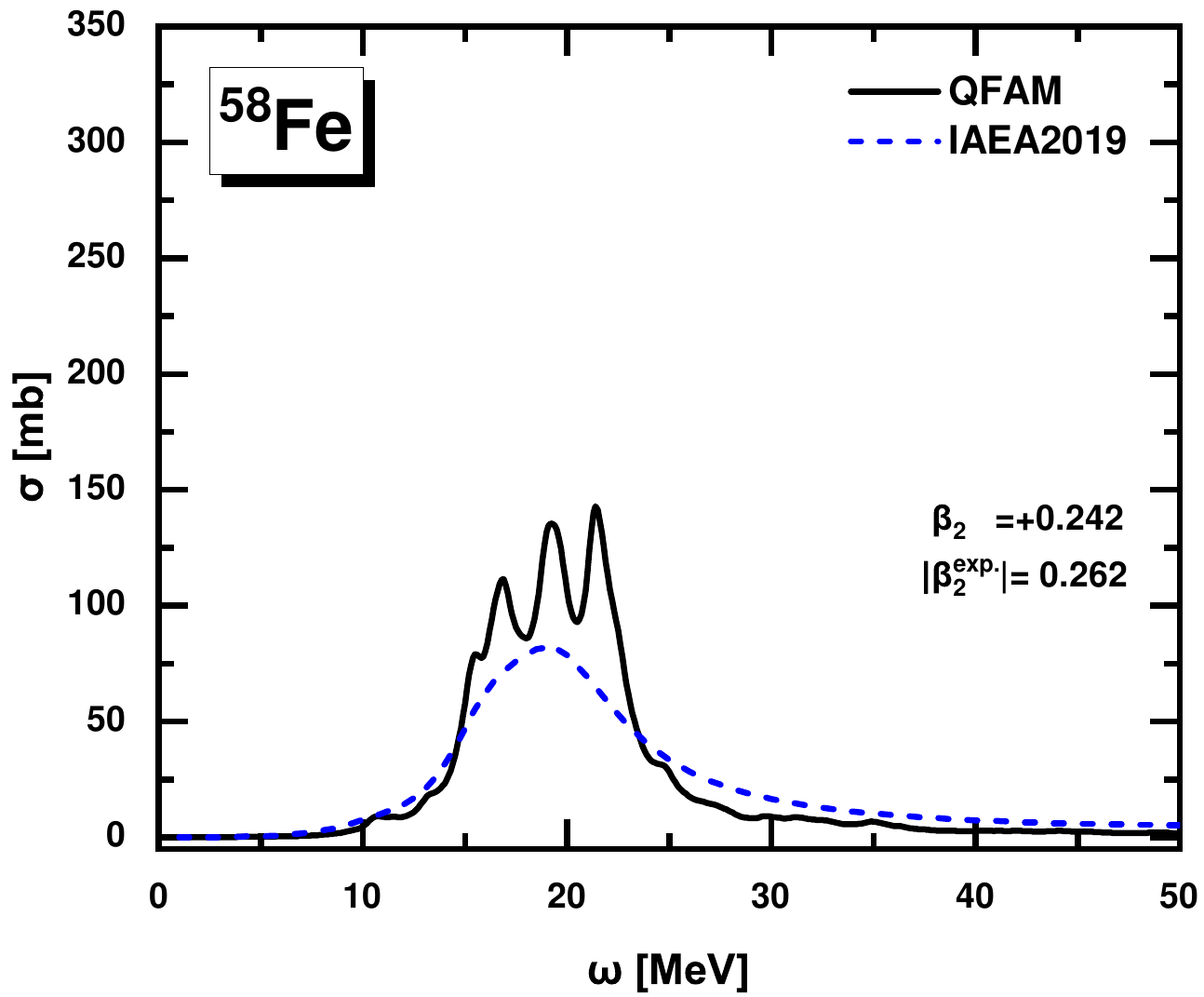}
    \includegraphics[width=0.35\textwidth]{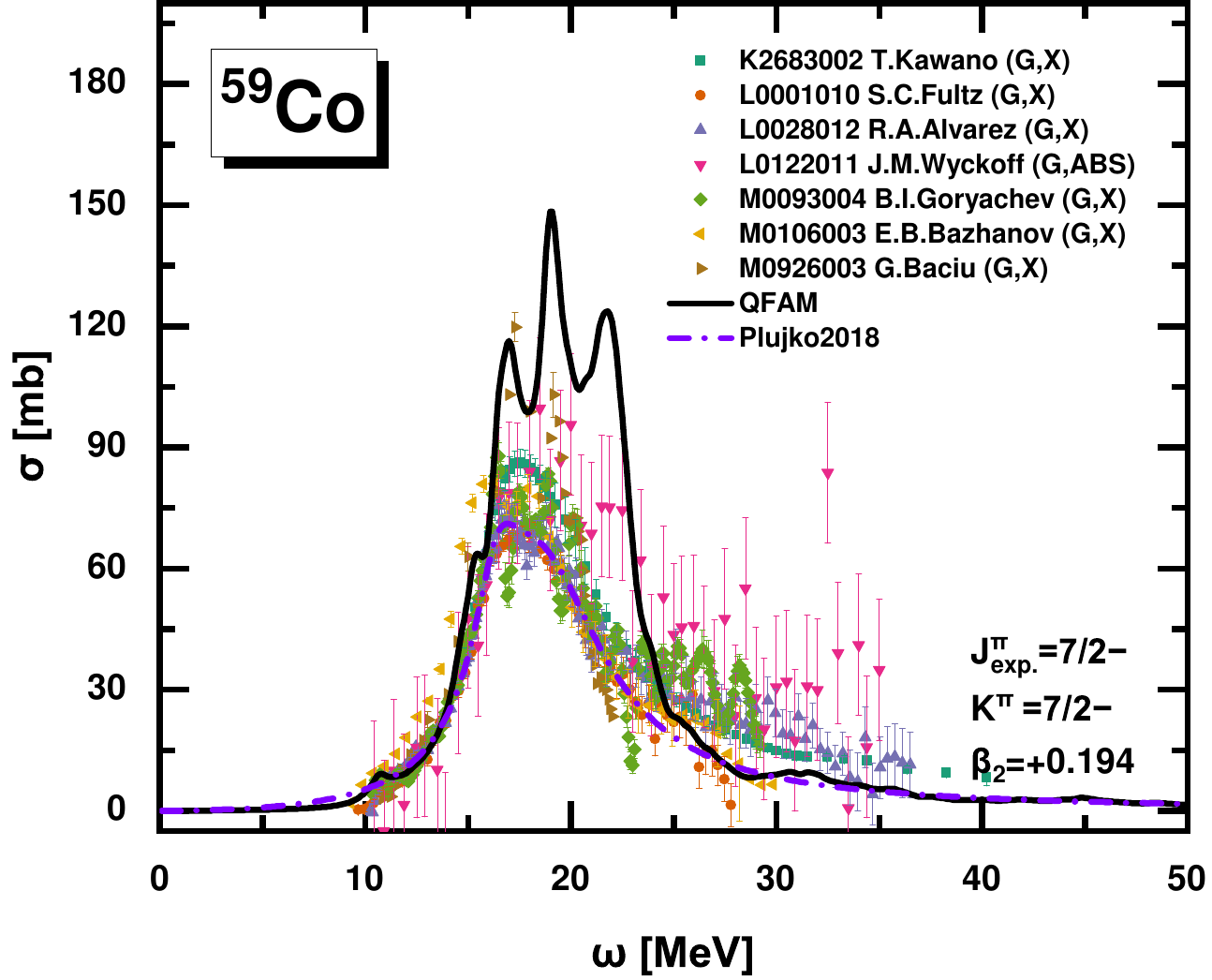}
    \includegraphics[width=0.35\textwidth]{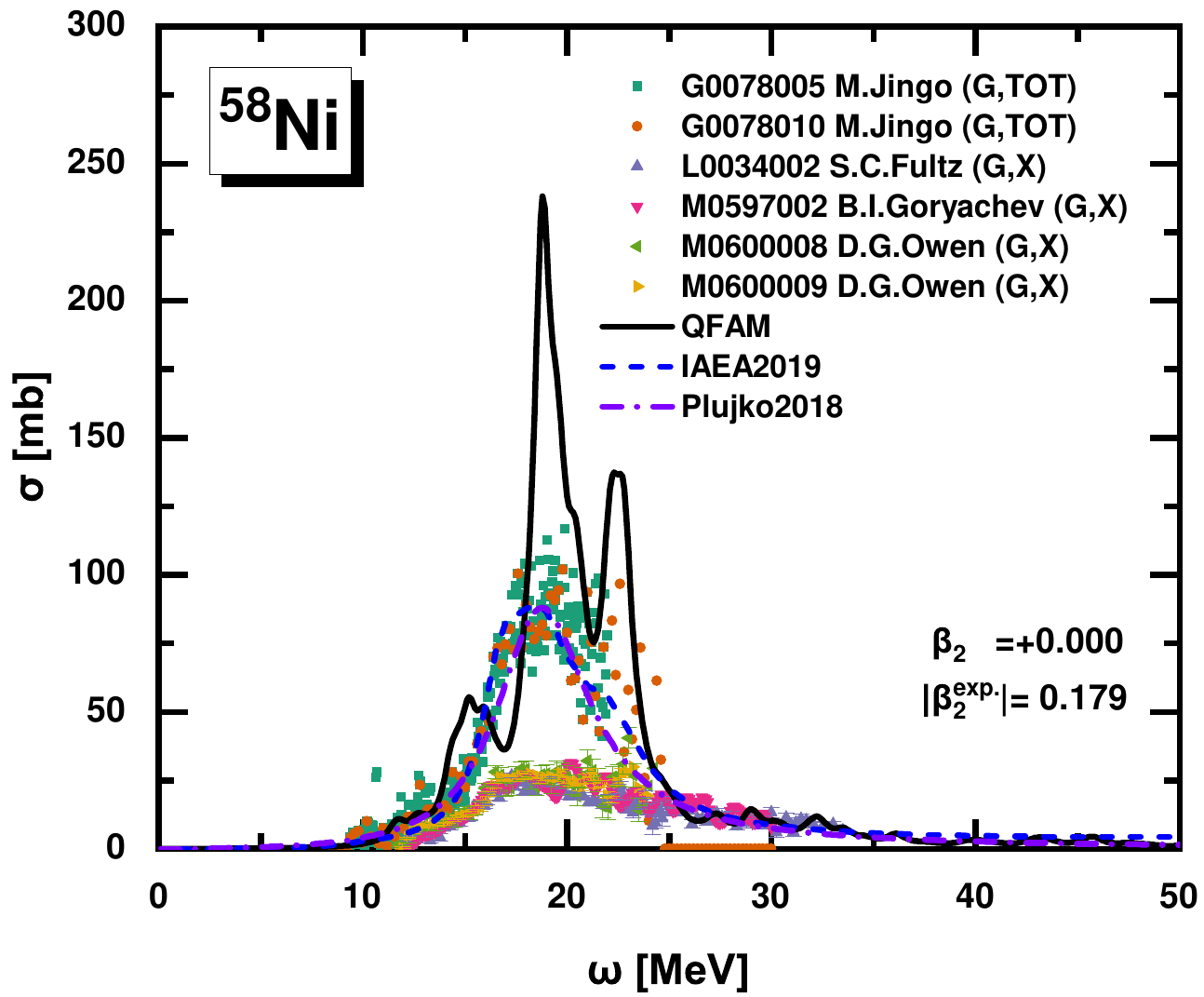}
    \includegraphics[width=0.35\textwidth]{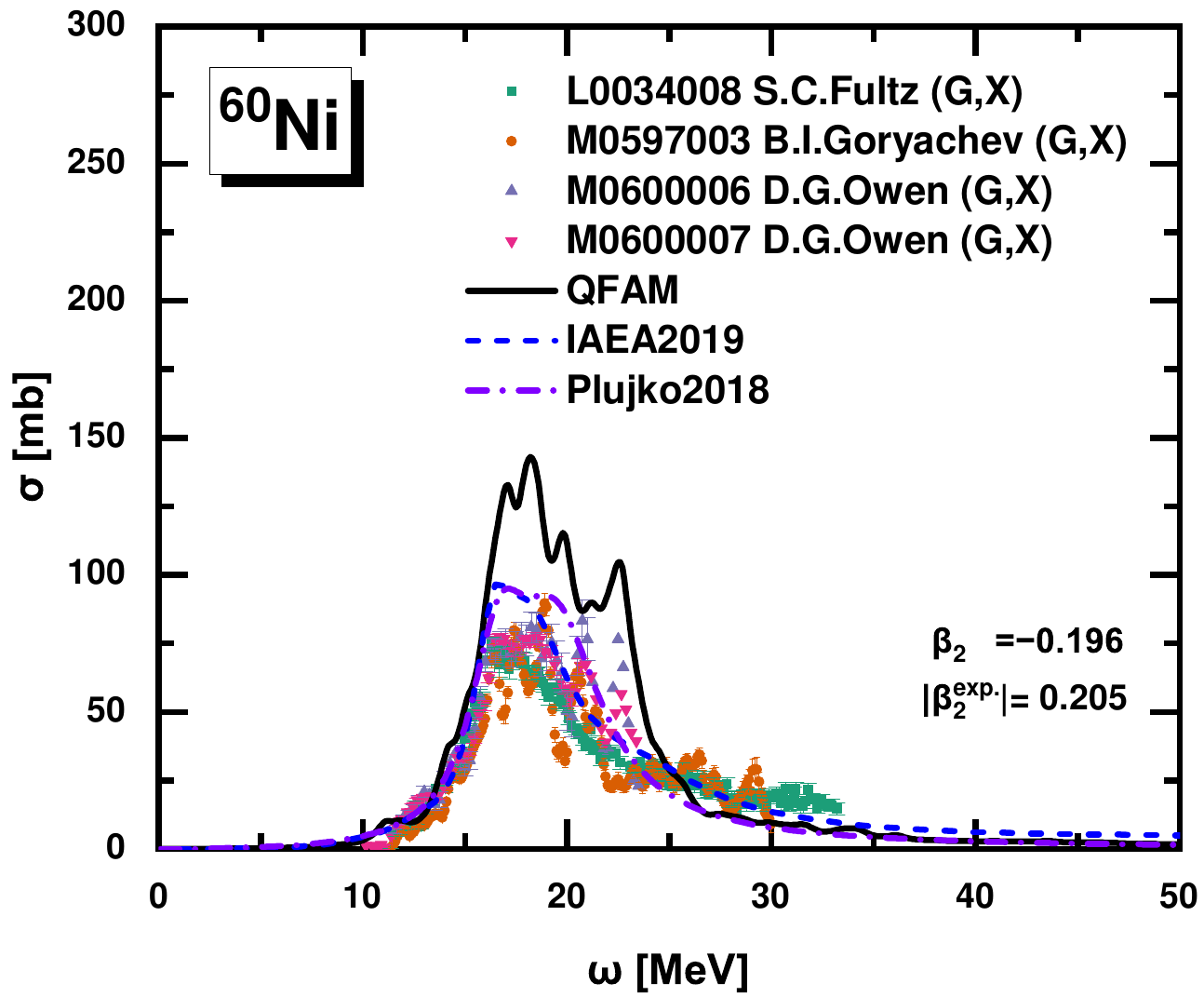}
\end{figure*}
\begin{figure*}\ContinuedFloat
    \centering
    \includegraphics[width=0.35\textwidth]{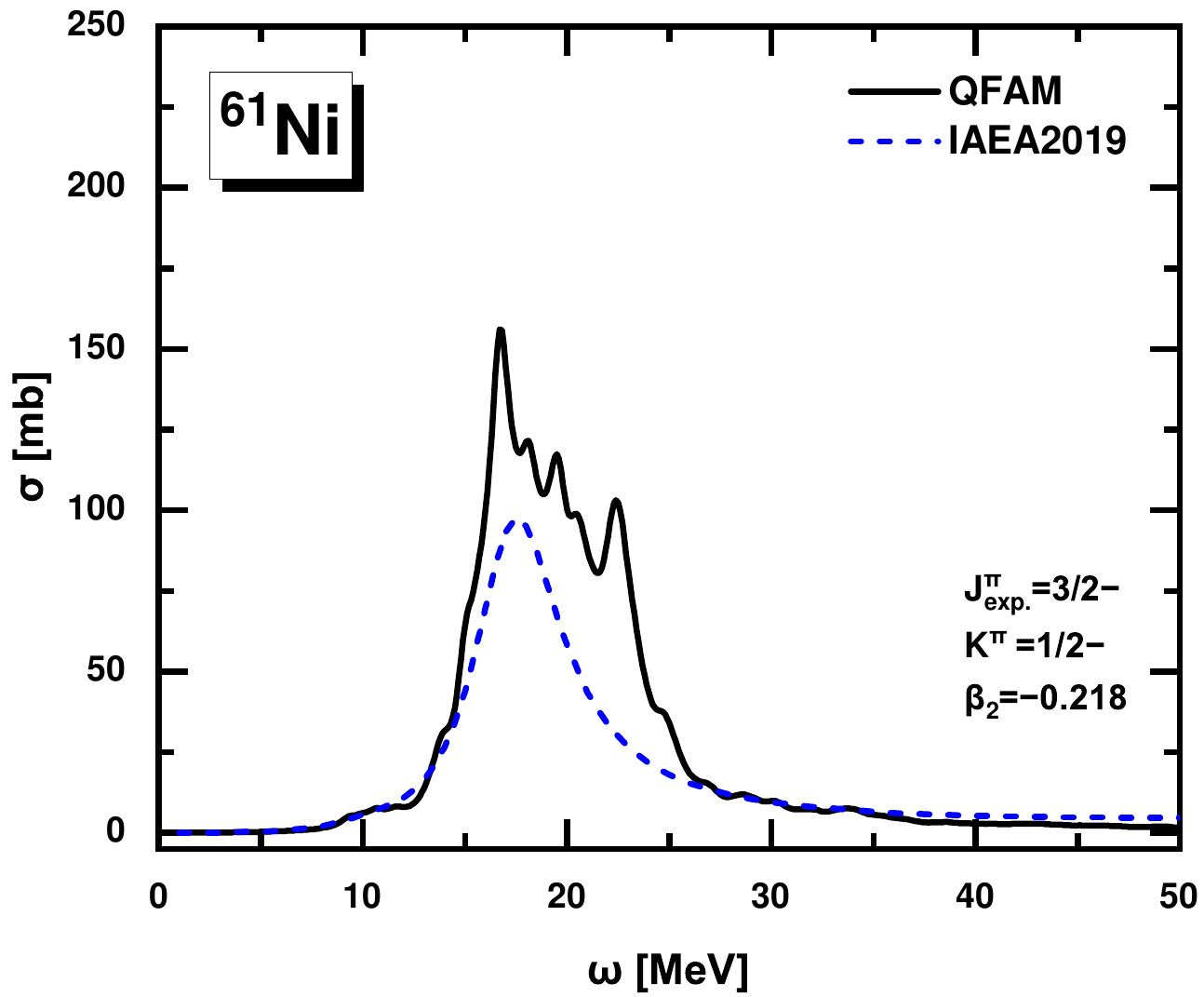}
    \includegraphics[width=0.35\textwidth]{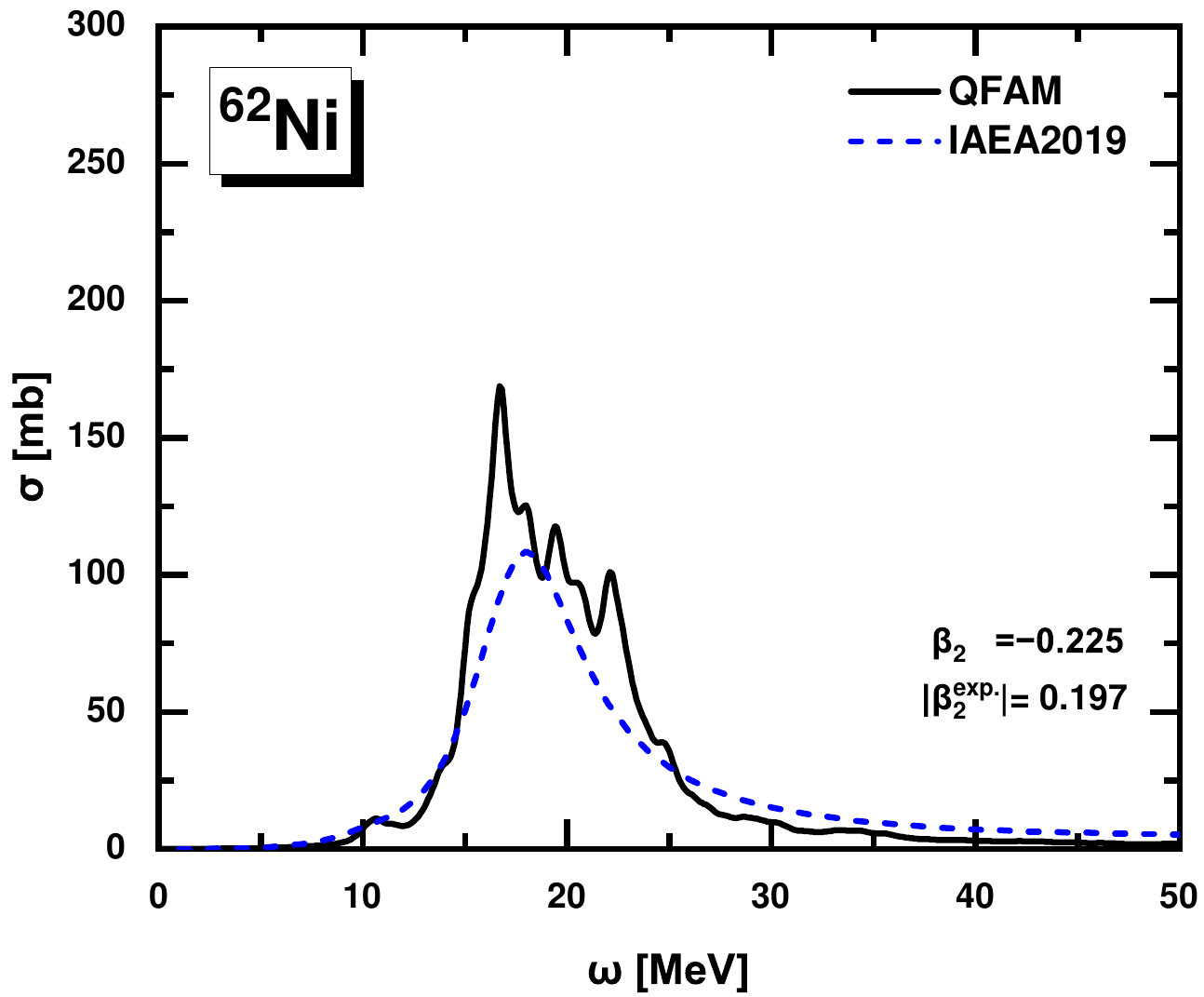}
    \includegraphics[width=0.35\textwidth]{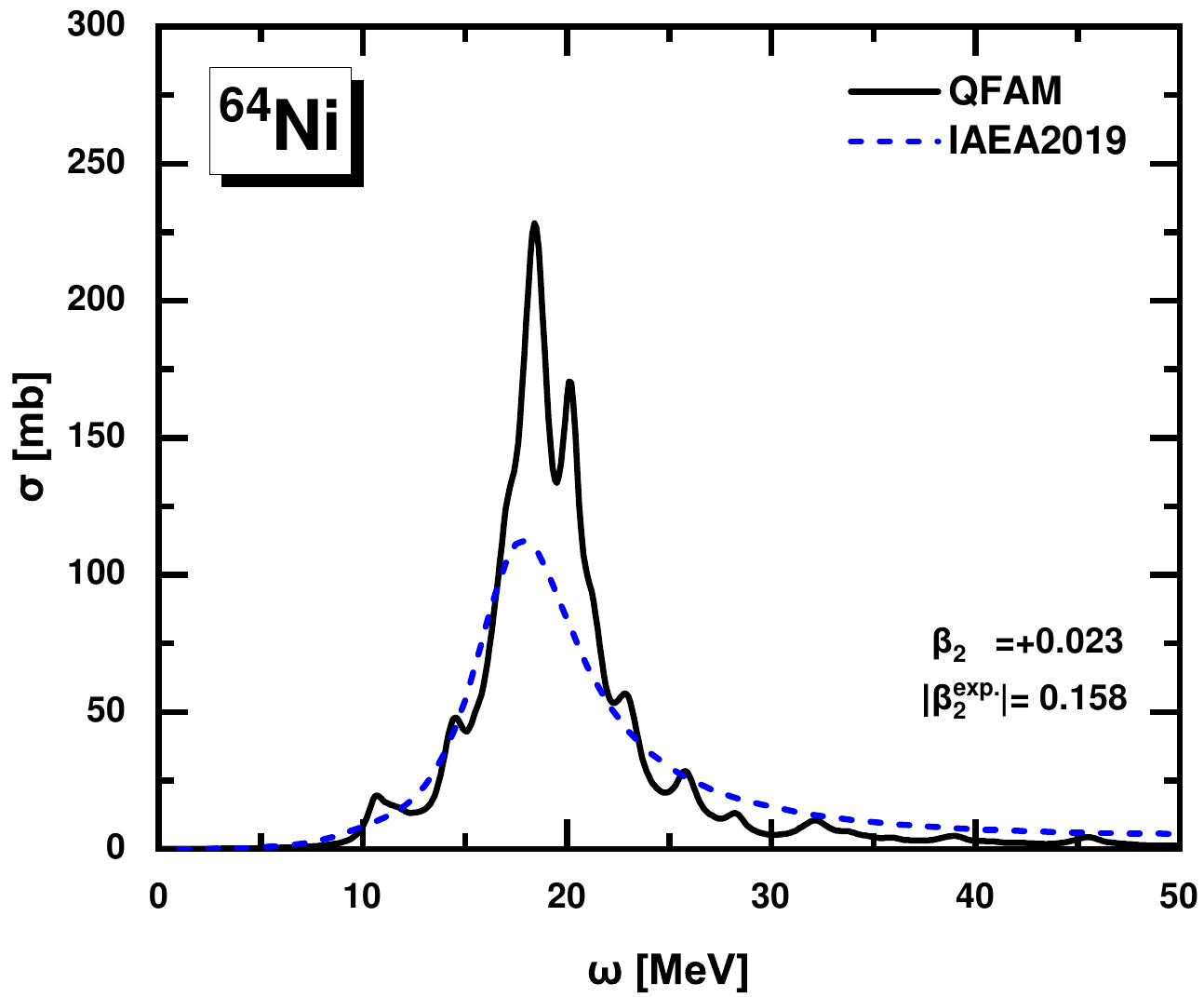}
    \includegraphics[width=0.35\textwidth]{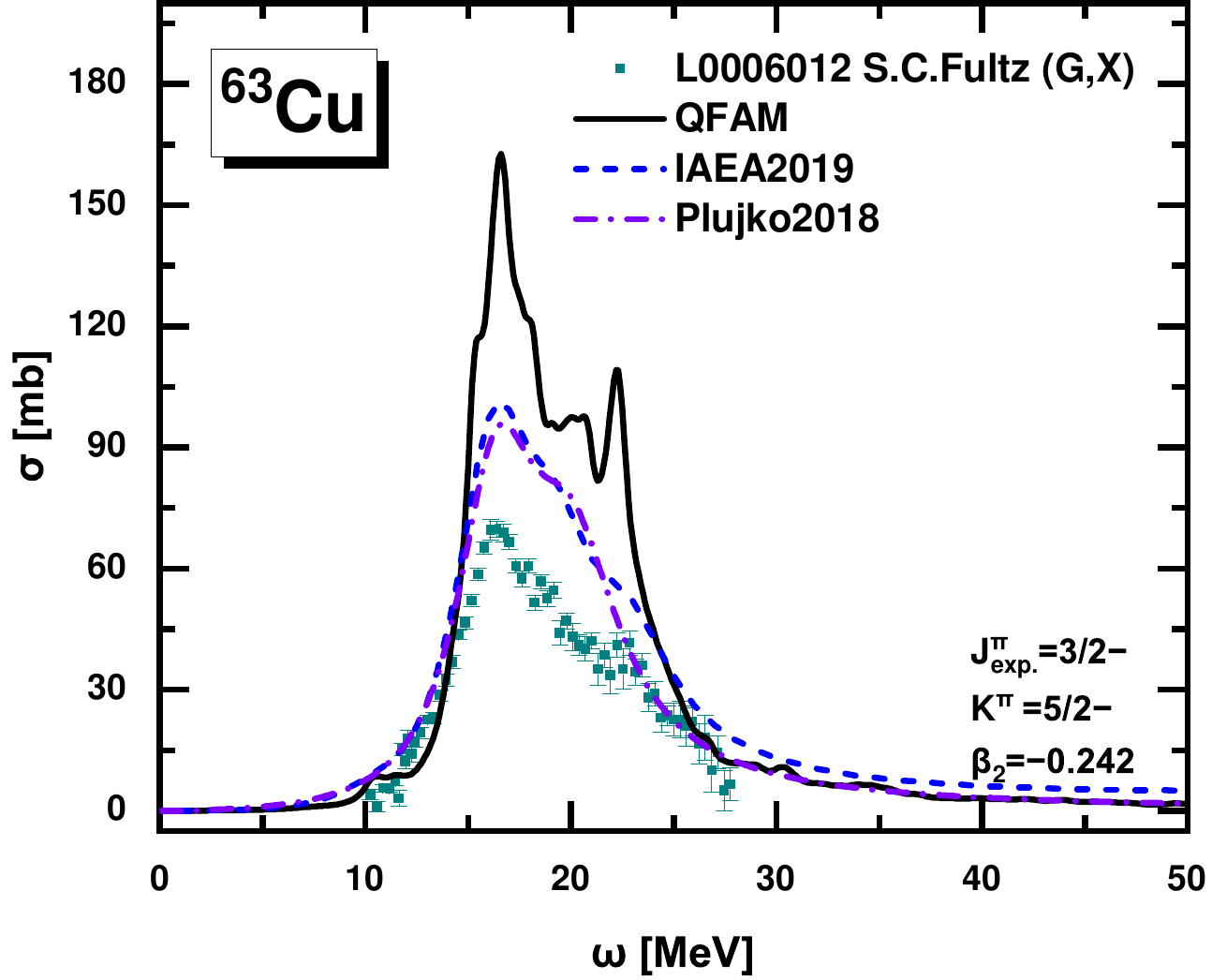}
    \includegraphics[width=0.35\textwidth]{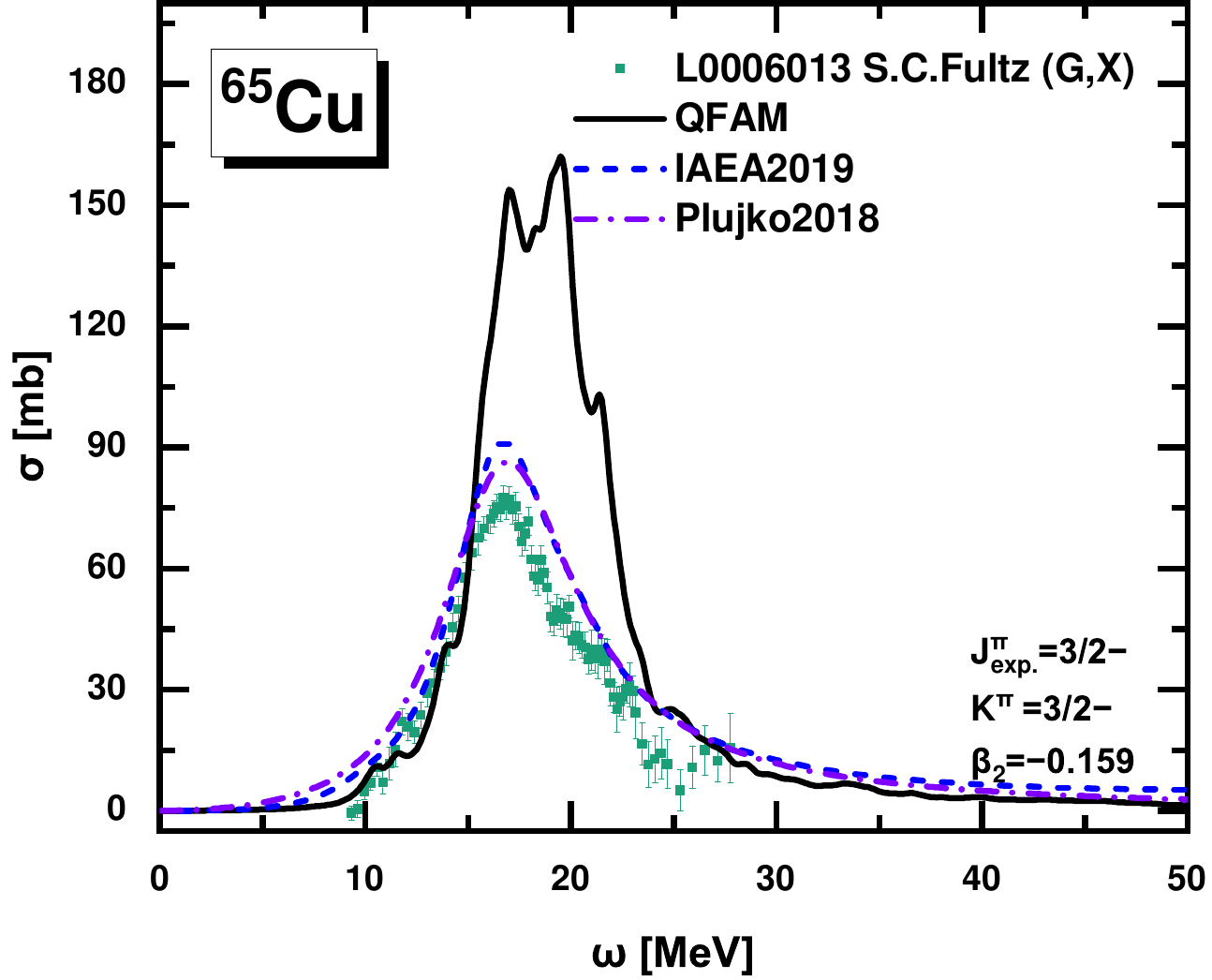}
    \includegraphics[width=0.35\textwidth]{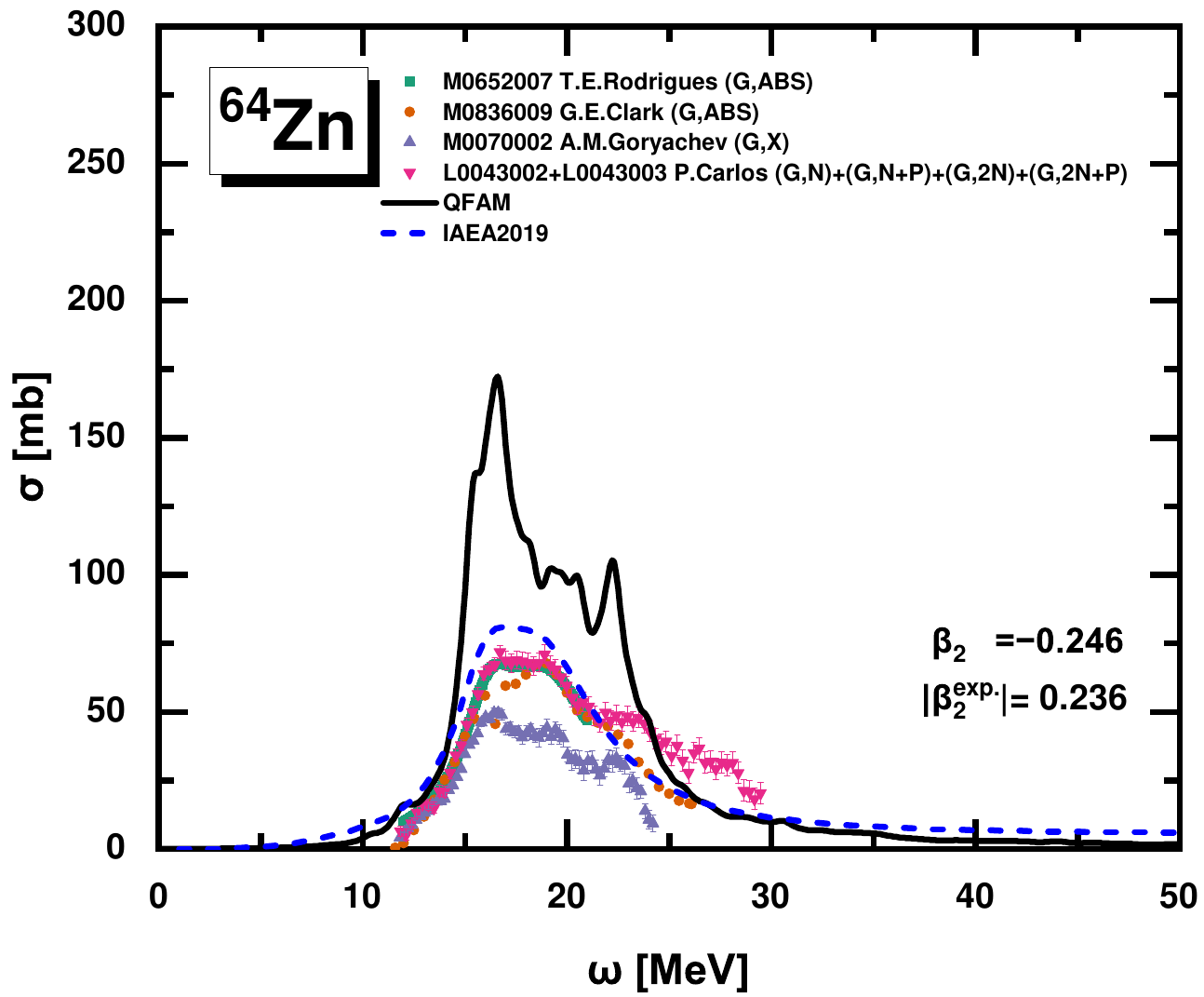}
    \includegraphics[width=0.35\textwidth]{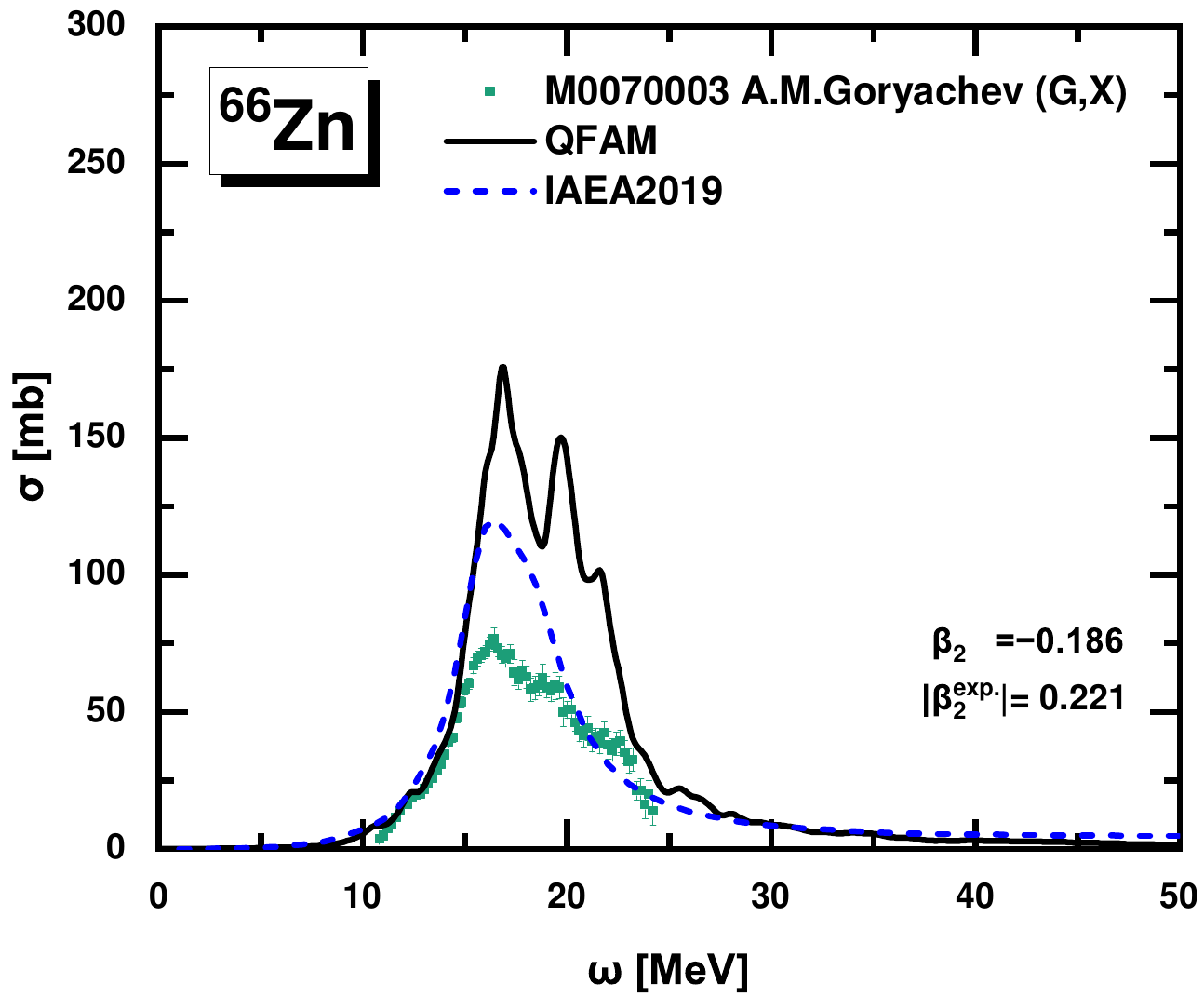}
    \includegraphics[width=0.35\textwidth]{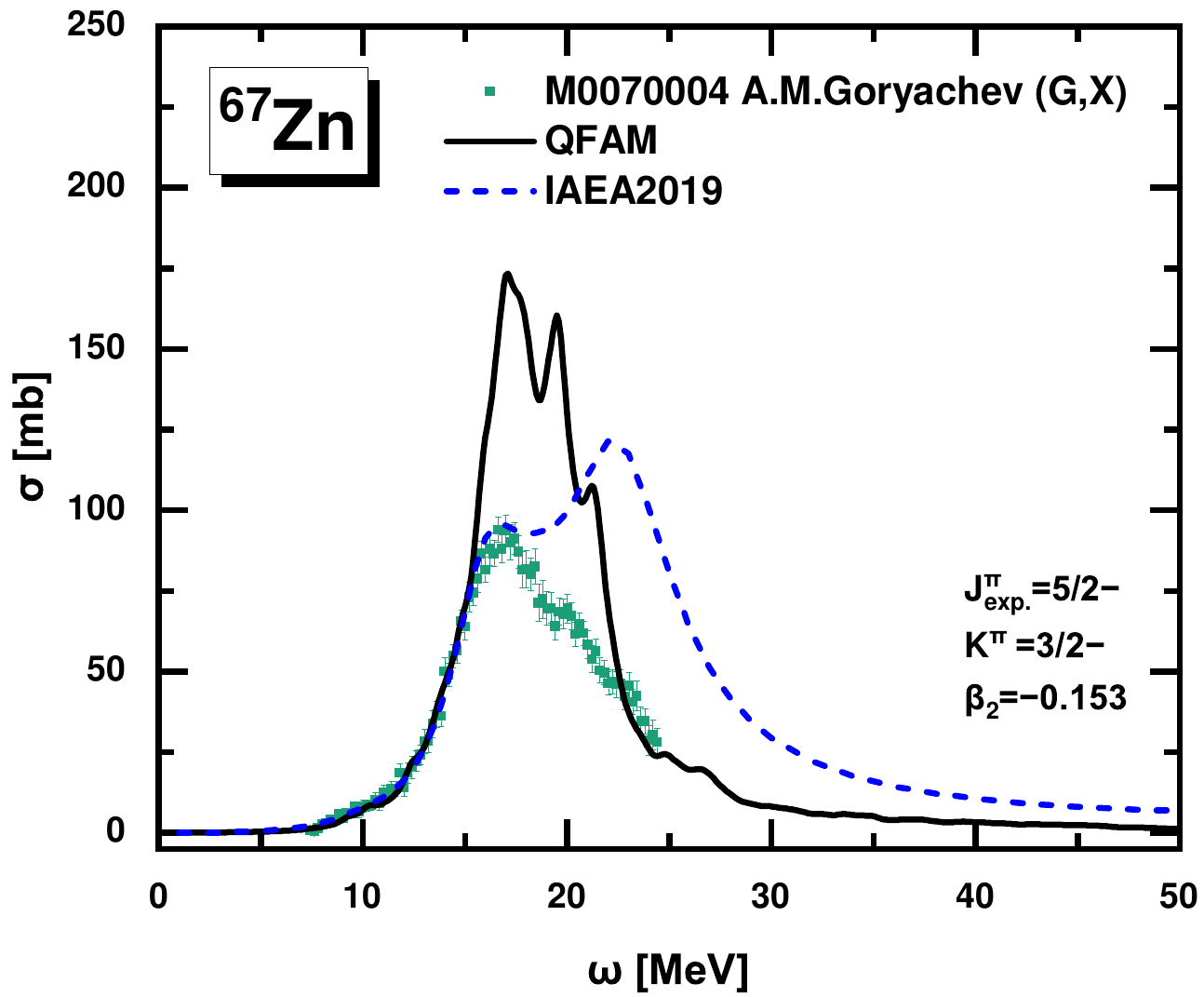}
\end{figure*}
\begin{figure*}\ContinuedFloat
    \centering
    \includegraphics[width=0.35\textwidth]{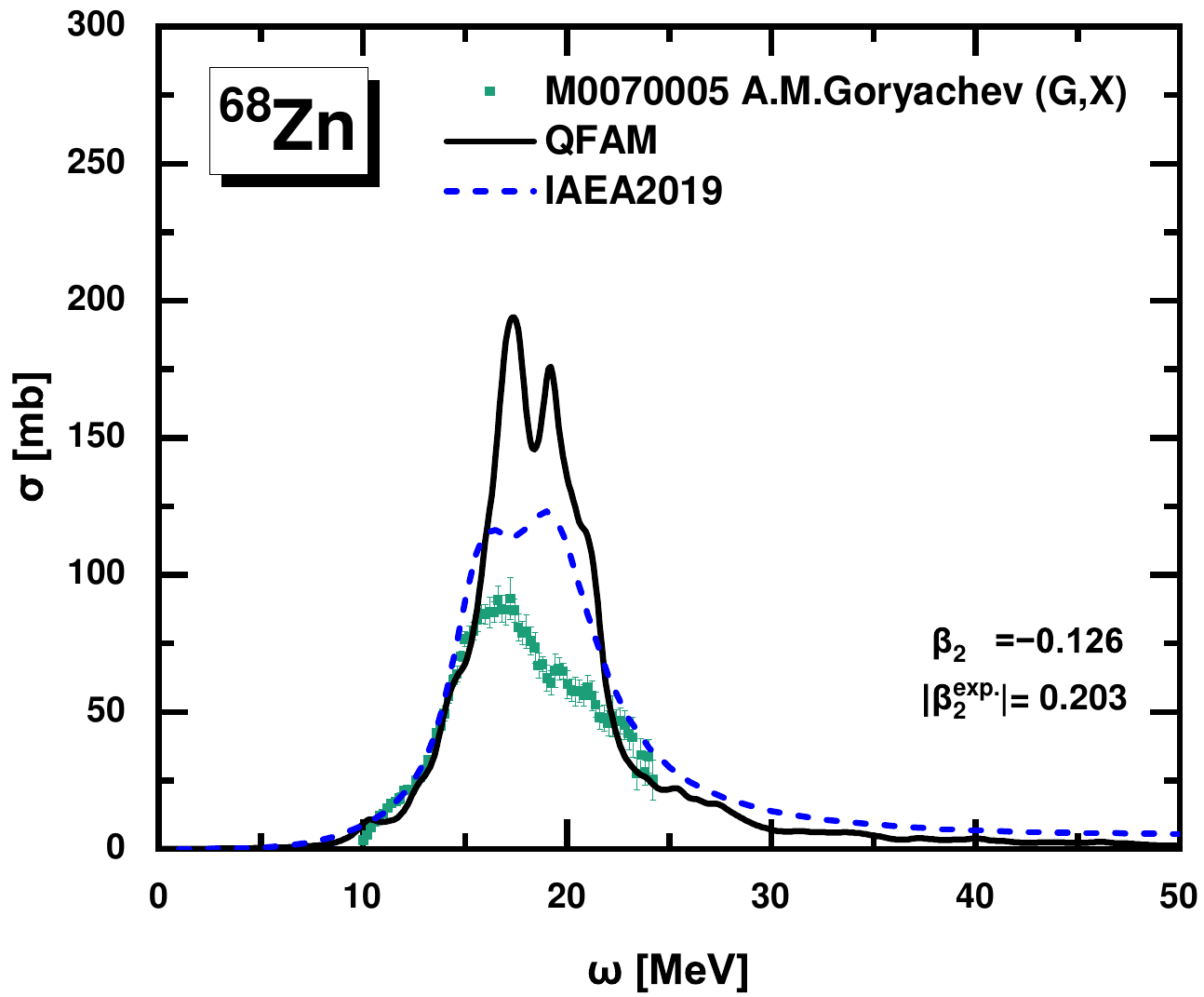}
    \includegraphics[width=0.35\textwidth]{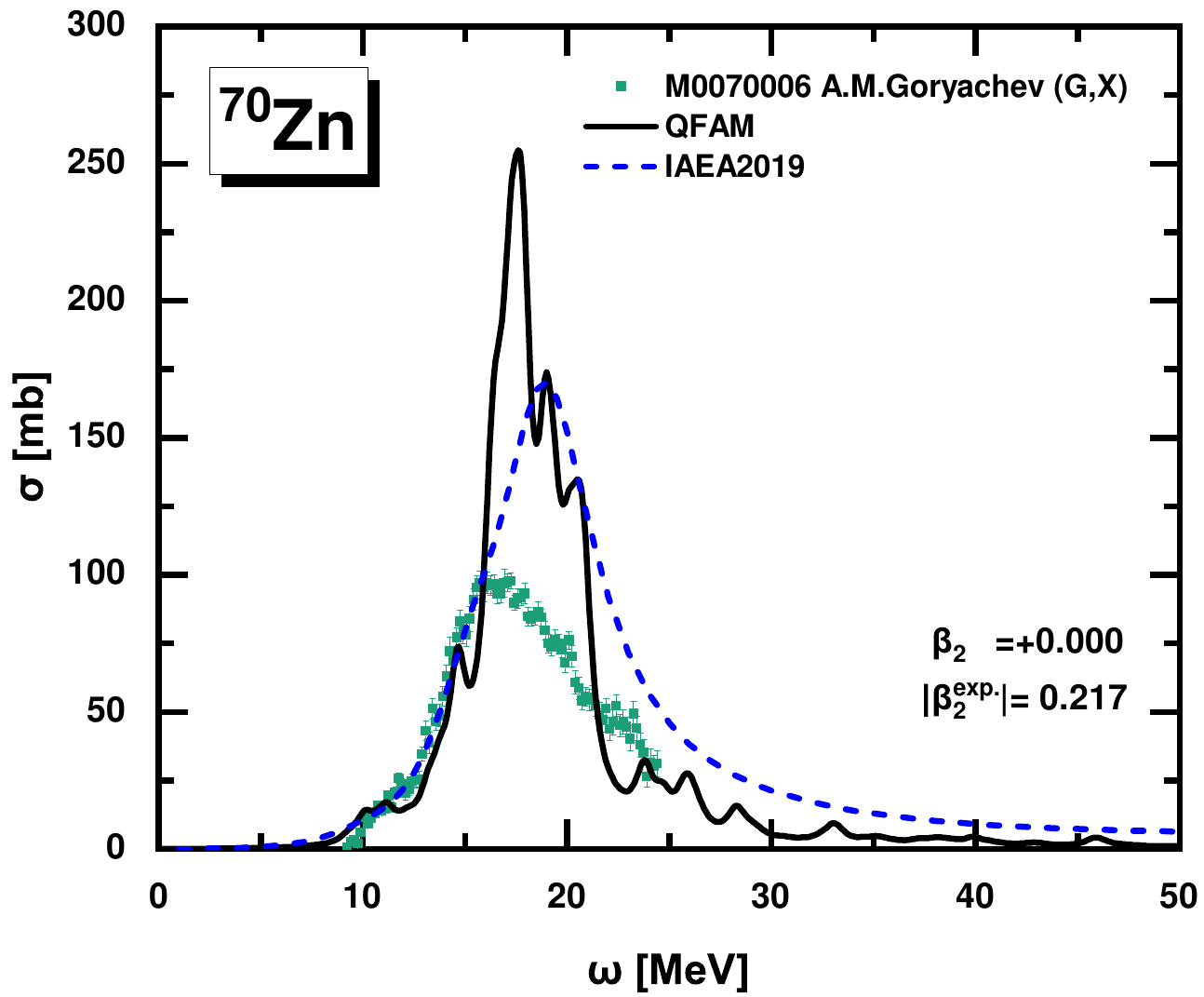}
    \includegraphics[width=0.35\textwidth]{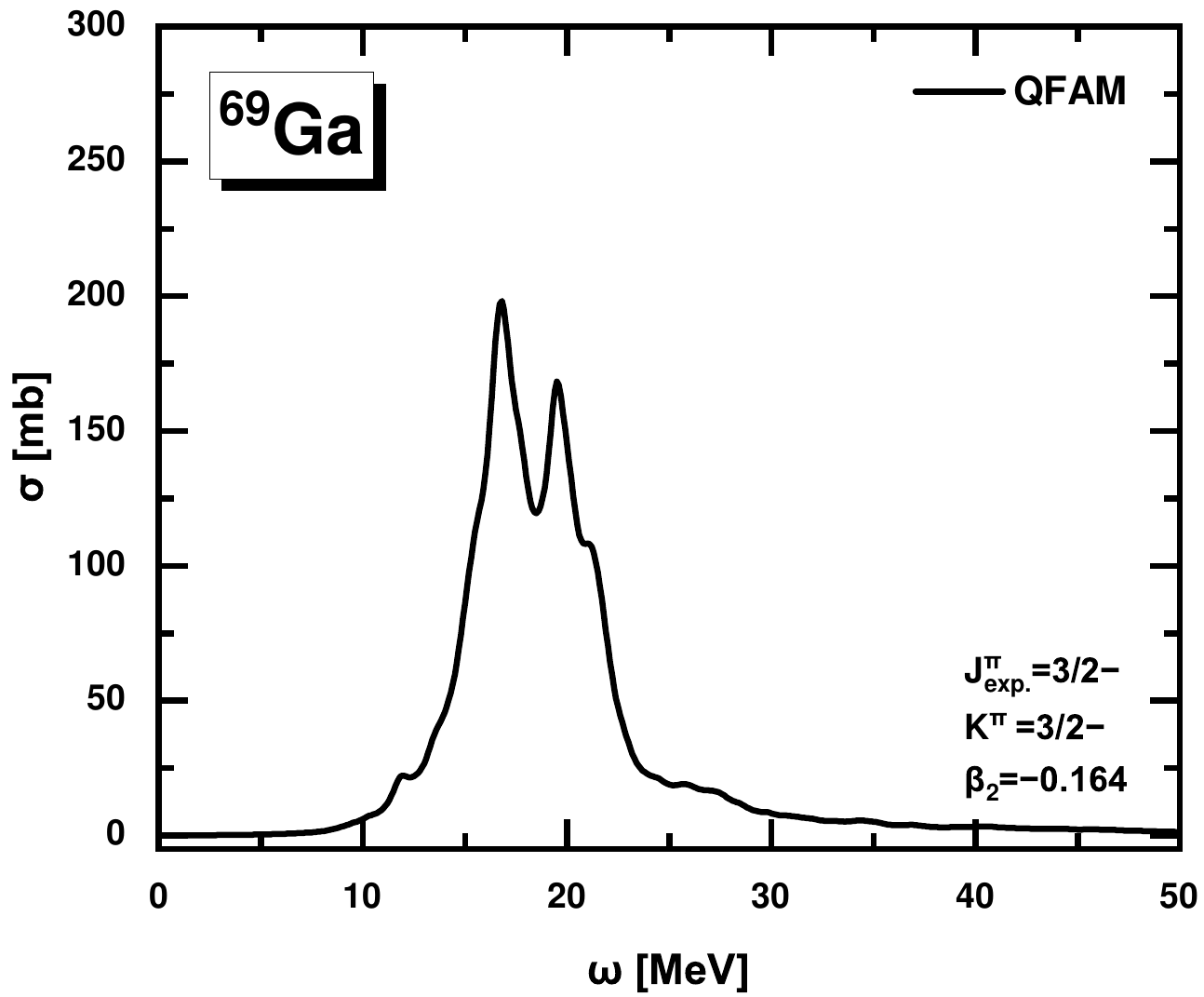}
    \includegraphics[width=0.35\textwidth]{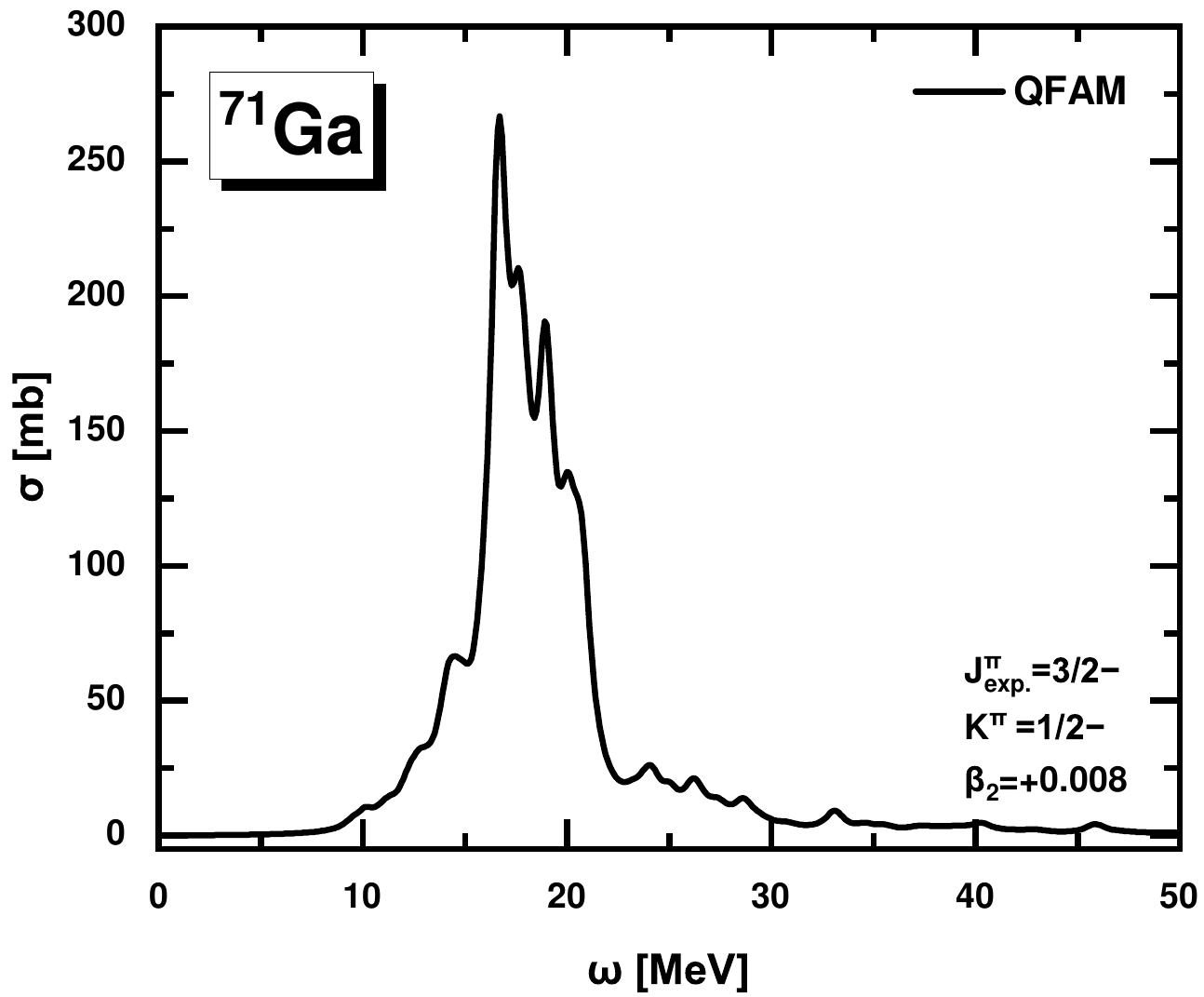}
    \includegraphics[width=0.35\textwidth]{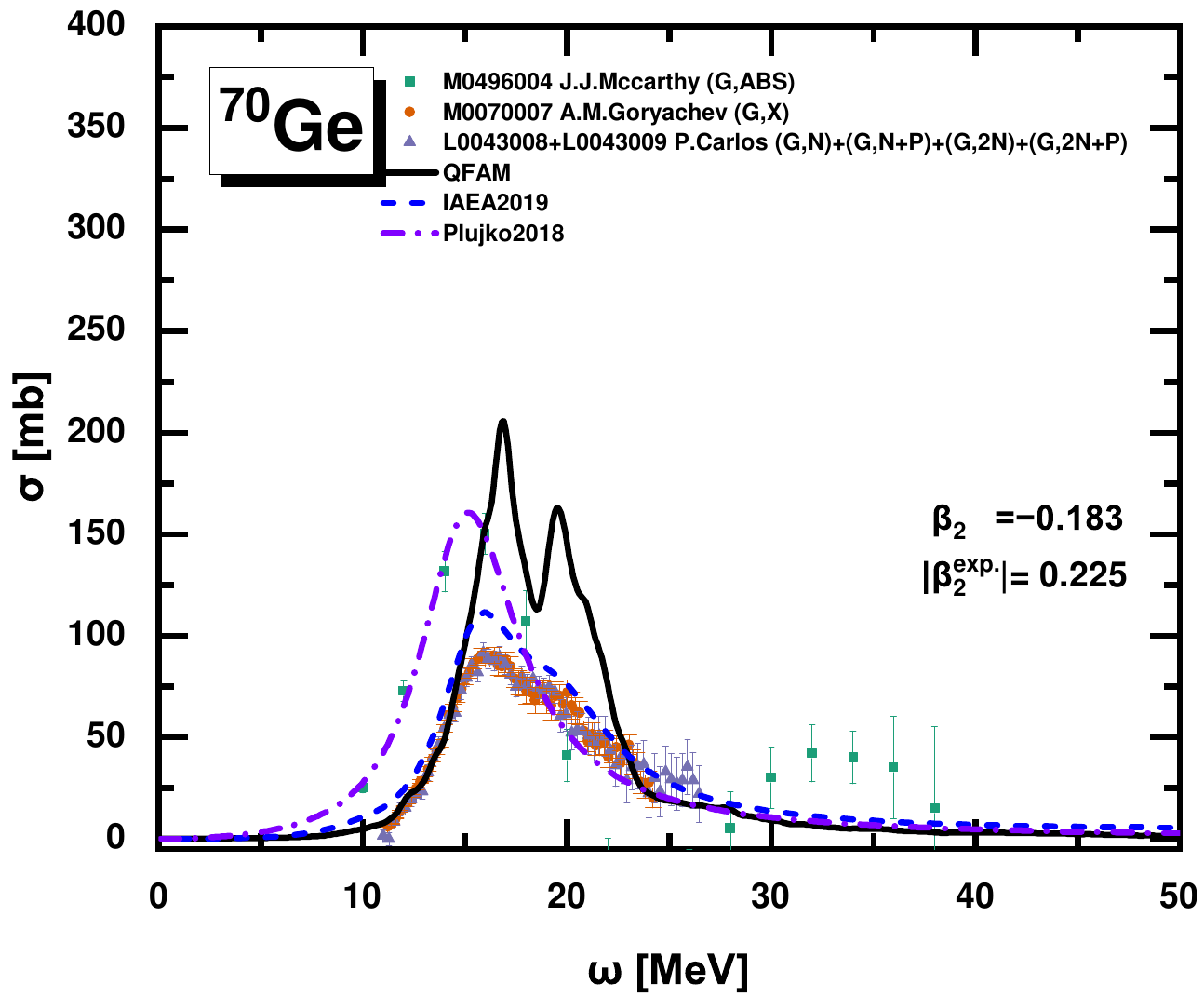}
    \includegraphics[width=0.35\textwidth]{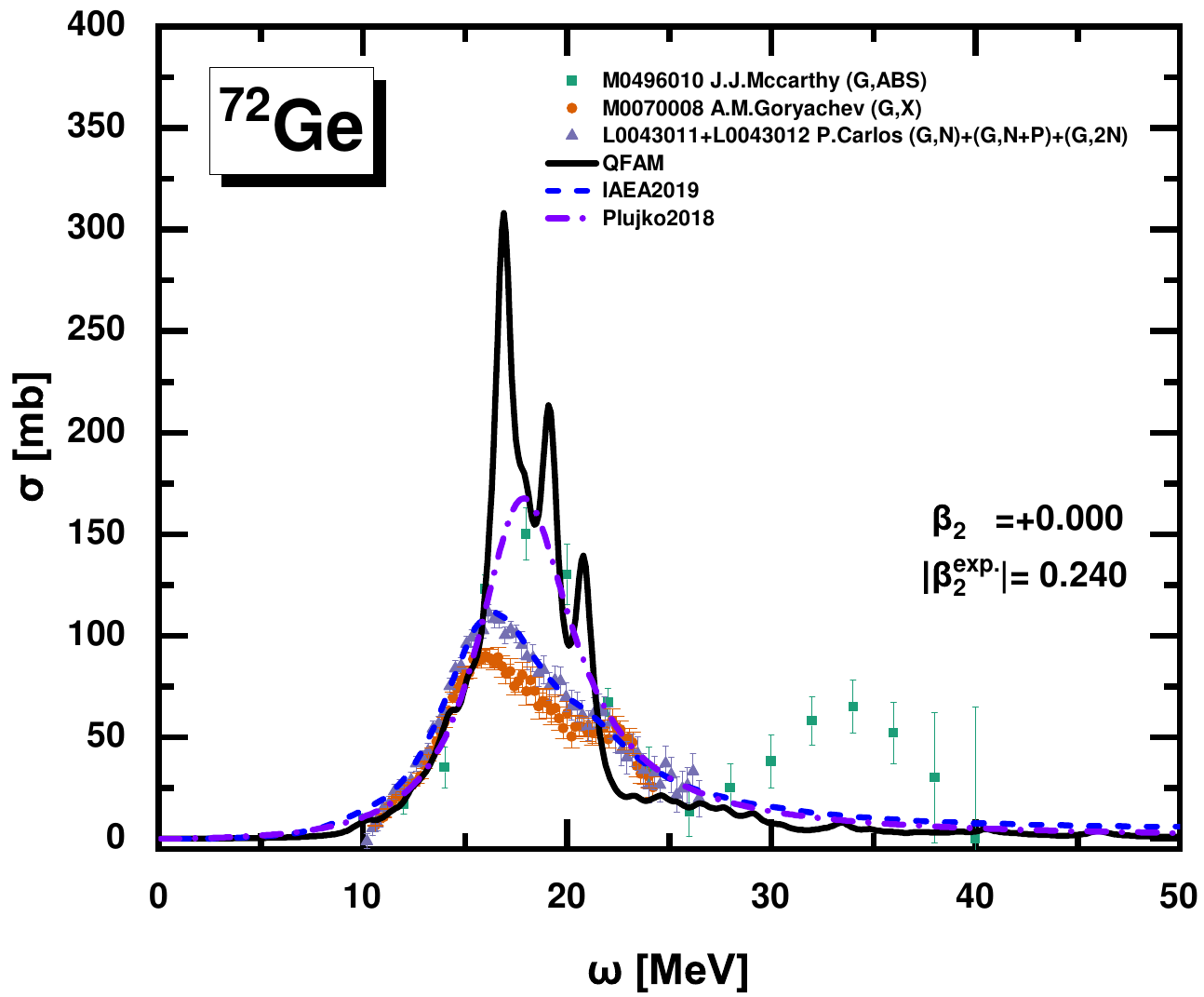}
    \includegraphics[width=0.35\textwidth]{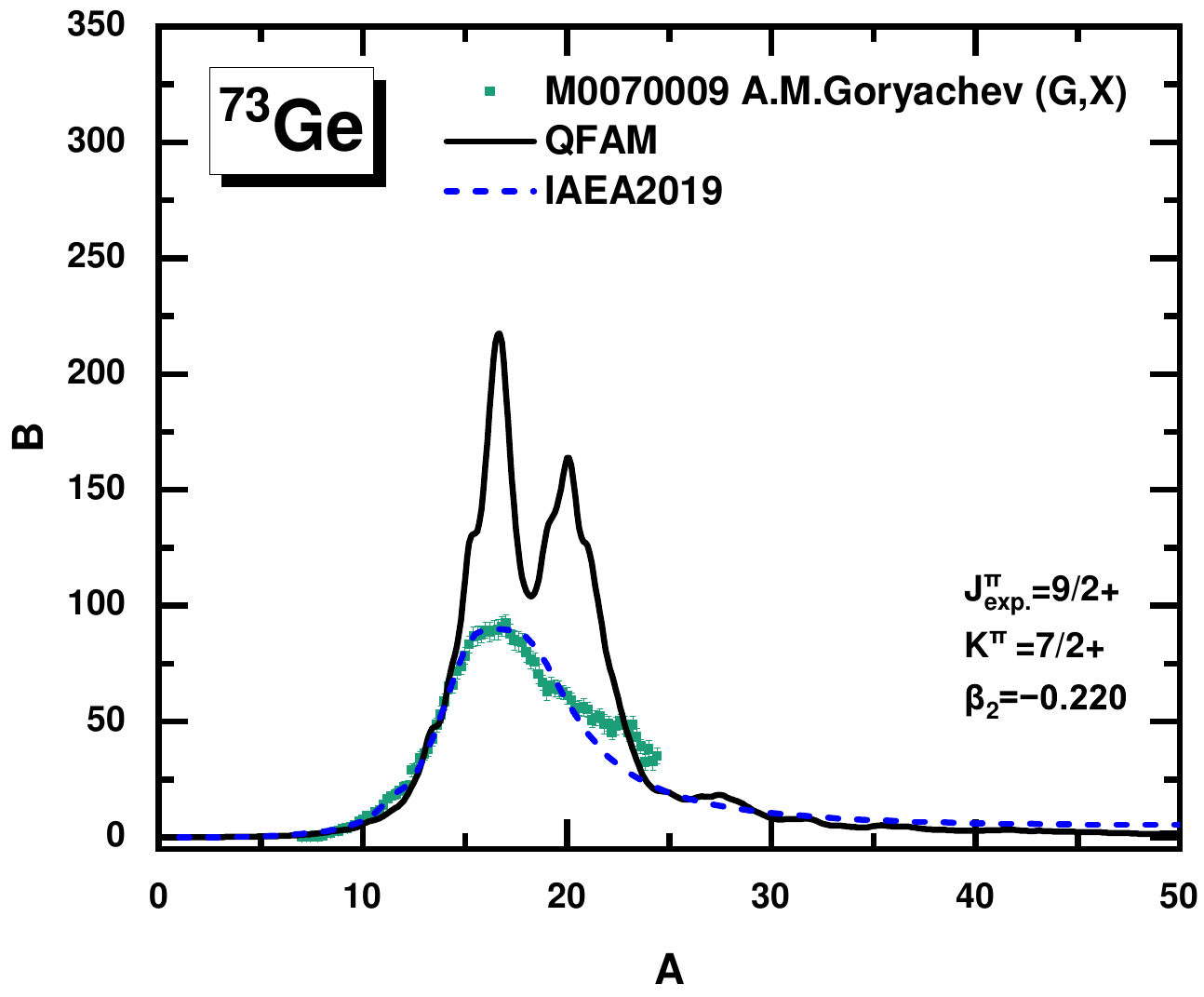}
    \includegraphics[width=0.35\textwidth]{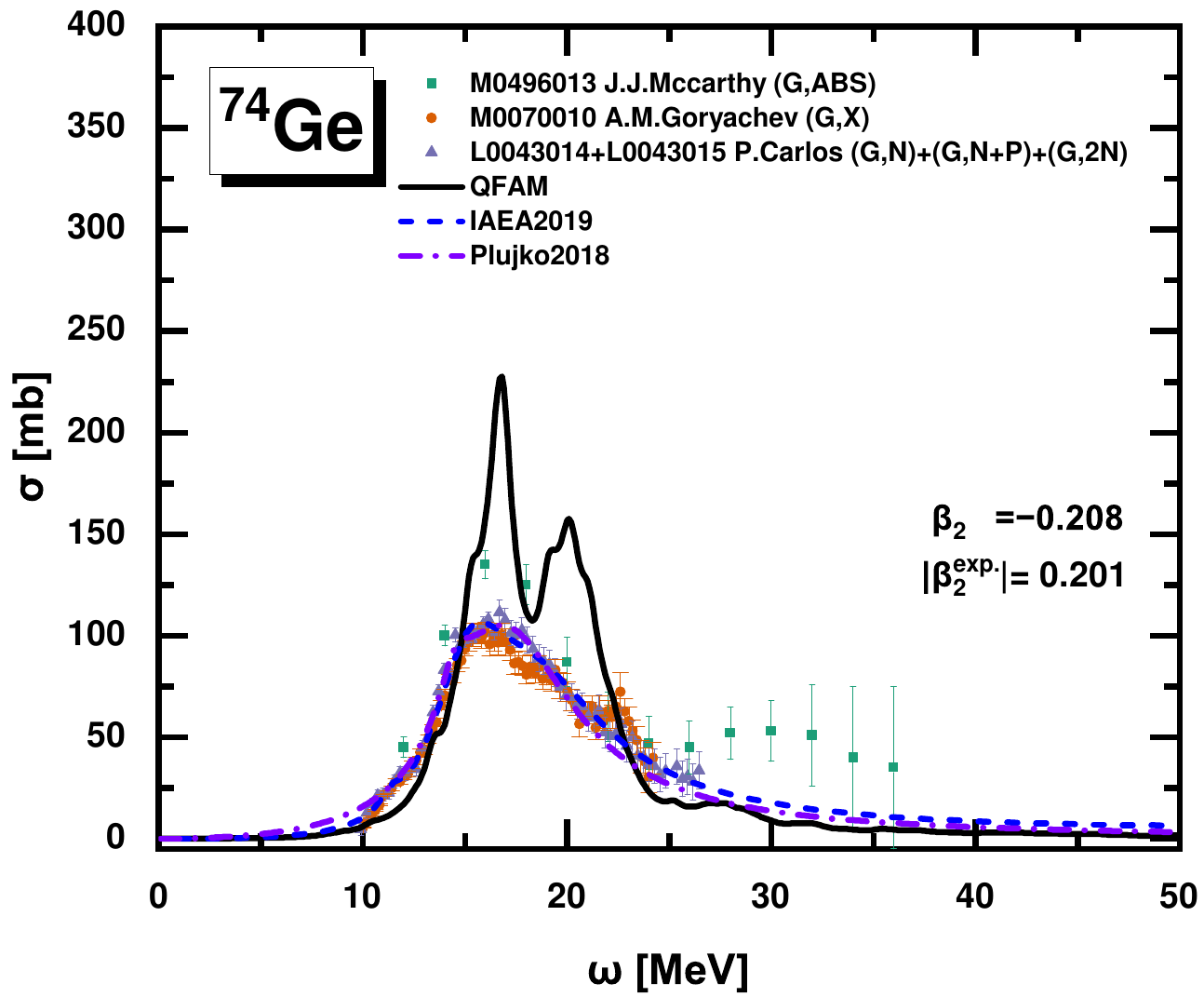}
\end{figure*}
\begin{figure*}\ContinuedFloat
    \centering
    \includegraphics[width=0.35\textwidth]{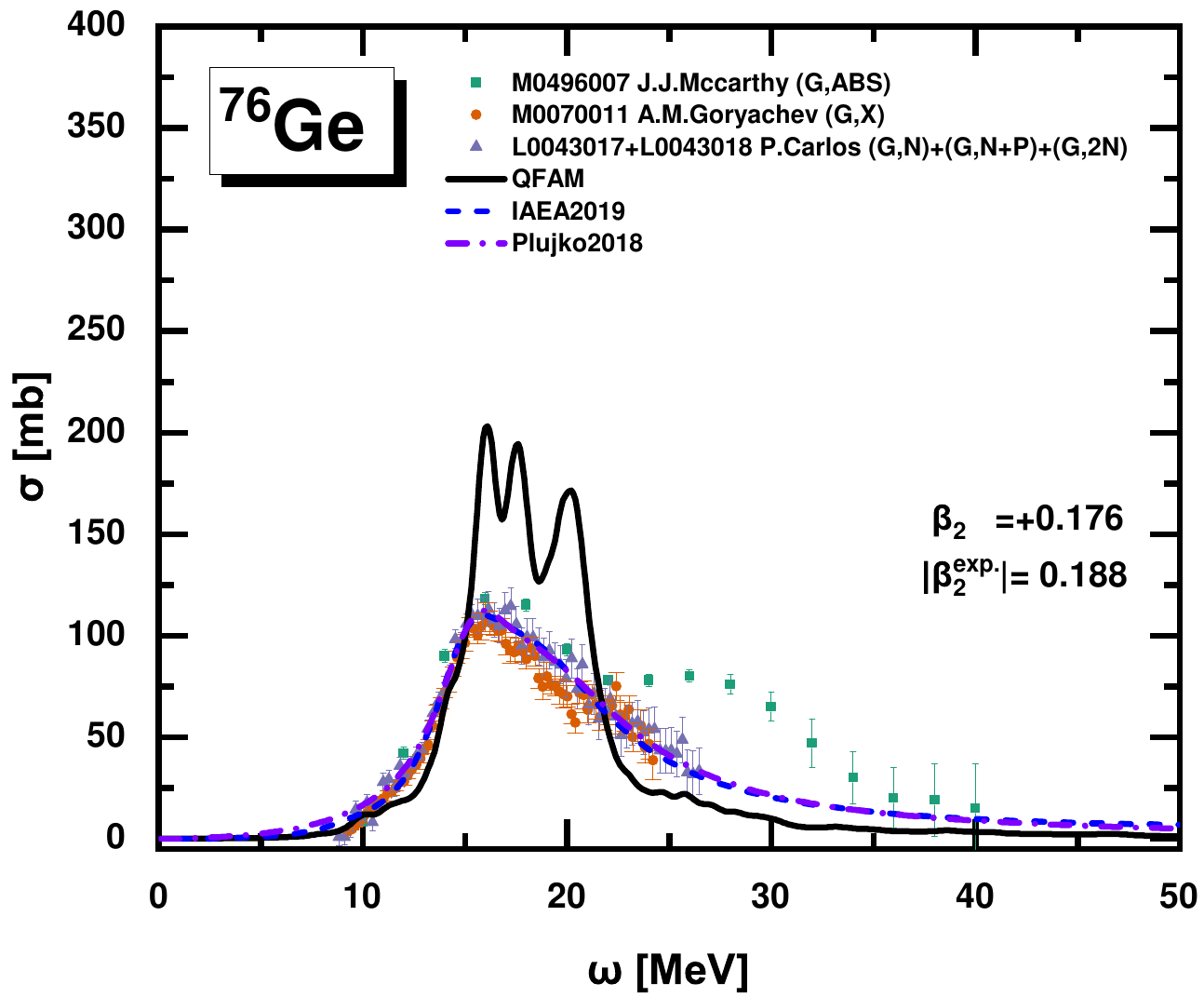}
    \includegraphics[width=0.35\textwidth]{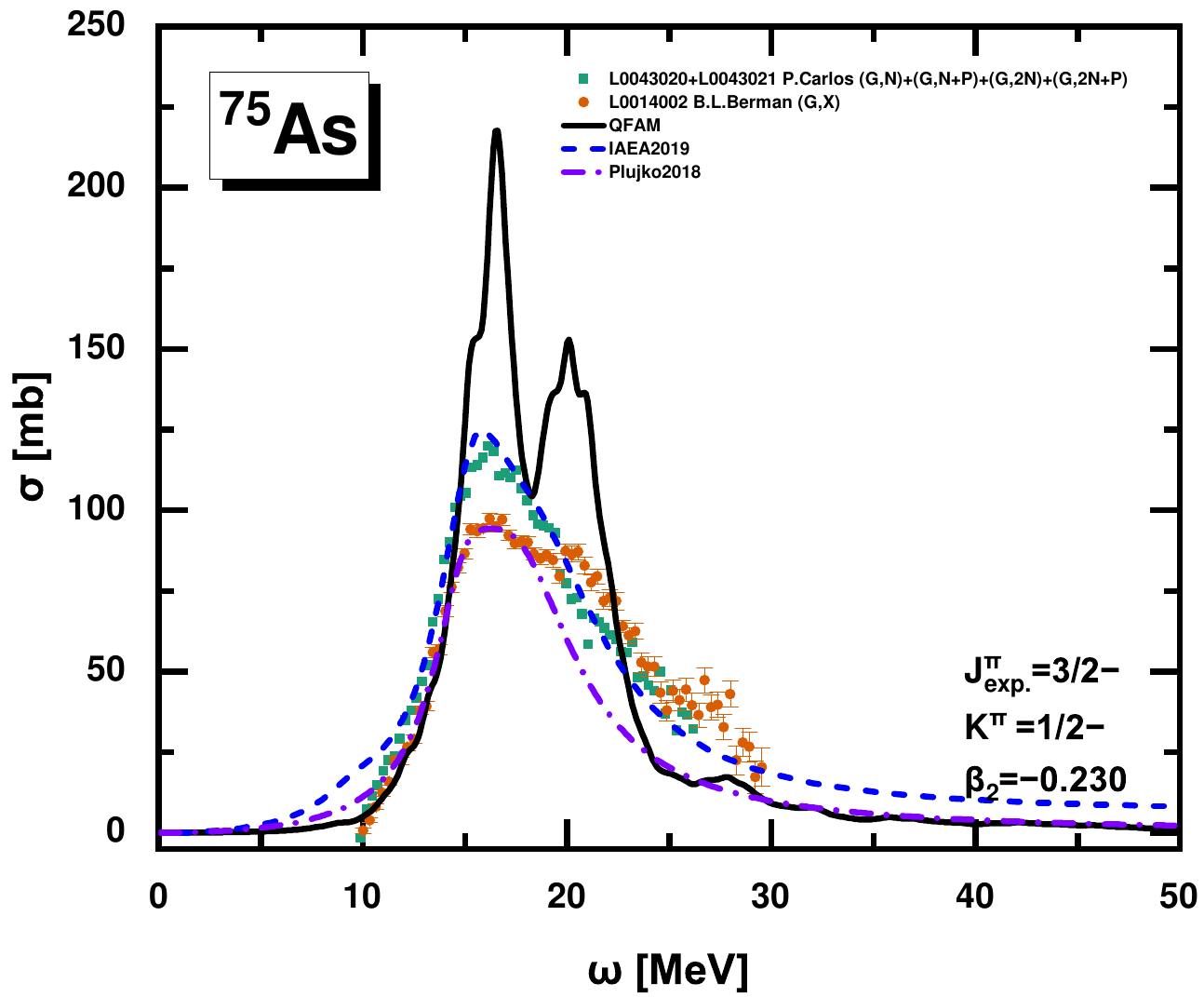}
    \includegraphics[width=0.35\textwidth]{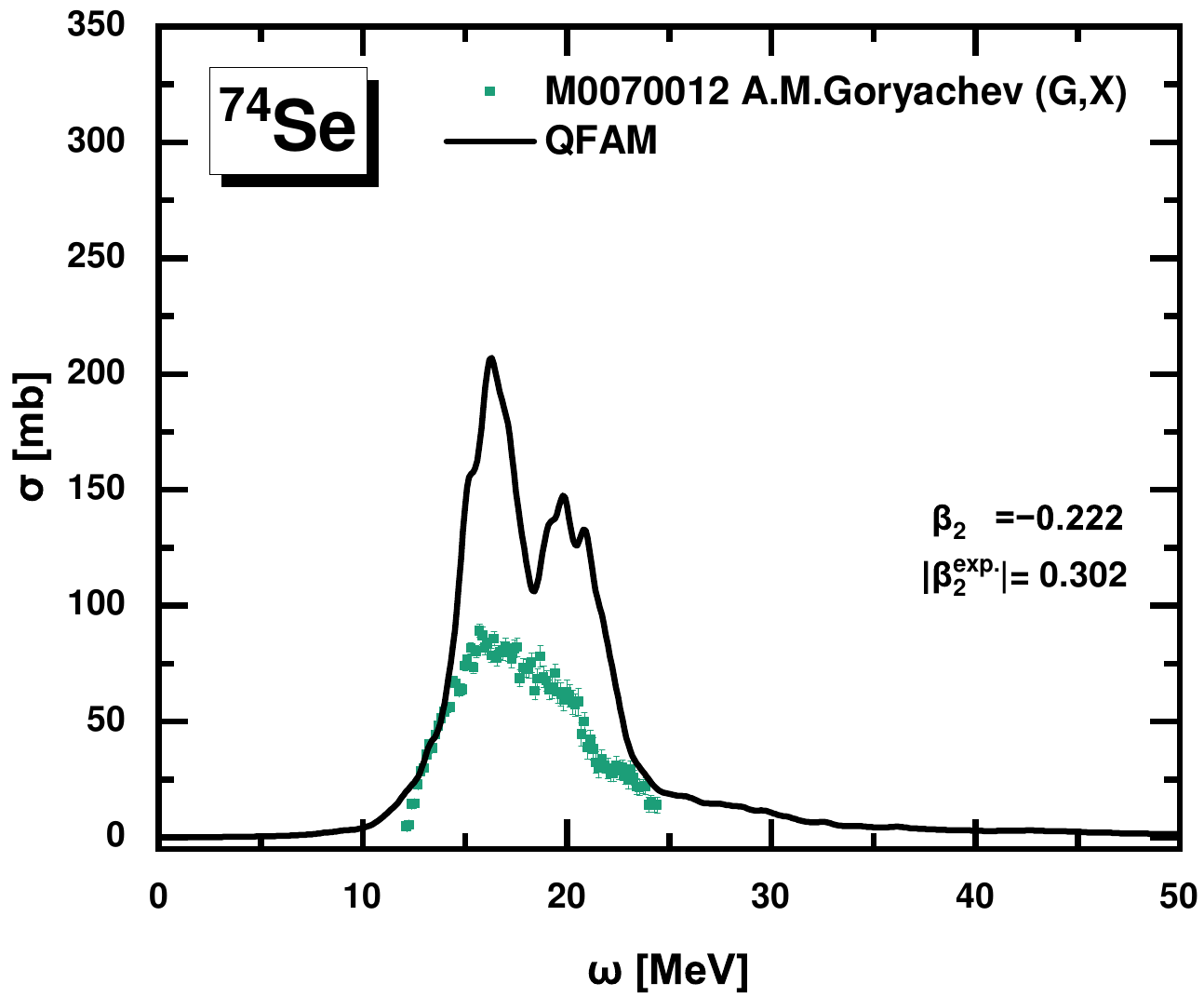}
    \includegraphics[width=0.35\textwidth]{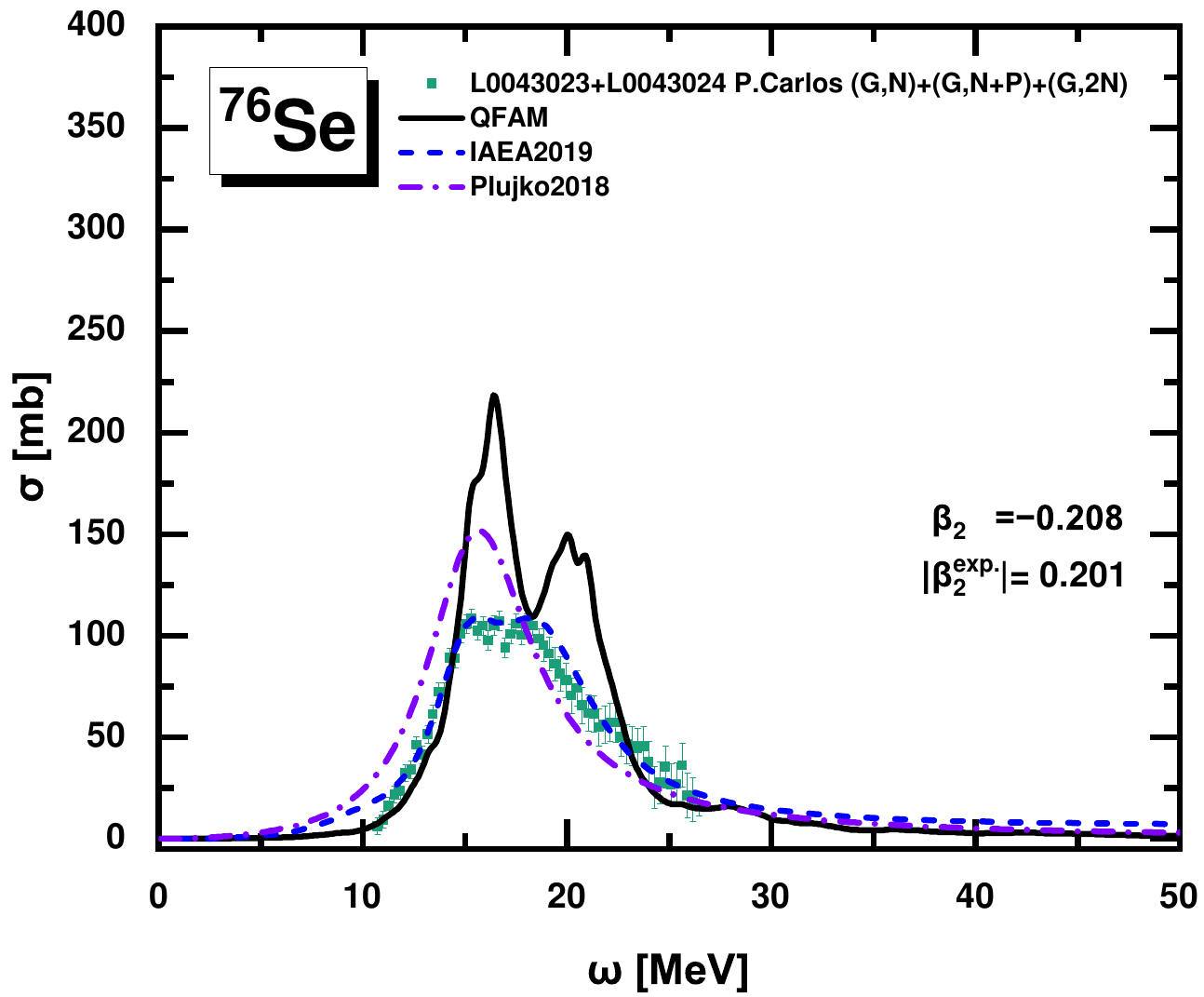}
    \includegraphics[width=0.35\textwidth]{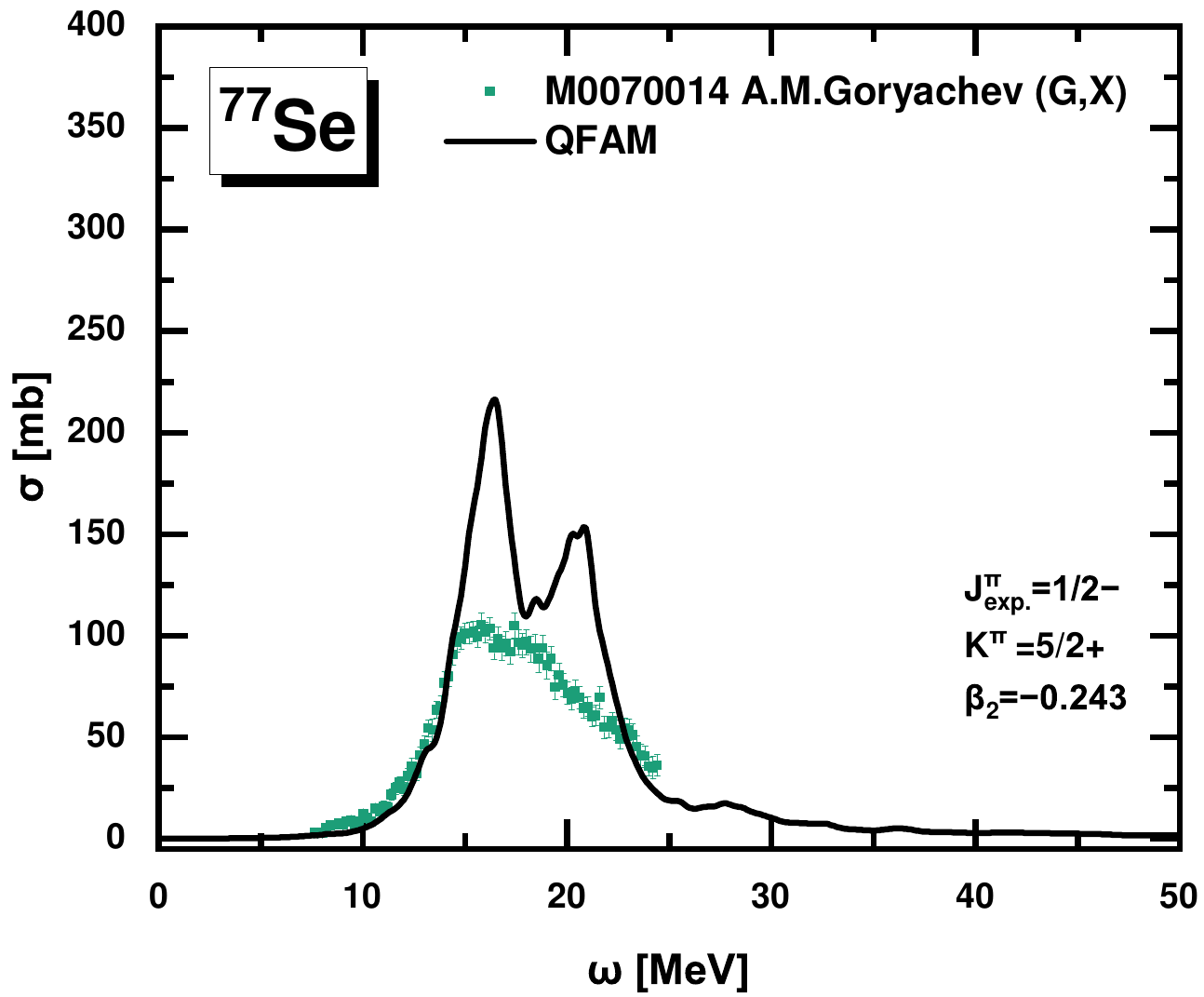}
    \includegraphics[width=0.35\textwidth]{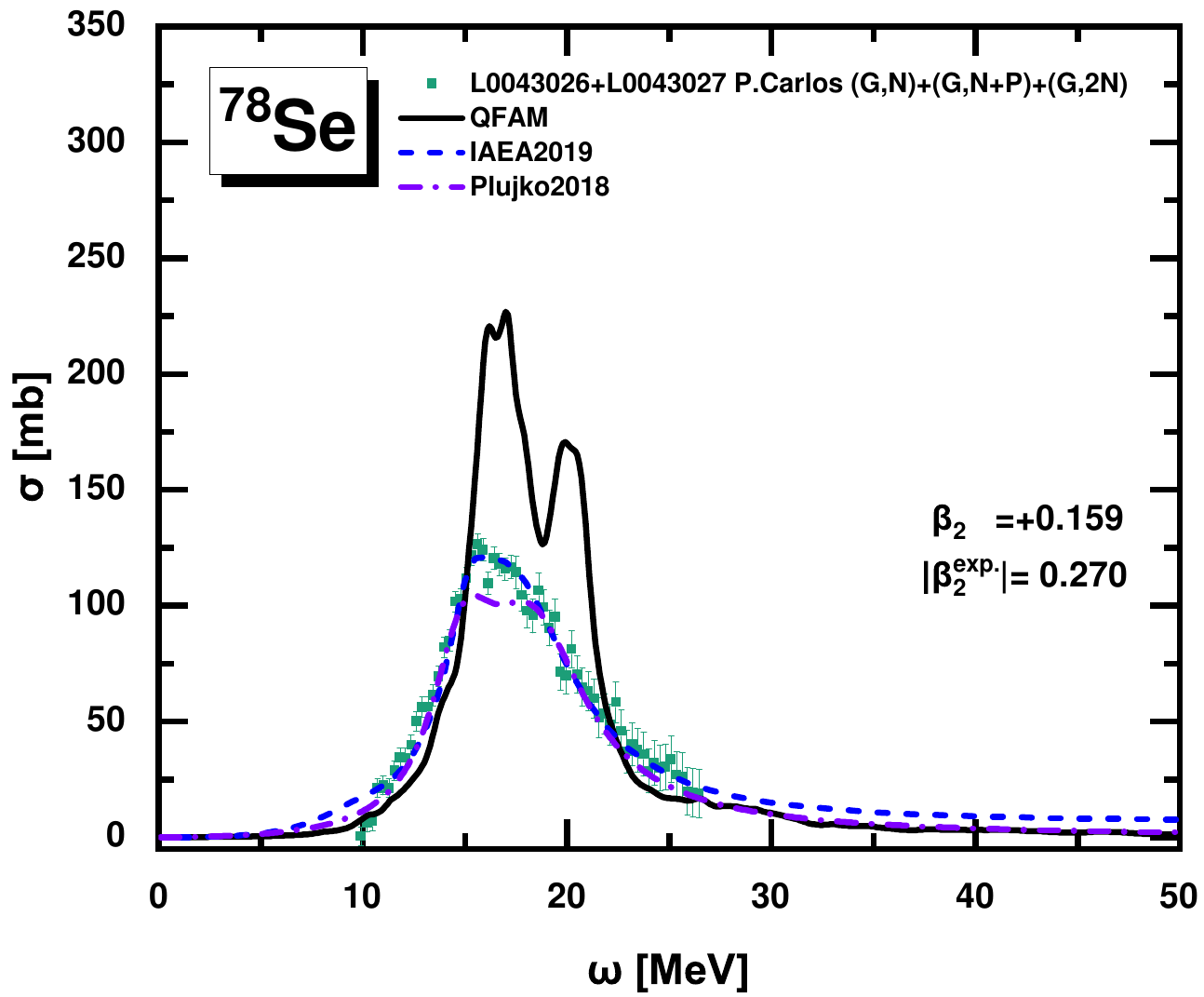}
    \includegraphics[width=0.35\textwidth]{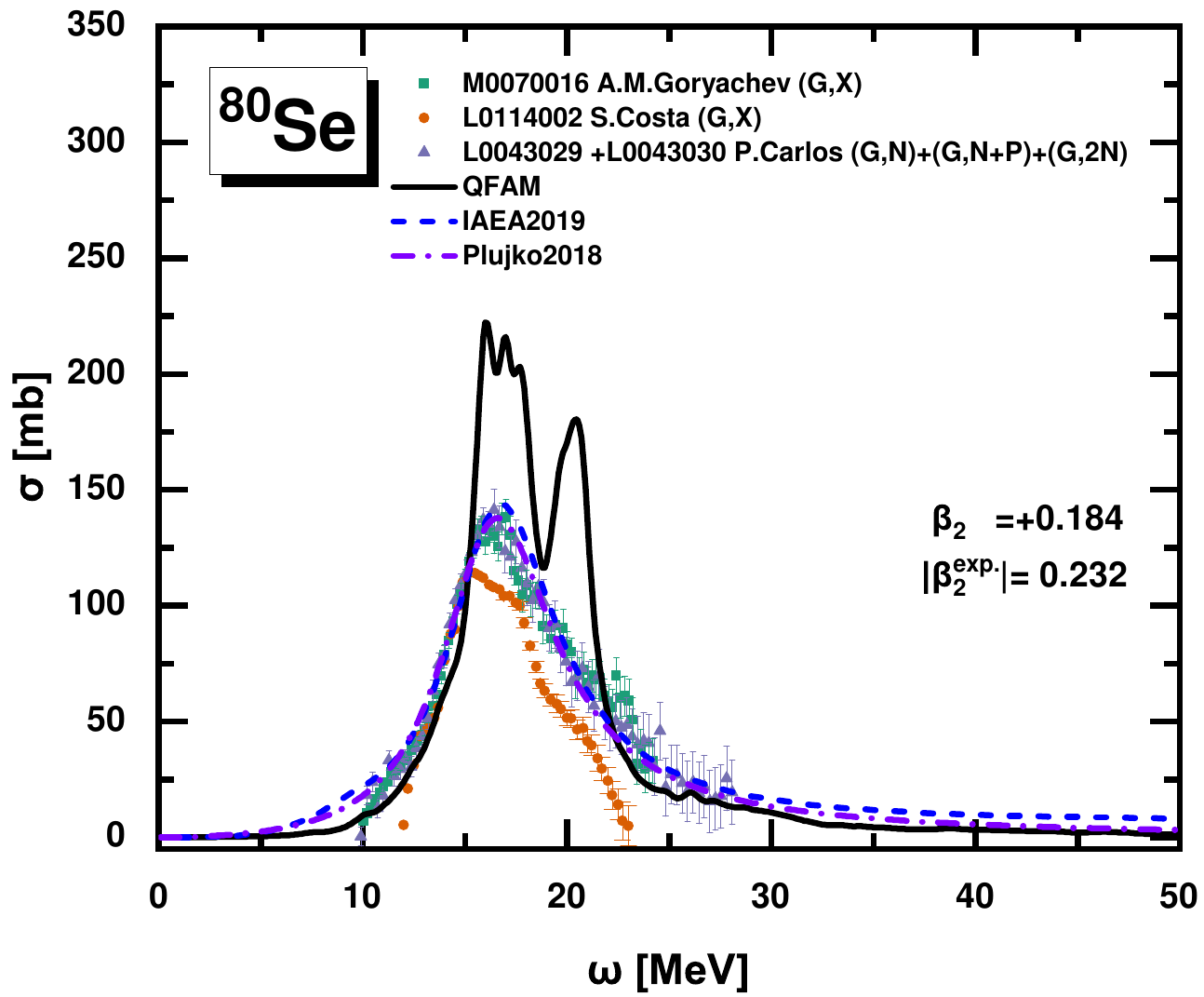}
    \includegraphics[width=0.35\textwidth]{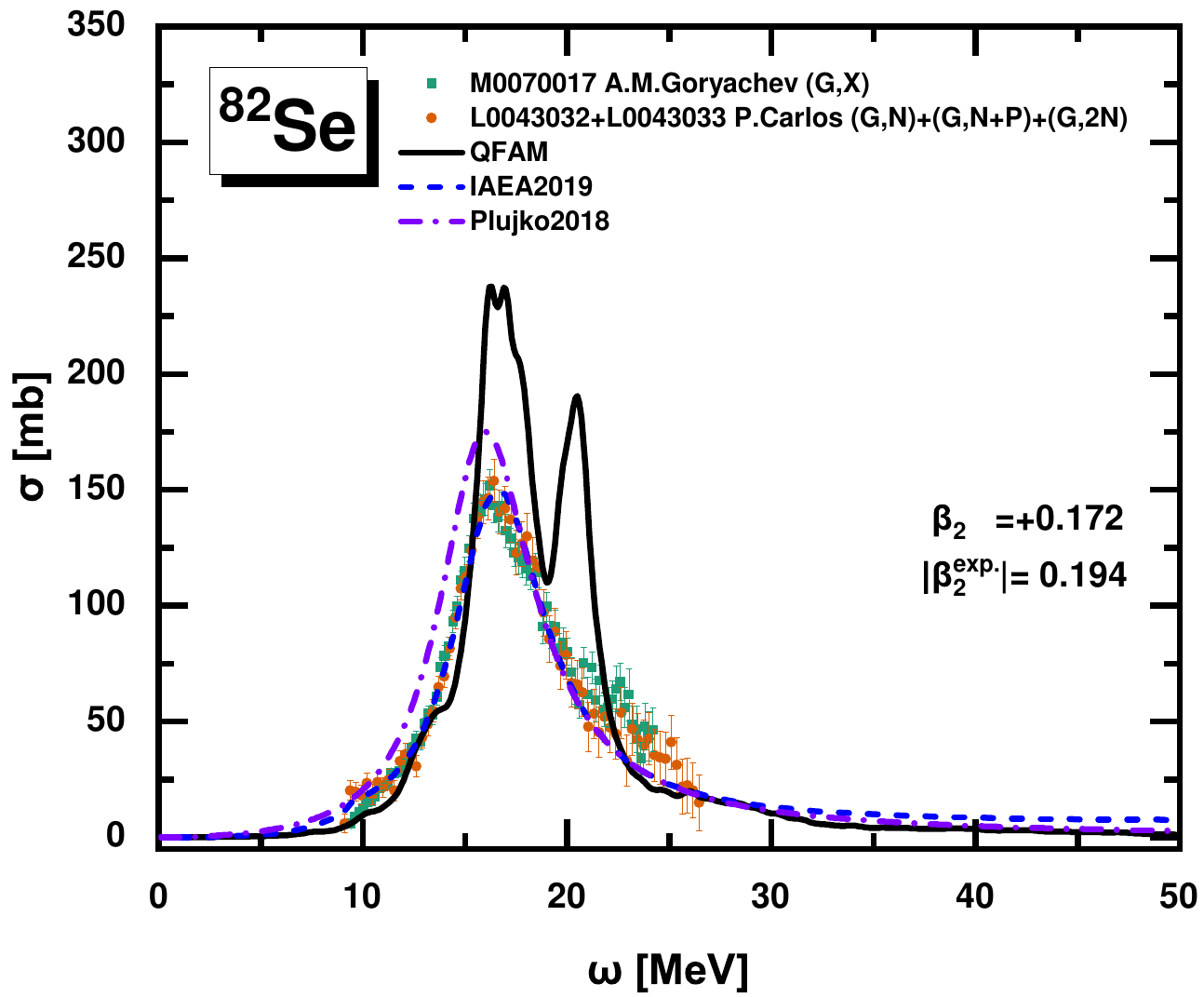}
\end{figure*}
\begin{figure*}\ContinuedFloat
    \centering
    \includegraphics[width=0.35\textwidth]{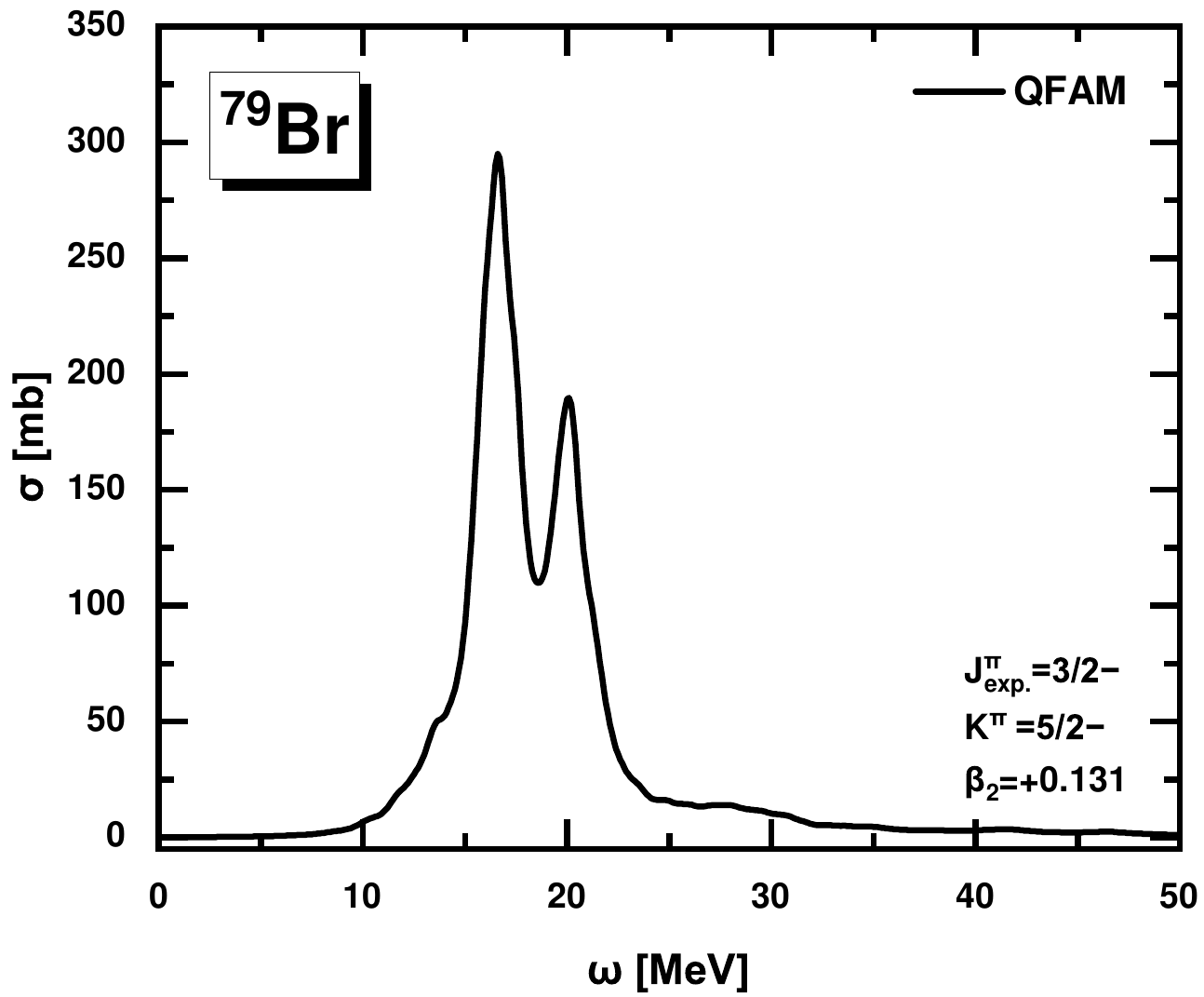}
    \includegraphics[width=0.35\textwidth]{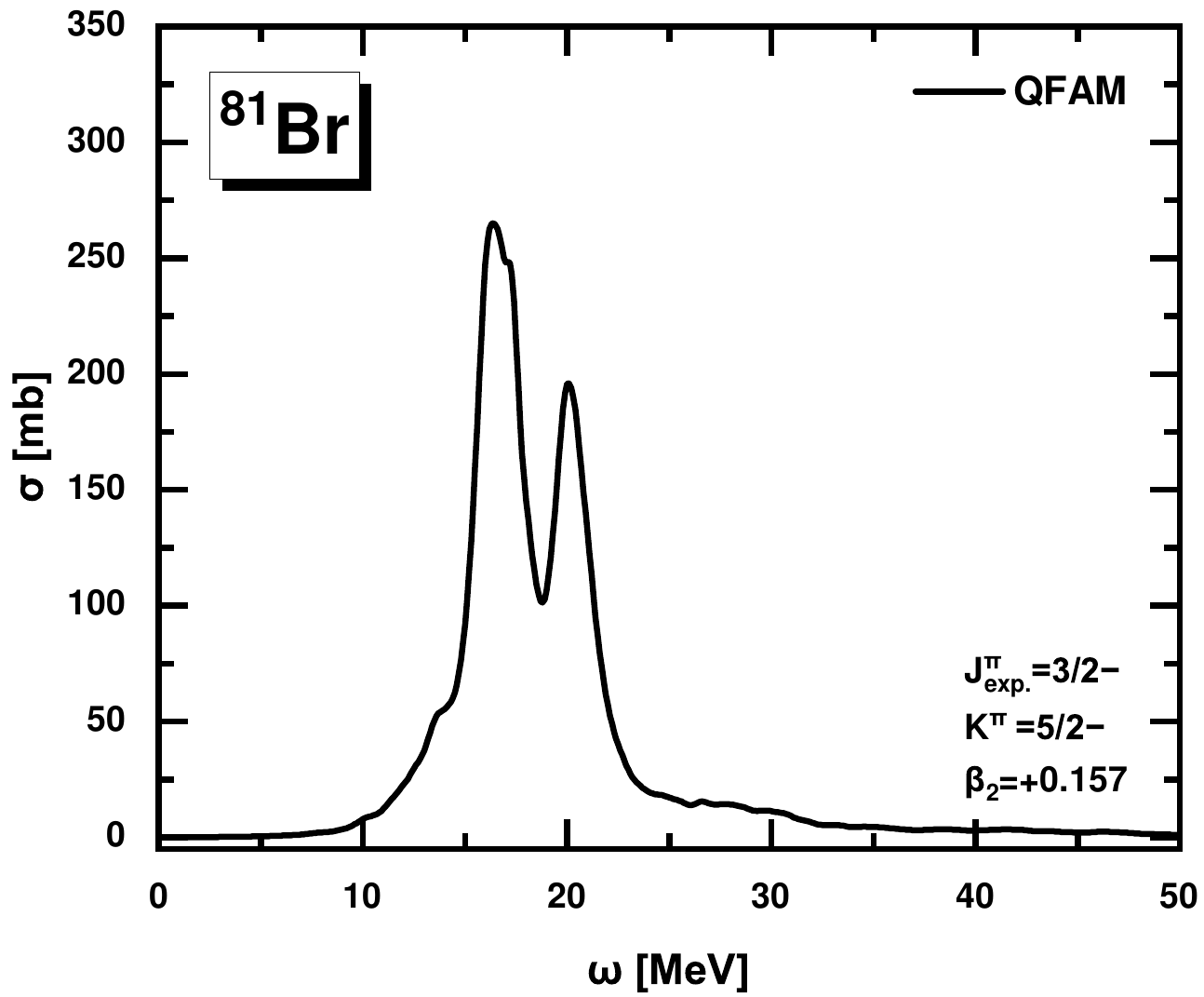}
    \includegraphics[width=0.35\textwidth]{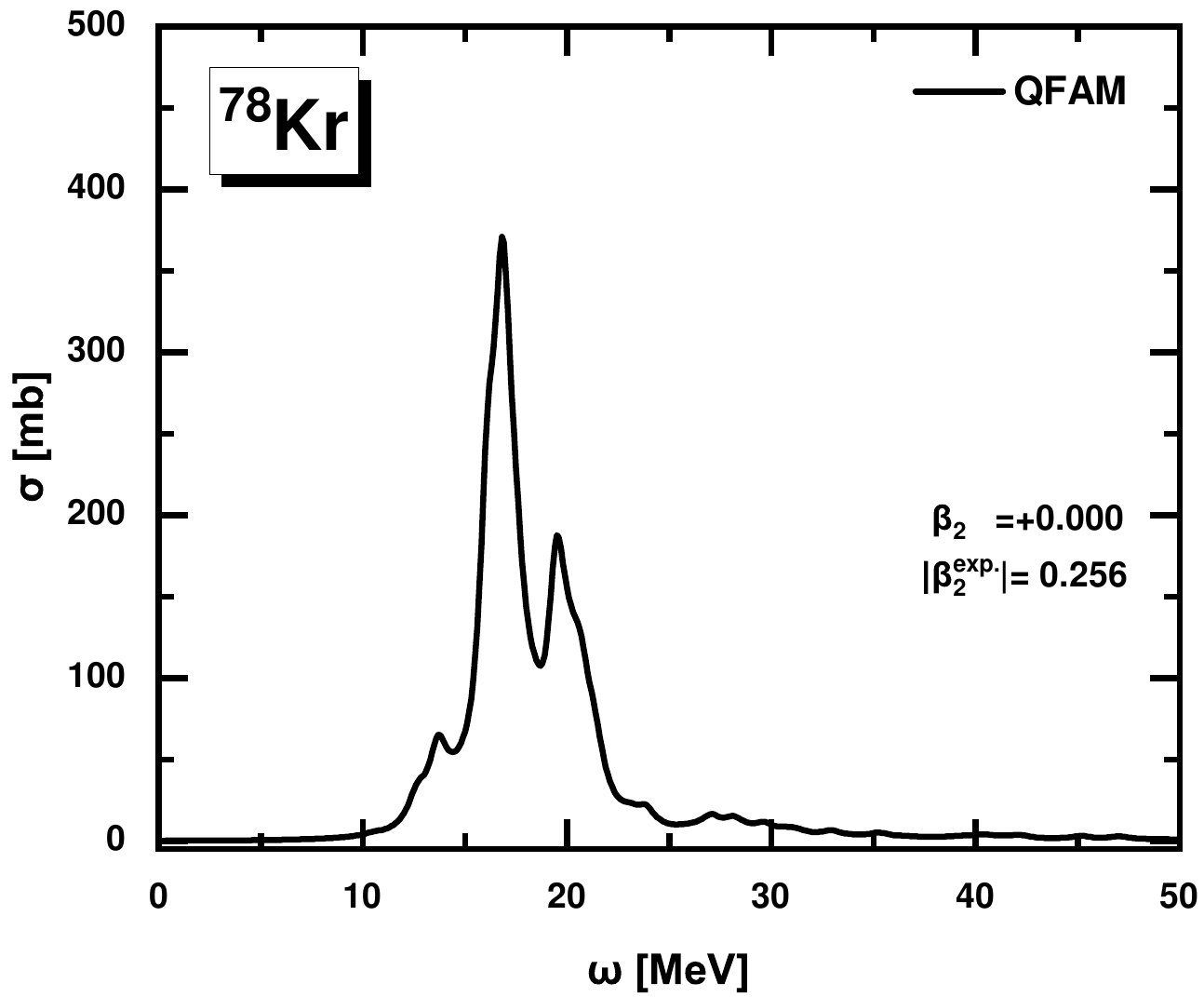}
    \includegraphics[width=0.35\textwidth]{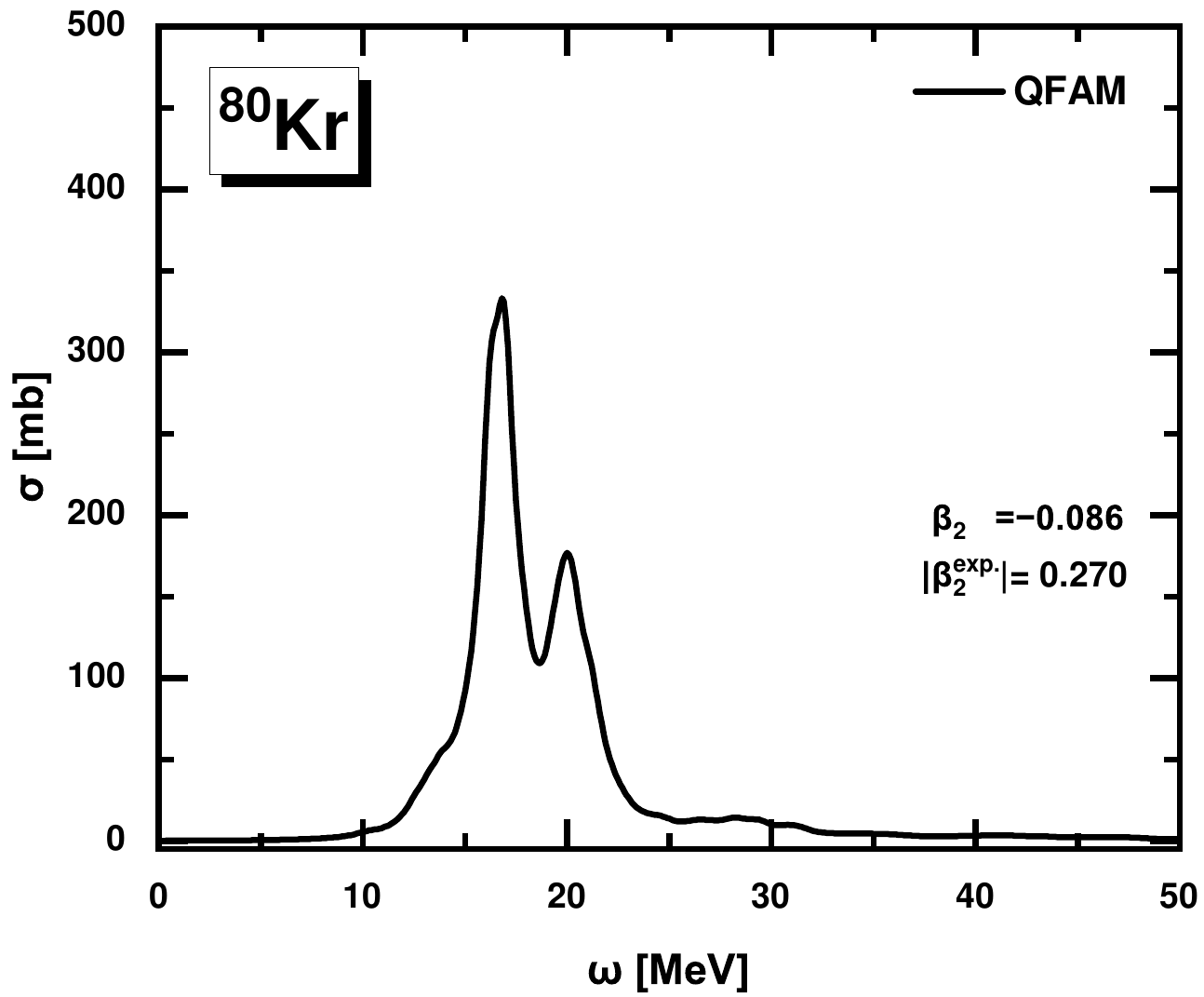}
    \includegraphics[width=0.35\textwidth]{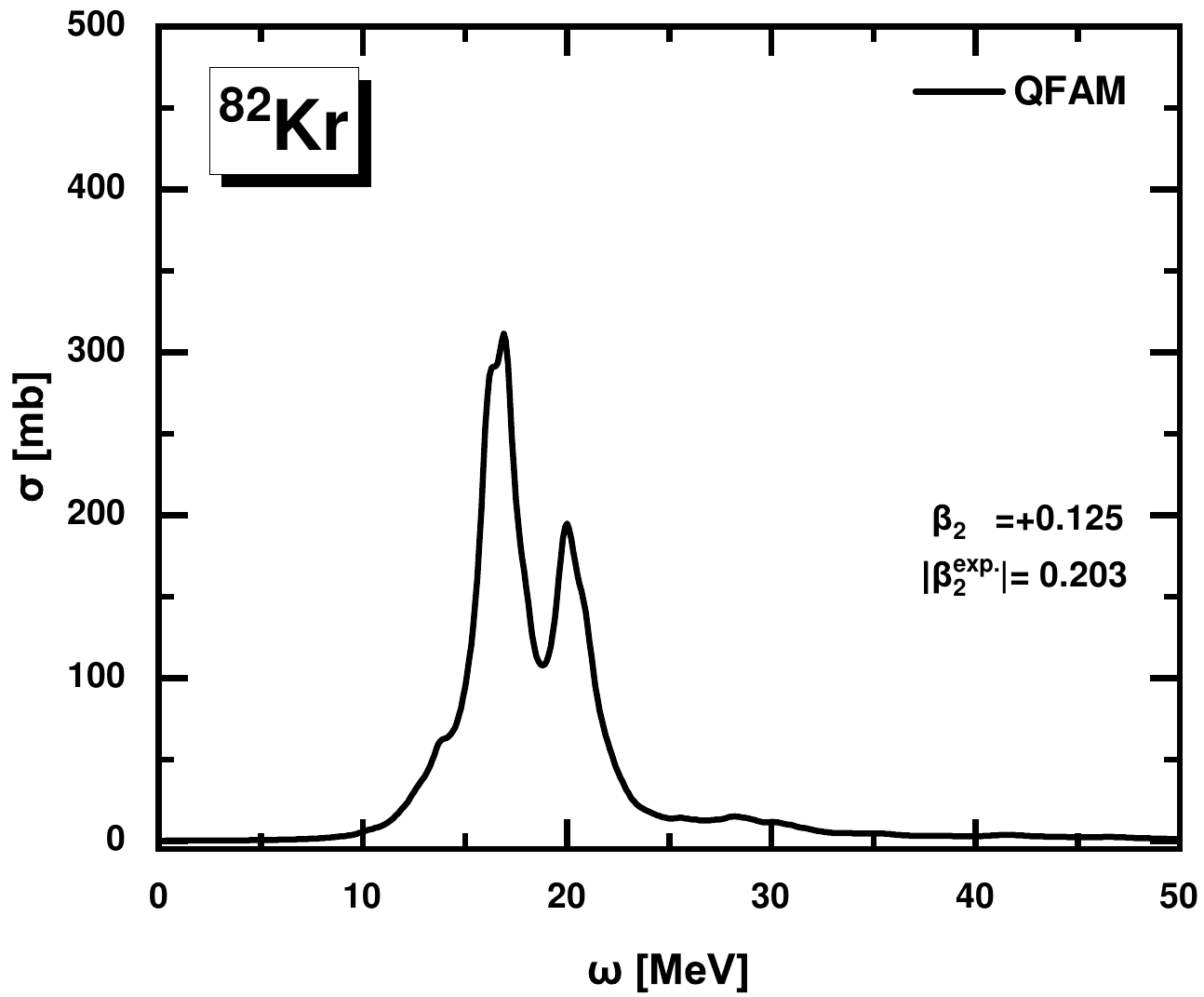}
    \includegraphics[width=0.35\textwidth]{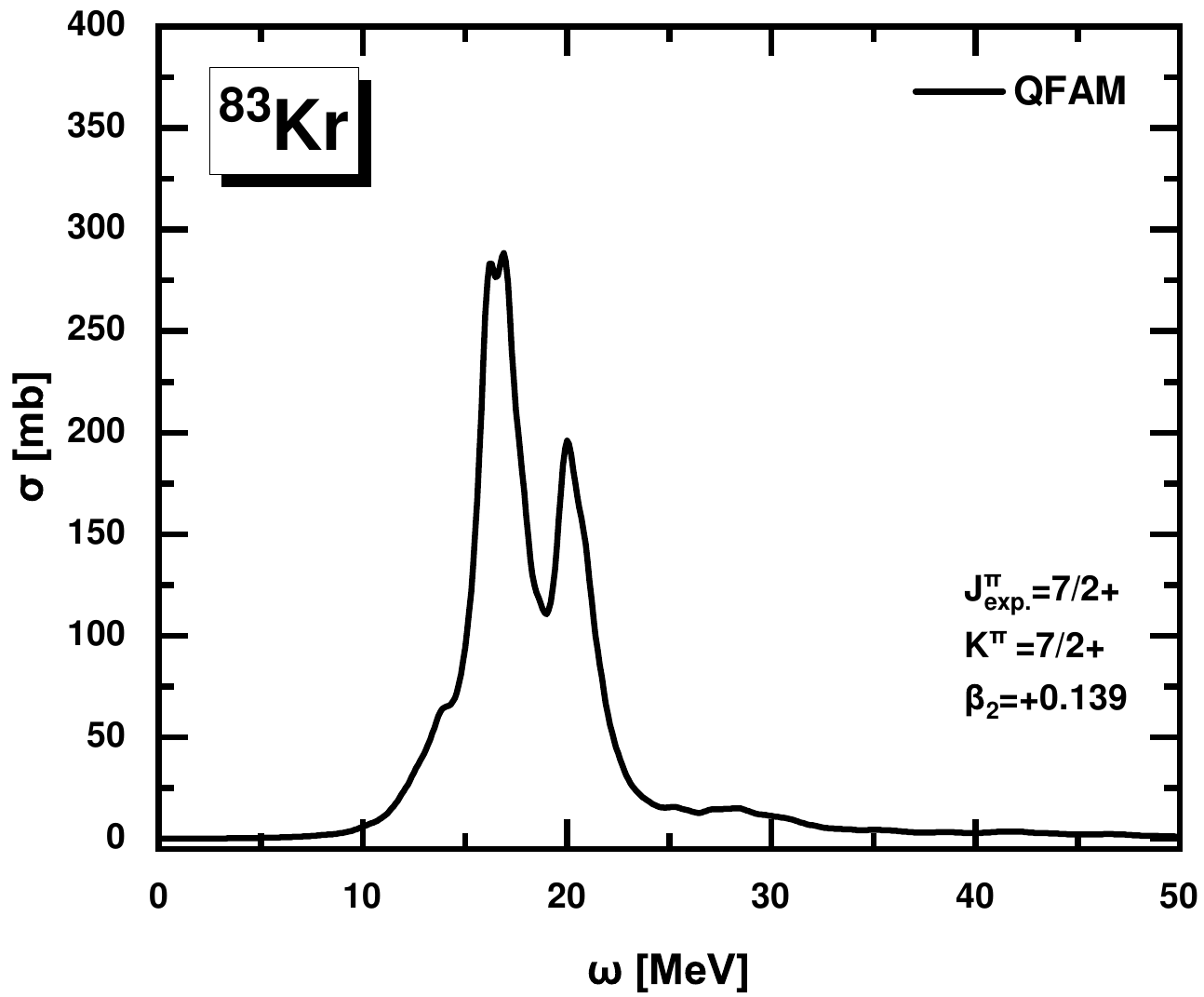}
    \includegraphics[width=0.35\textwidth]{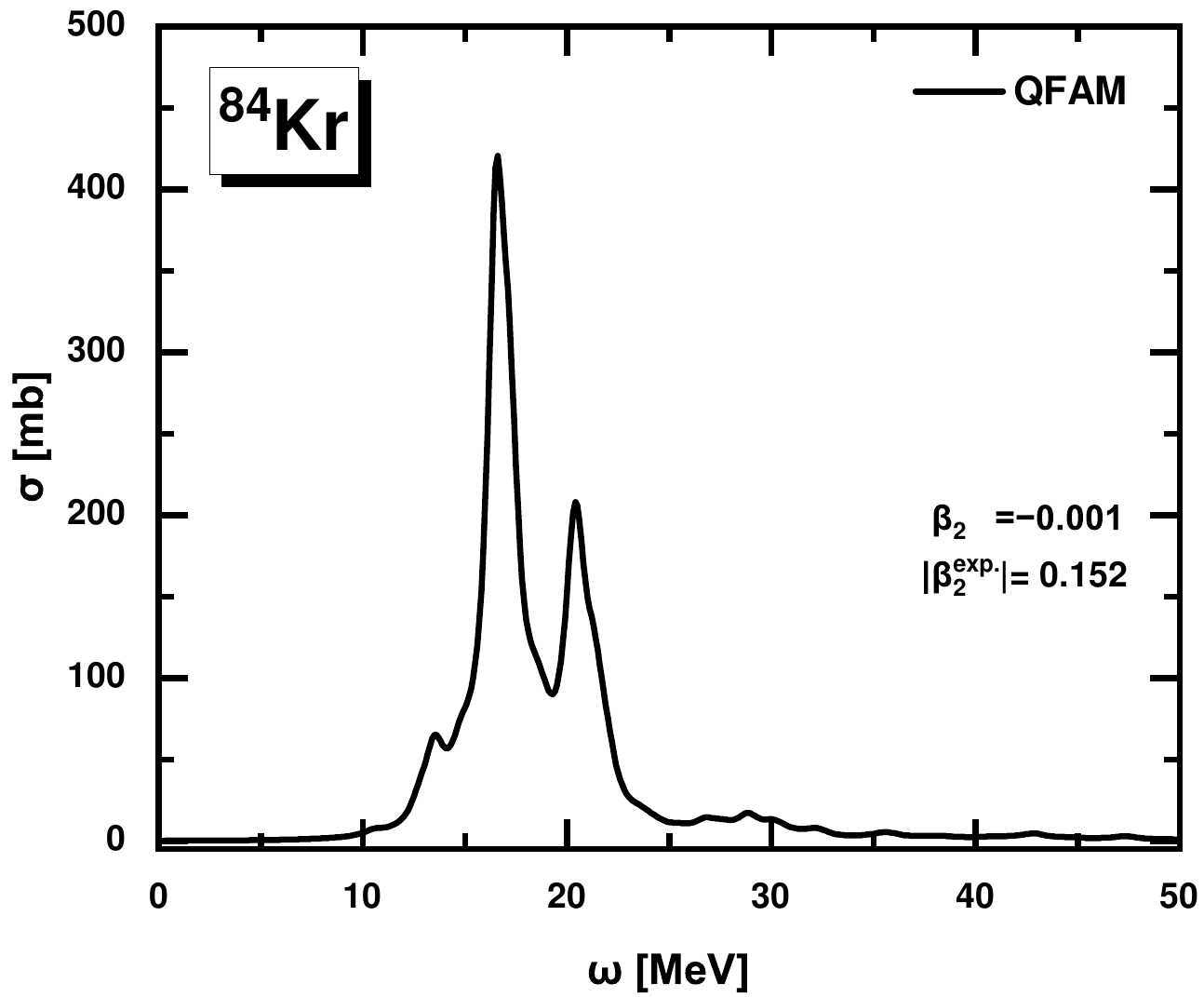}
    \includegraphics[width=0.35\textwidth]{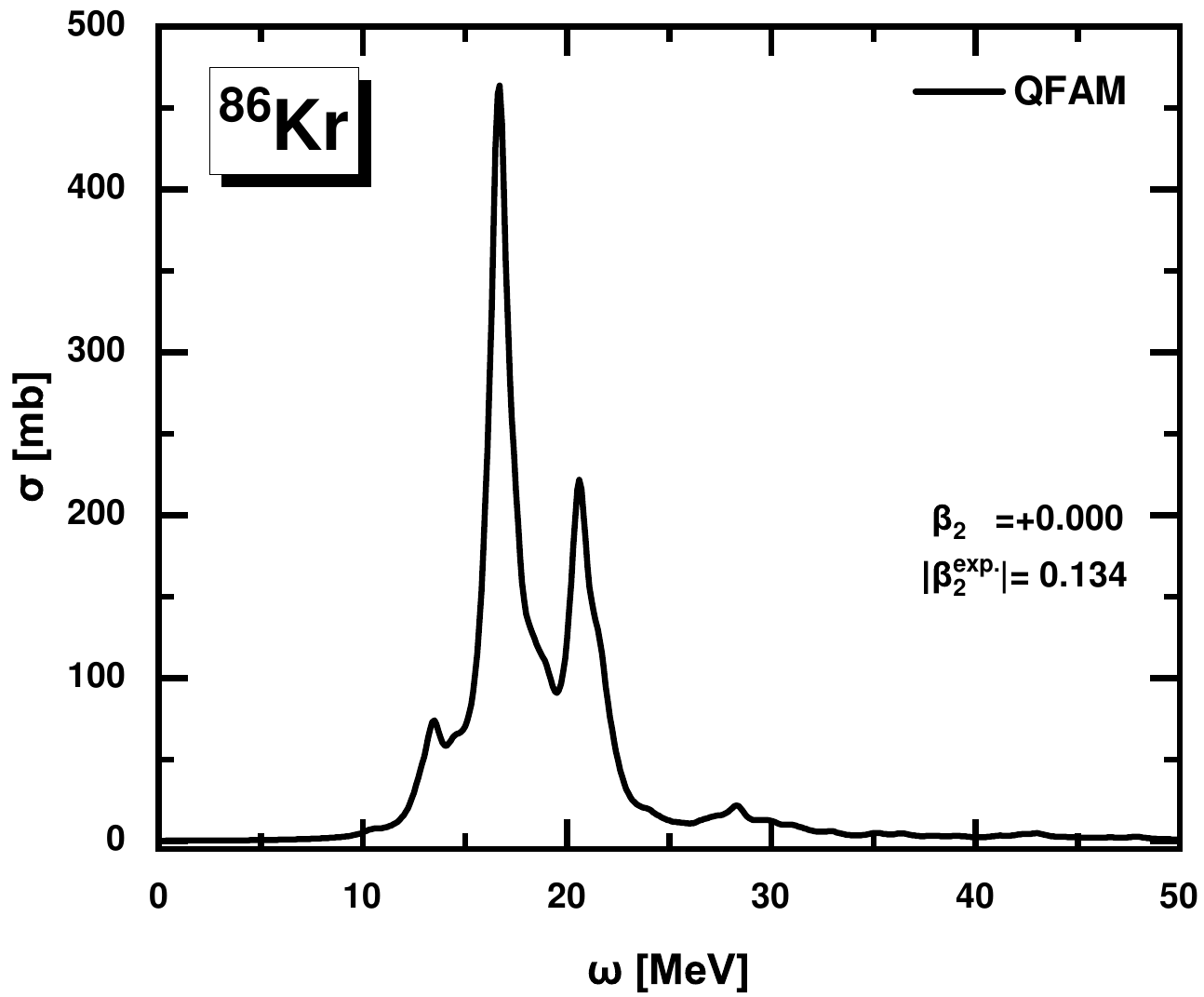}
\end{figure*}
\begin{figure*}\ContinuedFloat
    \centering
    \includegraphics[width=0.35\textwidth]{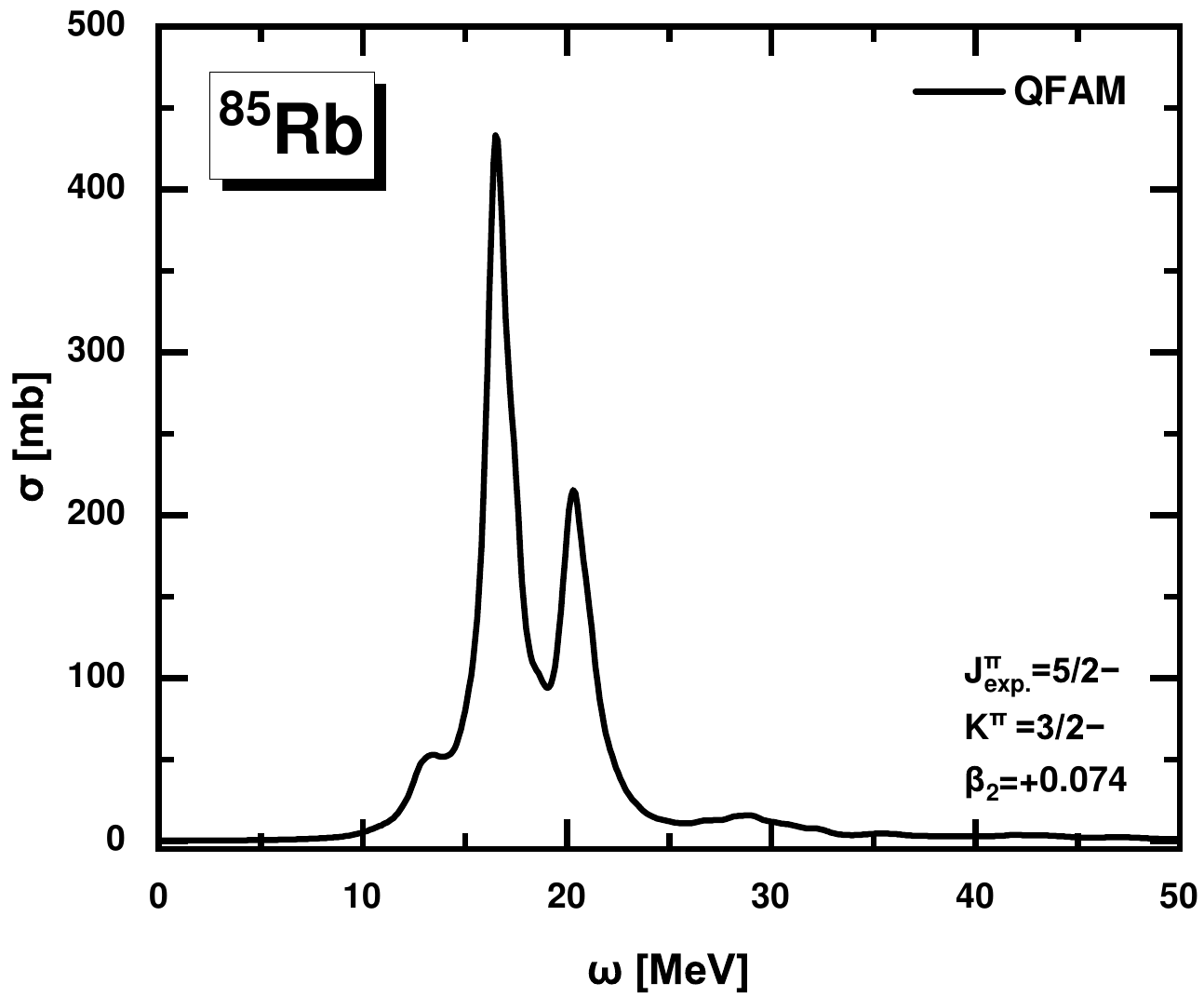}
    \includegraphics[width=0.35\textwidth]{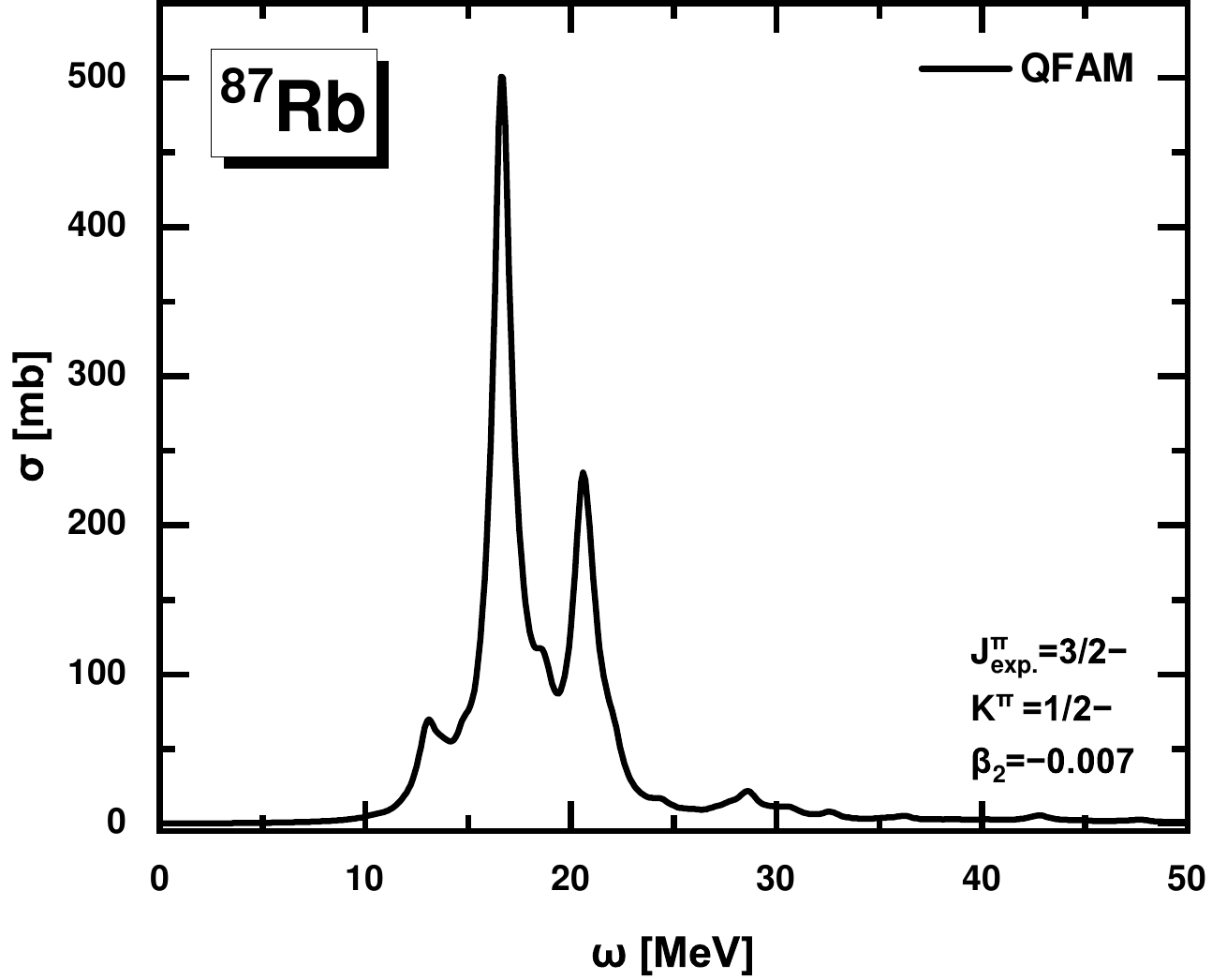}
    \includegraphics[width=0.35\textwidth]{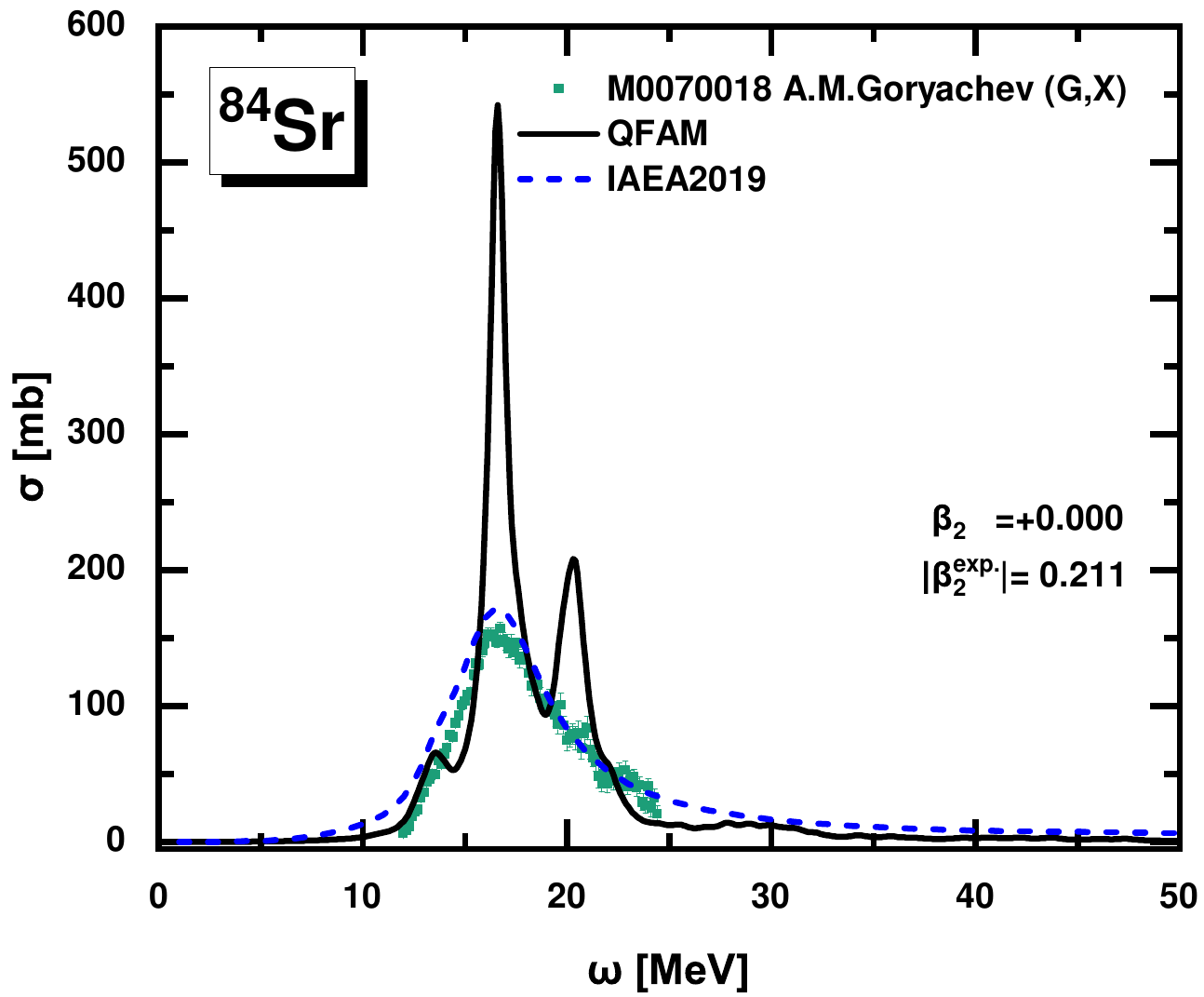}
    \includegraphics[width=0.35\textwidth]{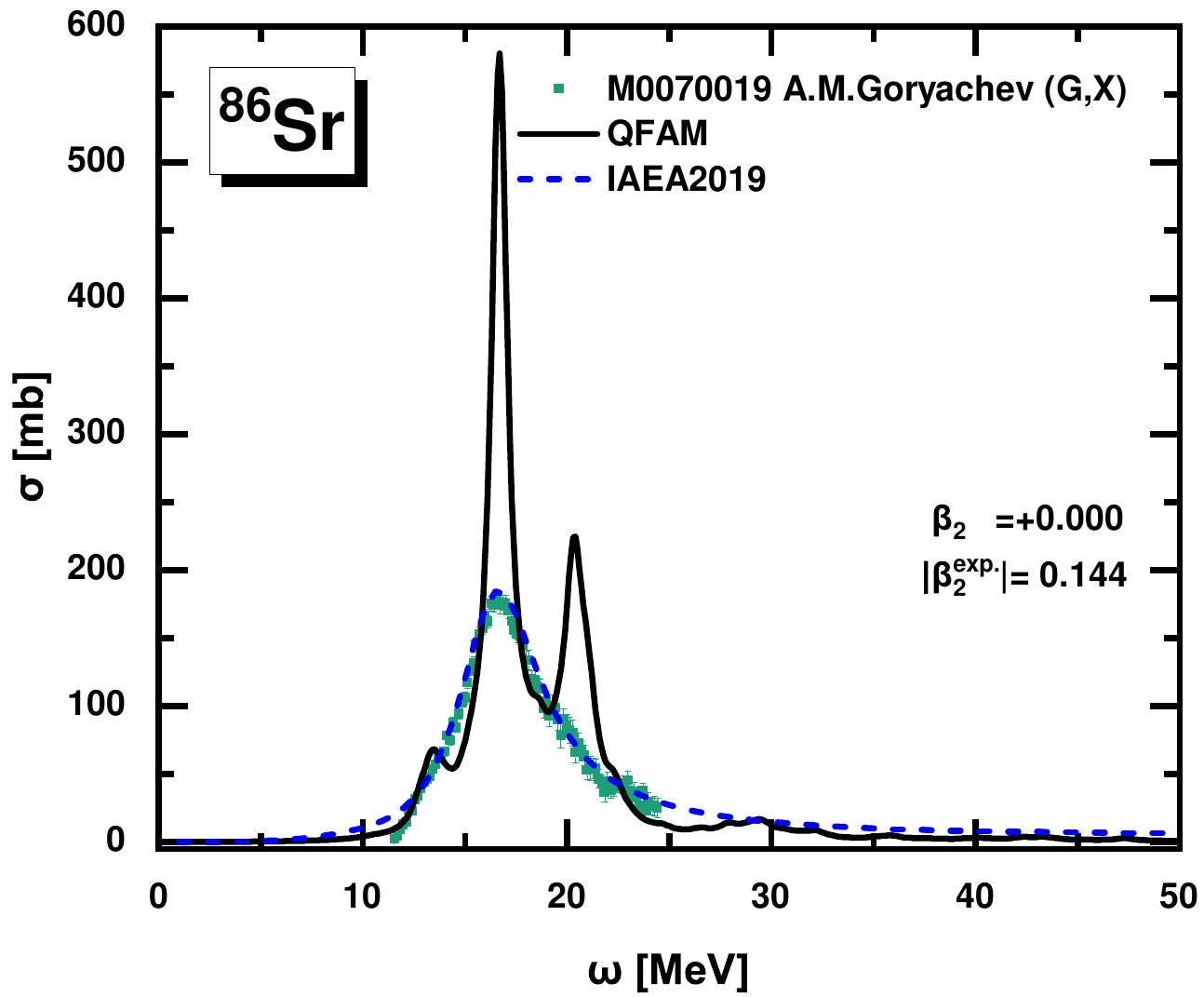}
    \includegraphics[width=0.35\textwidth]{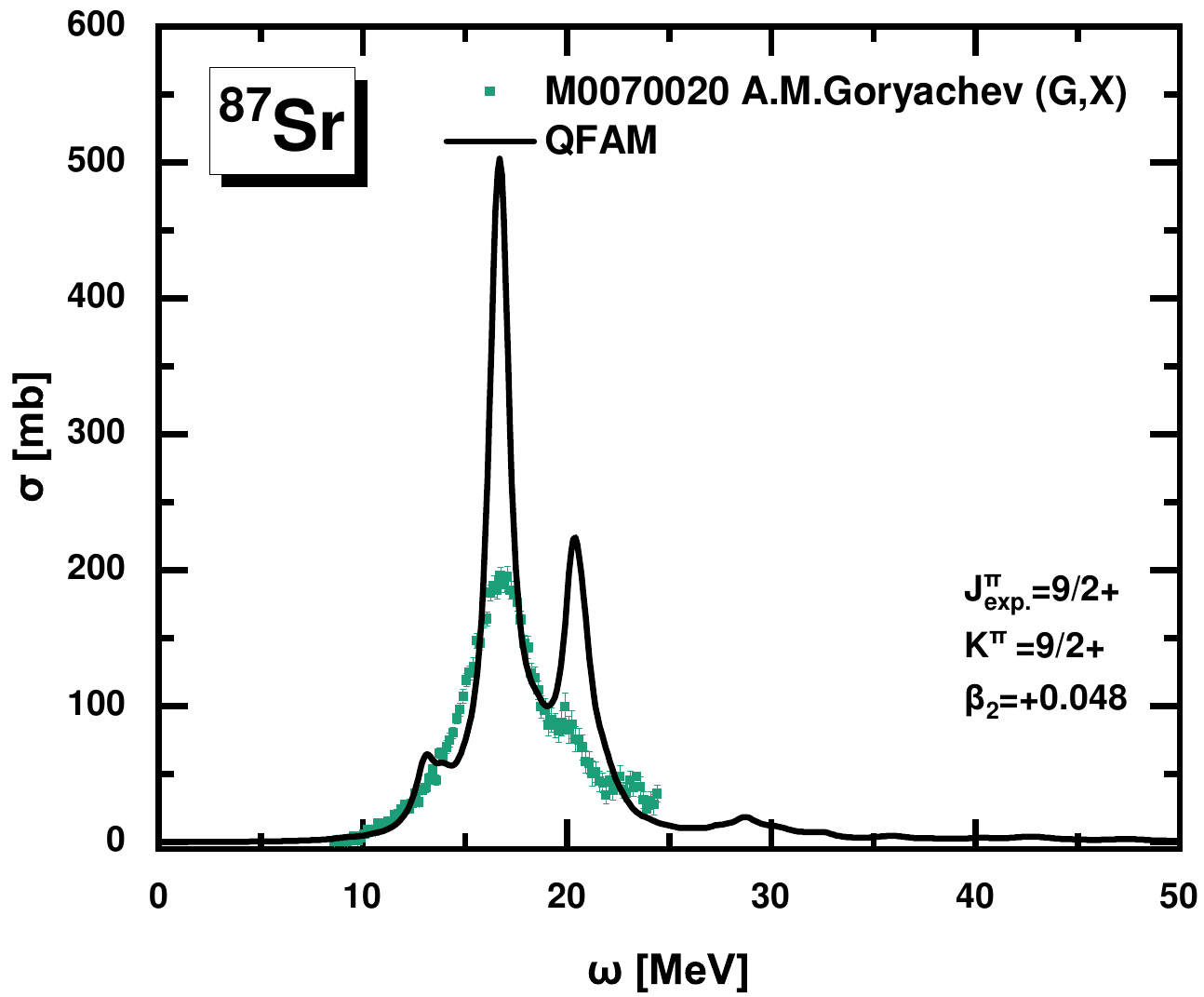}
    \includegraphics[width=0.35\textwidth]{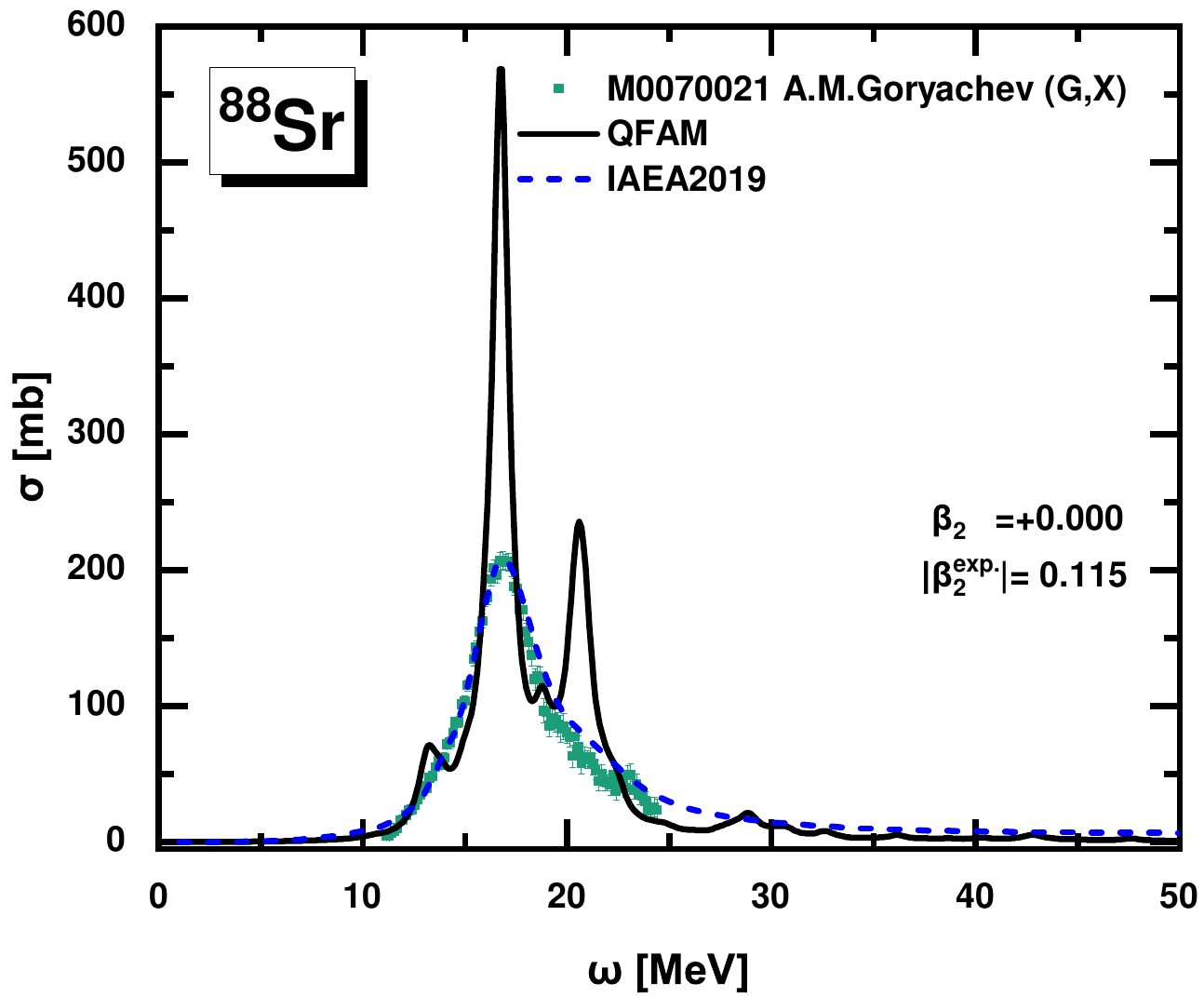}
    \includegraphics[width=0.35\textwidth]{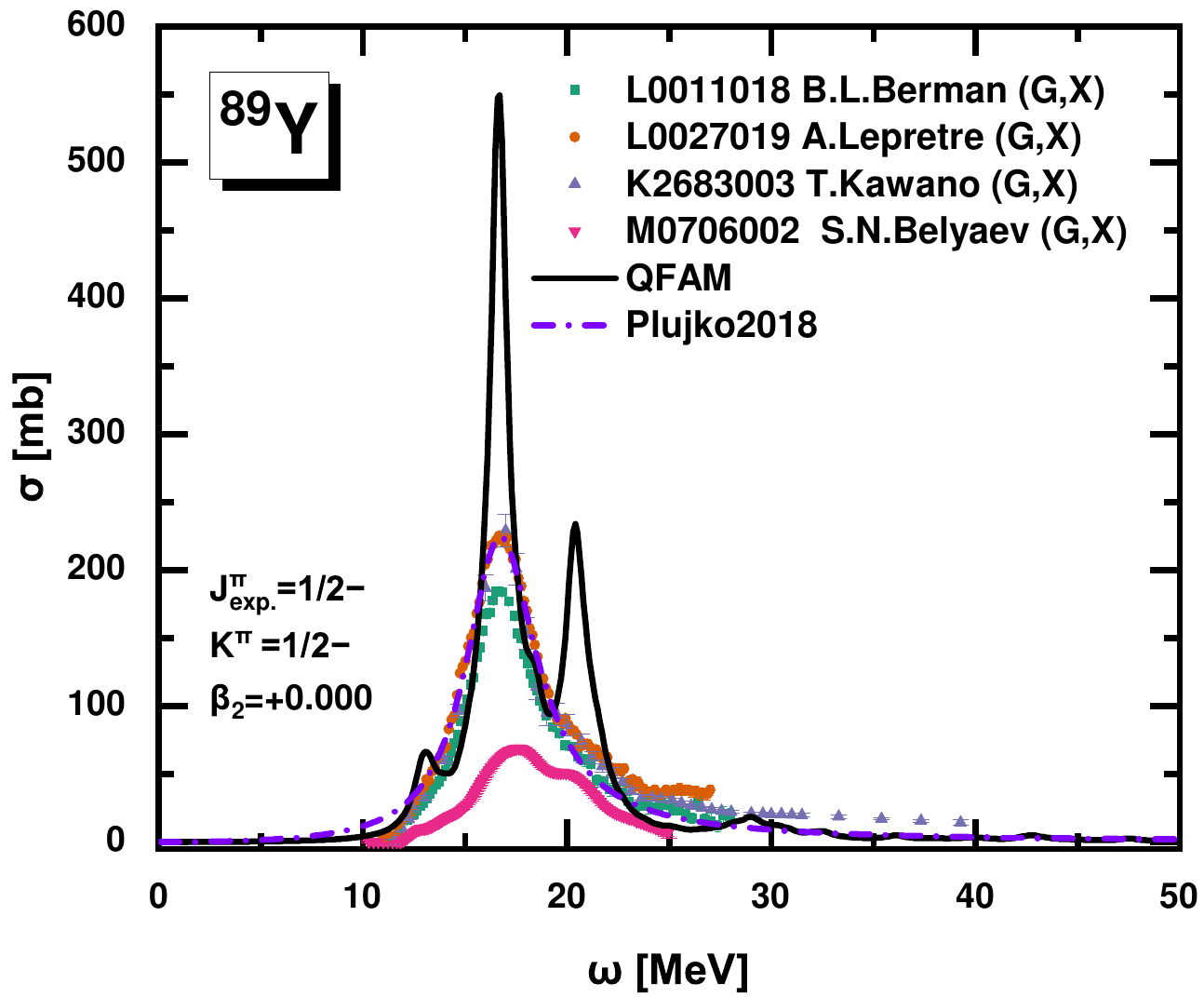}
    \includegraphics[width=0.35\textwidth]{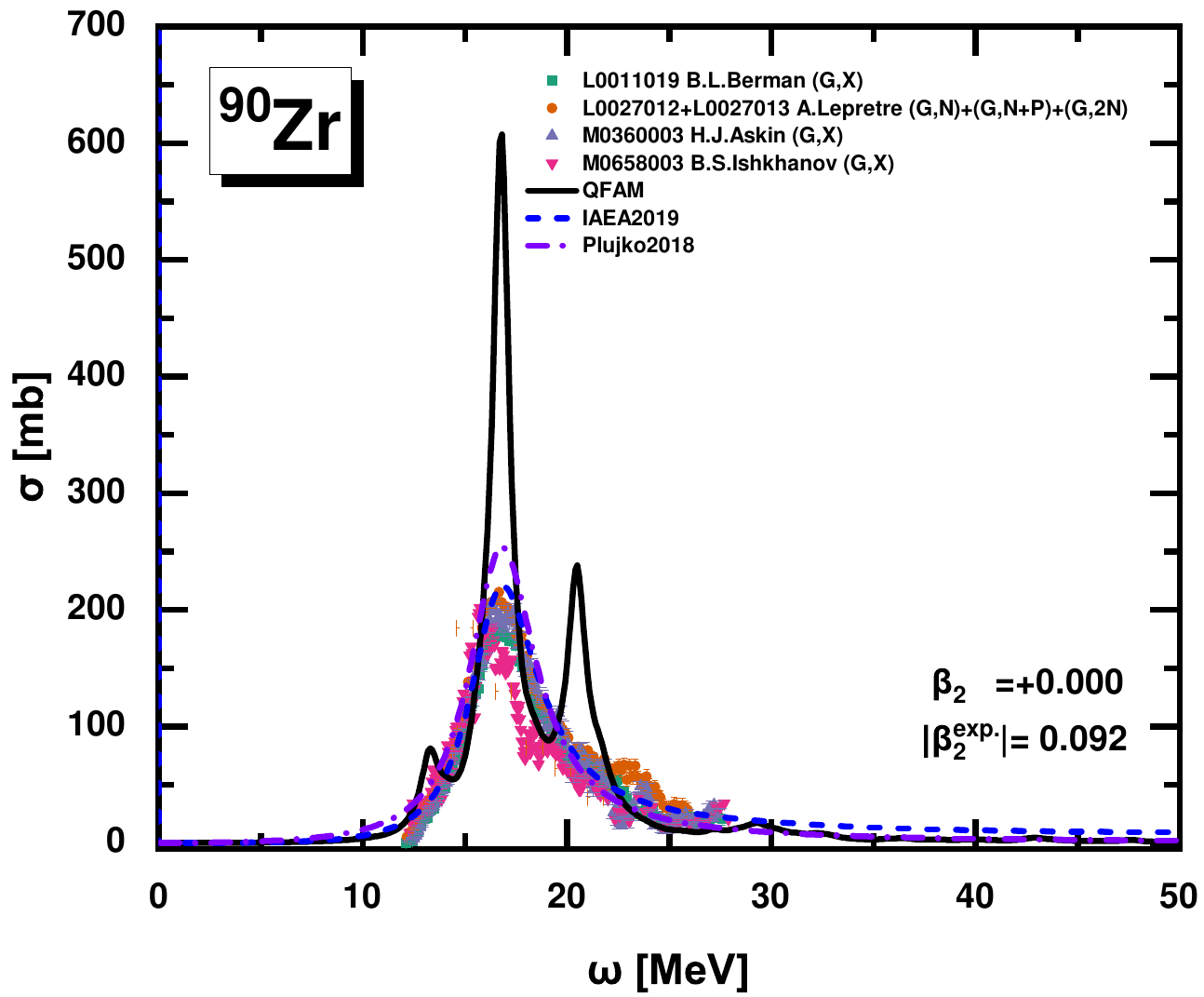}
\end{figure*}
\begin{figure*}\ContinuedFloat
    \centering
    \includegraphics[width=0.35\textwidth]{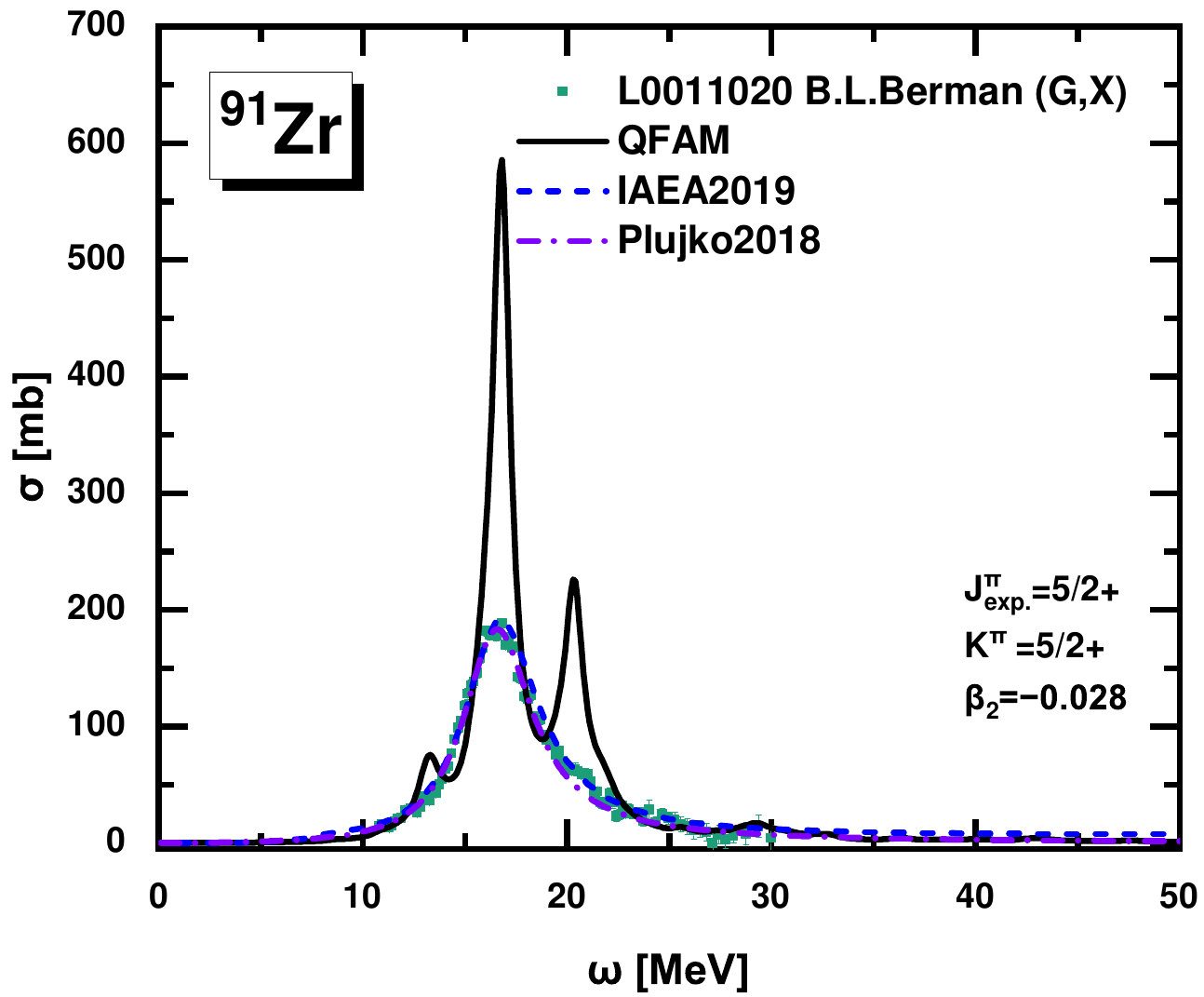}
    \includegraphics[width=0.35\textwidth]{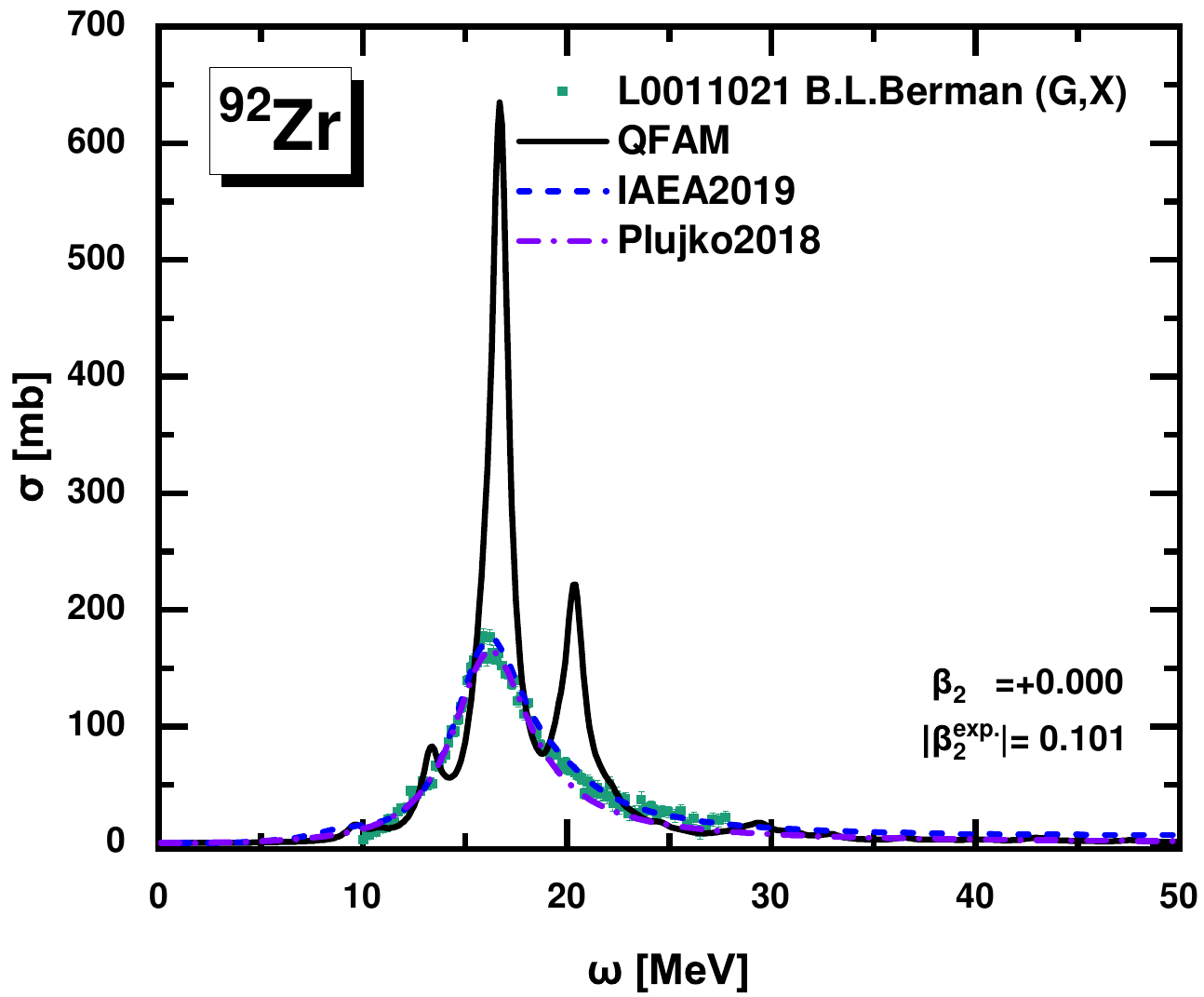}
    \includegraphics[width=0.35\textwidth]{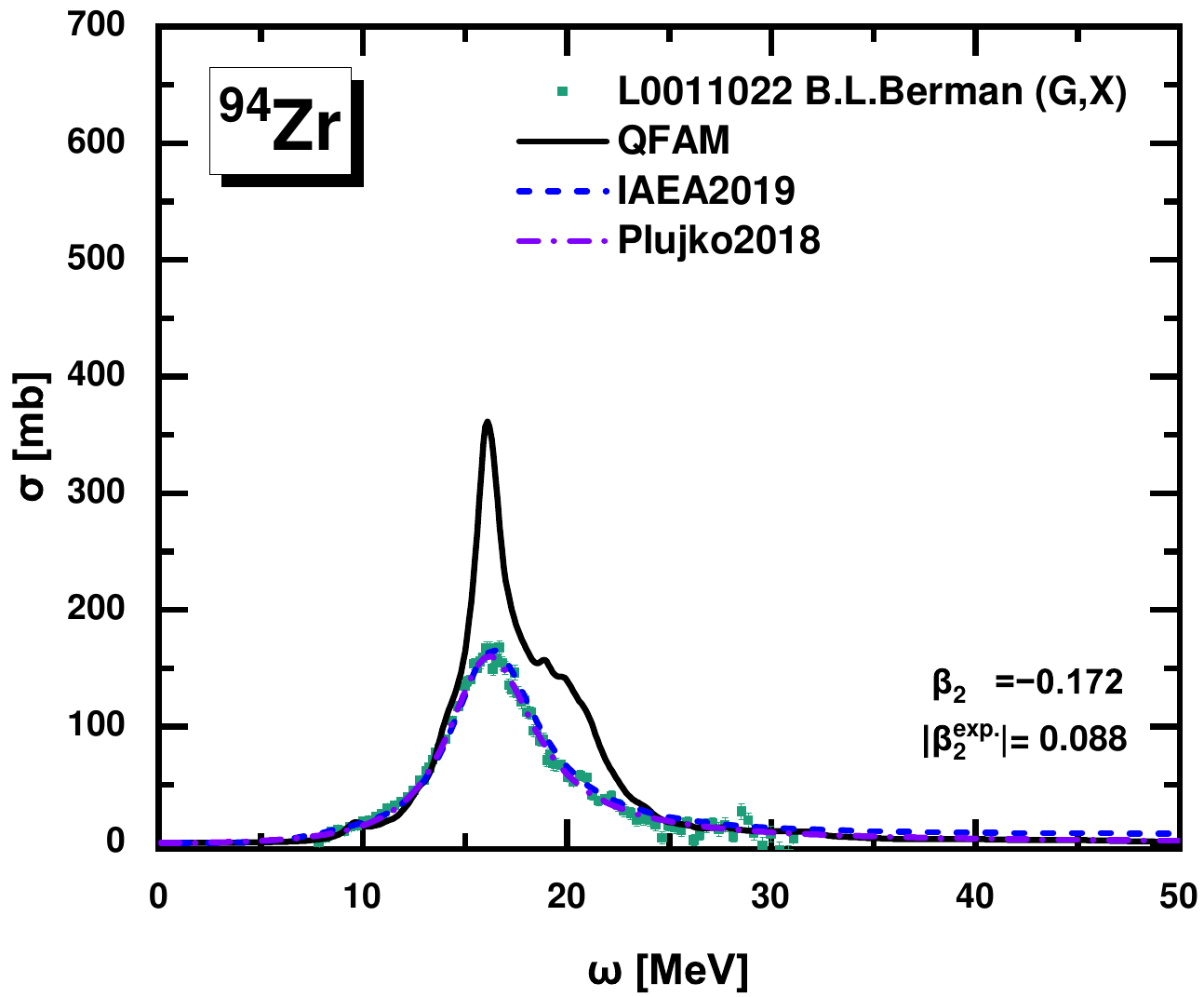}
    \includegraphics[width=0.35\textwidth]{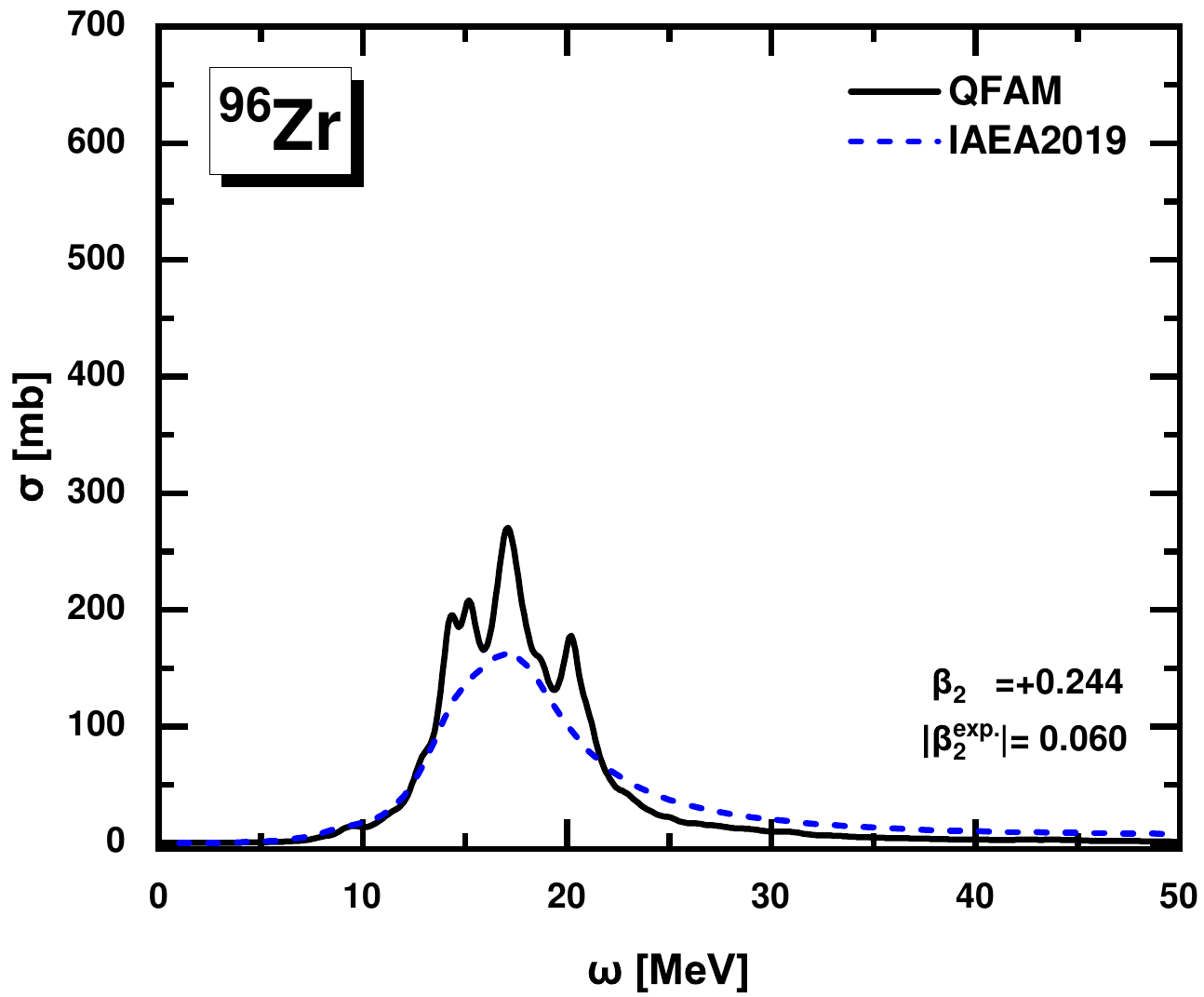}
    \includegraphics[width=0.35\textwidth]{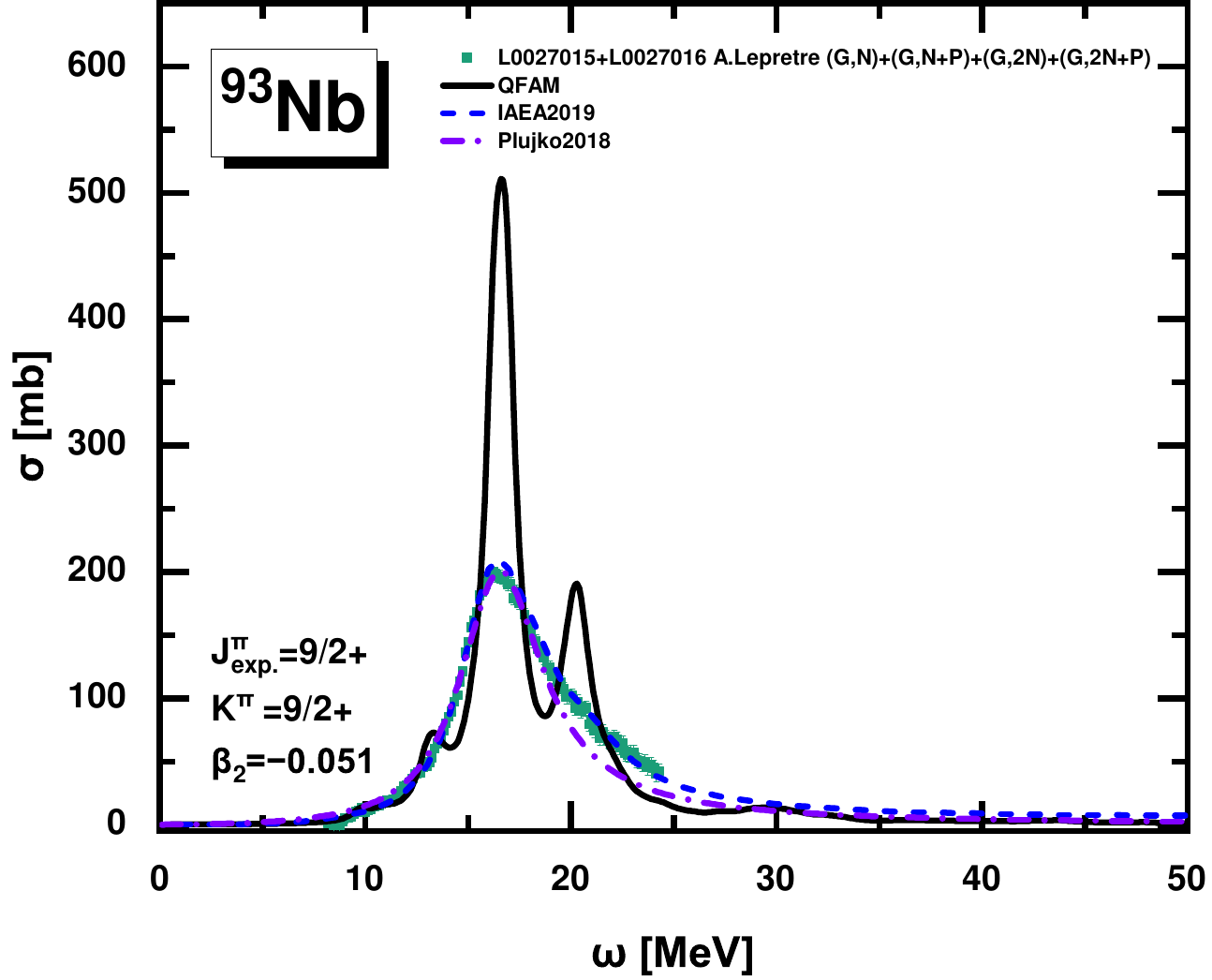}
    \includegraphics[width=0.35\textwidth]{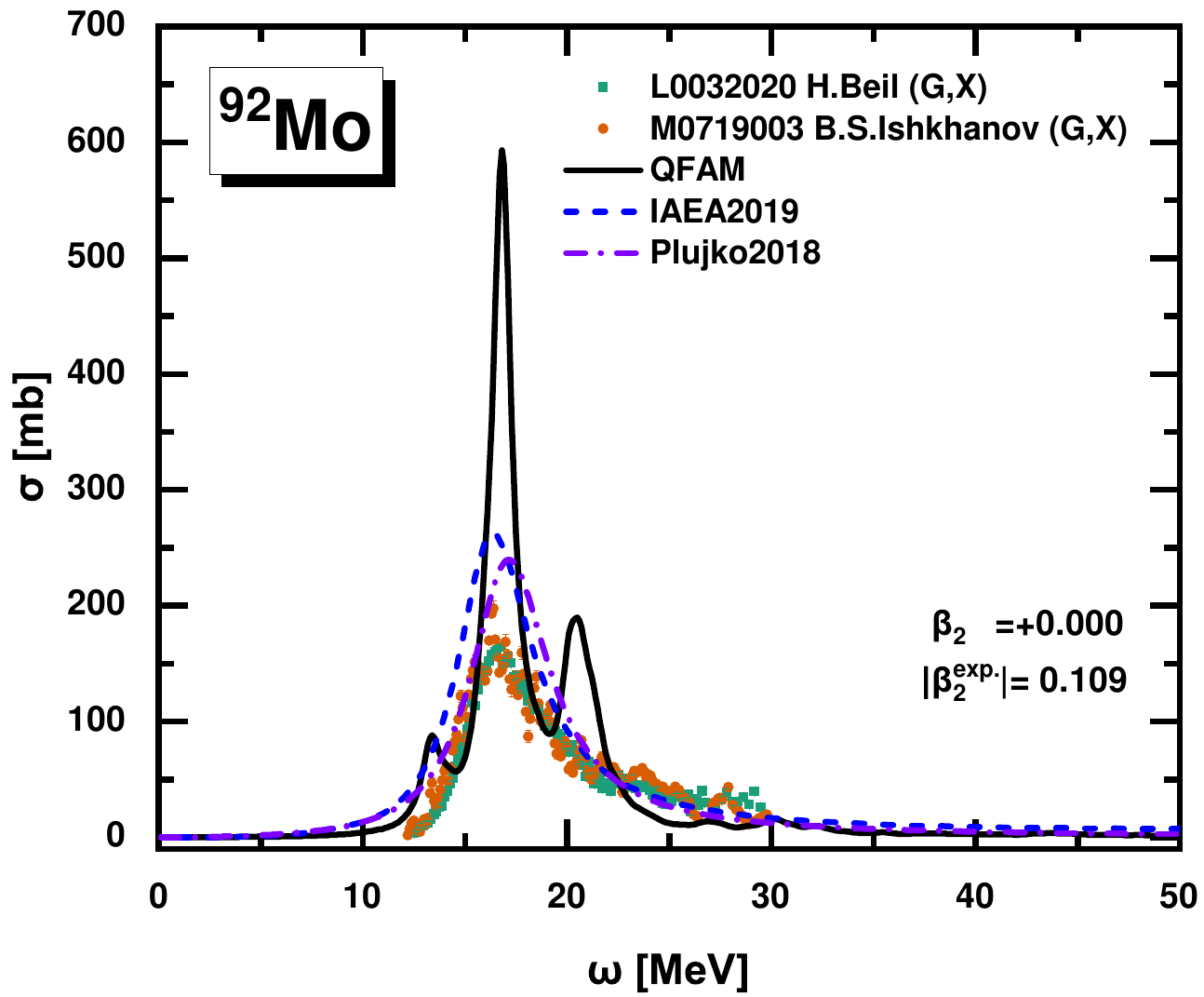}
    \includegraphics[width=0.35\textwidth]{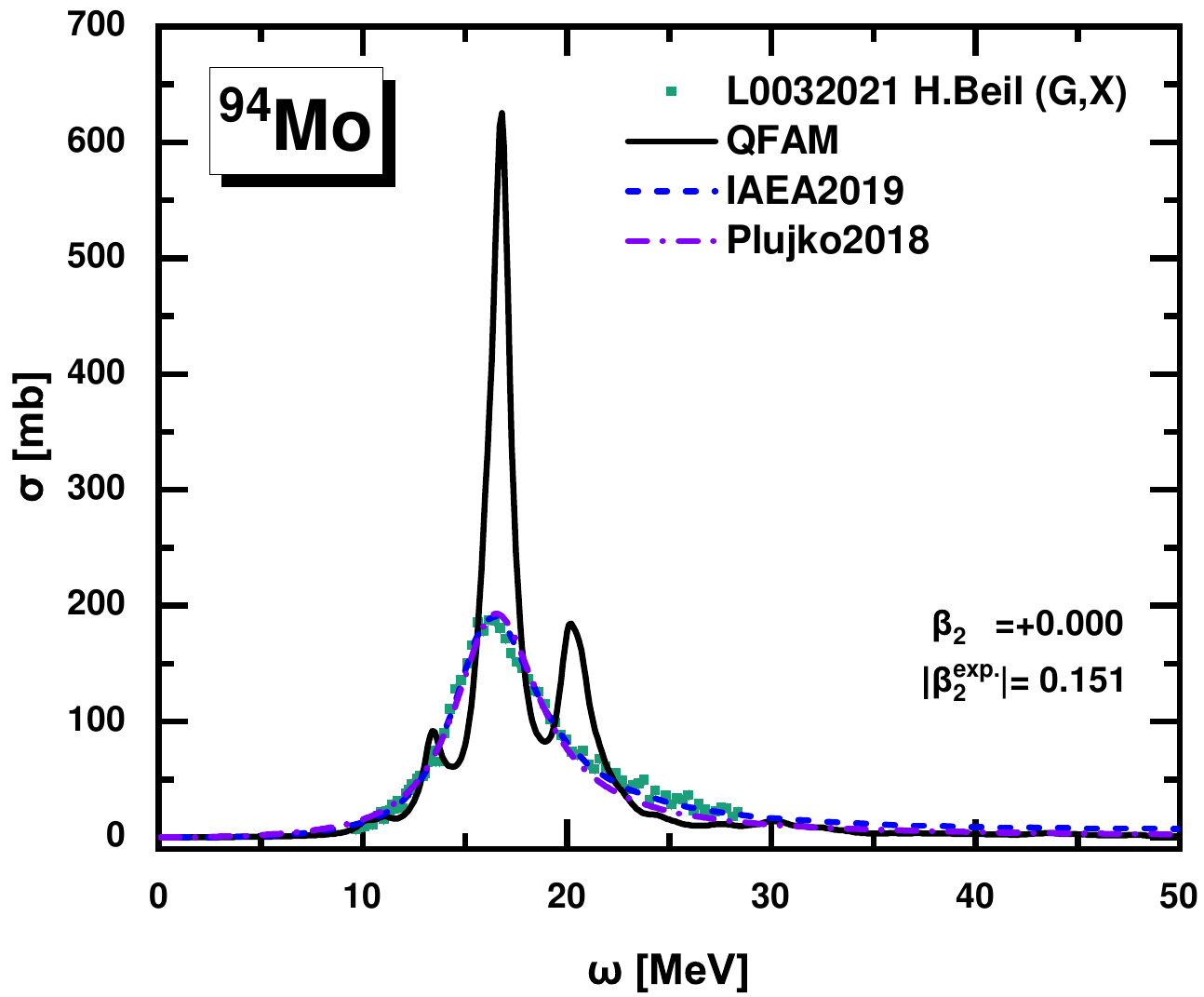}
    \includegraphics[width=0.35\textwidth]{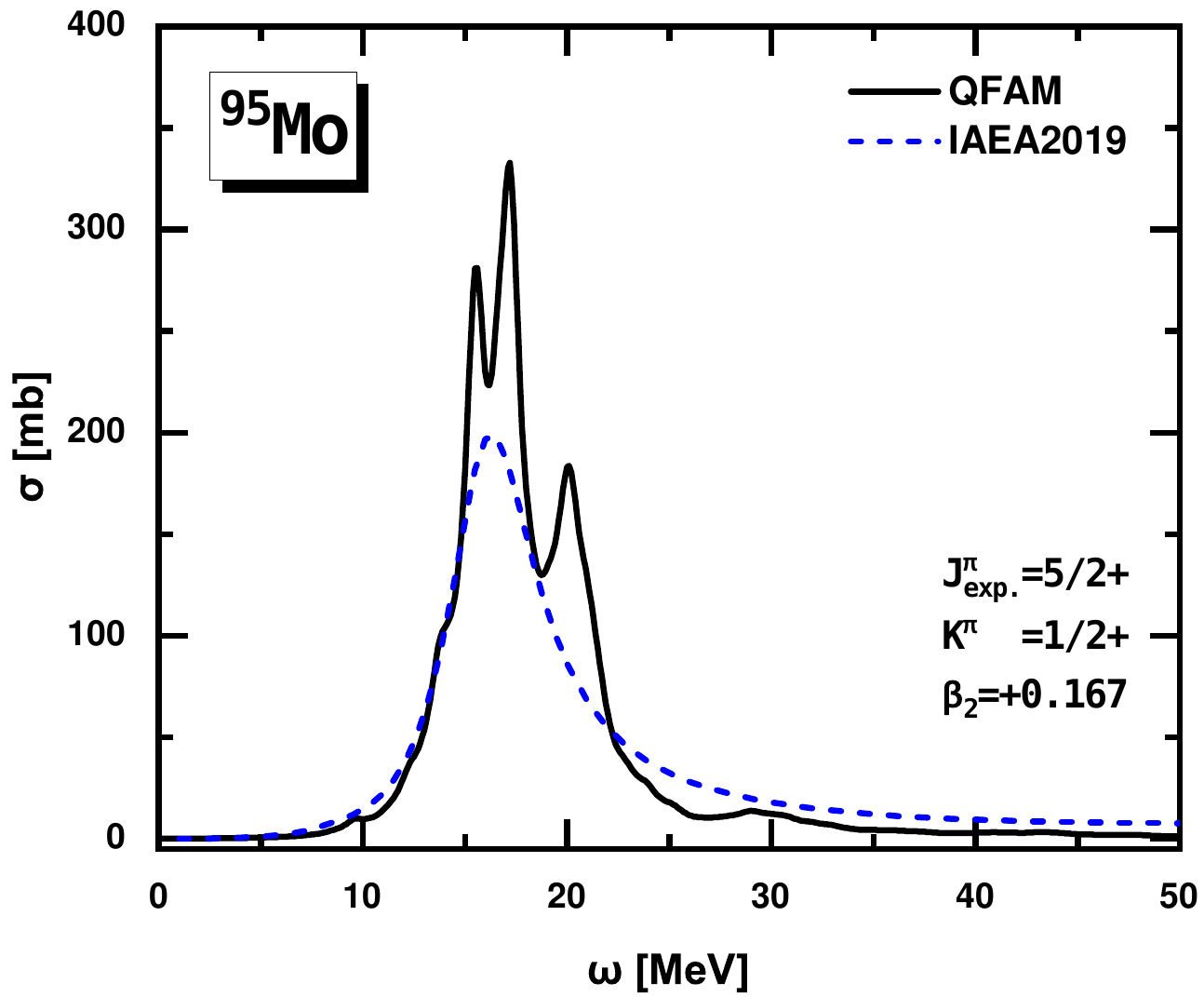}
\end{figure*}
\begin{figure*}\ContinuedFloat
    \centering
    \includegraphics[width=0.35\textwidth]{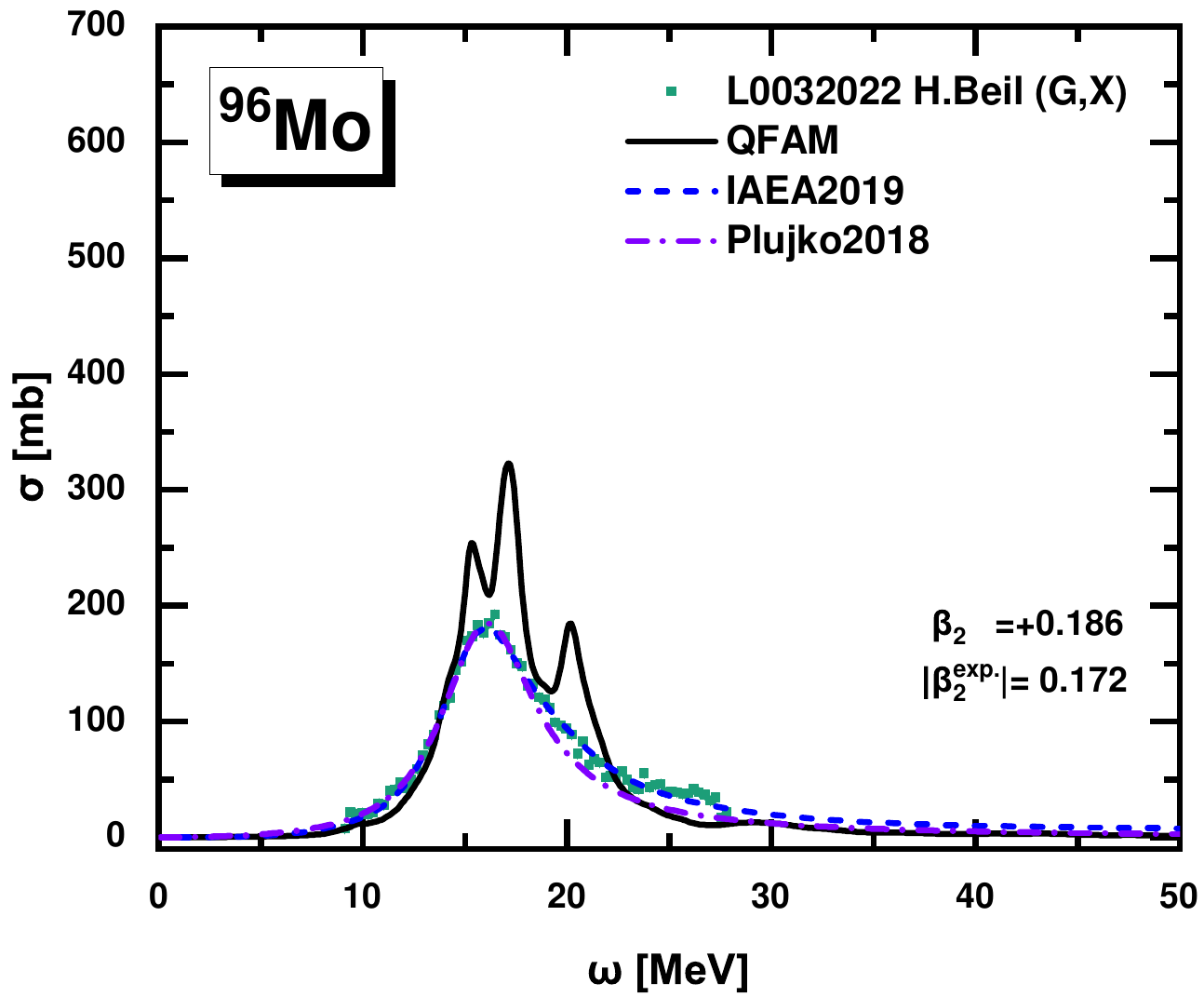}
    \includegraphics[width=0.35\textwidth]{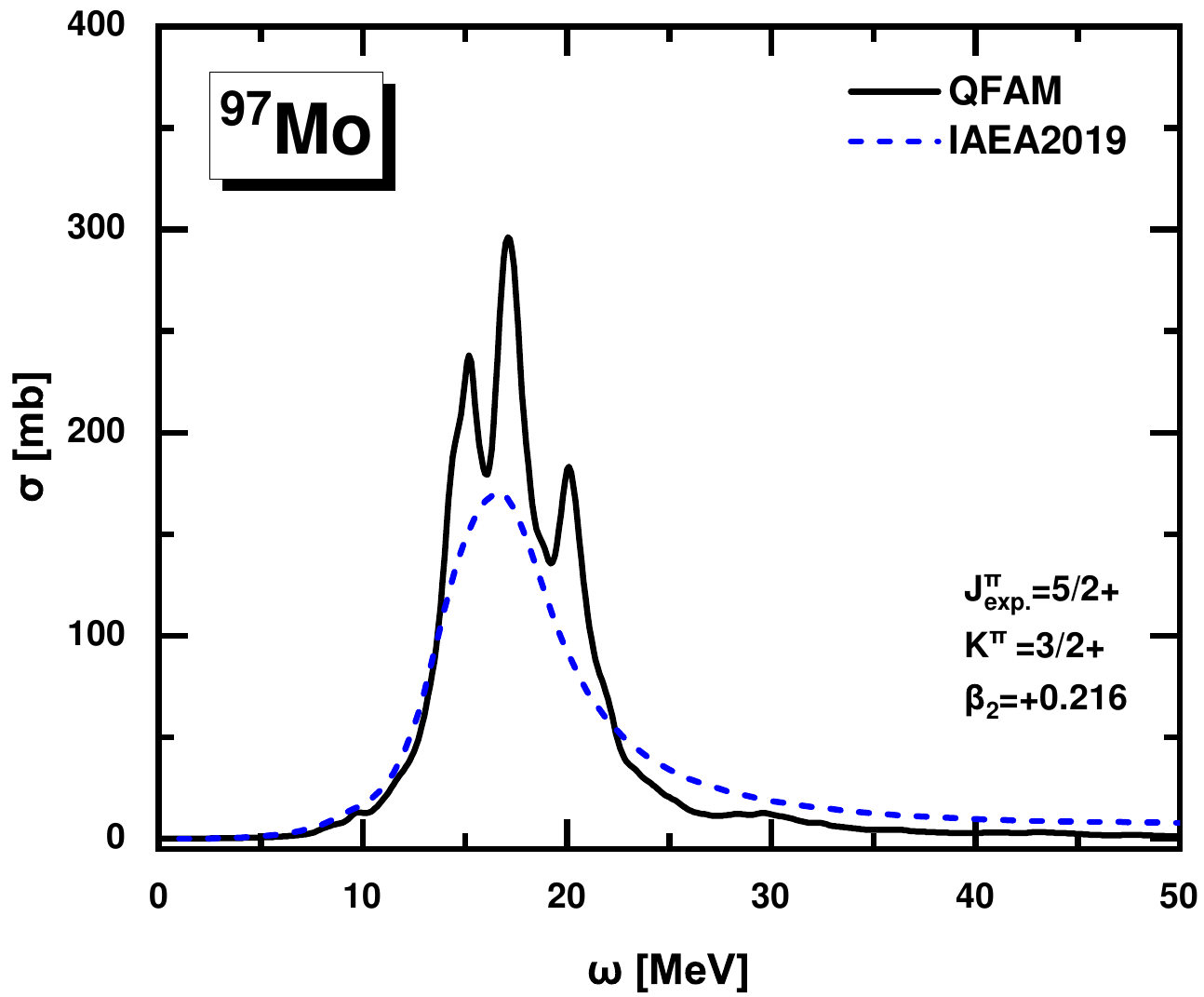}
    \includegraphics[width=0.35\textwidth]{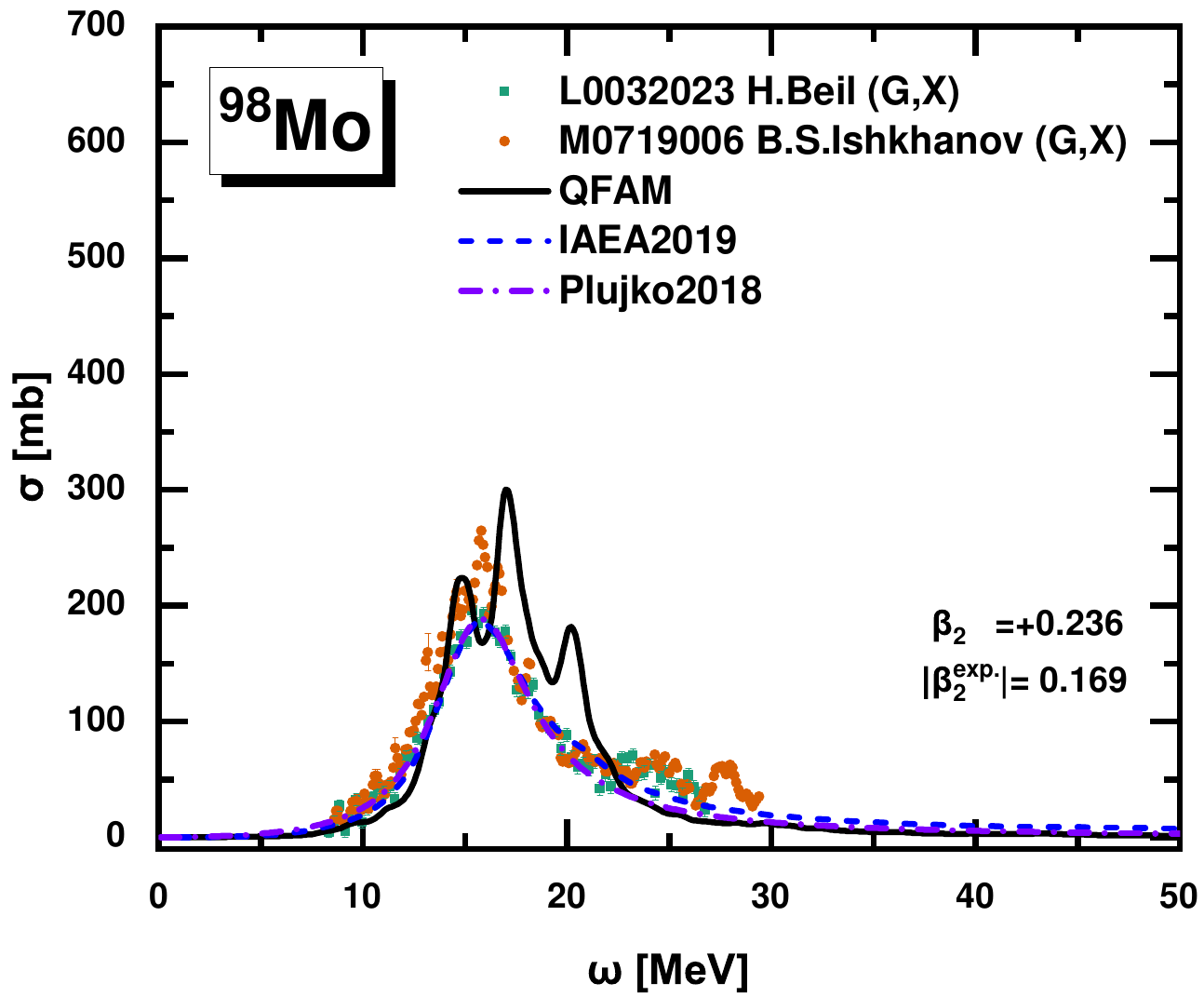}
    \includegraphics[width=0.35\textwidth]{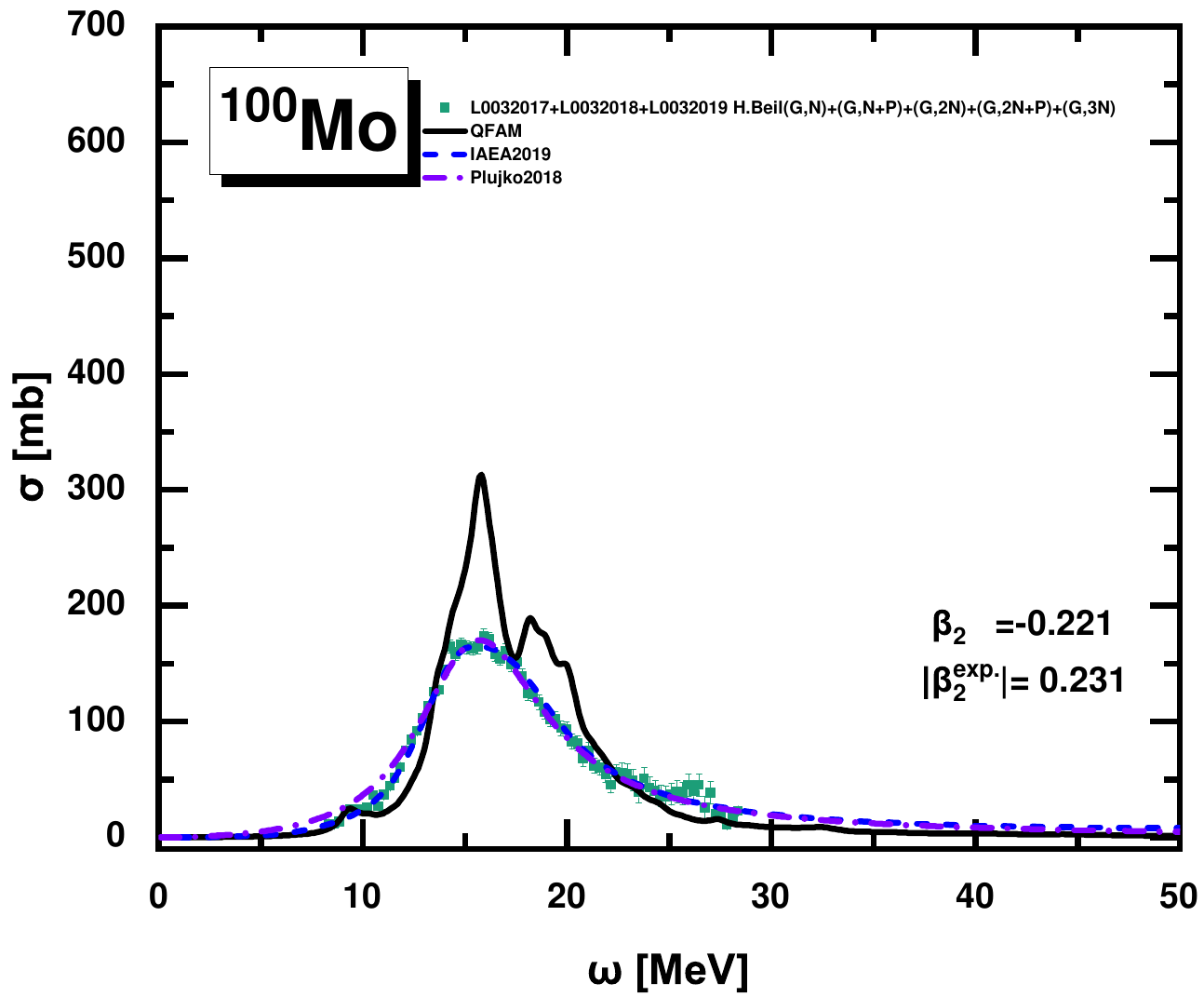}
    \includegraphics[width=0.35\textwidth]{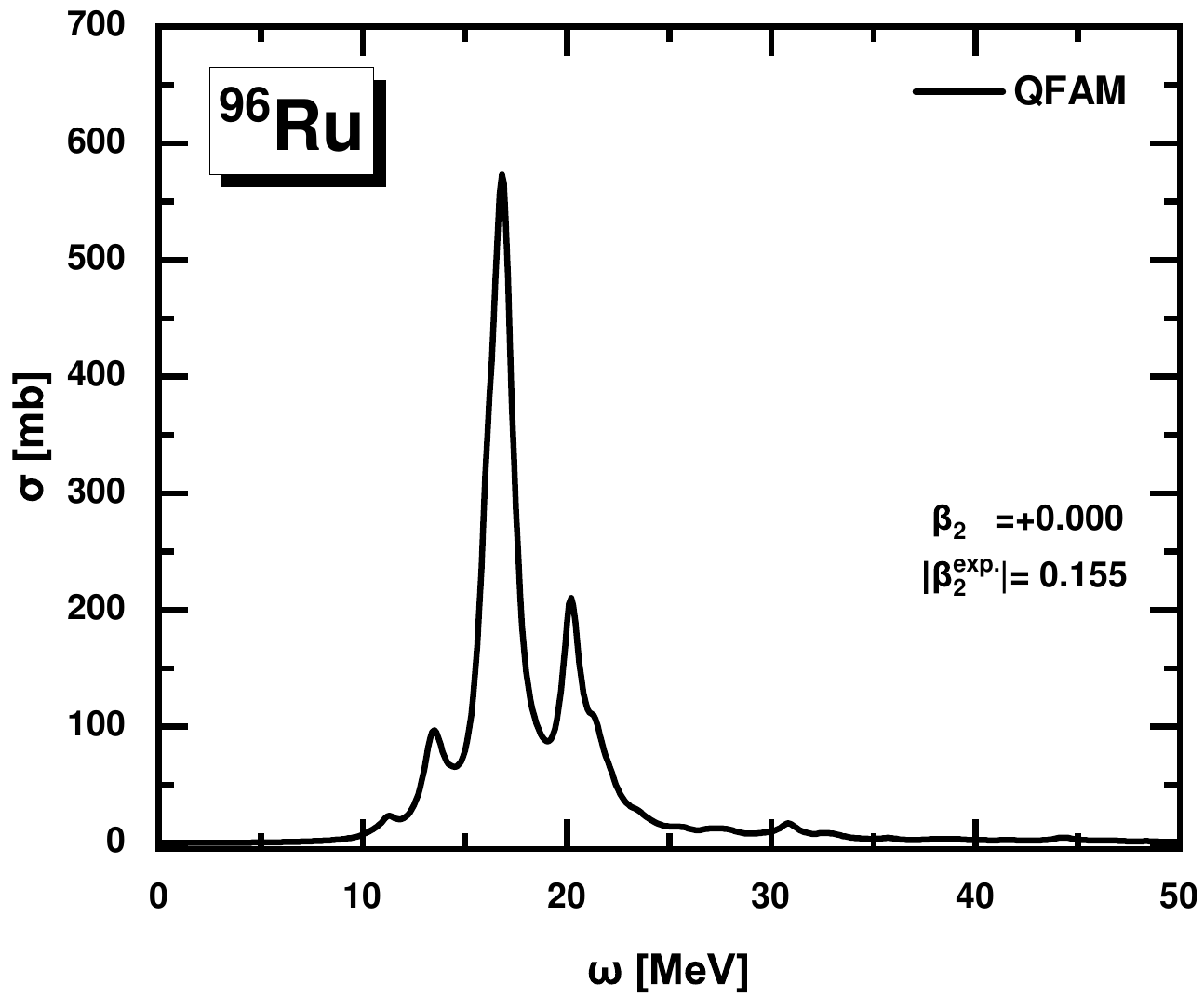}
    \includegraphics[width=0.35\textwidth]{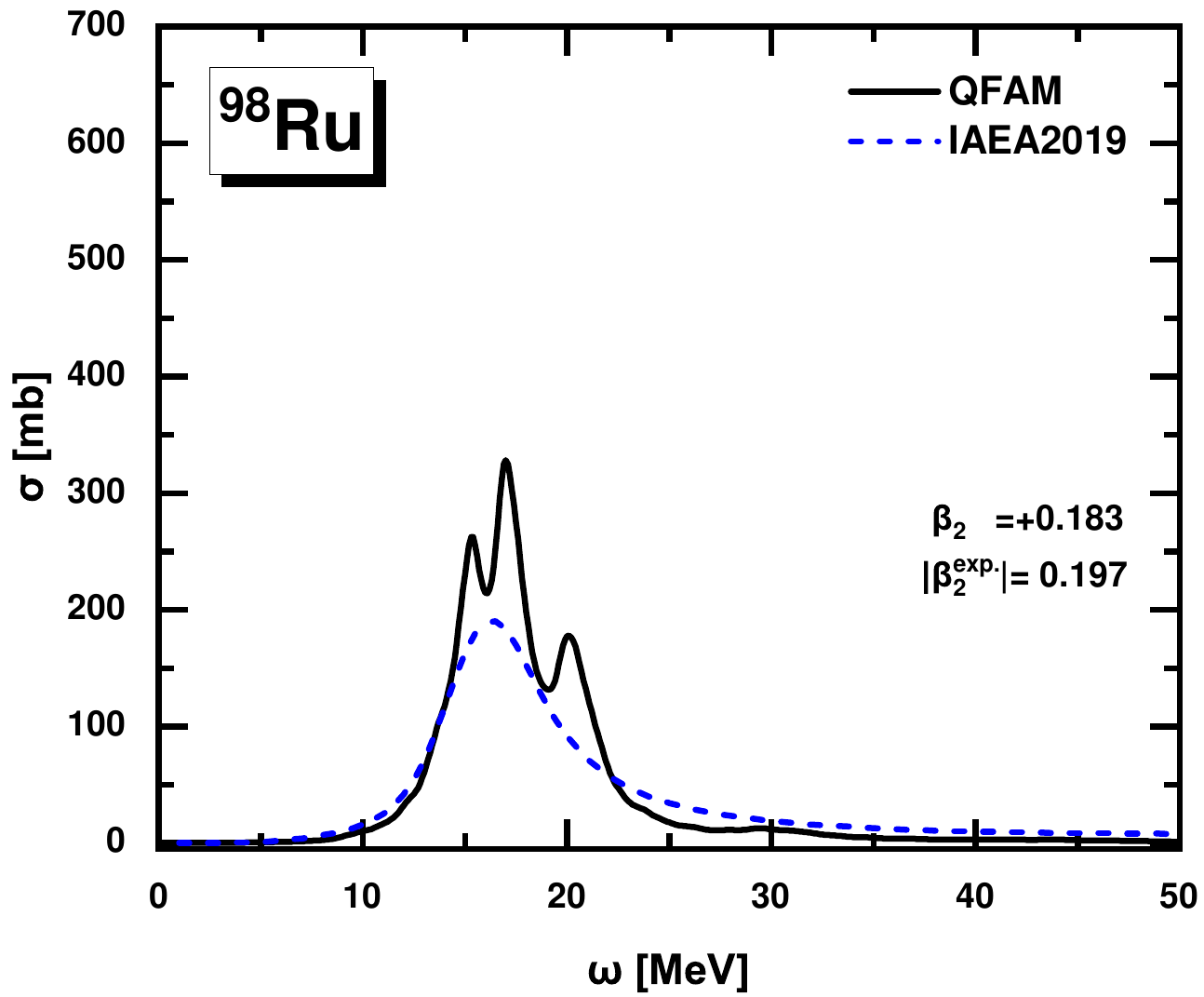}
    \includegraphics[width=0.35\textwidth]{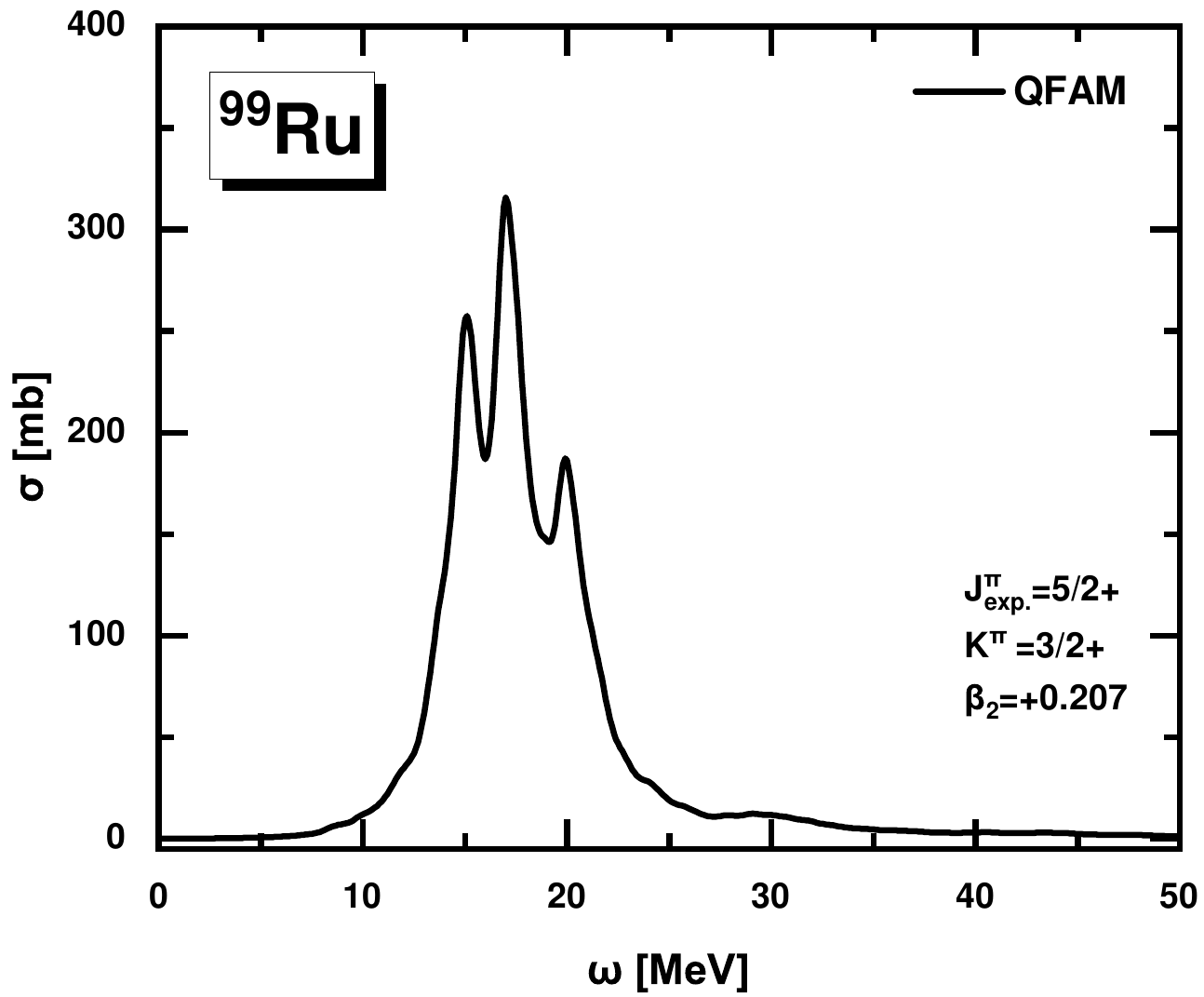}
    \includegraphics[width=0.35\textwidth]{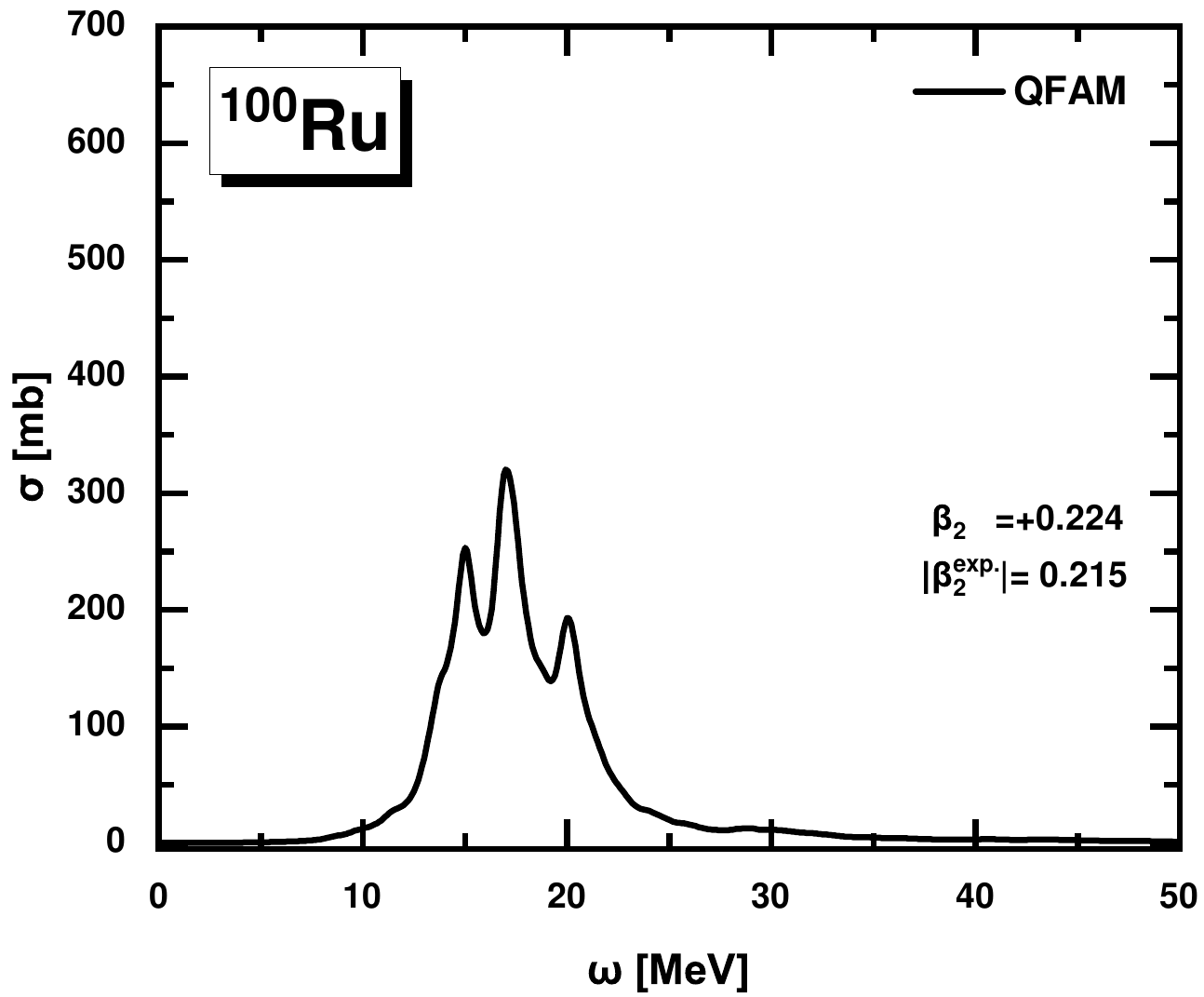}
\end{figure*}
\begin{figure*}\ContinuedFloat
    \centering
    \includegraphics[width=0.35\textwidth]{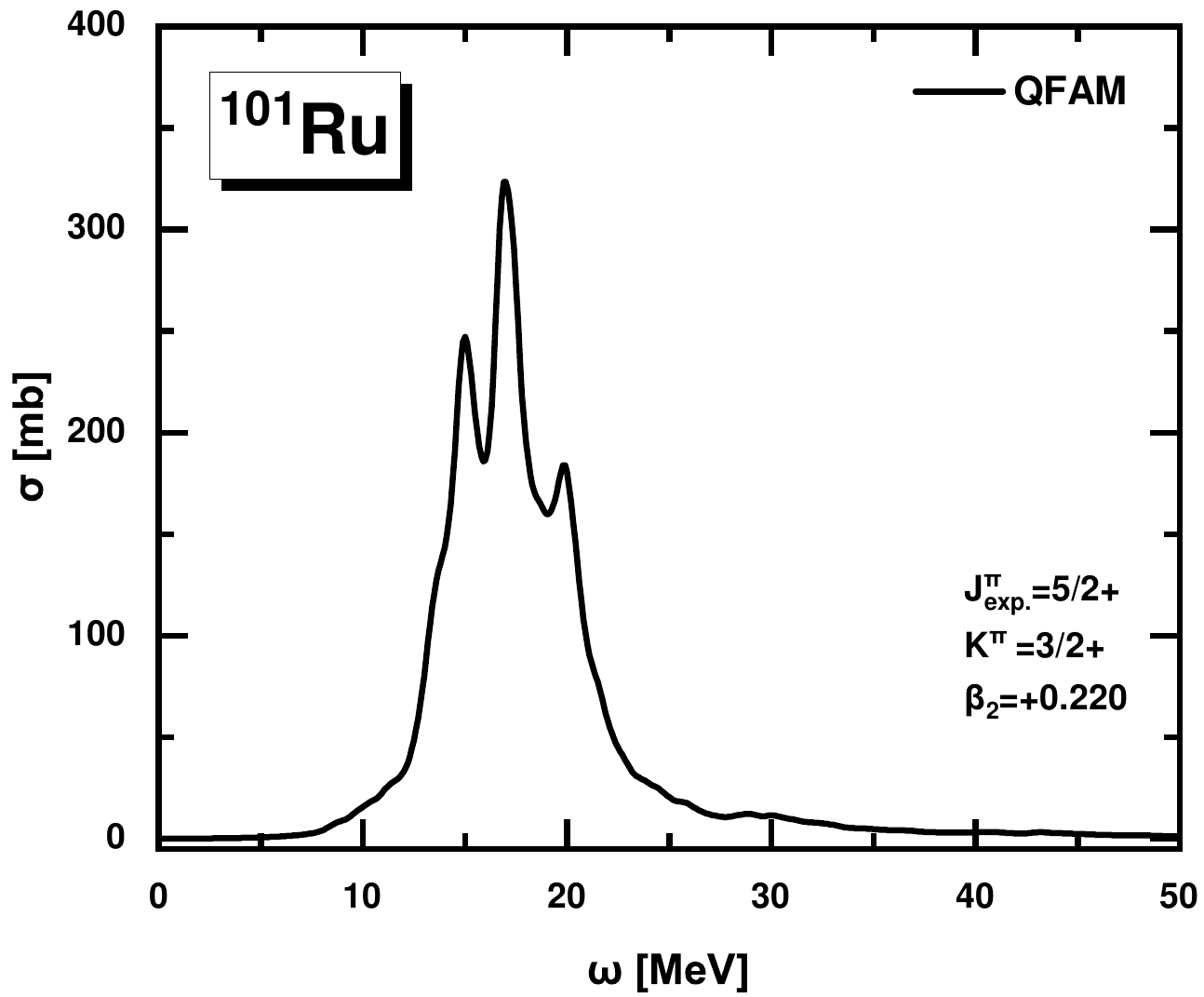}
    \includegraphics[width=0.35\textwidth]{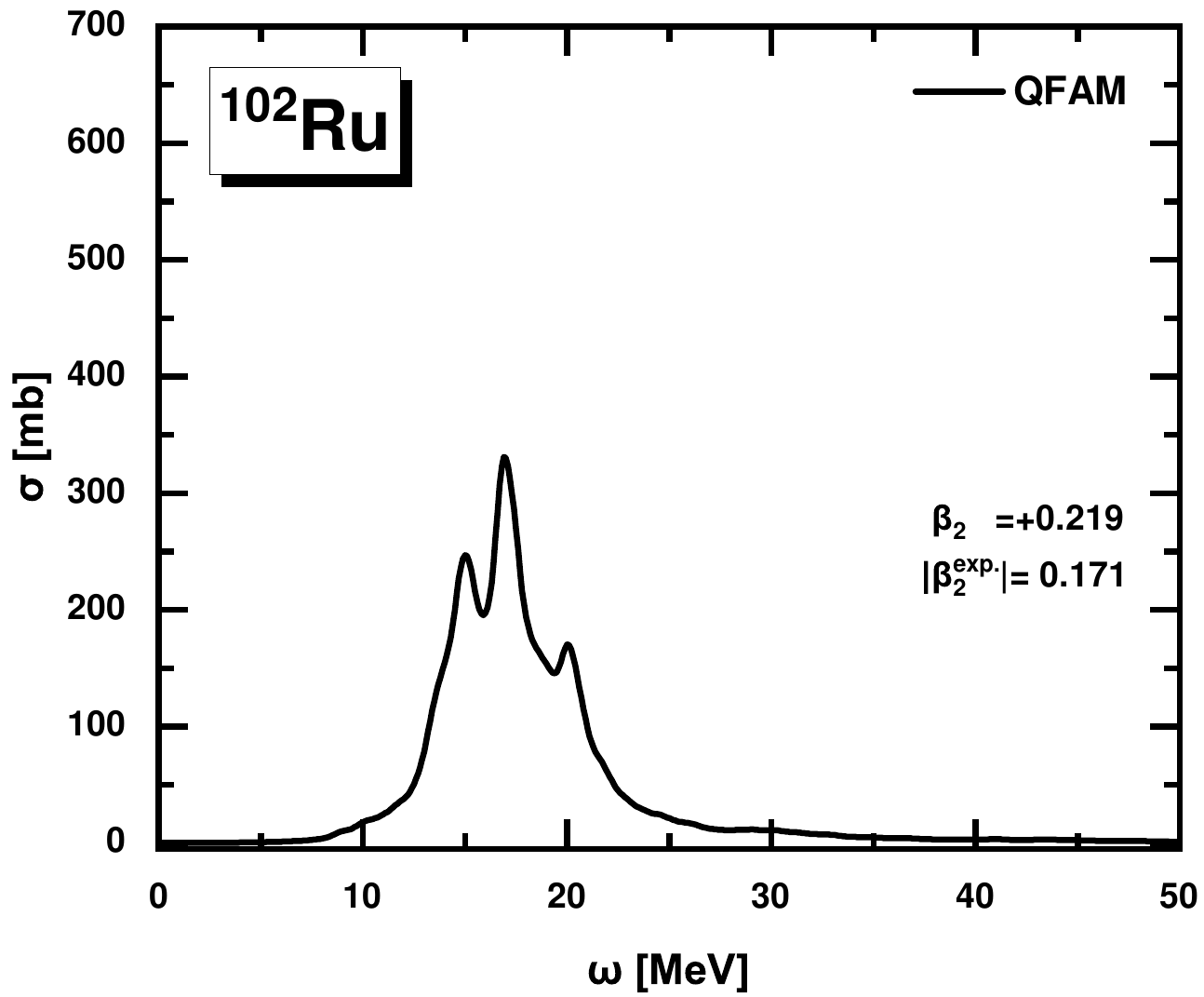}
    \includegraphics[width=0.35\textwidth]{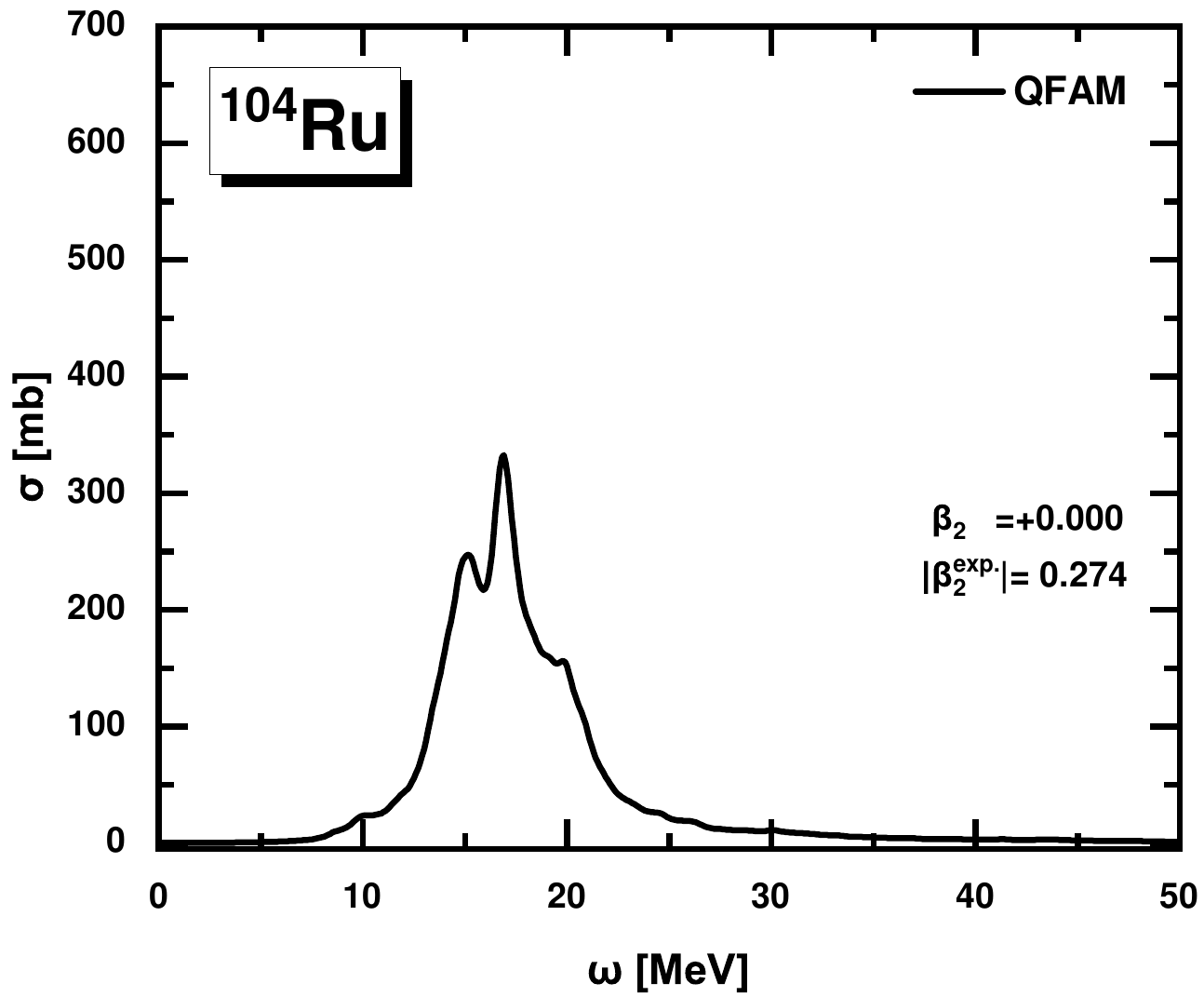}
    \includegraphics[width=0.35\textwidth]{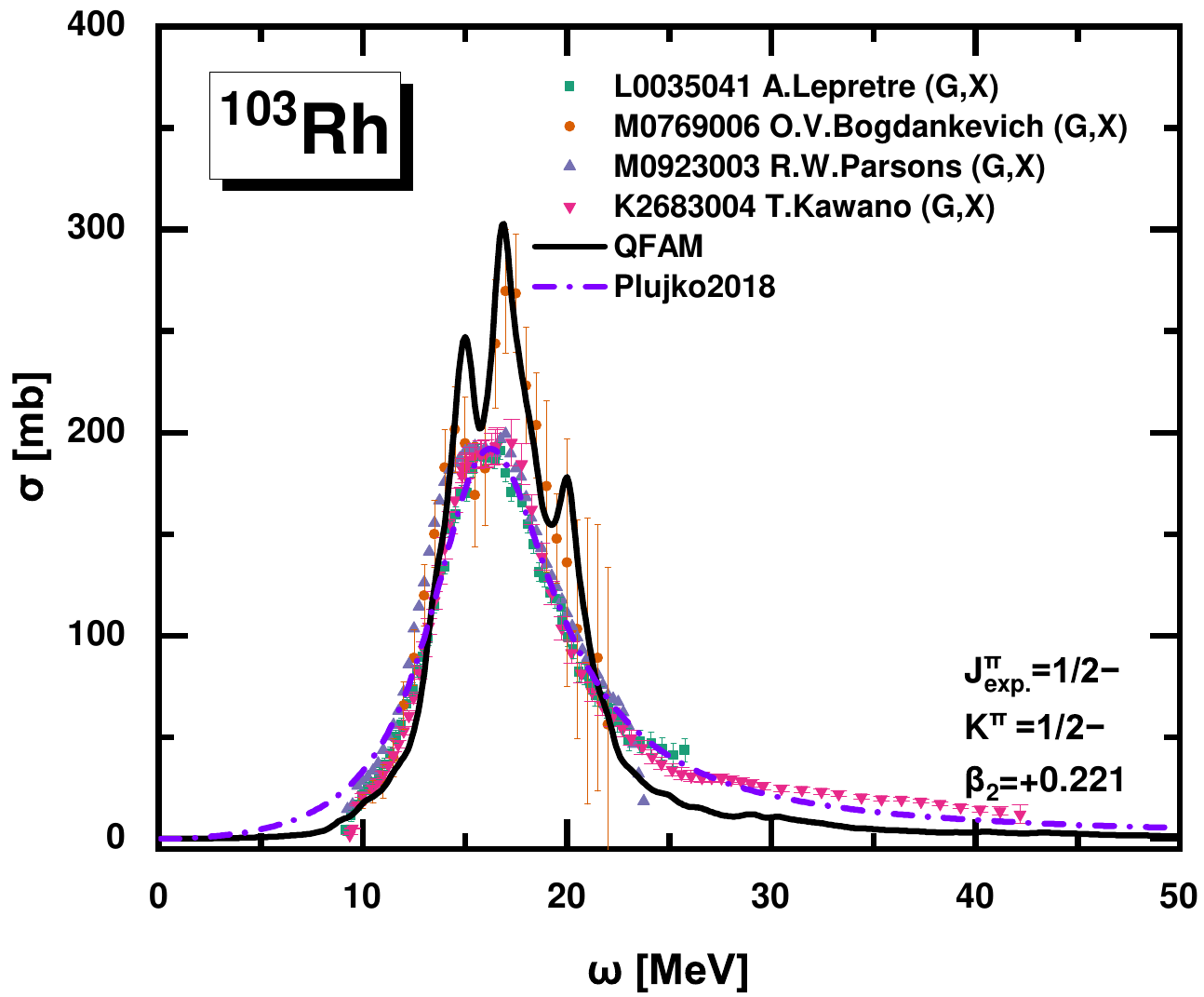}
    \includegraphics[width=0.35\textwidth]{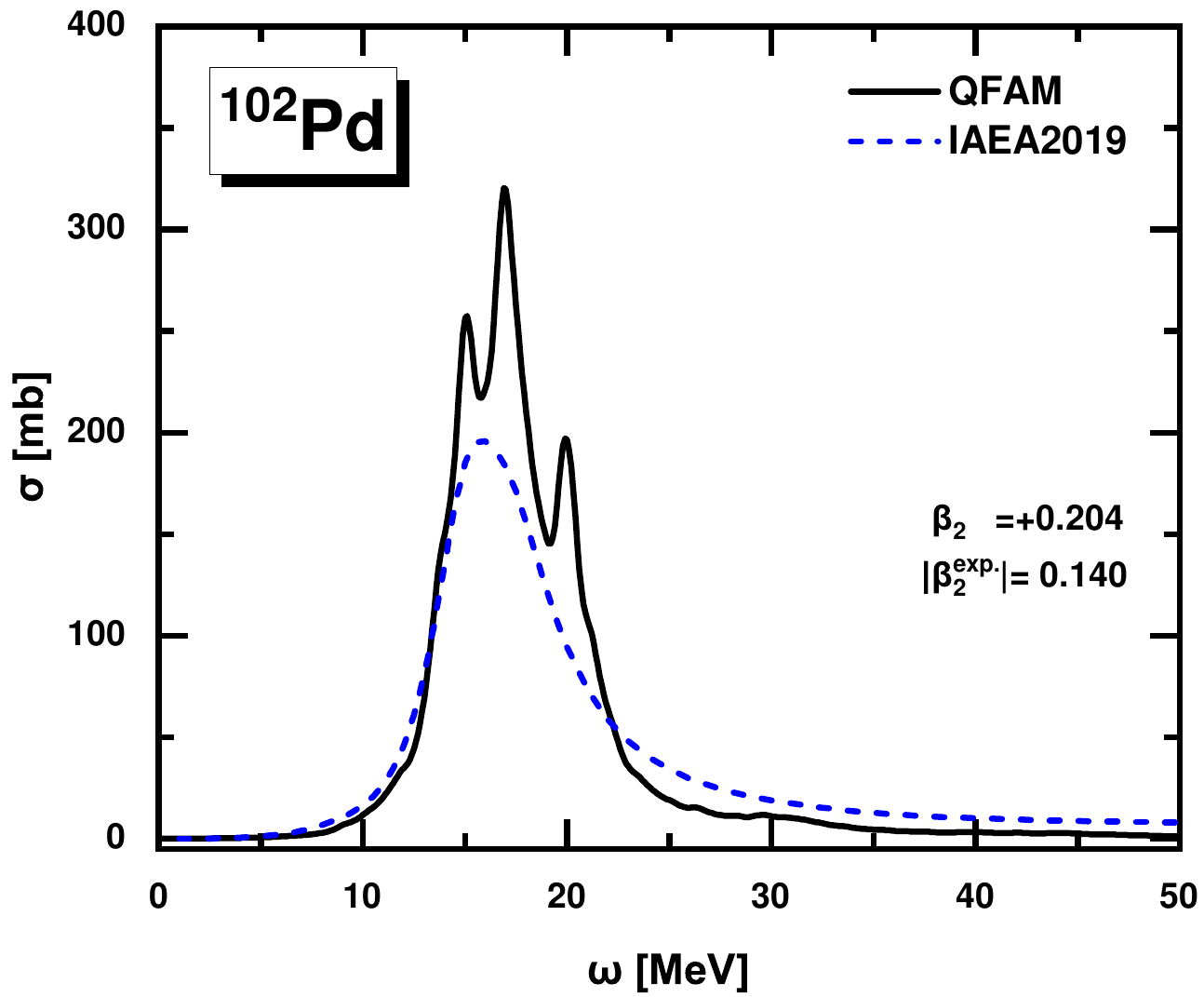}
    \includegraphics[width=0.35\textwidth]{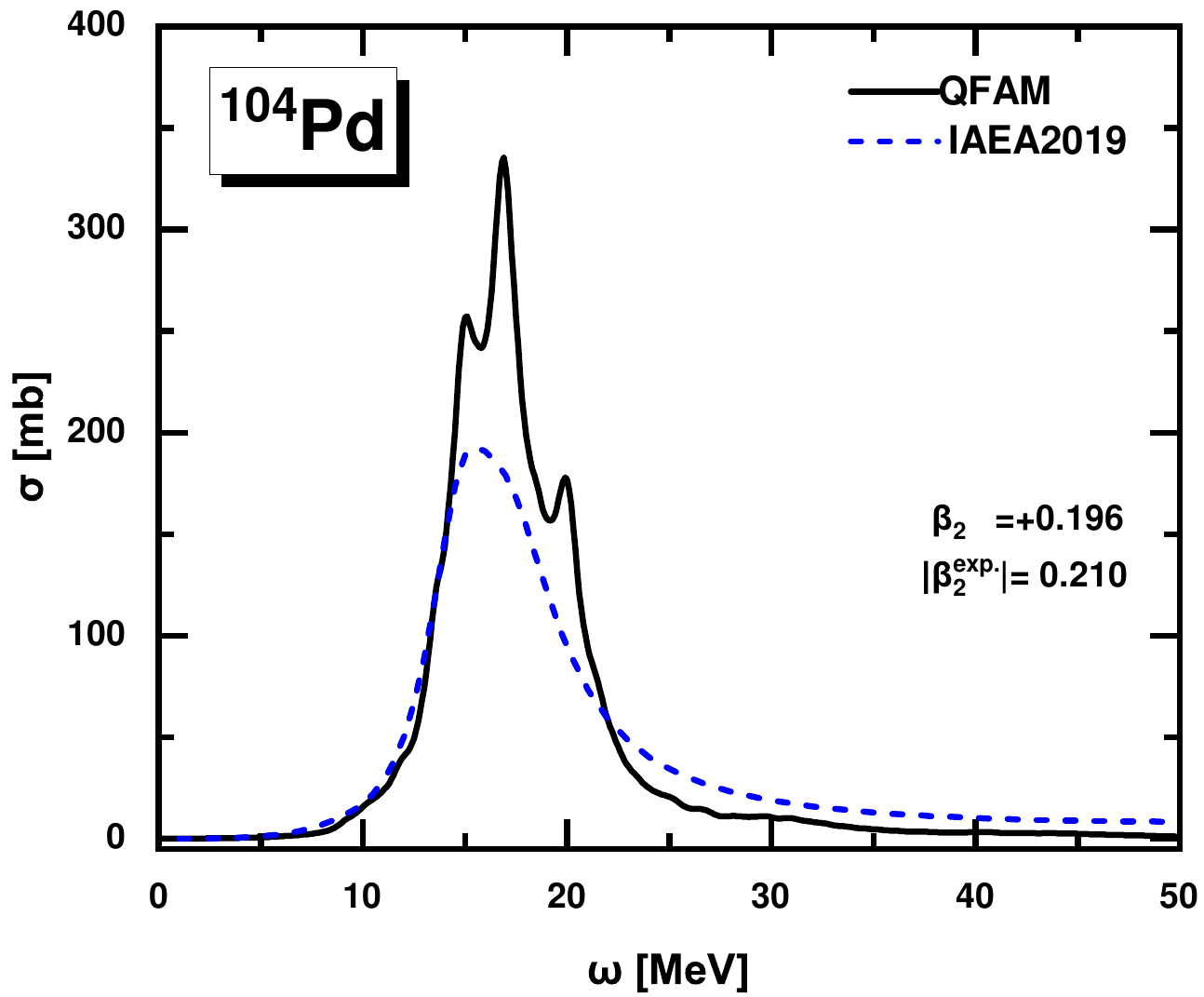}
    \includegraphics[width=0.35\textwidth]{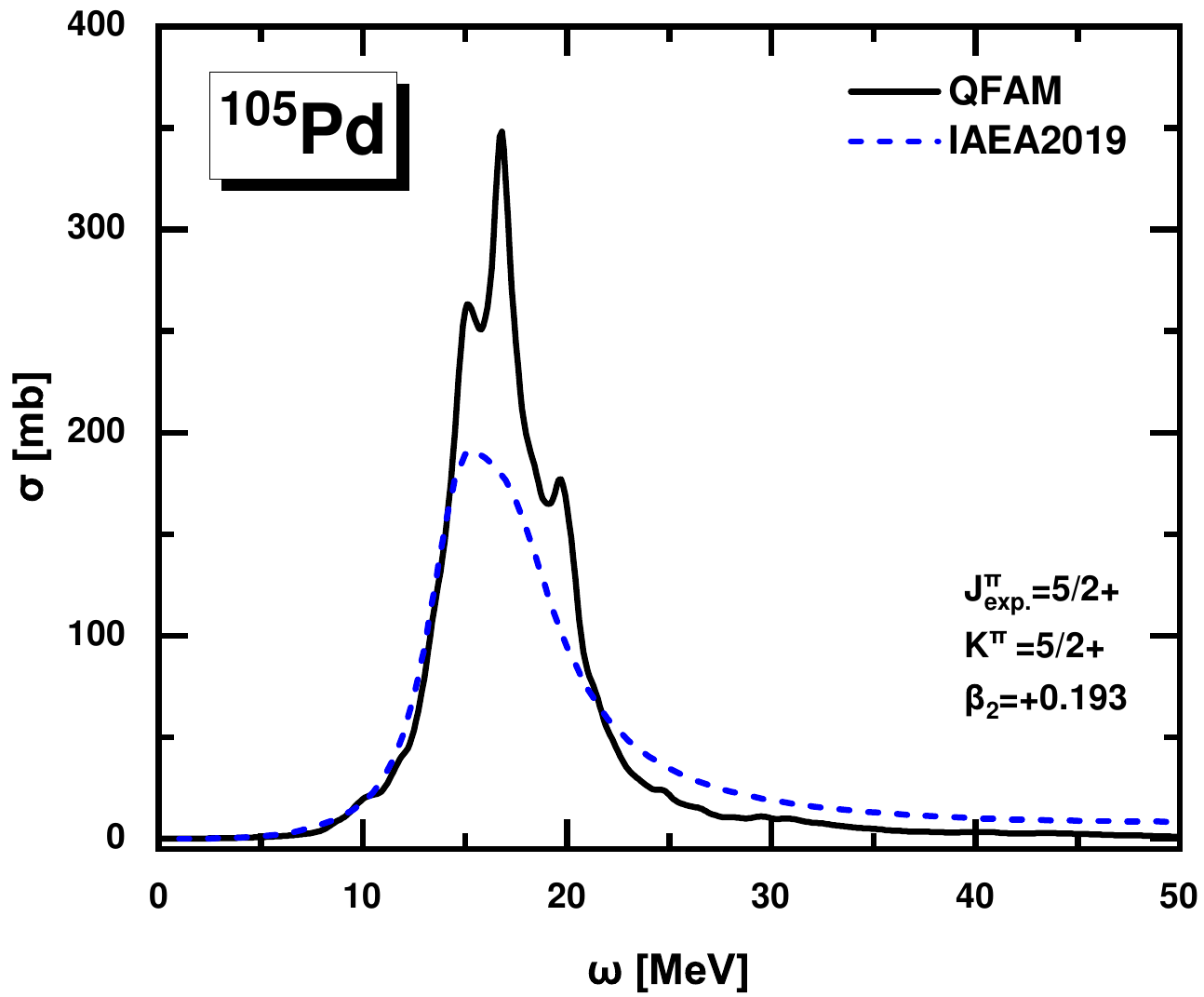}
    \includegraphics[width=0.35\textwidth]{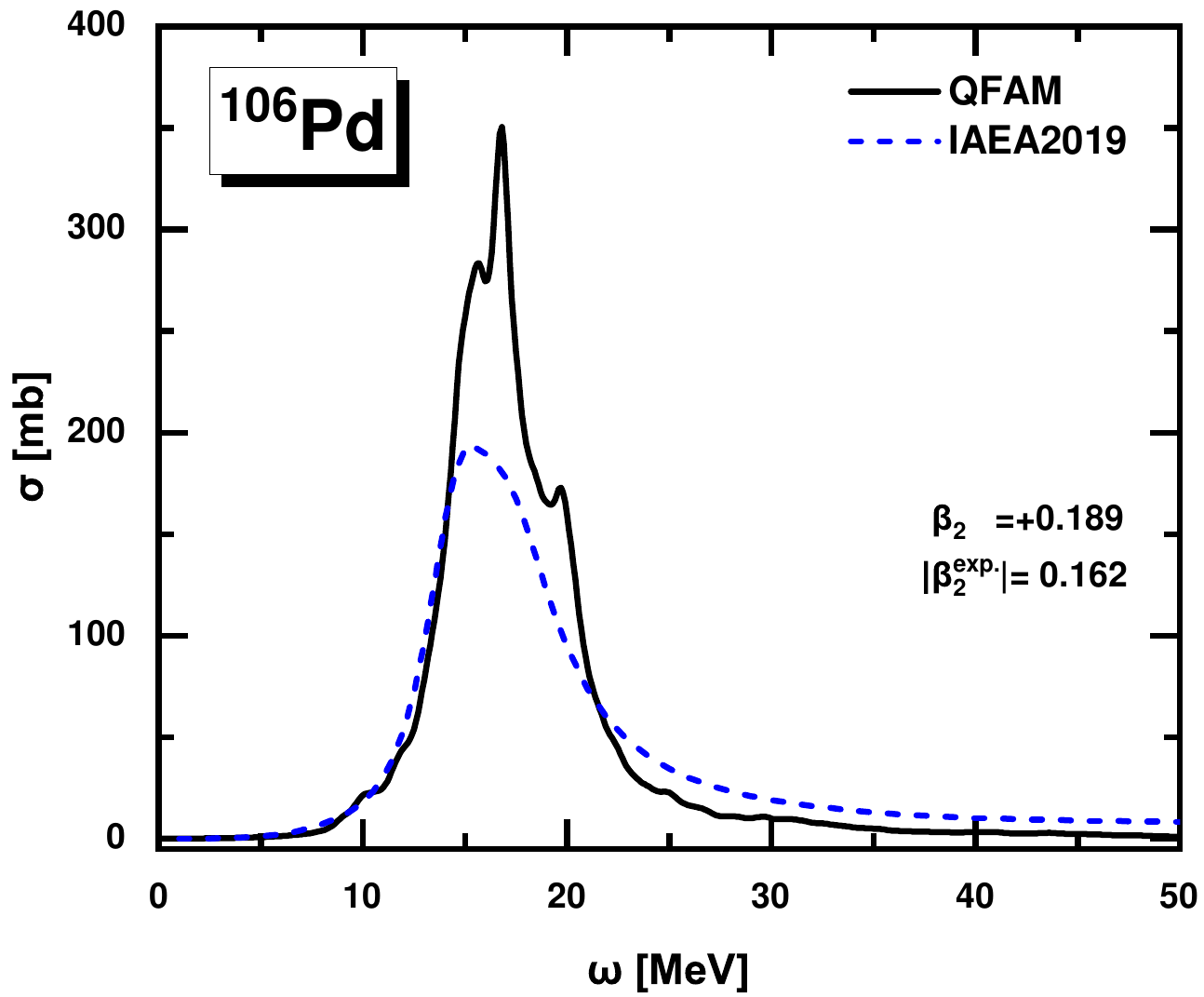}
\end{figure*}
\begin{figure*}\ContinuedFloat
    \centering
    \includegraphics[width=0.35\textwidth]{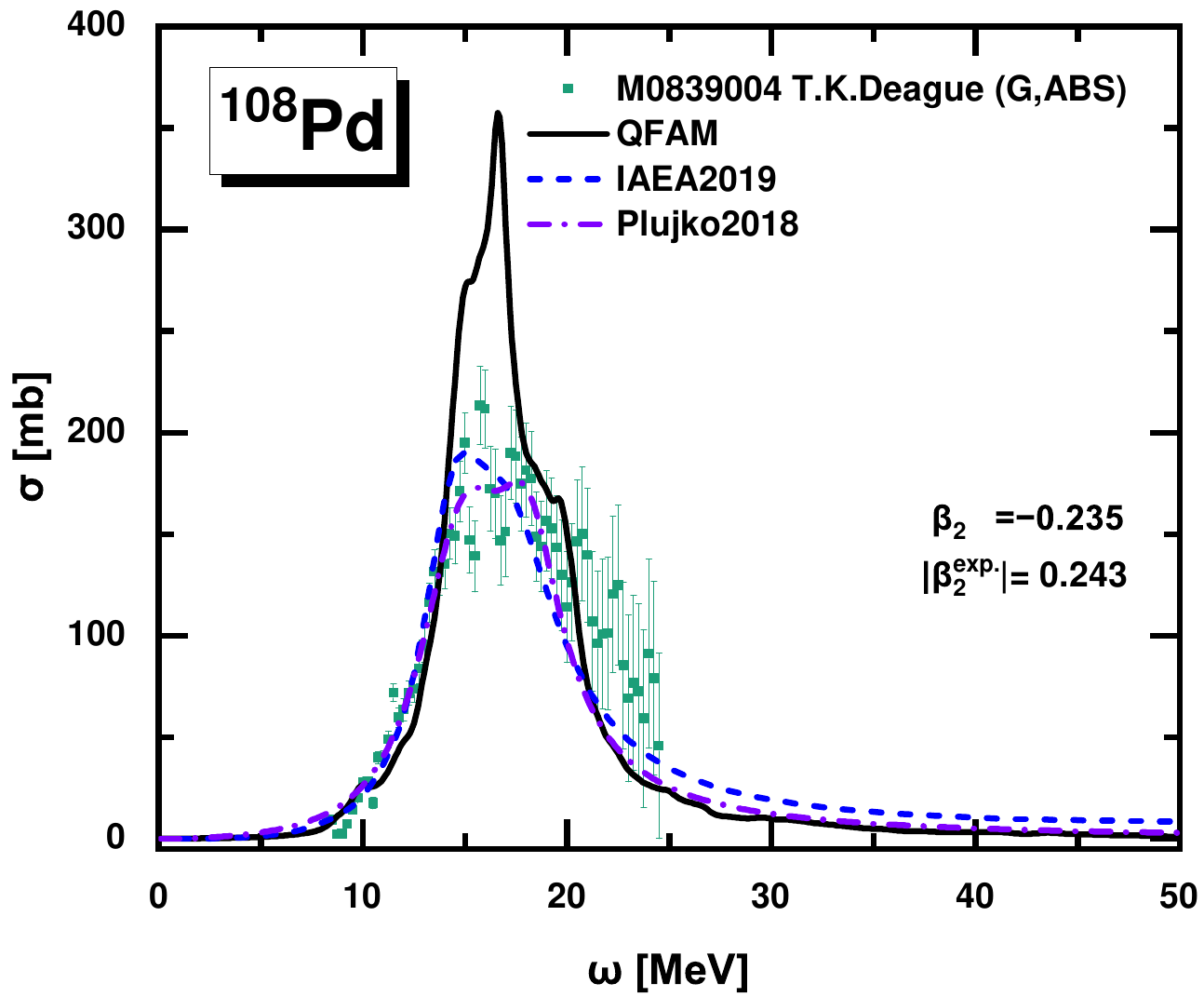}
    \includegraphics[width=0.35\textwidth]{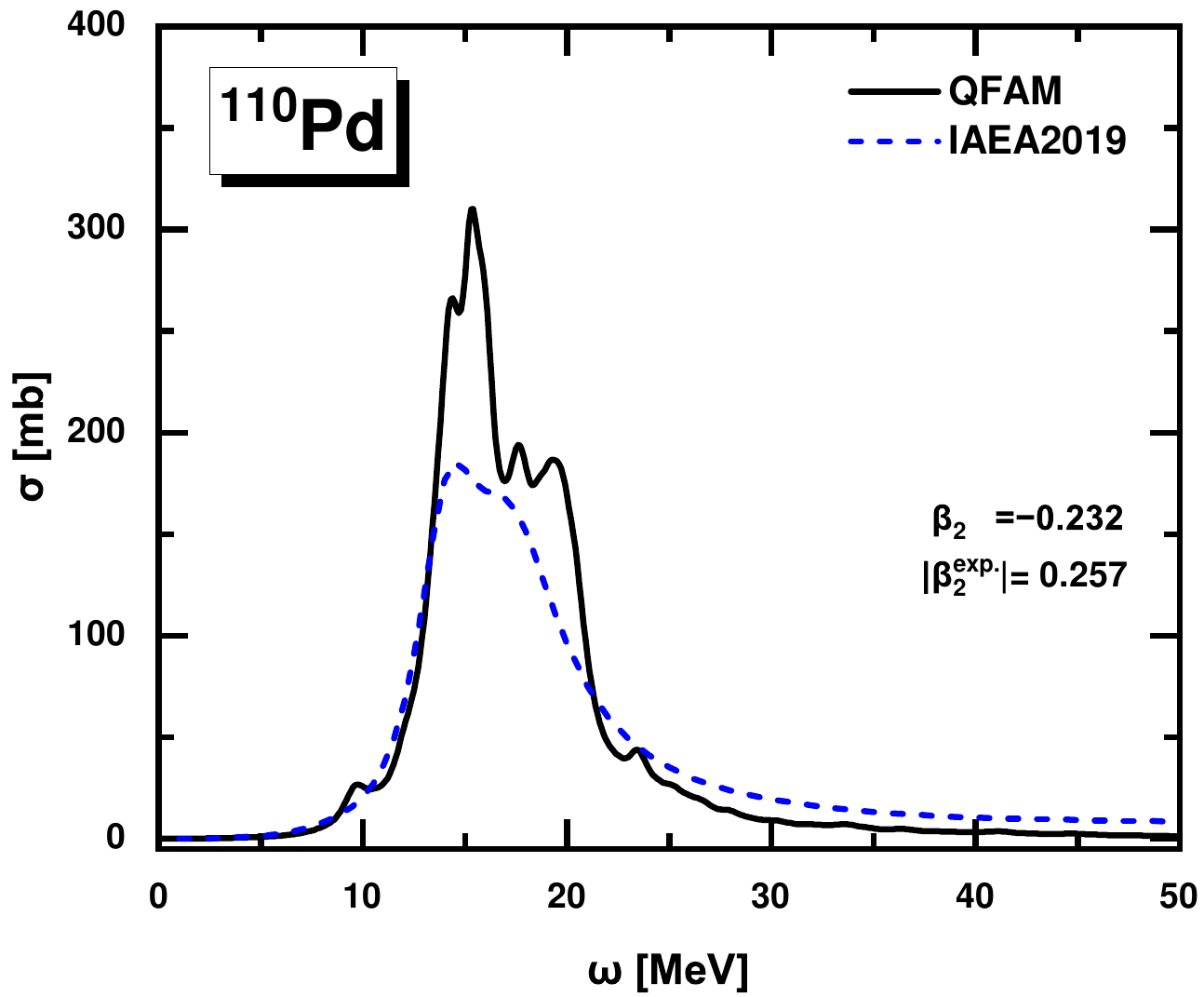}
    \includegraphics[width=0.35\textwidth]{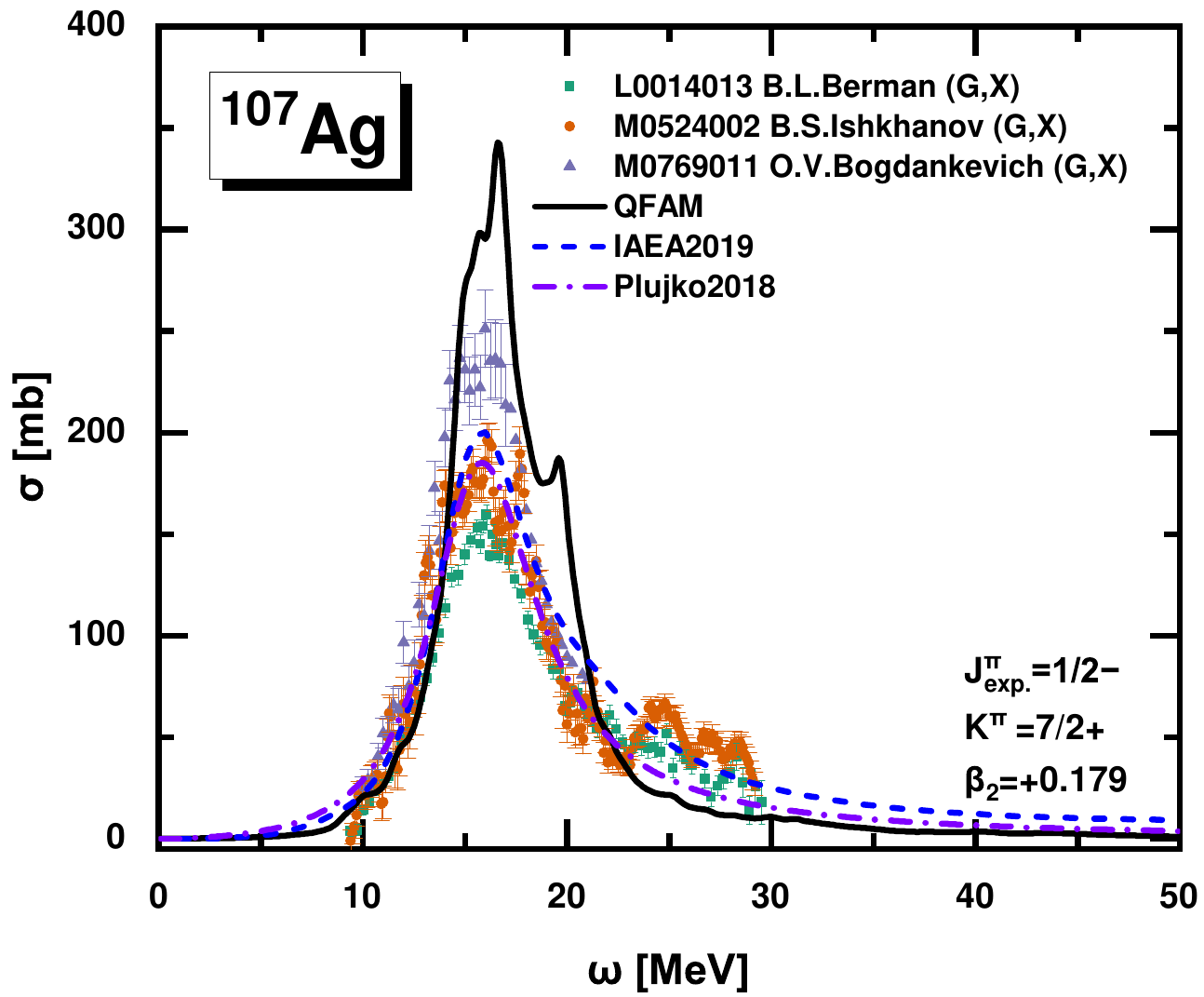}
    \includegraphics[width=0.35\textwidth]{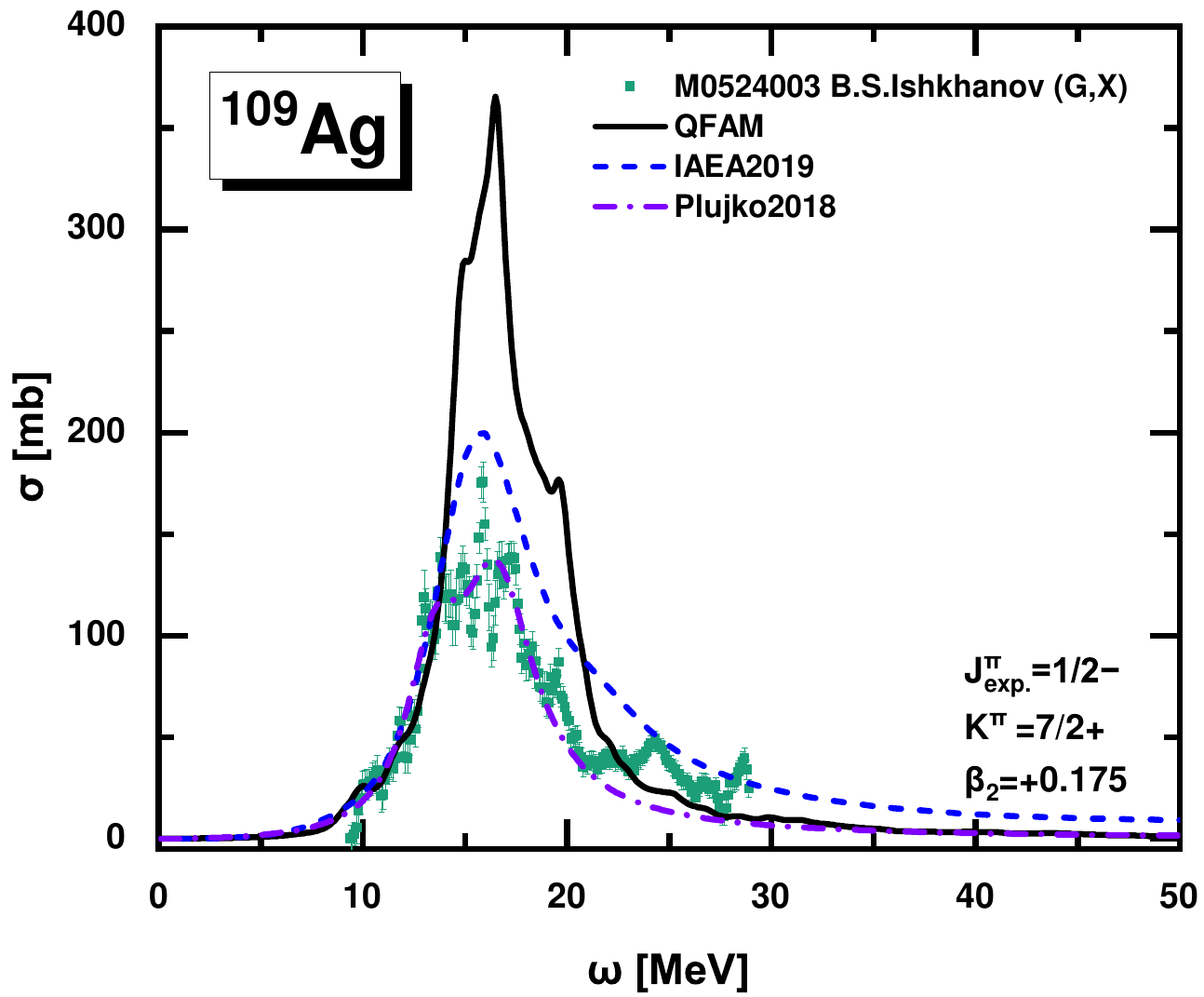}
    \includegraphics[width=0.35\textwidth]{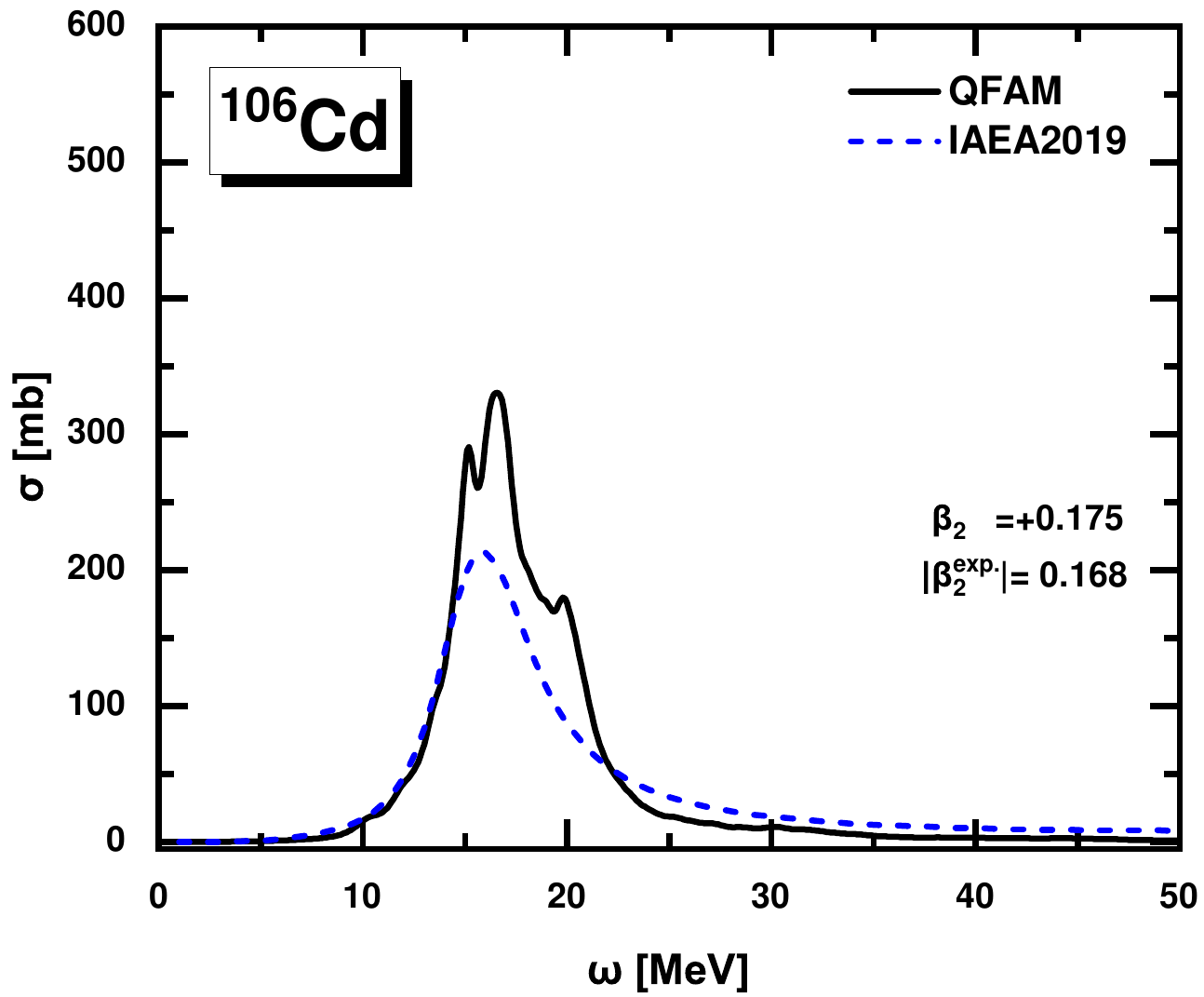}
    \includegraphics[width=0.35\textwidth]{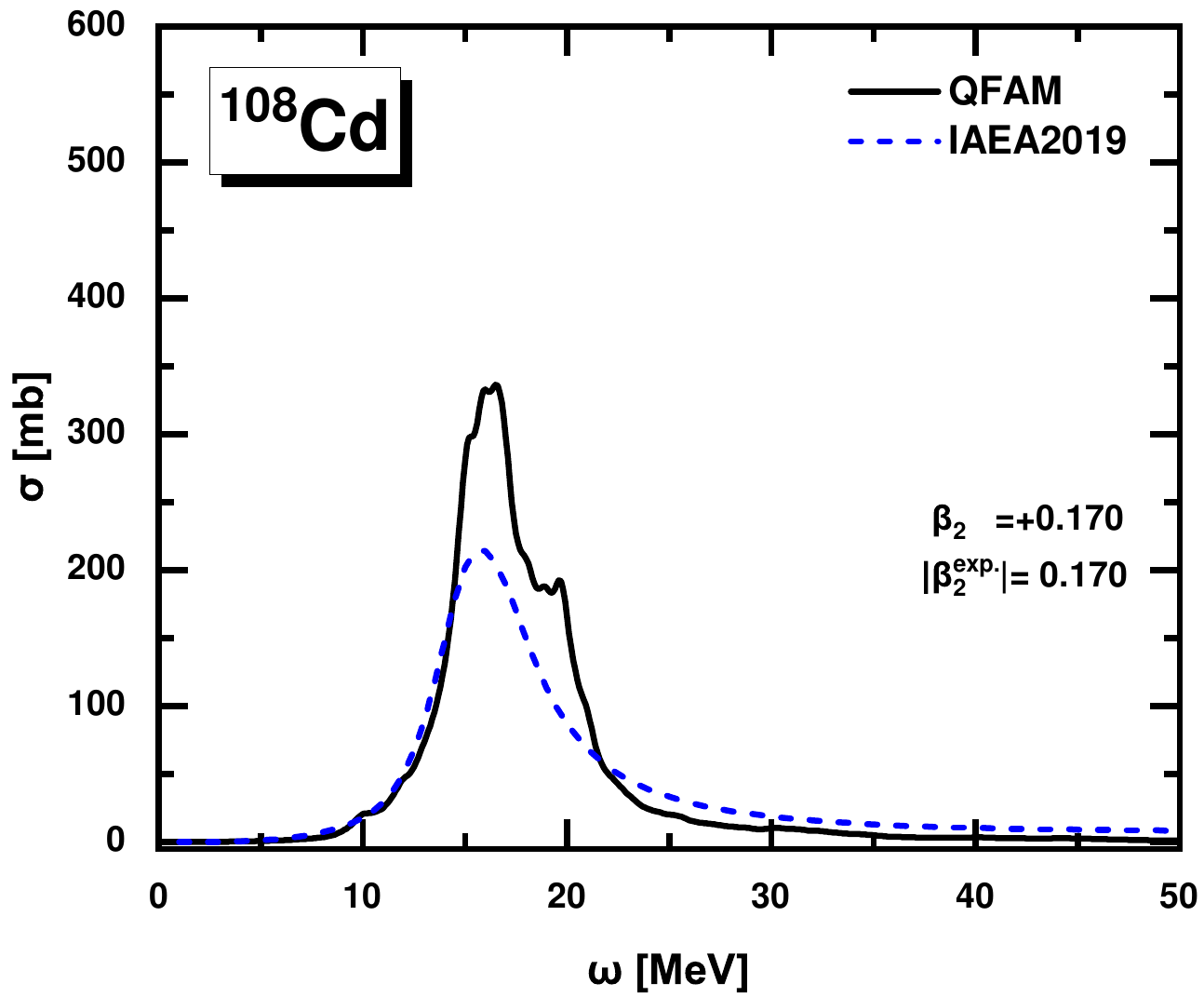}
    \includegraphics[width=0.35\textwidth]{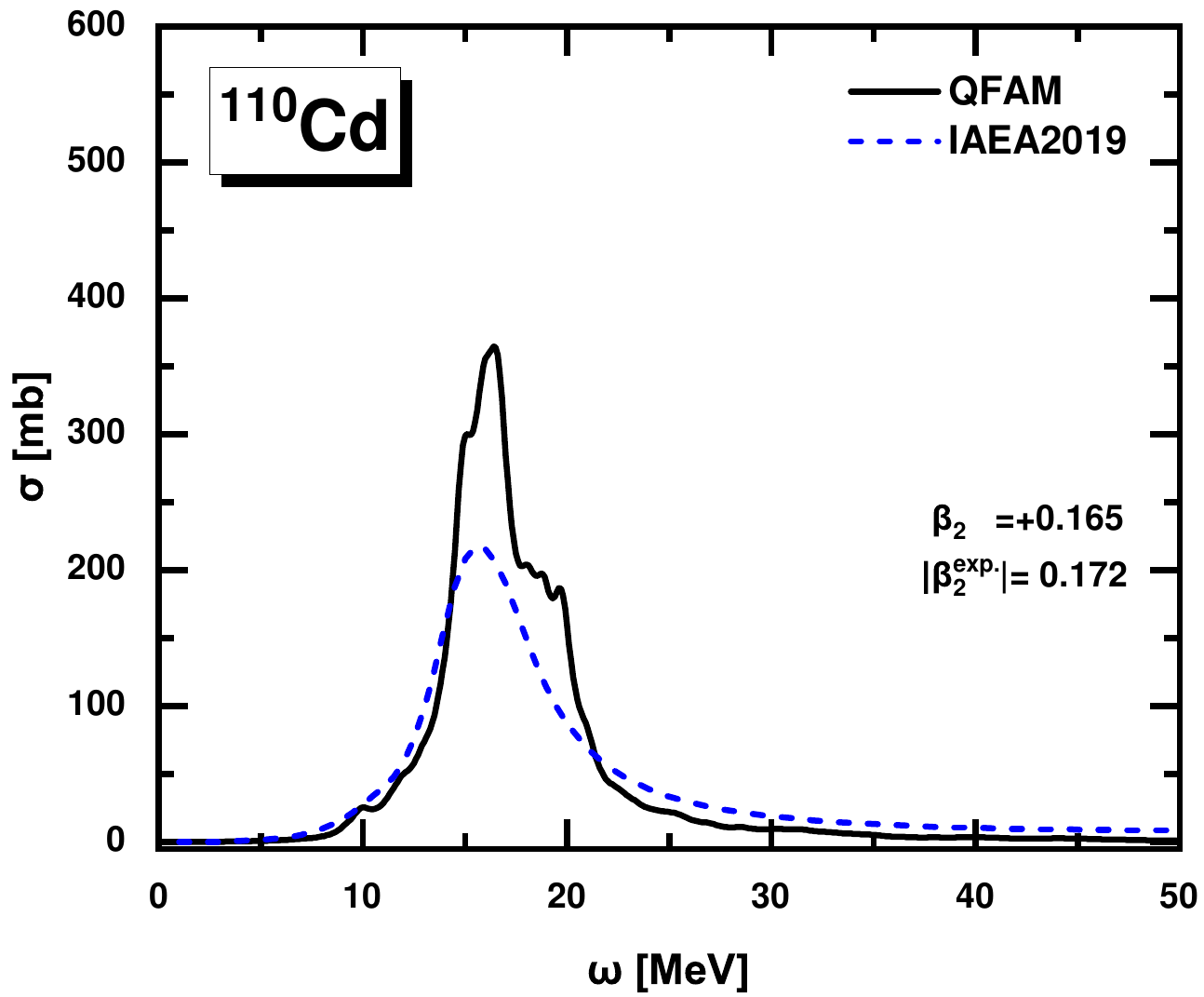}
    \includegraphics[width=0.35\textwidth]{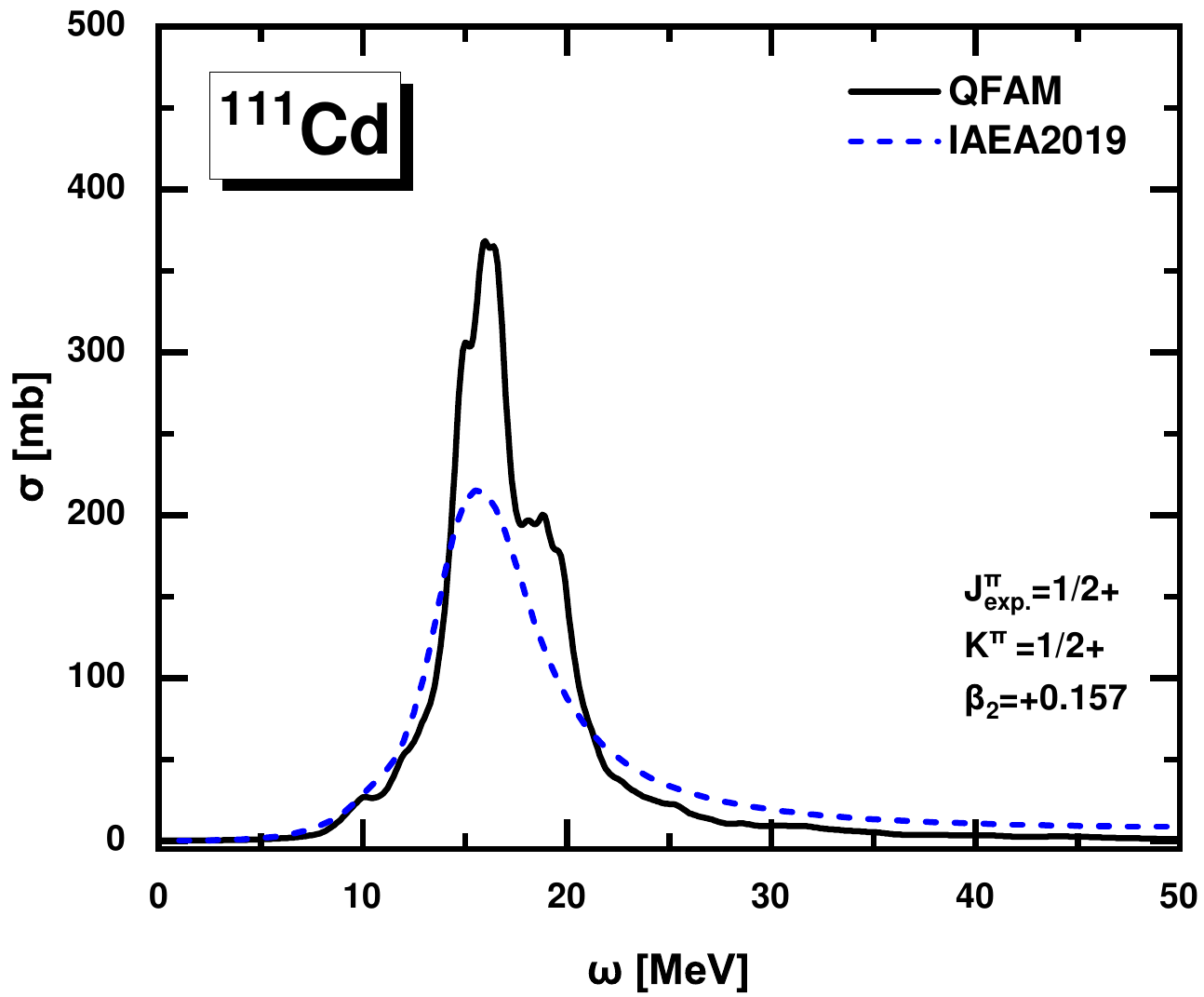}
\end{figure*}
\begin{figure*}\ContinuedFloat
    \centering
    \includegraphics[width=0.35\textwidth]{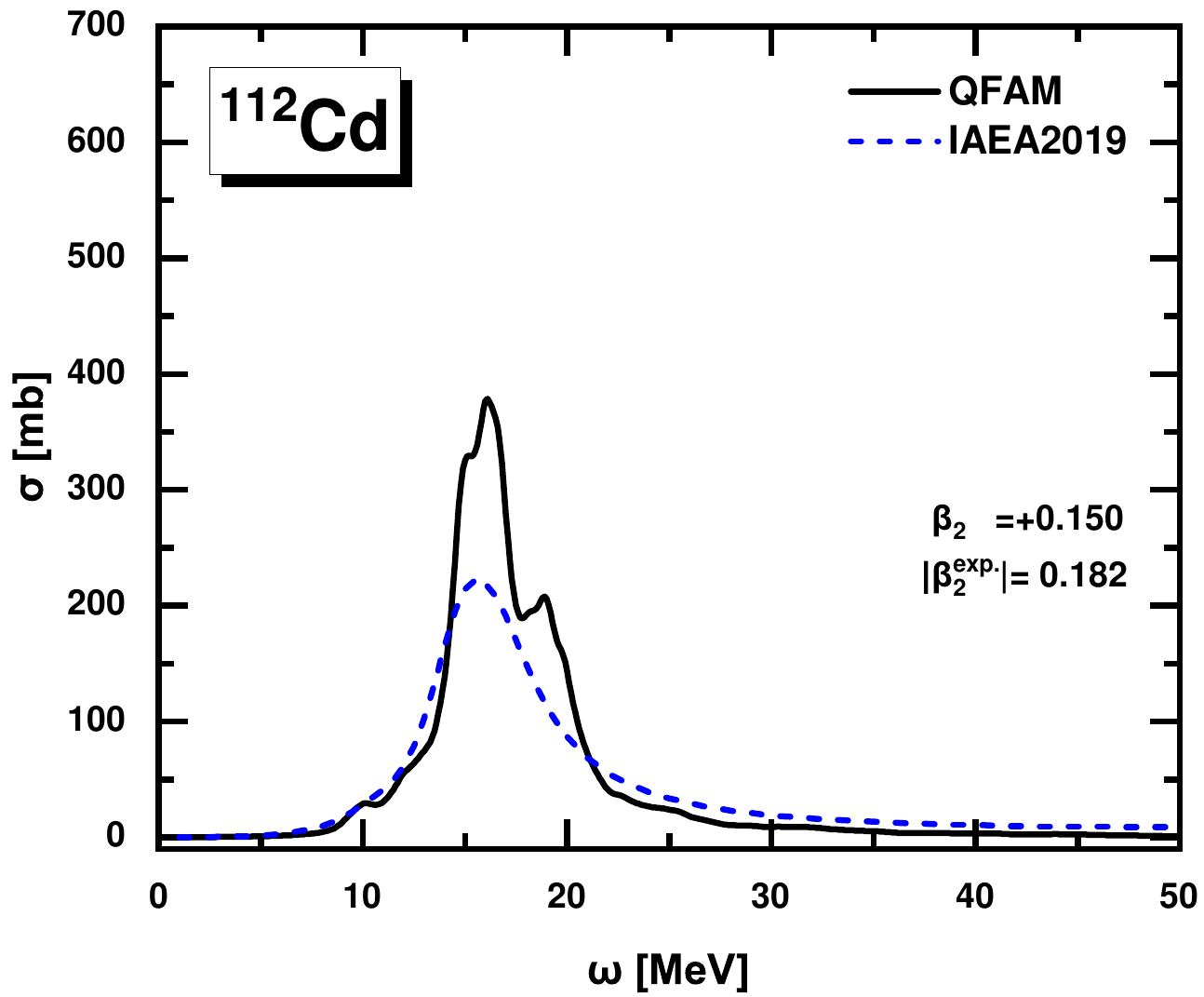}
    \includegraphics[width=0.35\textwidth]{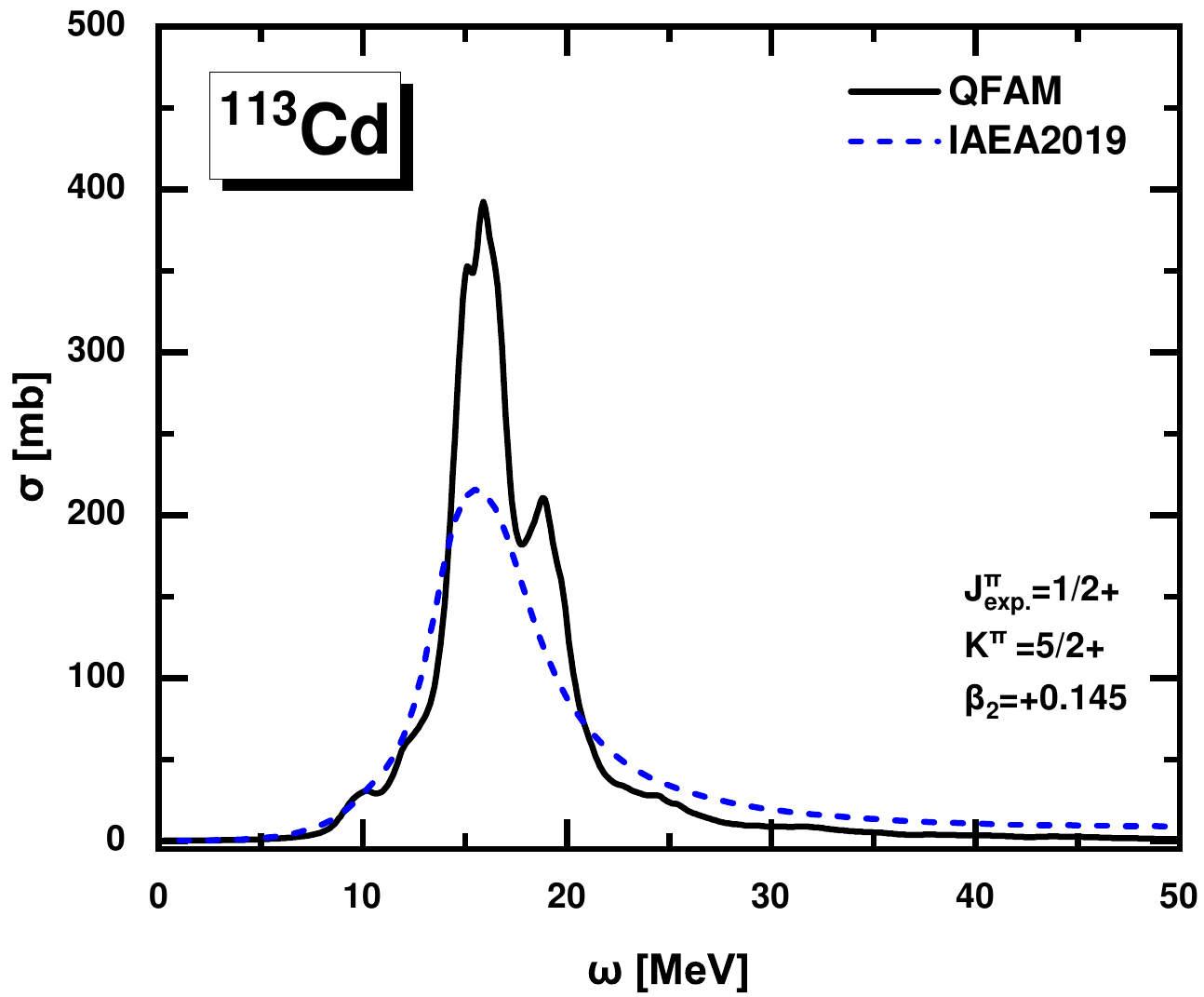}
    \includegraphics[width=0.35\textwidth]{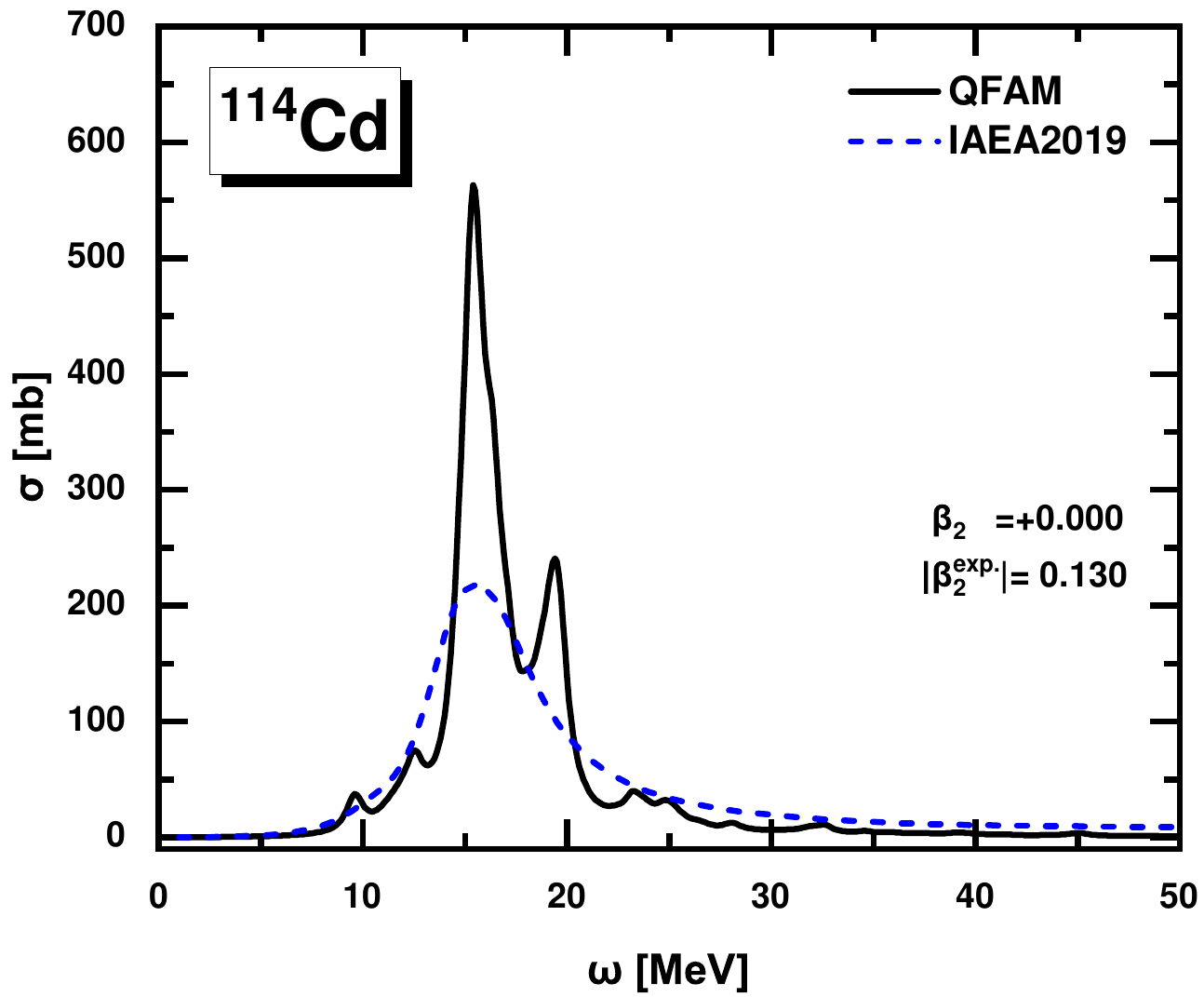}
    \includegraphics[width=0.35\textwidth]{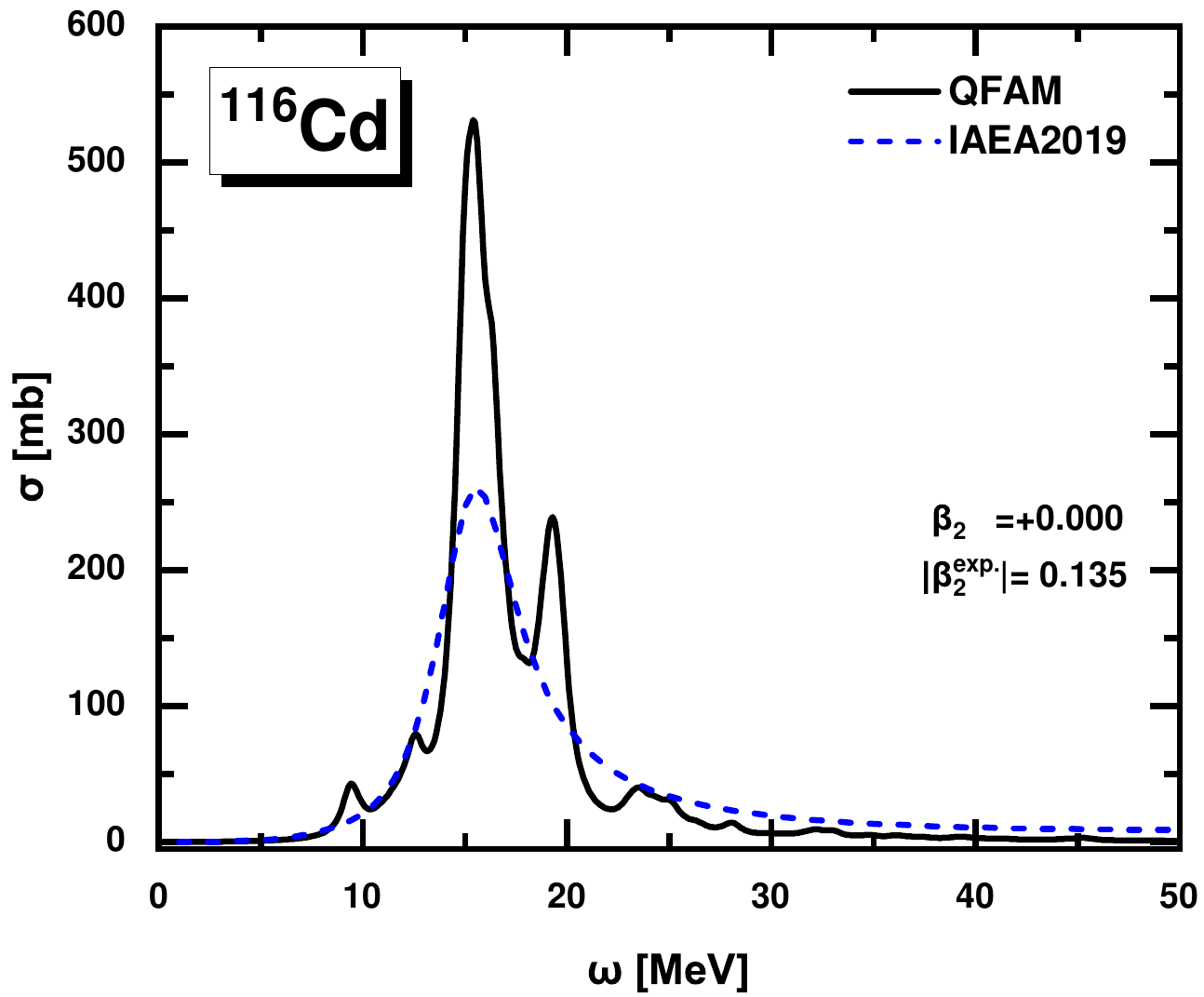}
    \includegraphics[width=0.35\textwidth]{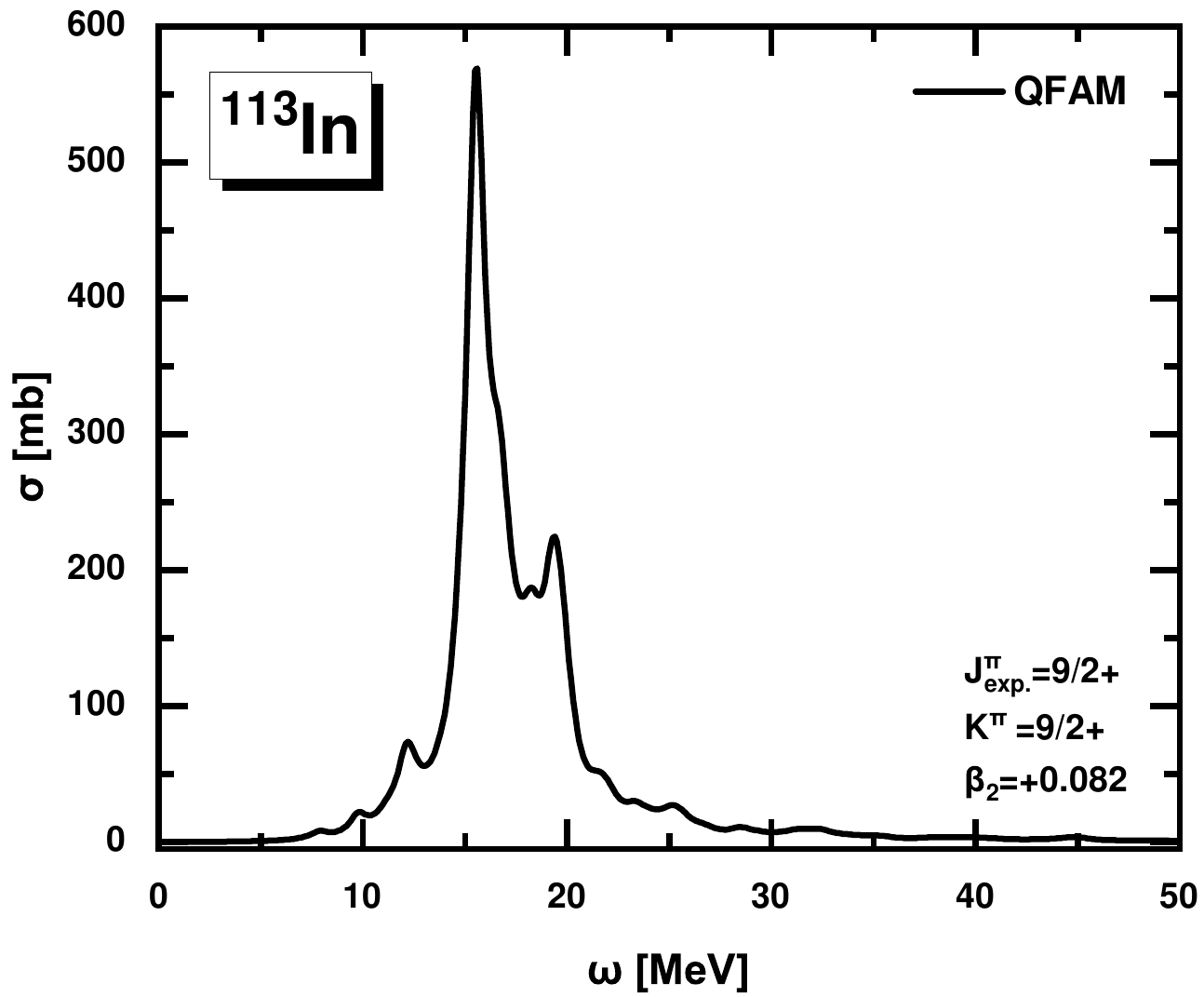}
    \includegraphics[width=0.35\textwidth]{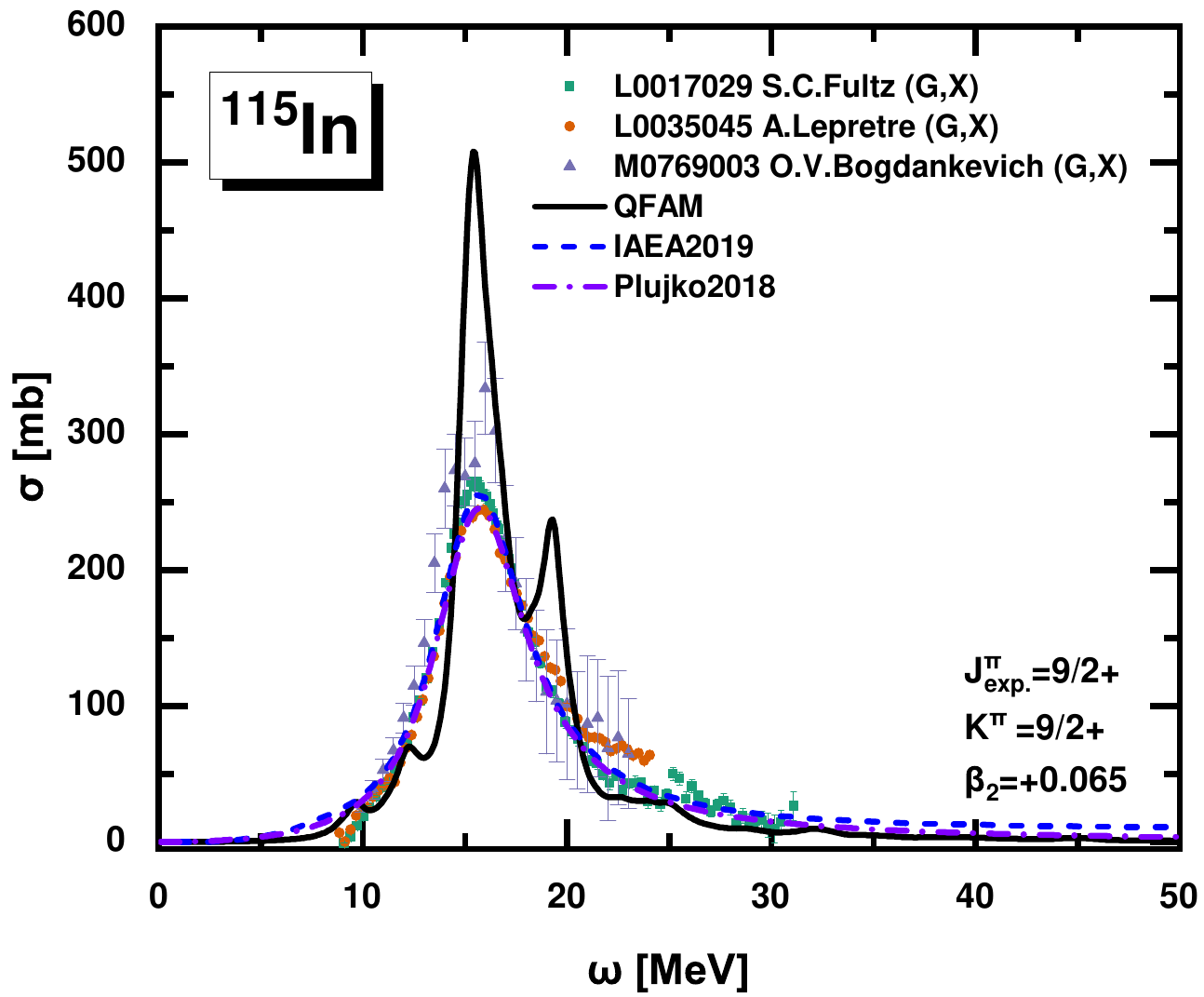}
    \includegraphics[width=0.35\textwidth]{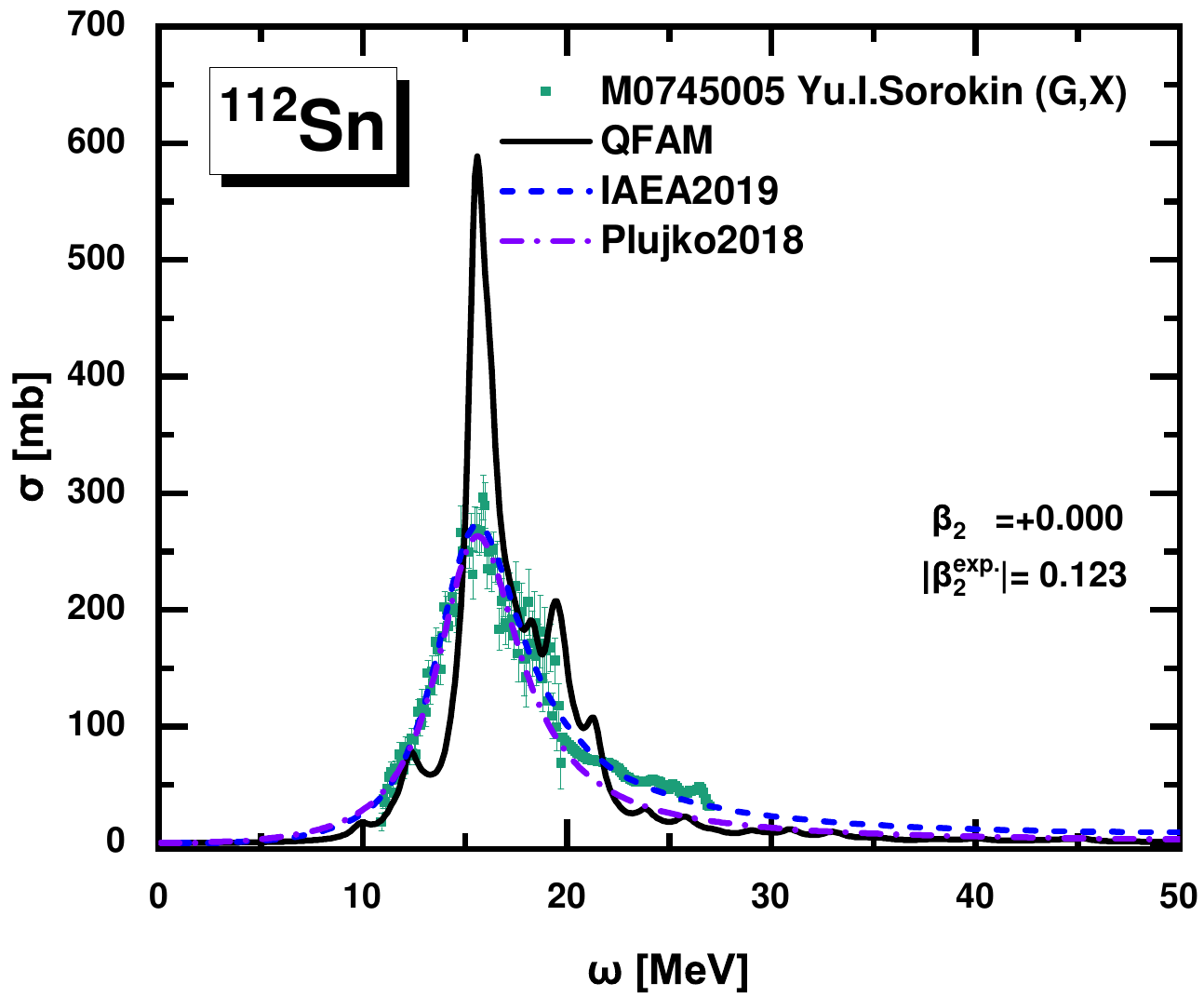}
    \includegraphics[width=0.35\textwidth]{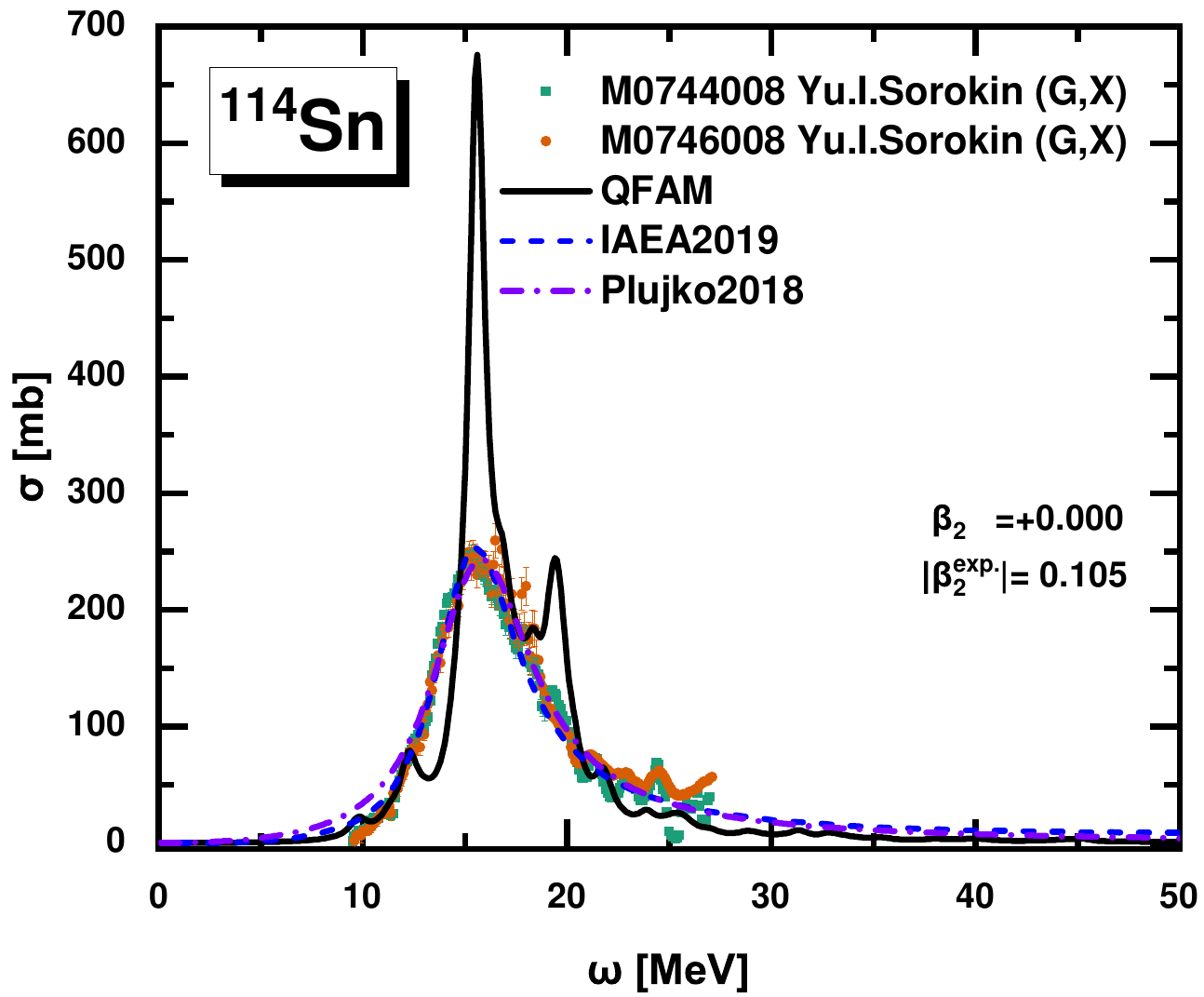}
\end{figure*}
\begin{figure*}\ContinuedFloat
    \centering
    \includegraphics[width=0.35\textwidth]{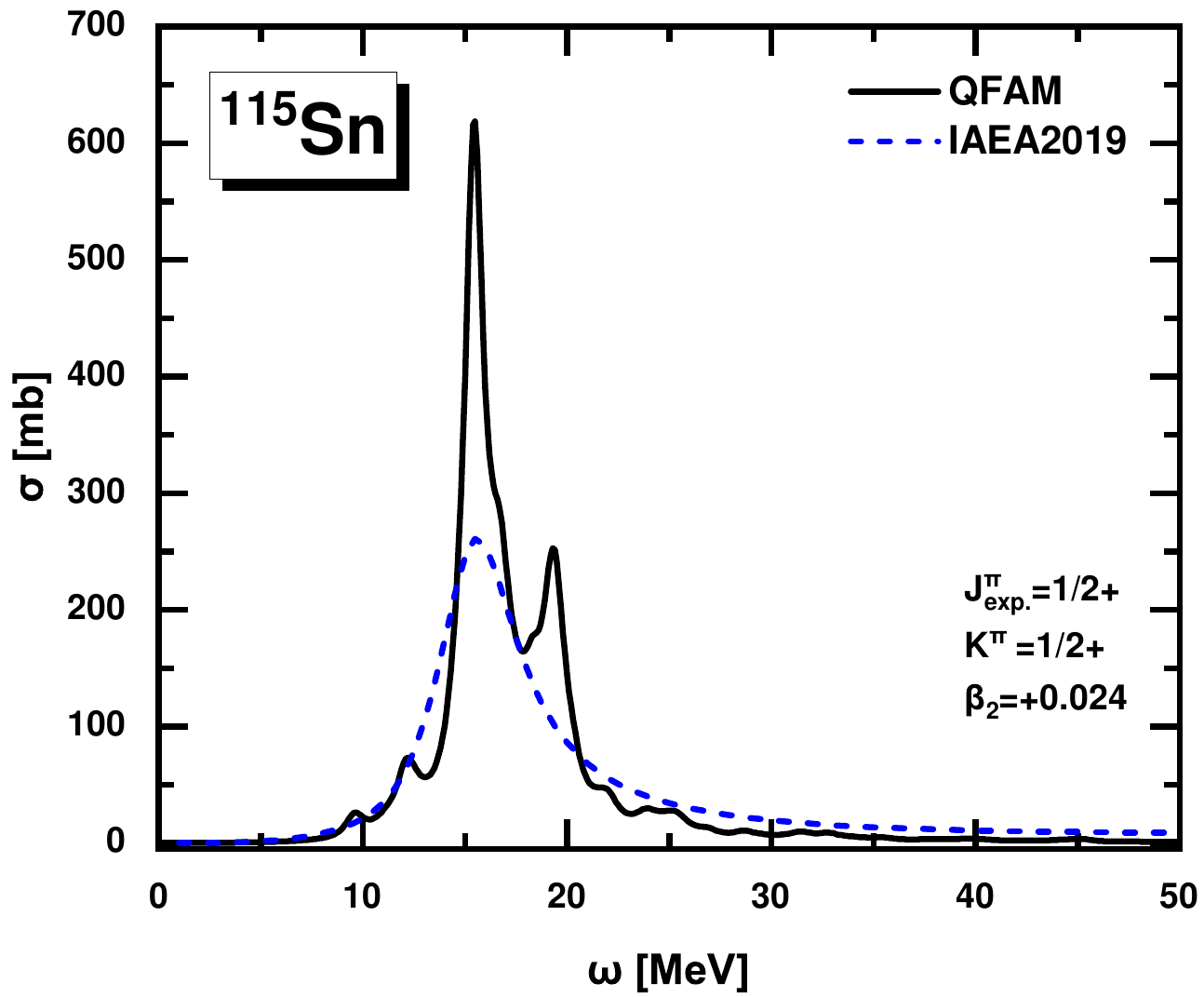}
    \includegraphics[width=0.35\textwidth]{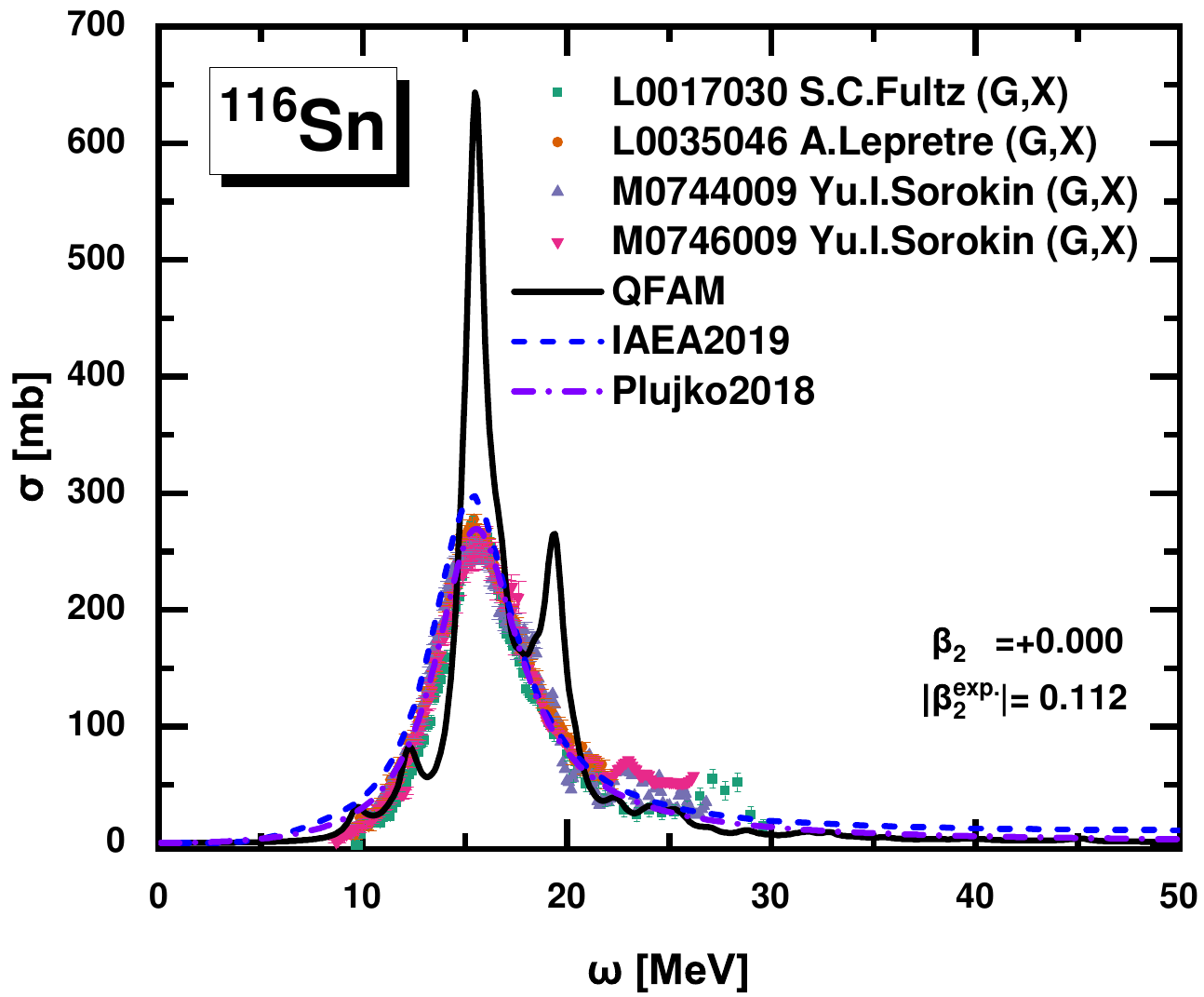}
    \includegraphics[width=0.35\textwidth]{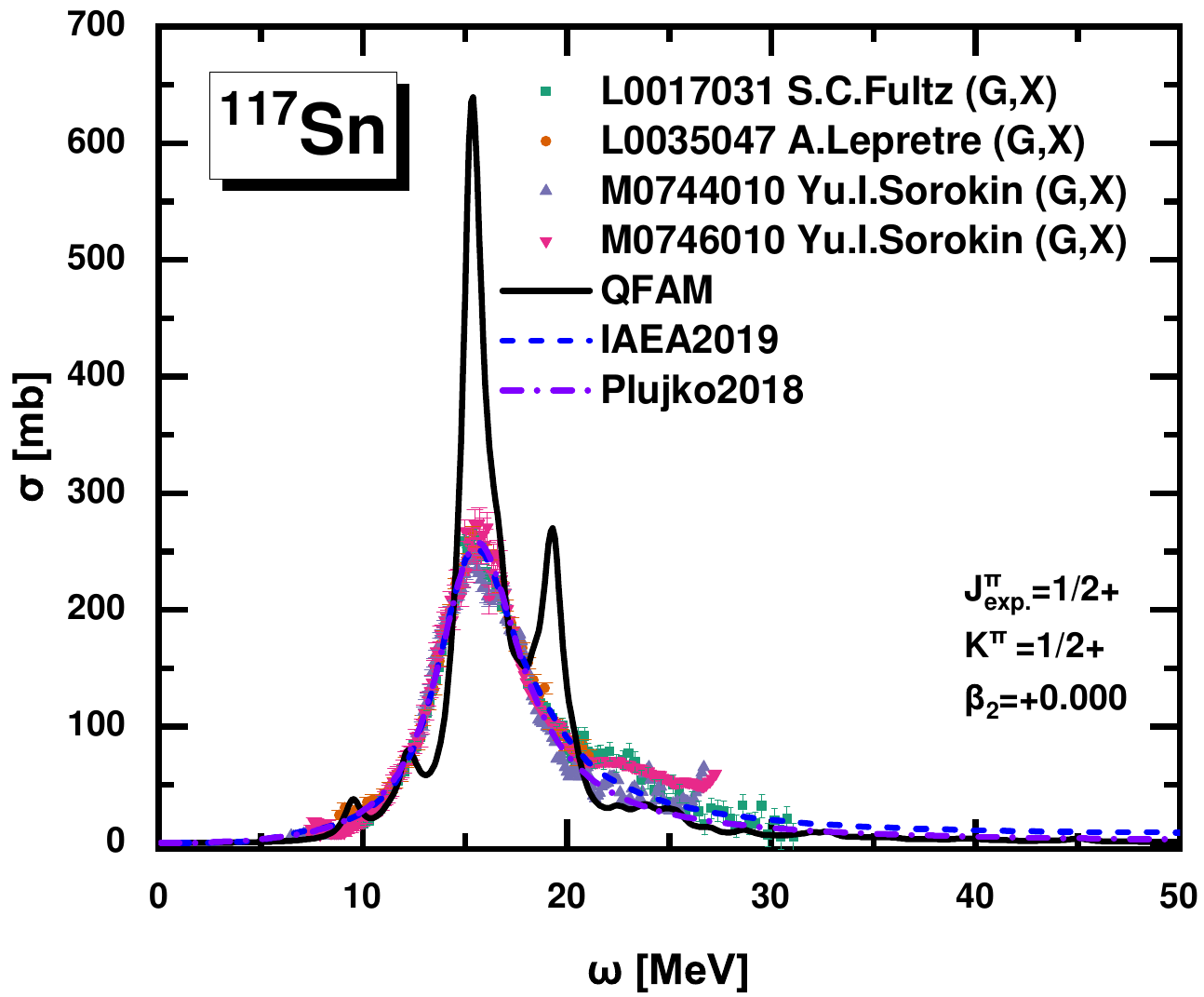}
    \includegraphics[width=0.35\textwidth]{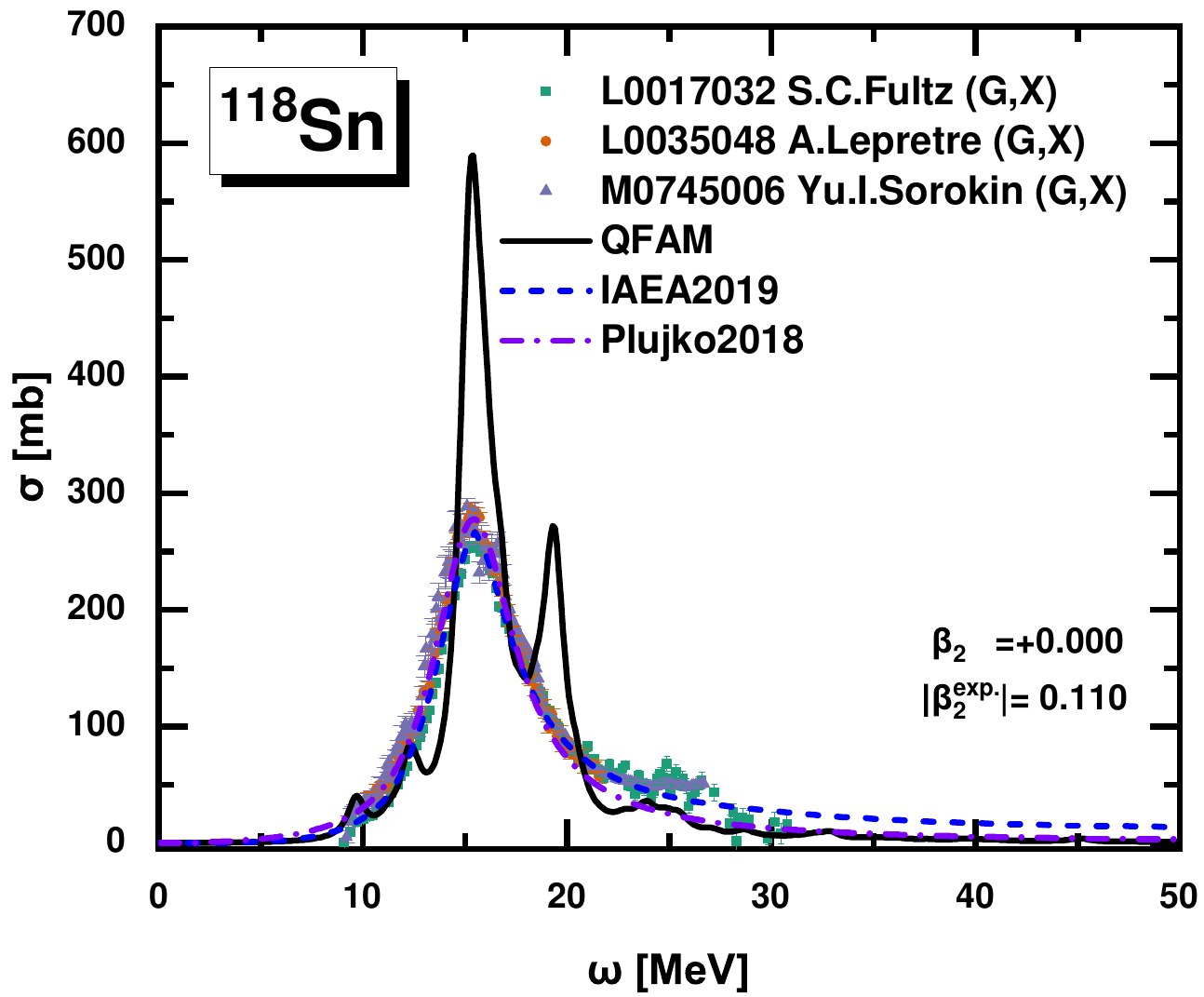}
    \includegraphics[width=0.35\textwidth]{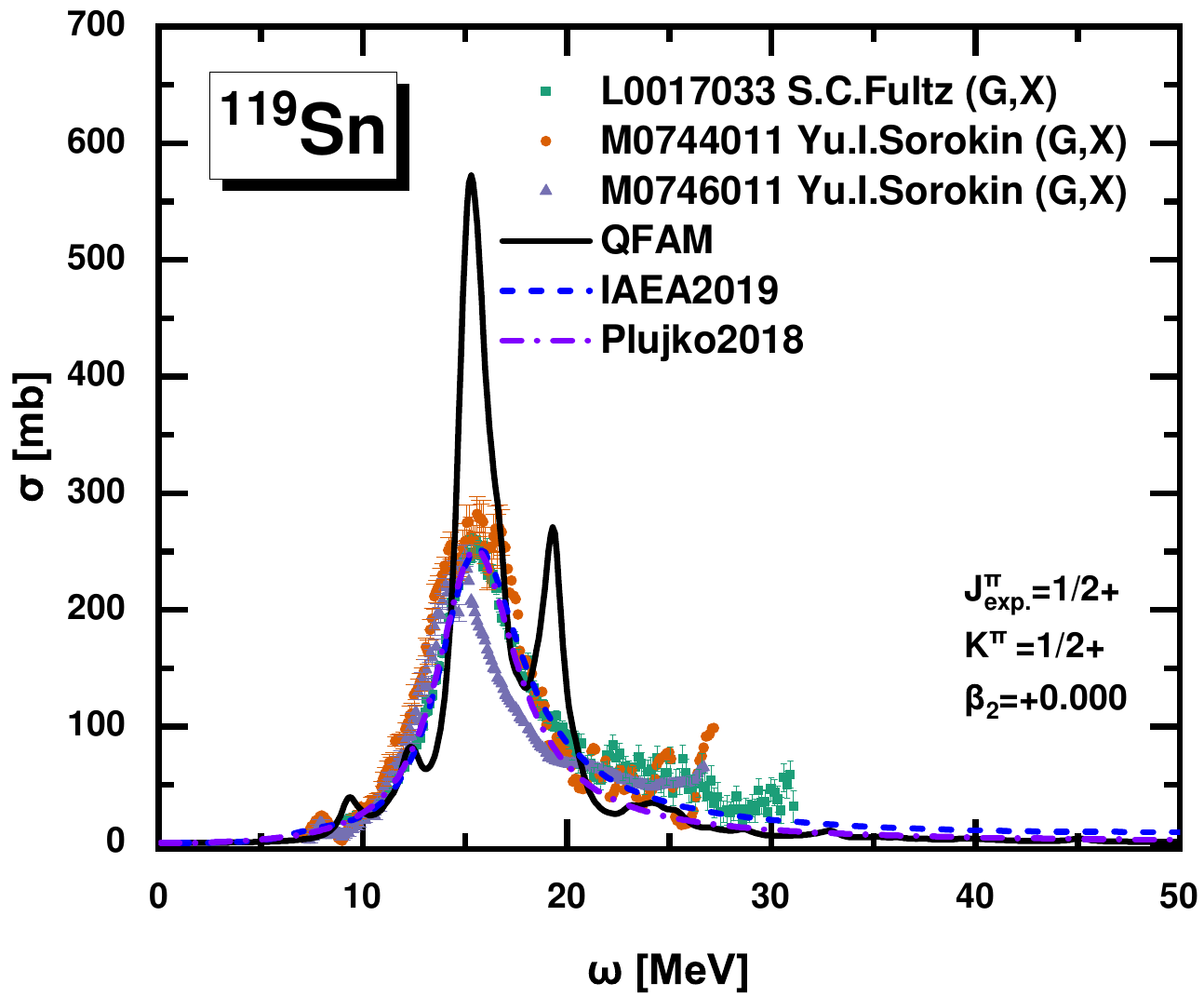}
    \includegraphics[width=0.35\textwidth]{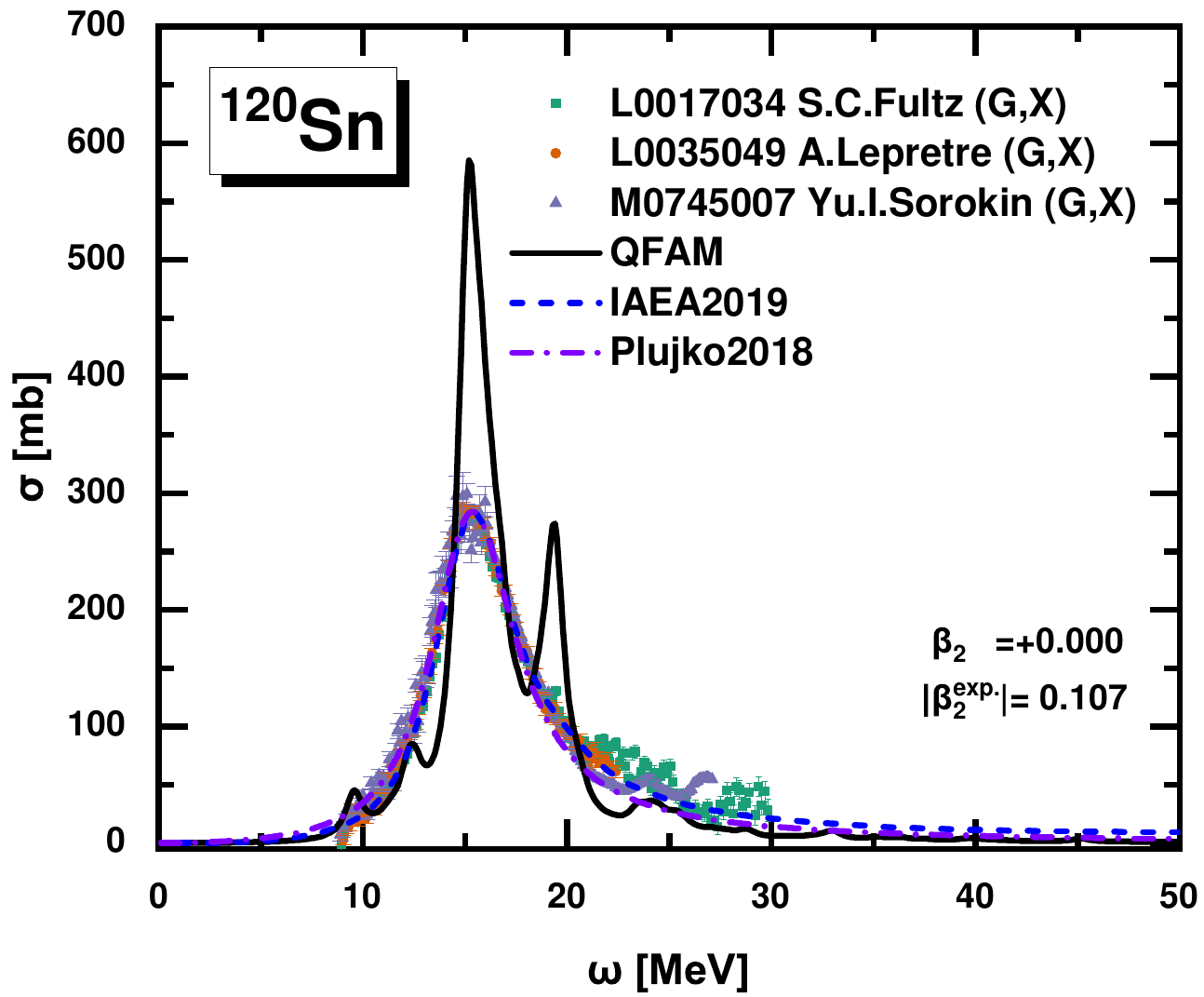}
    \includegraphics[width=0.35\textwidth]{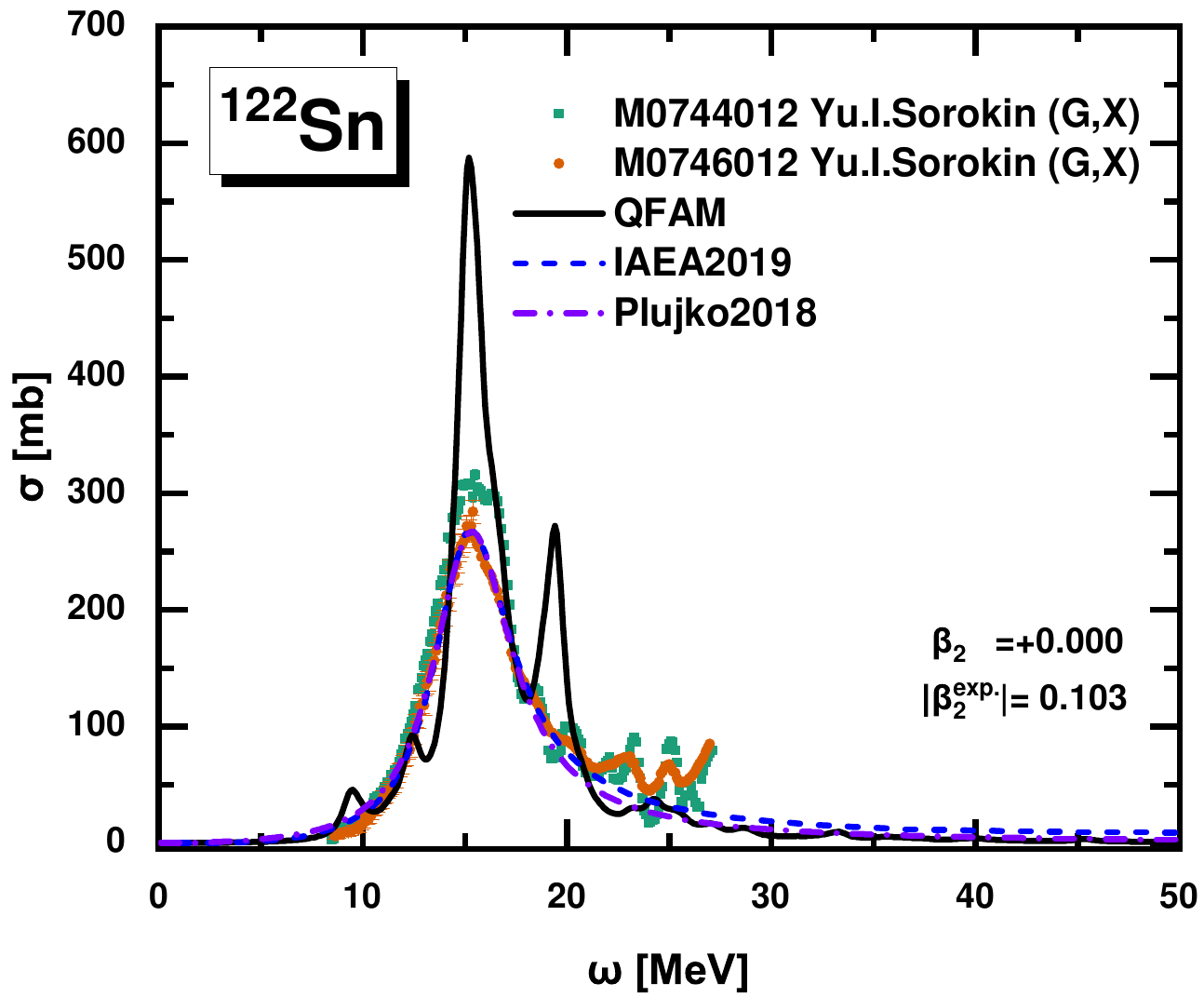}
    \includegraphics[width=0.35\textwidth]{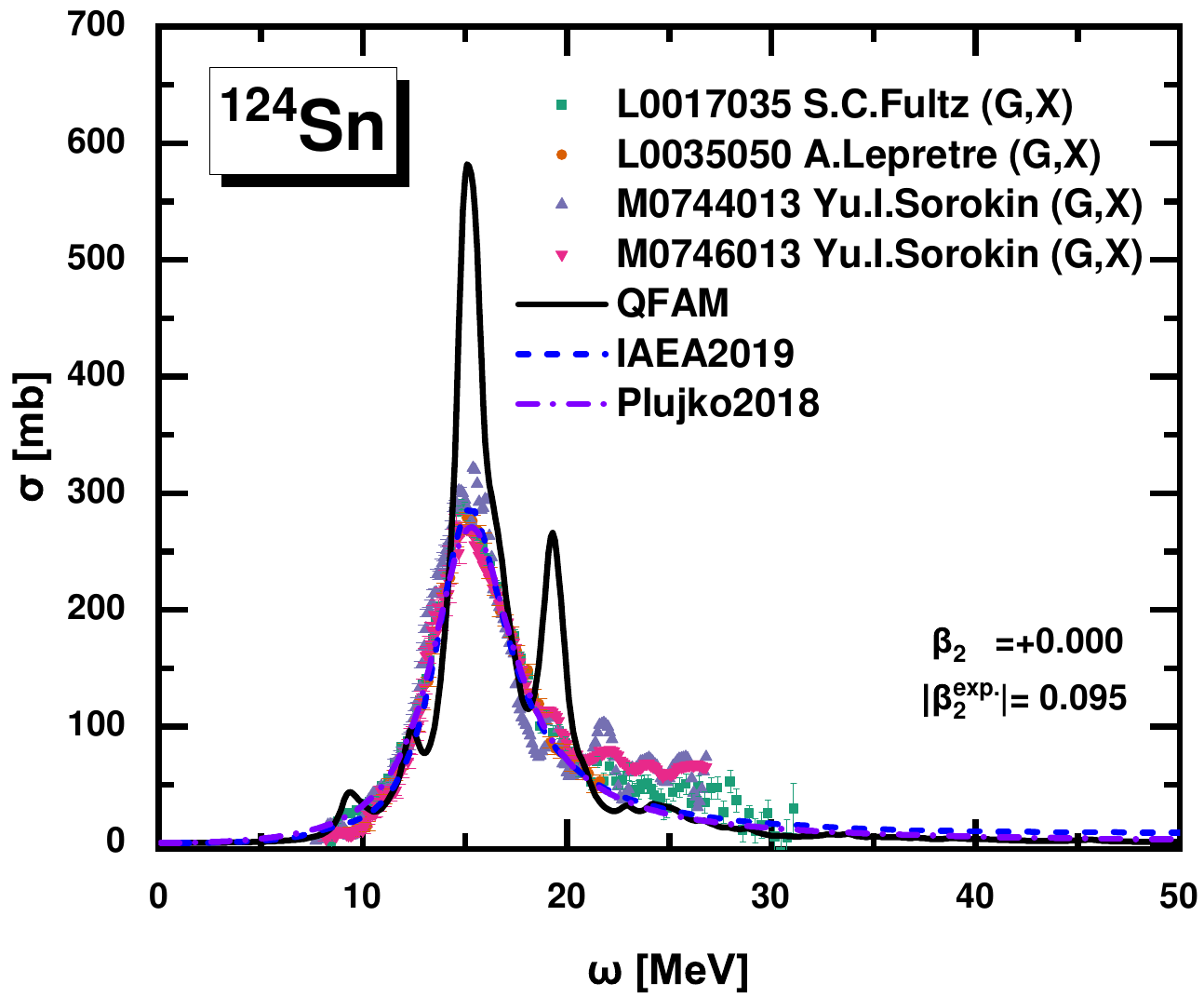}
\end{figure*}
\begin{figure*}\ContinuedFloat
    \centering
    \includegraphics[width=0.35\textwidth]{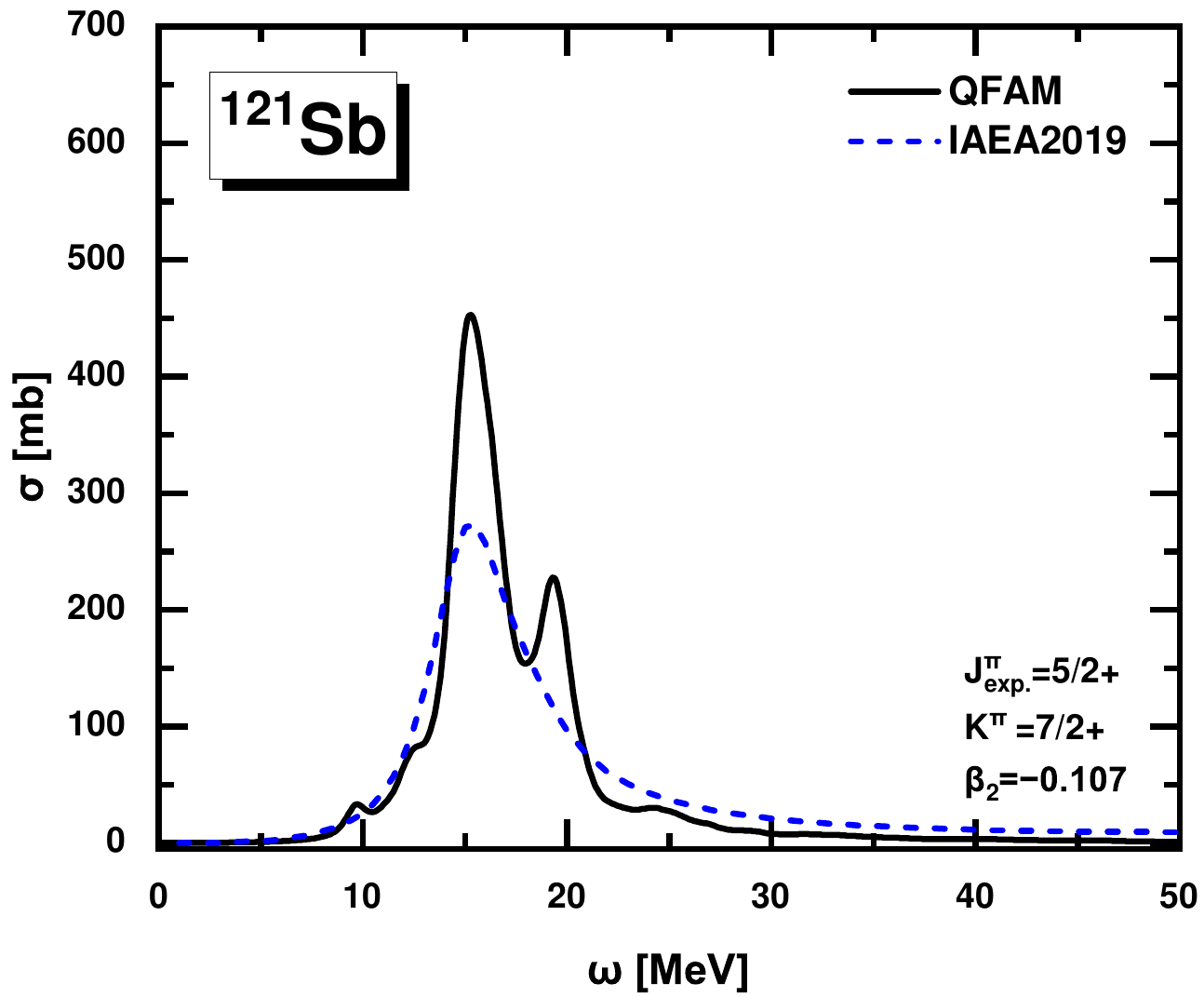}
    \includegraphics[width=0.35\textwidth]{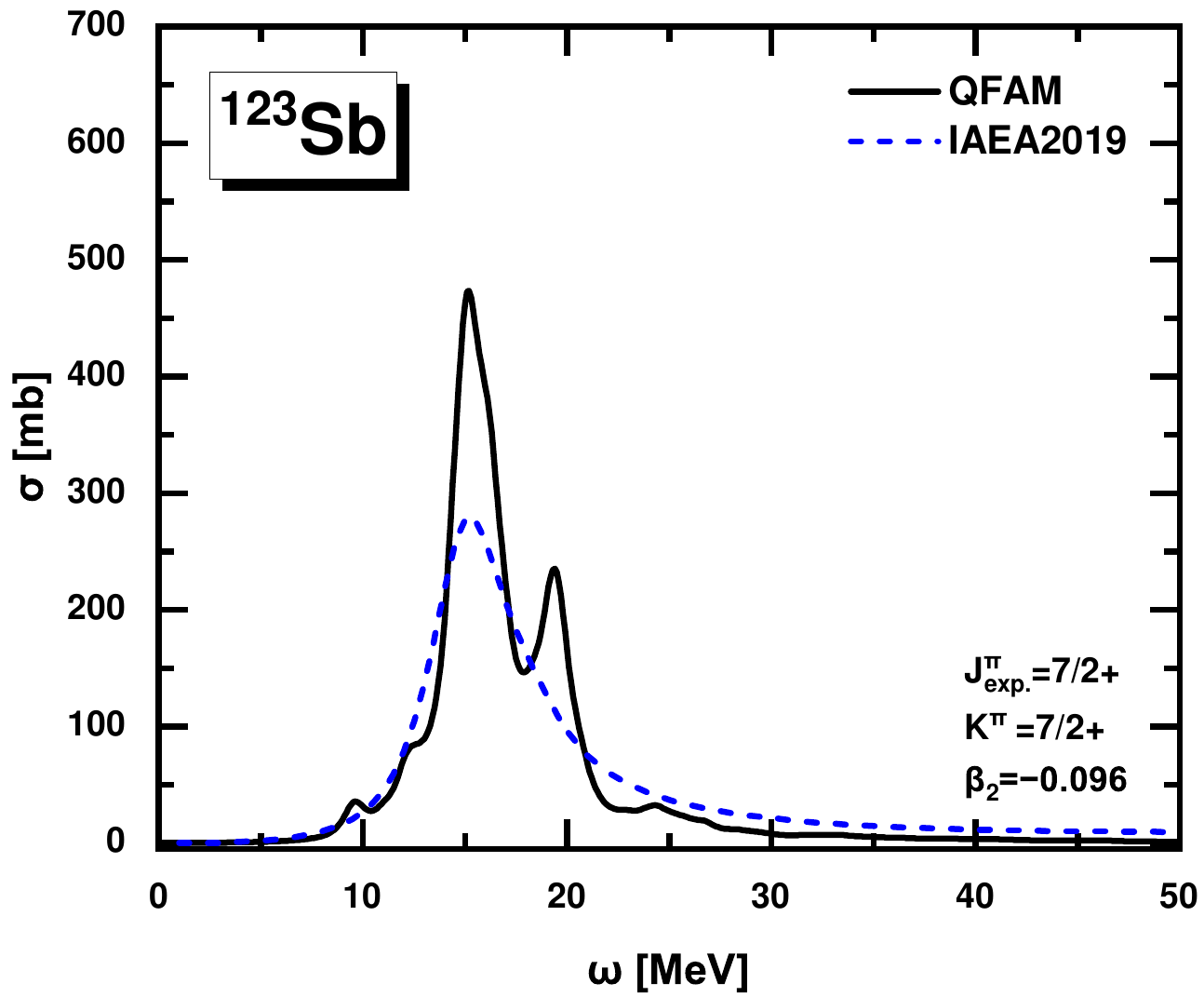}
    \includegraphics[width=0.35\textwidth]{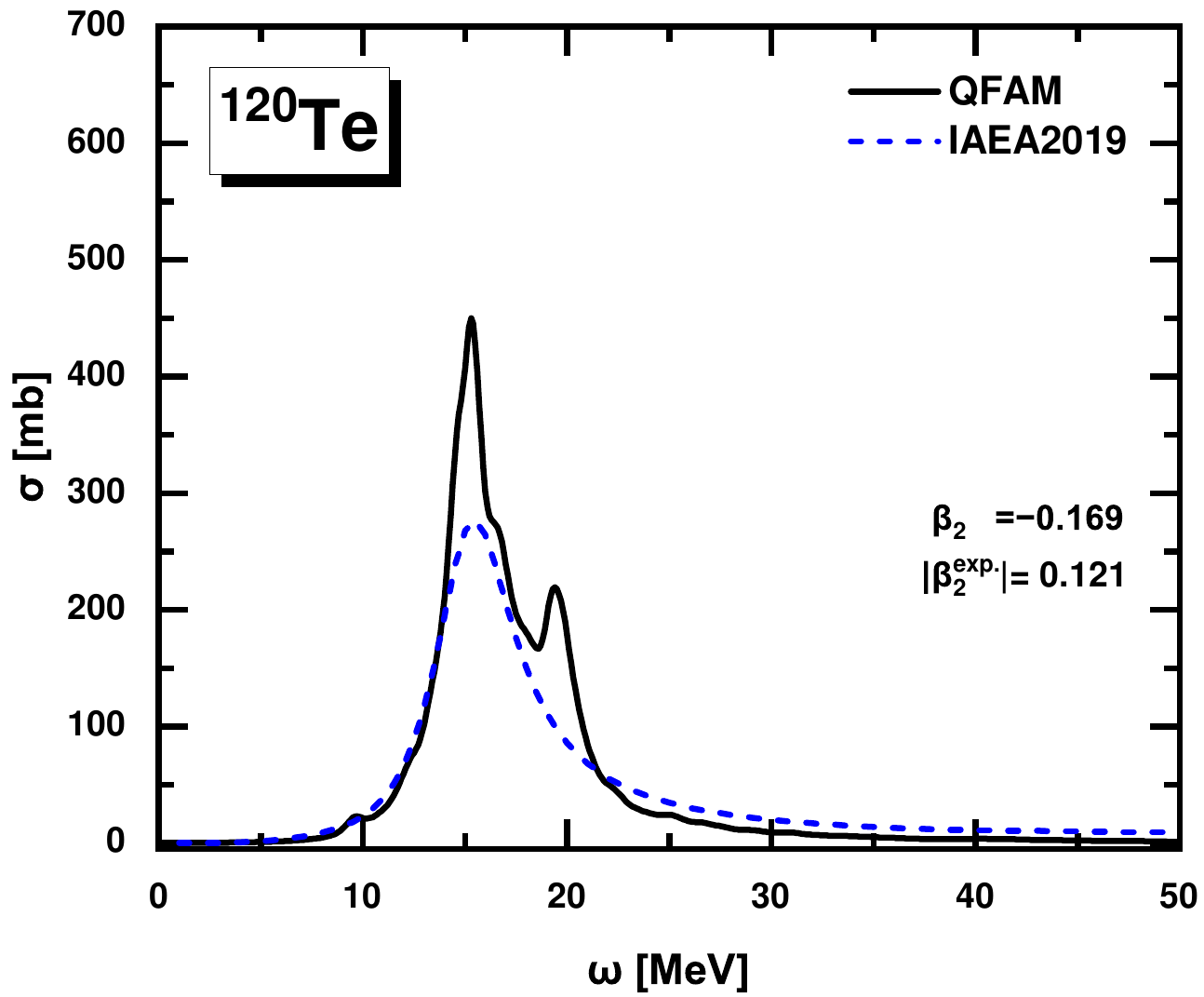}
    \includegraphics[width=0.35\textwidth]{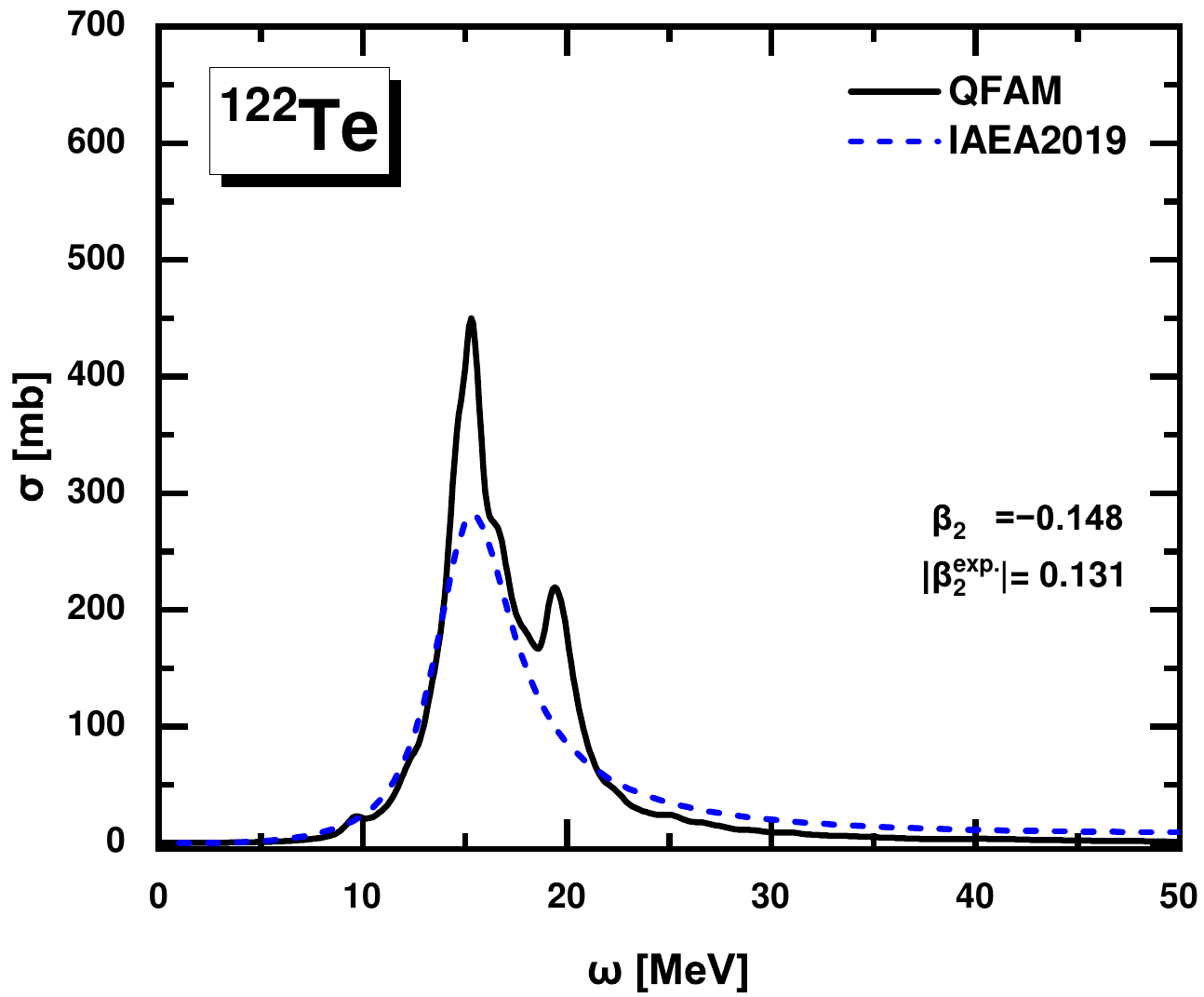}
    \includegraphics[width=0.35\textwidth]{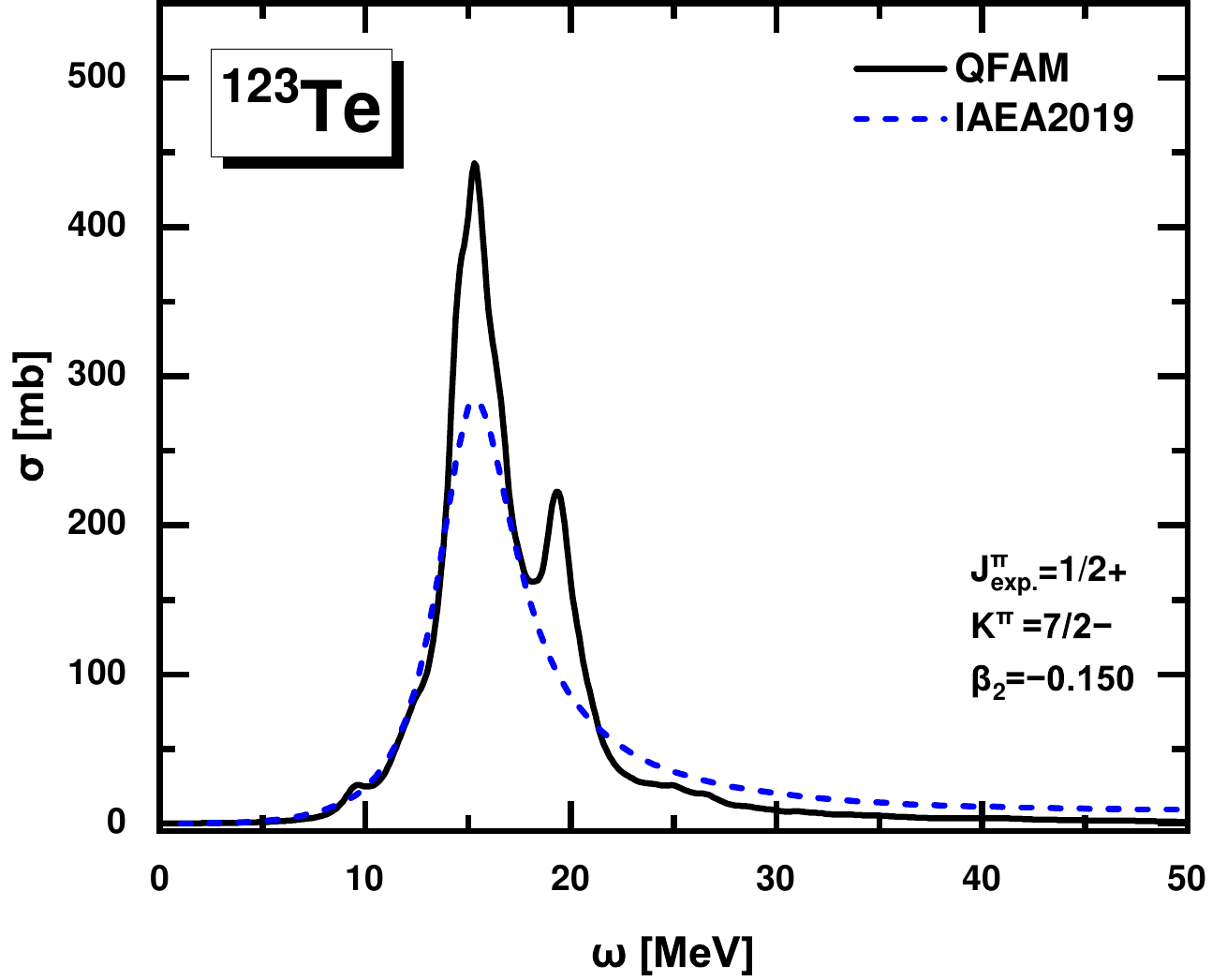}
    \includegraphics[width=0.35\textwidth]{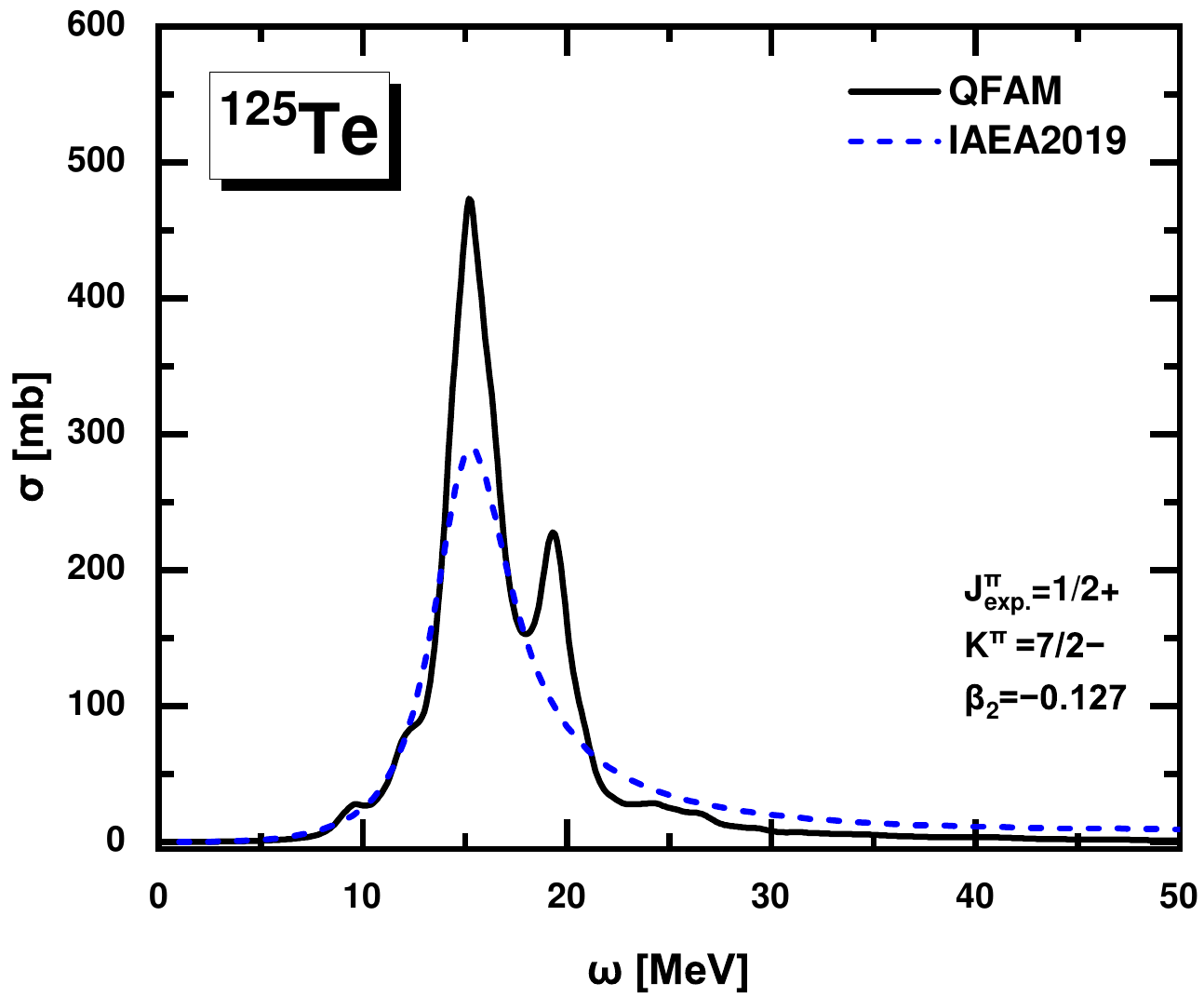}
    \includegraphics[width=0.35\textwidth]{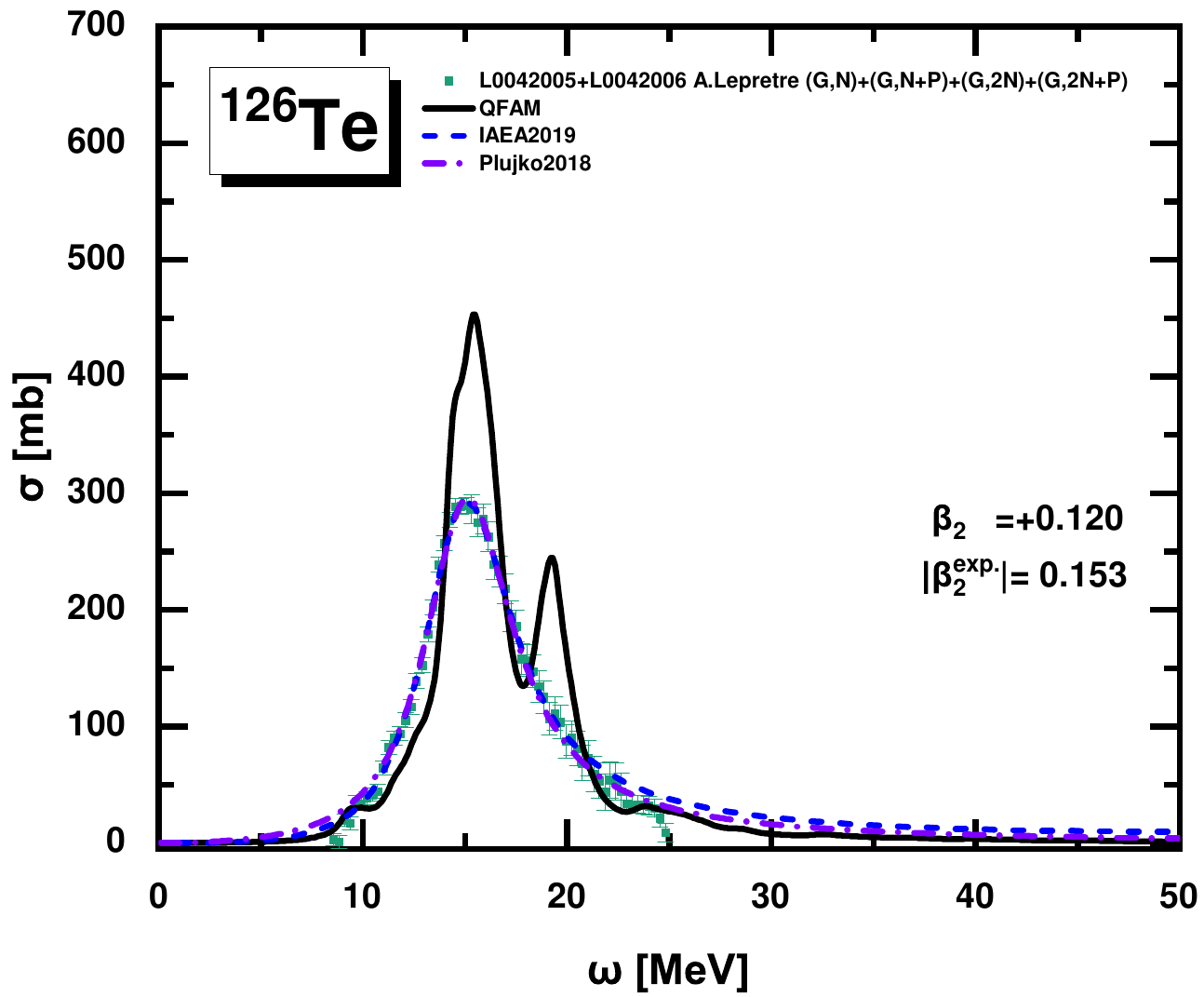}
    \includegraphics[width=0.35\textwidth]{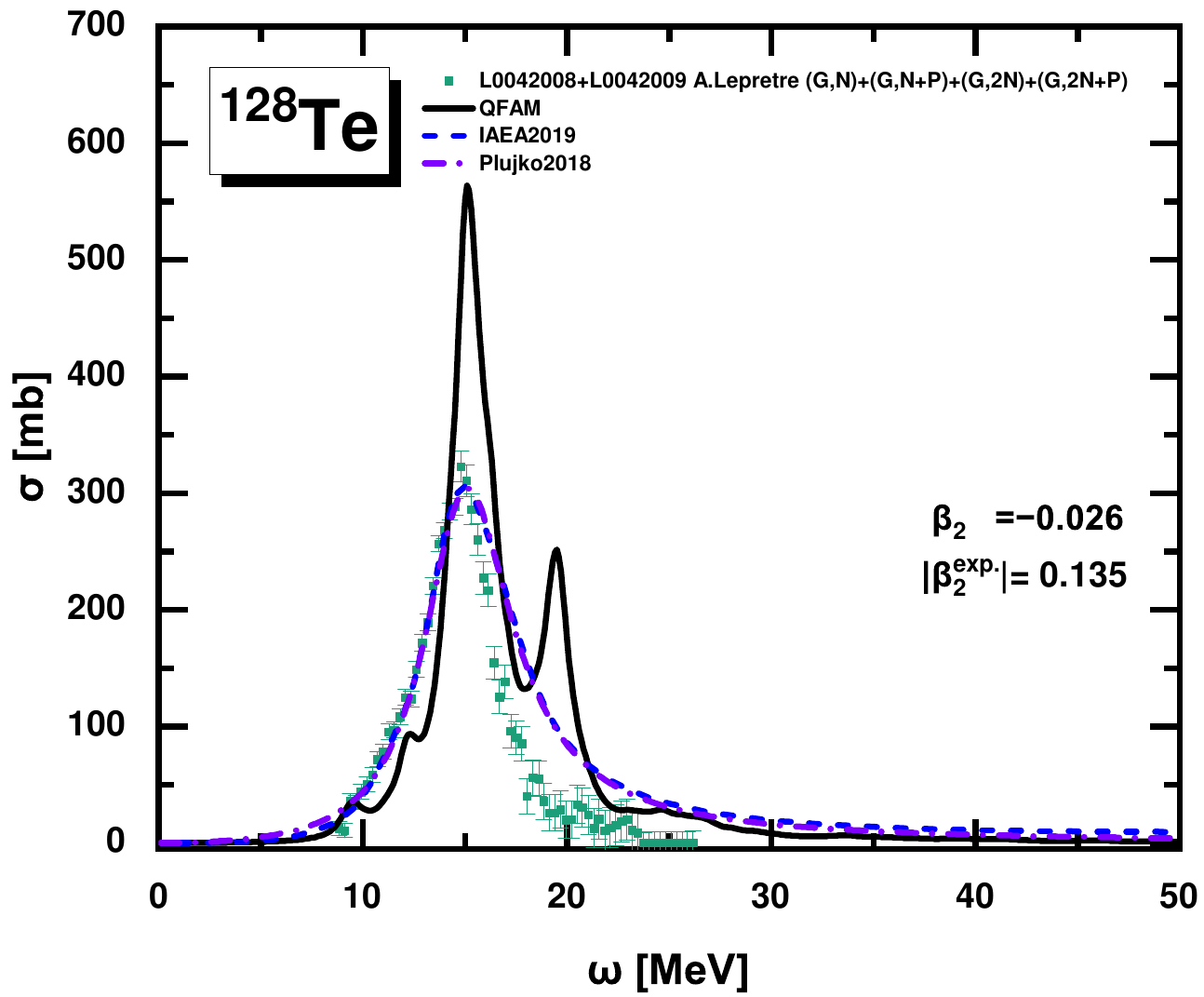}
\end{figure*}
\begin{figure*}\ContinuedFloat
    \centering
    \includegraphics[width=0.35\textwidth]{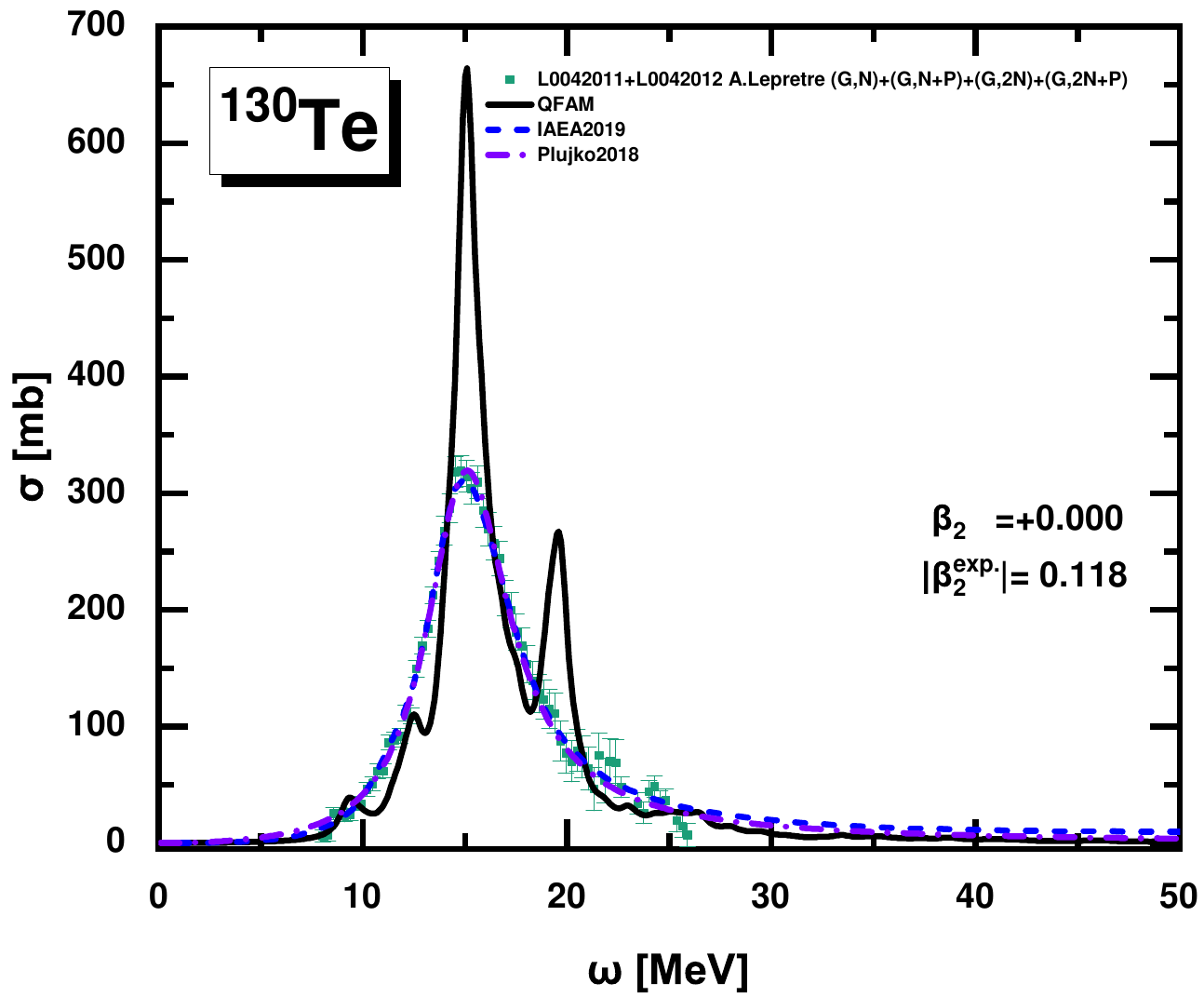}
    \includegraphics[width=0.35\textwidth]{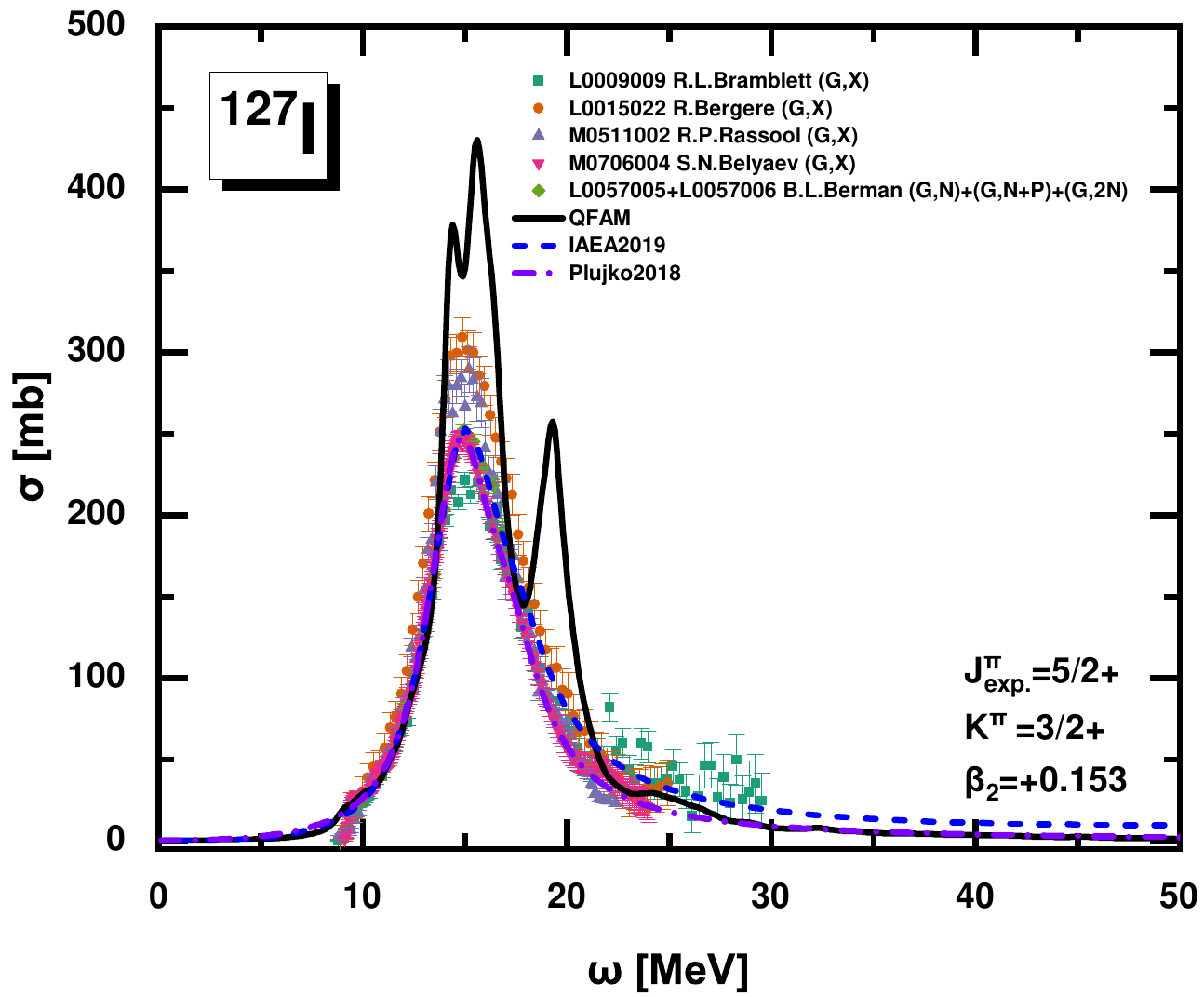}
    \includegraphics[width=0.35\textwidth]{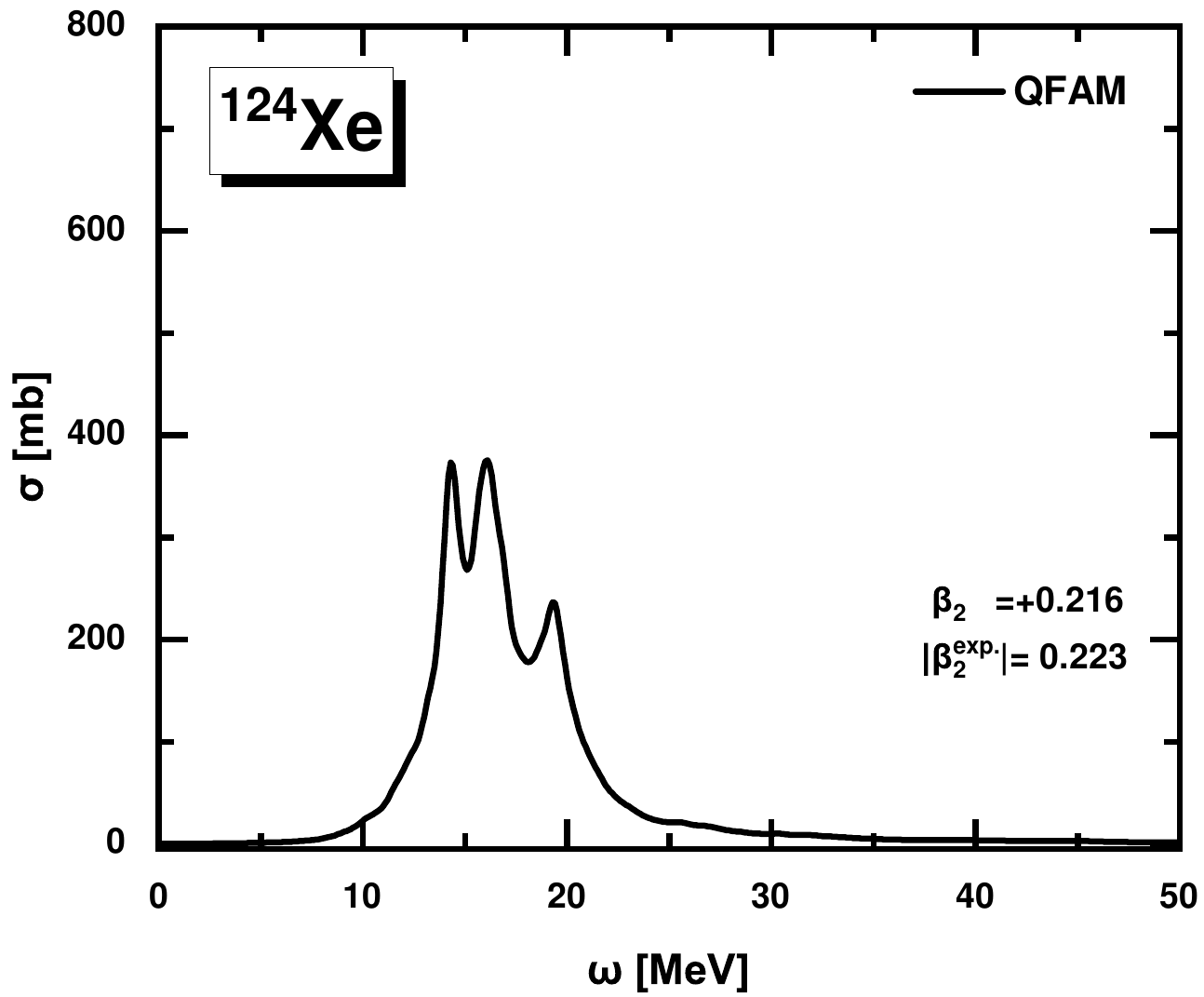}
    \includegraphics[width=0.35\textwidth]{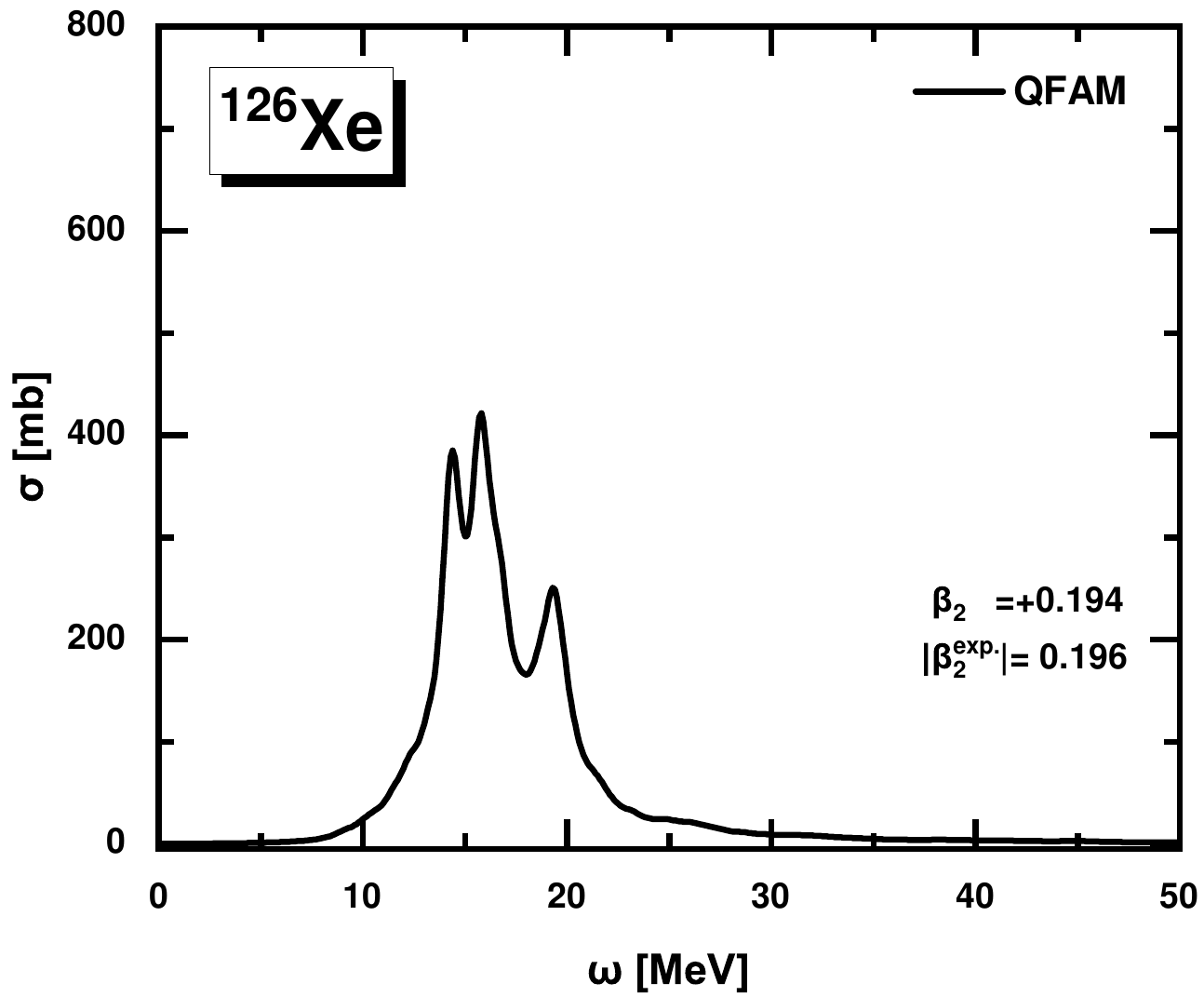}
    \includegraphics[width=0.35\textwidth]{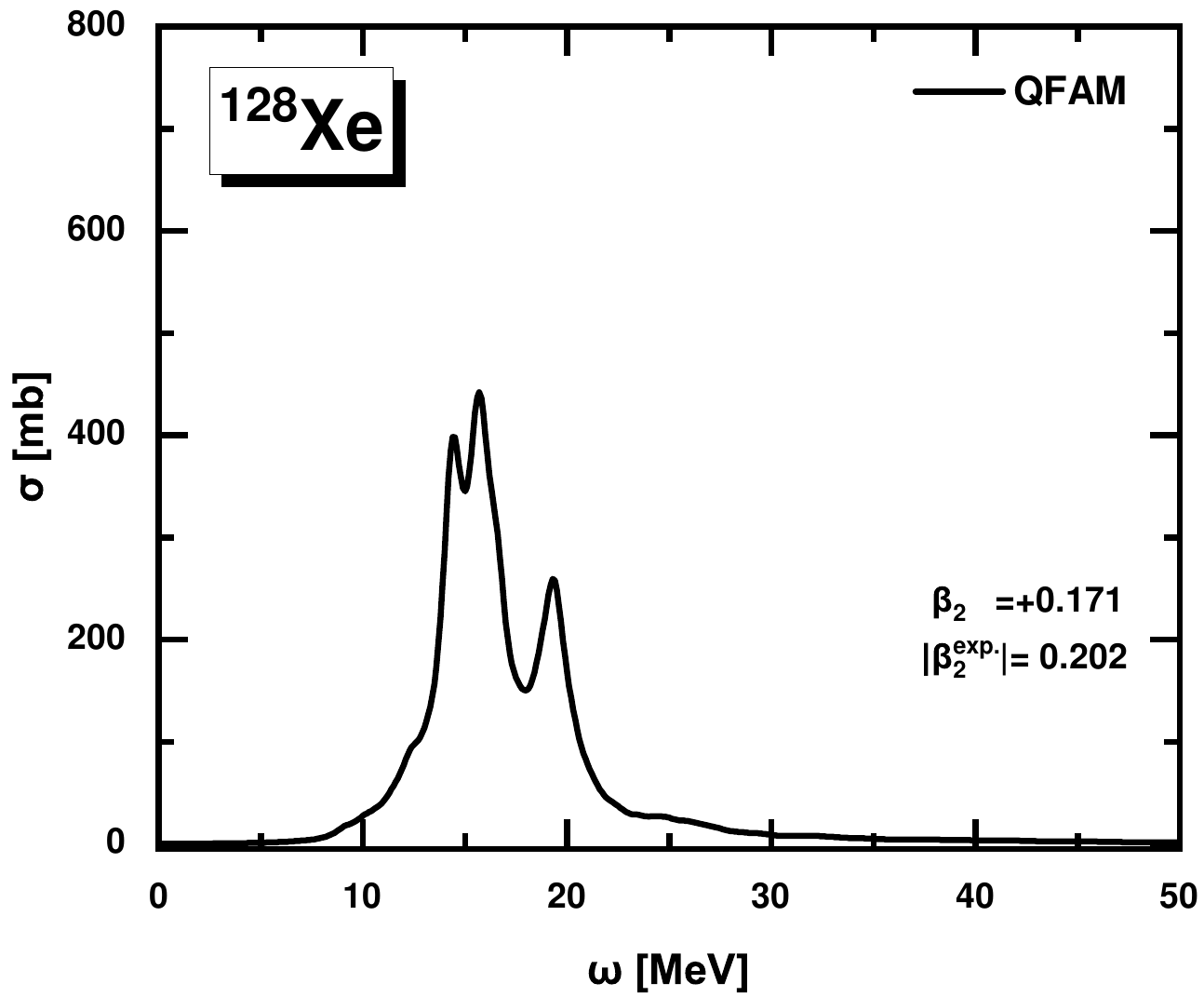}
    \includegraphics[width=0.35\textwidth]{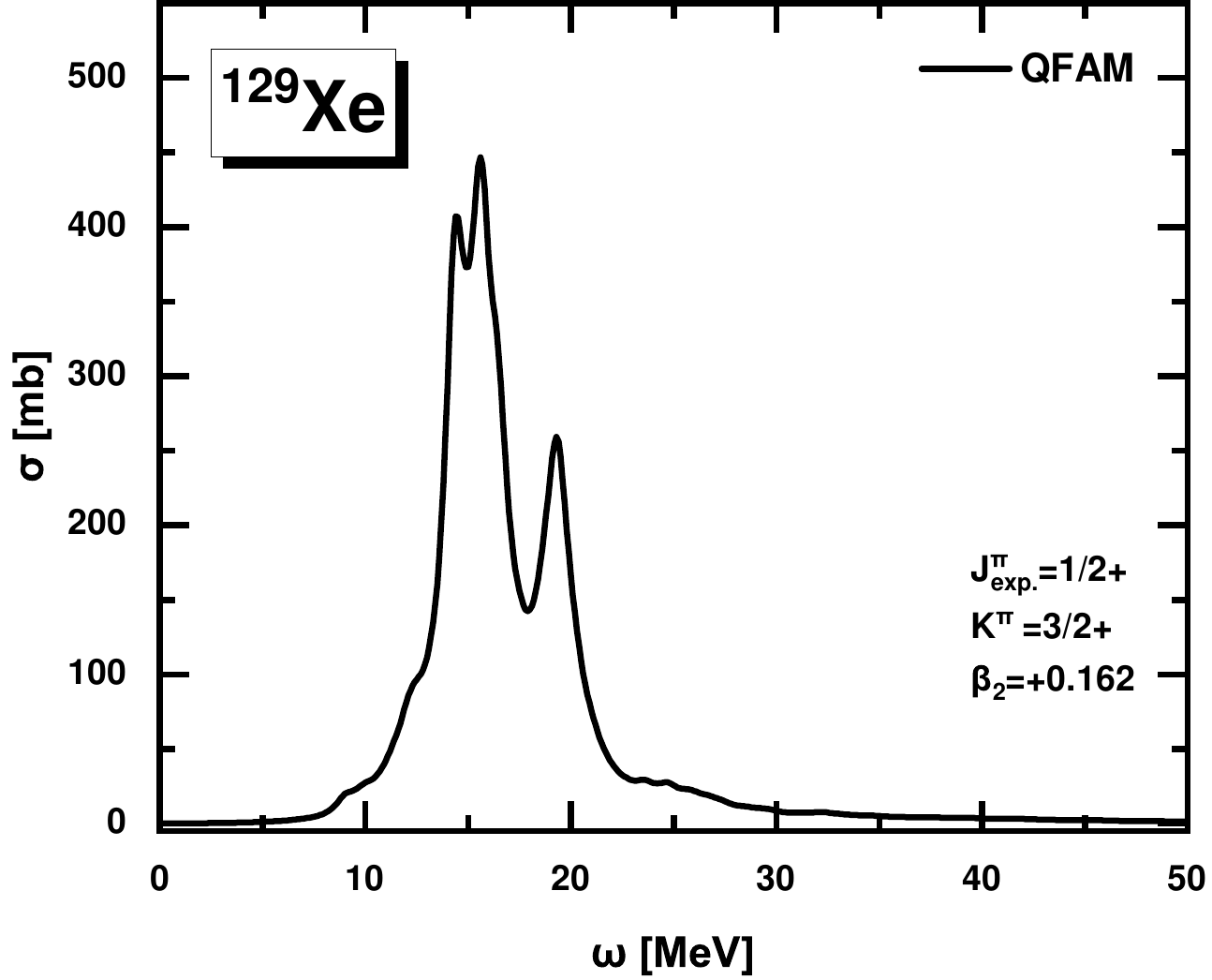}
    \includegraphics[width=0.35\textwidth]{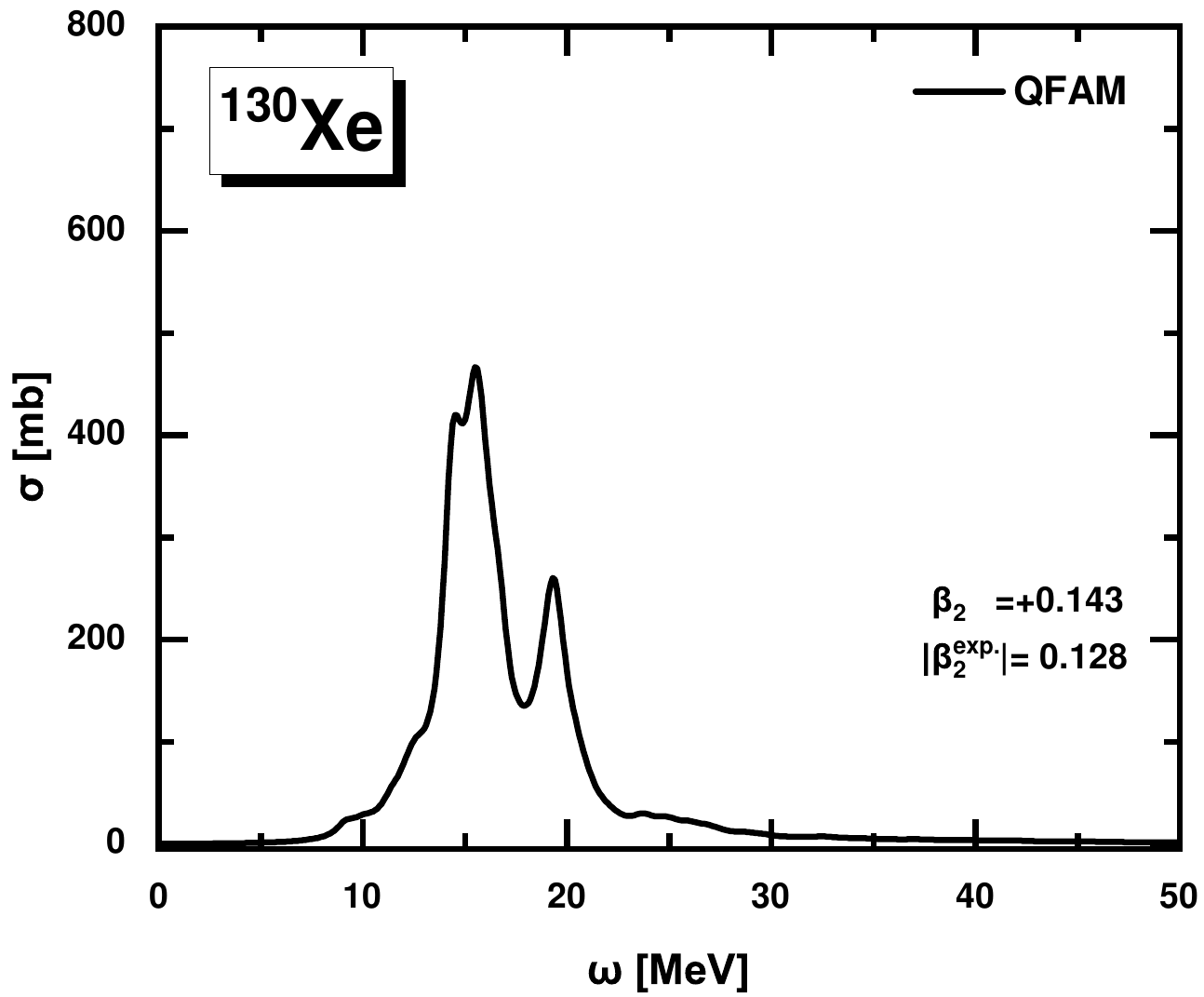}
    \includegraphics[width=0.35\textwidth]{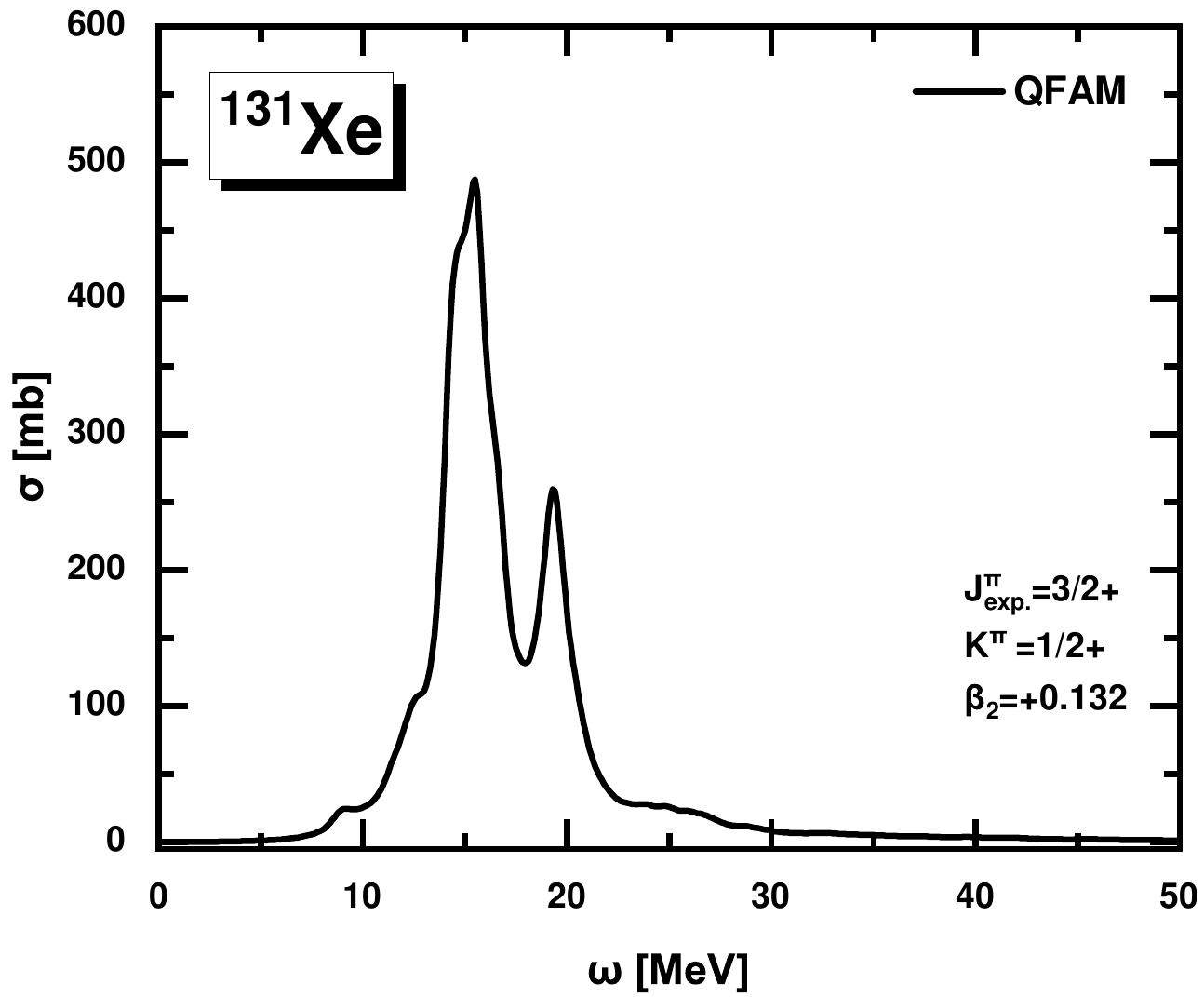}
\end{figure*}
\begin{figure*}\ContinuedFloat
    \centering
    \includegraphics[width=0.35\textwidth]{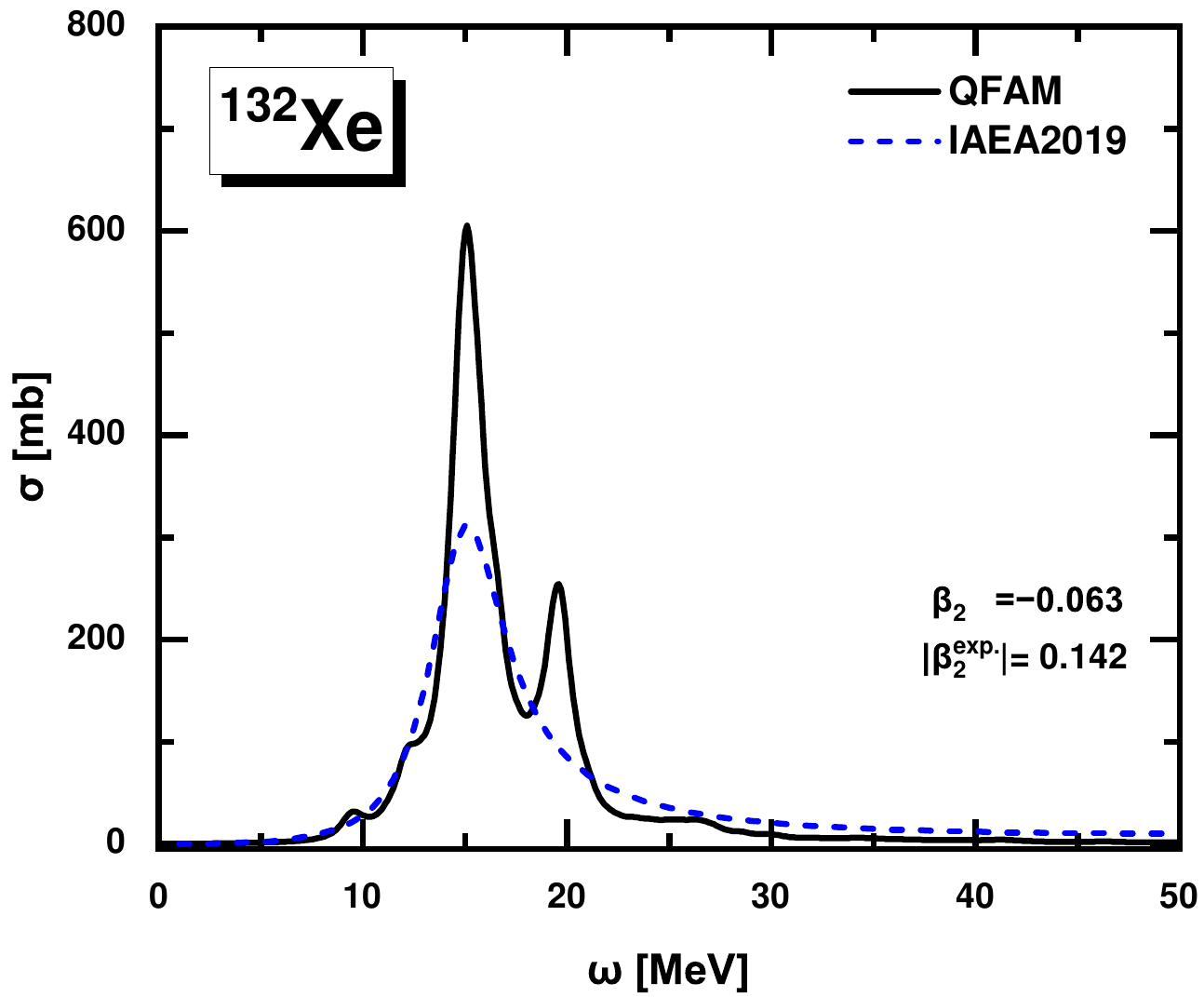}
    \includegraphics[width=0.35\textwidth]{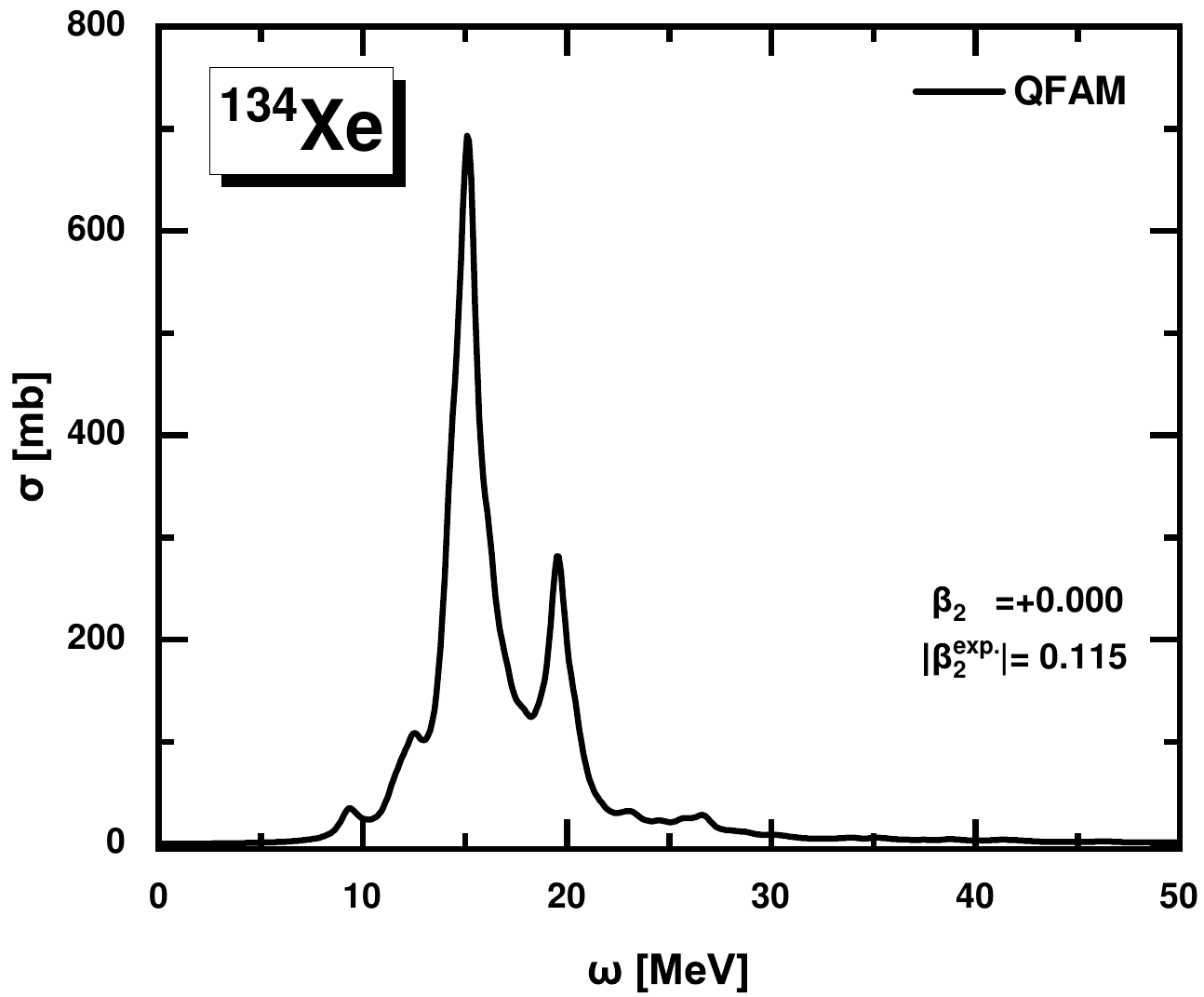}
    \includegraphics[width=0.35\textwidth]{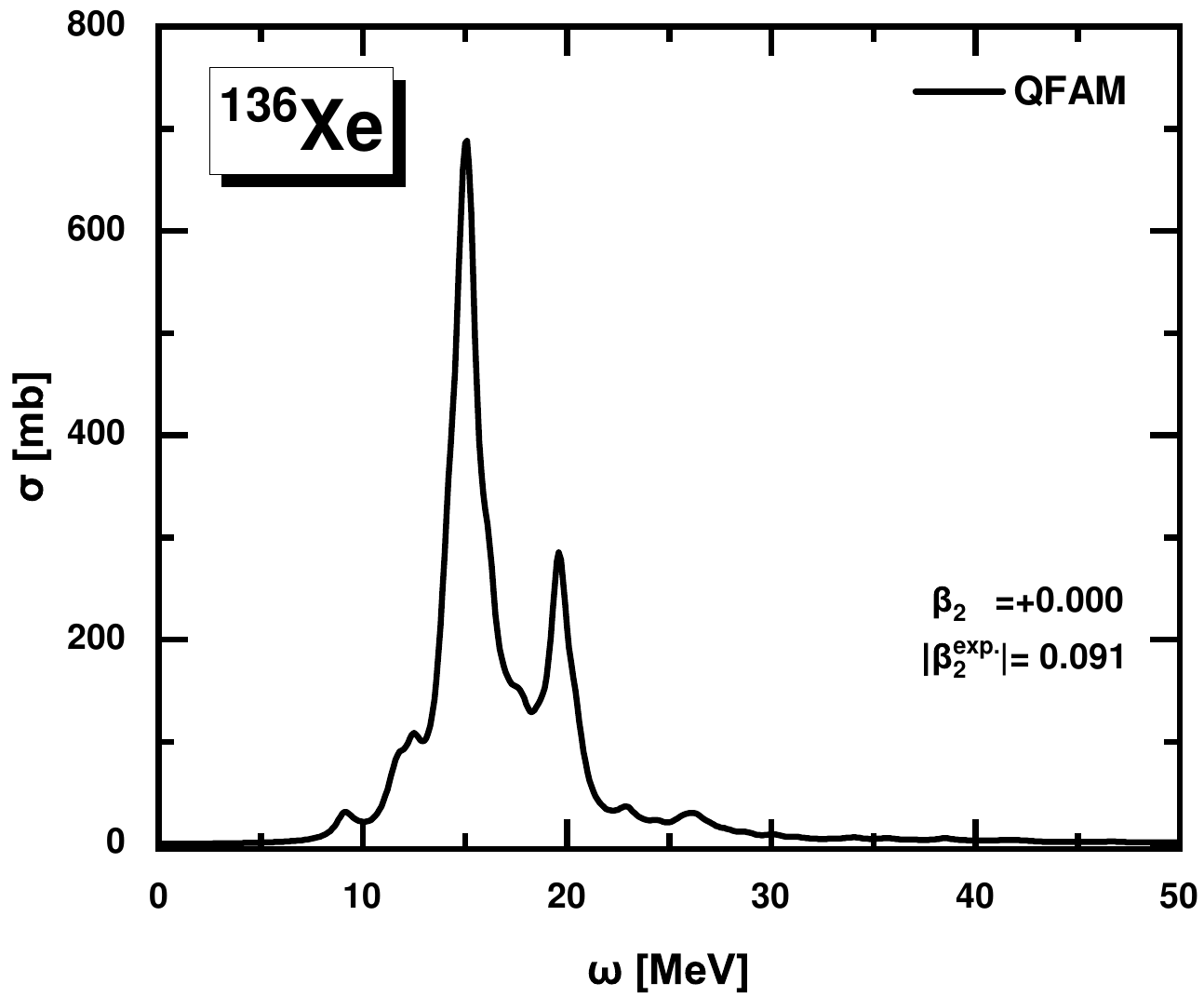}
    \includegraphics[width=0.35\textwidth]{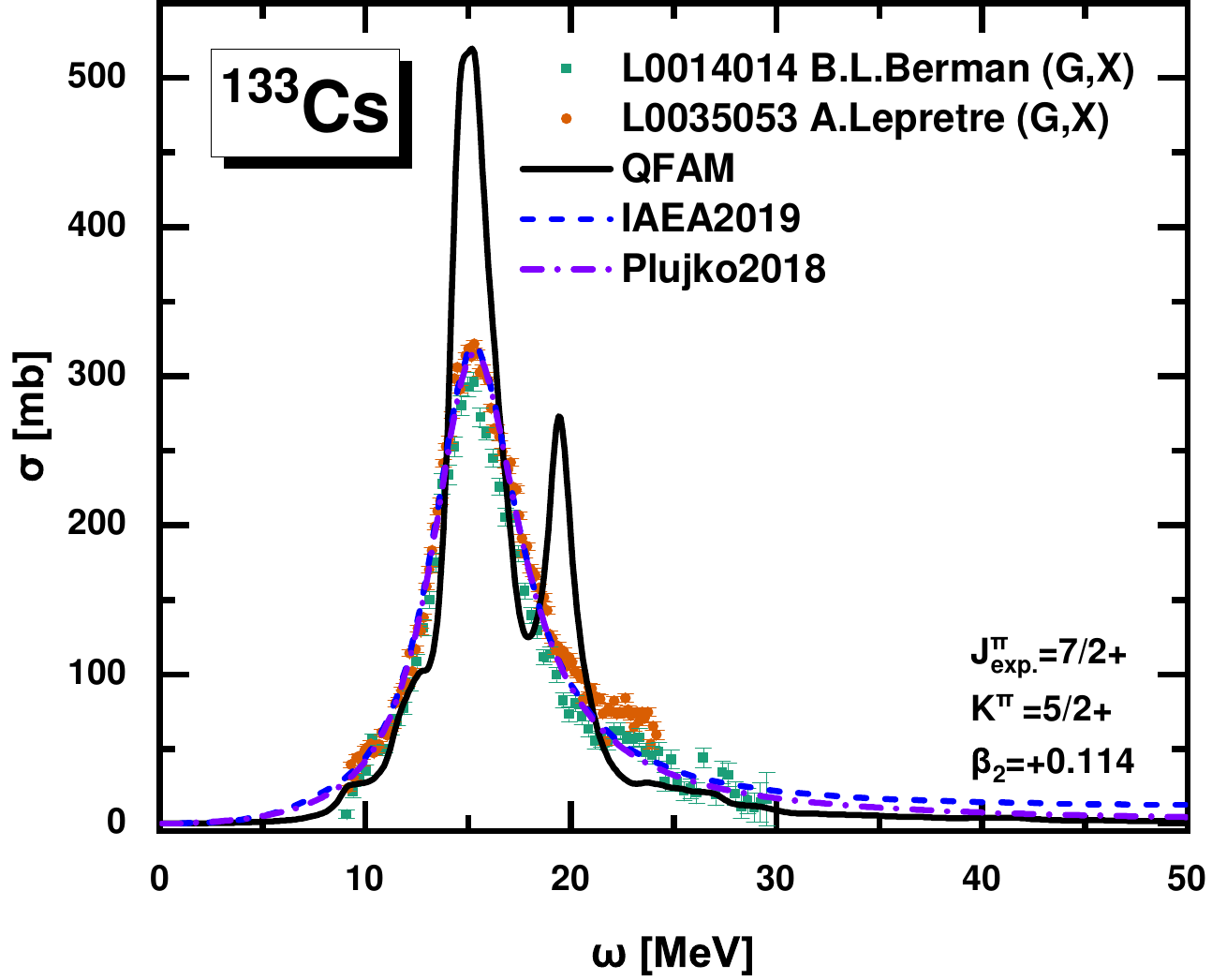}
    \includegraphics[width=0.35\textwidth]{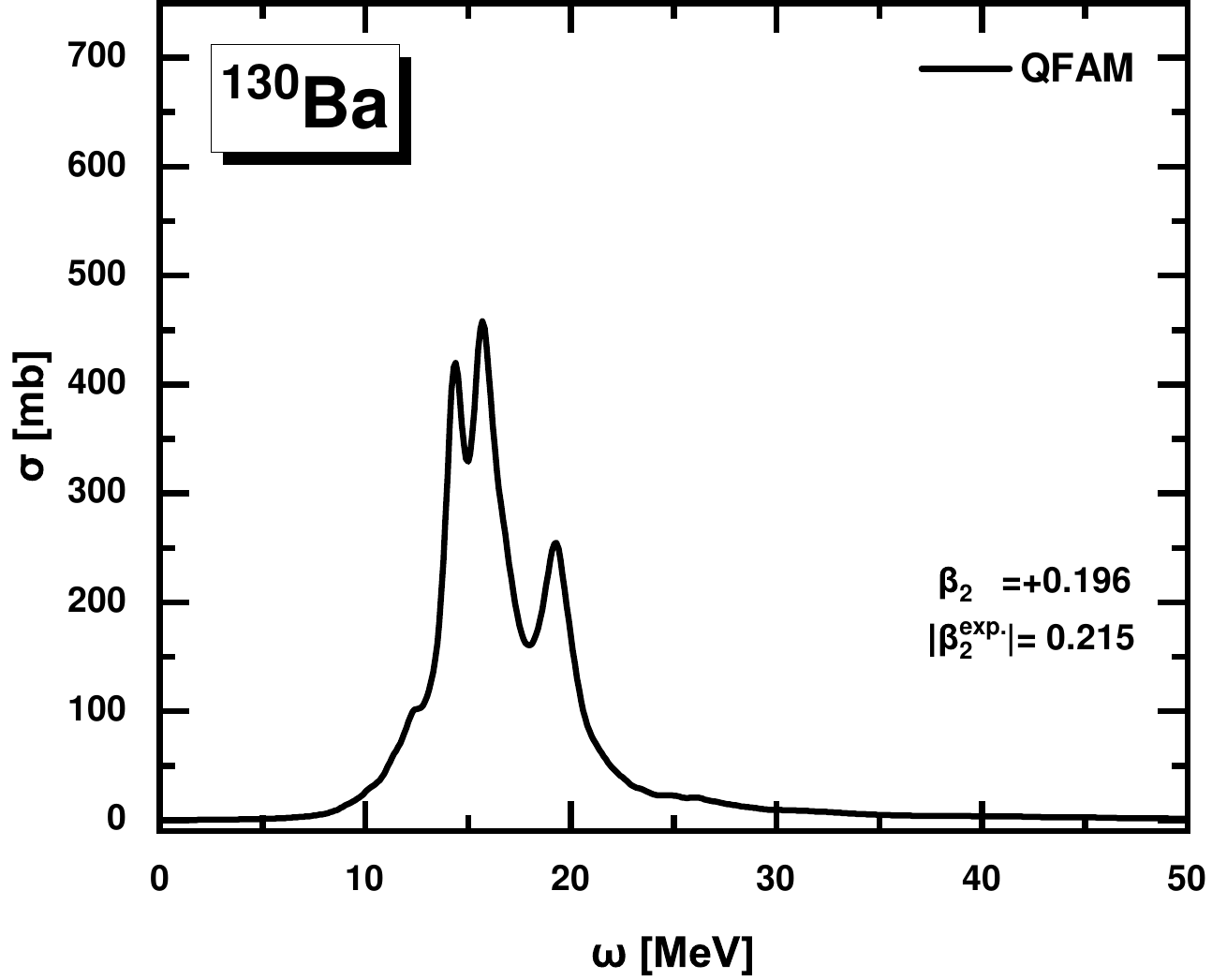}
    \includegraphics[width=0.35\textwidth]{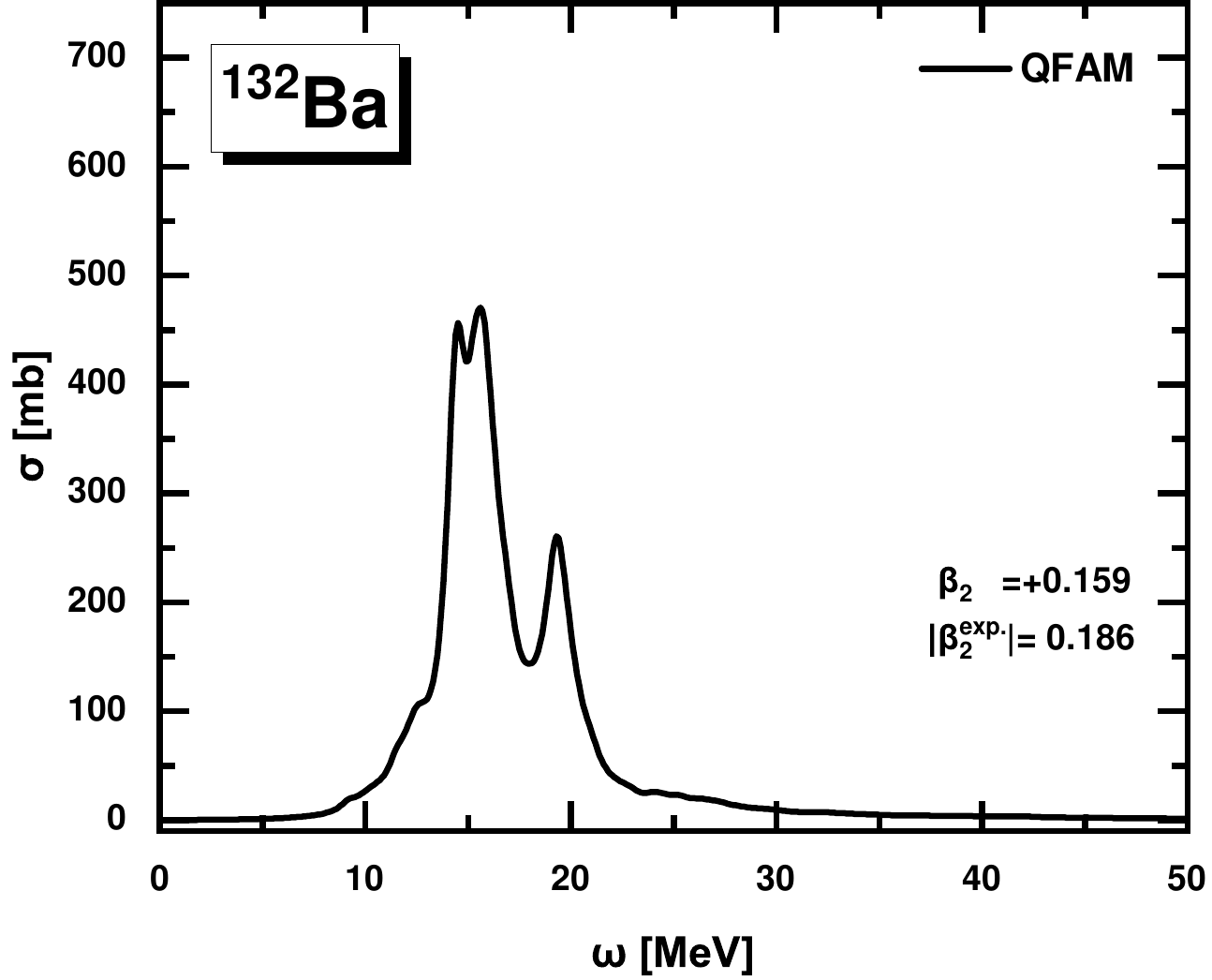}
    \includegraphics[width=0.35\textwidth]{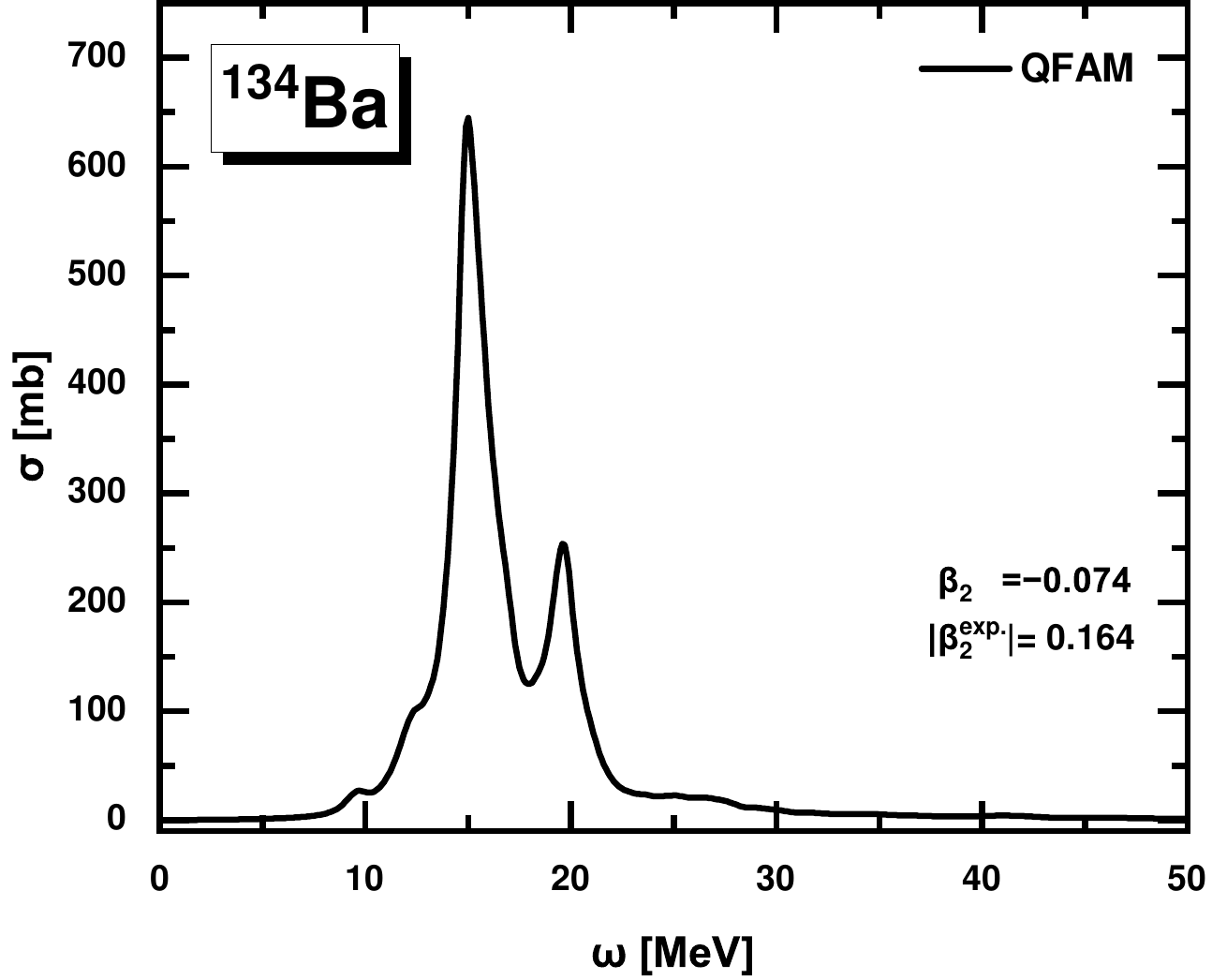}
    \includegraphics[width=0.35\textwidth]{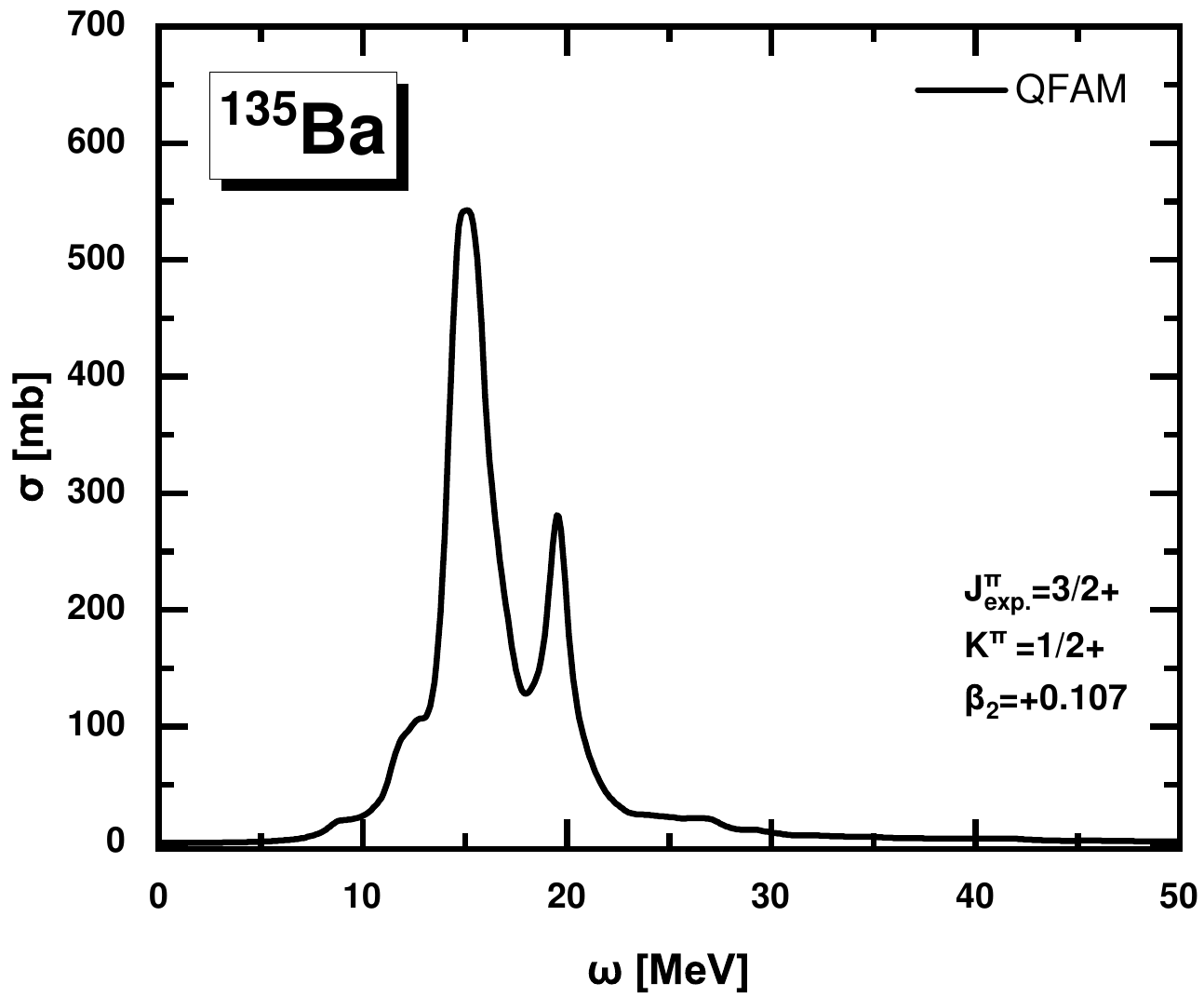}
\end{figure*}
\begin{figure*}\ContinuedFloat
    \centering
    \includegraphics[width=0.35\textwidth]{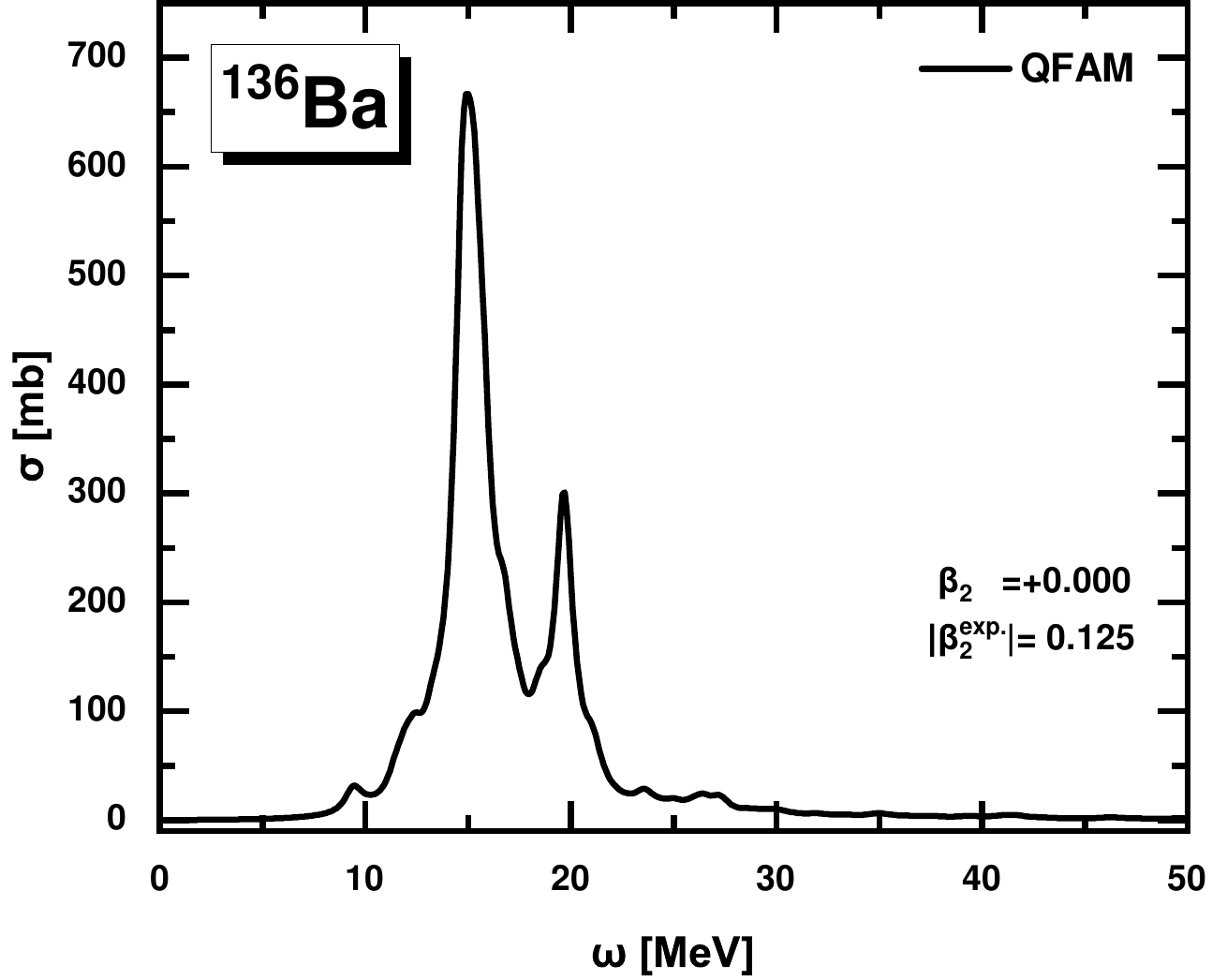}
    \includegraphics[width=0.35\textwidth]{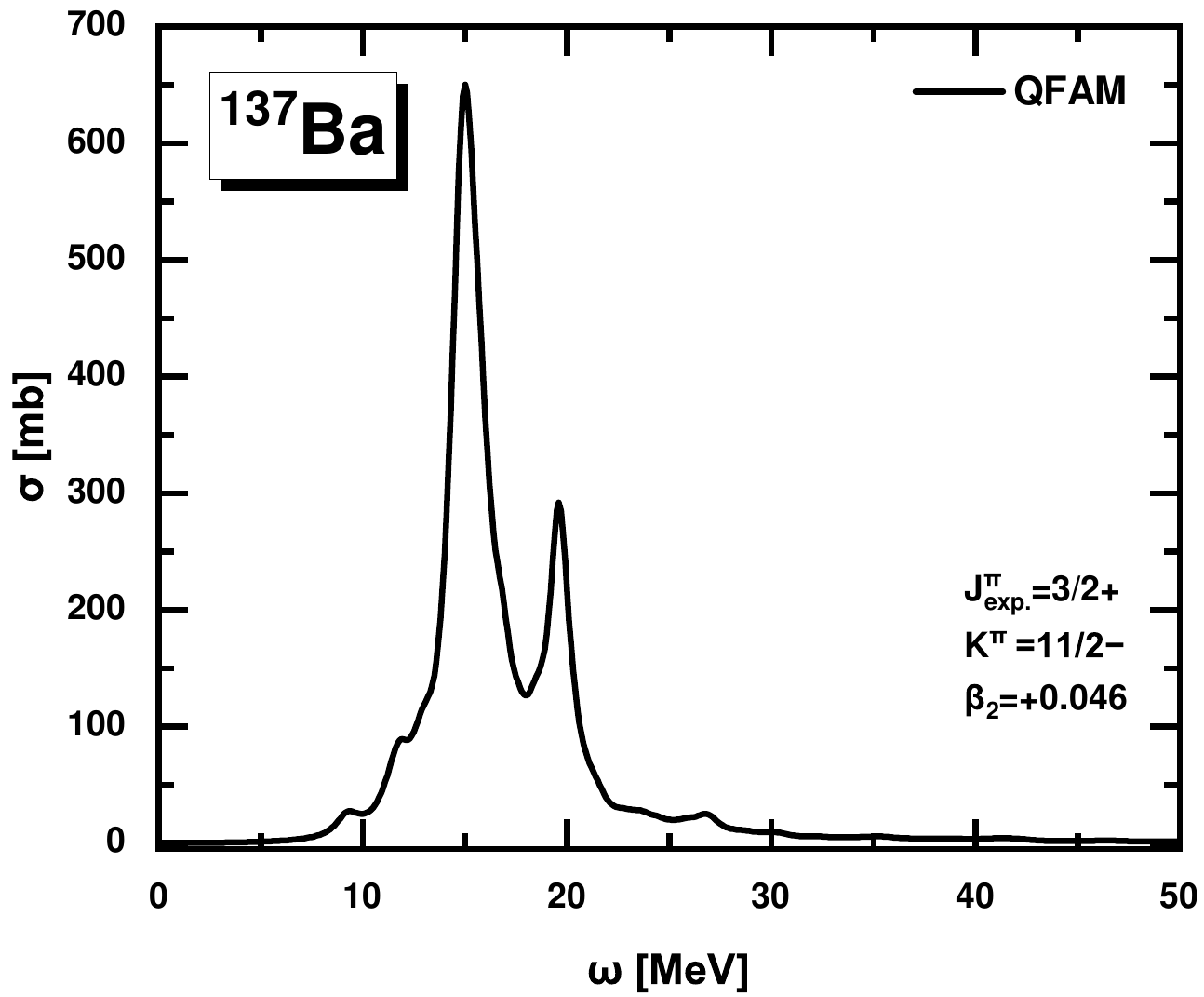}
    \includegraphics[width=0.35\textwidth]{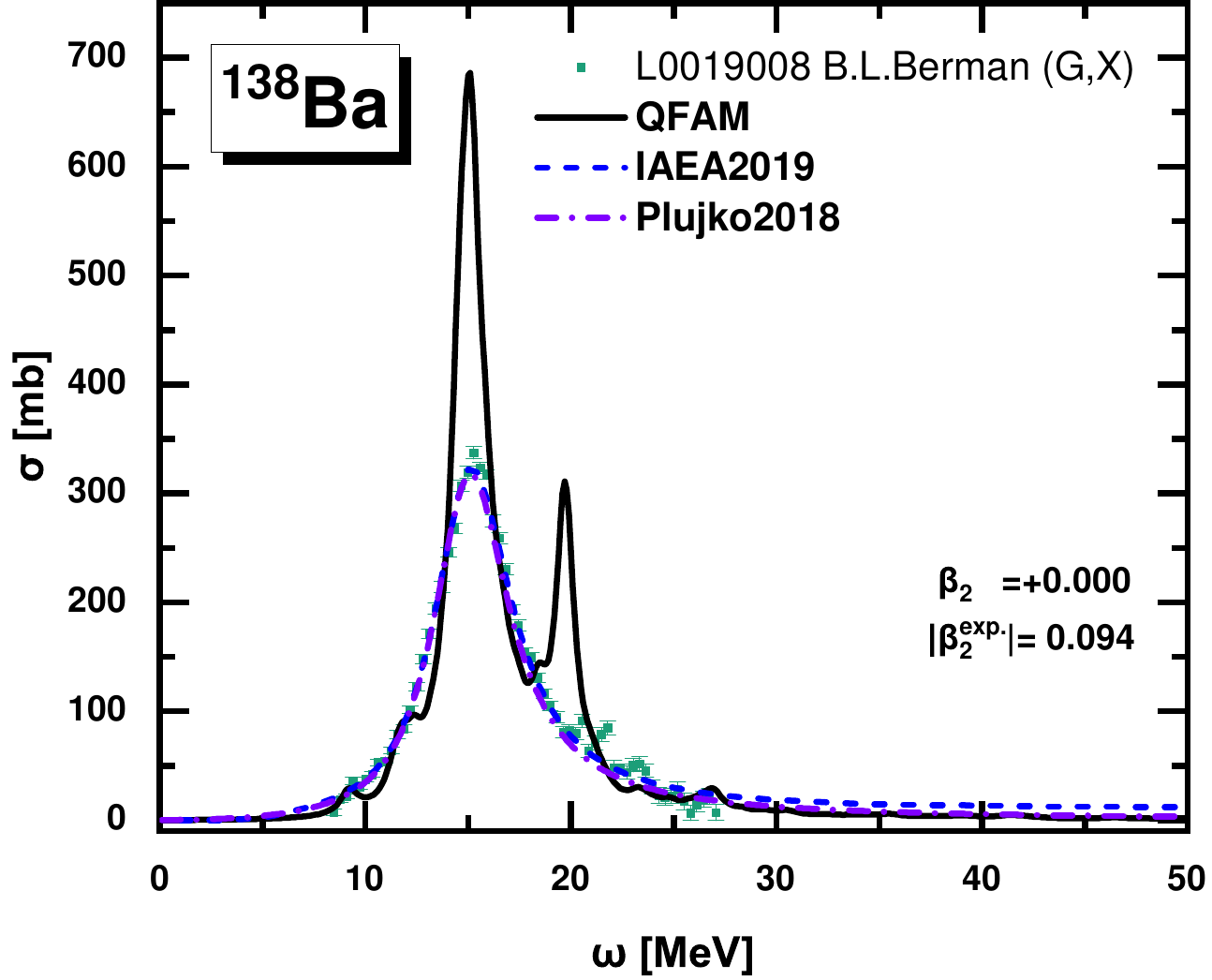}
    \includegraphics[width=0.35\textwidth]{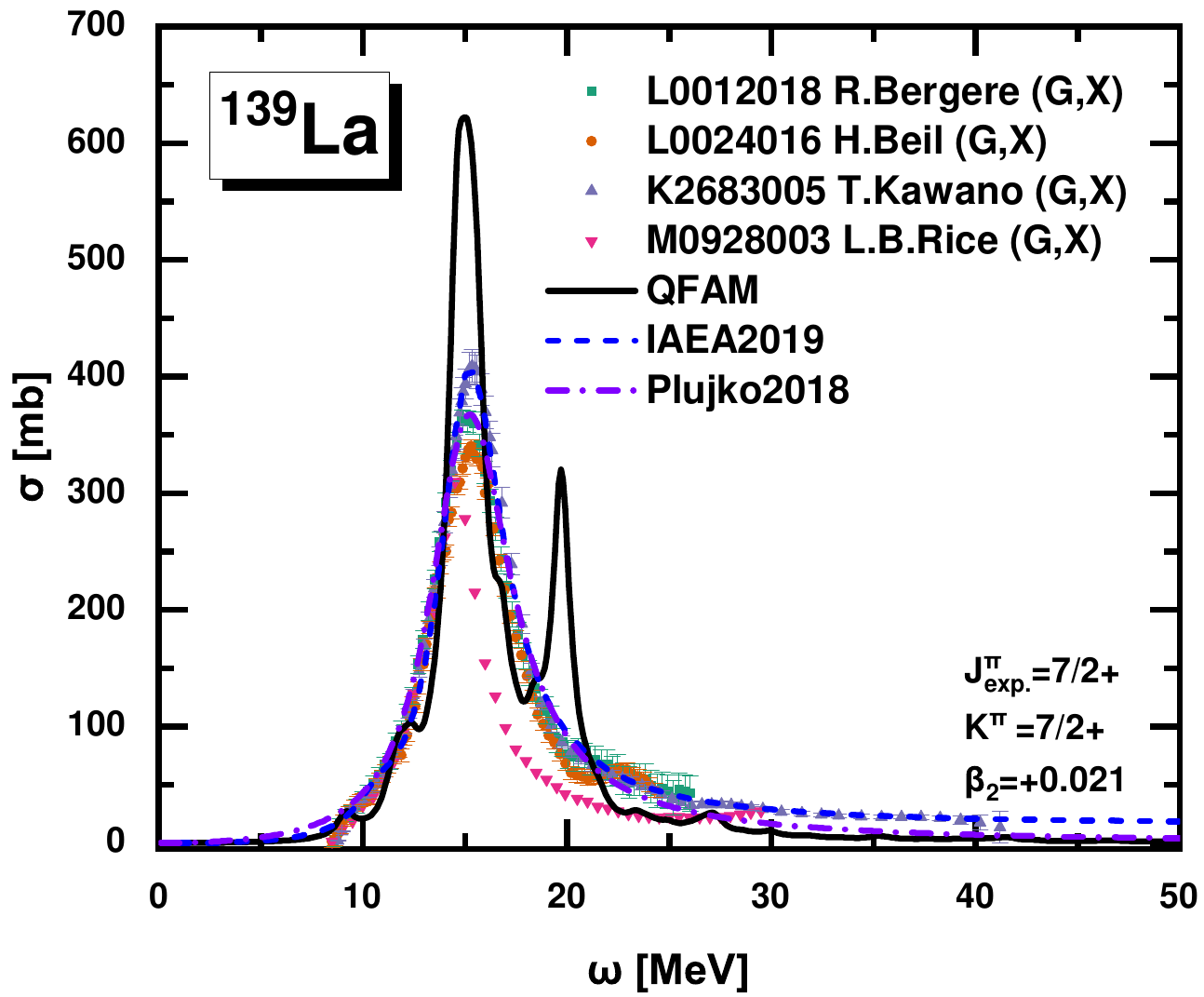}
    \includegraphics[width=0.35\textwidth]{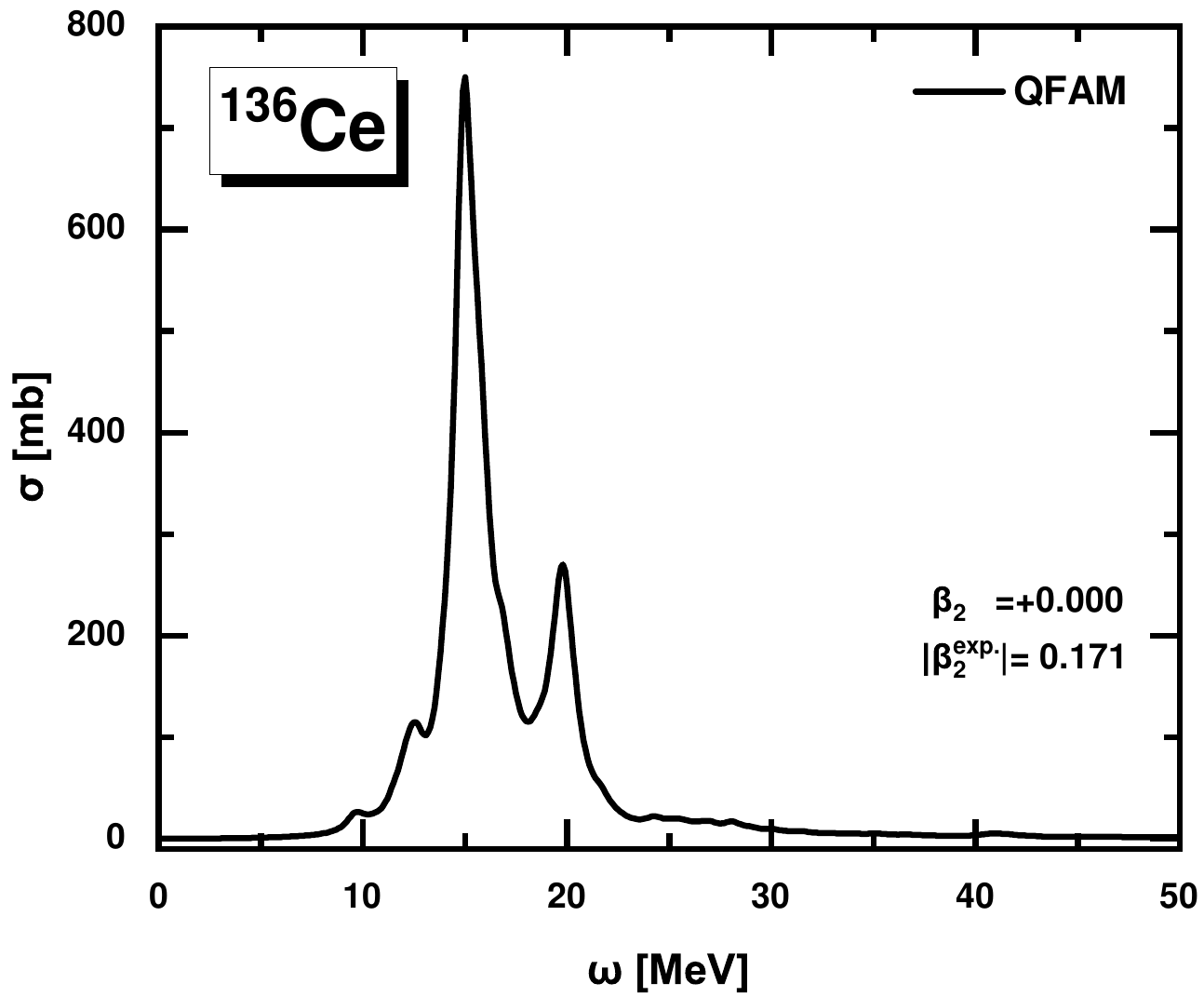}
    \includegraphics[width=0.35\textwidth]{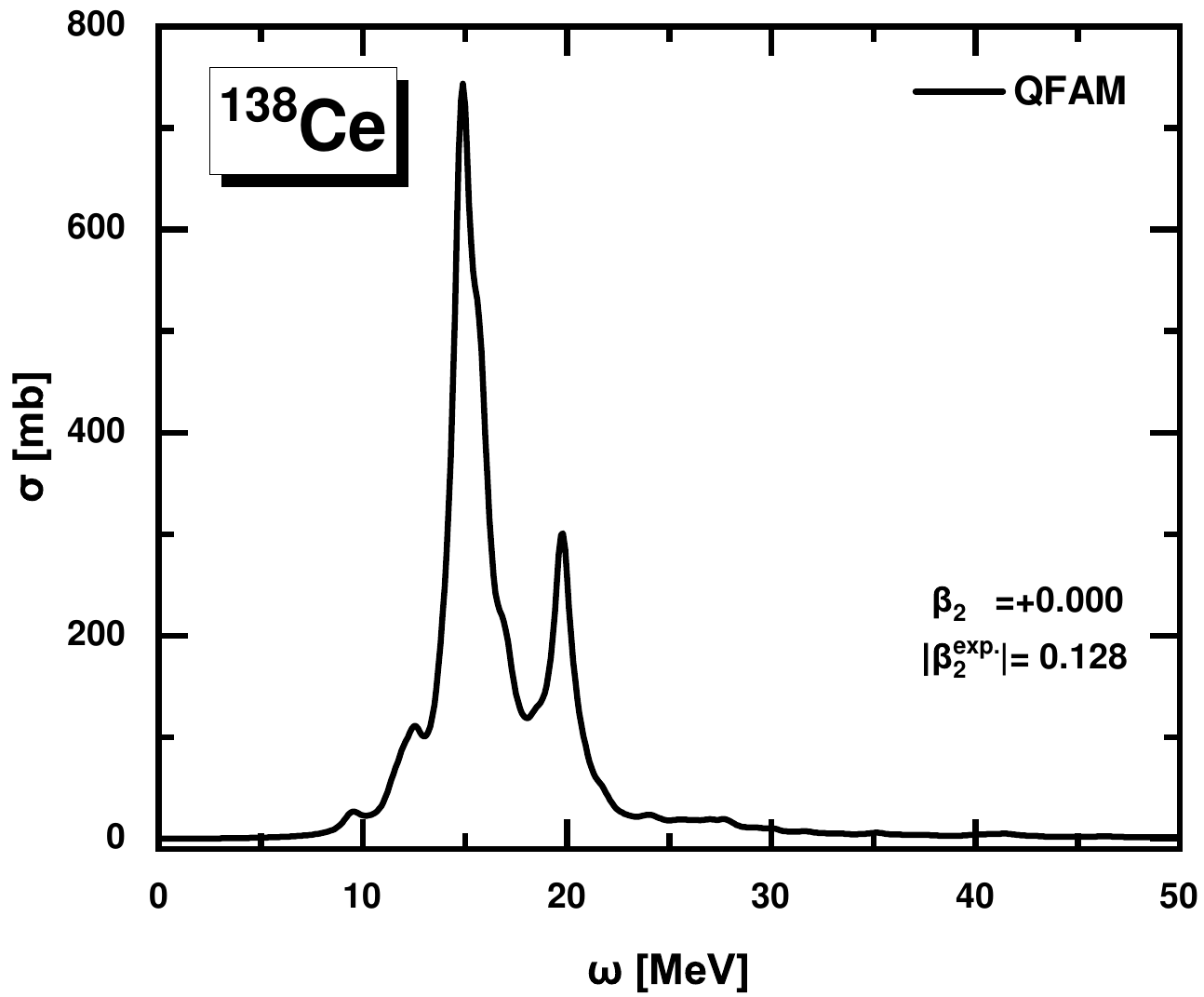}
    \includegraphics[width=0.35\textwidth]{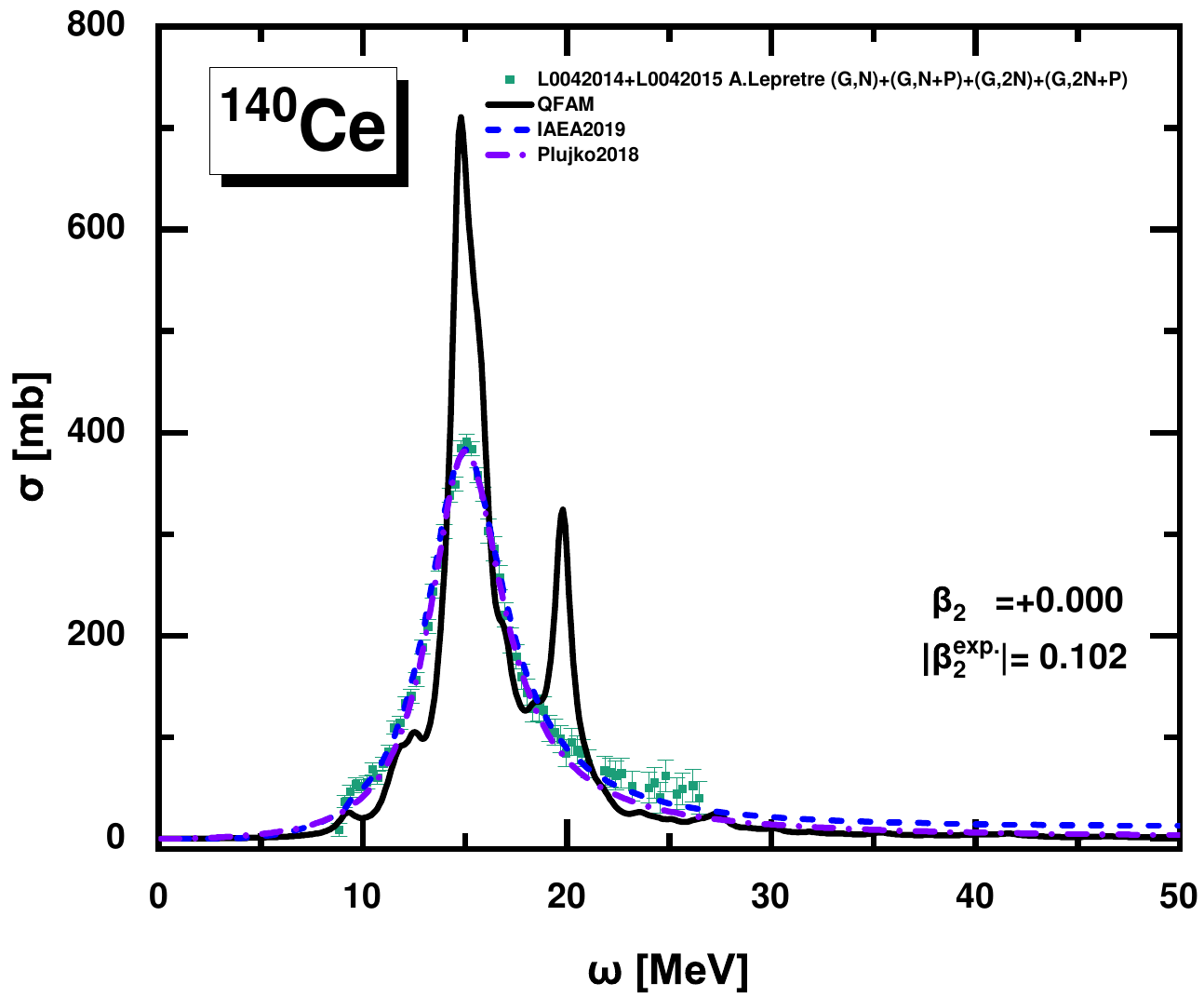}
    \includegraphics[width=0.35\textwidth]{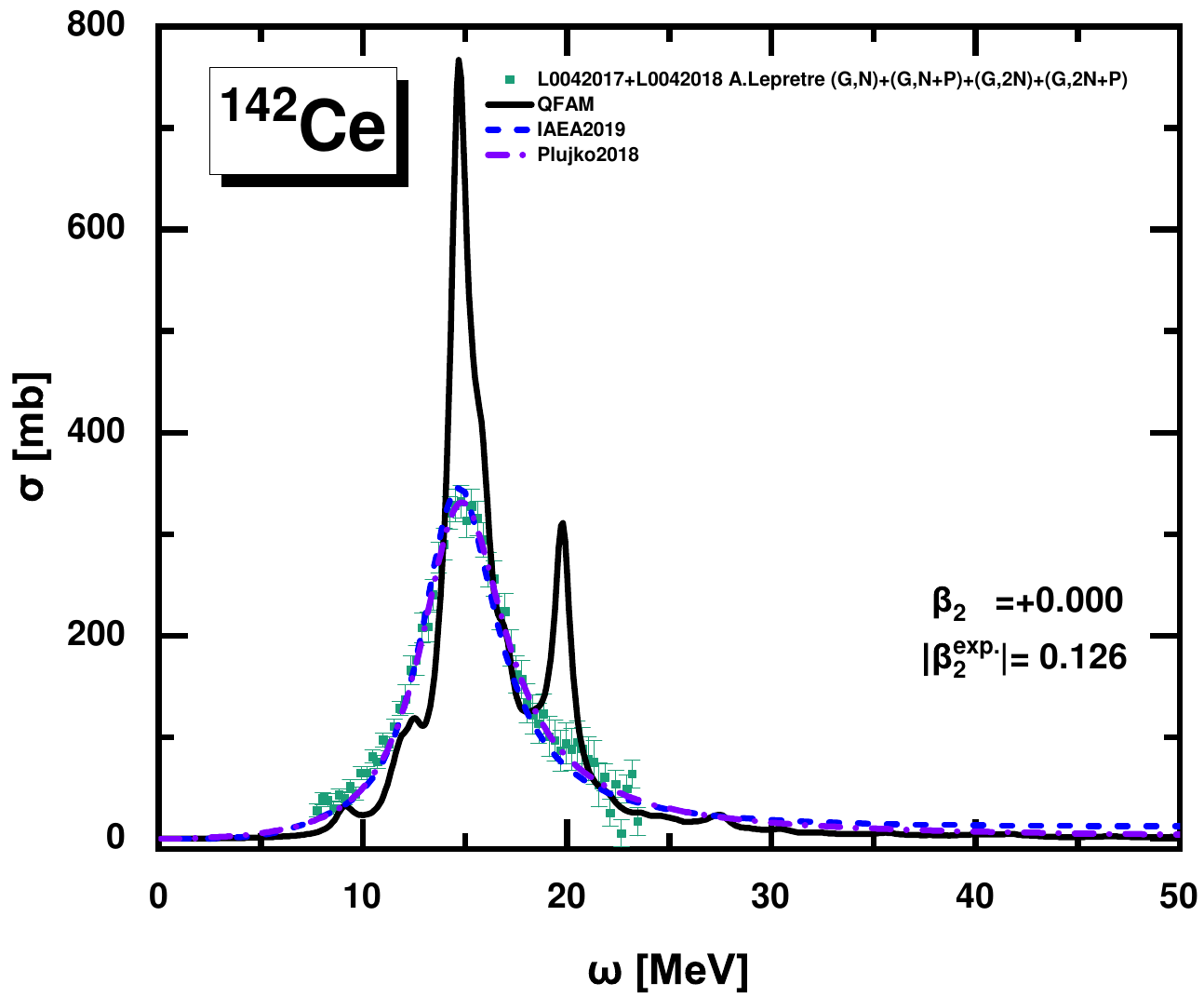}
\end{figure*}
\begin{figure*}\ContinuedFloat
    \centering
    \includegraphics[width=0.35\textwidth]{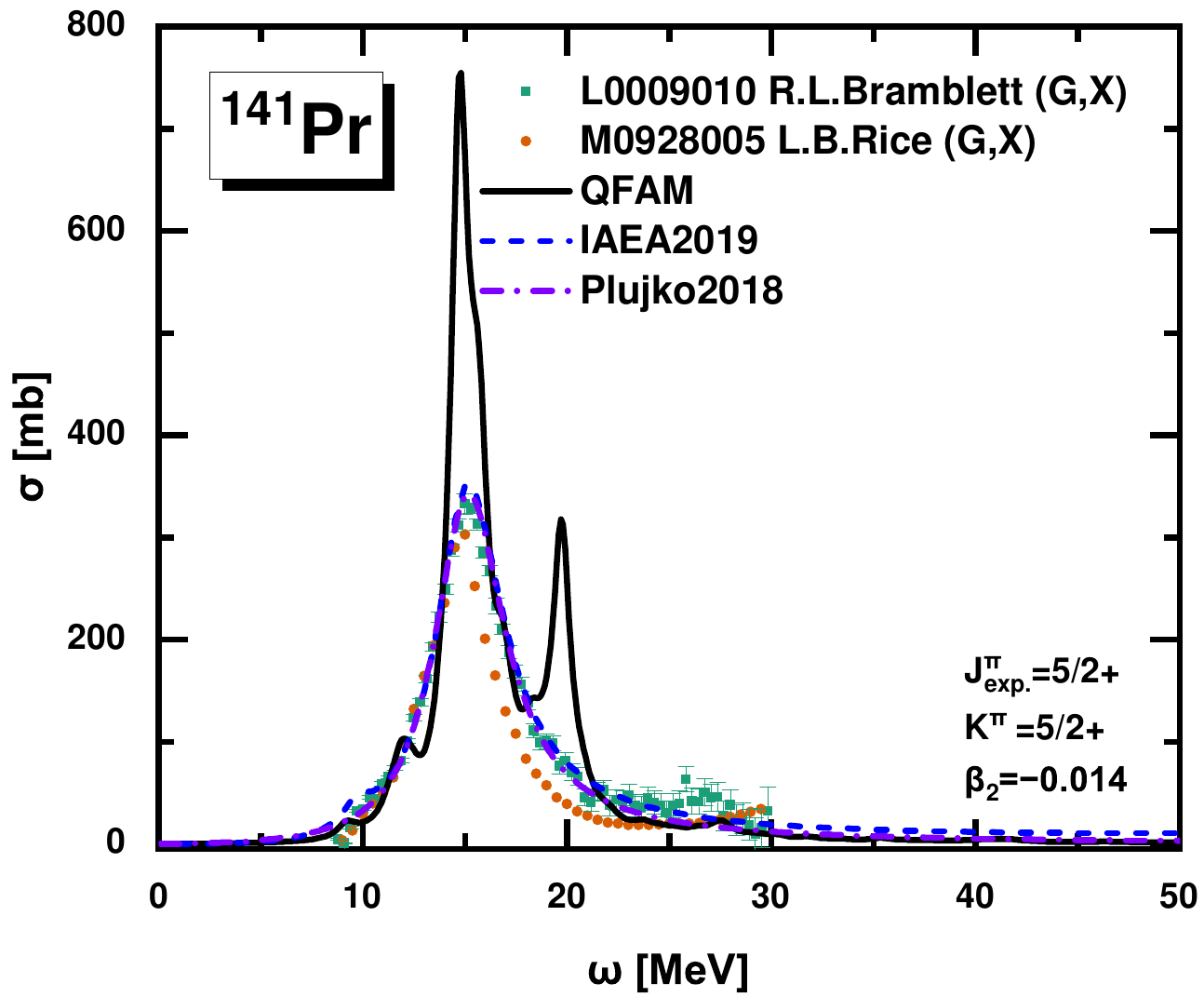}
    \includegraphics[width=0.35\textwidth]{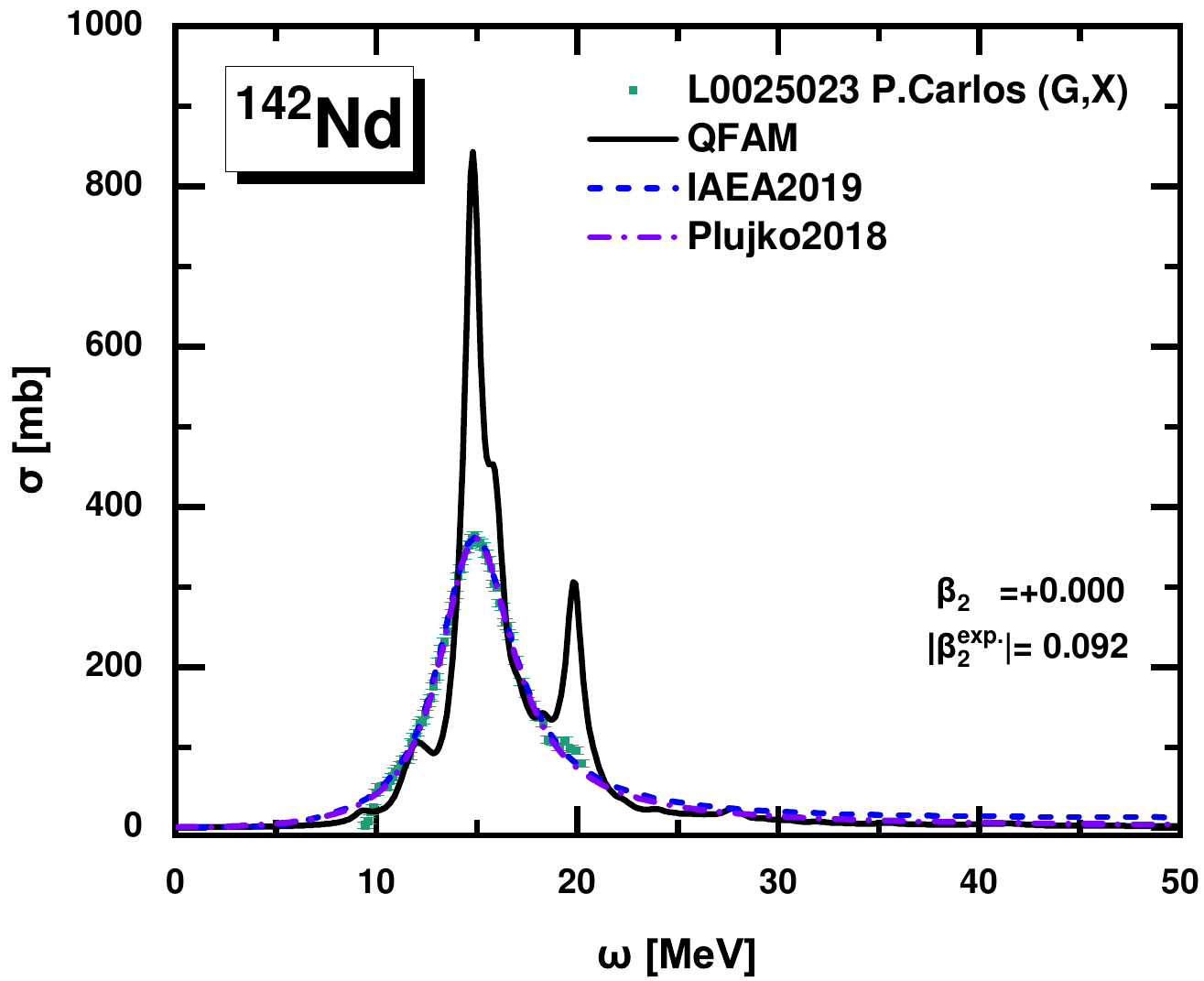}
    \includegraphics[width=0.35\textwidth]{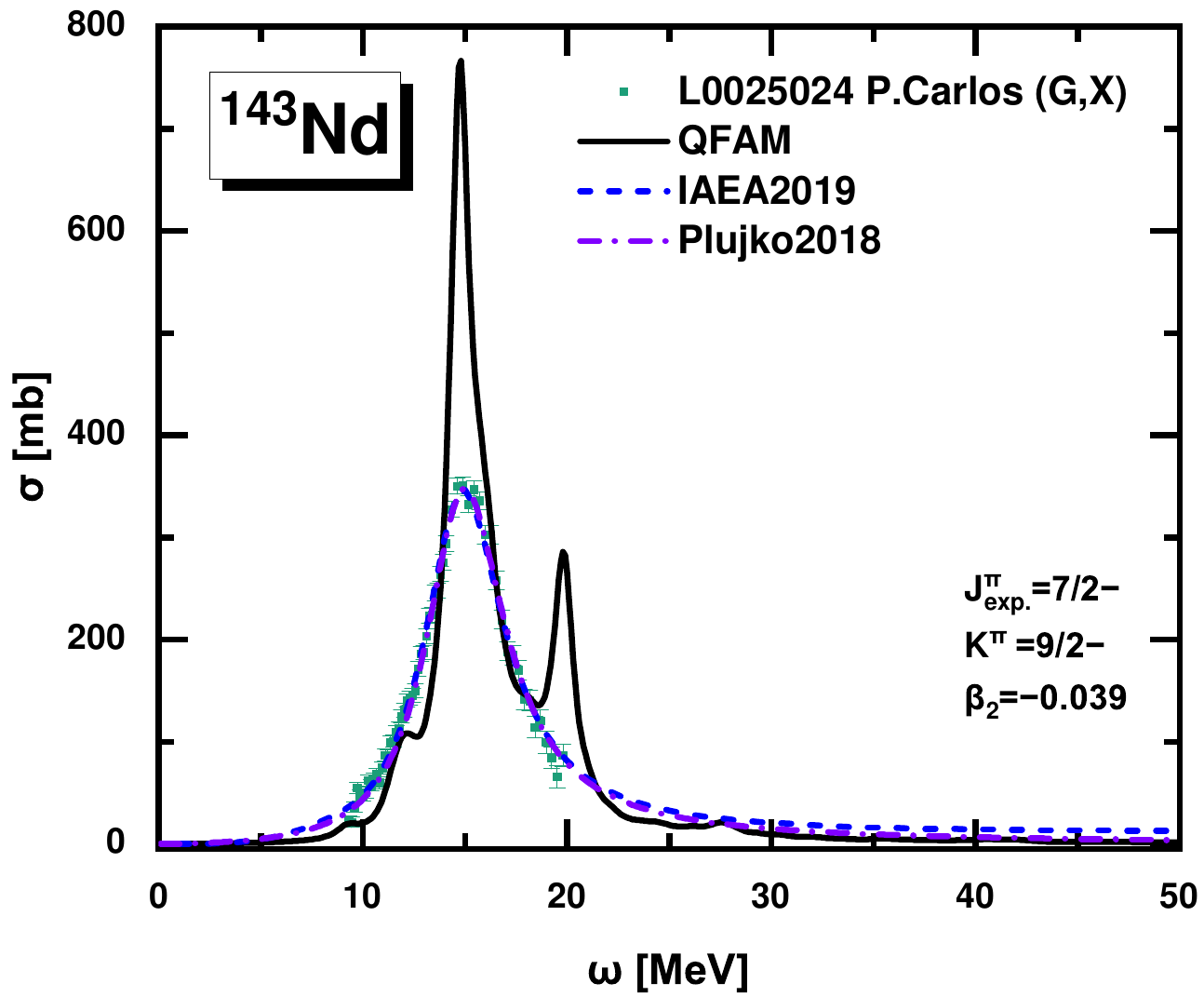}
    \includegraphics[width=0.35\textwidth]{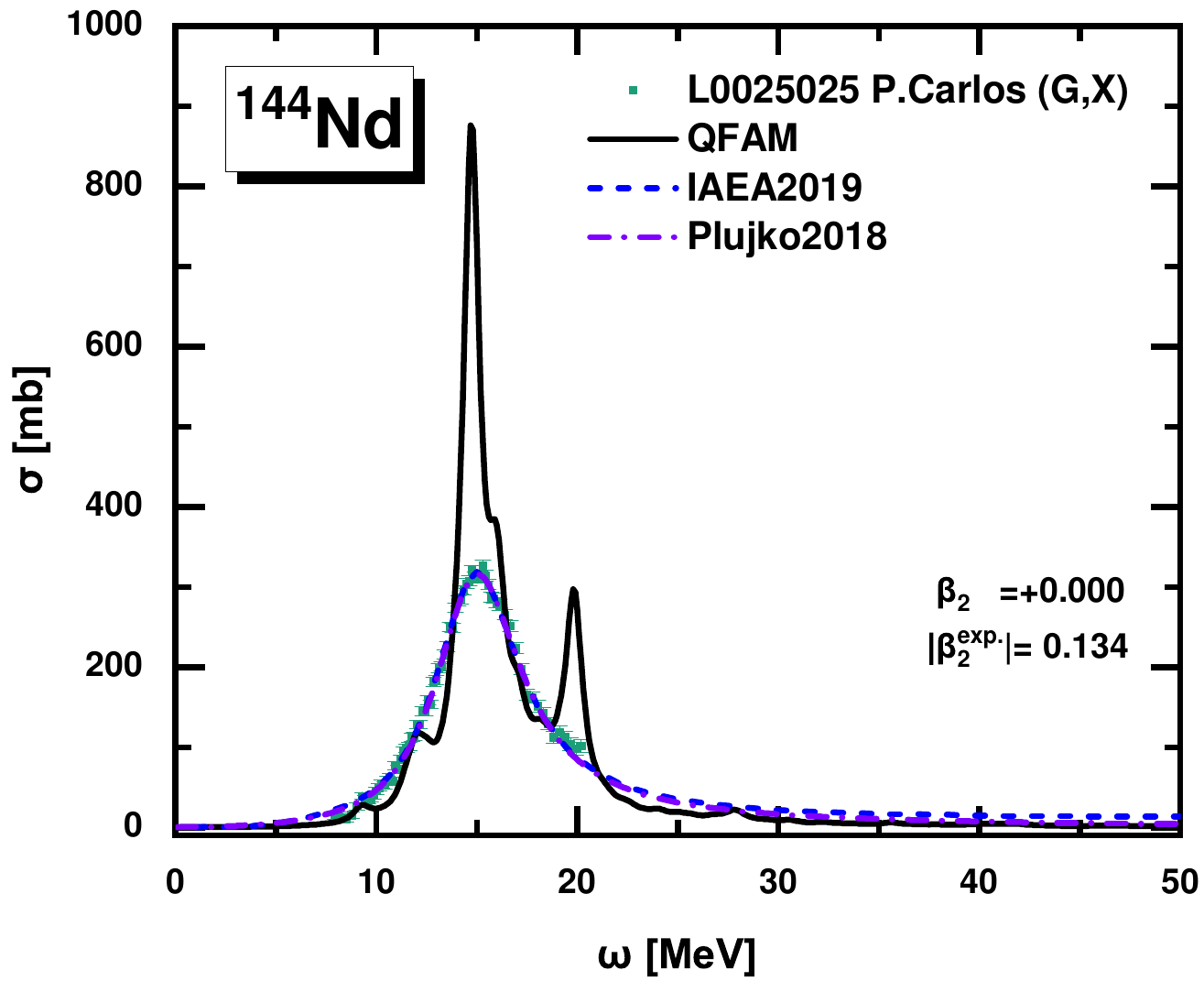}
    \includegraphics[width=0.35\textwidth]{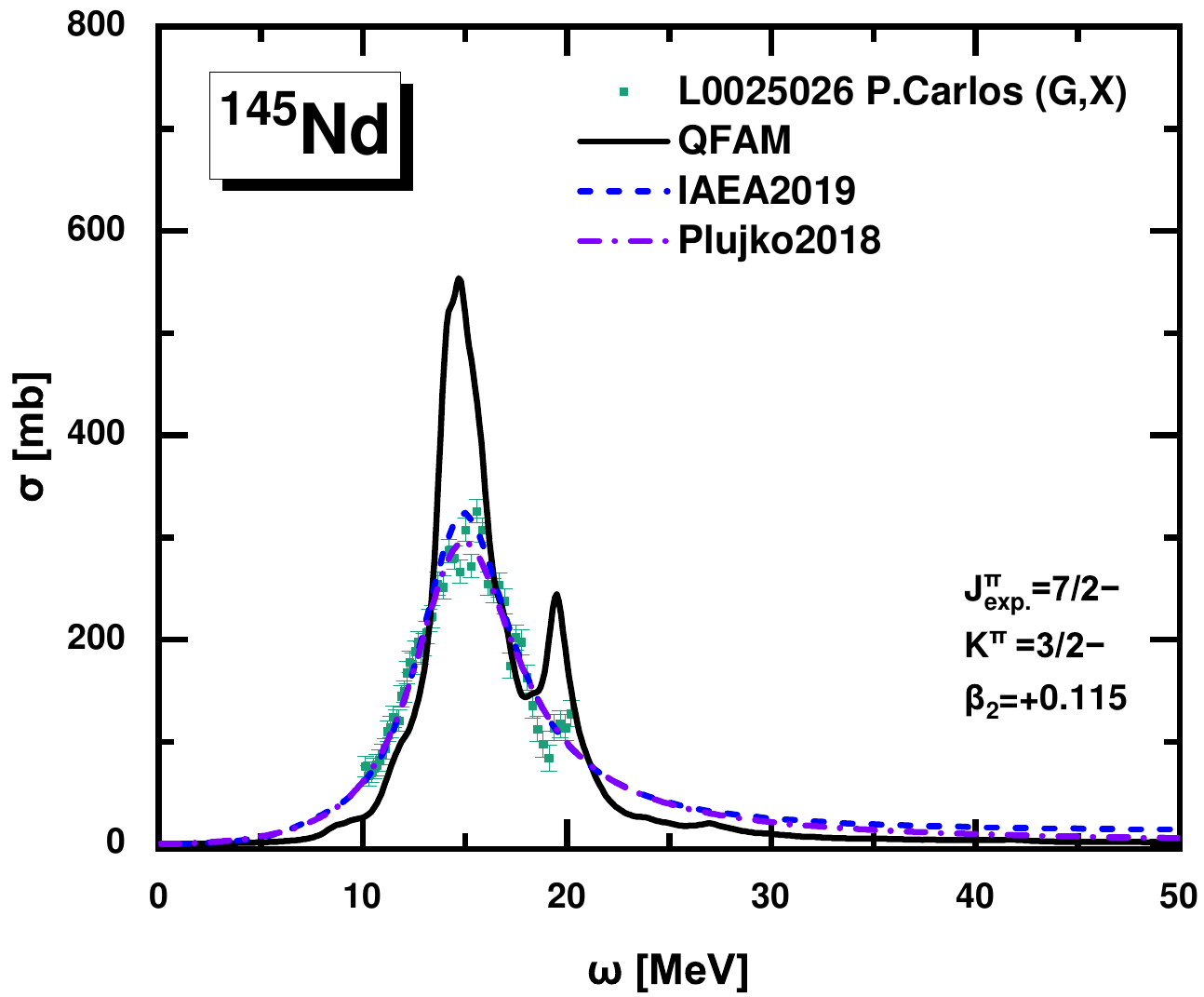}
    \includegraphics[width=0.35\textwidth]{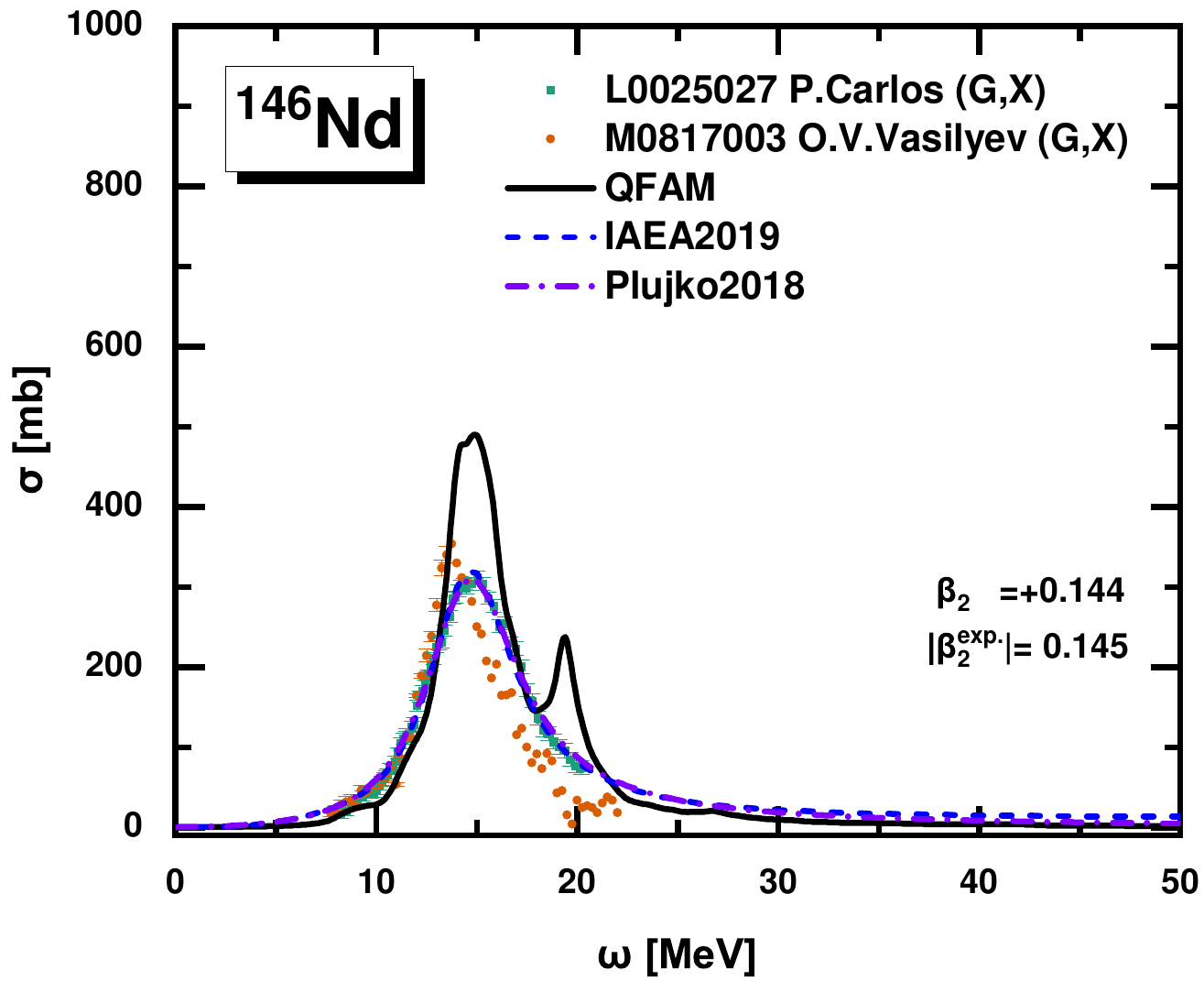}
    \includegraphics[width=0.35\textwidth]{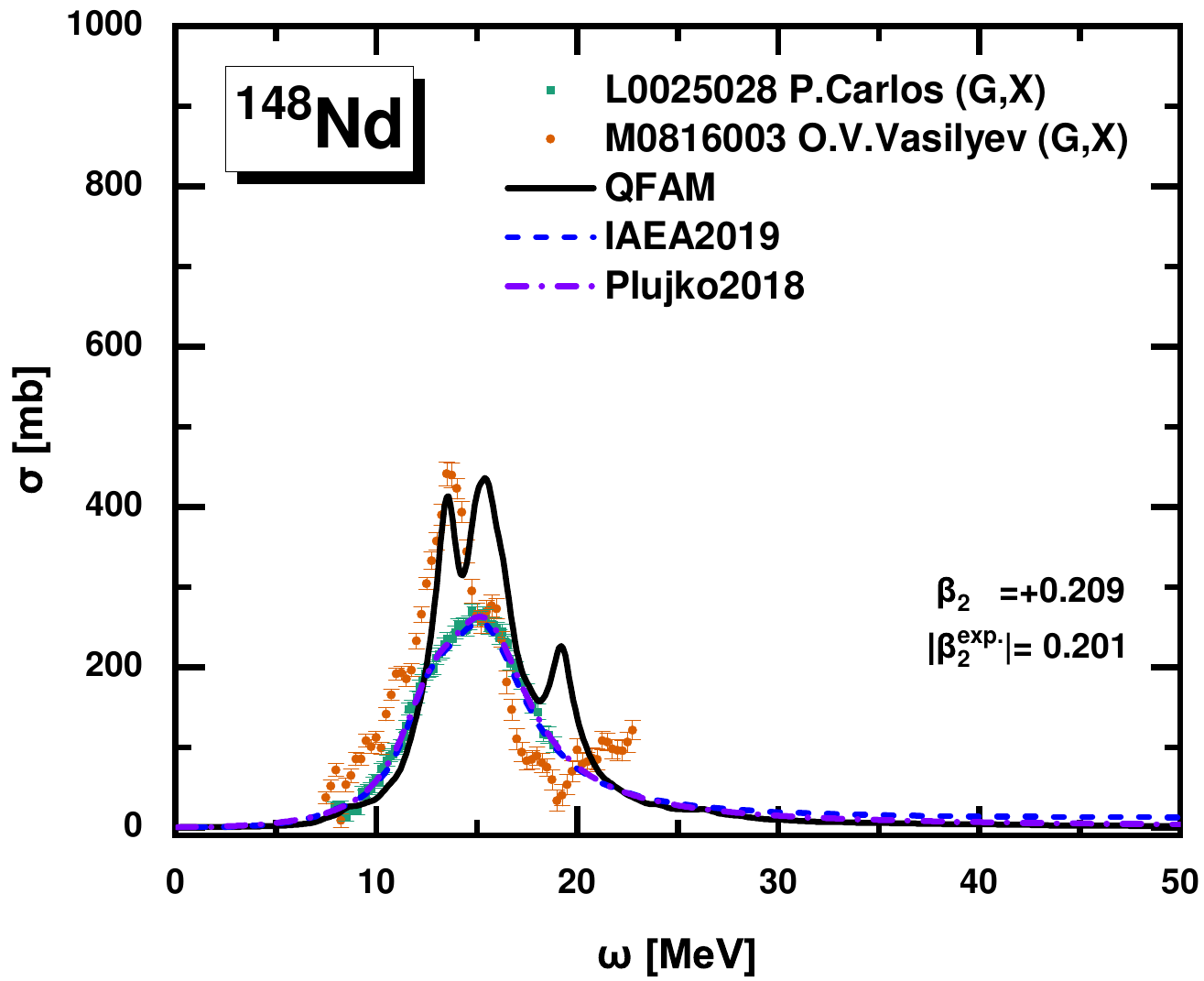}
    \includegraphics[width=0.35\textwidth]{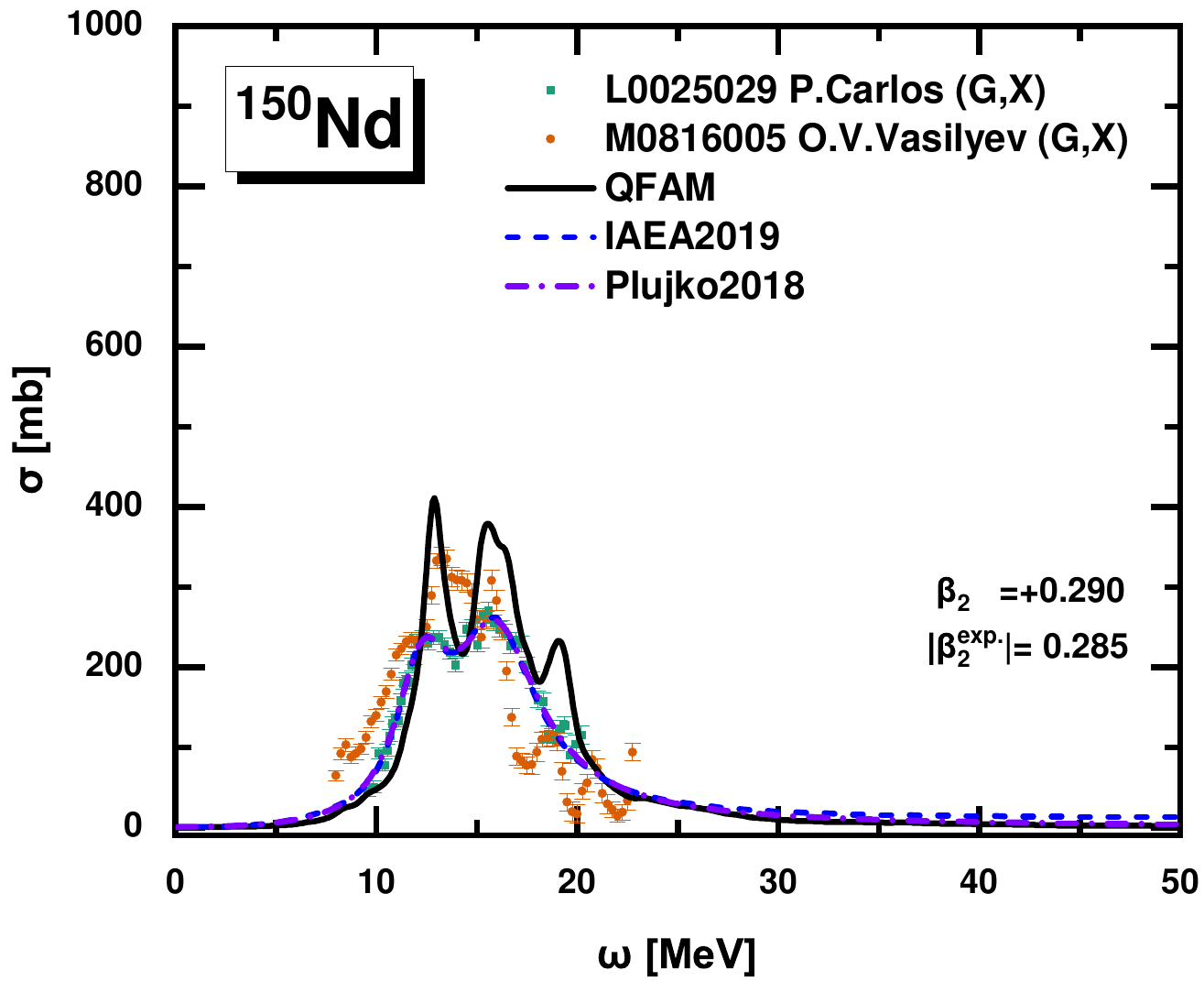}
\end{figure*}
\begin{figure*}\ContinuedFloat
    \centering
    \includegraphics[width=0.35\textwidth]{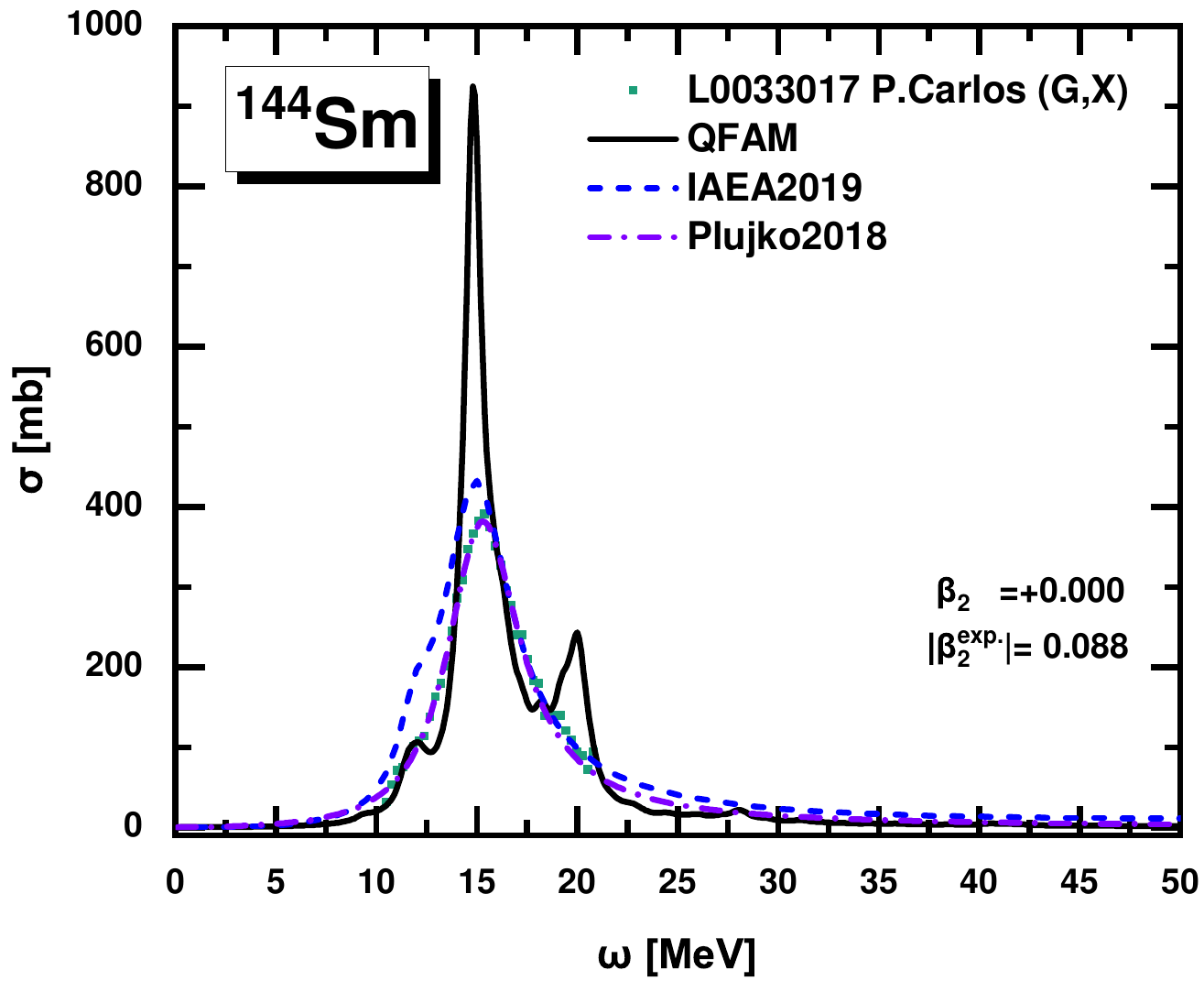}
    \includegraphics[width=0.35\textwidth]{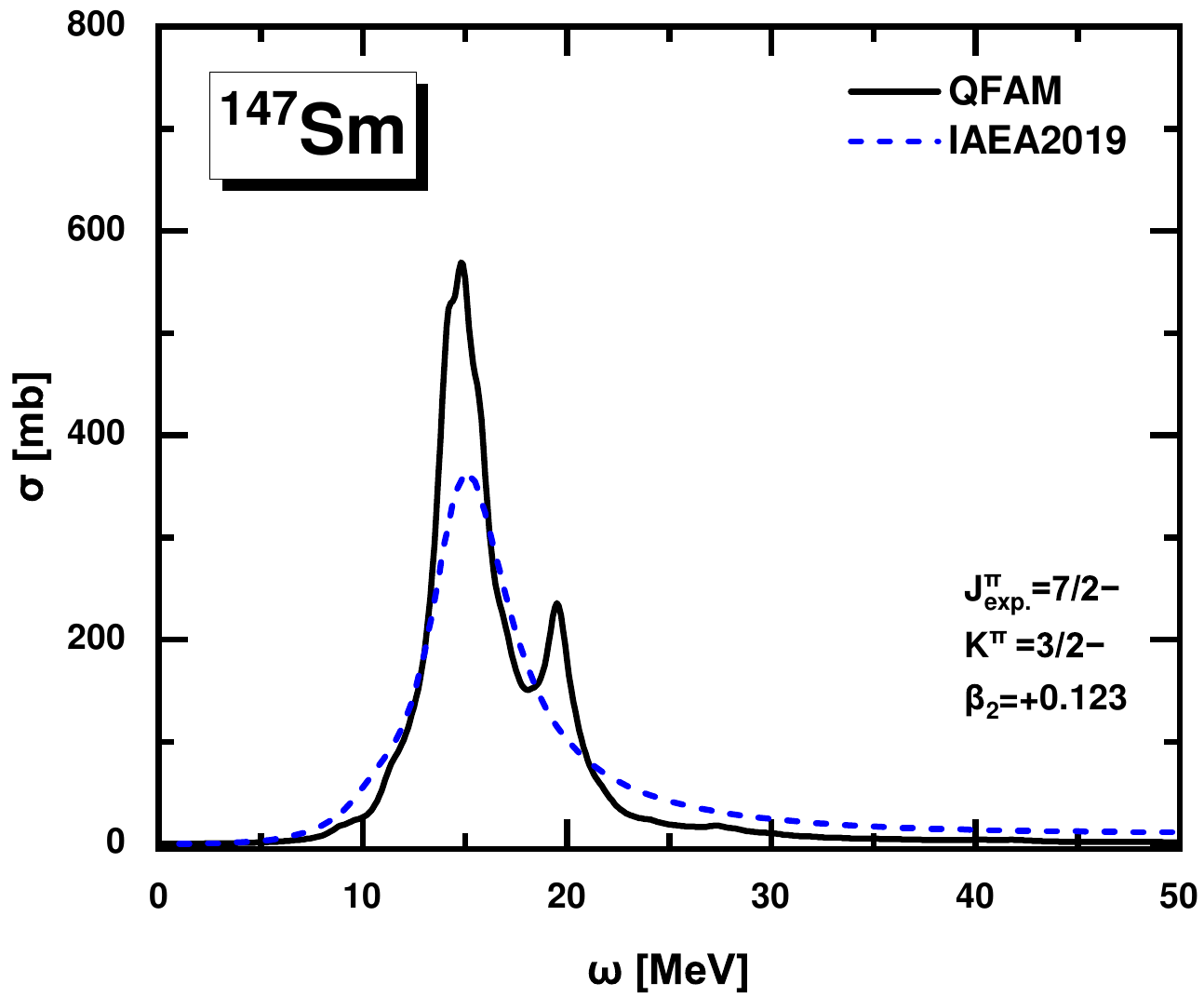}
    \includegraphics[width=0.35\textwidth]{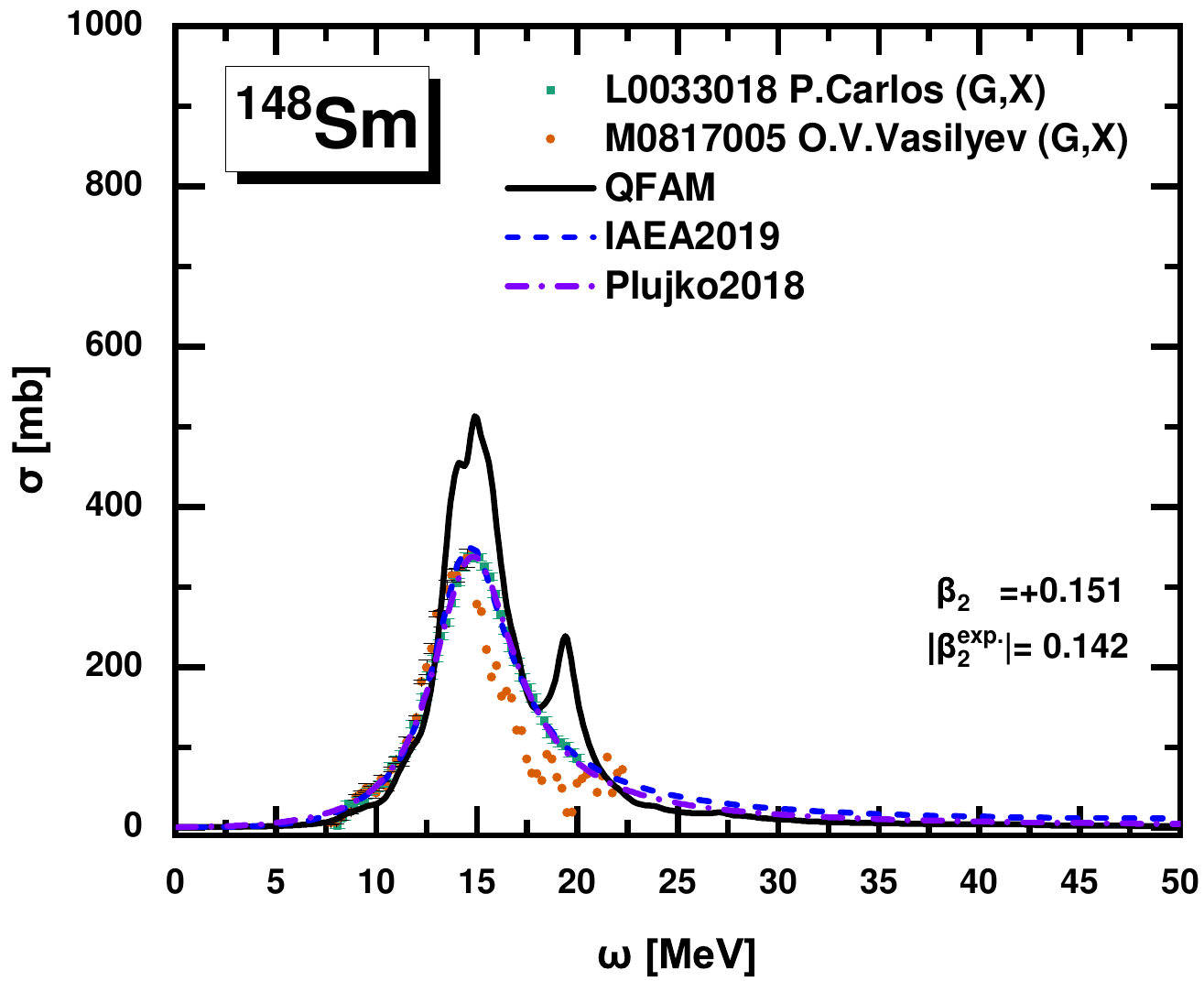}
    \includegraphics[width=0.35\textwidth]{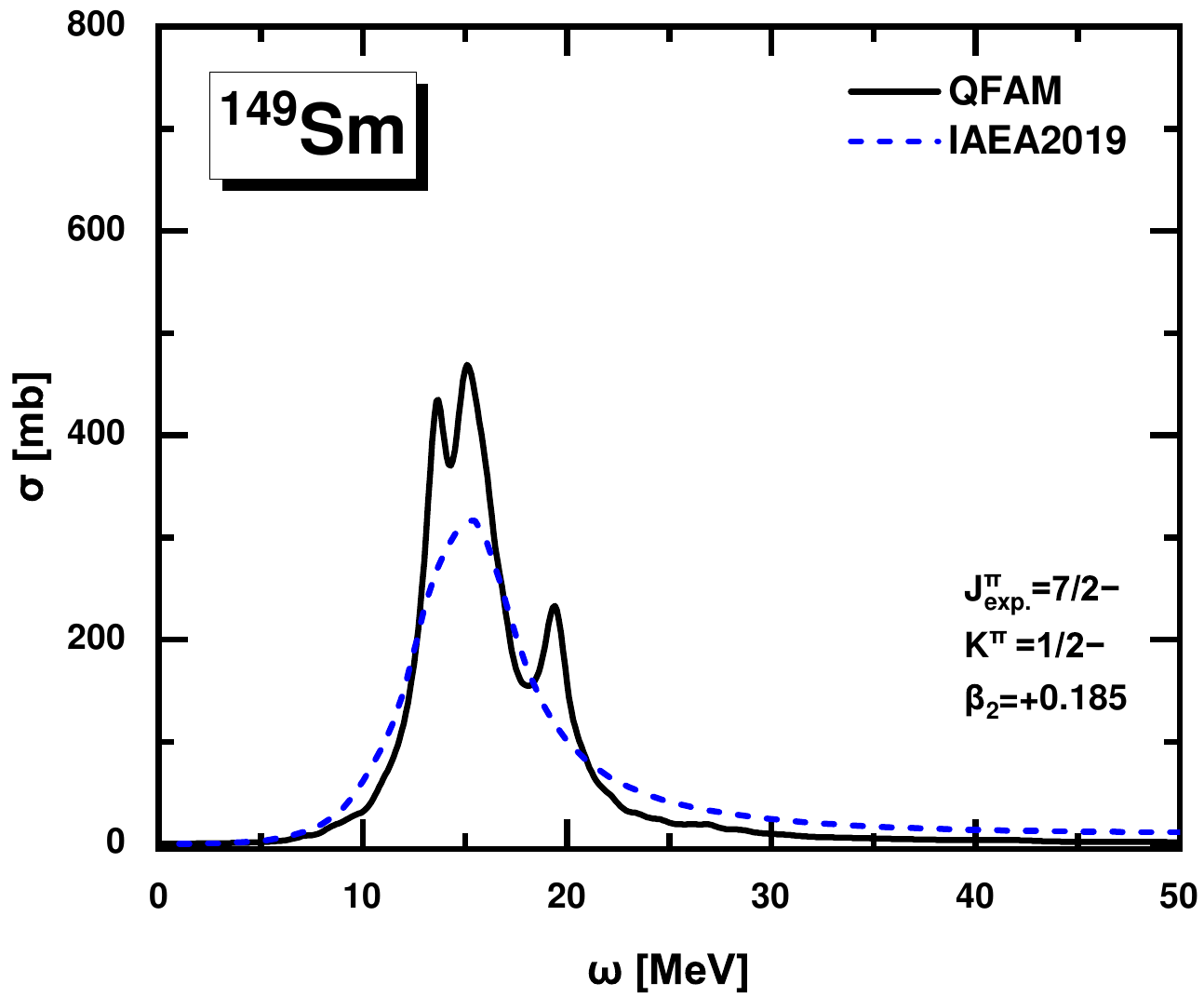}
    \includegraphics[width=0.35\textwidth]{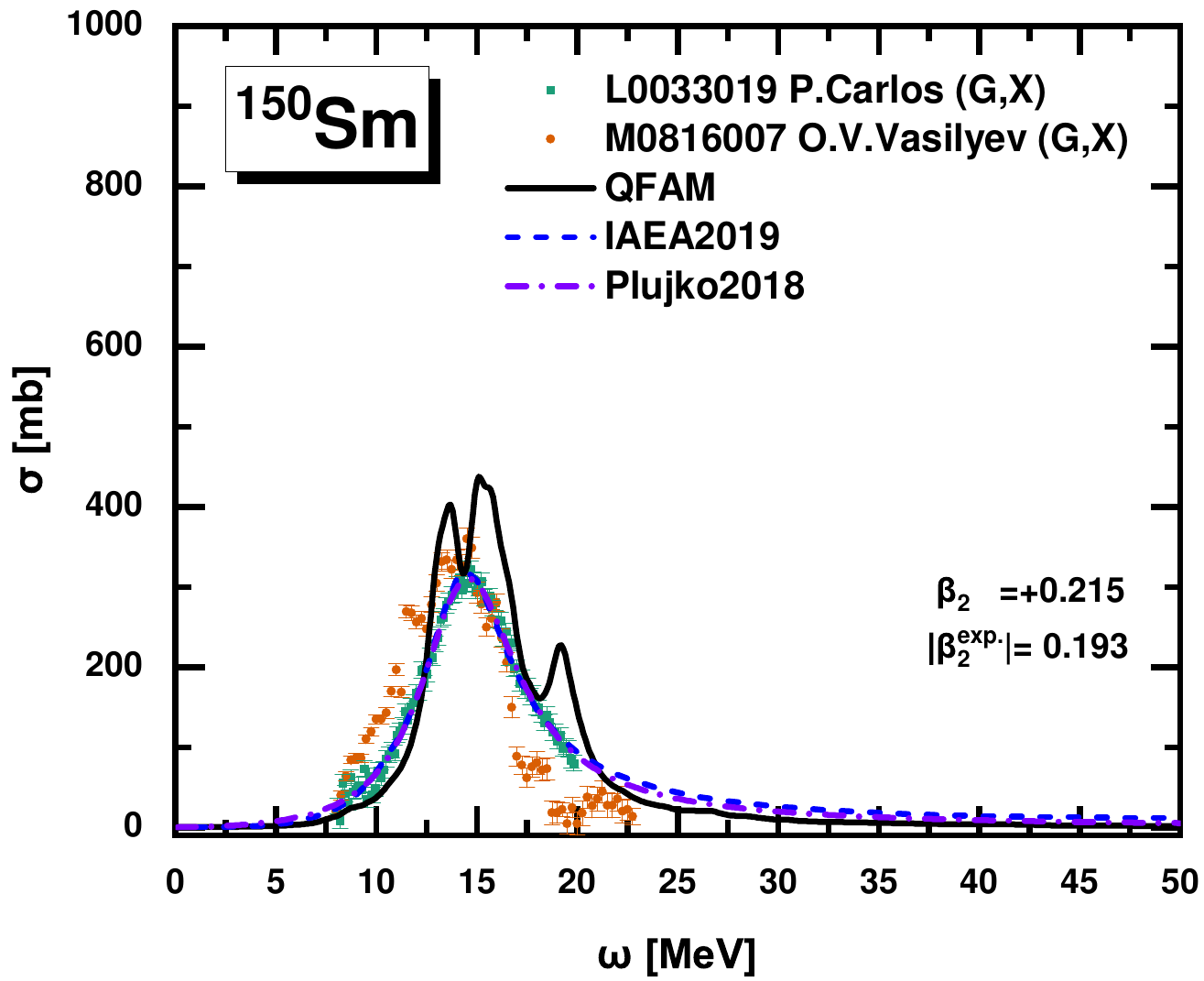}
    \includegraphics[width=0.35\textwidth]{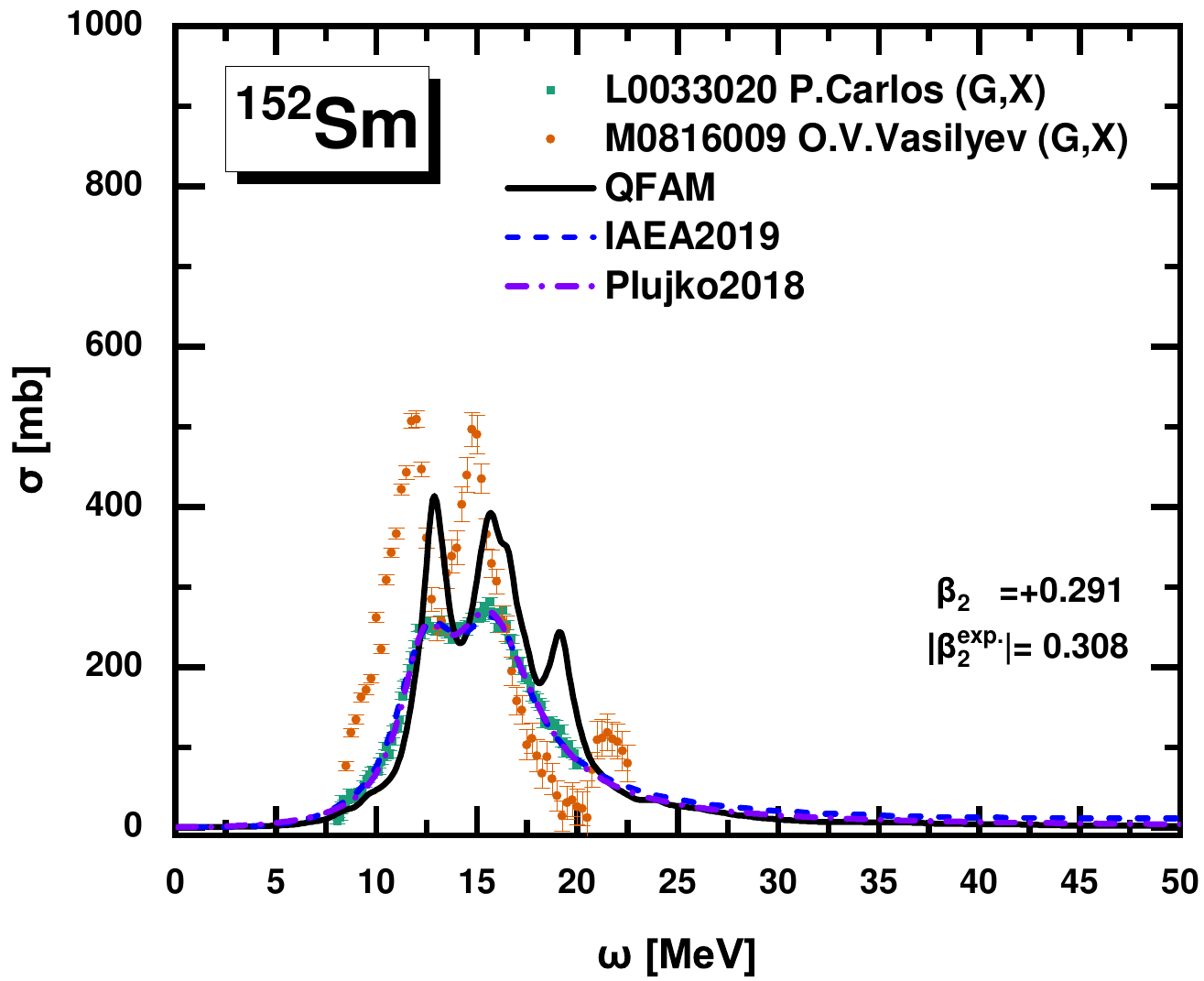}
    \includegraphics[width=0.35\textwidth]{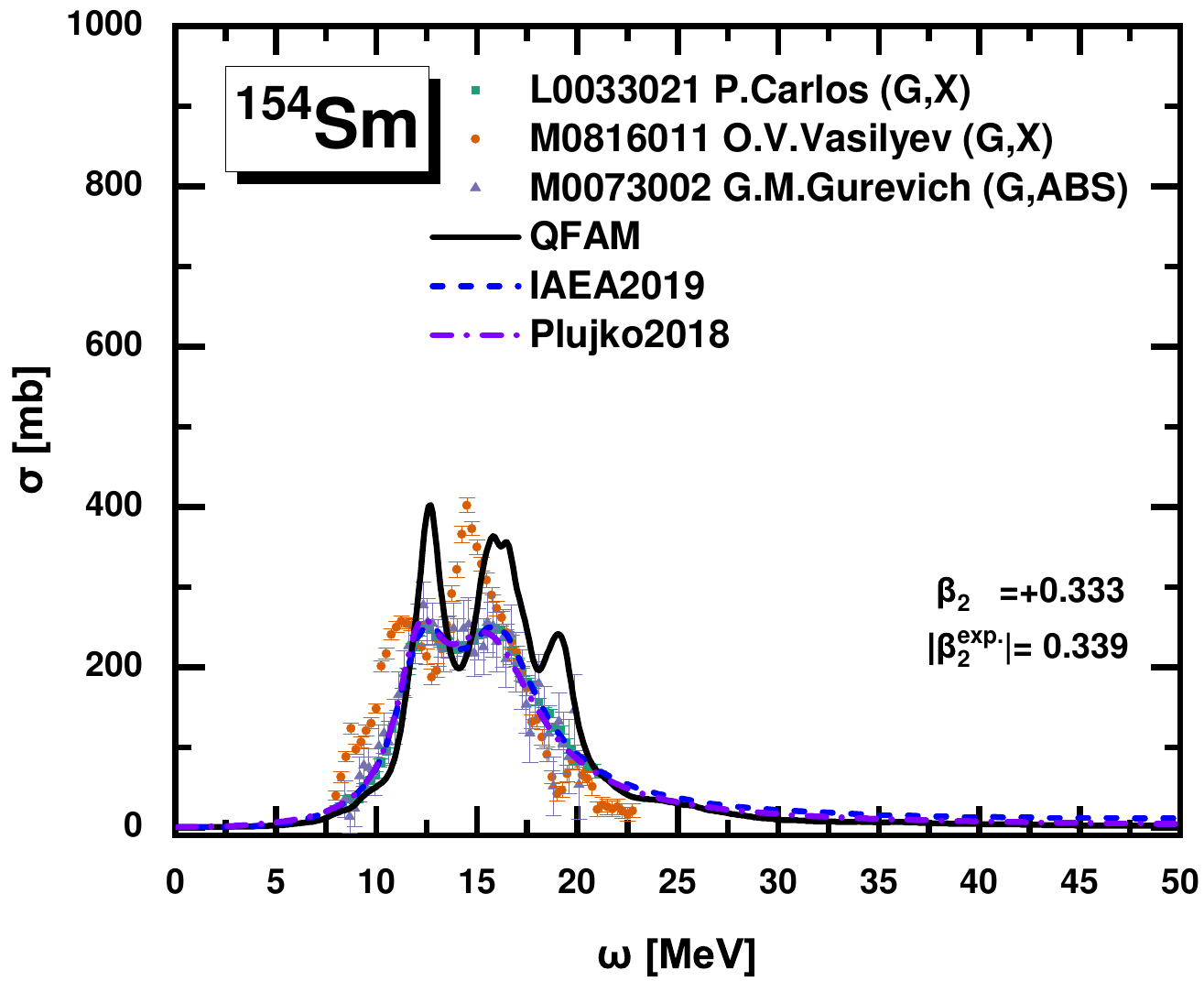}
    \includegraphics[width=0.35\textwidth]{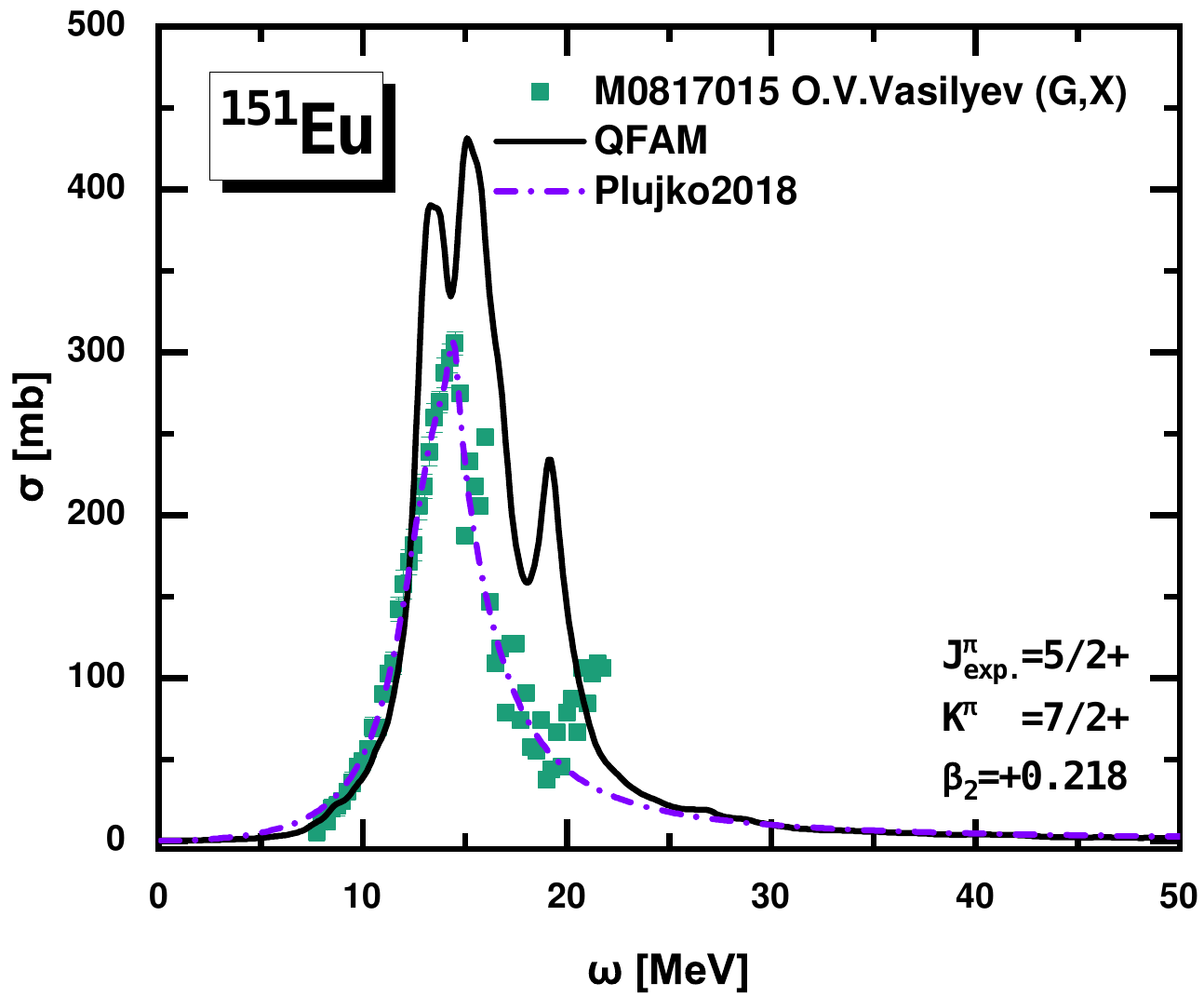}
\end{figure*}
\begin{figure*}\ContinuedFloat
    \centering
    \includegraphics[width=0.35\textwidth]{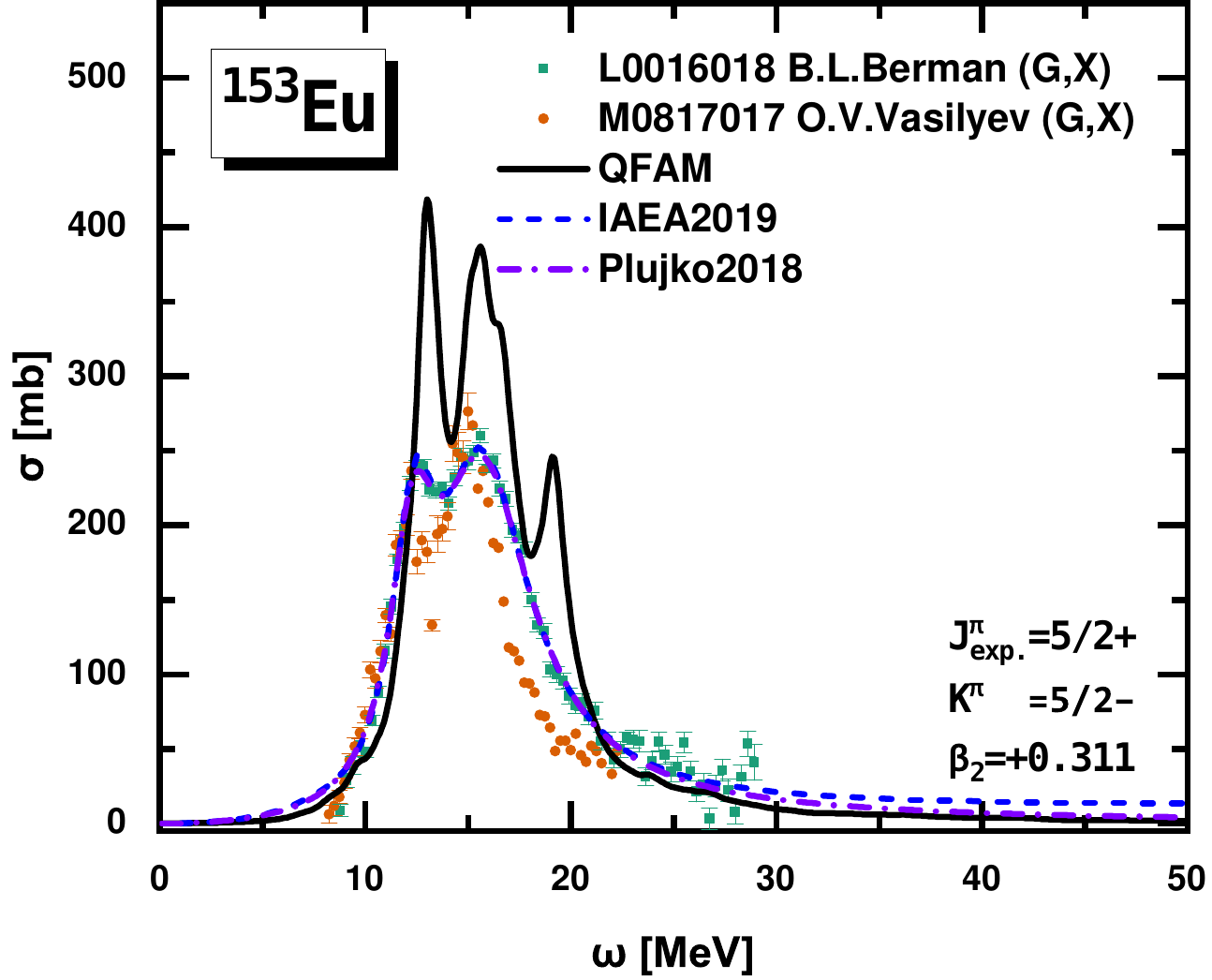}
    \includegraphics[width=0.35\textwidth]{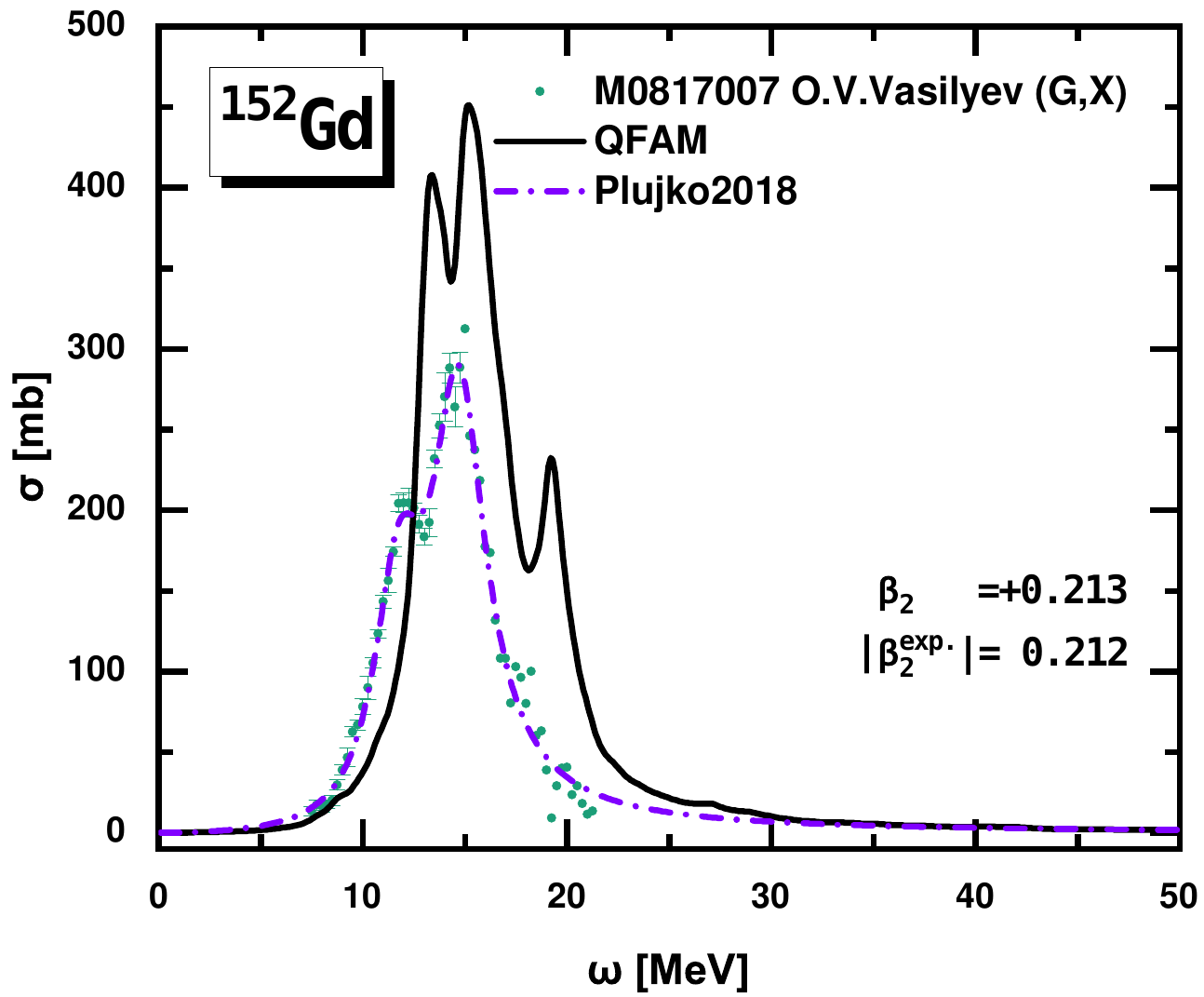}
    \includegraphics[width=0.35\textwidth]{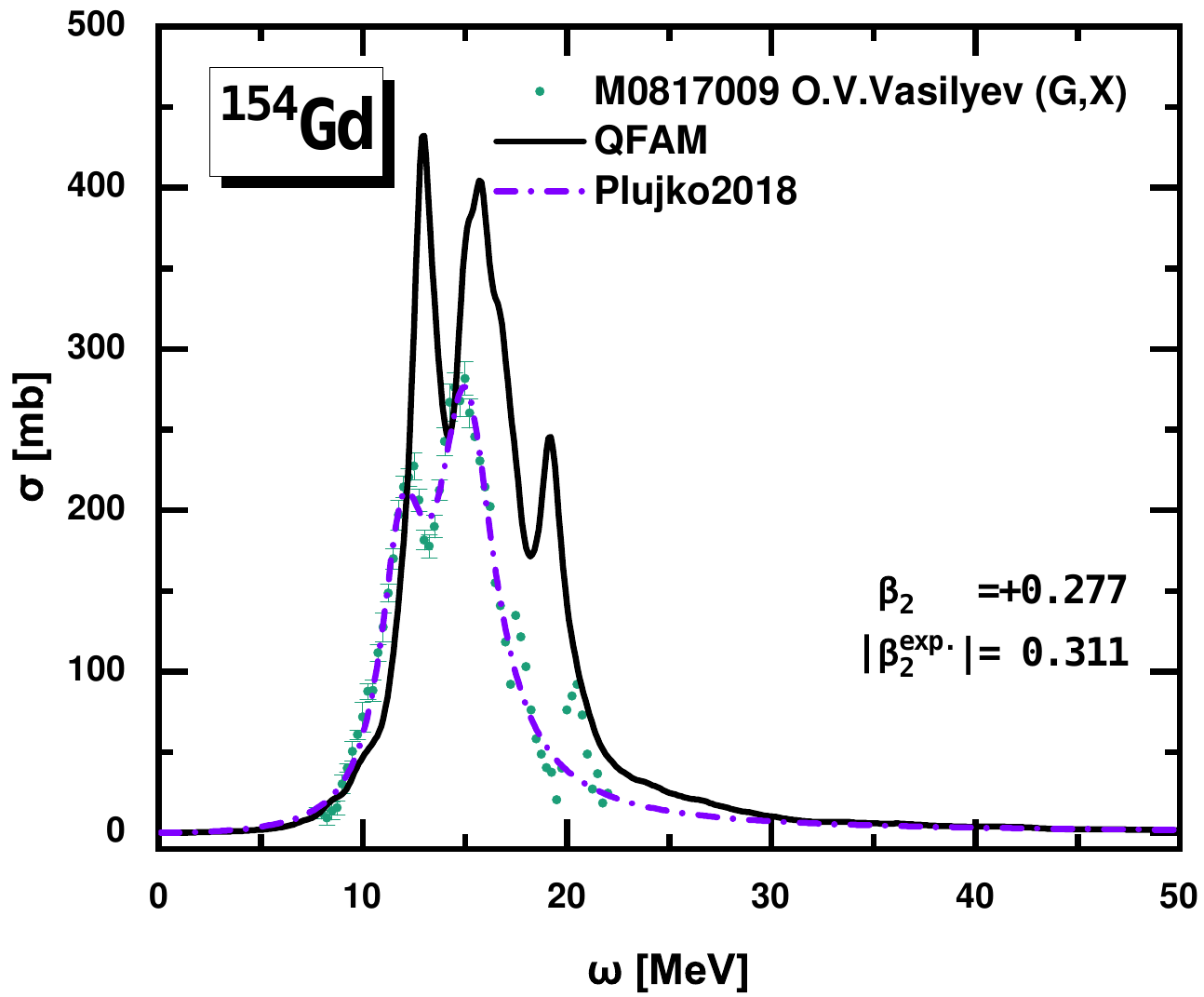}
    \includegraphics[width=0.35\textwidth]{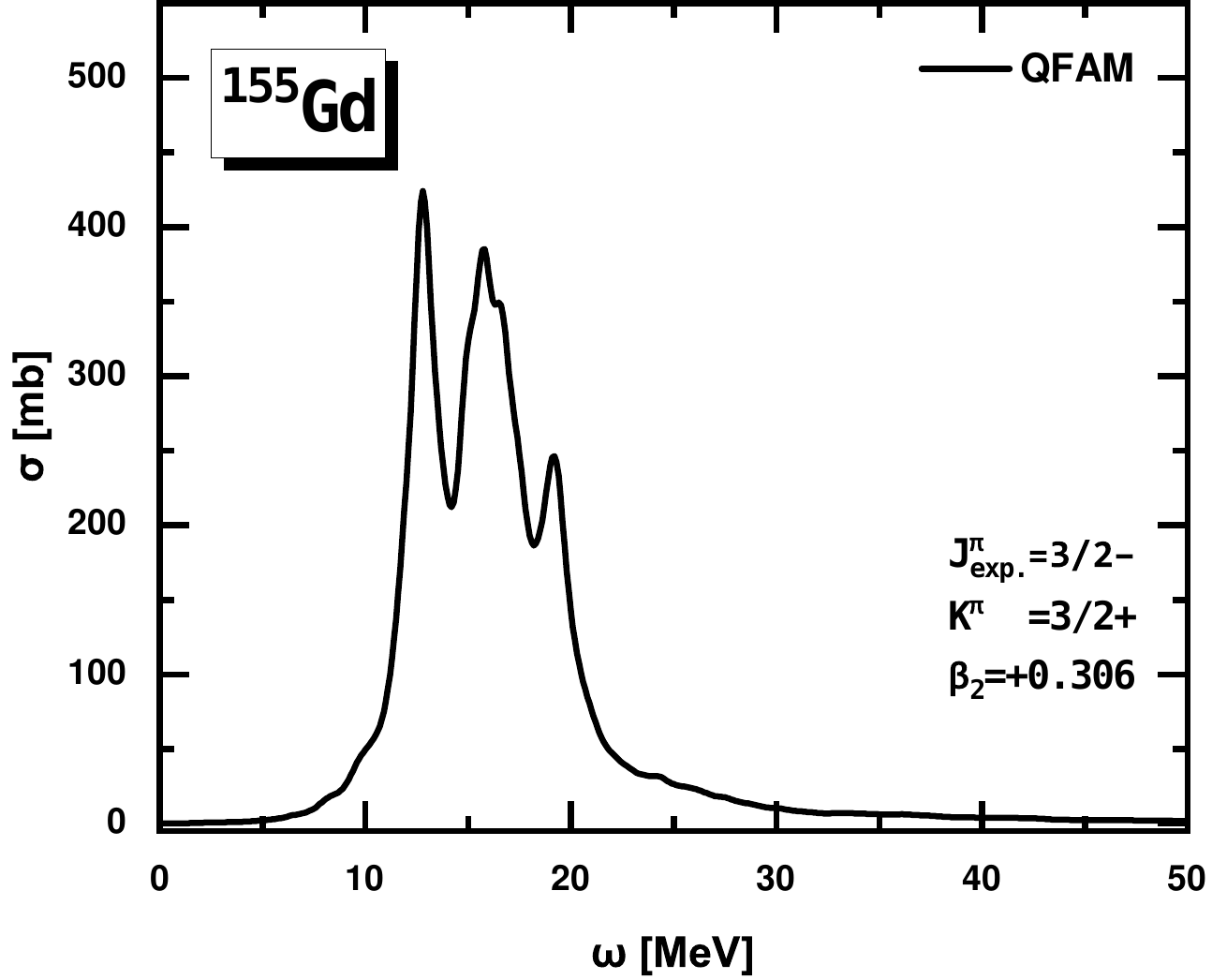}
    \includegraphics[width=0.35\textwidth]{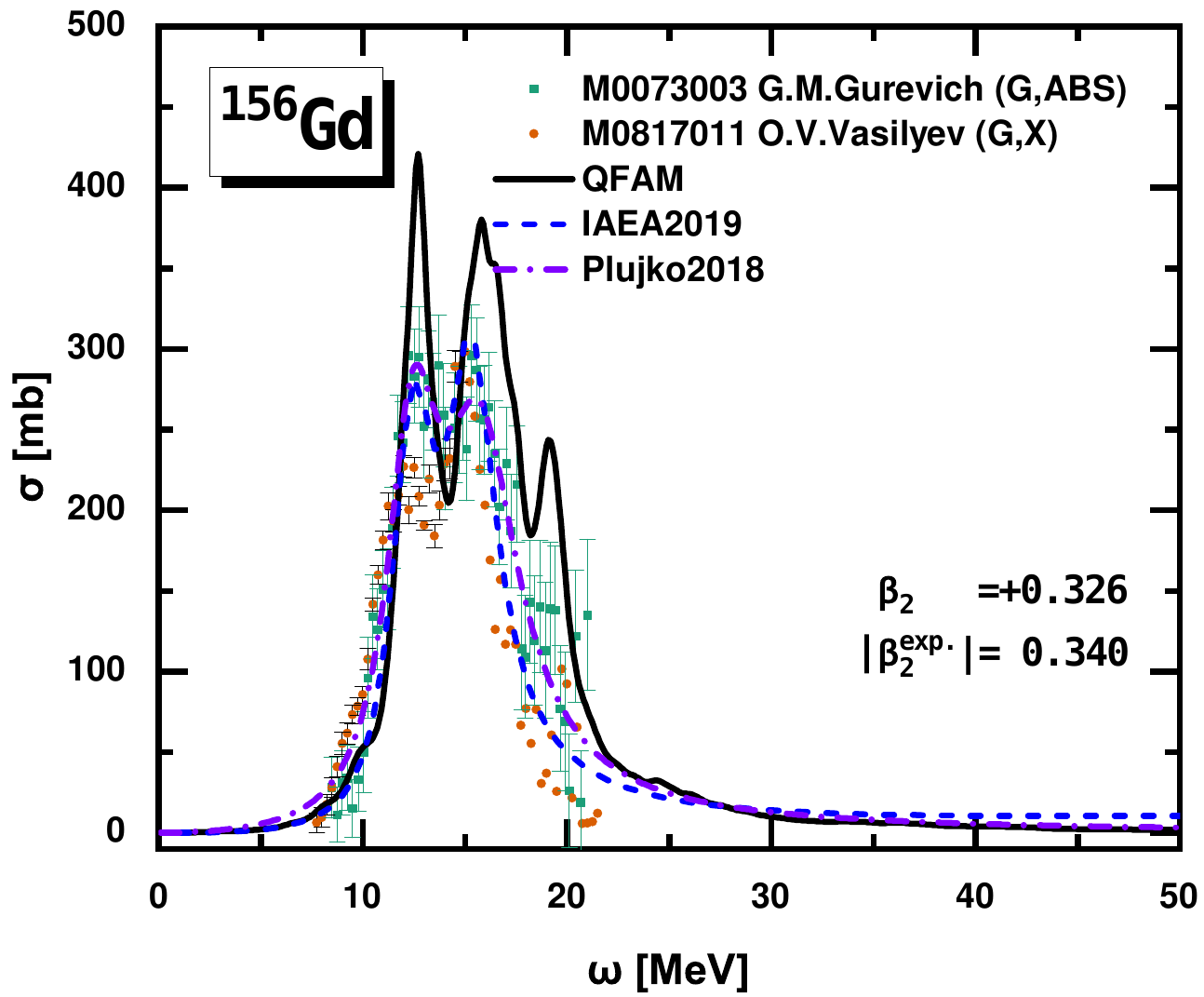}
    \includegraphics[width=0.35\textwidth]{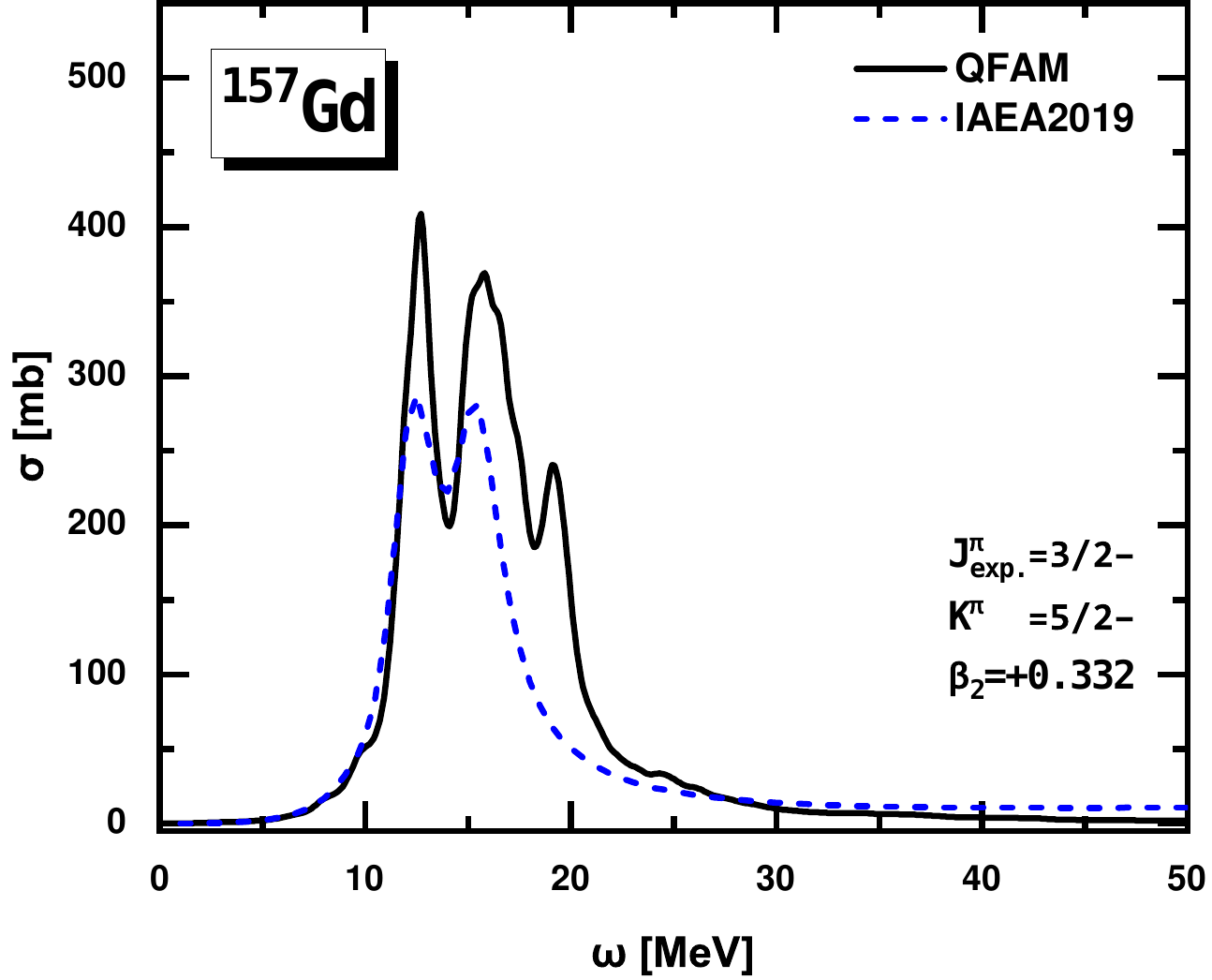}
    \includegraphics[width=0.35\textwidth]{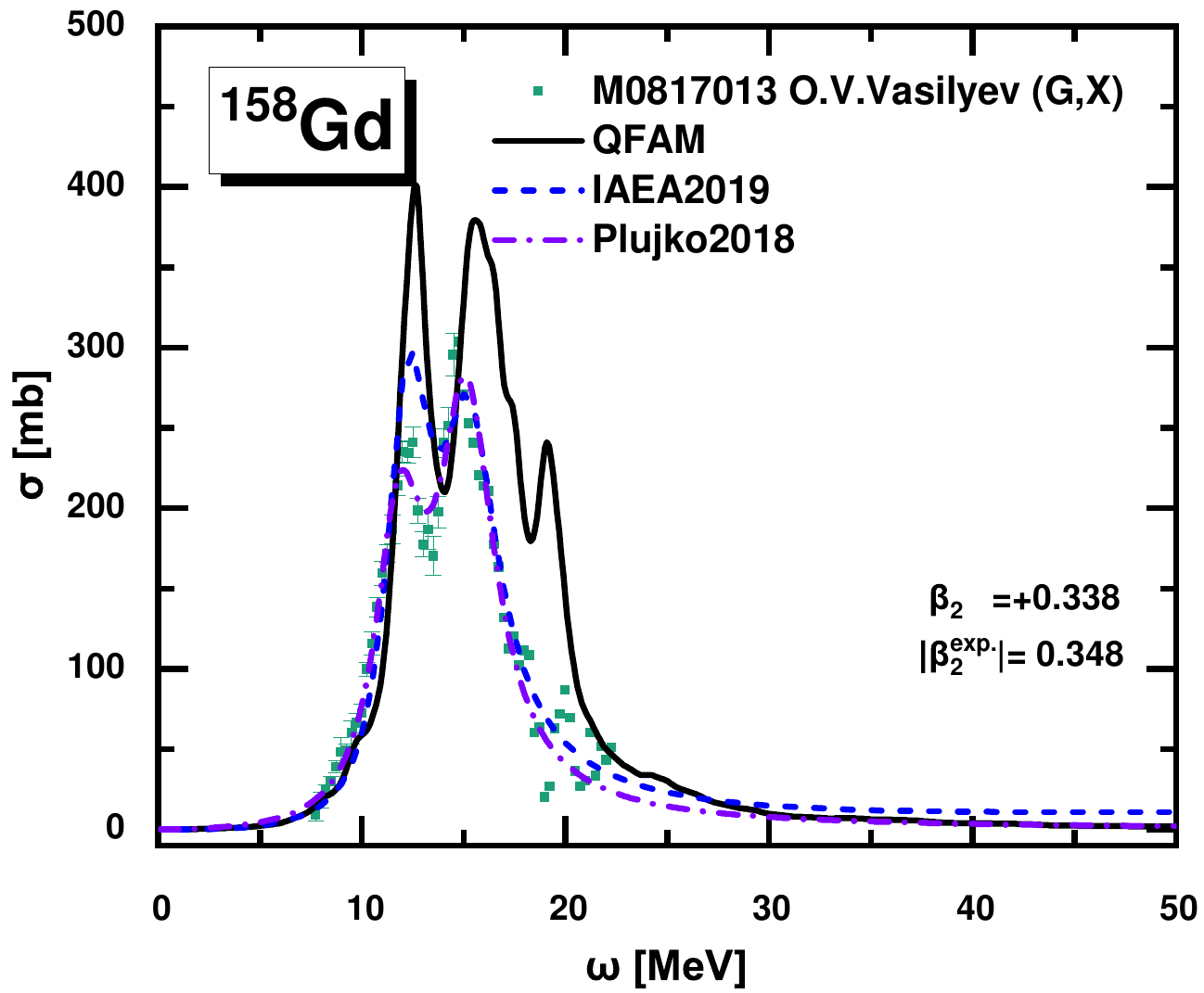}
    \includegraphics[width=0.35\textwidth]{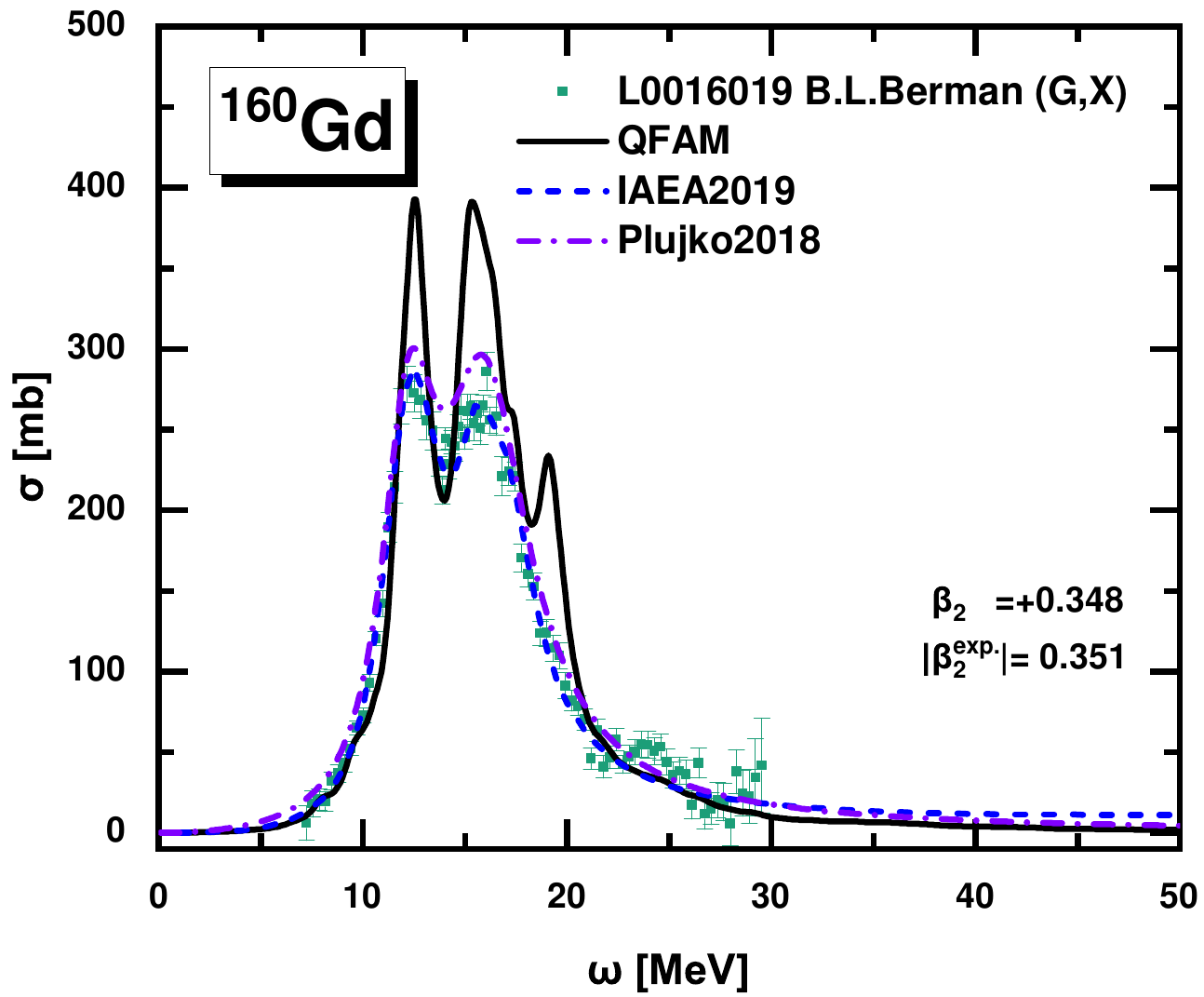}
\end{figure*}
\begin{figure*}\ContinuedFloat
    \centering
    \includegraphics[width=0.35\textwidth]{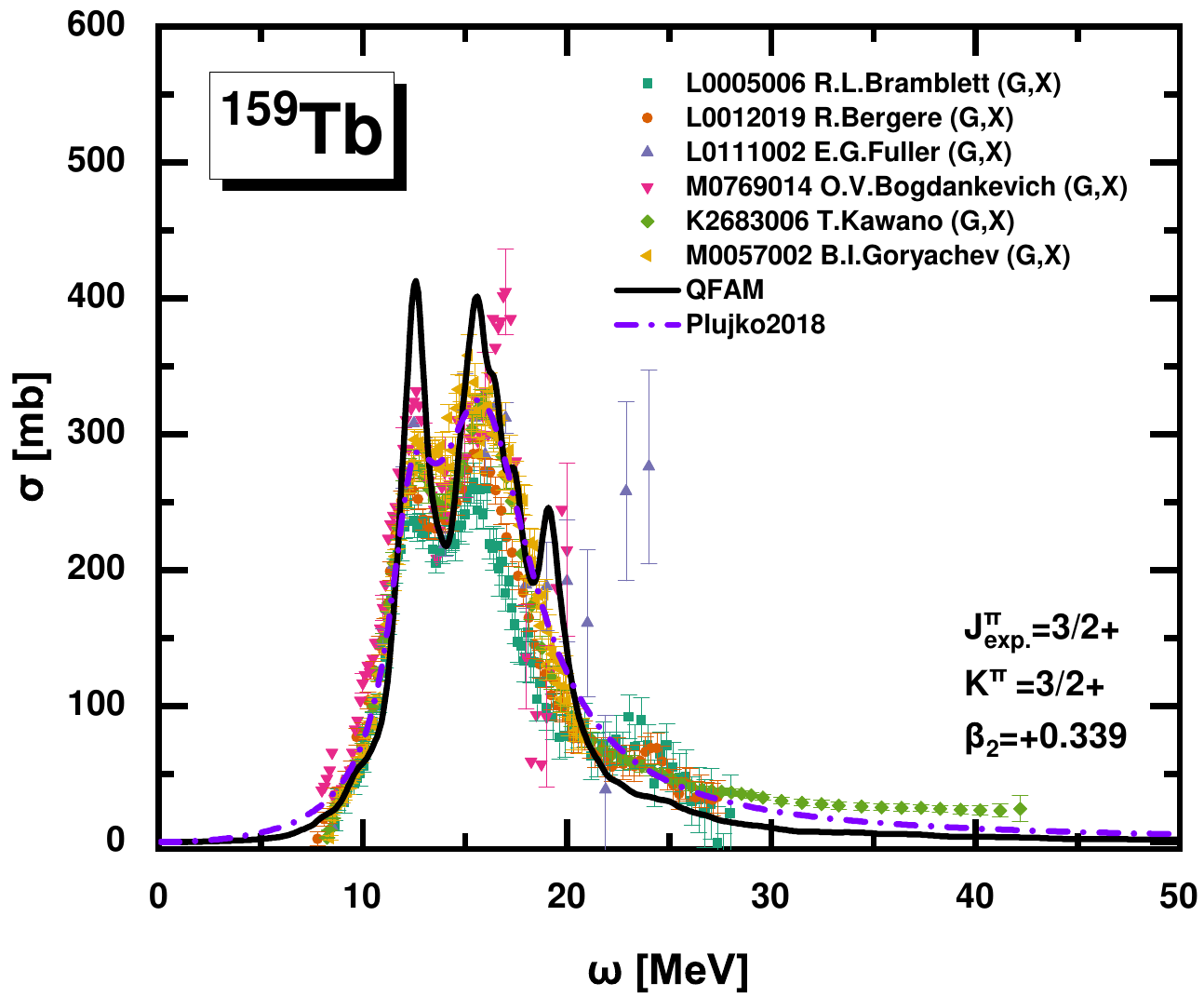}
    \includegraphics[width=0.35\textwidth]{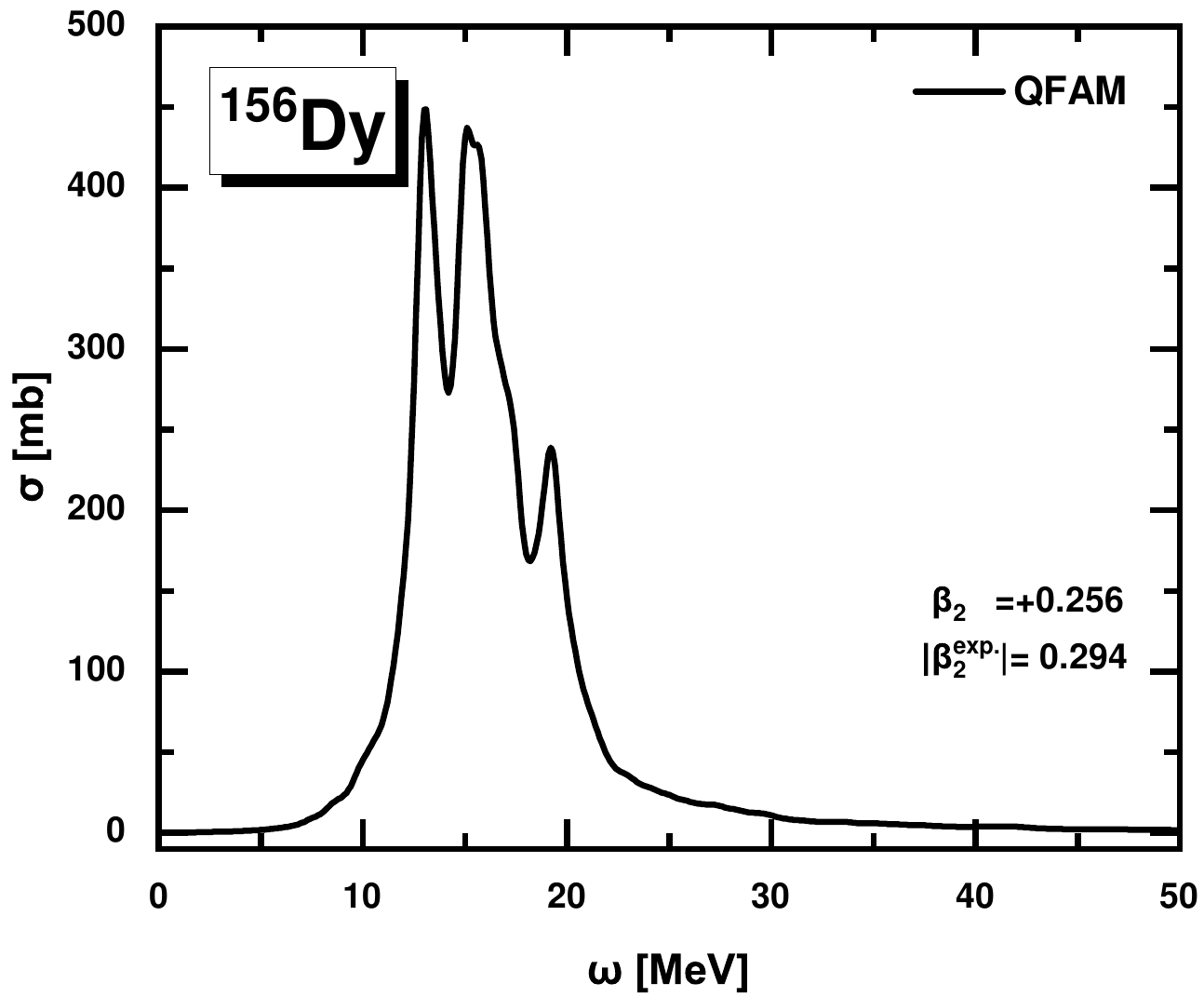}
    \includegraphics[width=0.35\textwidth]{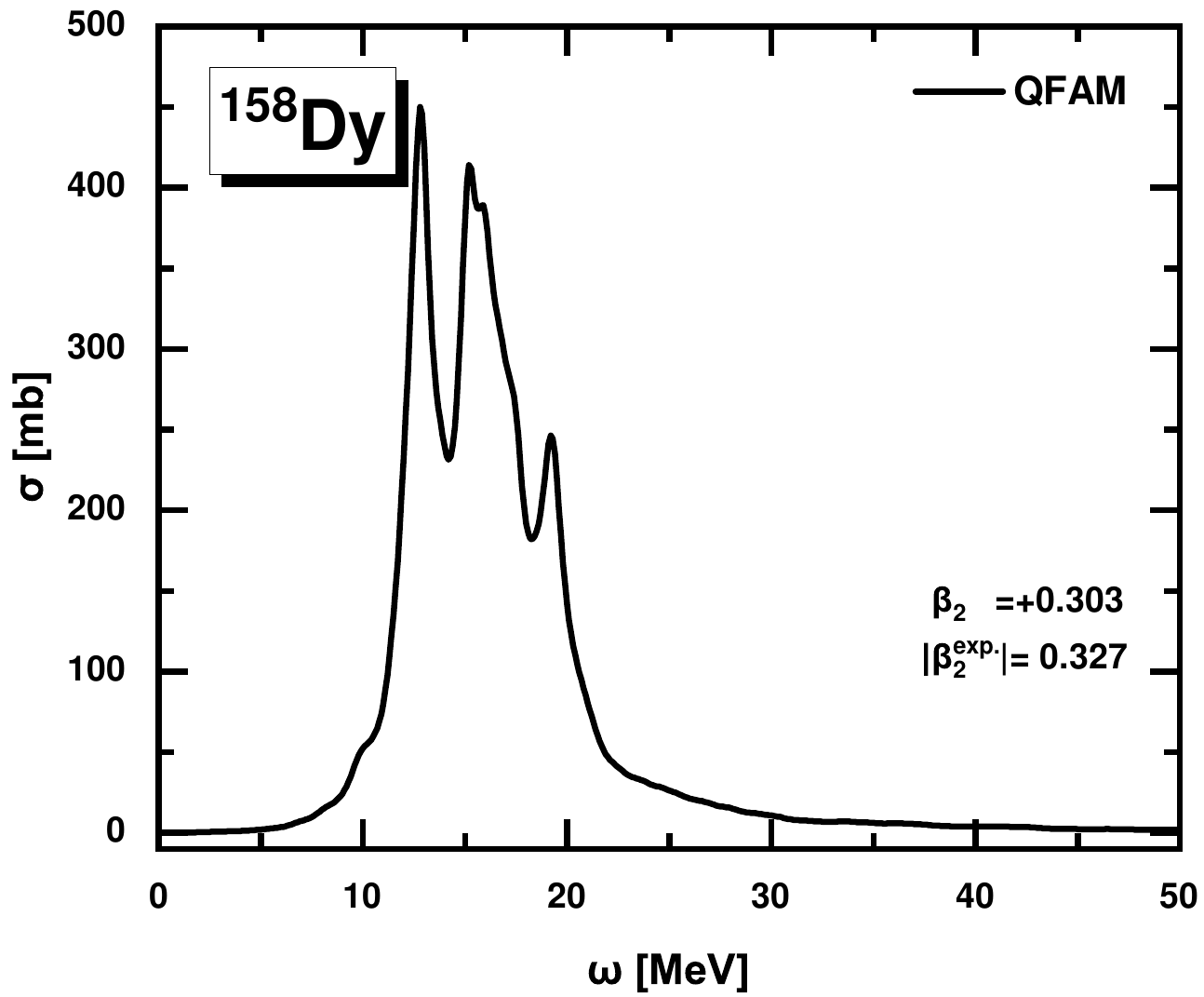}
    \includegraphics[width=0.35\textwidth]{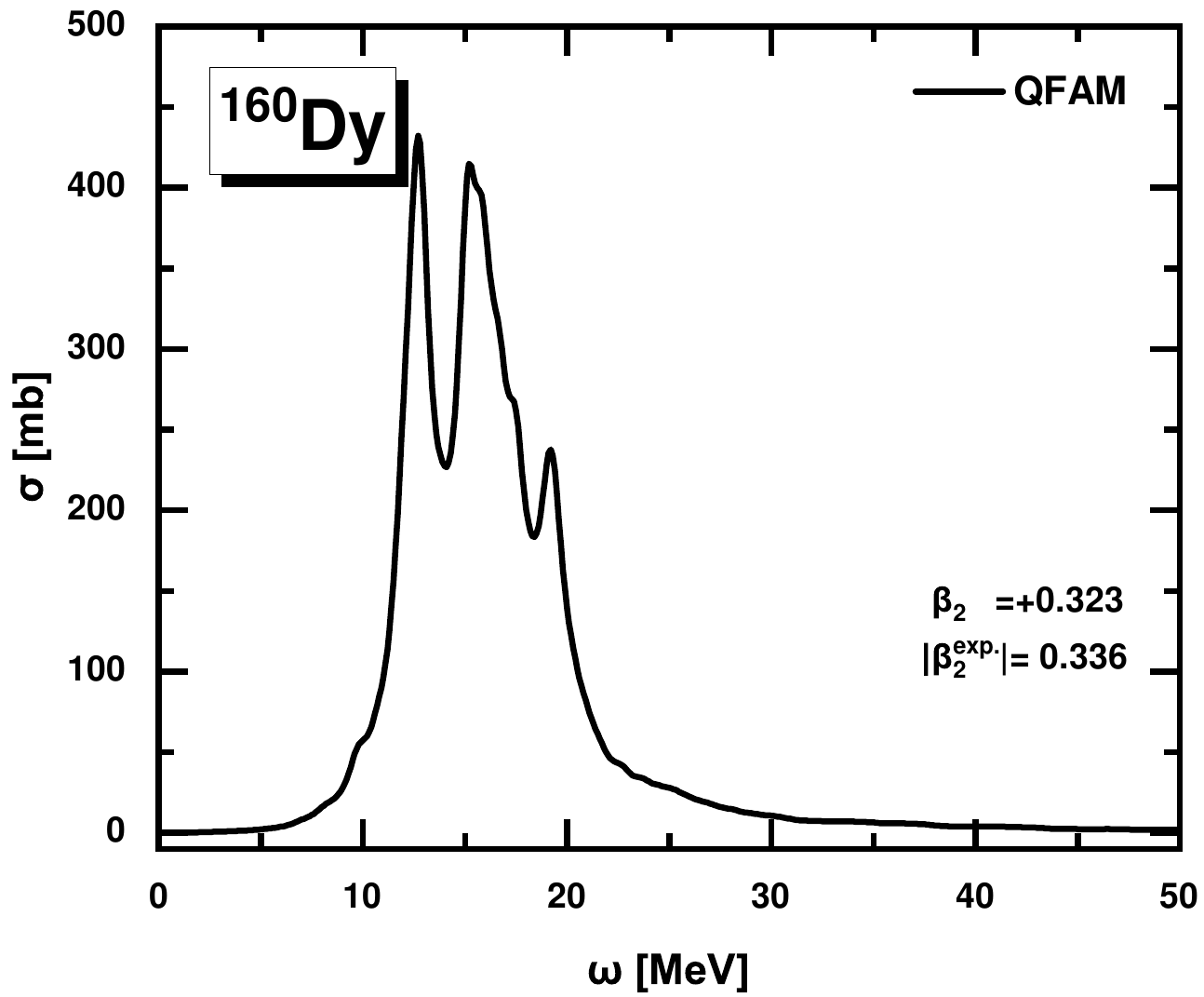}
    \includegraphics[width=0.35\textwidth]{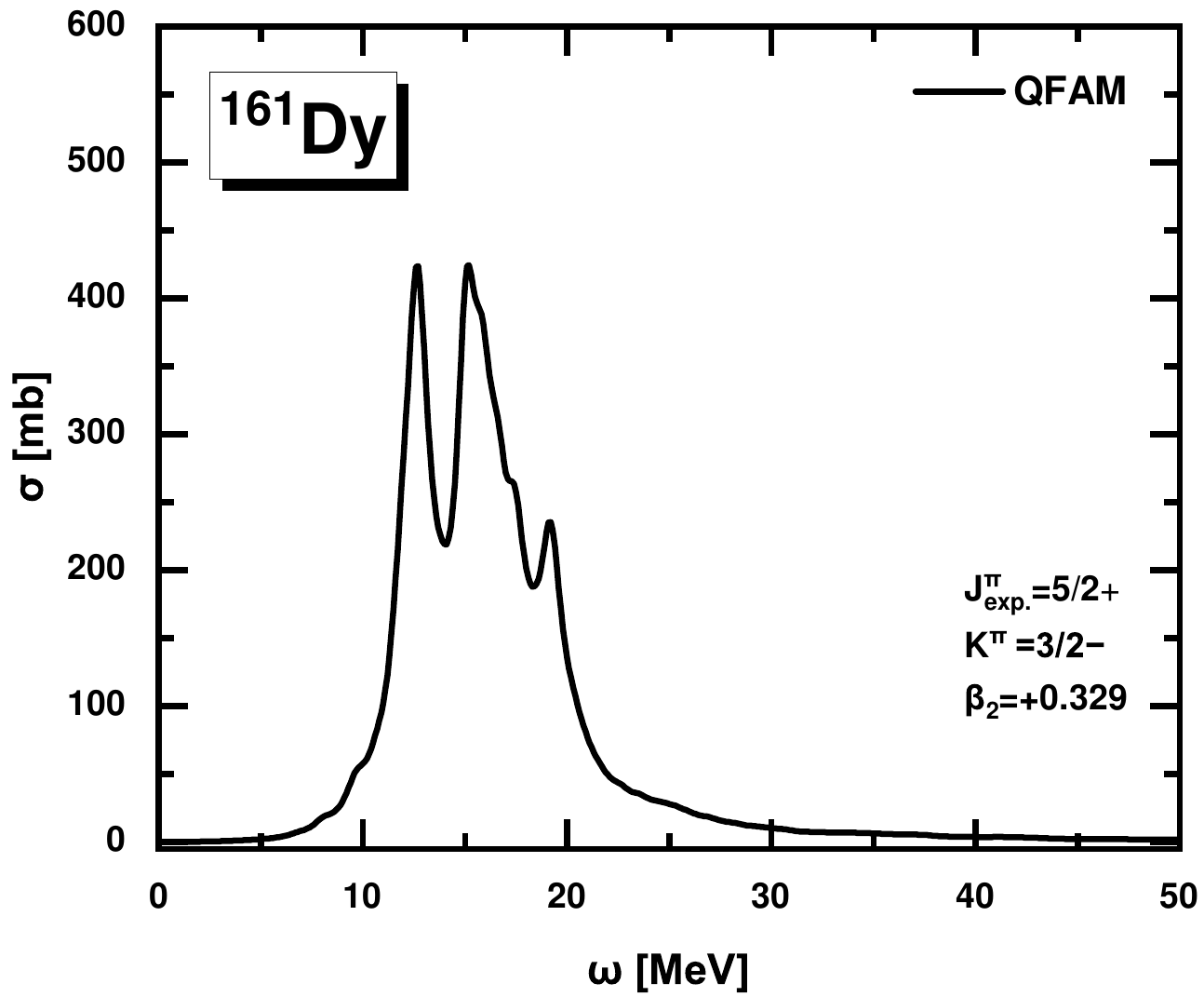}
    \includegraphics[width=0.35\textwidth]{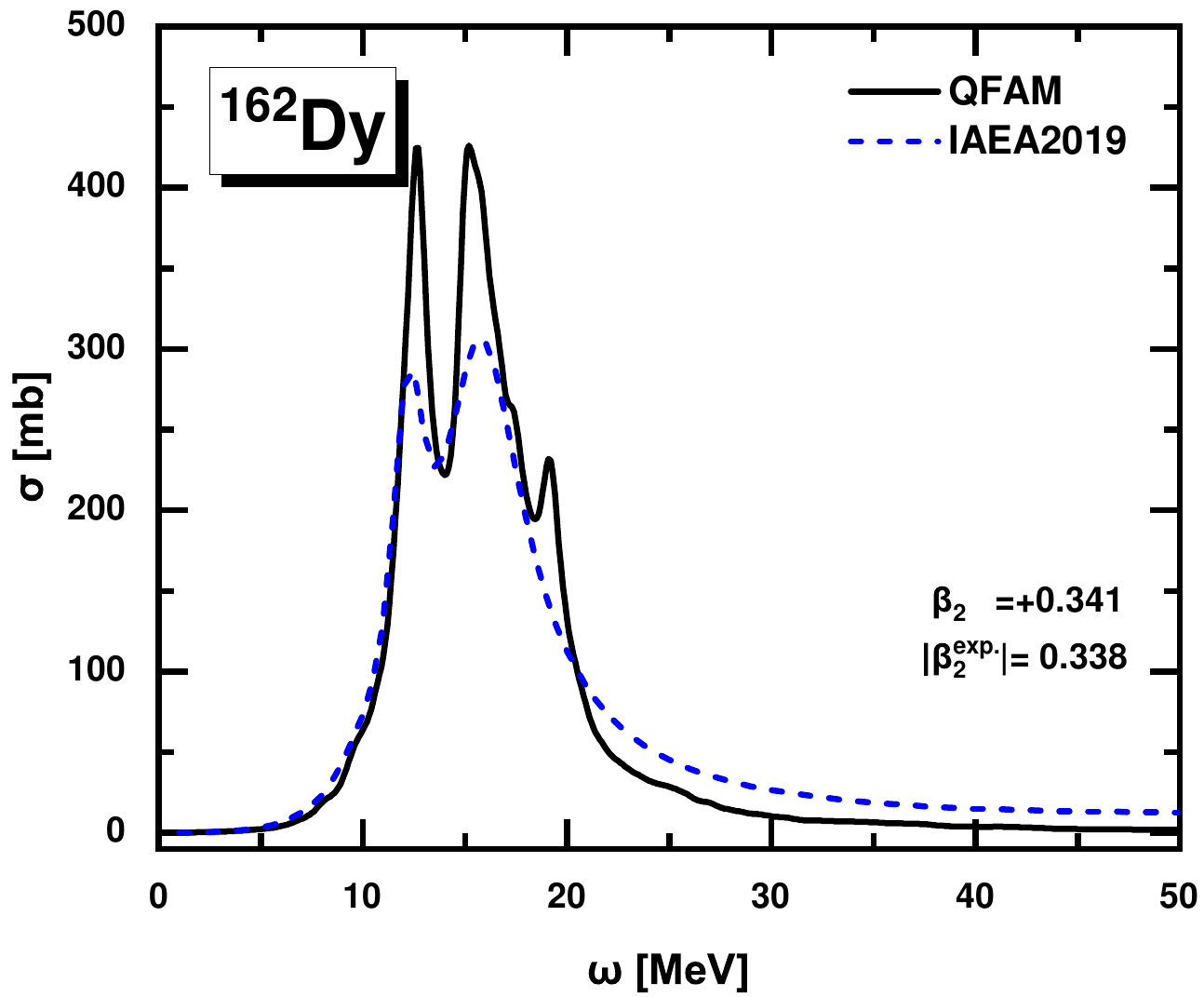}
    \includegraphics[width=0.35\textwidth]{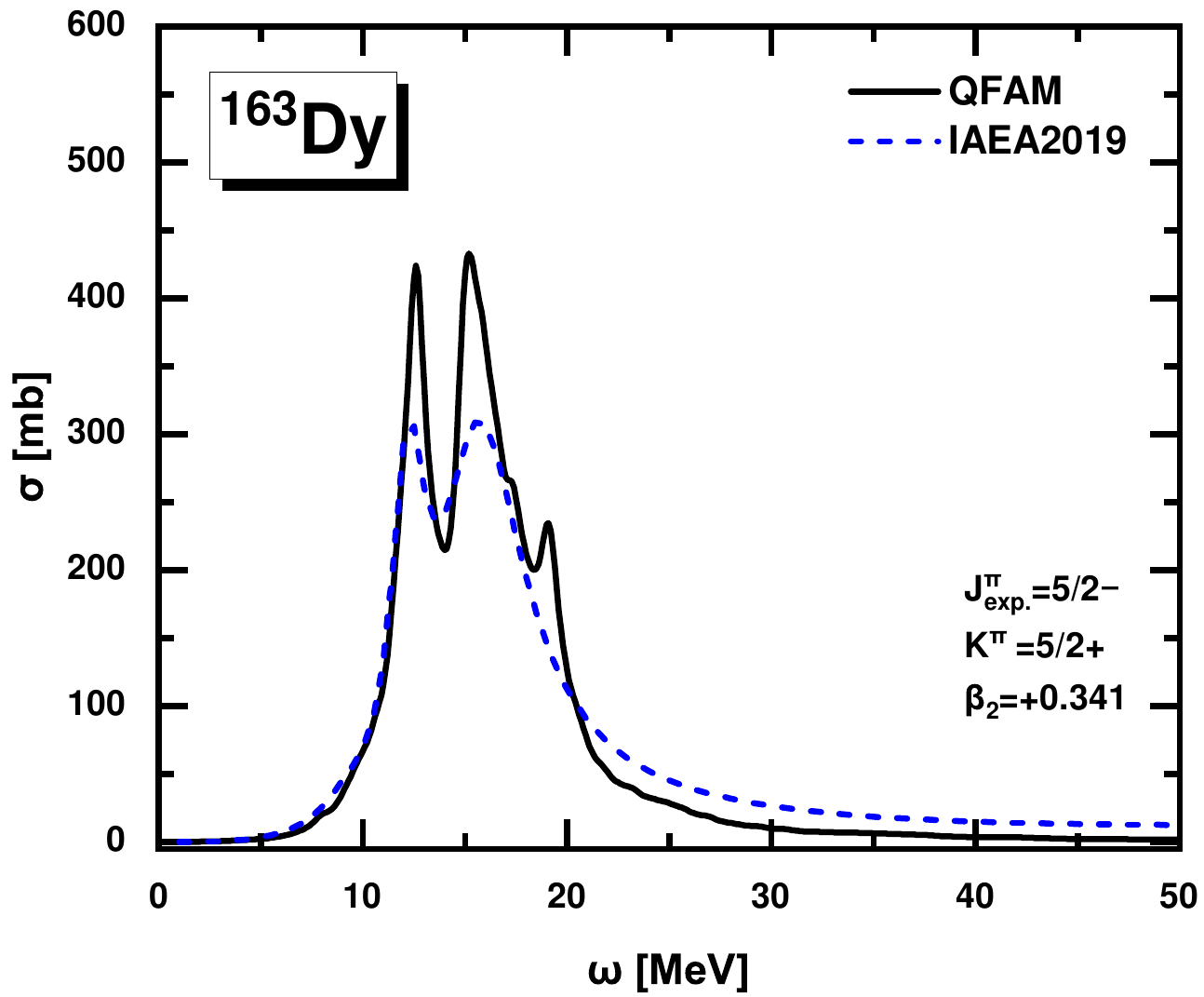}
    \includegraphics[width=0.35\textwidth]{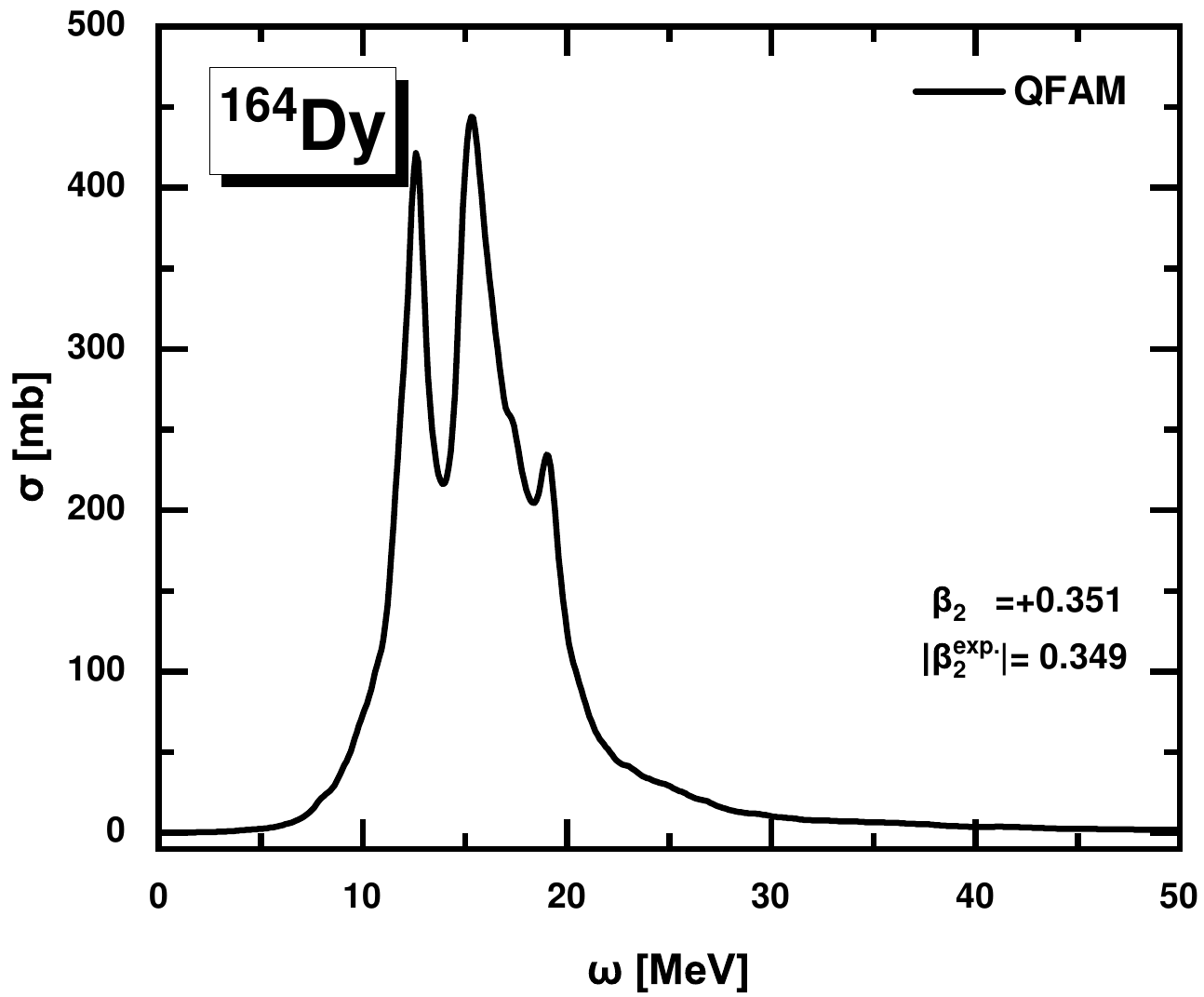}
\end{figure*}
\begin{figure*}\ContinuedFloat
    \centering
    \includegraphics[width=0.35\textwidth]{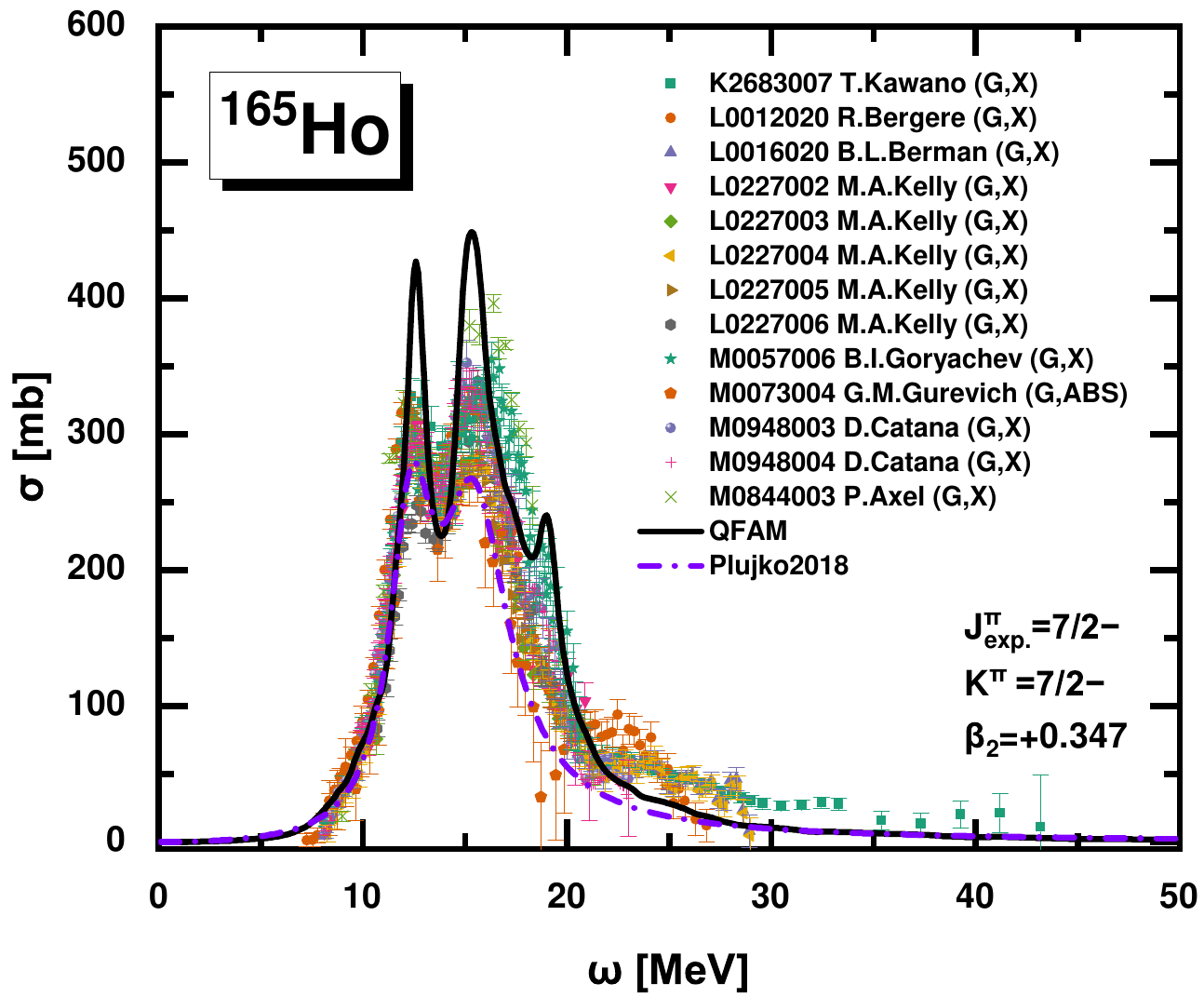}
    \includegraphics[width=0.35\textwidth]{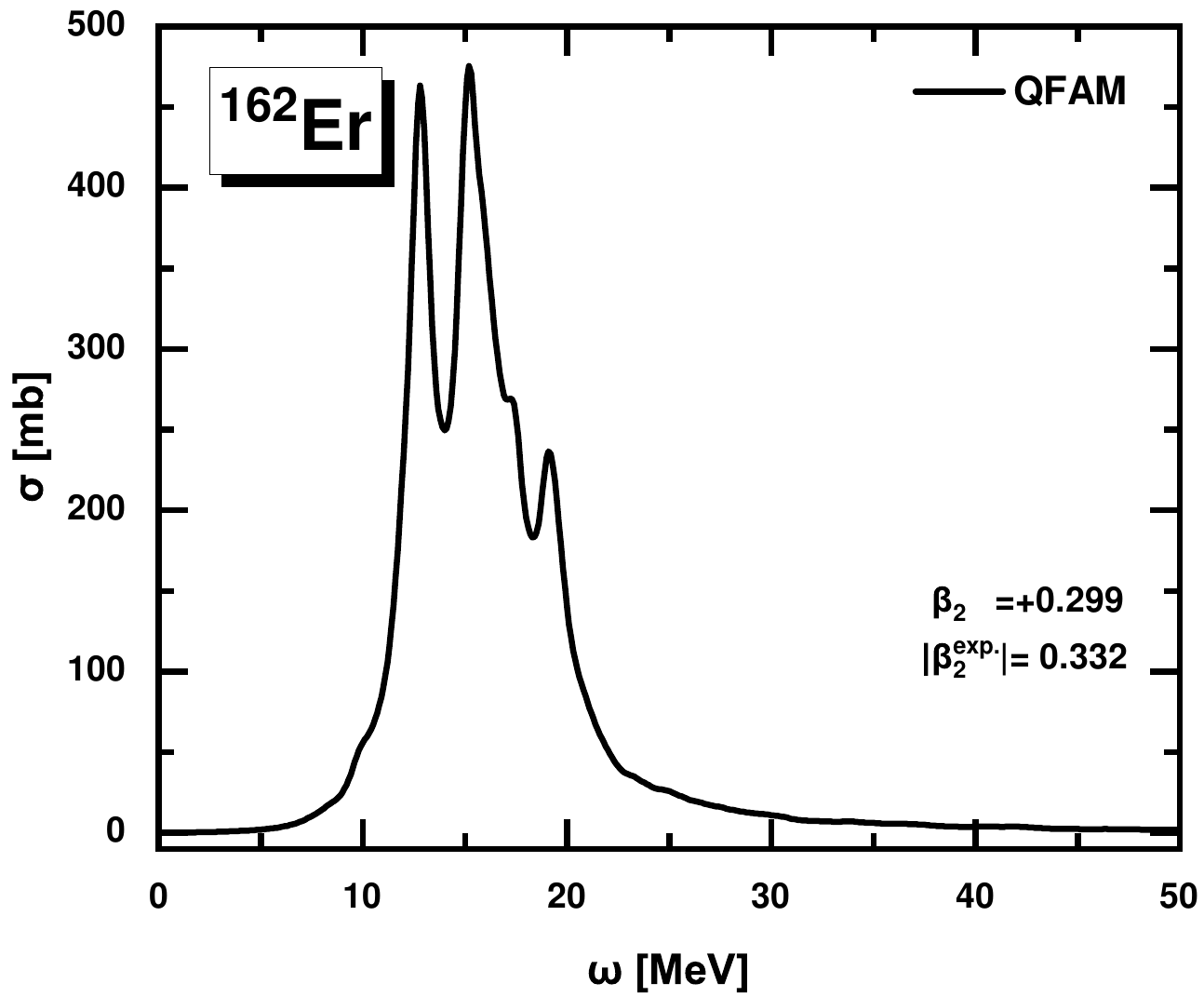}
    \includegraphics[width=0.35\textwidth]{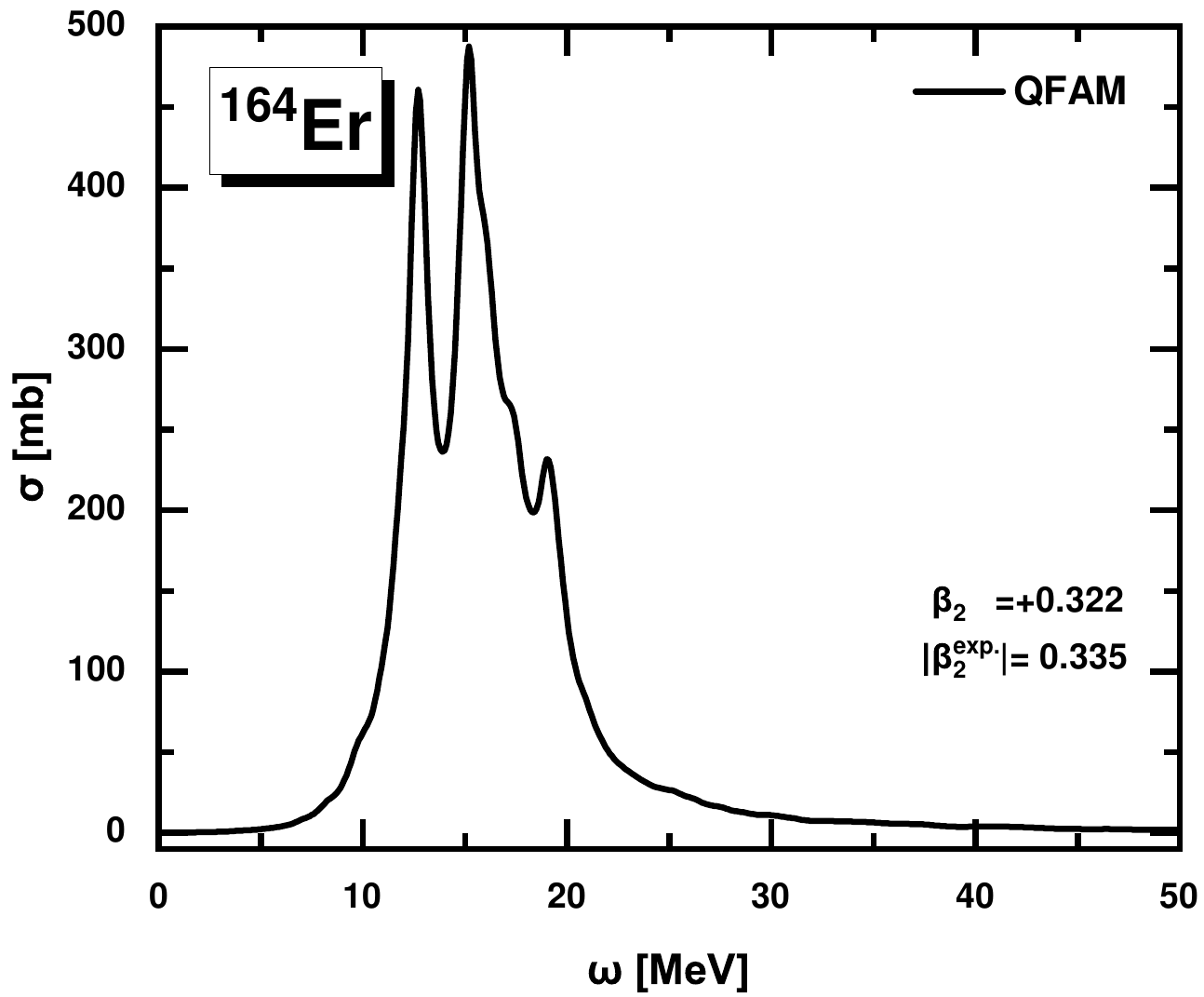}
    \includegraphics[width=0.35\textwidth]{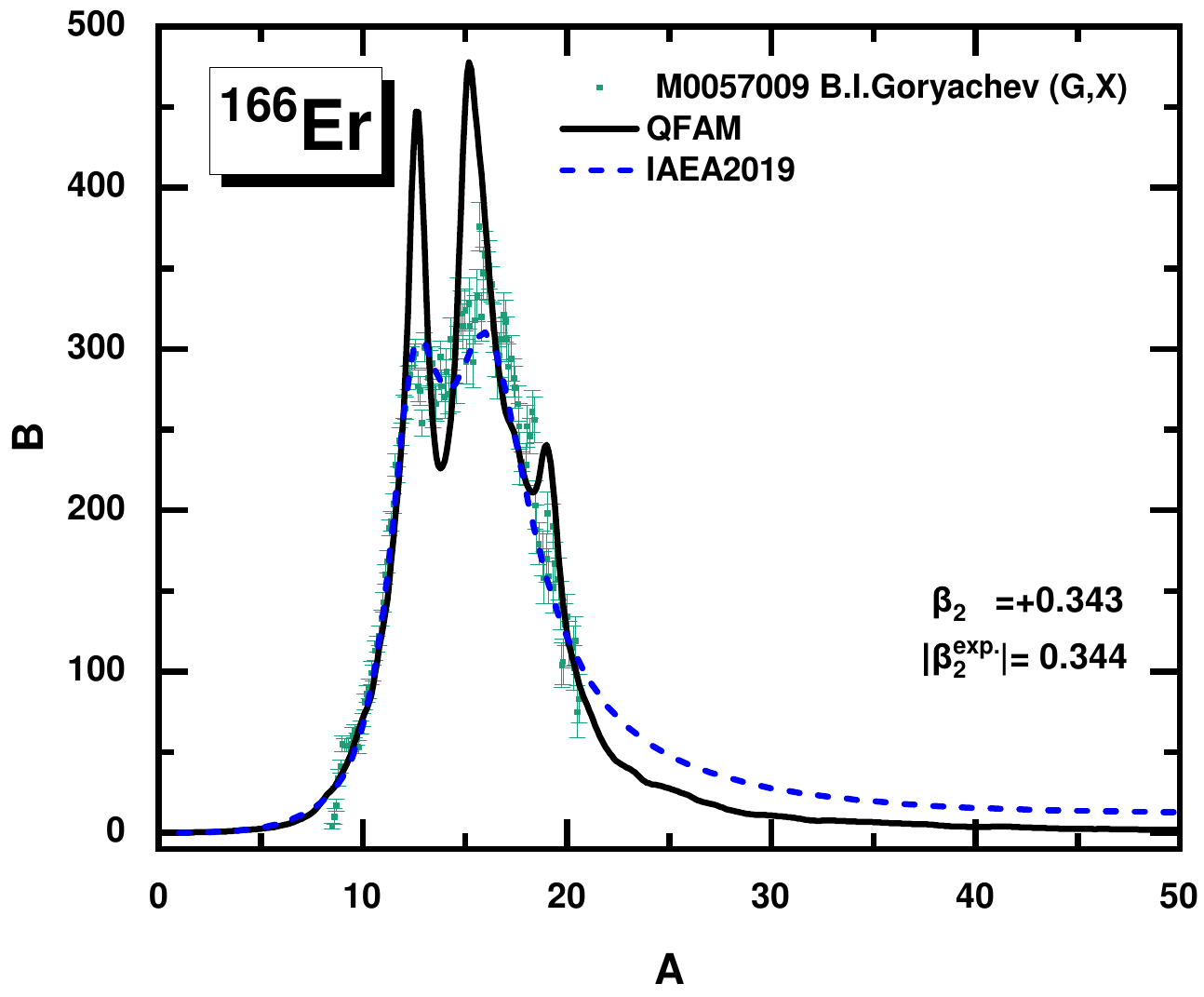}
    \includegraphics[width=0.35\textwidth]{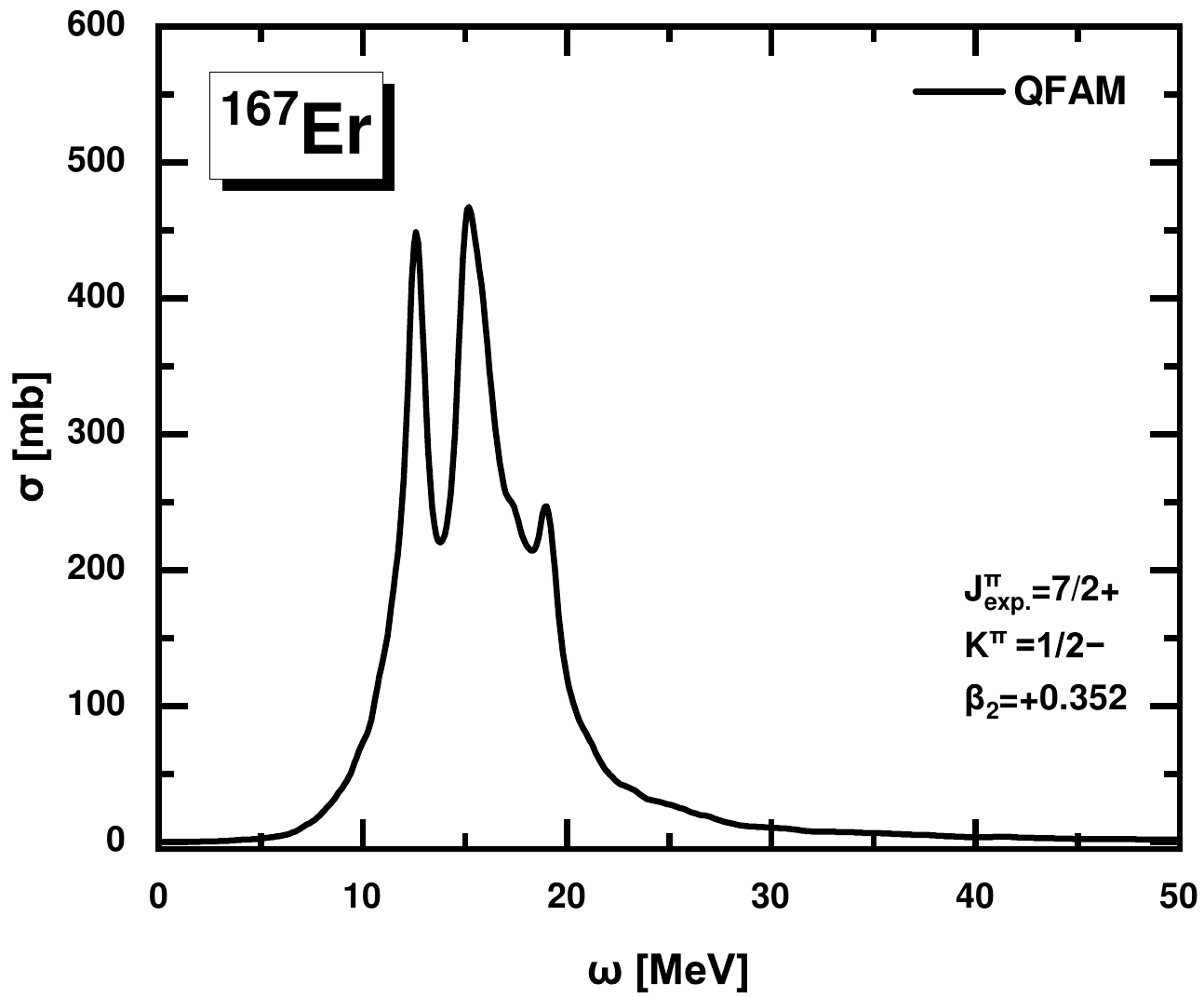}
    \includegraphics[width=0.35\textwidth]{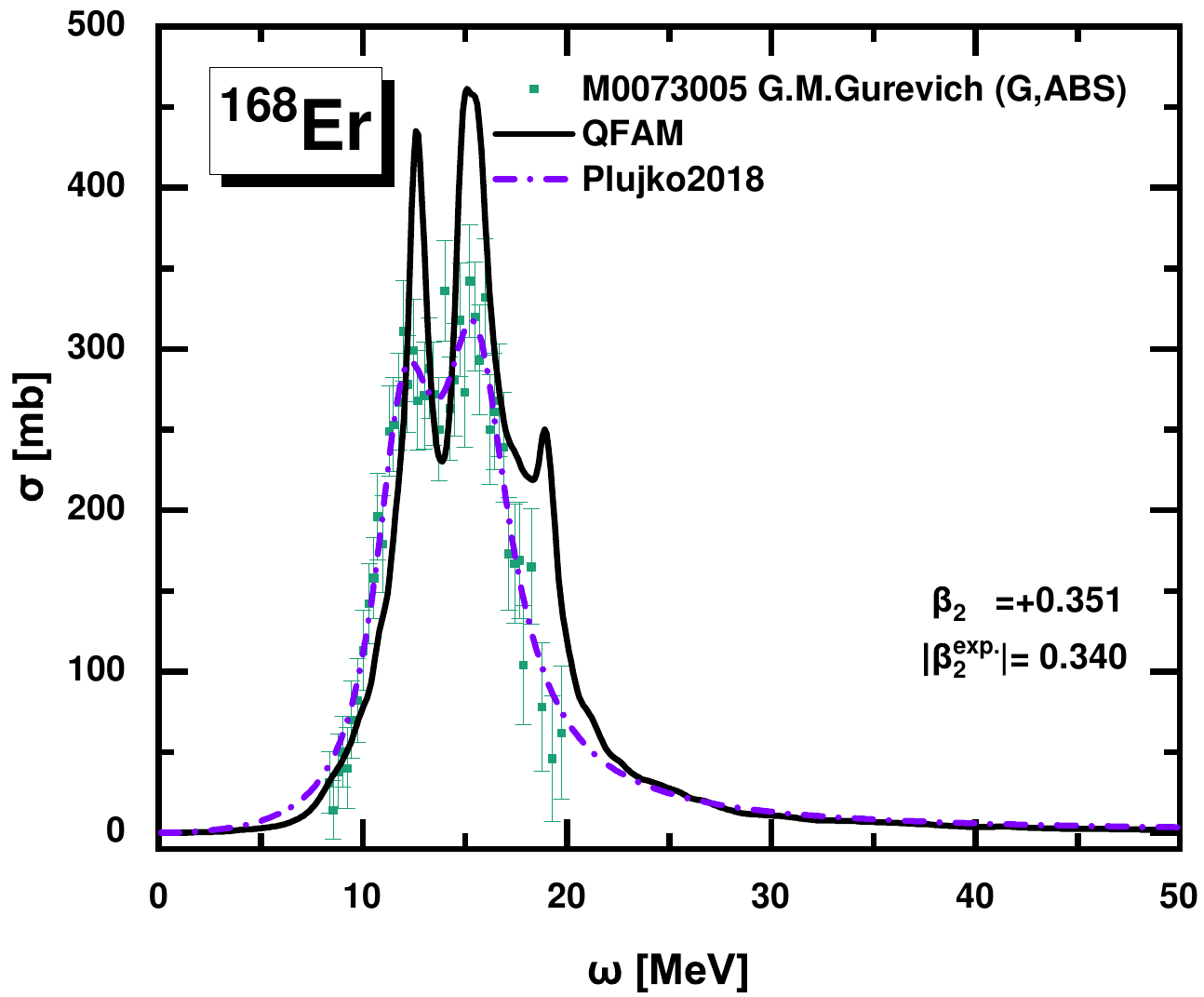}
    \includegraphics[width=0.35\textwidth]{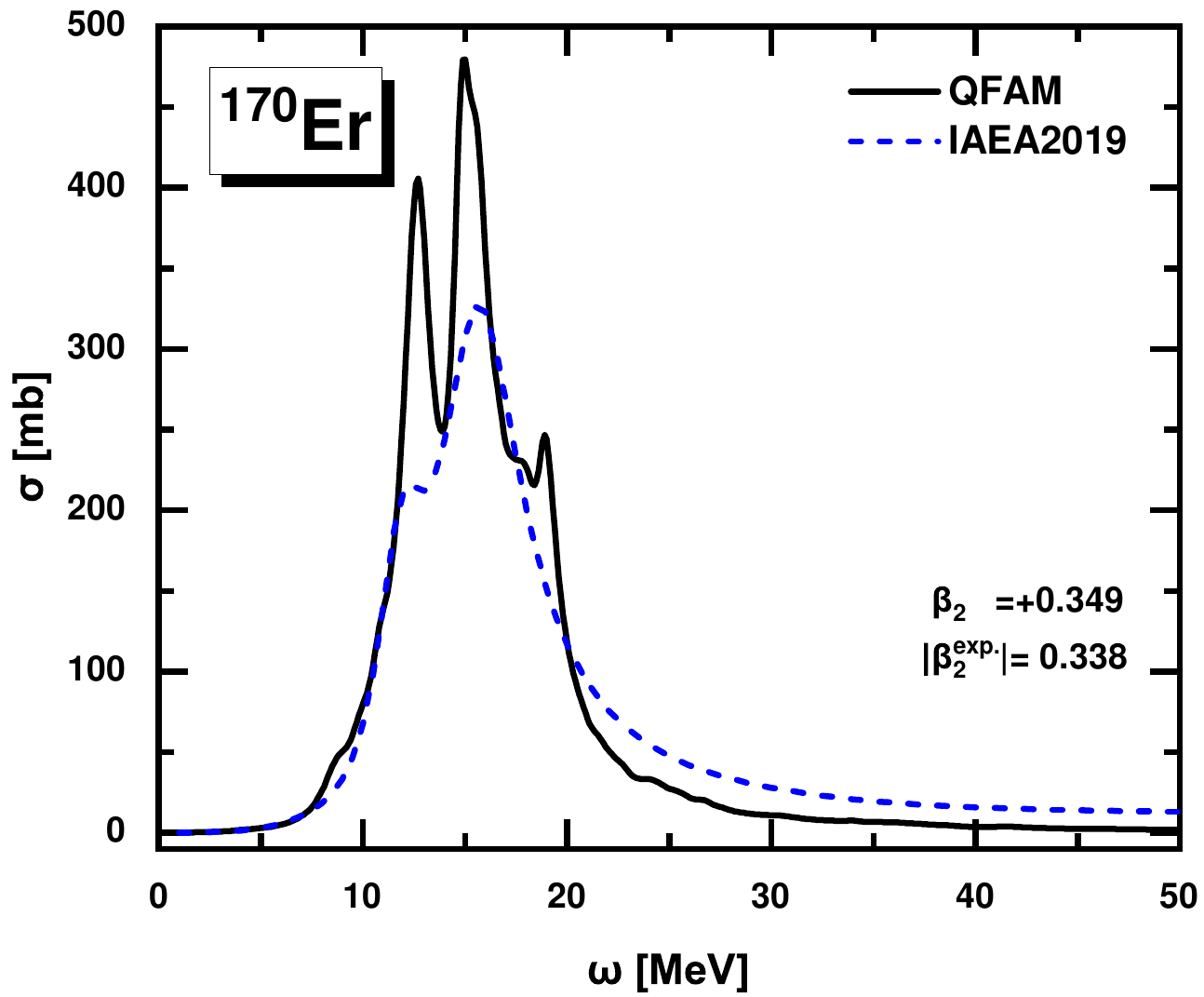}
    \includegraphics[width=0.35\textwidth]{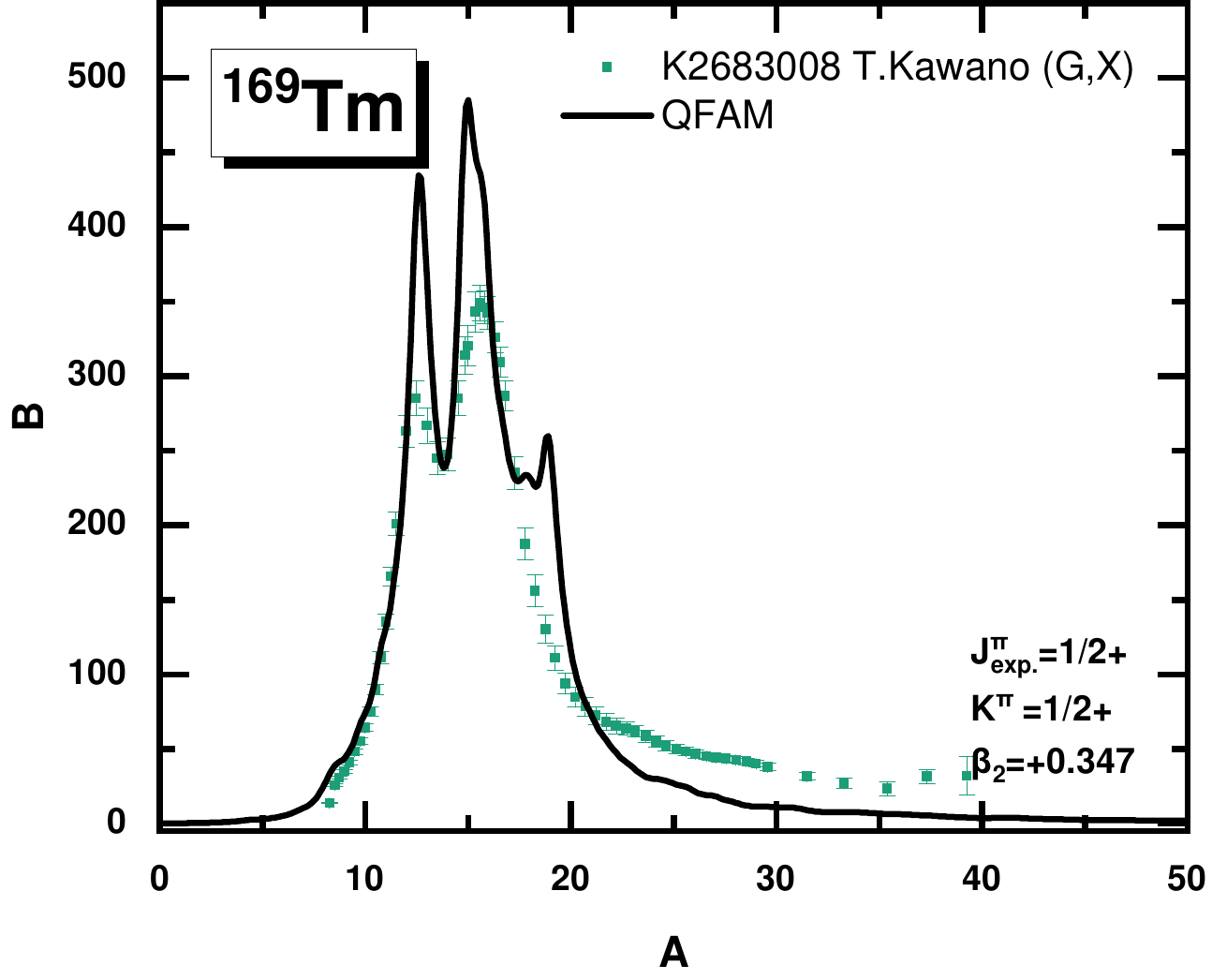}
\end{figure*}
\begin{figure*}\ContinuedFloat
    \centering
    \includegraphics[width=0.35\textwidth]{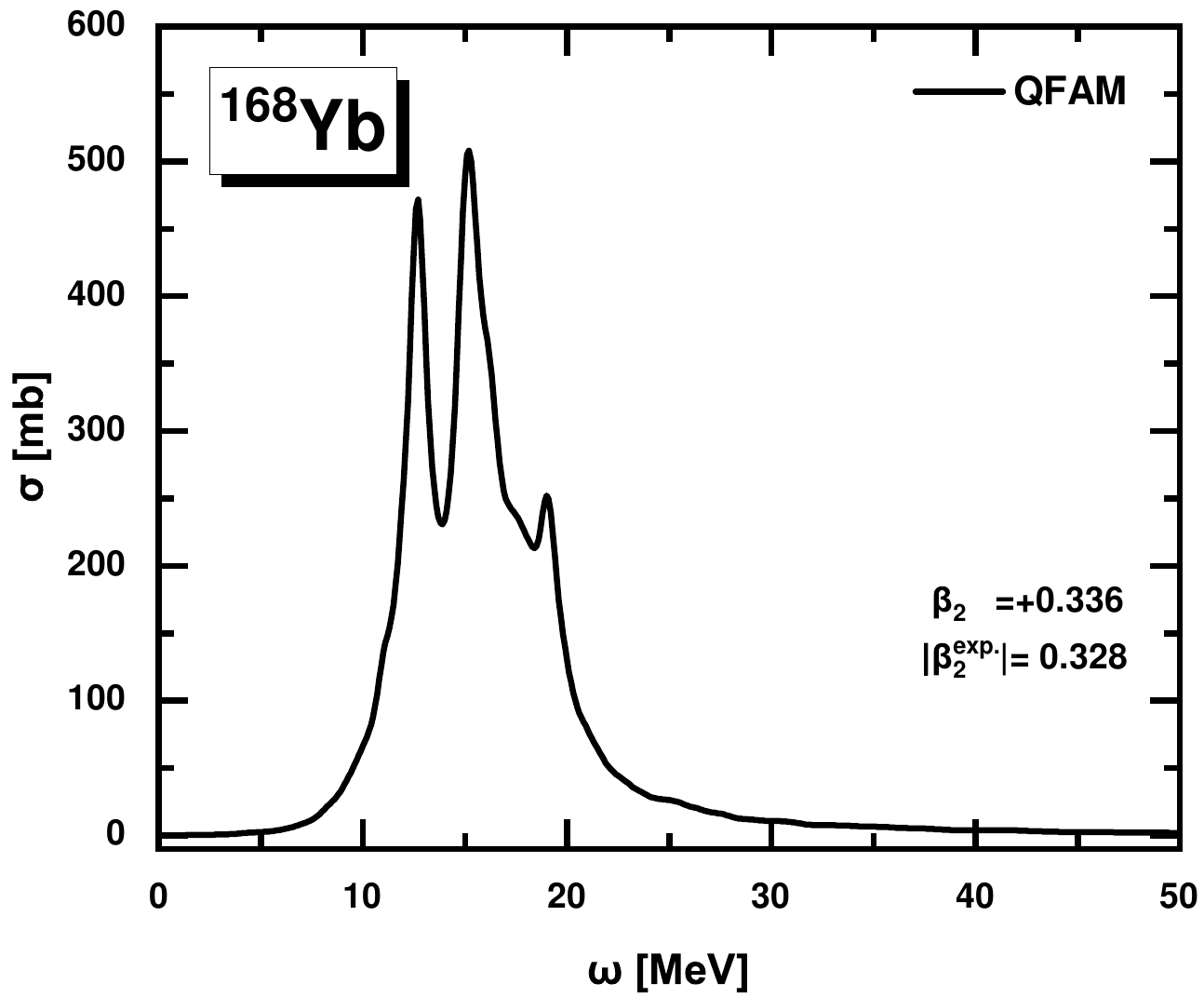}
    \includegraphics[width=0.35\textwidth]{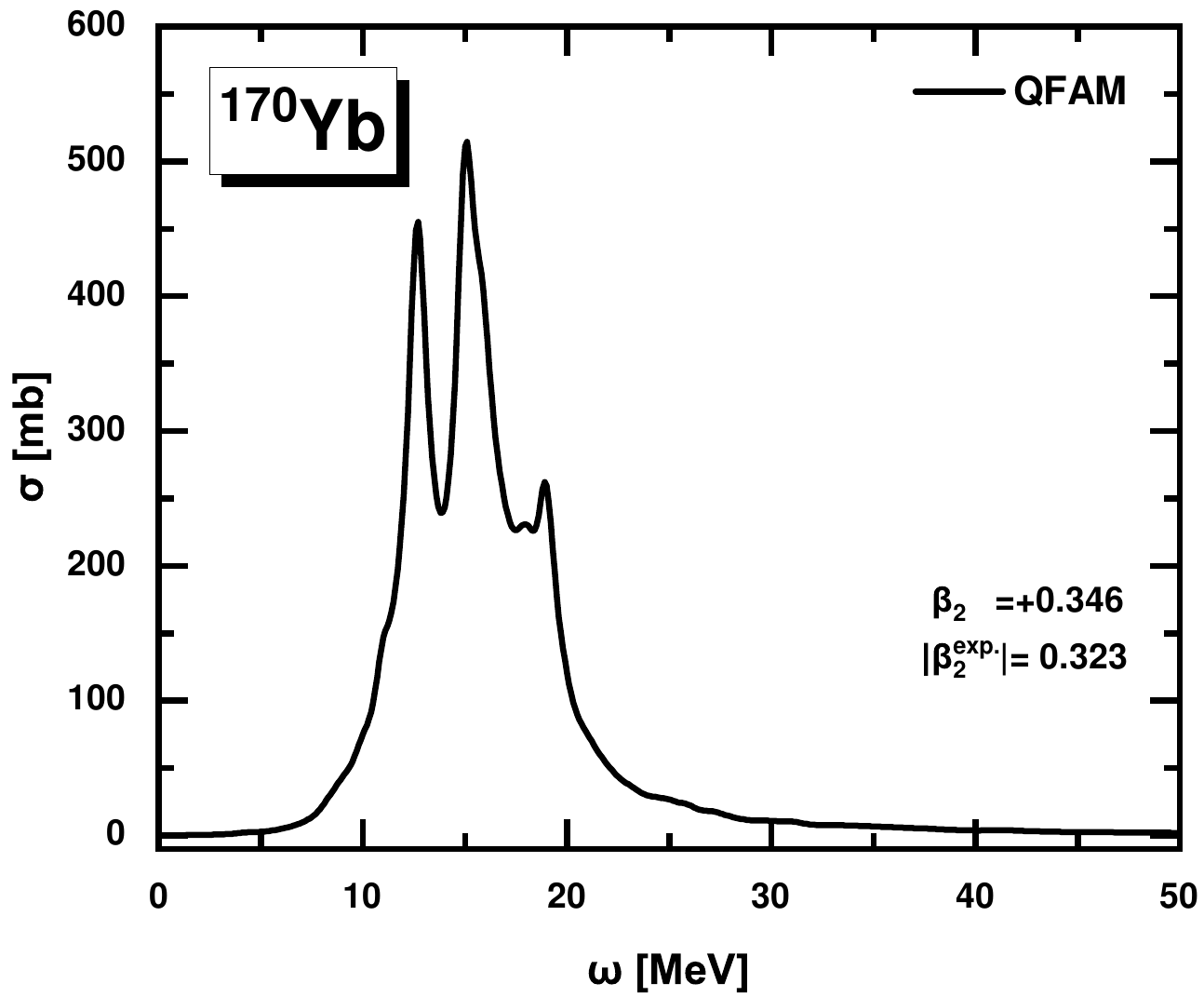}
    \includegraphics[width=0.35\textwidth]{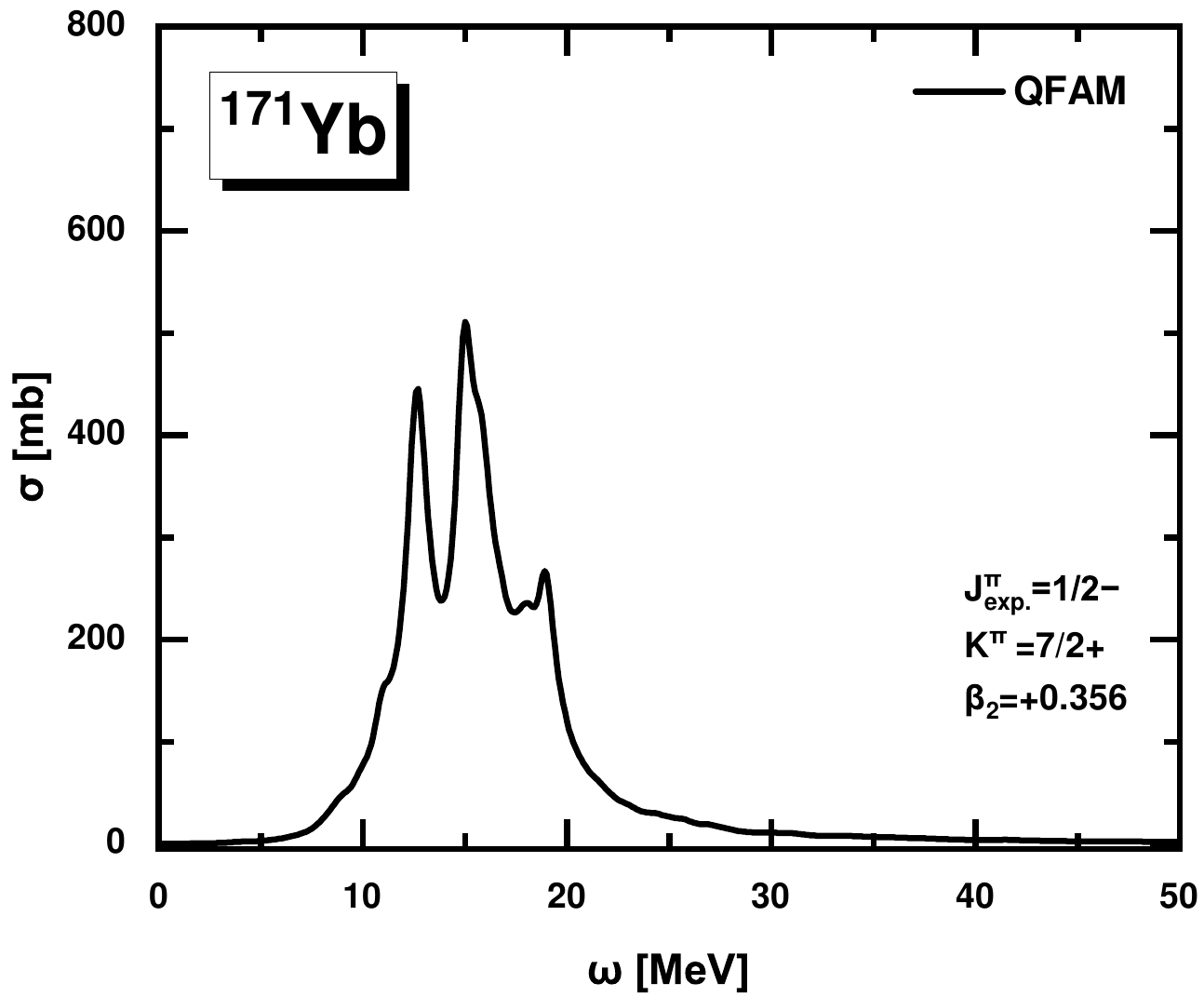}
    \includegraphics[width=0.35\textwidth]{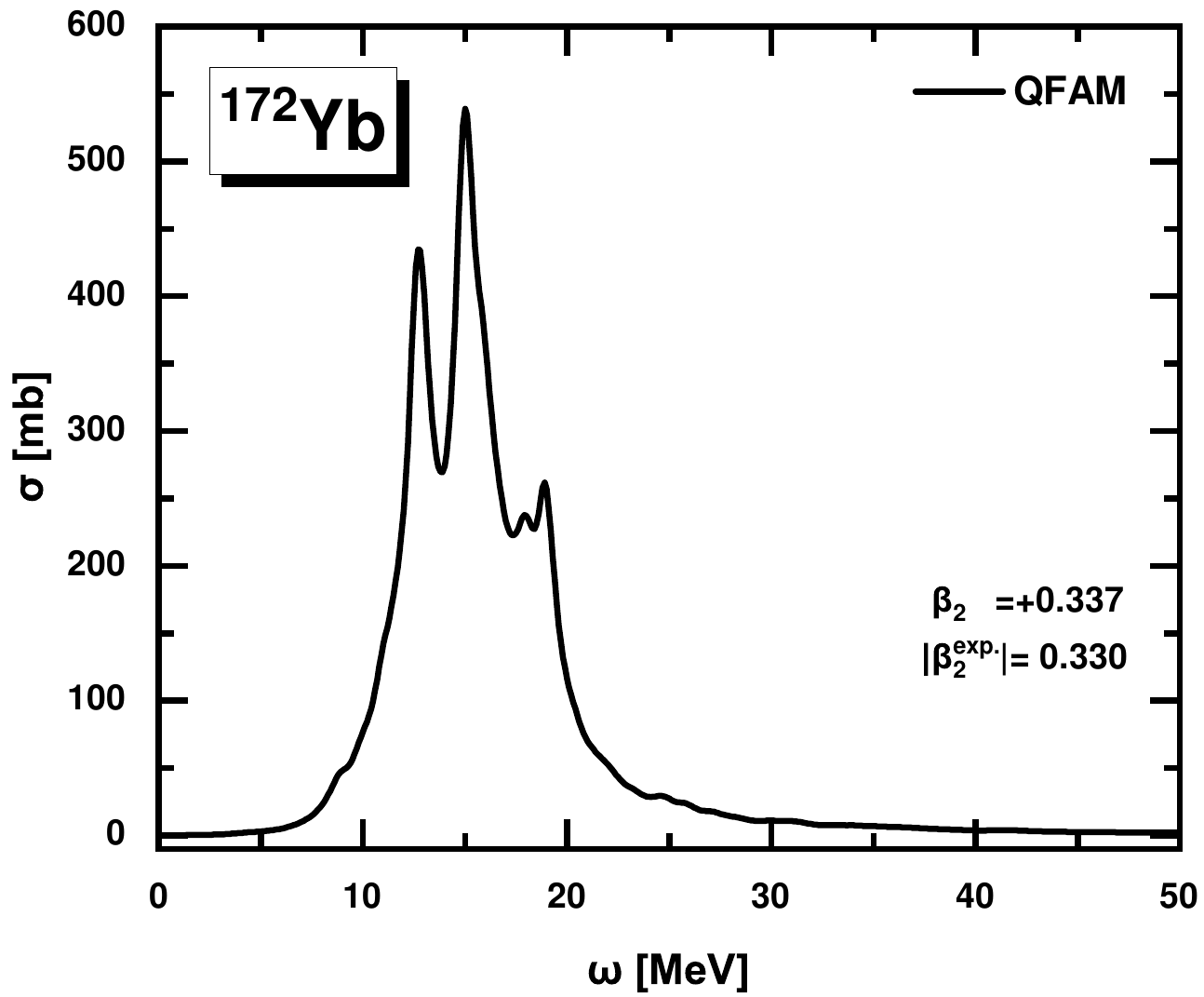}
    \includegraphics[width=0.35\textwidth]{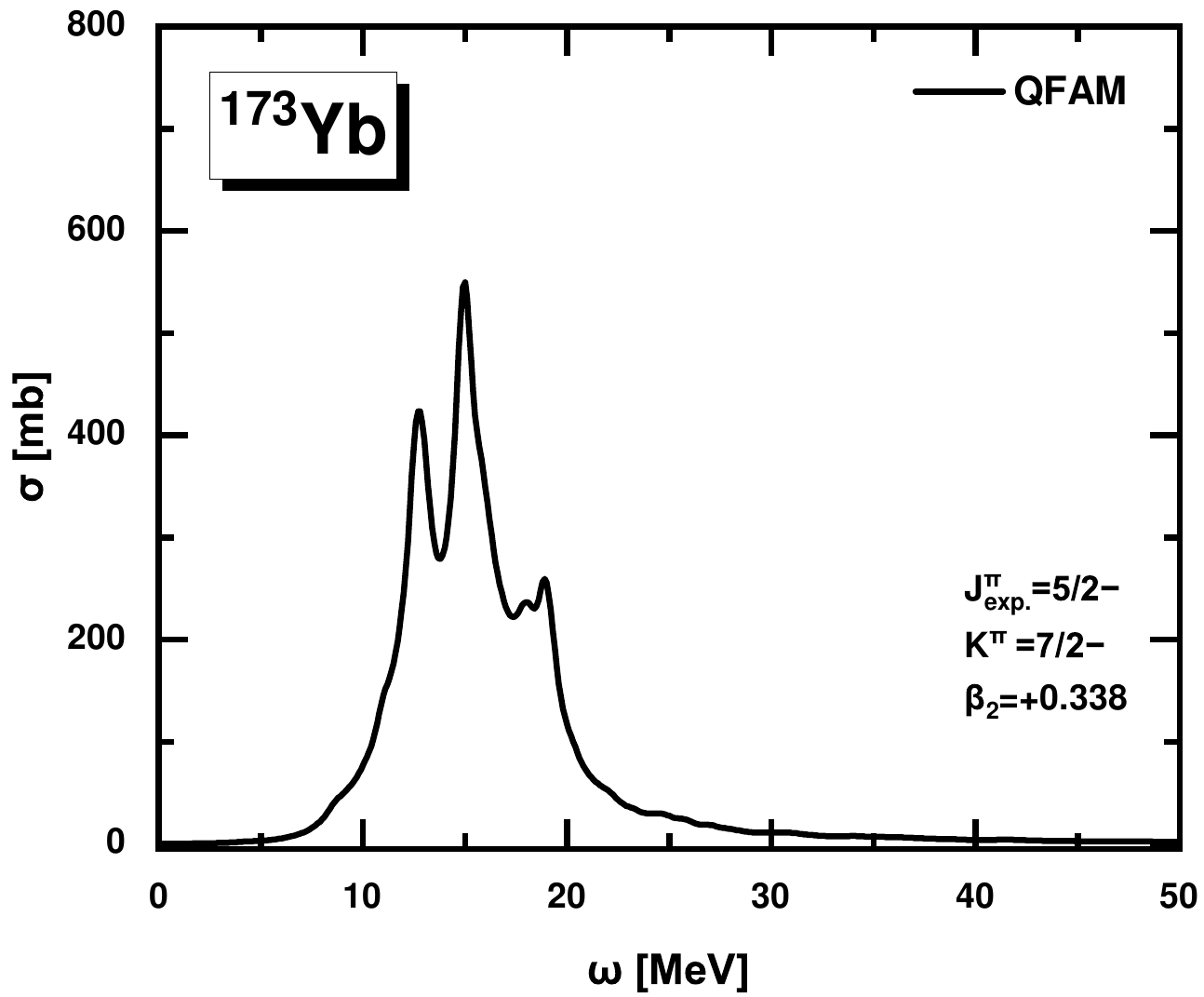}
    \includegraphics[width=0.35\textwidth]{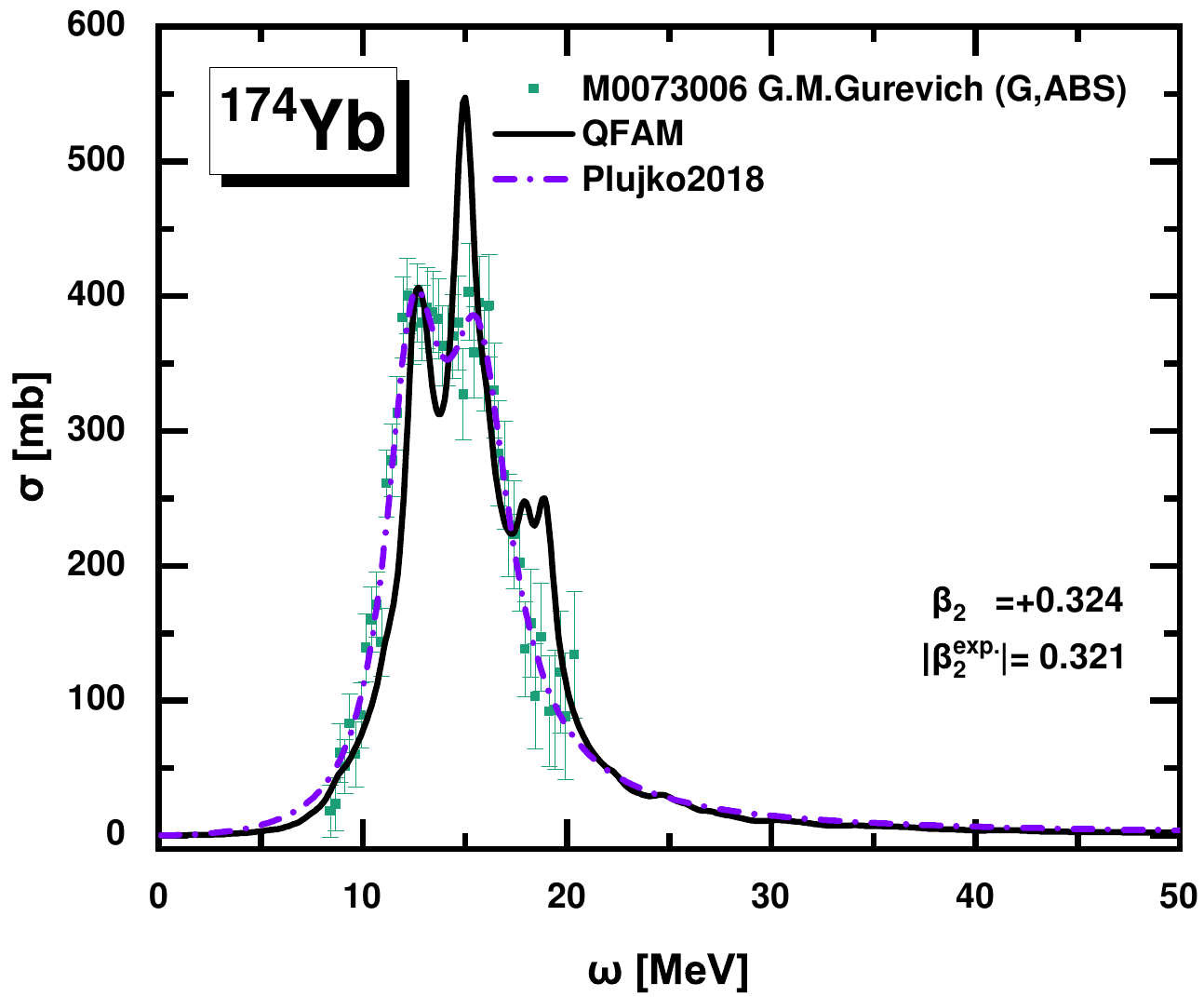}
    \includegraphics[width=0.35\textwidth]{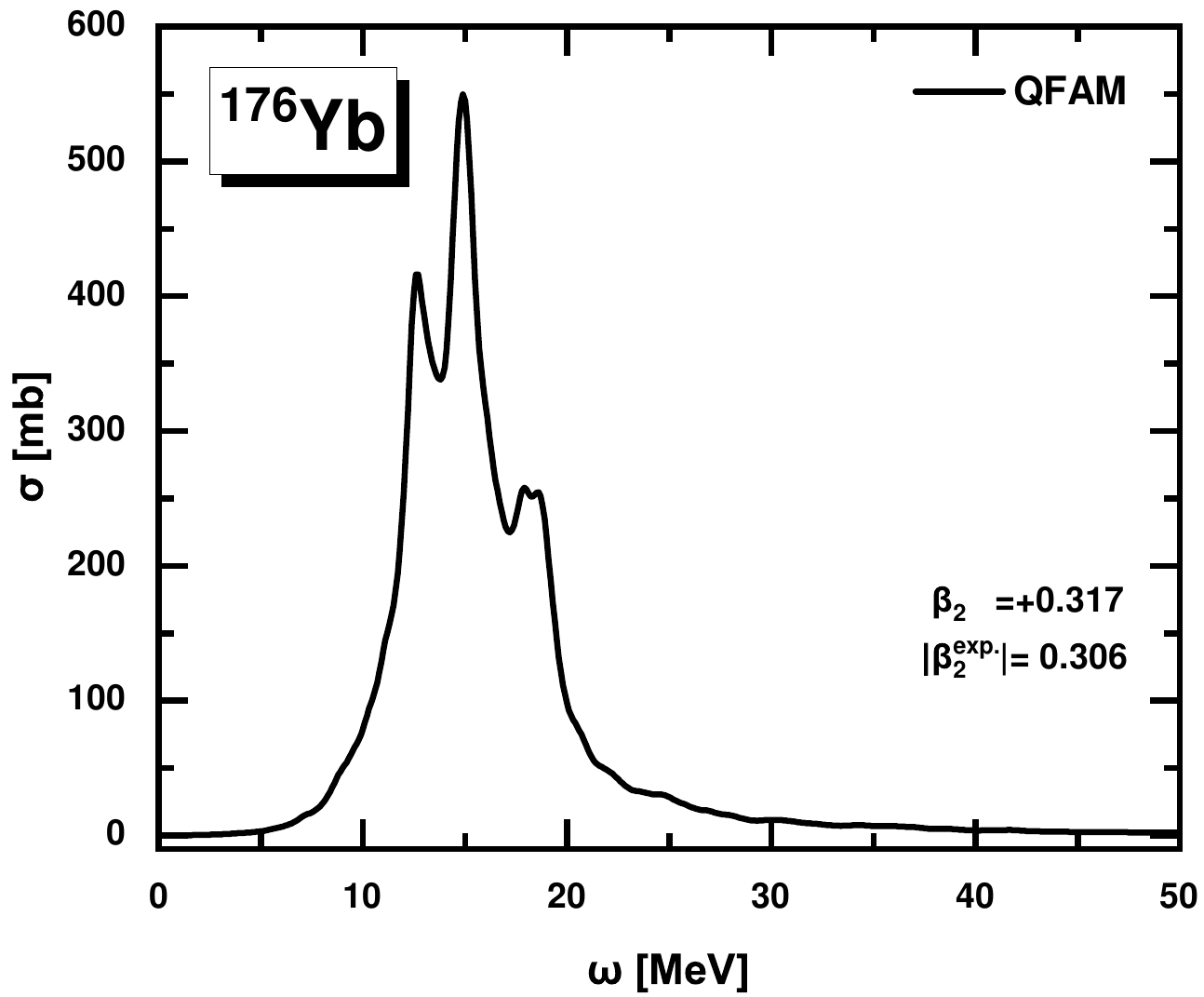}
    \includegraphics[width=0.35\textwidth]{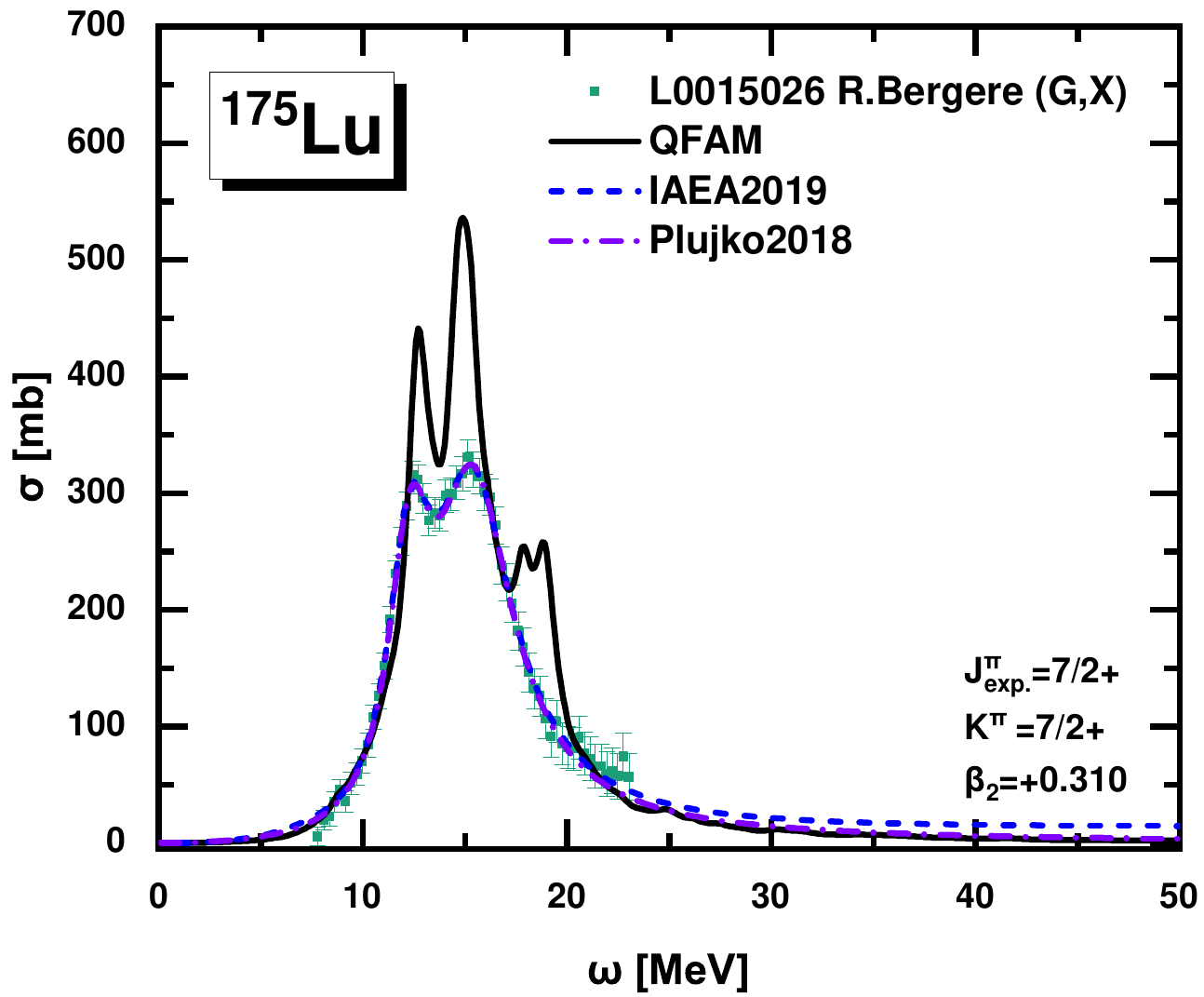}
\end{figure*}
\begin{figure*}\ContinuedFloat
    \centering
    \includegraphics[width=0.35\textwidth]{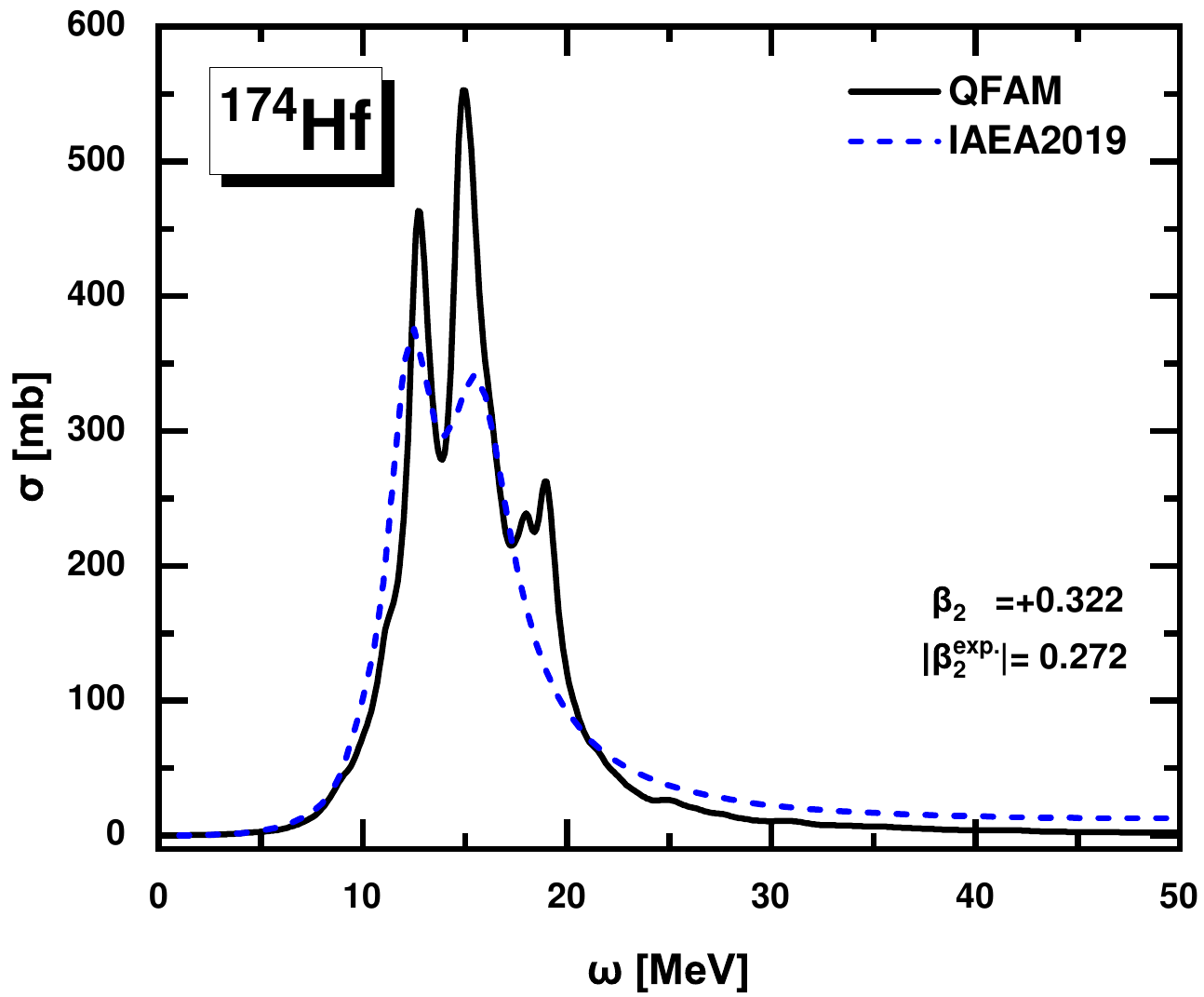}
    \includegraphics[width=0.35\textwidth]{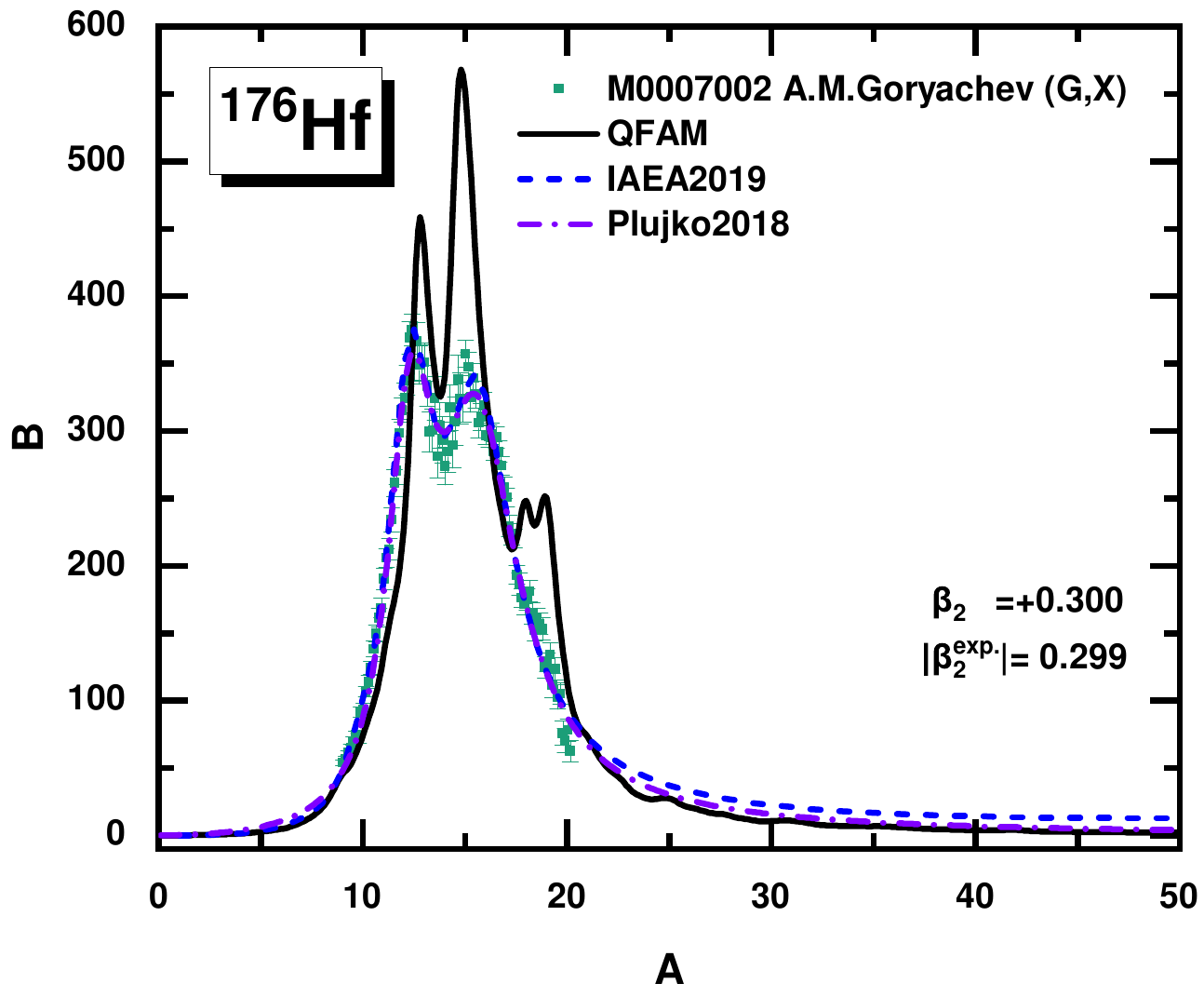}
    \includegraphics[width=0.35\textwidth]{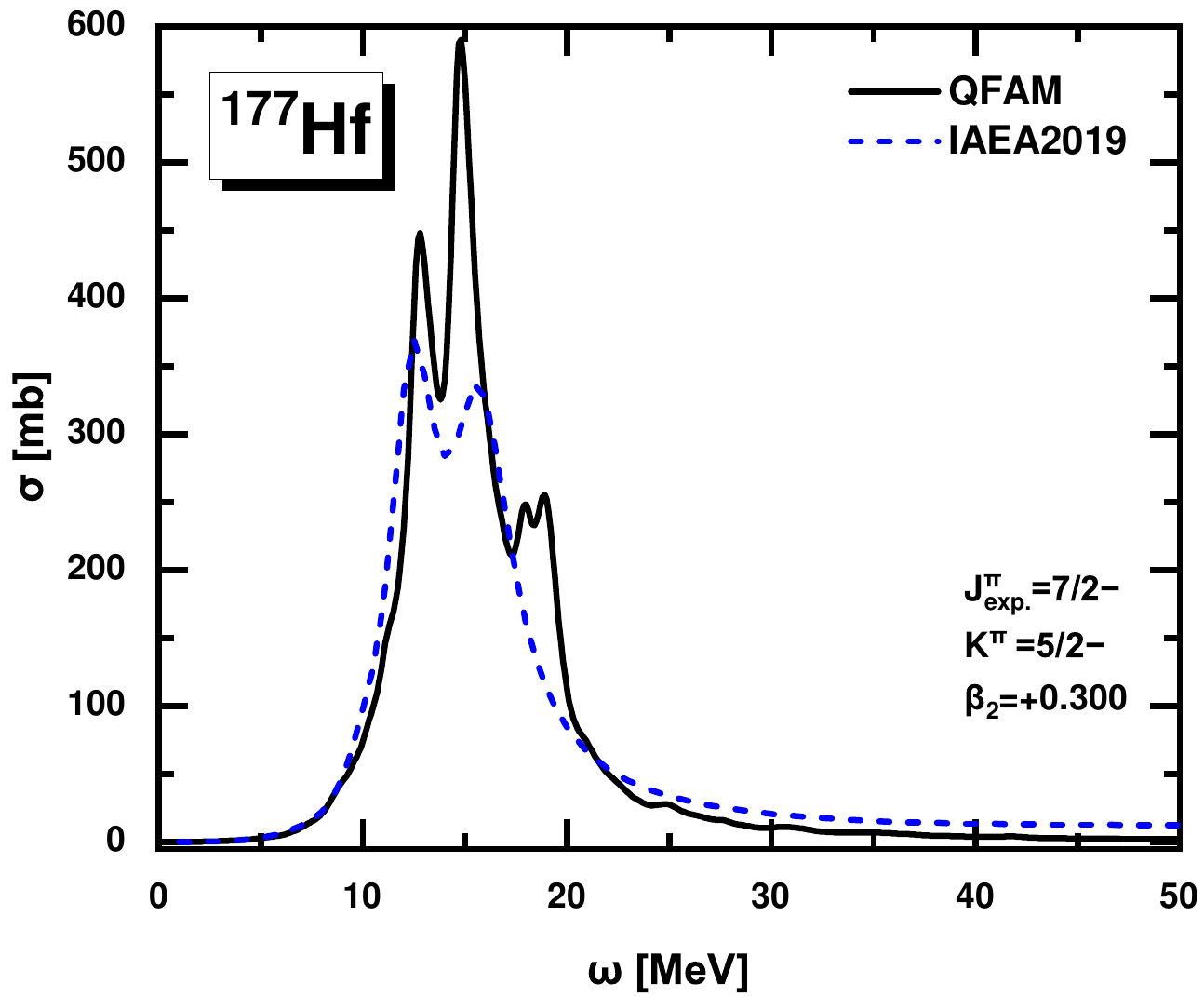}
    \includegraphics[width=0.35\textwidth]{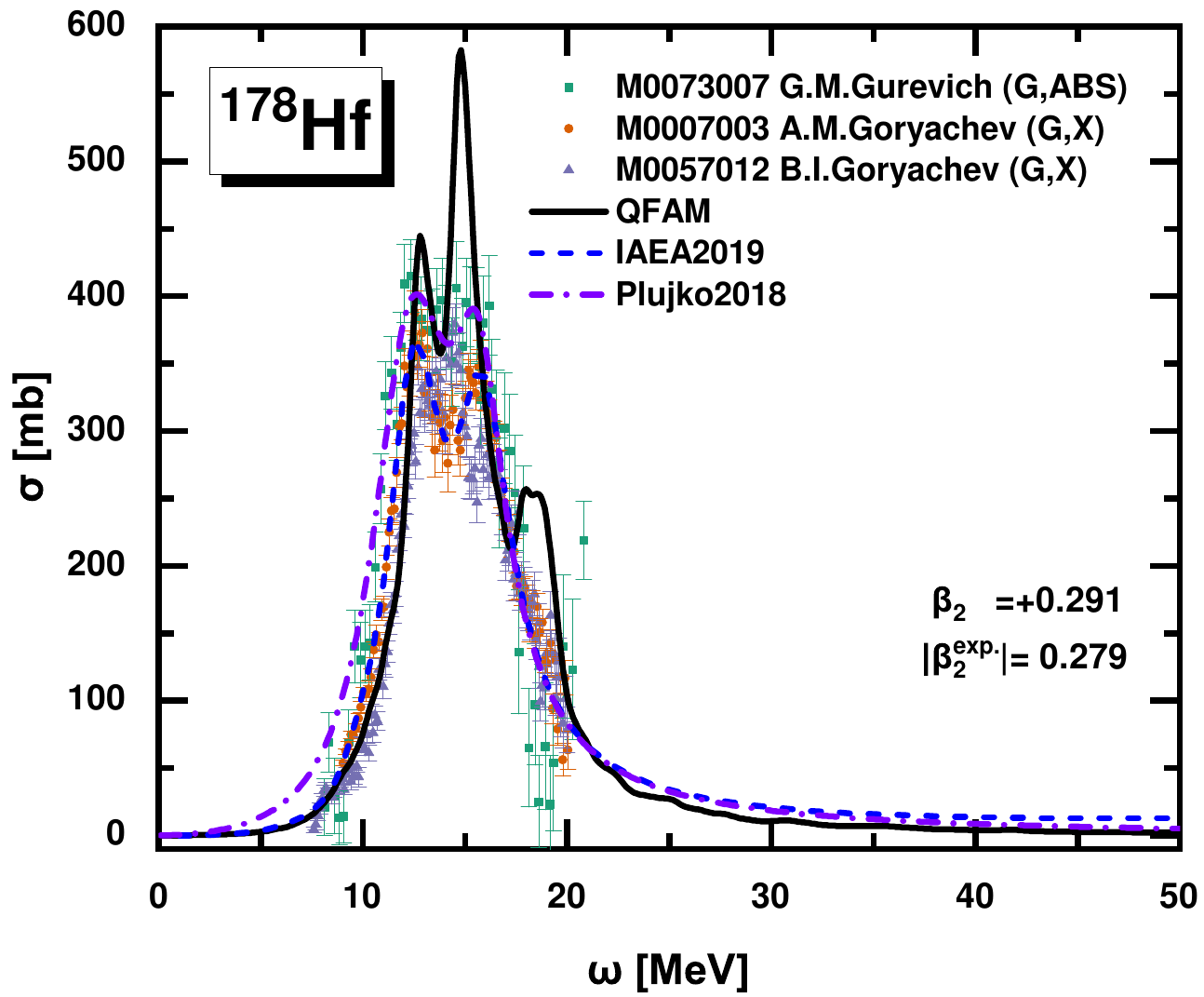}
    \includegraphics[width=0.35\textwidth]{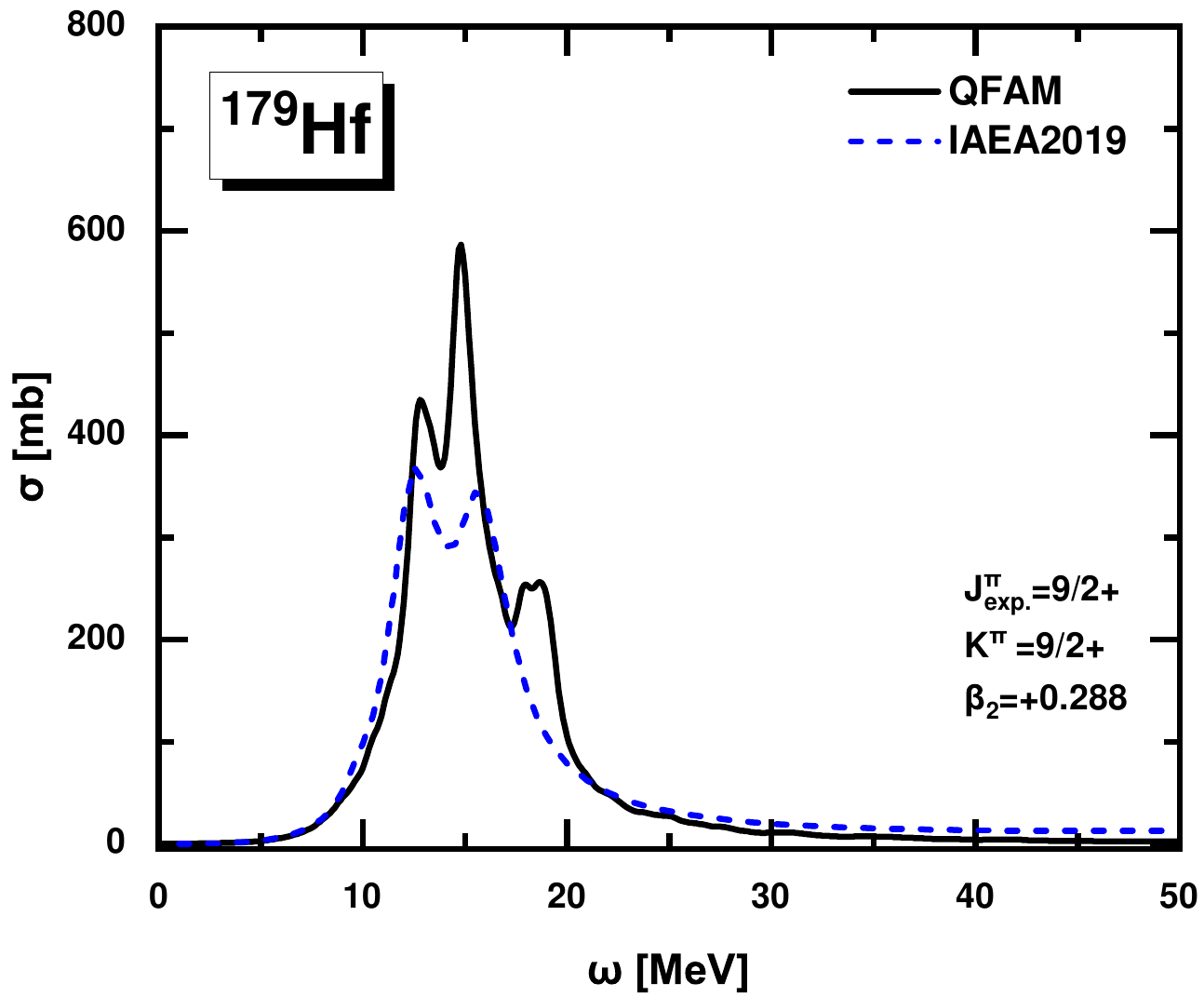}
    \includegraphics[width=0.35\textwidth]{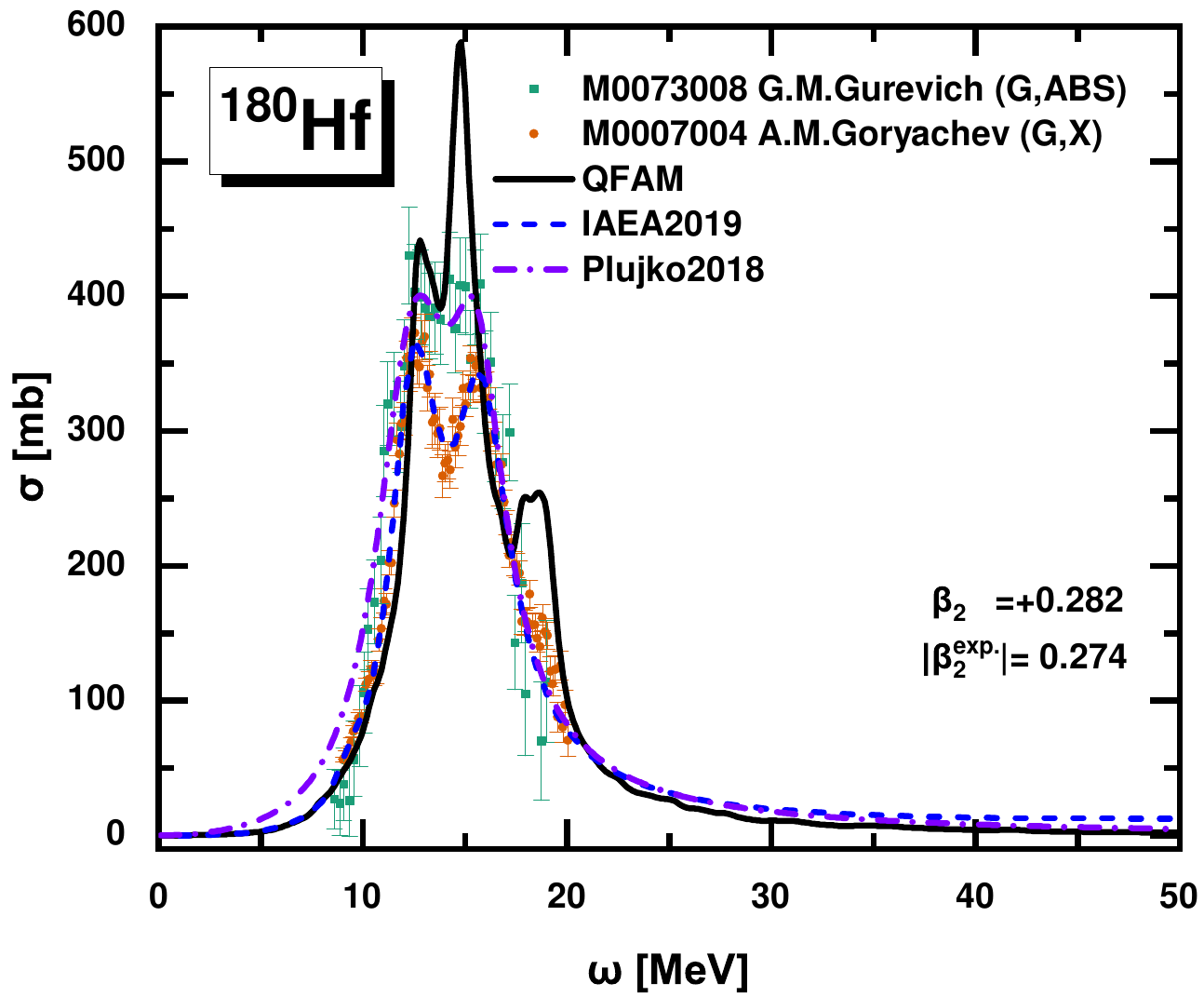}
    \includegraphics[width=0.35\textwidth]{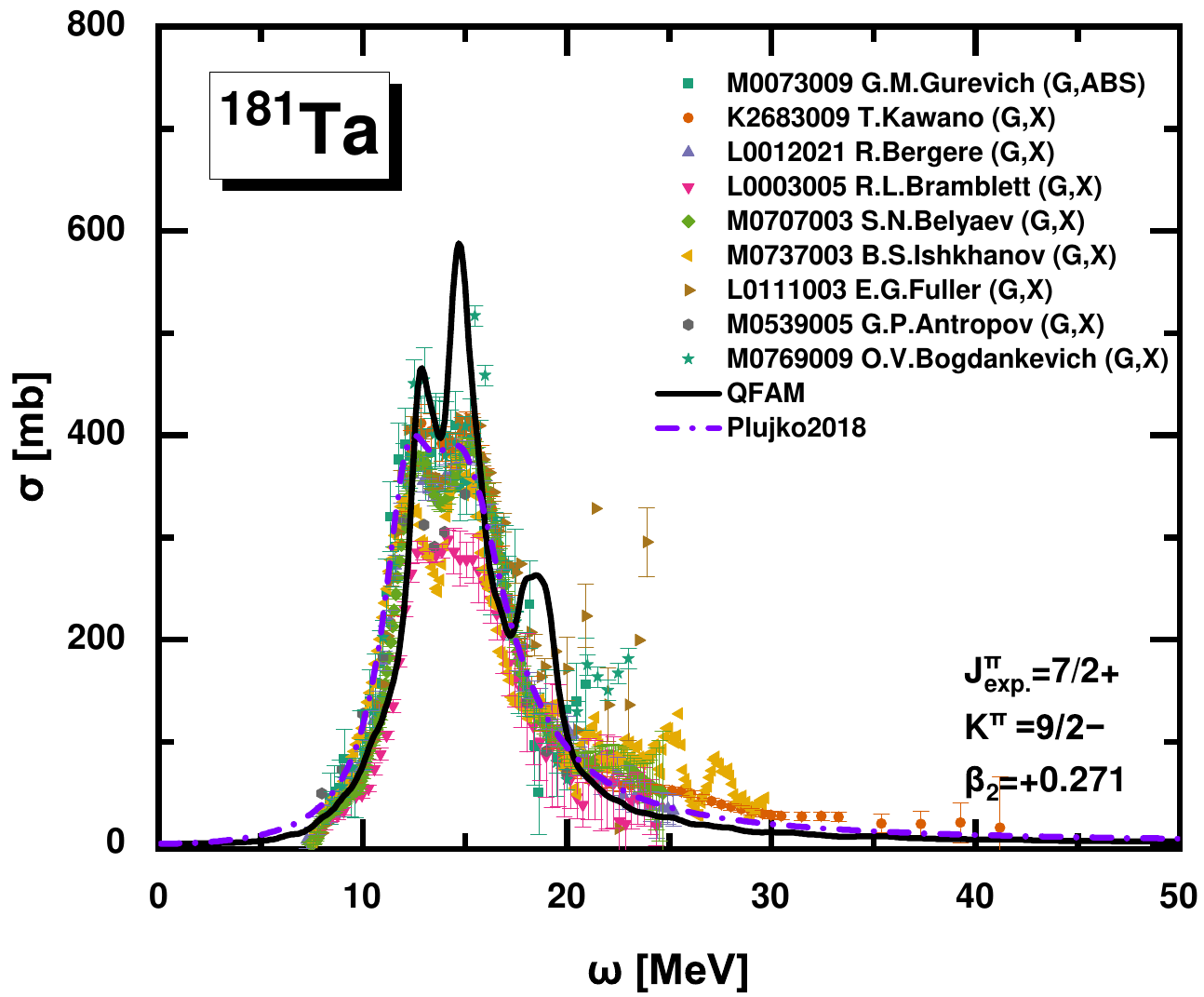}
    \includegraphics[width=0.35\textwidth]{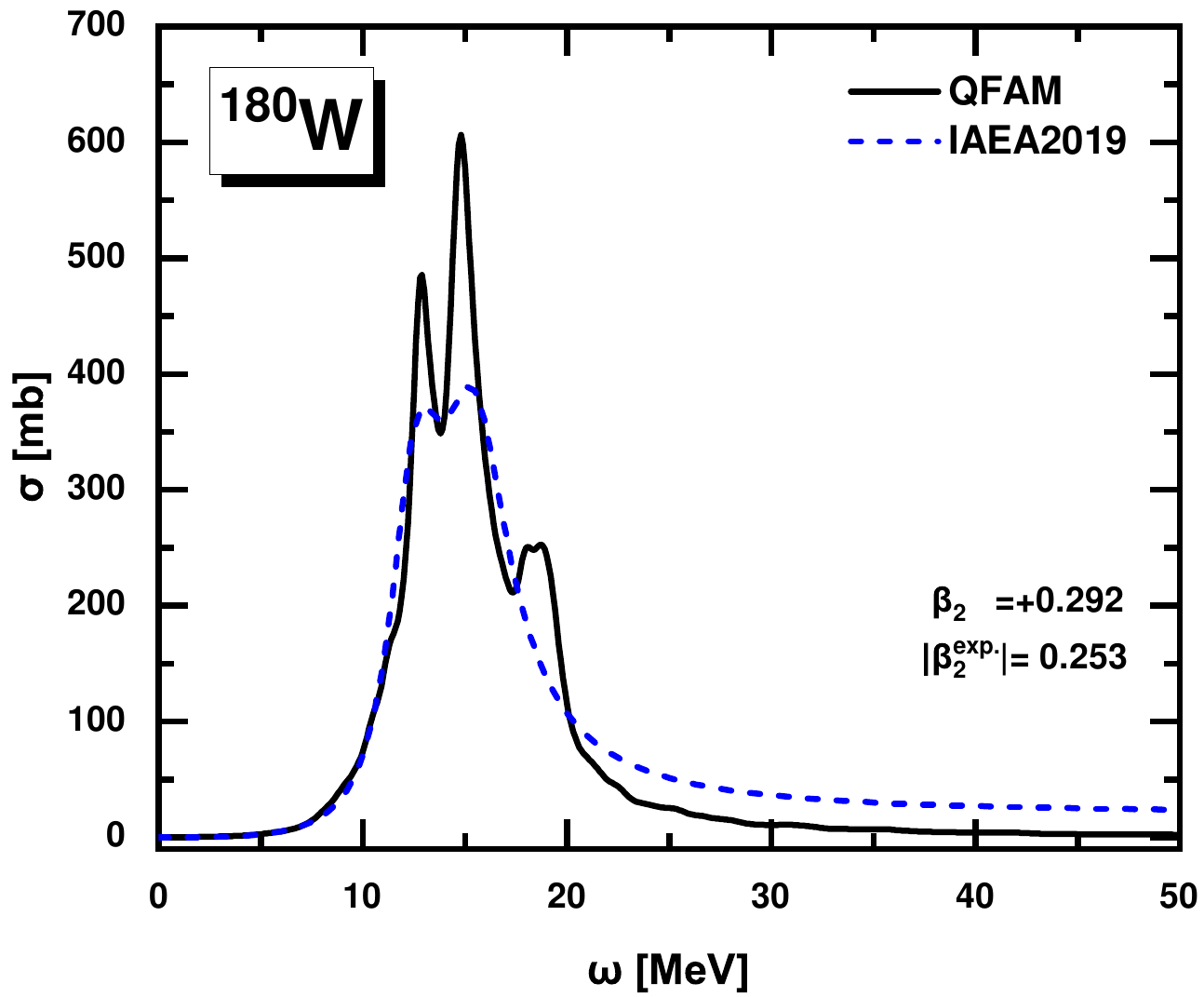}
\end{figure*}
\begin{figure*}\ContinuedFloat
    \centering
    \includegraphics[width=0.35\textwidth]{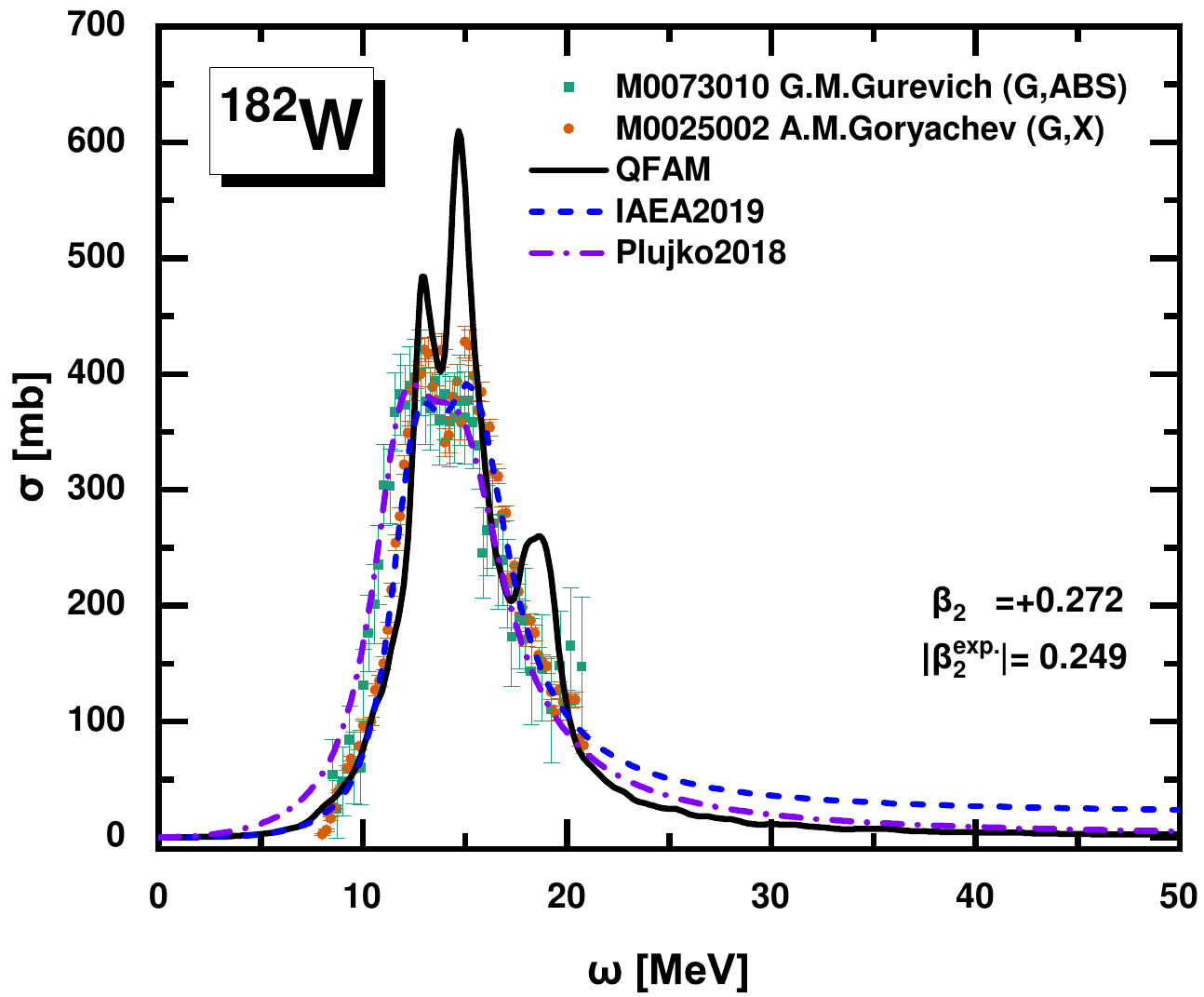}
    \includegraphics[width=0.35\textwidth]{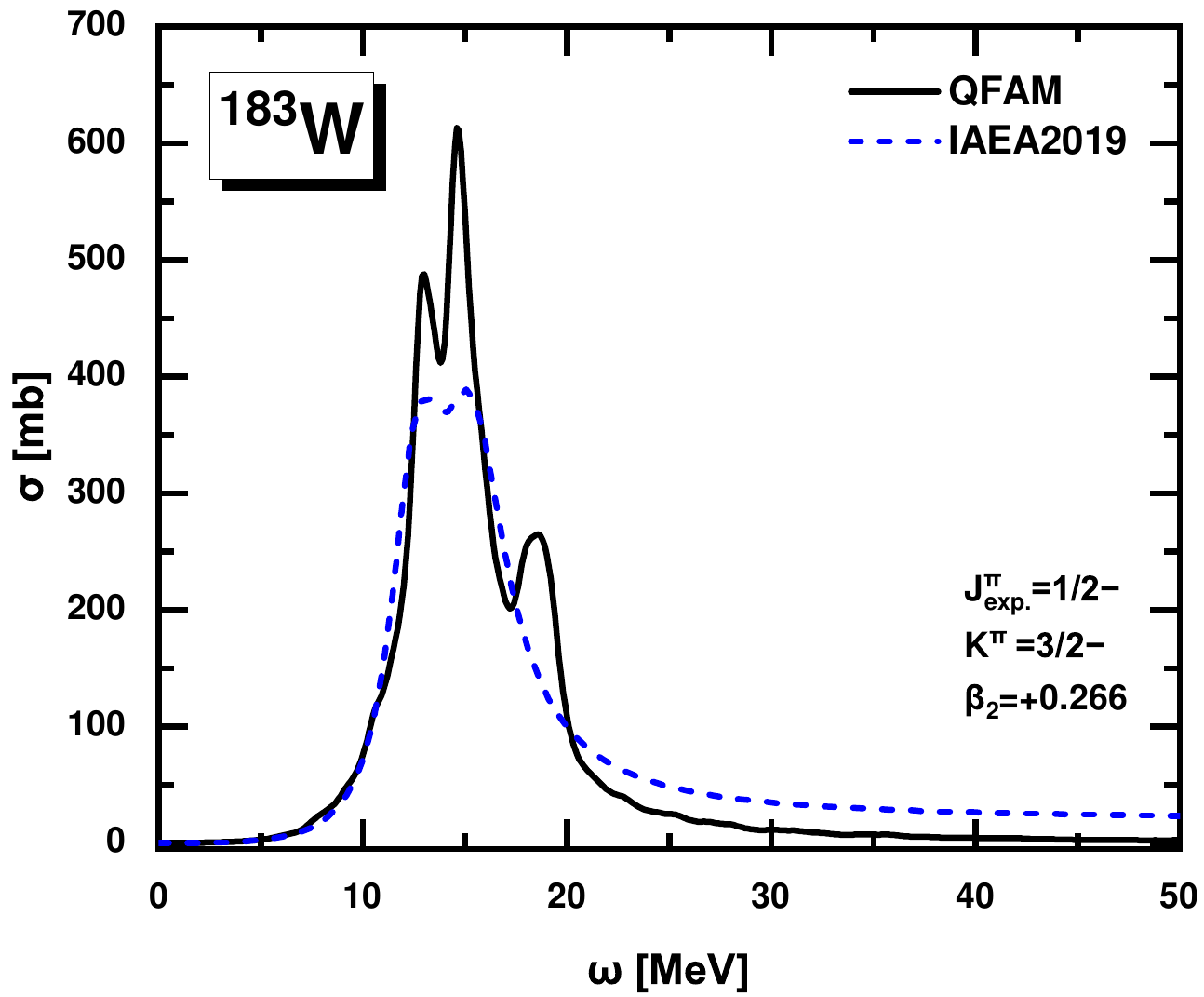}
    \includegraphics[width=0.35\textwidth]{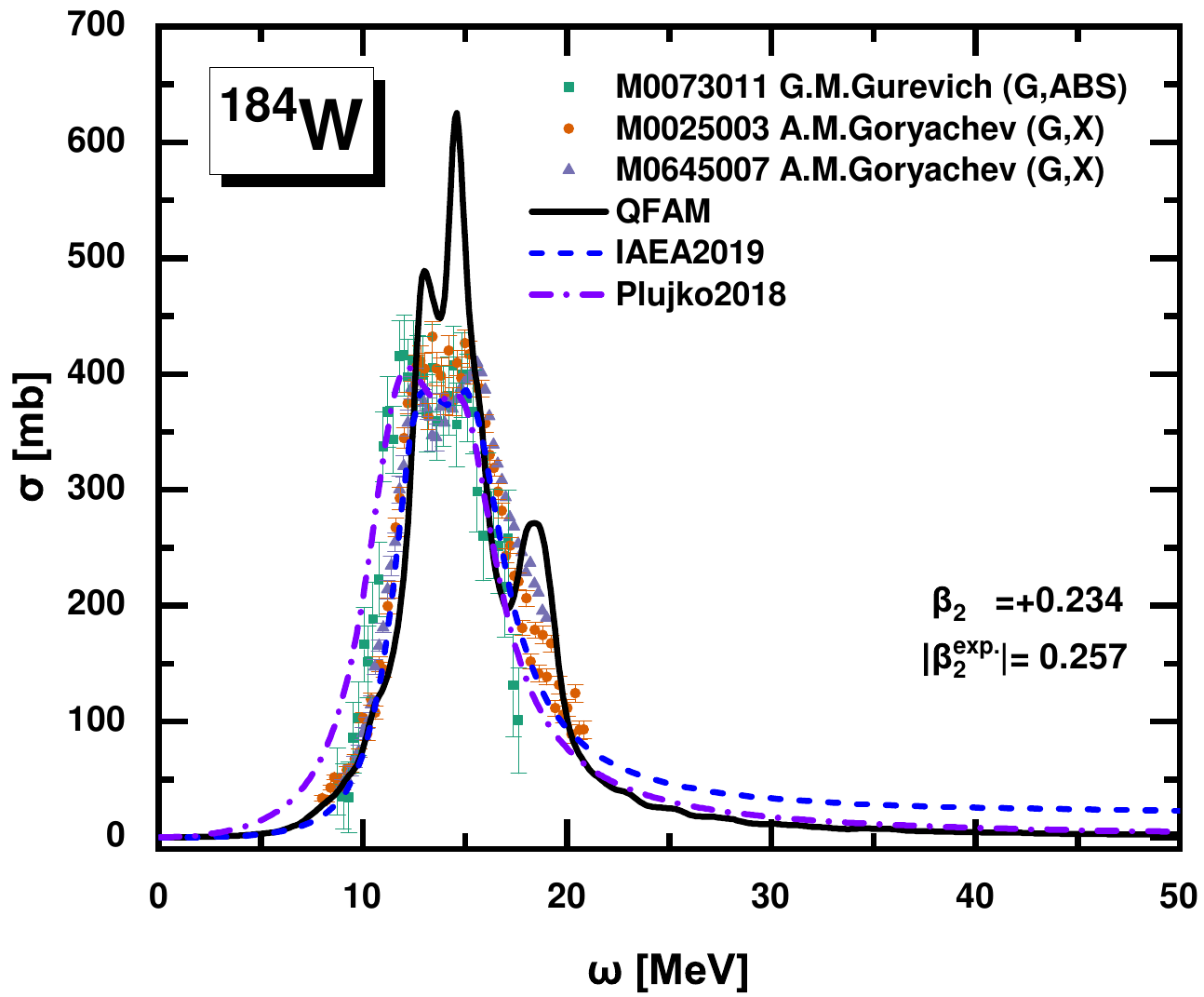}
    \includegraphics[width=0.35\textwidth]{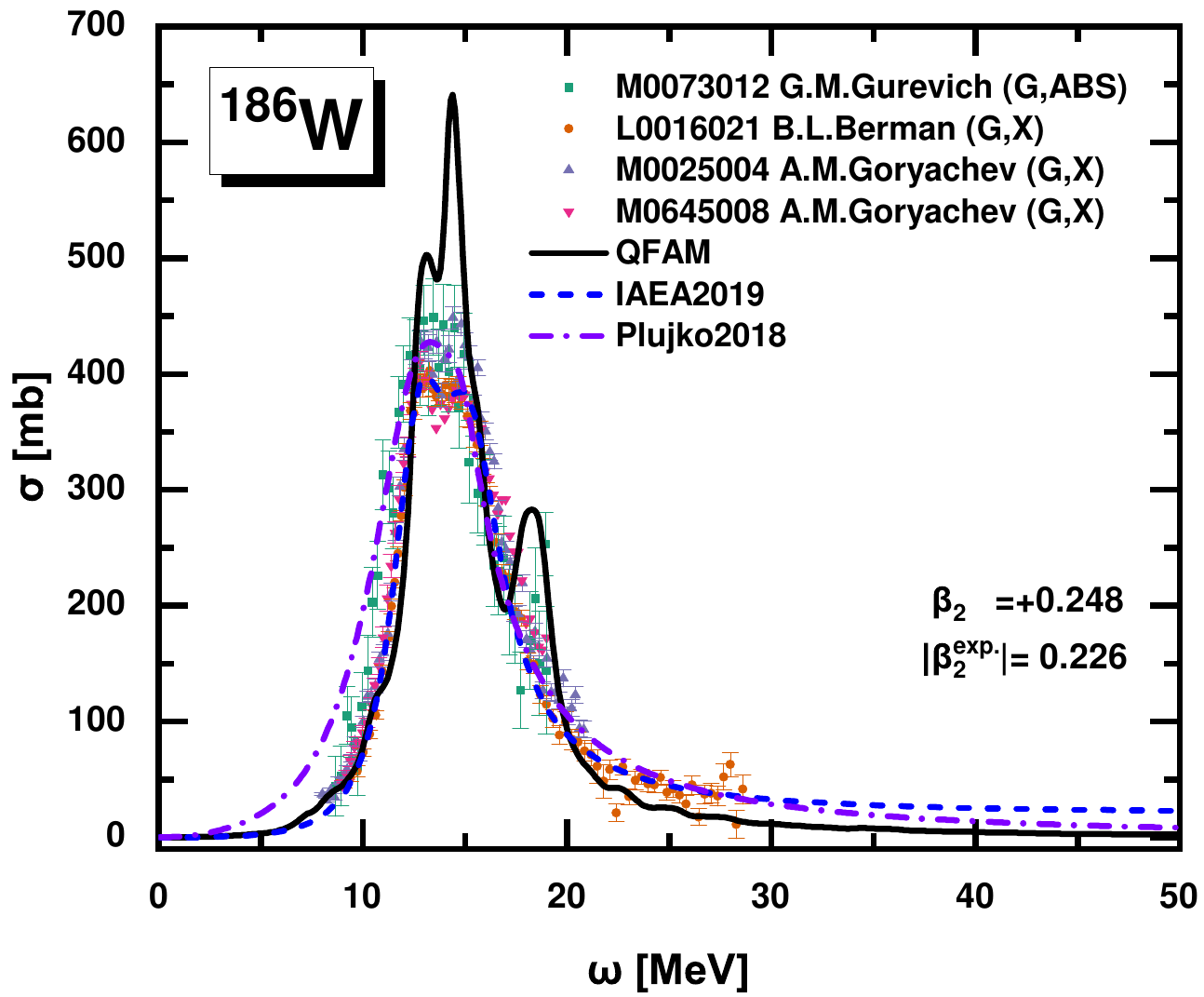}
    \includegraphics[width=0.35\textwidth]{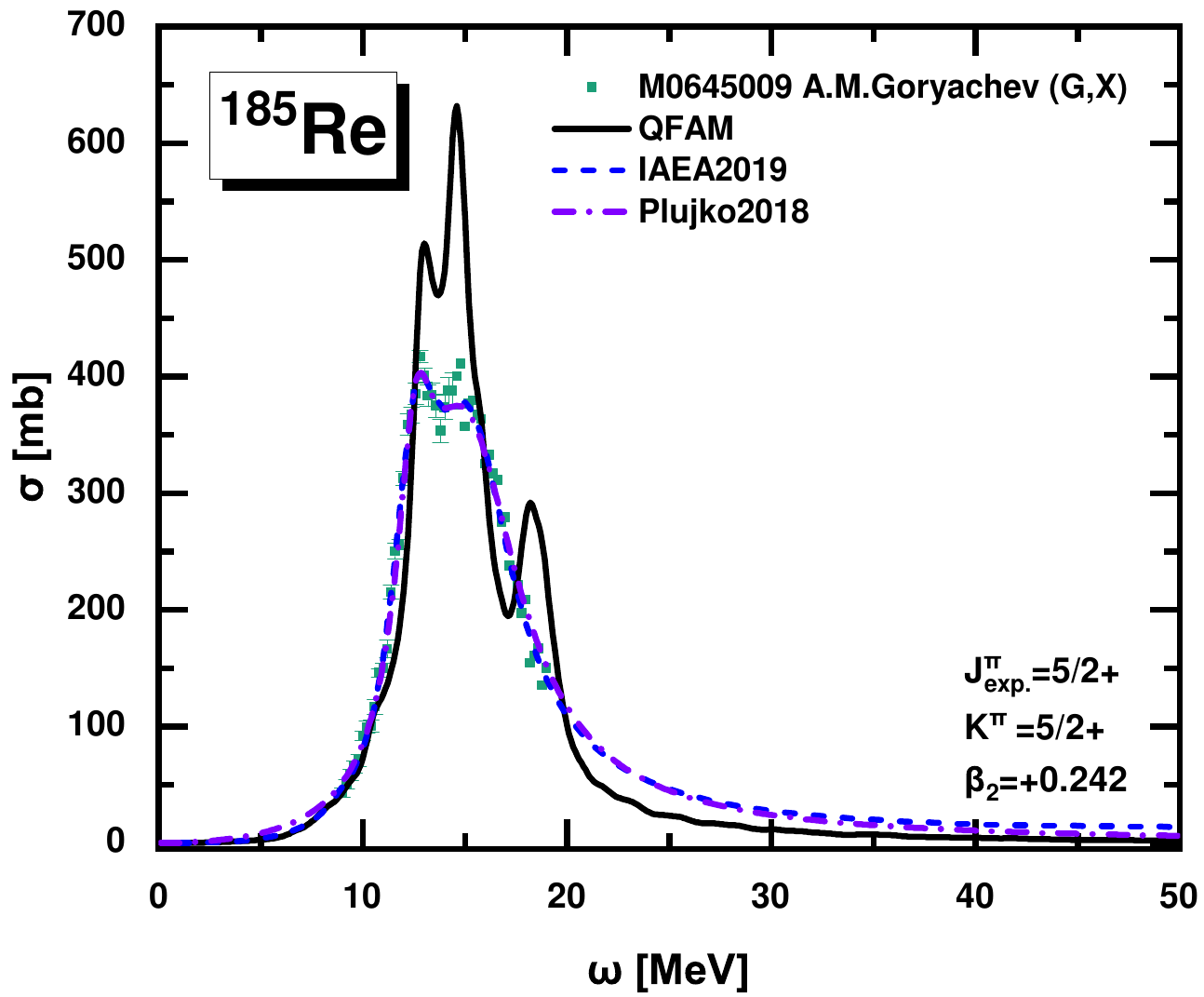}
    \includegraphics[width=0.35\textwidth]{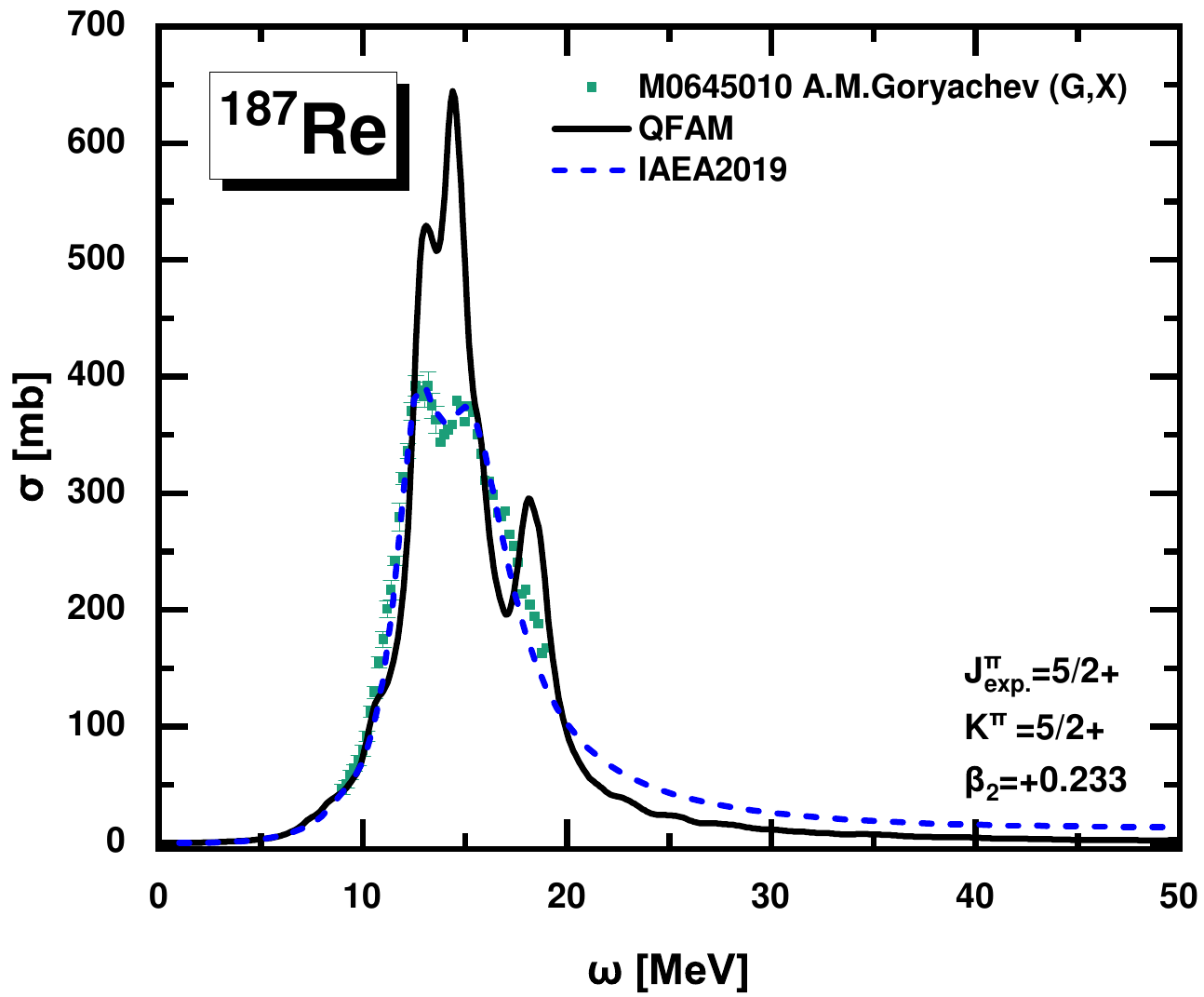}
    \includegraphics[width=0.35\textwidth]{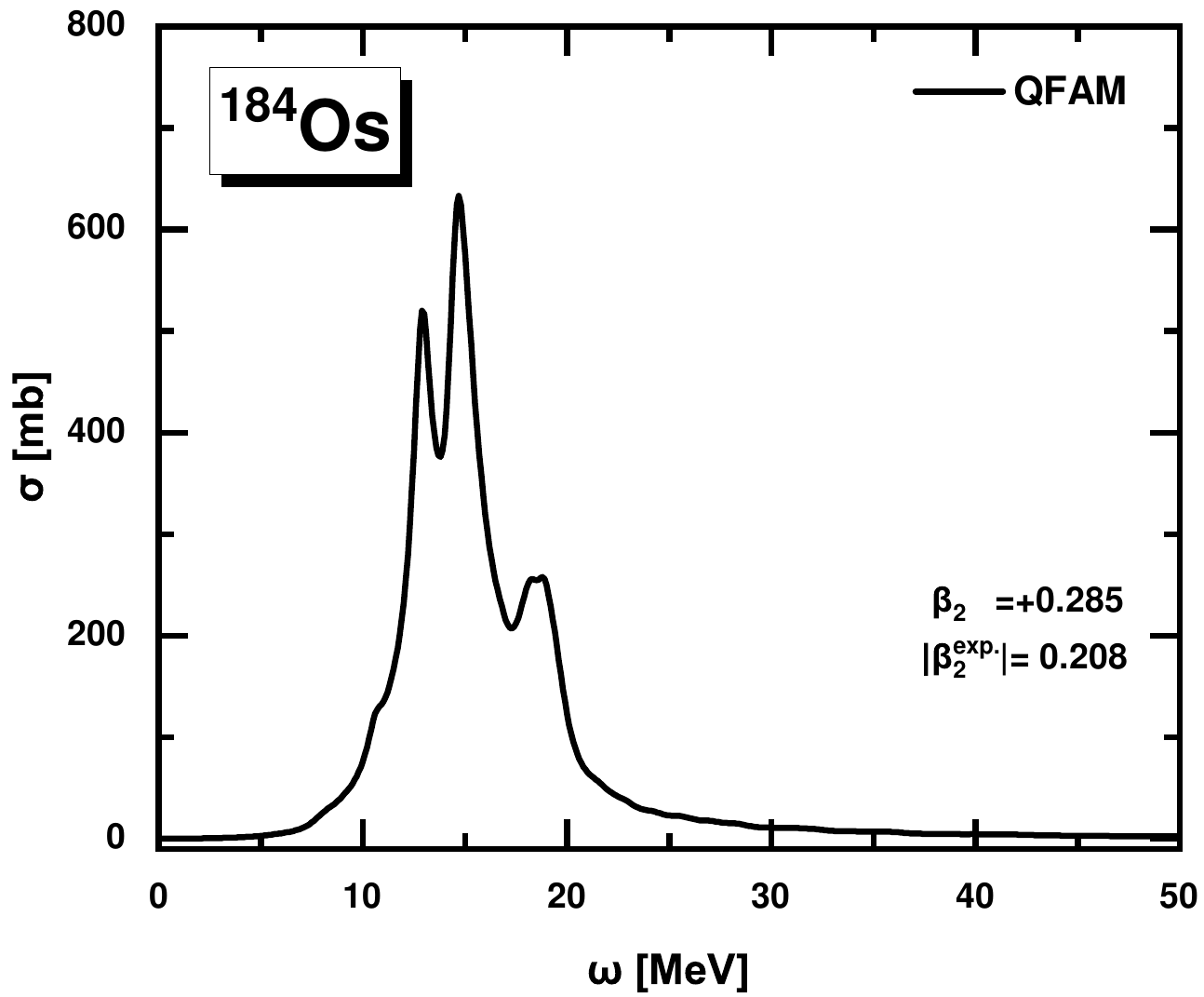}
    \includegraphics[width=0.35\textwidth]{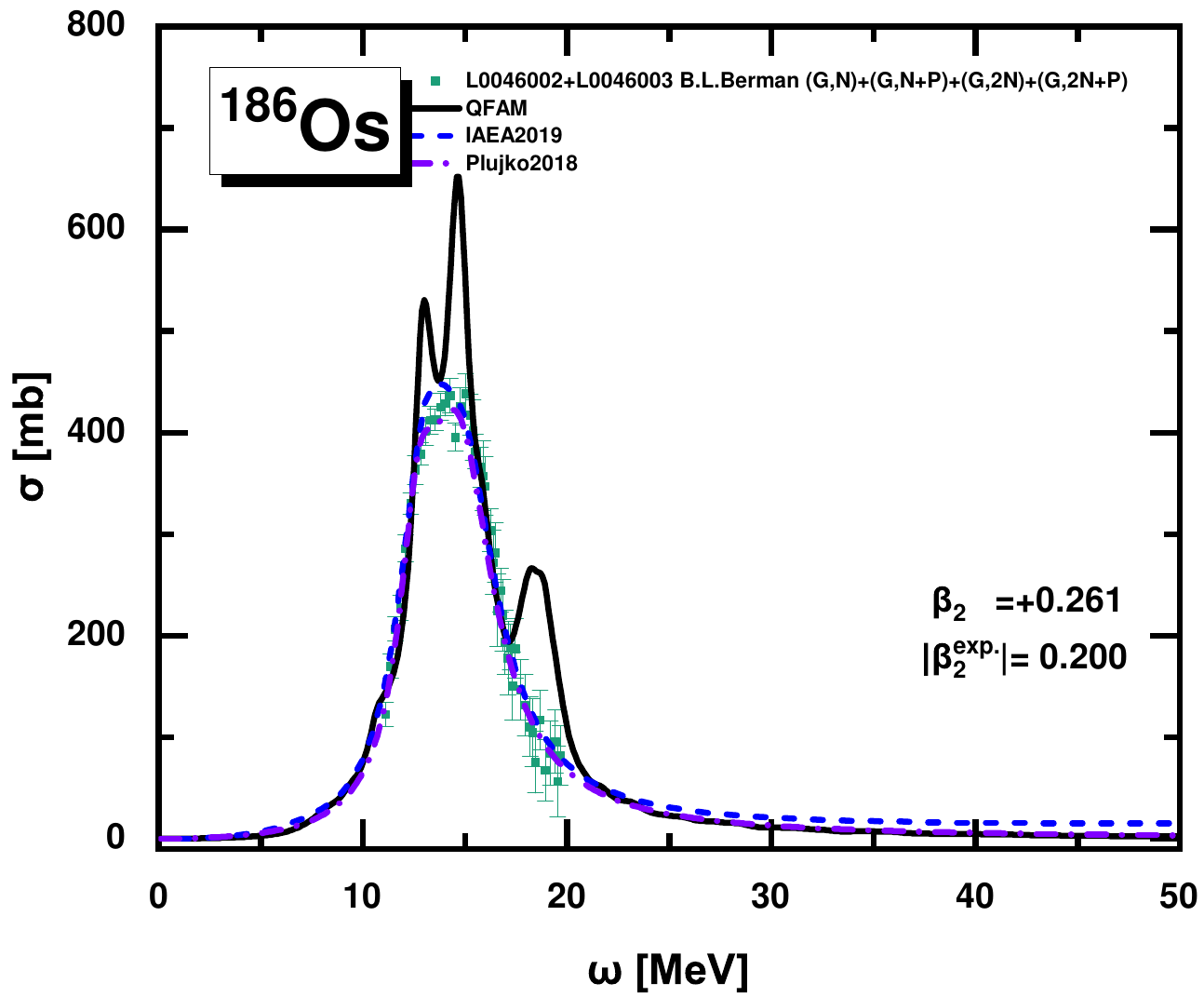}
\end{figure*}
\begin{figure*}\ContinuedFloat
    \centering
    \includegraphics[width=0.35\textwidth]{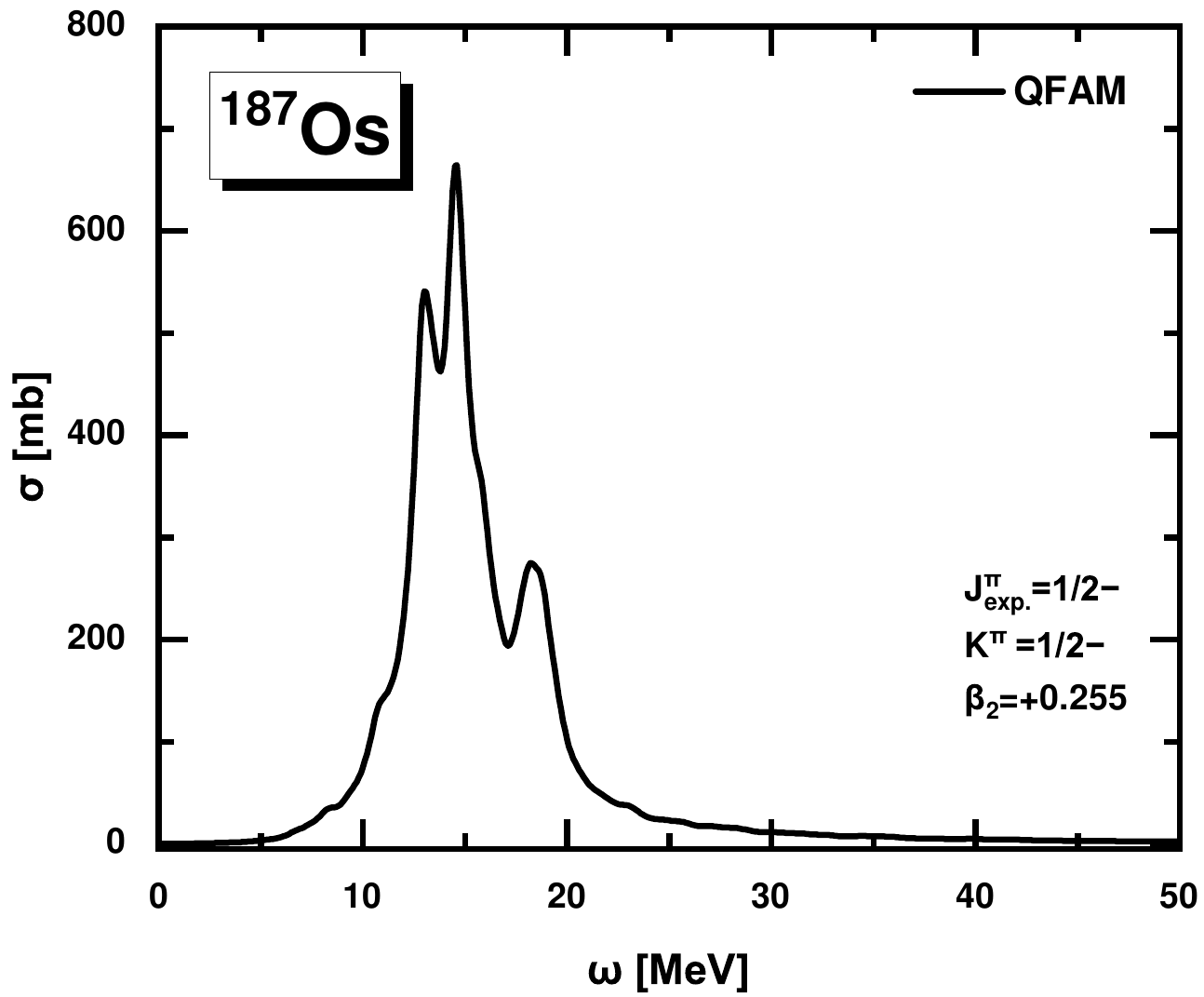}
    \includegraphics[width=0.35\textwidth]{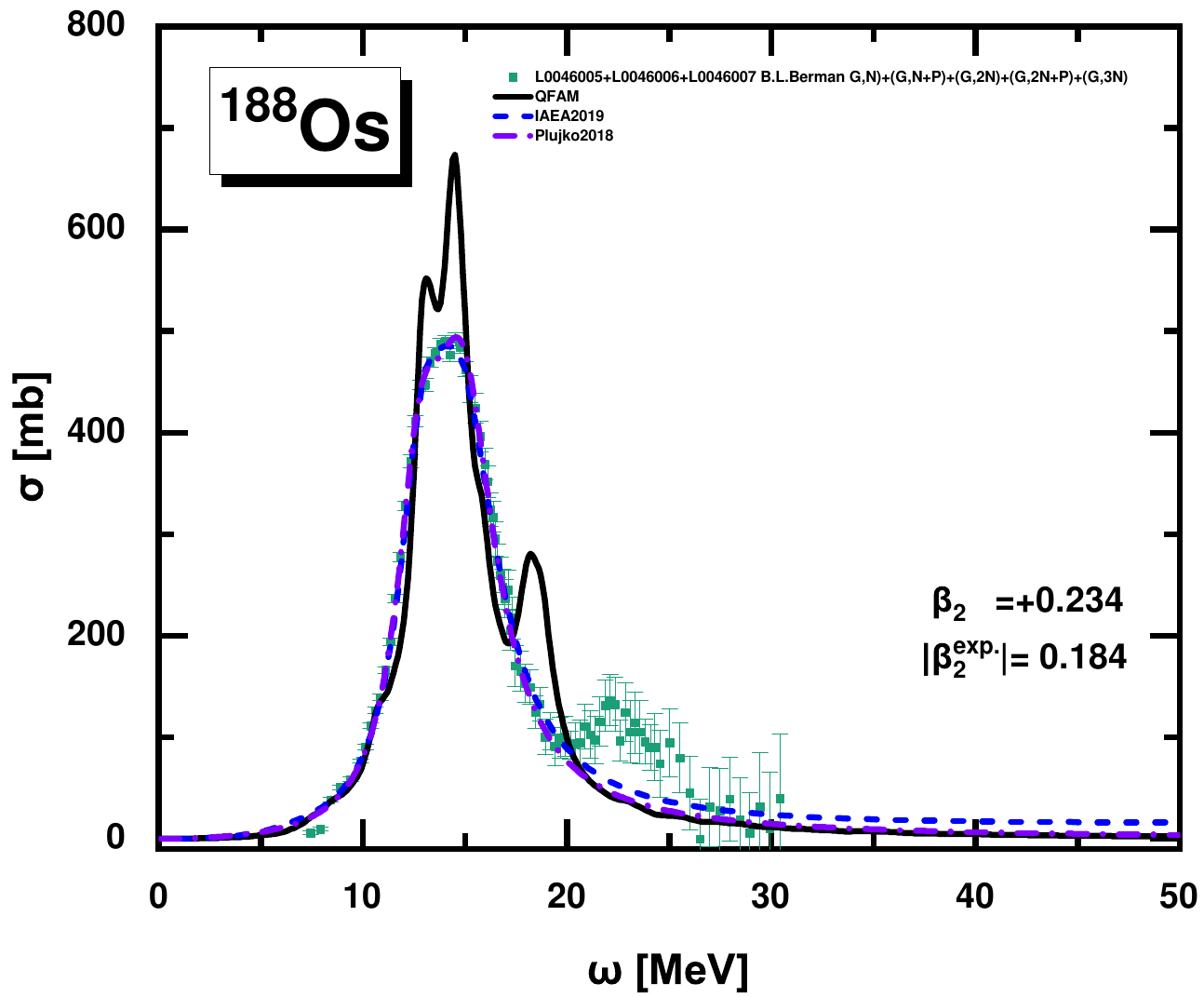}
    \includegraphics[width=0.35\textwidth]{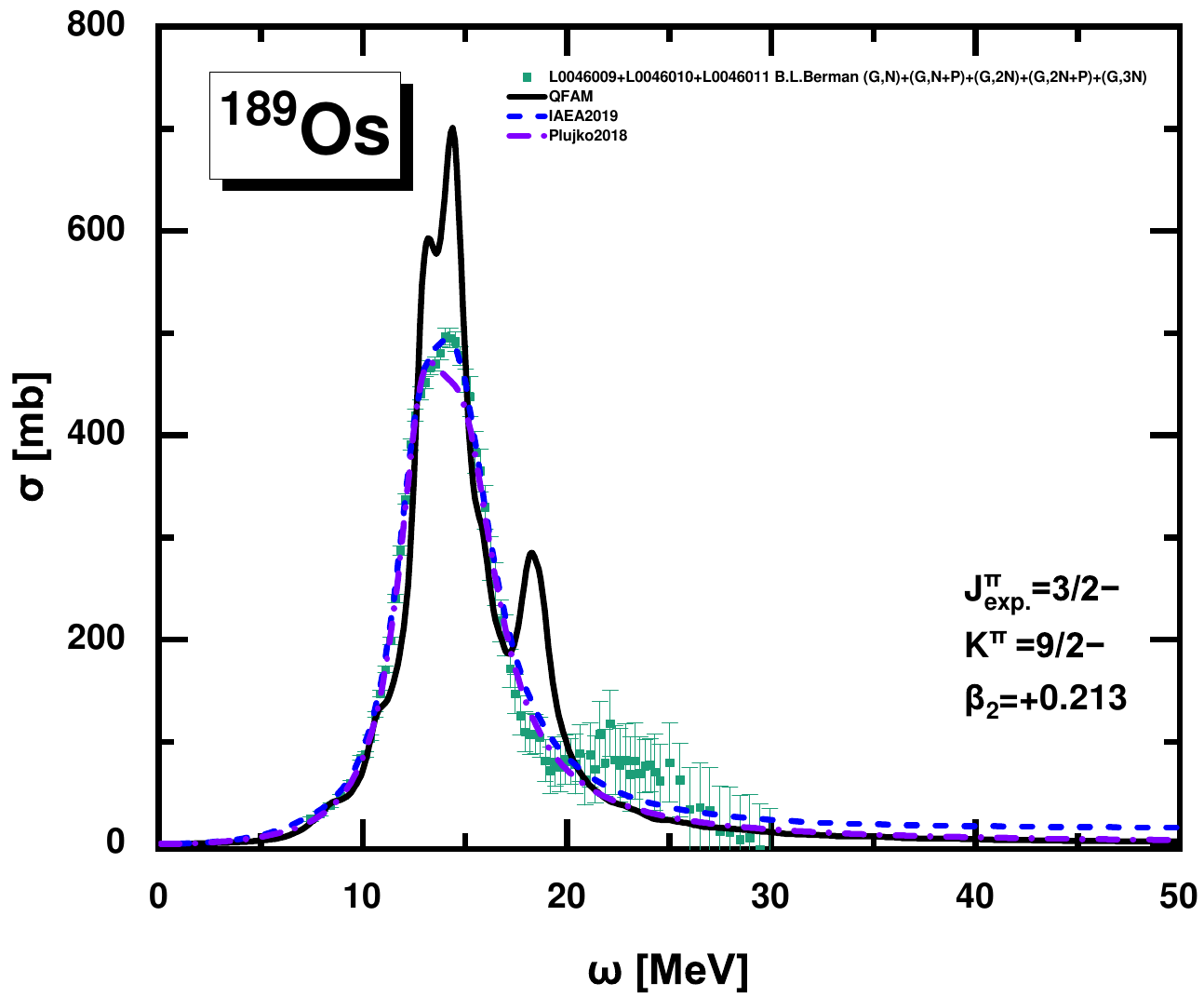}
    \includegraphics[width=0.35\textwidth]{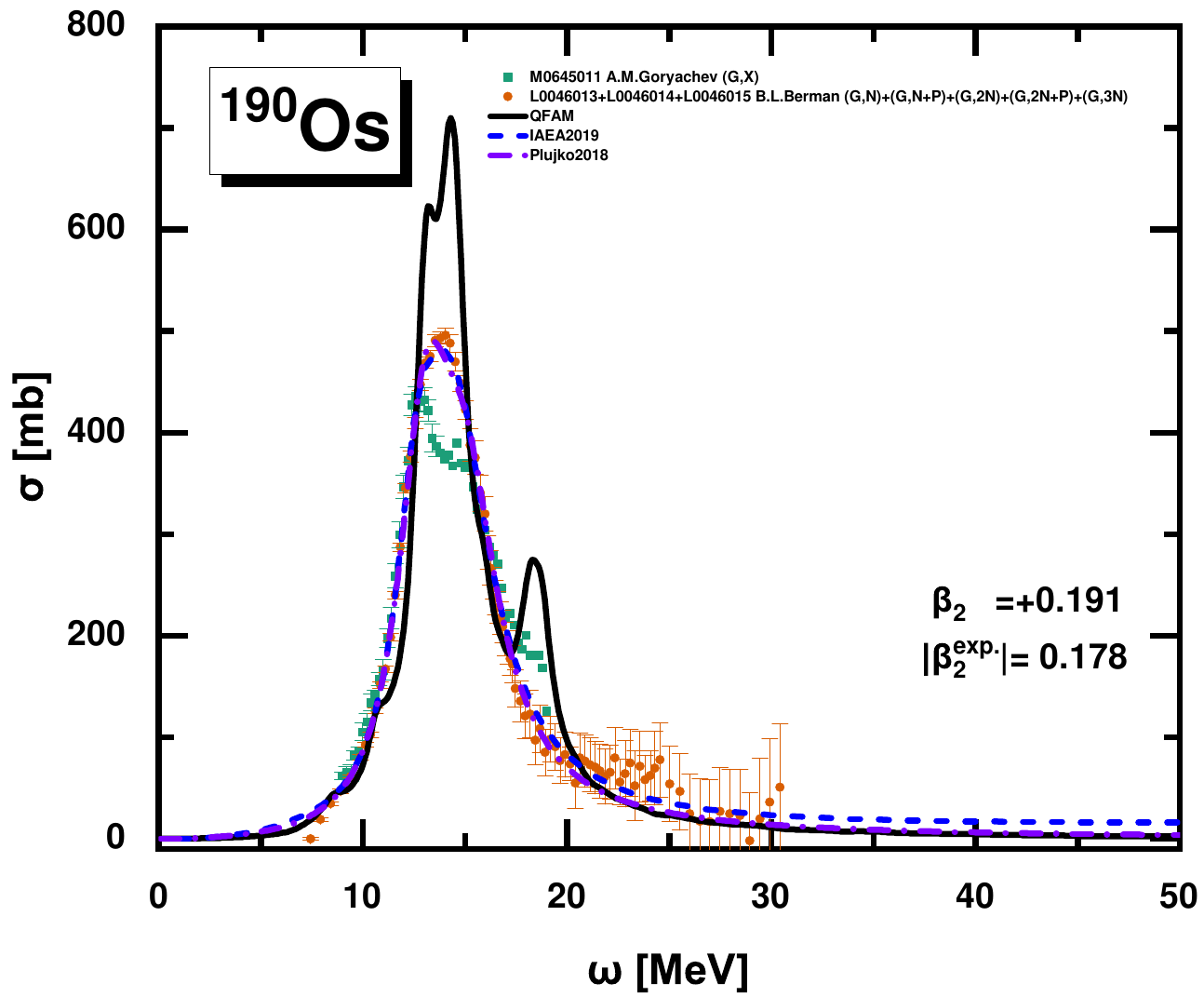}
    \includegraphics[width=0.35\textwidth]{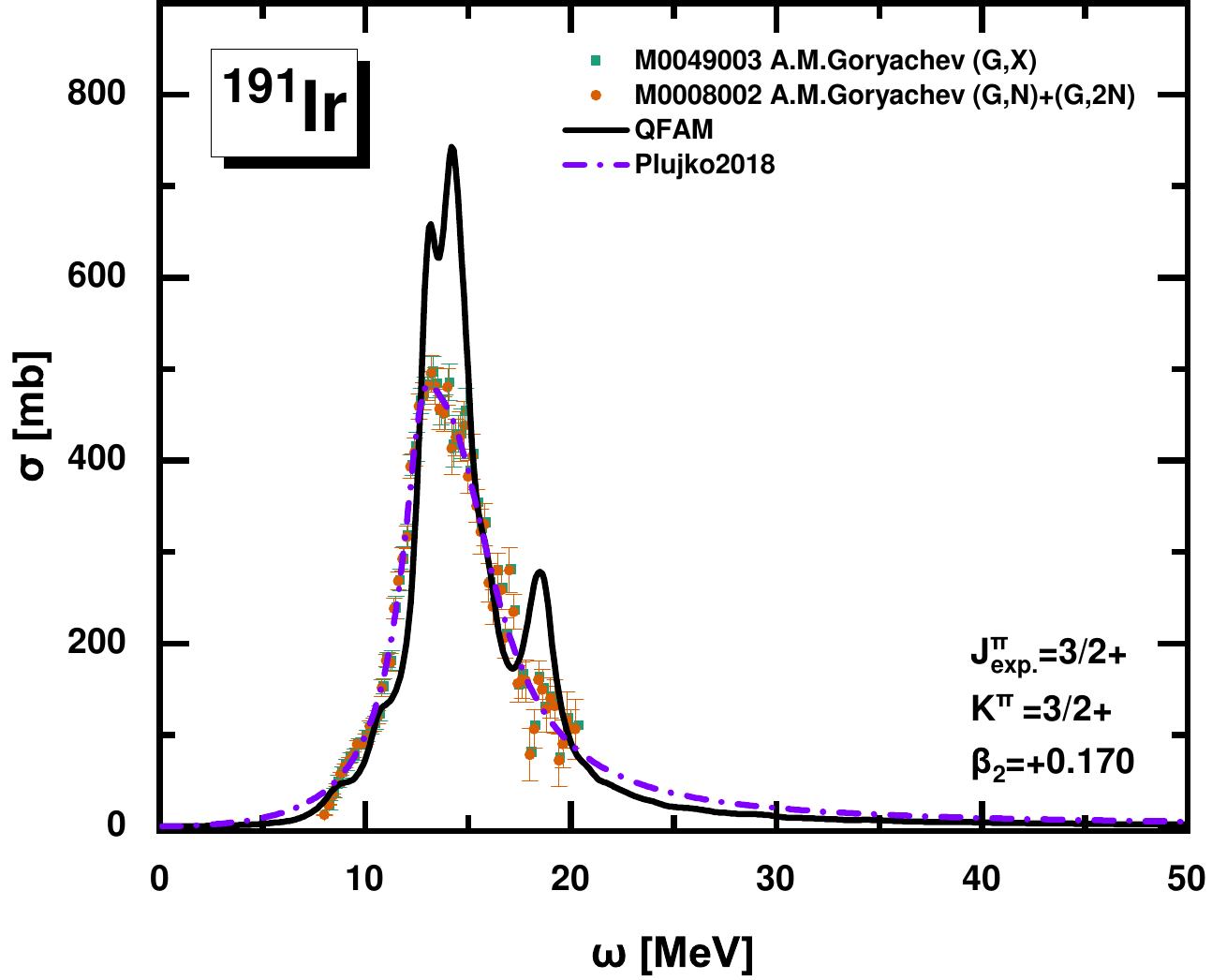}
    \includegraphics[width=0.35\textwidth]{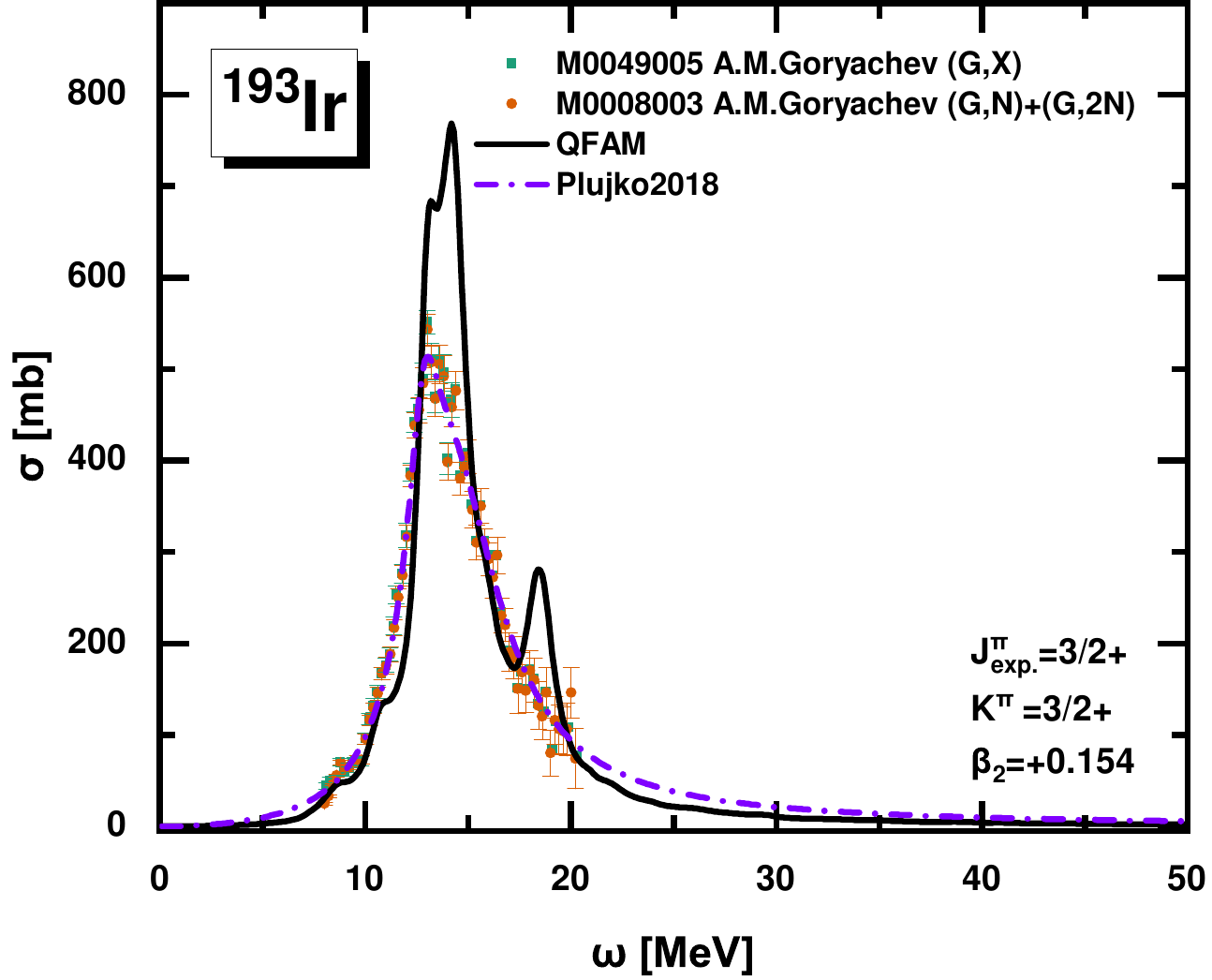}
    \includegraphics[width=0.35\textwidth]{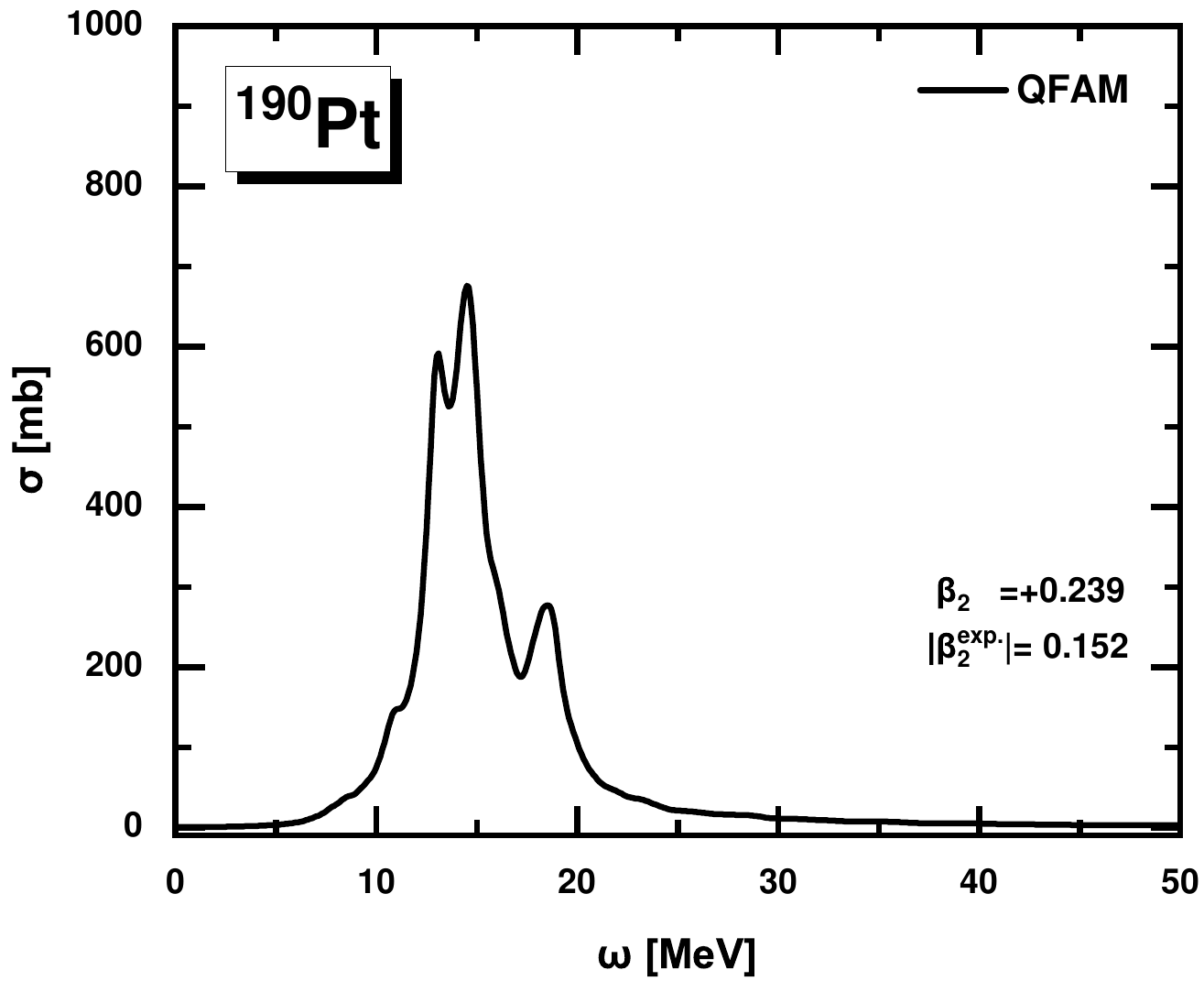}
    \includegraphics[width=0.35\textwidth]{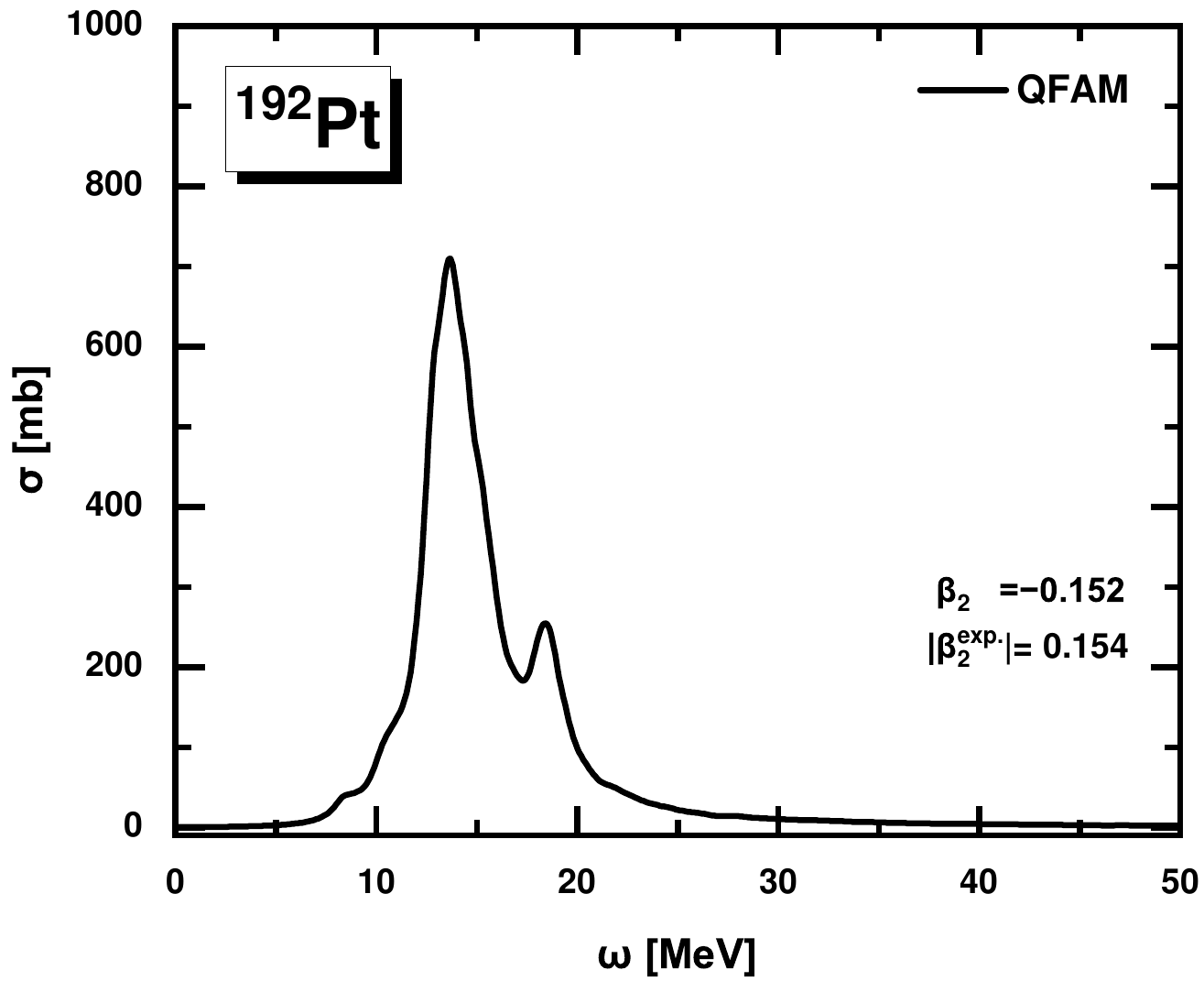}
\end{figure*}
\begin{figure*}\ContinuedFloat
    \centering
    \includegraphics[width=0.35\textwidth]{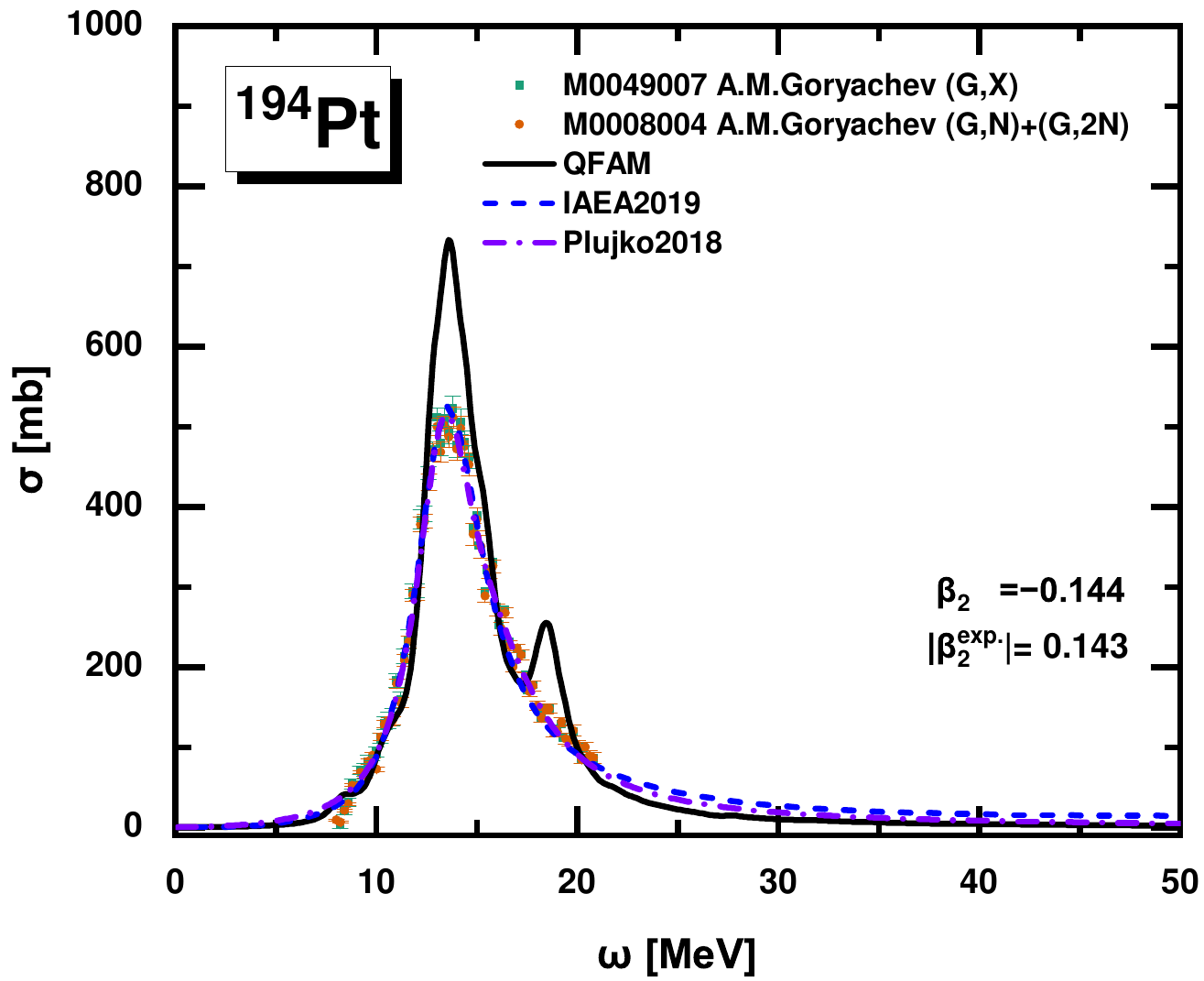}
    \includegraphics[width=0.35\textwidth]{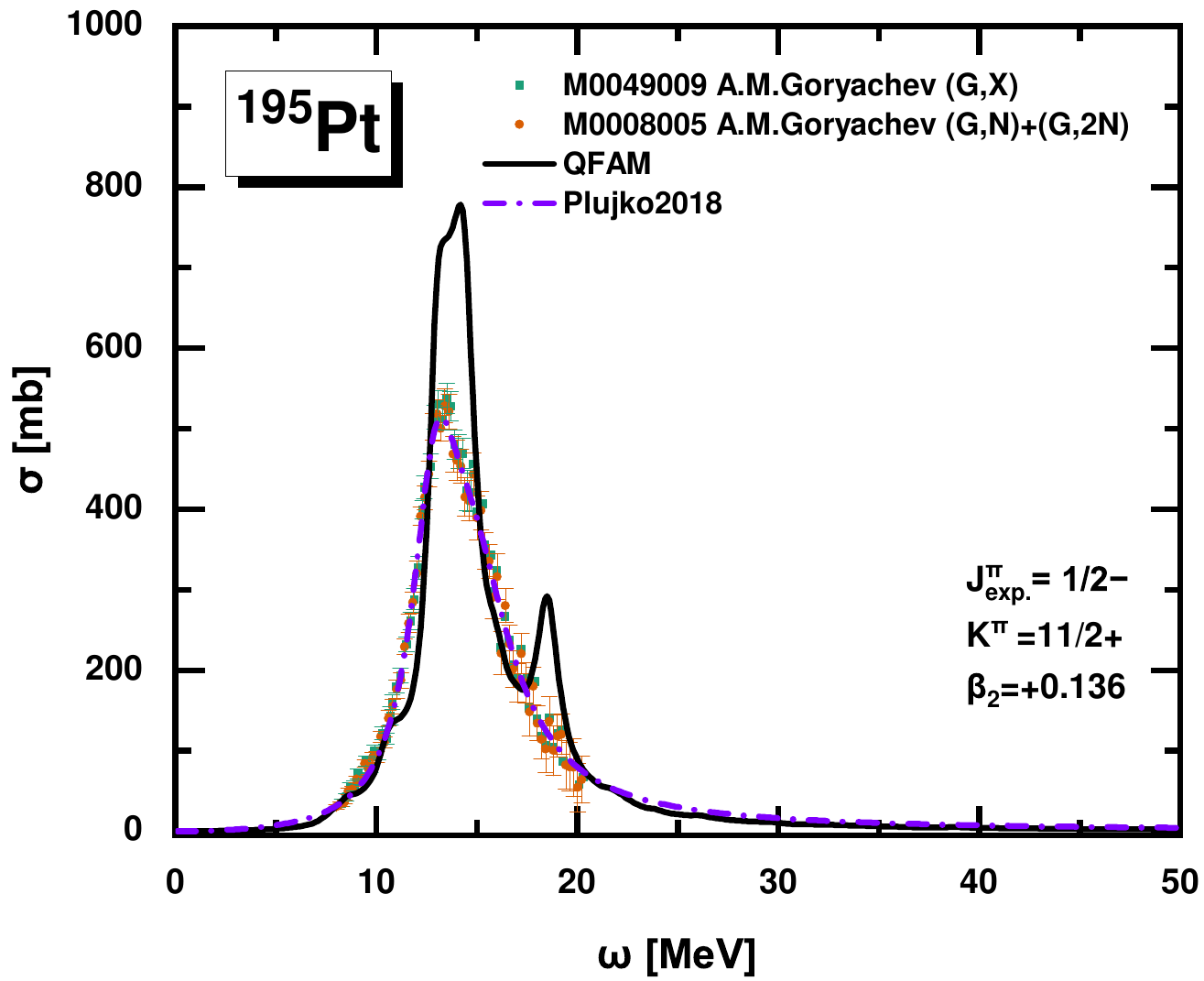}
    \includegraphics[width=0.35\textwidth]{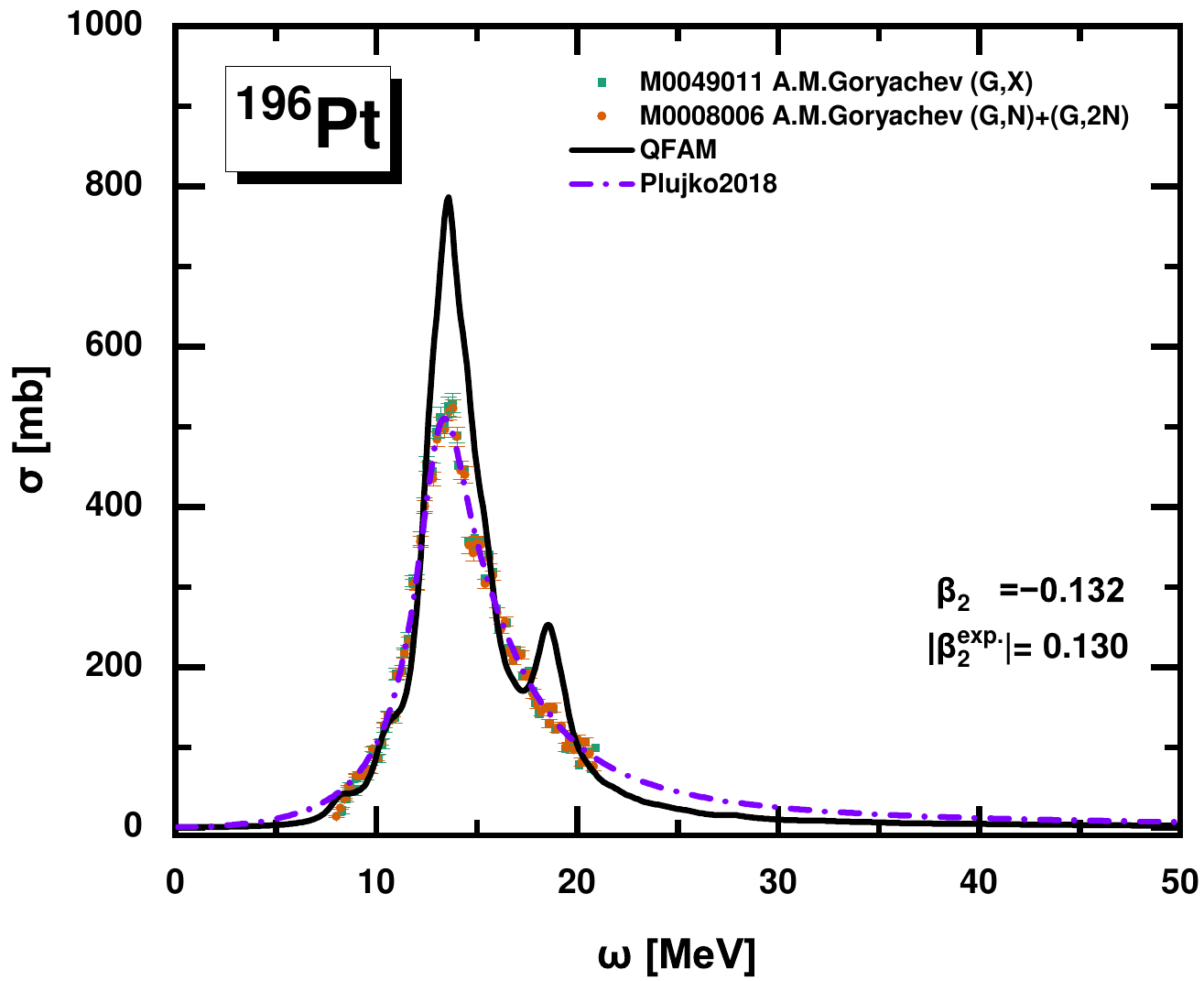}
    \includegraphics[width=0.35\textwidth]{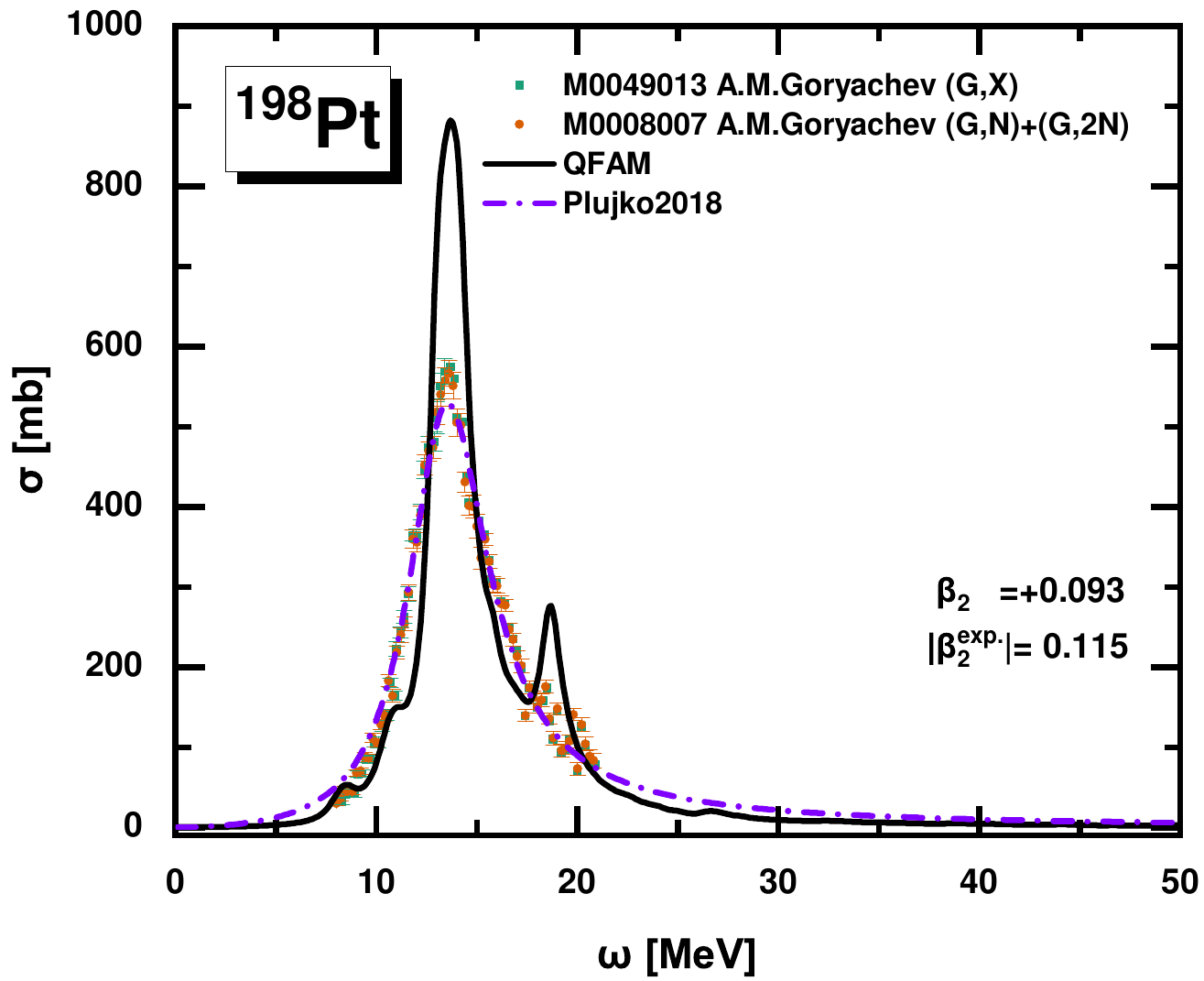}
    \includegraphics[width=0.35\textwidth]{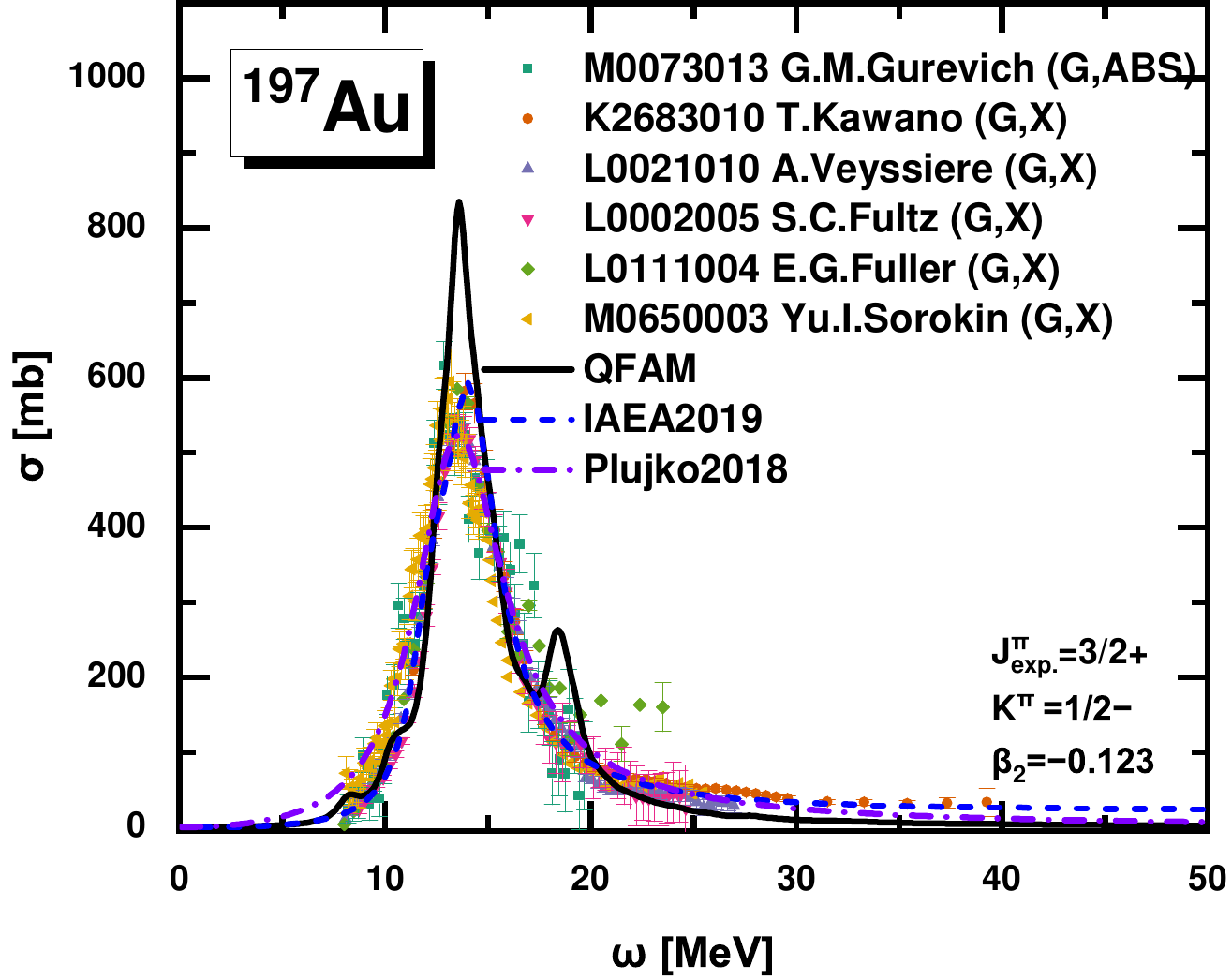}
    \includegraphics[width=0.35\textwidth]{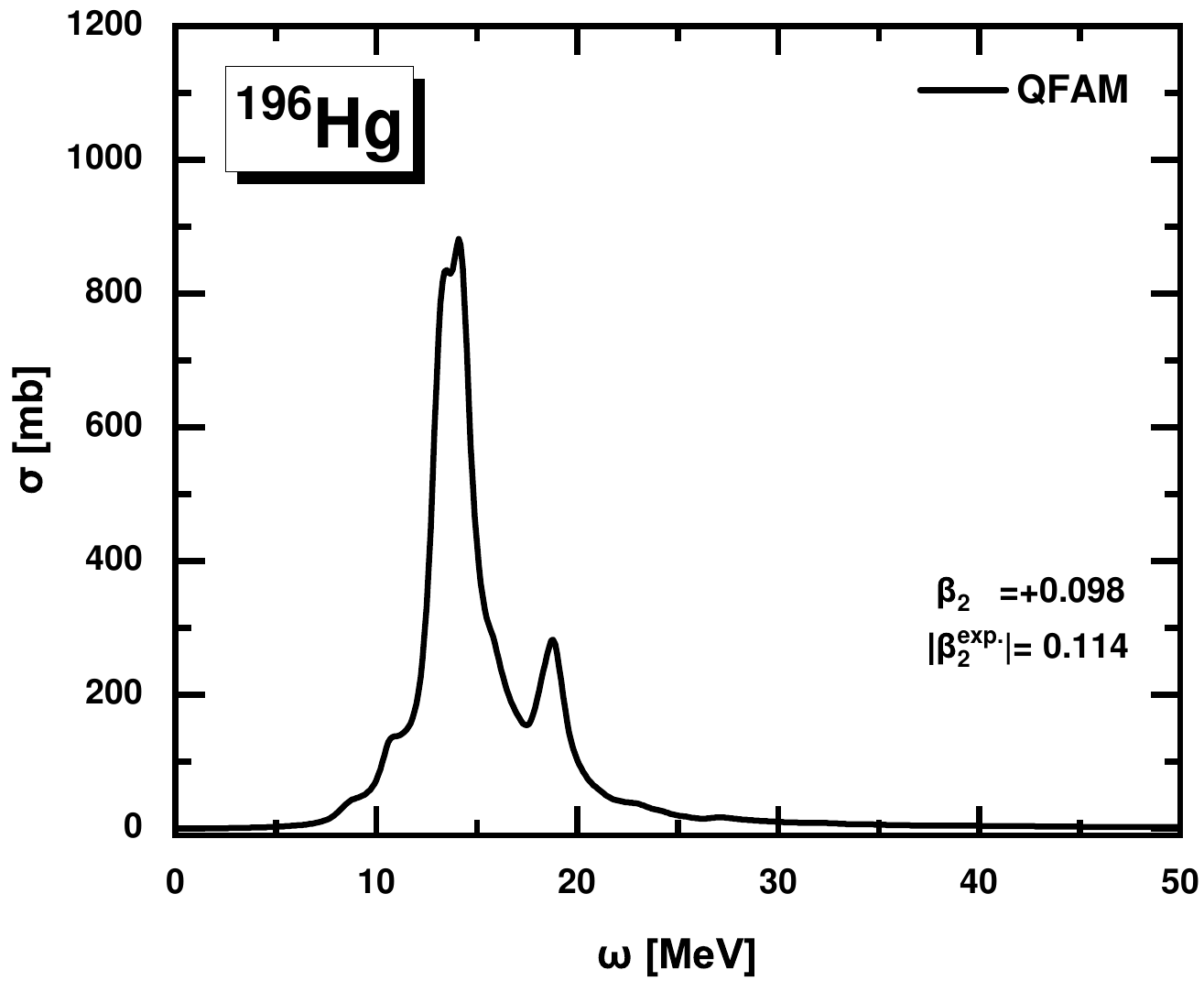}
    \includegraphics[width=0.35\textwidth]{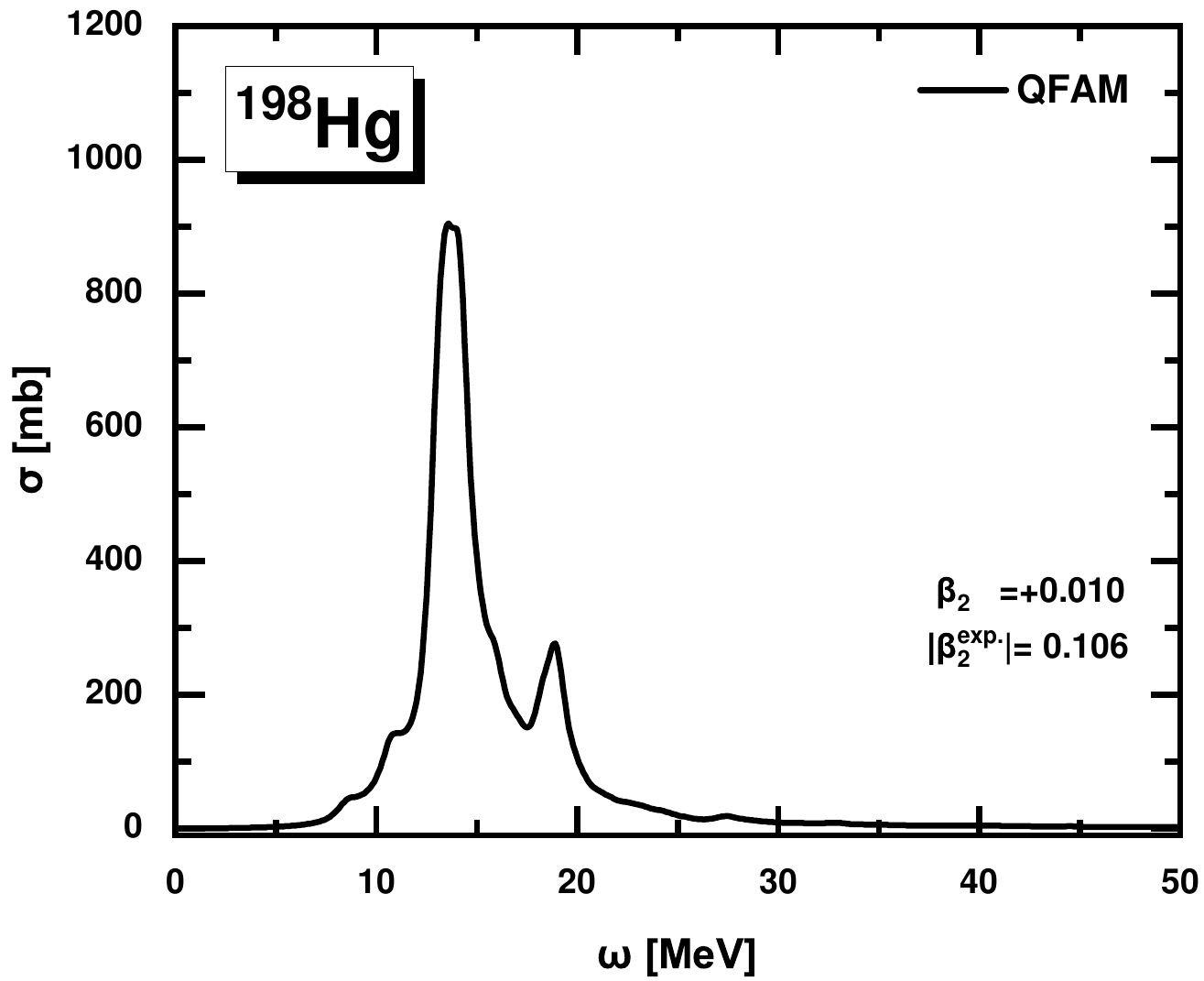}
    \includegraphics[width=0.35\textwidth]{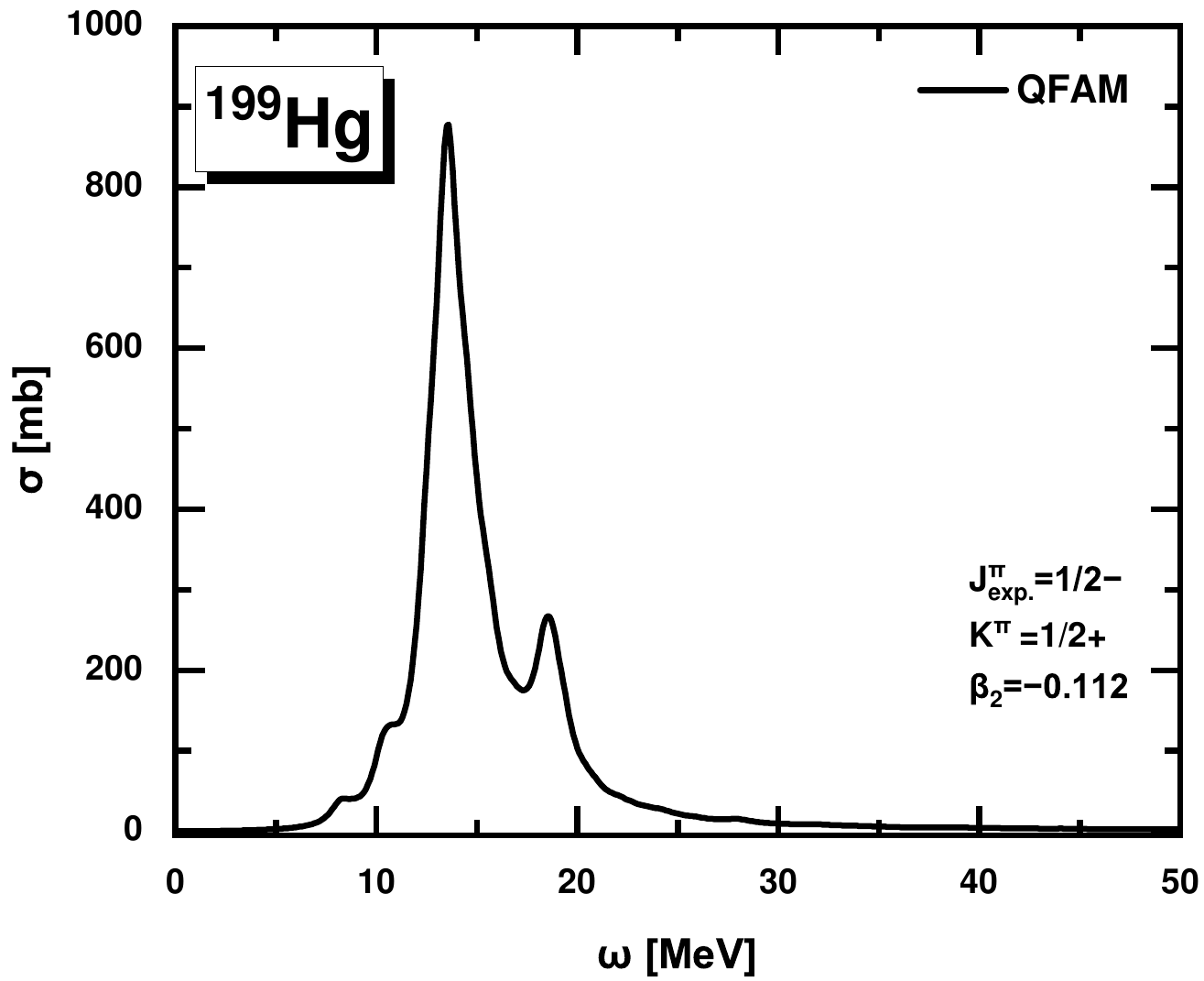}
\end{figure*}
\begin{figure*}\ContinuedFloat
    \centering
    \includegraphics[width=0.35\textwidth]{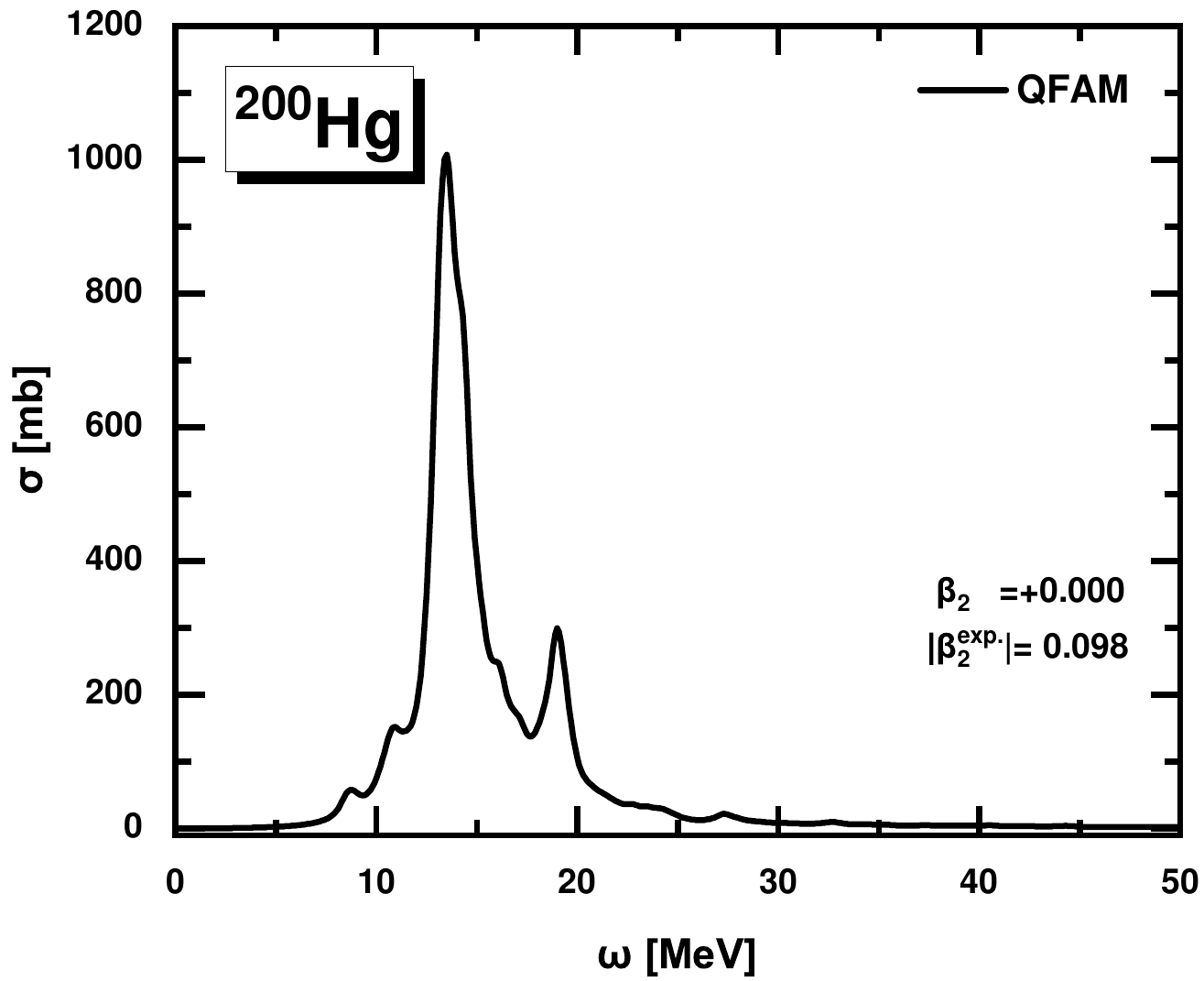}
    \includegraphics[width=0.35\textwidth]{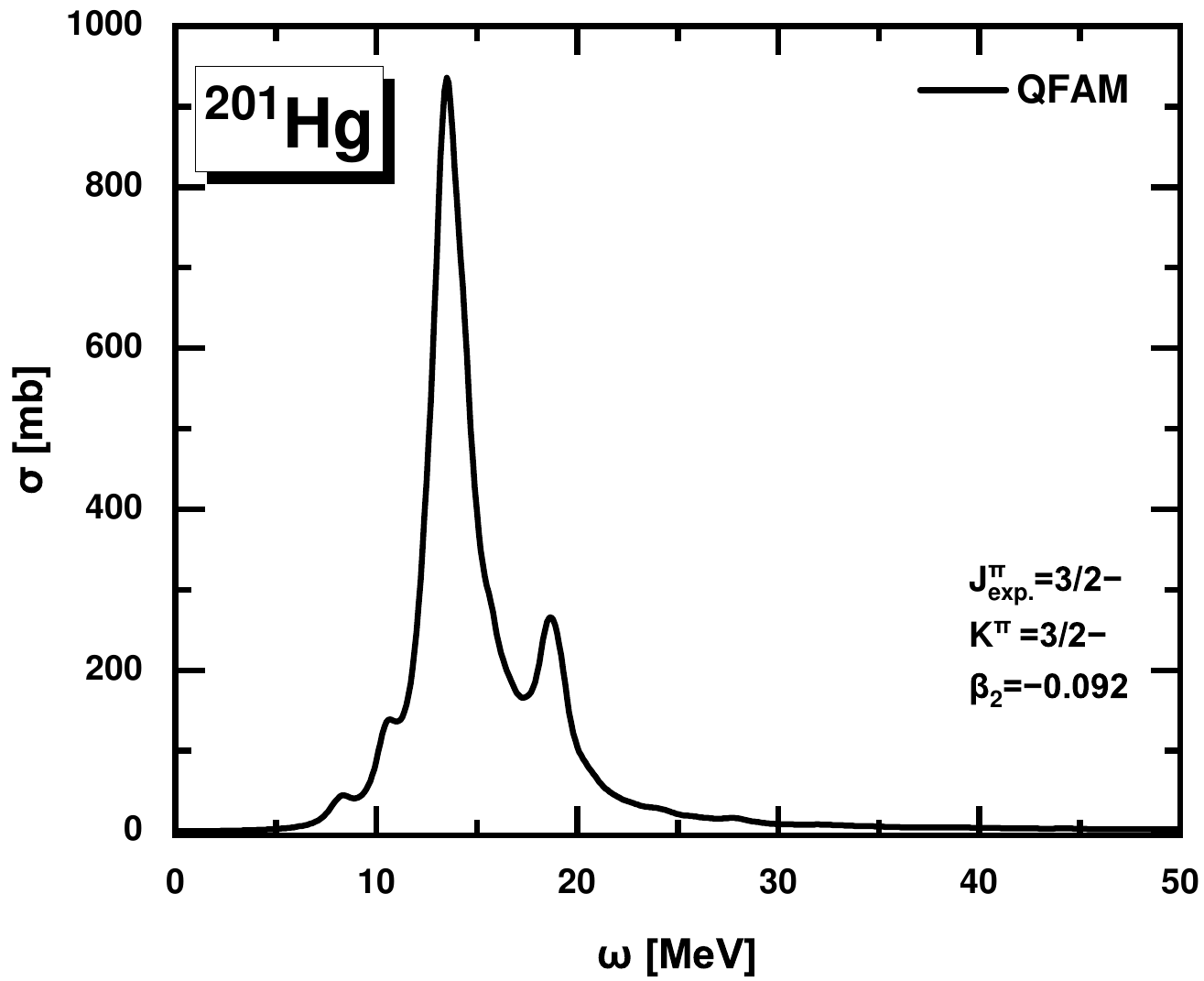}
    \includegraphics[width=0.35\textwidth]{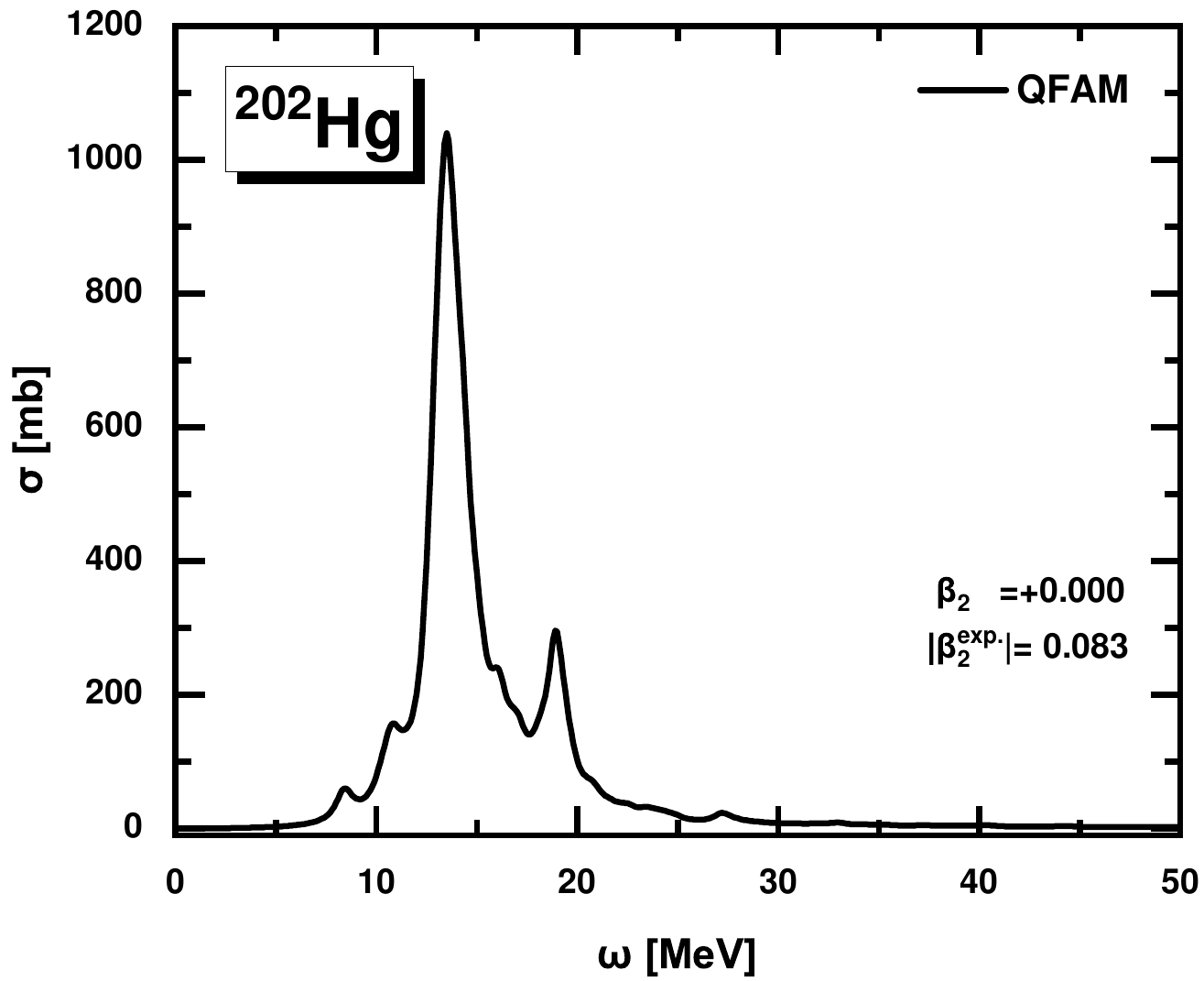}
    \includegraphics[width=0.35\textwidth]{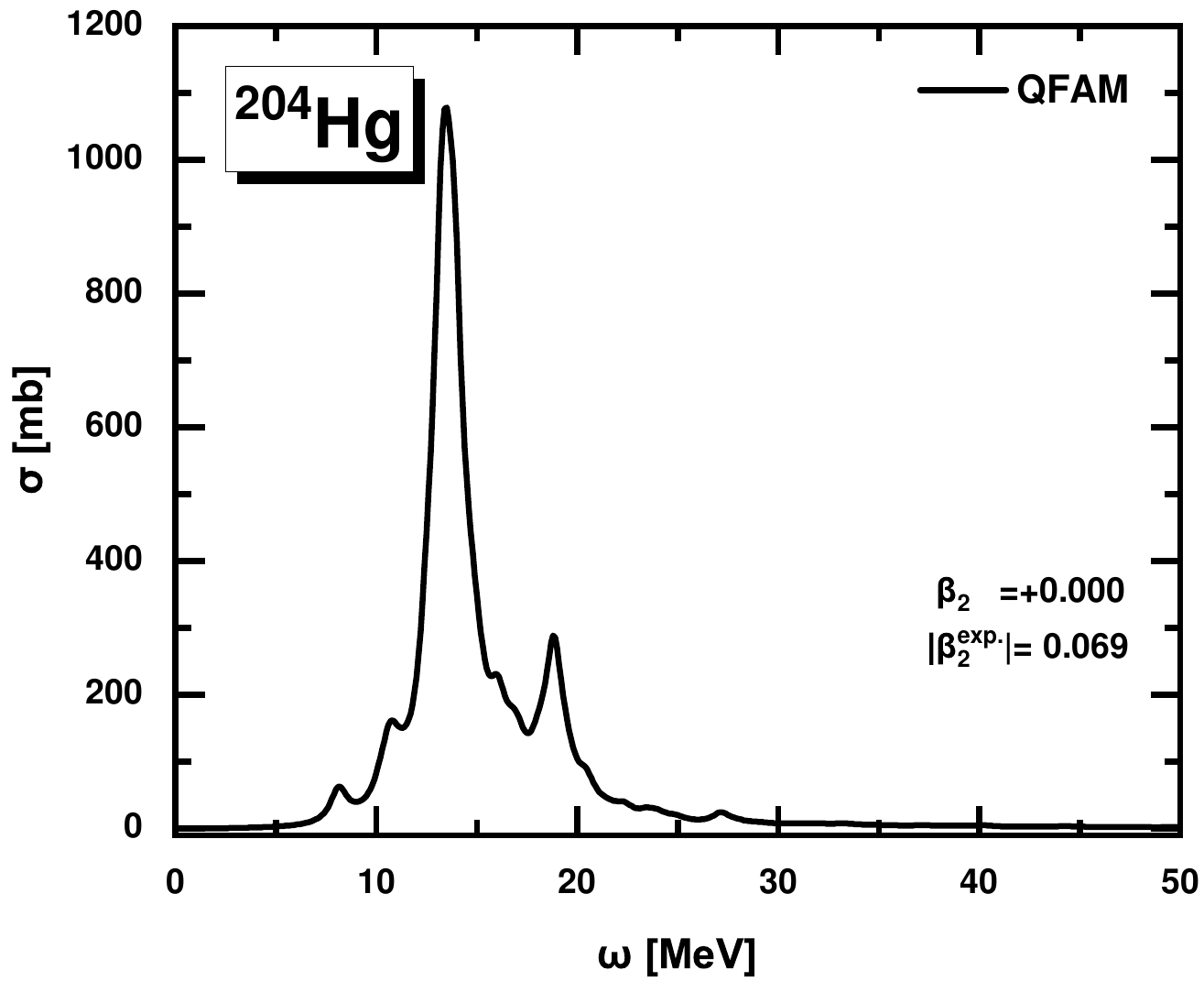}
    \includegraphics[width=0.35\textwidth]{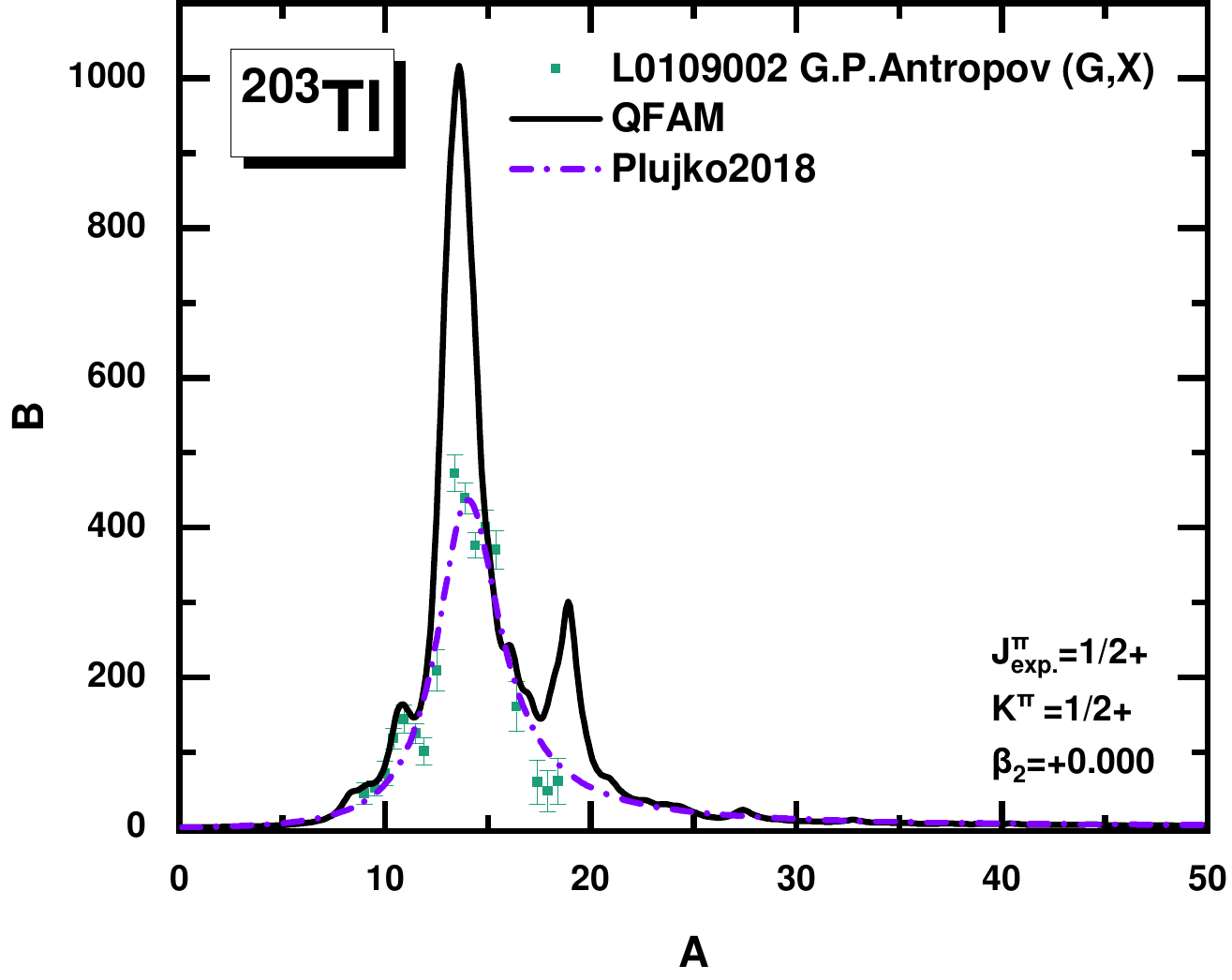}
    \includegraphics[width=0.35\textwidth]{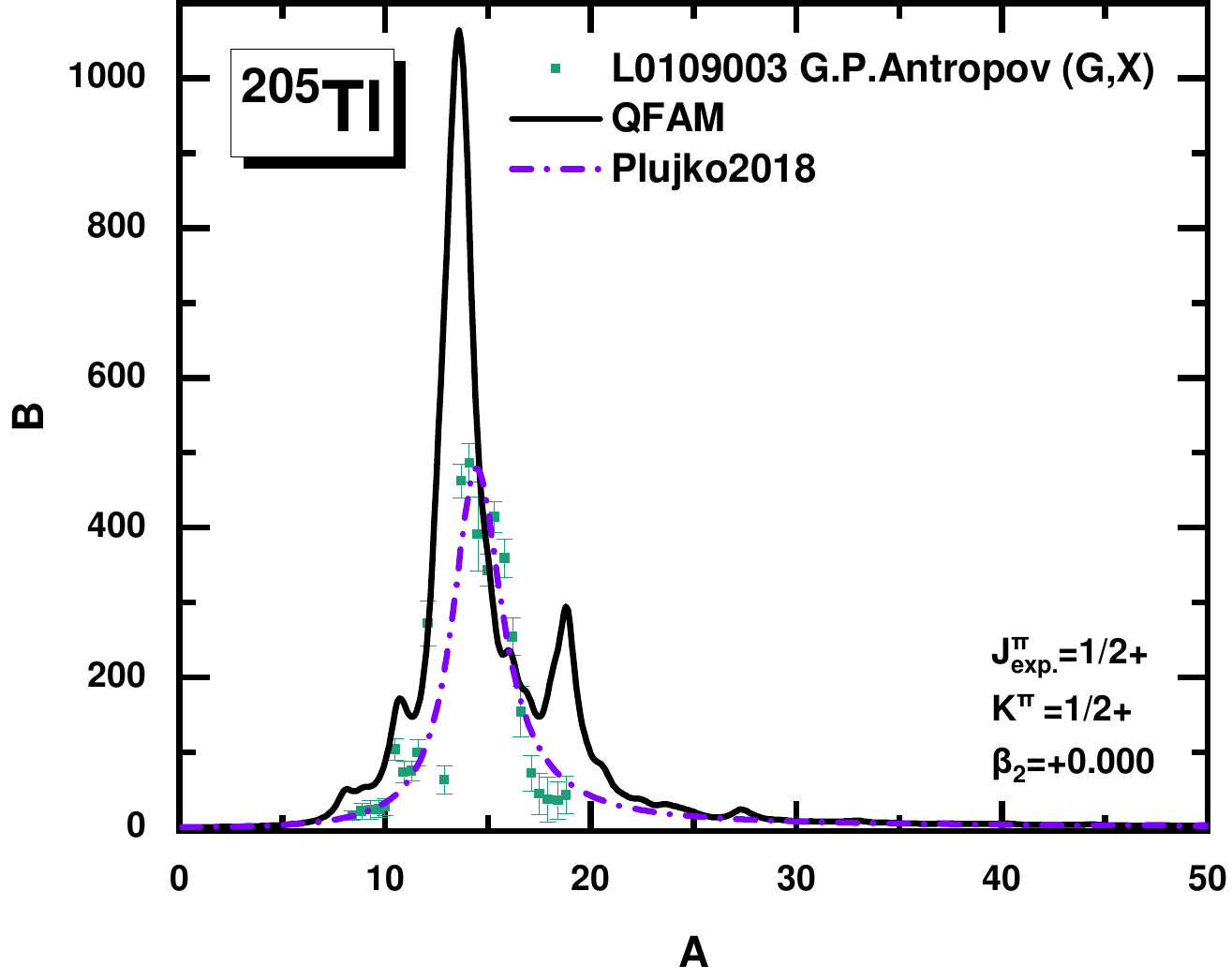}
    \includegraphics[width=0.35\textwidth]{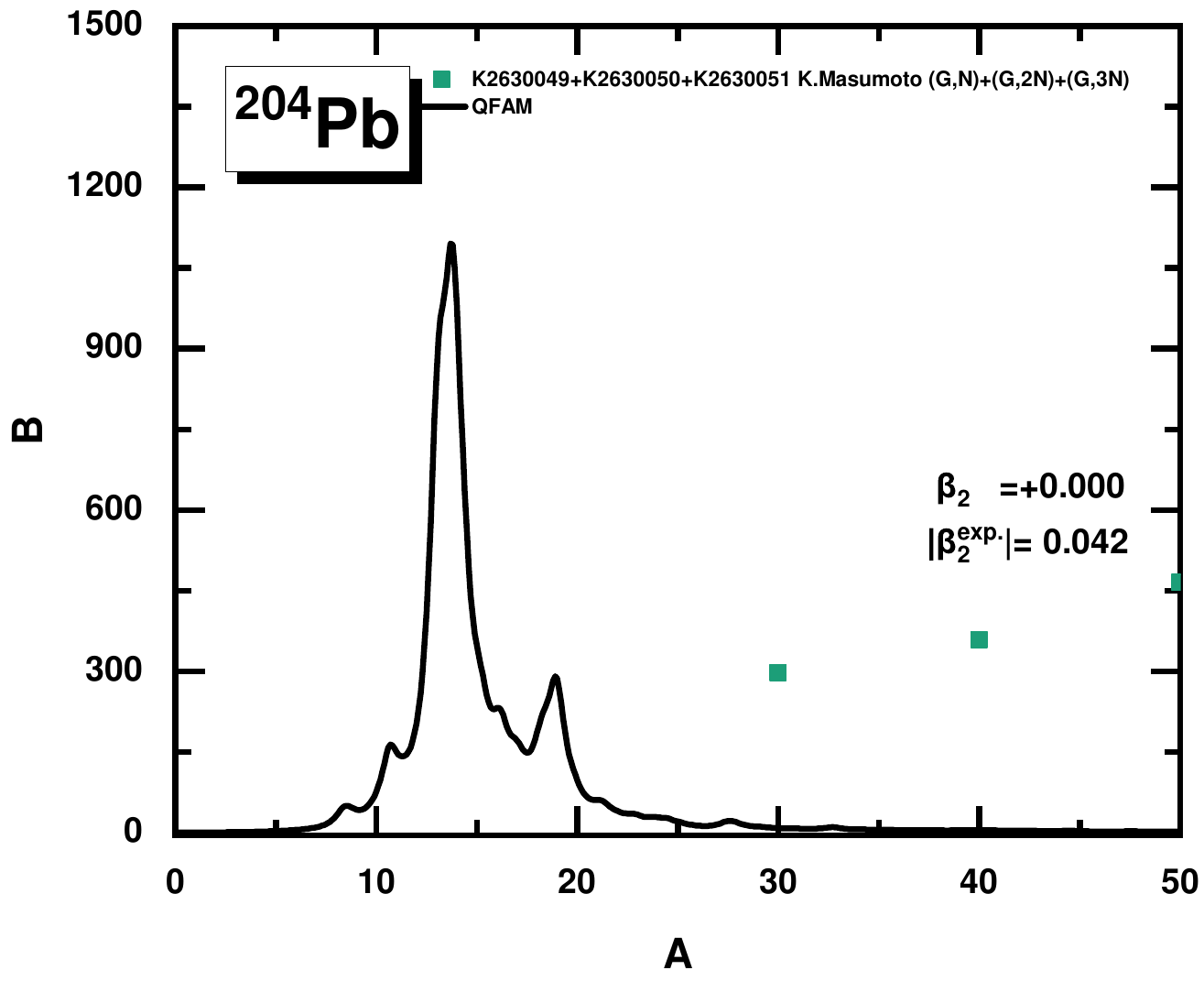}
    \includegraphics[width=0.35\textwidth]{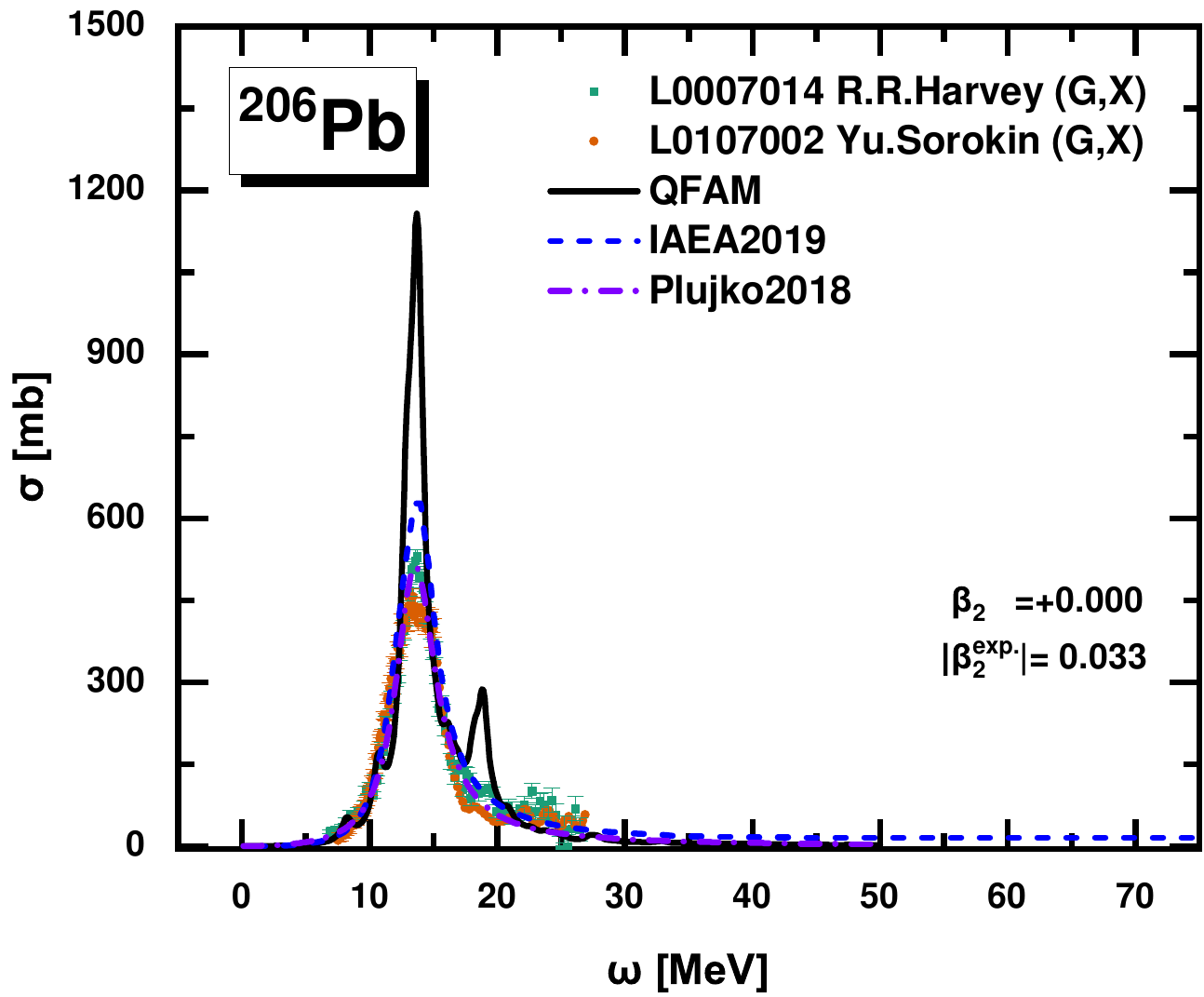}
\end{figure*}
\begin{figure*}\ContinuedFloat
    \centering
    \includegraphics[width=0.35\textwidth]{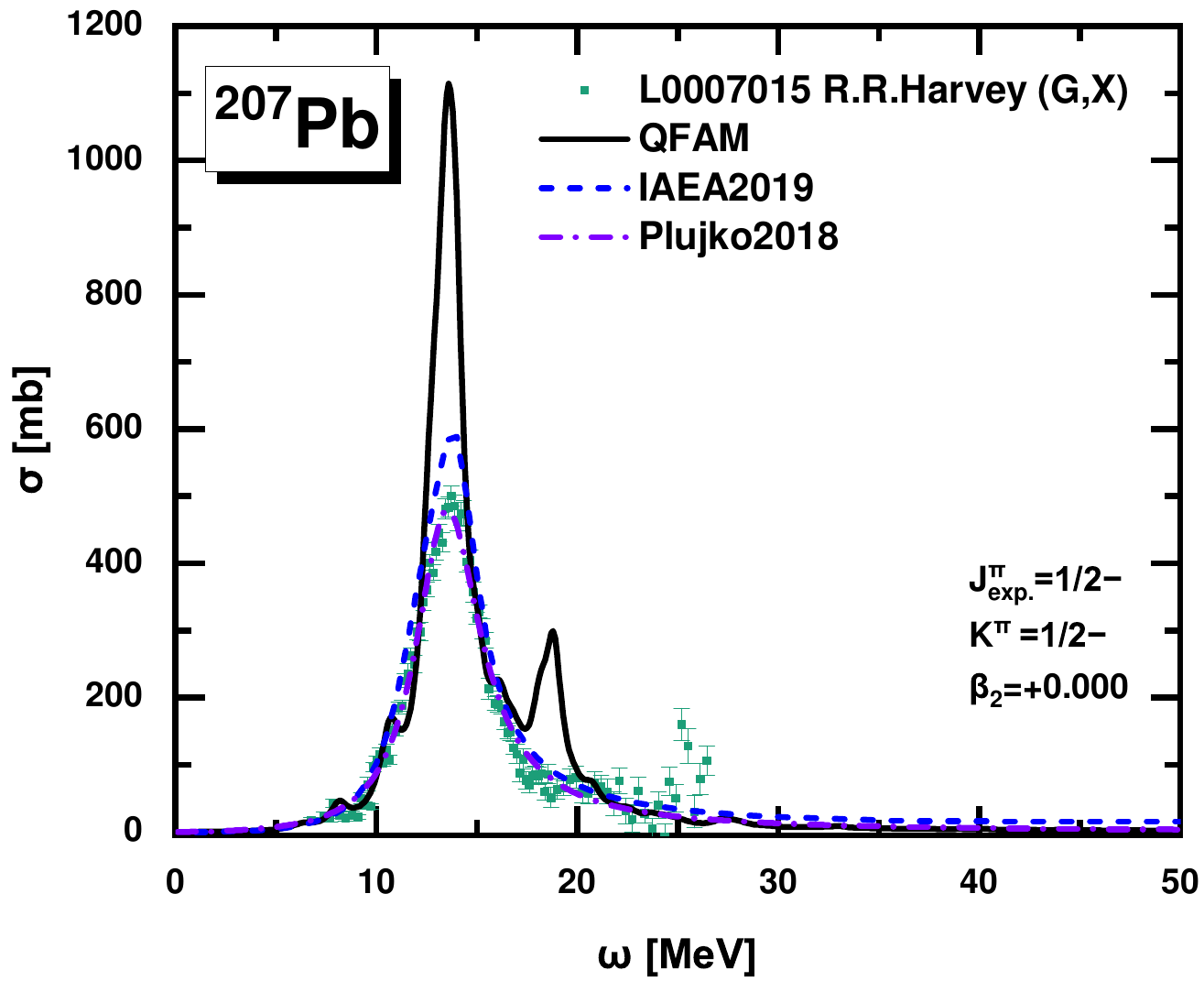}
    \includegraphics[width=0.35\textwidth]{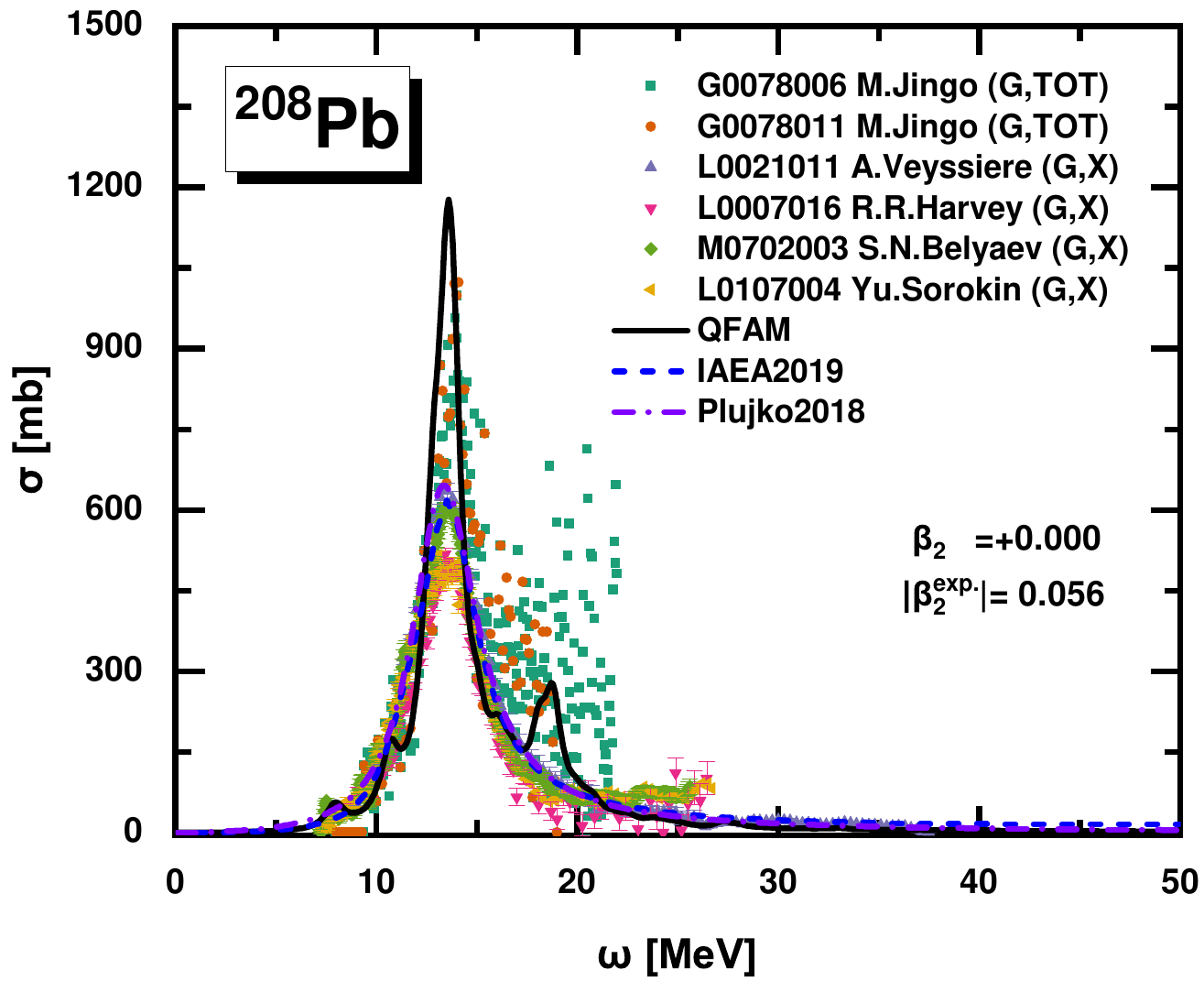}
    \includegraphics[width=0.35\textwidth]{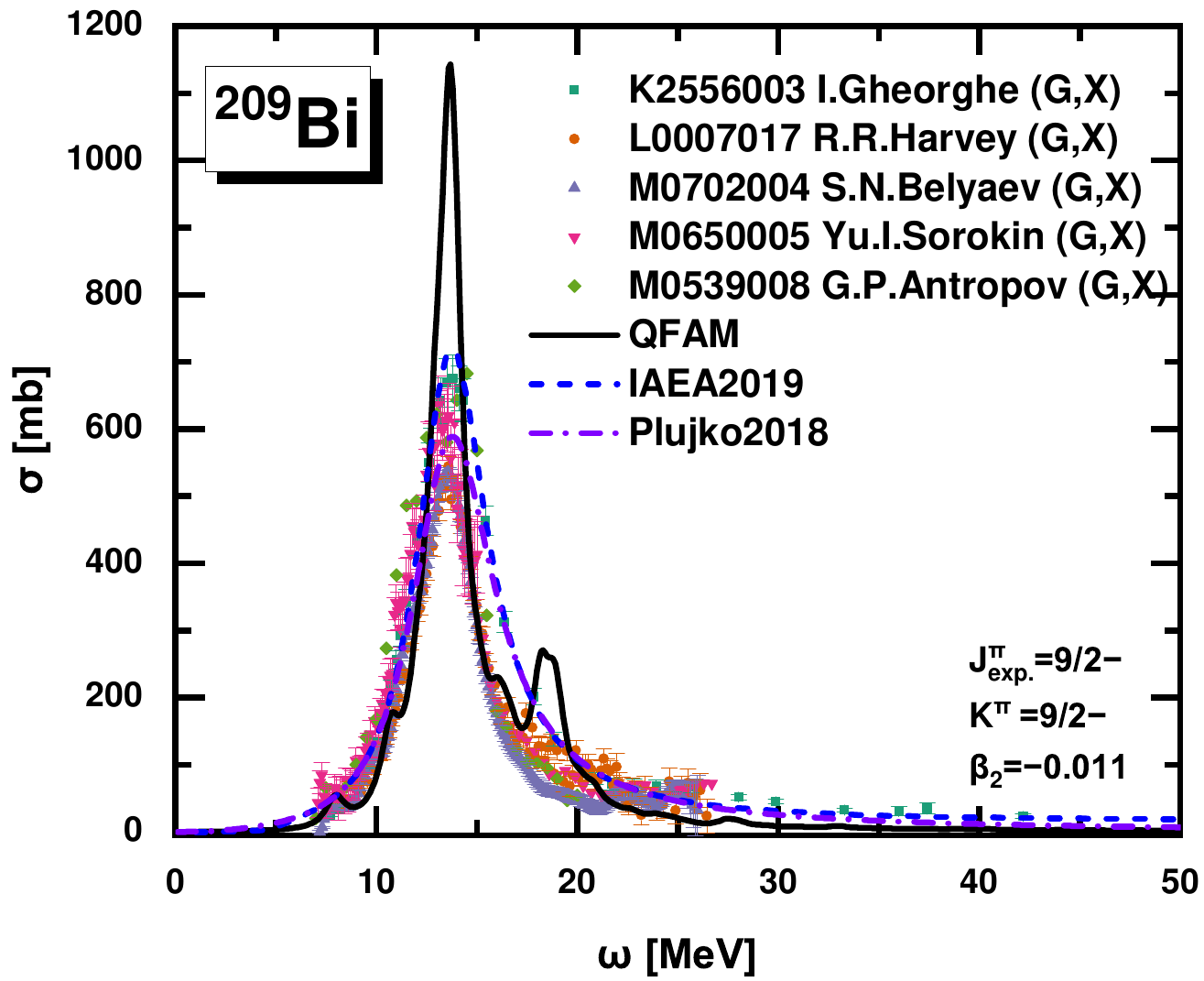}
\end{figure*}

\end{document}